\documentclass[12pt]{article}
\pdfoutput=1
\usepackage{color}
\usepackage{graphicx}
\usepackage{latexsym}
\usepackage{amssymb, amsmath, amsfonts}
\usepackage{indentfirst}
\usepackage{float,times}
\usepackage{bbm}
\usepackage{subfigure}
\usepackage{slashed}
\usepackage{ifthen}
\usepackage{mathrsfs}
\usepackage{setspace}
\def\s#1{\setbox0=\hbox{$#1$}%
\rlap{\ifdim\wd0>.7em\kern.22\wd0\else\kern.1\wd0\fi /}#1}
\def\circa#1{\,\raise.3ex\hbox{$#1$\kern-.75em\lower1ex\hbox{$\sim$}}\,}
\newcommand{\cm}{\,{\rm cm}}
\newskip\humongous \humongous=0pt plus 1000pt minus 1000pt

\newif\ifdtup

\def\oldreffmt#1{\rlap{[#1]} \hbox to 2\parindent{}}

\def\figfmt#1{\rlap{Figure {#1}} \hbox to 1in{}}

\def\beq{\begin{equation}}
\def\eeq{\end{equation}}
\def\bea{\begin{eqnarray}}
\def\eea{\end{eqnarray}}

\def\bq{\begin{quote}}
\def\eq{\end{quote}}

\relax
\def\be{\begin{equation}} 
\def\ee{\end{equation}} 
\def\bea{\begin{eqnarray}}
\def\eea{\end{eqnarray}}
\def\ba{\begin{array}}
\def\ea{\end{array}}
\def\kslash{\raise.15ex\hbox{/}\kern-.57em k}
\def\TeV{{\rm TeV}}

\newcommand{\vev}[1]{{\langle #1 \rangle}}

\newcommand{\bear}{\begin{eqnarray}}
\newcommand{\eear}{\end{eqnarray}}

\newcommand{\nn}{~\nonumber\\}
\newcommand{\sfr}[2]{{\textstyle\frac{#1}{#2}}}

\newcommand{\GeV}{\,\mbox{GeV}}

\newcommand{\matel}[3]{\langle #1|#2|#3\rangle}

\newcommand{\dU}{{d_{\cal U}}}
\newcommand{\cU}{{\cal U}}

\def\beq{\begin{equation}}
\def\eeq{\end{equation}}
\def\bal{\begin{align}}
\def\eal{\end{align}}

\newcommand{\drawsquare}[2]{\hbox{%
\rule{#2pt}{#1pt}\hskip-#2pt
\rule{#1pt}{#2pt}\hskip-#1pt
\rule[#1pt]{#1pt}{#2pt}}\rule[#1pt]{#2pt}{#2pt}\hskip-#2pt
\rule{#2pt}{#1pt}}
\newcommand{\Yfund}{\raisebox{-.5pt}{\drawsquare{6.5}{0.4}}}
\newcommand{\Yasymm}{\raisebox{-3.5pt}{\drawsquare{6.5}{0.4}}\hskip-6.9pt%
                     \raisebox{3pt}{\drawsquare{6.5}{0.4}}%
                    }
\newcommand{\Ysymm}{\Yfund\hskip-0.4pt%
                    \Yfund}
\def\symm{\Ysymm}
\def\bsymm{\overline{\Ysymm}}
\def\drawbox#1#2{\hrule height#2pt
        \hbox{\vrule width#2pt height#1pt \kern#1pt
              \vrule width#2pt}
              \hrule height#2pt}
\def\Fund#1#2{\vcenter{\vbox{\drawbox{#1}{#2}}}}
\def\Asym#1#2{\vcenter{\vbox{\drawbox{#1}{#2}
              \kern-#2pt 
              \drawbox{#1}{#2}}}}

\def\fund{\Fund{6.4}{0.3}}

\def\bfund{\overline{\fund}}

\newcommand{\Ythrees}{\raisebox{-.5pt}{\drawsquare{6.5}{0.4}}\hskip-0.4pt%
          \raisebox{-.5pt}{\drawsquare{6.5}{0.4}}\hskip-0.4pt%
          \raisebox{-.5pt}{\drawsquare{6.5}{0.4}}}
\newcommand{\Ythreea}{\raisebox{-3.5pt}{\drawsquare{6.5}{0.4}}\hskip-6.9pt%
        \raisebox{3pt}{\drawsquare{6.5}{0.4}}\hskip-6.9pt
        \raisebox{9.5pt}{\drawsquare{6.5}{0.4}}}

\newcommand{\Yadjoint}{\raisebox{-3.5pt}{\drawsquare{6.5}{0.4}}\hskip-6.9pt%
        \raisebox{3pt}{\drawsquare{6.5}{0.4}}\hskip-0.4pt
        \raisebox{3pt}{\drawsquare{6.5}{0.4}}}
\newcommand{\Om}{{\cal O}_{\widetilde{\psi}{\psi}}}
\newcommand{\dm}{ d_{\widetilde{\psi}{\psi}}}
\def\swsqeffl{\sin^2{\theta_\mathrm{eff}}}
\newcommand{\cw}{{w_T}}
\newcommand{\Cw}[1]{{w_T^{#1}}}
\newcommand{\hoch}{{M_\cU}}
\newcommand{\Hoch}[1]{{M_\cU^{#1}}}

\newcommand{\SU}{{\rm SU}}
\begin{document}
\begin{titlepage}
\begin{flushright}
{\it CP$^3$- Origins: 2009-20}
\end{flushright}
 \begin{center}
{{\LARGE {\color{red}
Conformal Dynamics for TeV Physics and Cosmology} 
 }} 
 \end{center}
 \par \vskip .2in \noindent
\begin{center}
{  \Large {\color{black}Francesco Sannino$^{\ast}$}  }
\end{center}
\begin{center}
  \par \vskip .1in \noindent
{ \large \it
CP$^3$-Origins, University of Southern Denmark, Odense~M. Denmark.
}
   \par \vskip .1in \noindent
\end{center}     
\begin{center}{\large Abstract}\end{center}
\begin{quote}

We introduce the topic of dynamical breaking of the electroweak symmetry and its link to unparticle physics and cosmology.  The knowledge of the phase diagram of strongly coupled theories plays a fundamental role when trying to construct viable extensions of the standard model (SM). Therefore  we present the state-of-the-art of the phase diagram for $SU$, $Sp$ and $SO$ gauge theories with fermionic matter transforming according to arbitrary representations of the underlying gauge group. We summarize several analytic methods used recently to acquire information about these gauge theories. We also provide new results for the phase diagram of the generalized Bars-Yankielowicz and Georgi-Glashow chiral gauge theories. These theories have been used for constructing grand unified models and have been the template for models of extended technicolor interactions. To gain information on the phase diagram of chiral gauge theories we will introduce a novel all orders beta function for chiral gauge theories. This permits the first unified study of all non-supersymmetric gauge theories with fermionic matter representation both chiral and non-chiral. To the best of our knowledge the phase diagram of these complex models appears here for the first time.  We will introduce recent extensions of the SM featuring minimal conformal gauge theories known as minimal walking models. Finally we will discuss the electroweak phase transition at nonzero temperature for models of dynamical electroweak symmetry breaking. \end{quote}
\par \vskip .1in
\vfill
{$^{\ast}$ \small Lectures presented at the 49$^{th}$ Cracow School of Theoretical Physics.}
\end{titlepage}
     
         \newpage

\def\baselinestretch{1.0}
\tiny
\normalsize

\tableofcontents

\newpage

\section{The Need to Go Beyond}

The energy scale at which the Large Hadron Collider experiment (LHC) will operate is determined by the need to complete the SM of particle interactions and, in particular, to understand the origin of mass of the elementary particle. Together with classical general relativity the SM constitutes one of the most successful models of nature.  We shall, however, argue that experimental results and theoretical arguments call for a more fundamental description of nature.

\begin{figure}[hb]
\label{figureSMplot1} 
\center{
\includegraphics[width=12cm]{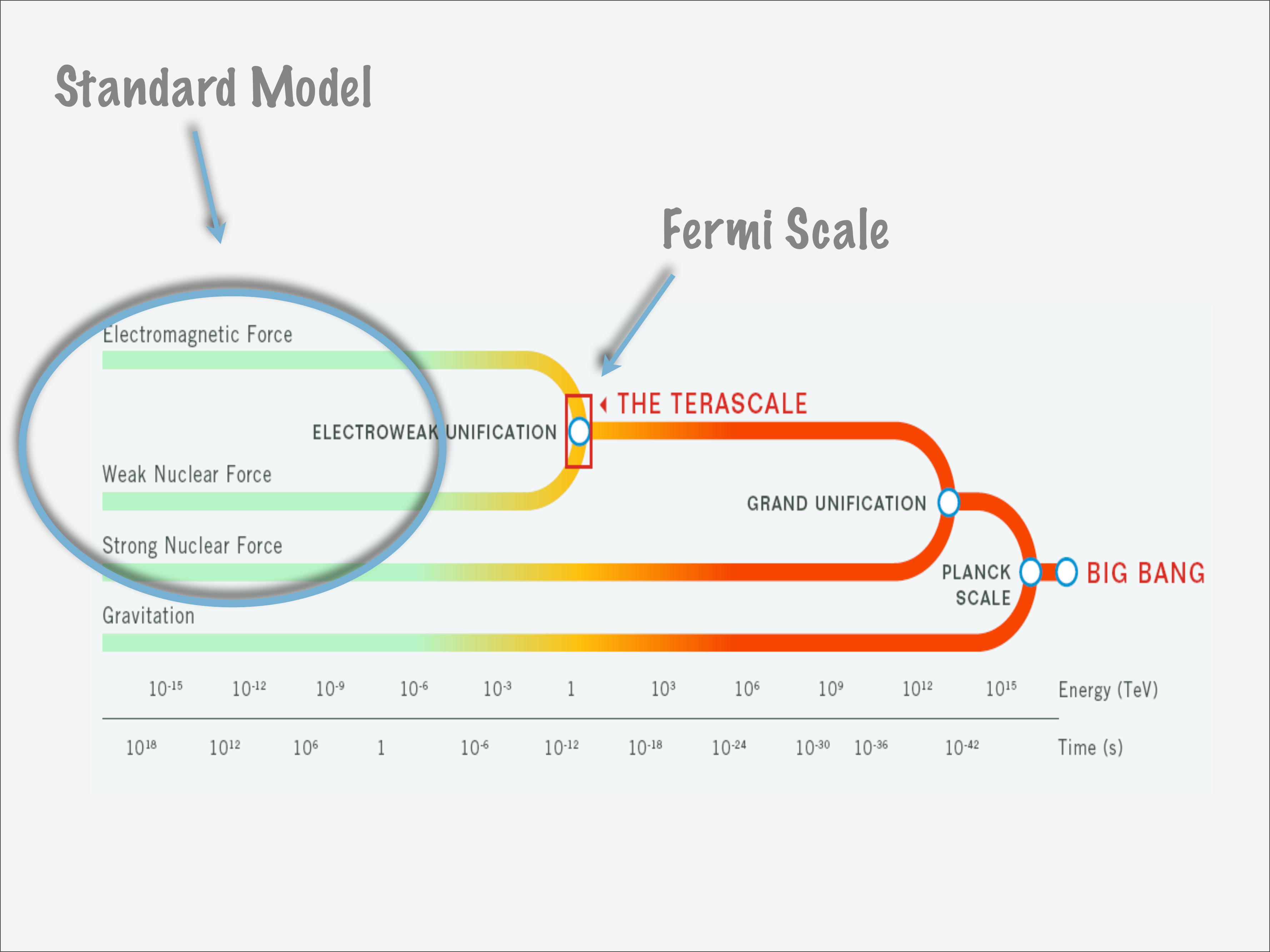}
\caption{Cartoon representing the various forces of  nature. At very high energies one may imagine that all the low-energy forces unify in a single force.}}
\end{figure}

In Figure \ref{figureSMplot1}, we schematically represent, in green, the known forces of  nature. The SM of particle physics describes the strong, weak and electromagnetic forces. The yellow region represents the energy scale around the TeV scale and will be explored directly at the LHC, while the red  part of the diagram is speculative.  

\begin{figure}[ht]
\begin{minipage}{16pc}
\vskip1.5cm\hskip -.5cm\includegraphics[width=18pc]{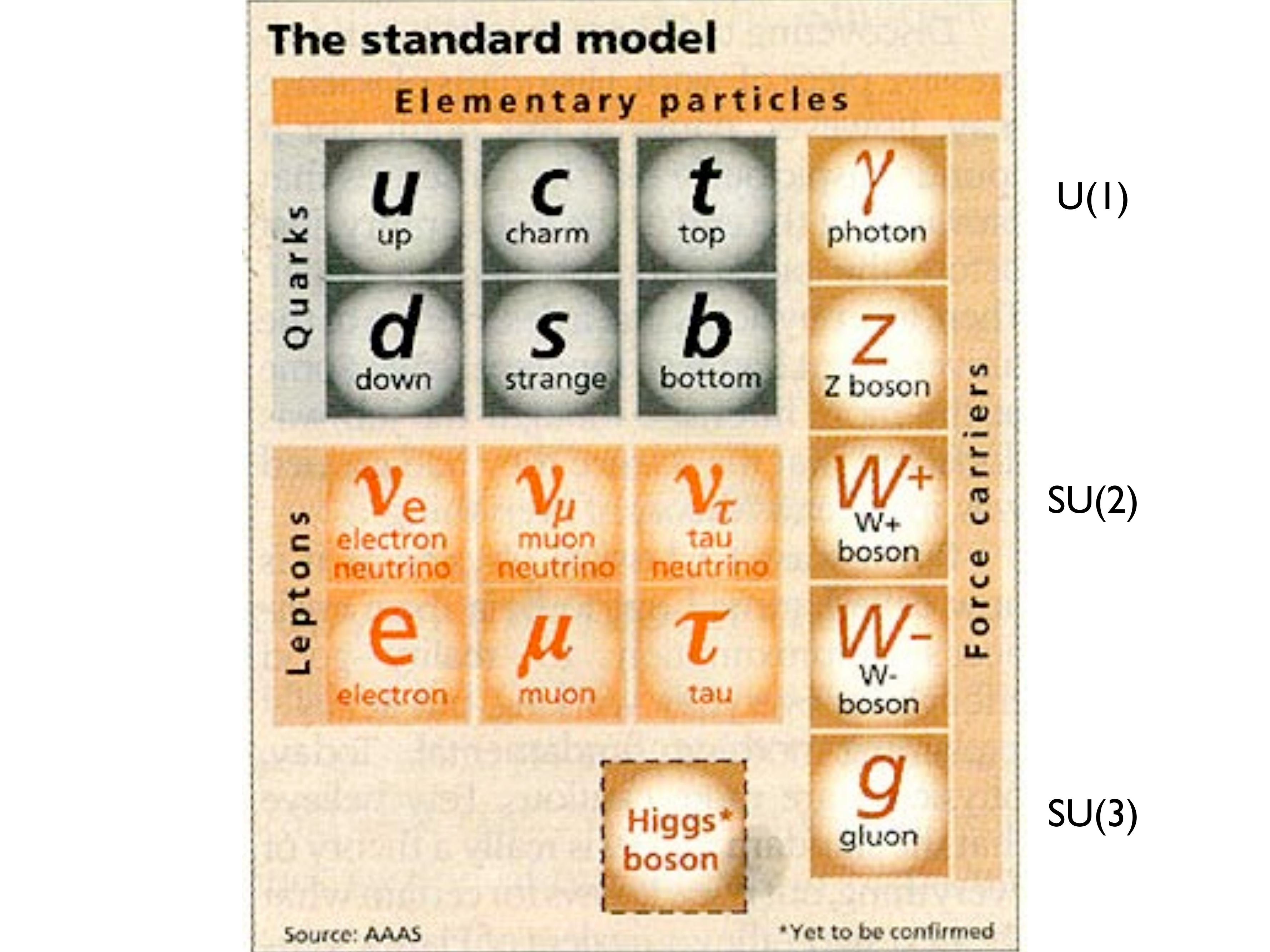}
\caption{Postage stamp representing all of the elementary particles which constitute the SM. The forces are mandated with the  $SU(3)\times SU(2) \times U(1)$ gauge group.}
\label{SMstamp}
\end{minipage}\hspace{3pc}%
\begin{minipage}{20pc}
\vskip -.7cm\hskip -1.3cm\includegraphics[width=28pc]{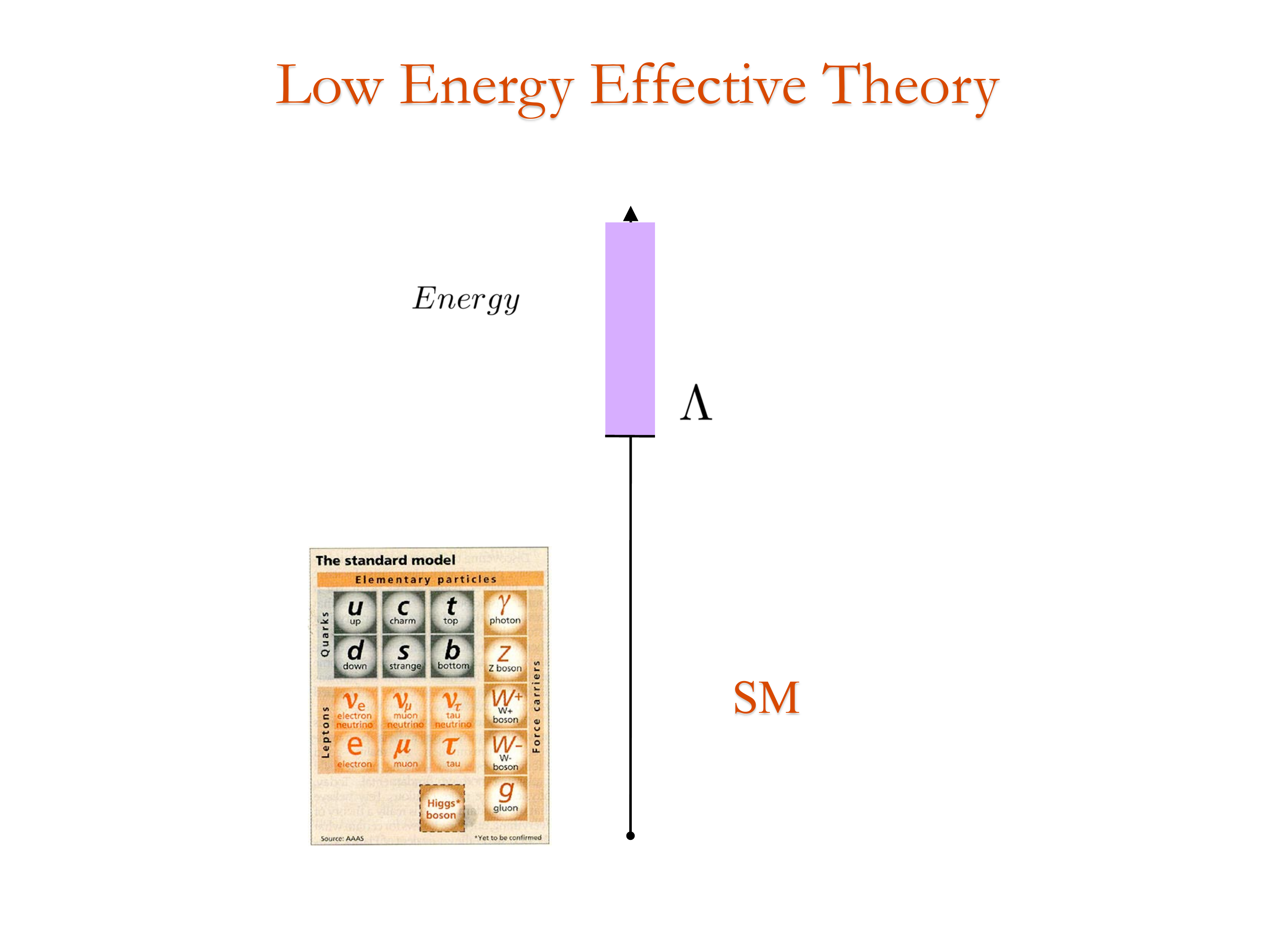}
\vskip -1cm\caption{The SM can be viewed as a low-energy theory valid up to a high energy scale $\Lambda$.}
\label{lowenergy}
\end{minipage} 
\end{figure}

All of the known elementary particles constituting the SM fit on the postage stamp shown in Fig.~\ref{SMstamp}. Interactions among quarks and leptons are carried by gauge bosons. Massless gluons mediate the strong force among quarks while the massive  gauge bosons, i.e. the $Z$ and $W$, mediate the weak force and interact with both quarks and leptons. Finally, the massless photon, the quantum of light, interacts with all of the electrically charged particles. The SM Higgs  does not feel strong interactions. The interactions emerge naturally by invoking a gauge principle. It is intimately linked with  the underlying symmetries relating the various particles of the SM. 
The asterisk on the Higgs boson in the postage stamp indicates that it has not yet been observed. Intriguingly the Higgs is the only fundamental scalar of the SM.

The SM can be viewed as a low-energy effective theory valid up to an energy scale $\Lambda$, as schematically represented in Fig.~\ref{lowenergy}. Above this scale new interactions, symmetries, extra dimensional worlds or any other extension could emerge. At sufficiently low energies with respect to this scale one expresses the existence of new physics via effective operators. The success of the SM is due to the fact that most of the corrections to its physical observables depend only logarithmically on this scale $\Lambda$. In fact, in the SM there exists only one operator which acquires corrections quadratic in $\Lambda$. This is the squared mass operator of the Higgs boson.  Since $\Lambda$ is expected to be the highest possible scale, in four dimensions the Planck scale, it is hard to explain {\it naturally}  why the mass of the Higgs is of the order of the electroweak scale. This is the hierarchy problem. Due to the occurrence of quadratic corrections in the cutoff this SM sector is most sensitive to the existence of new physics. 

\subsection{The Higgs}
It is a fact that the Higgs allows for a direct and economical way of spontaneously breaking the electroweak symmetry.  It generates simultaneously the masses of the quarks and leptons without introducing flavour changing neutral currents at the tree level.  
 The Higgs sector of the SM
possesses, when the gauge couplings are switched off, an
$SU_L(2)\times SU_R(2)$ symmetry. The full symmetry group can be made explicit when re-writing the Higgs
doublet field \begin{eqnarray} H=\frac{1}{\sqrt{2}}\left(%
\begin{array}{c}
  \pi_2 + i\, \pi_1 \\
  \sigma - i\, \pi_3 \\
\end{array}%
\right)\end{eqnarray} as the right column of the following two by two matrix:
\begin{eqnarray}
\frac{1}{\sqrt{2}}\left(\sigma + i\,
\vec{\tau}\cdot\vec{\pi} \right) \equiv M
 \ .
\end{eqnarray}
The first column can be identified with the column vector $\tau_2H^{\ast}$ while the second with $H$. We indicate this fact with $\left[i\,\tau_2H^{\ast}\
,\,H\right]= M$. $\tau^2$ is the second Pauli matrix. 
The $SU_L(2)\times SU_R(2)$ group acts linearly on $M$ according
to:
\begin{eqnarray}
M\rightarrow g_L M g_R^{\dagger} \qquad {\rm and} \qquad g_{L/R} \in SU_{L/R}(2)\ .
\end{eqnarray}
One can verify that:
\begin{eqnarray}
M\frac{\left(1-\tau^3\right)}{2} = \left[0\ , \, H\right] \ . \qquad
M\frac{\left(1+\tau^3\right)}{2} = \left[i\,\tau_2H^{\ast} \ , \, 0\right] \ .
\end{eqnarray}
The $SU_L(2)$ symmetry is gauged by introducing the weak gauge
bosons $W^a$ with $a=1,2,3$. The hypercharge generator is taken to
be the third generator of $SU_R(2)$. The ordinary covariant
derivative acting on the Higgs, in the present notation, is:
\begin{eqnarray}
D_{\mu}M=\partial_{\mu}M -i\,g\,W_{\mu}M + i\,g^{\prime}M\,B_{\mu} \ , \qquad {\rm
with}\qquad W_{\mu}=W_{\mu}^{a}\frac{\tau^{a}}{2} \ ,\quad
B_{\mu}=B_{\mu}\frac{\tau^{3}}{2} \ .
\end{eqnarray}
The Higgs Lagrangian is 
\begin{eqnarray}
{\cal L}&=&\frac{1}{2}{\rm Tr} \left[D_{\mu}M^{\dagger}
D^{\mu}M\right]-\frac{m^2}{2} {\rm
Tr}\left[M^{\dagger}M\right] - \frac{\lambda}{4}\,{\rm
Tr}\left[M^{\dagger}M\right]^2 \ .
\end{eqnarray}
At this point one {\it assumes} that the mass squared of
the Higgs field is negative and this leads to the electroweak
symmetry breaking. Except for the Higgs mass term the other SM operators have dimensionless couplings meaning that the natural scale for the SM is encoded in the Higgs mass\footnote{The mass of the proton is due mainly to strong interactions, however its value cannot be determined within QCD since the associated renormalization group invariant scale must be fixed to an hadronic observable.}  

At the tree level, when taking $m^2$ negative and the self-coupling $\lambda$ positive, one determines:
\begin{equation}
\langle \sigma \rangle^2 \equiv v_{weak}^2=\frac{|m^2|}{\lambda} \qquad {\rm and} \qquad \sigma = v_{weak} + h \ ,
\end{equation} 
where $h$ is the Higgs field. 
The global symmetry breaks to its diagonal subgroup:
\begin{eqnarray}
SU_L(2)\times SU_R(2) \rightarrow SU_V(2) \ .
\end{eqnarray}
To be more precise the $SU_R(2)$ symmetry is already broken explicitly by our choice of gauging only an $U_Y(1)$ subgroup of it and hence the actual symmetry breaking pattern is:
\begin{eqnarray}
SU_L(2)\times U_Y(1) \rightarrow U_Q(1) \ ,
\end{eqnarray}
with $U_Q(1)$ the electromagnetic abelian gauge symmetry. According to the Nambu-Goldstone's theorem three massless degrees of freedom appear, i.e. $\pi^{\pm}$ and $\pi^0$. In the unitary gauge these Goldstones become the longitudinal degree of freedom of the massive elecetroweak gauge-bosons. Substituting the vacuum value for $\sigma$ in the Higgs Lagrangian the gauge-bosons quadratic terms read:
\begin{equation}
\frac{v_{weak}^2}{8}\, \left[g^2\,\left(W_{\mu}^1
W^{\mu,1} +W_{\mu}^2 W^{\mu,2}\right)+ \left(g\,W_{\mu}^3 -
g^{\prime}\,B_{\mu}\right)^2\right]  \ . \end{equation}
 The $Z_{\mu}$ and the photon $A_{\mu}$ gauge bosons are:
\begin{eqnarray}
Z_{\mu}&=&\cos\theta_W\, W_{\mu}^3 - \sin\theta_{W}B_{\mu} \ ,\nonumber \\
A_{\mu}&=&\cos\theta_W\, B_{\mu} + \sin\theta_{W}W_{\mu}^3 \ ,
\end{eqnarray}
with $\tan\theta_{W}=g^{\prime}/g$ while the charged massive vector bosons are
$W^{\pm}_{\mu}=(W^1\pm i\,W^2_{\mu})/\sqrt{2}$. 
The bosons masses $M^2_W=g^2\,v_{weak}^2/4$ due to
the custodial symmetry satisfy the tree level relation $M^2_Z=M^2_W/\cos^2\theta_{W}$. 
Holding fixed the EW scale $v_{weak}$ the mass squared of the Higgs boson is $2\lambda v^2_{weak}$ and hence it increases with $\lambda$.  We recall that the Higgs Lagrangian has a familiar form since it is identical to the linear $\sigma$ Lagrangian which was introduced long ago to describe chiral symmetry breaking in QCD with two light flavors.

Besides breaking the electroweak symmetry dynamically the ordinary Higgs serves also the purpose to provide mass to all of the SM particles via the Yukawa terms of the type:
\beq -Y_d^{ij}\bar{Q}_L^i H d_R^j  - Y_u^{ij}\bar{Q}_L^i (i\tau_2 H^{\ast}) u_R^j  + {\rm h.c.}\ , \eeq
where $Y_{q}$ is the Yukawa coupling constant, $Q_L$ is the
left-handed Dirac spinor of quarks, $H$ the Higgs doublet and
 $q$ the right-handed Weyl spinor for the quark and $i,j$ the flavor indices. The $SU_L(2)$ weak and spinor indices are suppressed. 

When considering quantum corrections the Higgs mass acquires large quantum corrections  proportional to the scale of the cut-off squared. 
\begin{equation}
m^2_{\rm ren} - m^2 \propto \Lambda^2 \ .
\end{equation}
 $\Lambda$ is the highest energy above which the SM is no longer a valid description of Nature and a large fine tuning of the parameters of the Lagrangian is needed to offset the effects of the cut-off. This large fine tuning is needed because there are no symmetries protecting the Higgs mass operator from large corrections which would hence destabilize the Fermi scale (i.e. the electroweak scale). This problem is the one we referred above as the hierarchy problem of the SM.

The constant value of the Higgs field evaluated on the ground state is determined by the measured mass of the $W$ boson. On the other hand, the value of the SM Higgs mass  ($m_H$ ) is constrained only indirectly by the electroweak precision data. The preferred value of the Higgs mass is $m_H=87^{+36}_{-27}$~GeV at 68\% confidence level (CL) with a 95\% CL upper limit $m_H <160$~GeV. This value increases to $m_H <190$~GeV when including the LEP-2 direct lower limit $m_H >114$~GeV, as reported by the Electroweak Working Group ({http://lepewwg.web.cern.ch})\footnote{All the plots we use in this section are reported by the Electroweak Working Group and can be found at the web-address: {\rm http://lepewwg.web.cern.ch}.}.

\begin{figure}[t]
\begin{minipage}{16pc}
\includegraphics[width=16pc]{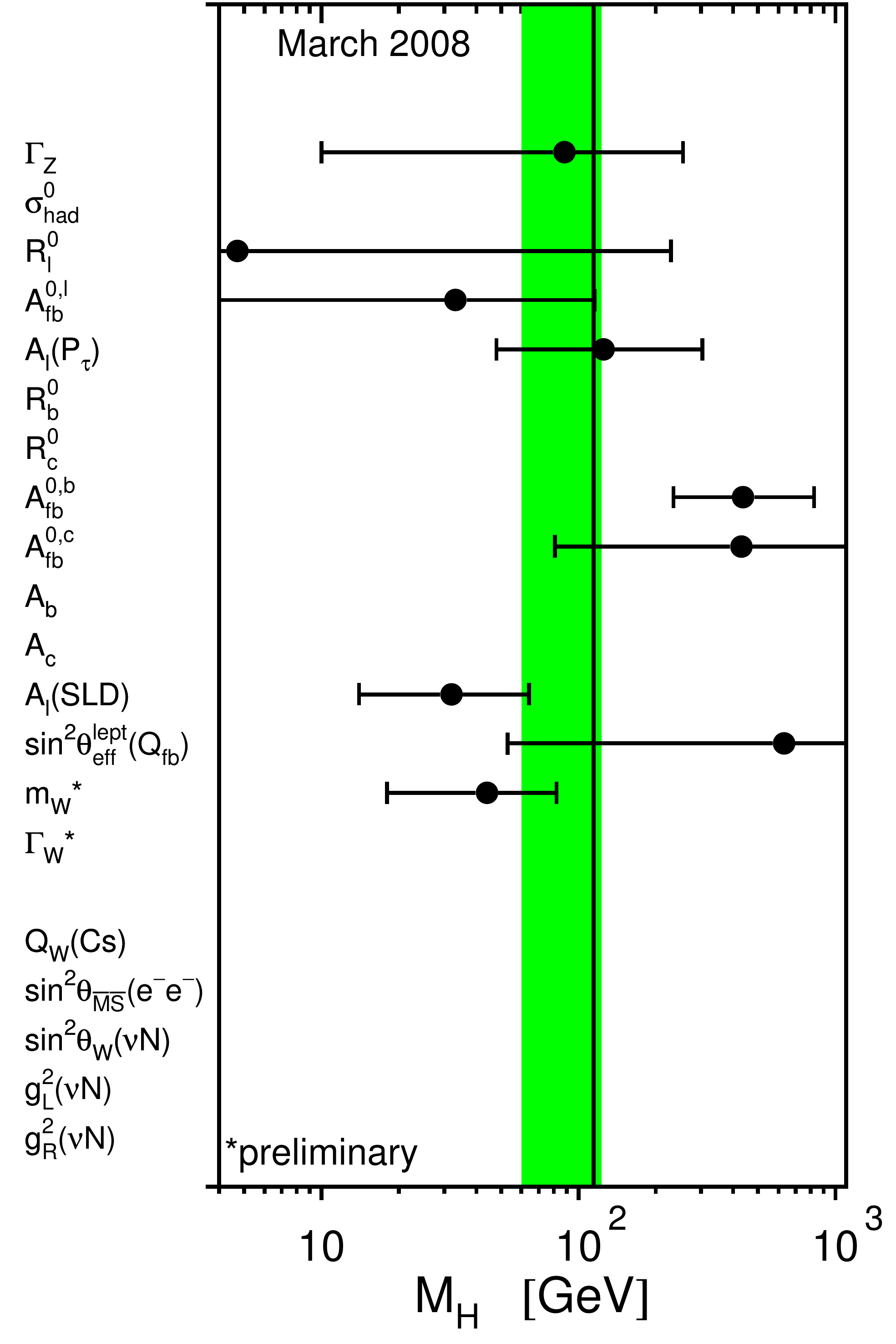}
\caption{Values of the Higgs mass extracted from different electroweak observables. The average is shown as a green band.}
\label{fig1}
\end{minipage}\hspace{1pc}%
\begin{minipage}{20pc}
\includegraphics[width=20pc]{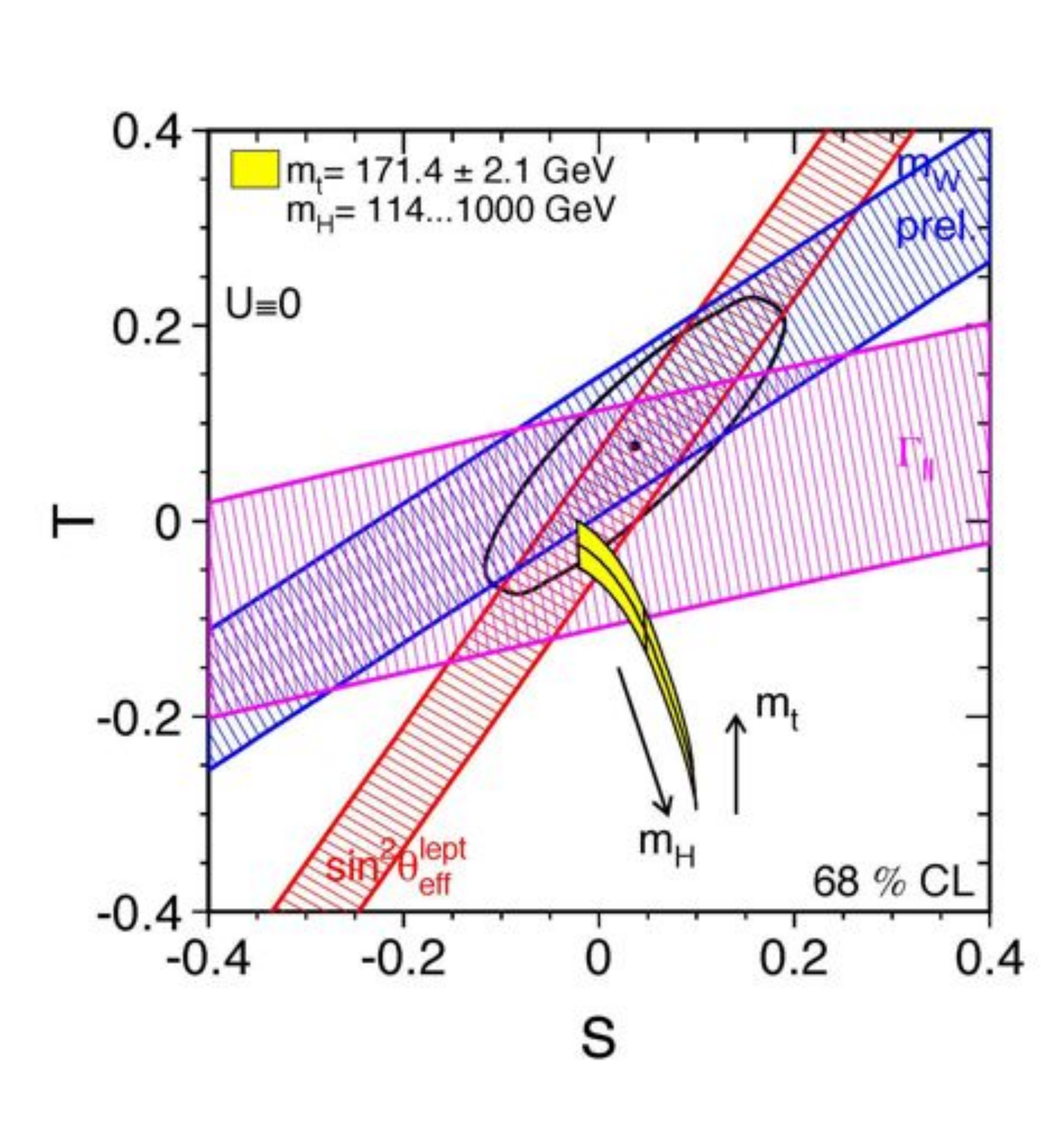}
\caption{The 1-$\sigma$ range of the electroweak parameters $S$ and $T$ determined from different observables. The ellipsis shows the 68\% probability from combined data. The yellow area gives the SM prediction with $m_t$ and $m_H$ varied as shown.}
\label{s06_stu_contours}
\end{minipage} 
\end{figure}
It is instructive to look separately at the various measurements influencing the fit for the SM Higgs mass. The final result of the average of all of the measures, however, has a Pearson's chi-square ($\chi^2$) test of 11.8 for 5 degrees
of freedom. This relatively high value of $\chi^2$ is due to the two most precise measurements of $\swsqeffl$, namely
those derived from the measurements of the lepton left-right asymmetries $A_l$ by SLD and of the forward-backward asymmetry
measured in $b \bar b$ production at LEP, $A^b_{FB}$. The two measurements differ by about
3 $\sigma$'s.  
\begin{figure}[ht]
\begin{minipage}{19pc}
\vskip -1cm\hskip -1cm\includegraphics[width=19pc]{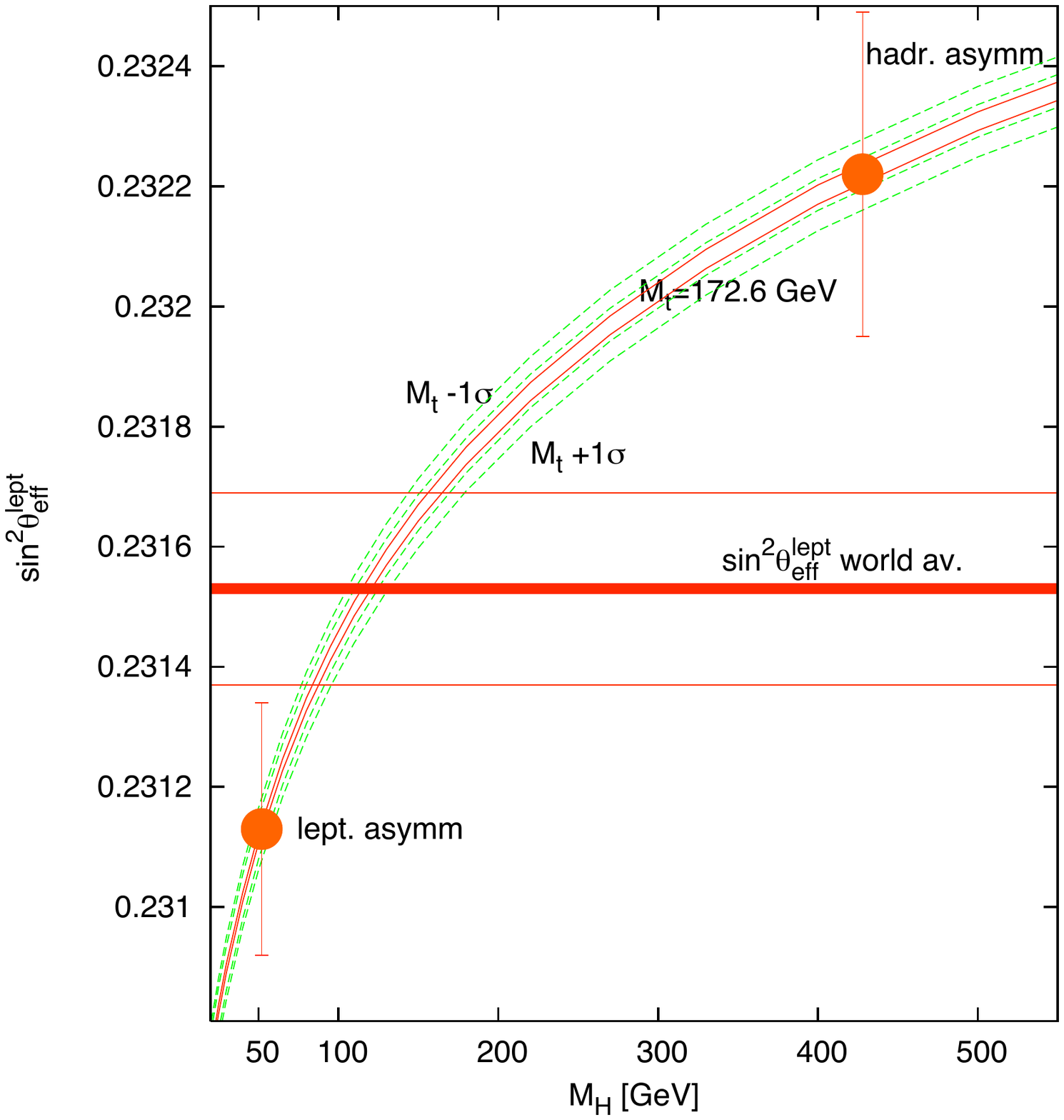}
\caption{The data for $\sin^2\theta_{\rm eff}^{\rm lept}$ are
plotted vs $m_H$. For presentation purposes the measured points are
shown each at the $m_H$ value that would ideally correspond to it
given the central value of $m_t$. \label{s2mH}}
\end{minipage}\hspace{2pc}%
\begin{minipage}{19pc}
\vskip -2.2cm \hskip -1.1cm\includegraphics[width=19pc]{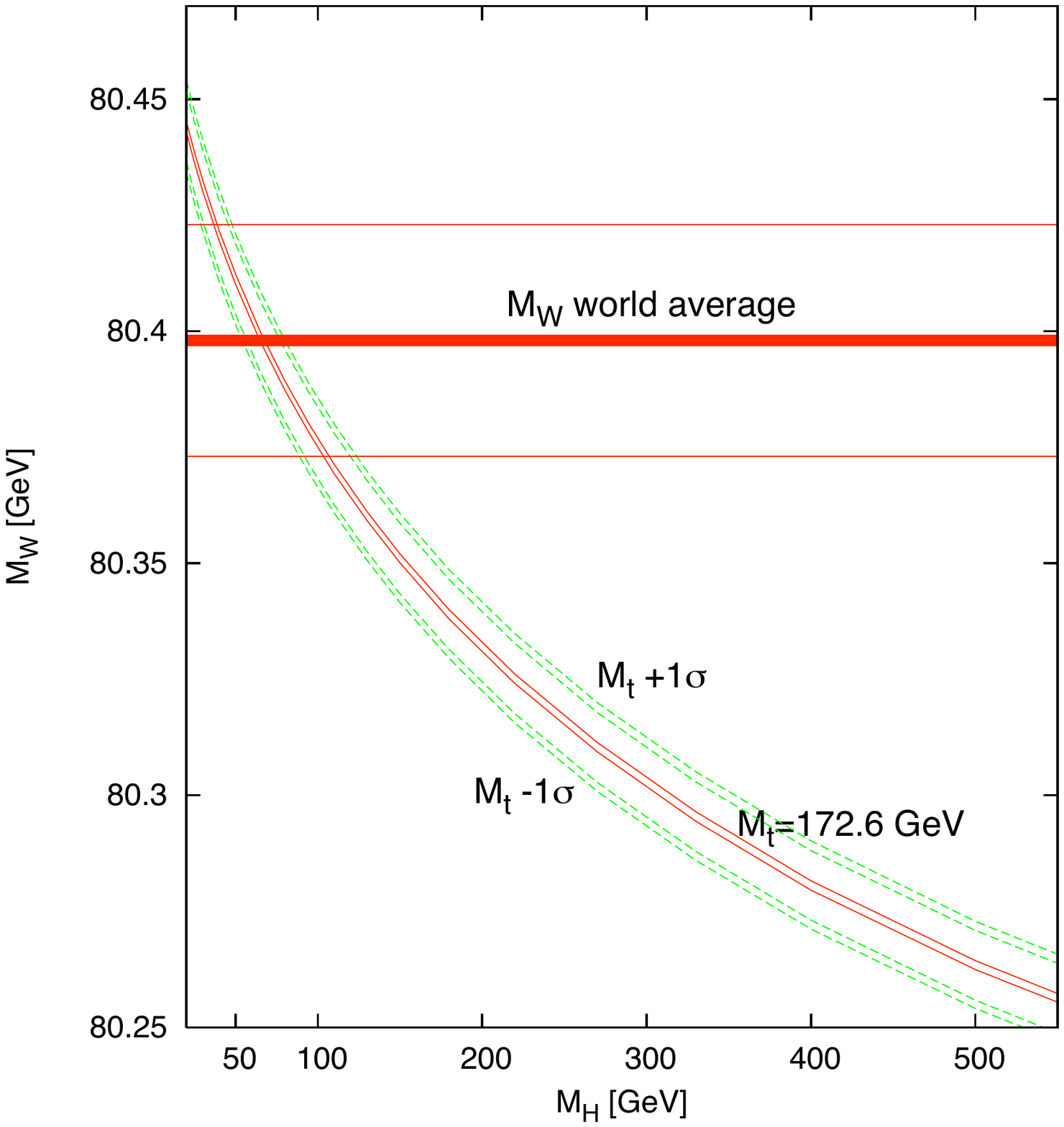}
\caption{The world average for $m_W$ is
plotted vs $m_H$. \label{wmH}}
\end{minipage} 
\end{figure}
The situation is shown in Fig.~\ref{s2mH}~ (updated values of \cite{Gambino:2003xc}).
The values of $\sin^2\theta_{\rm eff}$ and their average are shown each at the preferred value of $m_H$
 corresponding to a given central value of $m_t$.
The implications for the value of the mass of the Higgs are interesting. 
The $b\bar b$ forward-backward asymmetry leads to the prediction of a relatively heavy Higgs with $m_H=420^{+420}_{-190}$~GeV. On the other hand, the lepton left-right asymmetry corresponds to $m_H=31^{+33}_{-19}$~GeV, in conflict with the lower limit $m_H>114$~GeV from direct LEP searches. Moreover, the world average of the $W$ mass, $m_W=80.398\pm 0.025$~GeV (see Fig.~\ref{wmH}), is still larger than the value extracted from a SM fit, again requiring $m_H$ to be smaller than what is allowed by the LEP Higgs searches. 
This tension may be due to new physics, to a statistical fluctuation or to an unknown experimental problem. The overall situation is summarized in Fig.~\ref{fig1}, where the predicted values of $m_H$ from the different observables are shown.  A very light SM Higgs is deduced only when averaging over the whole set of data.

Summarising, the experimental window for the SM Higgs mass $
114~{\rm GeV}<m_H<190~{\rm GeV}$ coincides with the theoretical range $125~{\rm GeV} <m_H<175~{\rm GeV}$ of allowed values of the SM Higgs mass  naturally compatible with a high cutoff scale\footnote{The scale associated with unification in four dimensions is typically of the order of $10^{15}$~GeV.} and the stability of the ground state of the SM. This fact may be a coincidence or may be an argument in favour of the naturality of the SM Higgs.

A Higgs heavier than $190$~GeV is compatible with precision tests if we allow simultaneously new physics to compensate for the effects of the heavier value of the mass. The precision measurements of direct interest for the Higgs sector are often reported using the $S$ and $T$ parameters as shown in Fig.~\ref{s06_stu_contours}. {}From this graph one deduces that a heavy Higgs is compatible with data at the expense of a large value of the $T$ parameter.  Actually, even the lower direct experimental limit on the Higgs mass can be evaded with suitable extensions of the SM Higgs sector.

Many more questions need an answer if the Higgs is found at the LHC: Is it composite? How many Higgs fields are there in nature? Are there hidden sectors?

\subsection{Riddles}

Why do we expect that there is new physics awaiting to be discovered? Of course, we still have to observe the Higgs, but this cannot be everything. Even with the Higgs discovered, the SM has both conceptual problems and phenomenological shortcomings. In fact, theoretical arguments indicate that the SM is not the ultimate description of nature: 
 \begin{itemize}
\item{{\bf Hierarchy Problem:} The Higgs sector is highly fine-tuned. We have no natural separation between the Planck and the electroweak scale.}
\item{{\bf Strong CP Problem:} There is no natural explanation for the smallness of the electric dipole moment of the neutron within the SM. This problem is also known as the strong CP problem.}
\item{{\bf Origin of Patterns:} The SM can fit, but cannot explain the number of matter generations and their mass texture. }
\item{{\bf Unification of the Forces:} Why do we have so many different interactions? It is appealing to imagine that the SM forces could unify into a single Grand Unified Theory (GUT). We could imagine that at very high energy scales gravity also becomes part of a unified description of nature.}
\end{itemize}
There is no doubt that the SM is incomplete since we cannot even account for a number of basic observations:  
\begin{itemize}

\item{{\bf Neutrino Physics:}
Only recently it has been possible to have some definite answers about properties of neutrinos. We now know that they have a tiny mass, which can be naturally accommodated in extensions of the SM, featuring for example a  Òsee-sawÓ  mechanism. We do not yet know if the neutrinos have a Dirac or a Majorana nature.}

\item{{\bf Origin of Bright and Dark Mass:}
Leptons, quarks and the gauge bosons mediating the weak interactions possess a rest mass. Within the SM this mass can be accounted for by the Higgs mechanism, 
which constitutes the electroweak symmetry breaking sector of the SM. However, the associated 
Higgs particle has not yet been discovered. Besides, the SM cannot account for the observed large fraction of ÒdarkÓ mass of the universe. What is interesting is that in the universe the dark matter is about five times more abundant than the known baryonic matter, i.e. ÒbrightÓ matter. We do not know why the ratio of dark to bright matter is of order unity.}

\item{{\bf Matter-Antimatter Asymmetry:} 
 From our everyday experience we know that there is very little ÒbrightÓ antimatter in the universe. The SM fails to predict the observed excess of matter. }
 
 \end{itemize}
 
These arguments do not imply that the SM is necessarily wrong, but it must certainly be extended to answer any of the questions raised above. The truth is that we do not have an answer to the basic question: What lies beneath the SM?

A number of possible generalizations of the SM have been
conceived (see \cite{Ellis:2009pz,Altarelli:2009bz,Giudice:2007qj,Mangano:2008ag,Altarelli:2008yi,Barbieri:2008zz,Grojean:2009fd} for reviews). Such extensions are introduced on the base of one or more guiding principles or prejudices. Two  technical reviews are \cite{DeRoeck:2009id,Accomando:2006ga}. 

In the models we will consider here the electroweak symmetry breaks via a
fermion bilinear condensate. The Higgs sector of the SM
becomes an effective description of a more fundamental fermionic
theory. This is similar to the Ginzburg-Landau theory of
superconductivity. If the force underlying the fermion condensate driving electroweak symmetry
breaking is due to a strongly interacting gauge theory these models are termed
technicolor. 

Technicolor, in brief, is an additional non-abelian and strongly interacting gauge theory augmented
with (techni)fermions transforming under a given representation of the gauge group.
The Higgs Lagrangian is replaced by a suitable
new fermion sector interacting strongly via a new gauge interaction (technicolor). Schematically:
\begin{eqnarray}
L_{Higgs} \rightarrow -\frac{1}{4}F_{\mu\nu}F^{\mu\nu} + i \bar{Q} \gamma_{\mu}D^{\mu} Q  + \cdots\ ,
\end{eqnarray}
where, to be as general as possible, we have left unspecified the underlying nonabelian gauge group and the associated technifermion ($Q$) representation. The dots represent new sectors which may even be needed to avoid, for example, anomalies introduced by the technifermions. 
The intrinsic scale of the new theory is expected to be less or of the order of a few TeVs.
The chiral-flavor symmetries of this theory, as for ordinary QCD,
break spontaneously when the technifermion condensate $\bar{Q} Q $ forms. It is
possible to choose the fermion charges in such a way that there is,
at least, a weak left-handed doublet of technifermions and the
associated right-handed one which is a weak singlet. The covariant derivative contains the new gauge field as well as the electroweak ones. The condensate
spontaneously breaks the electroweak symmetry down to the
electromagnetic and weak interactions.
The Higgs is now interpreted as the lightest scalar field with the same quantum numbers of the fermion-antifermion composite field. The Lagrangian part responsible for the mass-generation of the ordinary fermions will also be modified since the Higgs particle is no longer an elementary object. 

Models of electroweak symmetry breaking via new strongly interacting
theories of technicolor type \cite{Weinberg:1979bn,Susskind:1978ms}
are a mature subject (for recent reviews see
\cite{Hill:2002ap,Sannino:2008ha,Lane:2002wv}). One of the main difficulties in
constructing such extensions
of the SM is the very
limited knowledge about generic strongly interacting theories. This
has led theorists to consider specific models of technicolor which
resemble ordinary quantum chromodynamics and for which the large
body of experimental data at low energies can be directly exported
to make predictions at high energies. Unfortunately the simplest version of this type of models are at odds with electroweak precision measurements. New strongly coupled theories with dynamics very different from the one featured by a scaled up version of QCD are needed \cite{Sannino:2004qp}. 

We will review models of dynamical electroweak symmetry breaking making use of new type of four dimensional gauge theories \cite{Sannino:2004qp,Dietrich:2005jn,Dietrich:2005wk} and their low energy effective description 
\cite{Foadi:2007ue} useful for collider phenomenology. The phase structure of a large number of  strongly interacting nonsupersymmetric theories, as function of number of underlying colors will be uncovered with traditional nonperturbative methods \cite{Dietrich:2006cm} as well as novel ones \cite{{Ryttov:2007cx}}. We will discuss possible applications to cosmology as well. These lectures should be integrated with earlier reviews  \cite{Hill:2002ap,Sannino:2008ha,Shrock:2007km,Lane:2002wv,Sarkar:1995dd,Chanowitz:1988ae,Farhi:1980xs,Kaul:1981uk,Chivukula:2000mb} on the various subjects treated here. 

\newpage
\section{Dynamical Electroweak Symmetry Breaking}

It is a fact that the SM does not fail, when experimentally tested, to describe all of the known forces to a very high degree of experimental accuracy. This is true even if we include gravity. Why is it so successful?

The SM is a low energy effective theory valid up to a scale $\Lambda$ above which new interactions, symmetries, extra dimensional worlds or any possible extension can emerge. At sufficiently low energies with respect to the cutoff scale $\Lambda$ one expresses the existence of new physics via effective operators. The success of the SM is due to the fact that most of the corrections to its physical observable depend only logarithmically on the cutoff scale $\Lambda$. 

Superrenormalizable operators are very sensitive to the cut off scale. In the SM there exists only one operator with naive mass dimension two which acquires corrections quadratic in $\Lambda$. This is the squared mass operator of the Higgs boson.  Since $\Lambda$ is expected to be the highest possible scale, in four dimensions the Planck scale, it is hard to explain {\it naturally}  why the mass of the Higgs is of the order of the electroweak scale. The Higgs is also the only particle predicted in the SM yet to be directly produced in experiments. Due to the occurrence of quadratic corrections in the cutoff this is the SM sector highly sensitve to the existence of new physics. 

In Nature we have already observed Higgs-type mechanisms. Ordinary superconductivity and chiral symmetry breaking in QCD are two time-honored examples. In both cases the mechanism has an underlying dynamical origin with the Higgs-like particle being a composite object of fermionic fields.

\subsection{Superconductivity versus Electroweak Symmetry Breaking }
The breaking of the electroweak theory is a relativistic screening effect. It is useful to parallel it to ordinary superconductivity which is also a screening phenomenon albeit non-relativistic. The two phenomena happen at a temperature lower than a critical one. In the case of superconductivity one defines a density of superconductive electrons $n_s$ and to it one associates a macroscopic wave function $\psi$ such that  its modulus squared
\begin{eqnarray}
|\psi|^2 = n_C = \frac{n_s}{2} \ ,
\end{eqnarray}
is the density of Cooper's pairs. That we are describing a nonrelativistic system is manifest in the fact that the macroscopic wave function squared, in natural units, has mass dimension three while the modulus squared of the Higgs wave function evaluated at the minimum is equal to $<|H|^2> = v_{weak}^2/2$ and has mass dimension two, i.e. is a relativistic wave function. One can adjust the units by considering, instead of the wave functions, the Meissner-Mass of the photon in the superconductor which is
\begin{equation}
 M^2=q^2n_s/(4m_e) \ ,
 \end{equation}
  with $q=-2e$ and $2m_e$ the charge and the mass of a Cooper pair which is constituted by two electrons. In the electroweak theory the Meissner-Mass of the photon  is compared  with the relativistic mass of the $W$ gauge boson
  \begin{equation}
  M^2_W=g^2{v_{weak}^2}/{4}\ ,
  \end{equation}
  with $g$ the weak coupling constant and $v_{weak}$ the electroweak scale.  In a superconductor the relevant scale is given by the density of superconductive electrons typically of the order of $n_s\sim 4\times 10^{28}m^{-3}$ yielding a screening length of the order of $
\xi = 1/M\sim 10^{-6}{\rm cm}$. In the weak interaction case we measure directly the mass of the weak gauge boson which is of the order of $80$~GeV yielding a weak screening length $\xi_W=1/M_W\sim 10^{-15}{\rm cm}$.  

{}For a superconductive system it is clear from the outset that the wave function $\psi$ is not a fundamental degree of freedom, however for the Higgs we are not yet sure about its origin. The Ginzburg-Landau effective theory in terms of $\psi$ and the photon degree of freedom describes the spontaneous breaking of the $U_Q(1)$ electric symmetry and it is the equivalent of the Higgs Lagrangian.  

If the Higgs is due to a macroscopic relativistic screening phenomenon we expect it to be an effective description of a more fundamental system with possibly an underlying new strong gauge dynamics replacing the role of the phonons in the superconductive case. A dynamically generated Higgs system solves the problem of the quadratic divergences by replacing the cutoff $\Lambda$ with the weak energy scale itself, i.e. the scale of  compositness.  An underlying strongly coupled asymptotically free gauge theory, a la QCD,  is an example. 
 
 \subsection{From Color to Technicolor}
In fact even in complete absence of the Higgs sector in the SM the electroweak symmetry breaks \cite{Farhi:1980xs} due to the condensation of the following quark bilinear in QCD: 
\beq \langle\bar u_Lu_R + \bar d_Ld_R\rangle \neq 0 \ . \label{qcd-condensate}\eeq
  This mechanism, however, cannot account for the whole contribution to the weak gauge bosons masses. If QCD would be the only source contributing to the spontaneous breaking of the electroweak symmetry one would have
\beq M_W = \frac{gF_\pi}{2} \sim 29 {\rm MeV}\ , \eeq 
with $F_{\pi}\simeq 93$~MeV the pion decay constant. This contribution is very small with respect to the actual value of the $M_W$ mass that one typically neglects it. 

According to the original idea of technicolor \cite{Weinberg:1979bn,Susskind:1978ms} one augments the SM with another gauge interaction similar to QCD but with a new dynamical scale of the order of the electroweak one. It is sufficient that the new gauge theory is asymptotically free and has global symmetry able to contain the SM $SU_L(2)\times U_Y(1)$ symmetries. It is also required that the new global symmetries break dynamically in such a way that the embedded $SU_L(2)\times U_Y(1)$  breaks to the electromagnetic abelian charge $U_{Q}(1)$ . The dynamically generated scale will then be fit to the electroweak one. 

Note that, excepet in certain cases, dynamical behaviors are typically nonuniversal which means that different gauge groups and/or matter representations will, in general, posses very different dynamics. 

The simplest example of technicolor theory is the scaled up version of QCD, i.e. an $SU(N_{TC})$  nonabelian gauge theory with two Dirac Fermions transforming according to the fundamental representation or the gauge group. We need at least two Dirac flavors  to realize the $SU_L(2) \times SU_R(2)$ symmetry of the SM discussed in the SM Higgs section. One simply chooses the scale of the theory to be such that the new pion decaying constant is: \beq F_\pi^{TC} = v_{\rm weak} \simeq 246~ {\rm GeV} \ . \eeq 
The flavor symmetries, for any $N_{TC}$ larger than 2 are $SU_L(2) \times SU_R(2)\times U_V(1)$ which spontaneously break to $SU_V(2)\times U_V(1)$. It is natural to embed the electroweak symmetries within the present technicolor model in a way that the hypercharge corresponds to the third generator of $SU_R(2)$. This simple dynamical model correctly accounts for the electroweak symmetry breaking. The new technibaryon number $U_V(1)$ can break due to not yet specified new interactions. 
In order to get some indication on the dynamics and spectrum of this theory one can use the 't Hooft large N limit  \cite{'tHooft:1973jz,Witten:1979kh,'tHooft:1980xb}.  {}For example the intrinsic scale of the theory is related to the QCD one via:
\beq \Lambda_{\rm TC} \sim
\sqrt{\frac{3}{N_{TC}}}\frac{F_\pi^{TC}}{F_\pi}\Lambda_{\rm QCD} \ . \eeq
At this point it is straightforward to use the QCD phenomenology for describing the experimental signatures and dynamics of a composite Higgs.  

\subsection{Constraints from Electroweak Precision Data}
\label{5}

The relevant corrections due to the presence of new physics trying to modify the electroweak breaking sector of the SM appear in the vacuum polarizations of the electroweak gauge bosons. These can be parameterized in terms of the three 
quantities $S$, $T$, and $U$ (the oblique parameters) 
\cite{Peskin:1990zt,Peskin:1991sw,Kennedy:1990ib,Altarelli:1990zd}, and confronted with the electroweak precision data. Recently, due to the increase precision of the measurements reported by LEP II, the list of interesting parameters to compute has been extended \cite{hep-ph/9306267,Barbieri:2004qk}.  We show below also the relation with the traditional one \cite{Peskin:1990zt}.  Defining with  $Q^2\equiv -q^2$ the Euclidean transferred momentum entering in a generic two point function vacuum polarization associated to the electroweak gauge bosons, and denoting derivatives with respect to $-Q^2$ with a prime we have \cite{Barbieri:2004qk}: 
\begin{eqnarray}
\hat{S} &\equiv & g^2 \ \Pi_{W^3B}^\prime (0) \ , \\
\hat{T} &\equiv & \frac{g^2}{M_W^2}\left[ \Pi_{W^3W^3}(0) -
\Pi_{W^+W^-}(0) \right] \ , \\
W &\equiv & \frac{g^2M_W^2}{2} \left[\Pi^{\prime\prime}_{W^3W^3}(0)\right] \ , \\
Y &\equiv & \frac{g'^2M_W^2}{2} \left[\Pi^{\prime\prime}_{BB}(0)\right] \ , \\
\hat{U} &\equiv & -g^2 \left[\Pi^\prime_{W^3W^3}(0)-
\Pi^\prime_{W^+W^-}(0)\right]\ , \\
V &\equiv & \frac{g^2 \, M^2_W}{2}\left[\Pi^{\prime\prime}_{W^3W^3}(0)-
\Pi^{\prime\prime}_{W^+W^-}(0)\right] \ , \\
X &\equiv & \frac{g g'\,M_W^2}{2} \ \Pi_{W^3B}^{\prime\prime}(0) \ .
\end{eqnarray}
Here $\Pi_V(Q^2)$ with $V=\{W^3B,\, W^3W^3,\, W^+W^-,\, BB\}$ represents the
self-energy of the vector bosons. Here the
electroweak couplings are the ones associated to the physical electroweak gauge bosons:
\begin{eqnarray}
\frac{1}{g^2} \equiv  \Pi^\prime_{W^+W^-}(0)
 \ , \qquad \frac{1}{g'^2}
\equiv  \Pi^\prime_{BB}(0) \ ,
\end{eqnarray}
while $G_F$ is
\begin{eqnarray}
\frac{1}{\sqrt{2}G_F}=-4\Pi_{W^+W^-}(0) \ ,
\end{eqnarray}
as in \cite{Chivukula:2004af}. $\hat{S}$ and $\hat{T}$ lend their name
from the well known Peskin-Takeuchi parameters $S$ and $T$ which are related to the new ones via
\cite{Barbieri:2004qk,Chivukula:2004af}:
\begin{eqnarray}
\frac{\alpha S}{4s_W^2} =  \hat{S} - Y - W  \ , \qquad 
\alpha T = \hat{T}- \frac{s_W^2}{1-s_W^2}Y \ .
\end{eqnarray}
Here $\alpha$ is the electromagnetic structure constant and $s_W=\sin \theta_W $
is the weak mixing angle. Therefore in the case where $W=Y=0$ we
have the simple relation
\begin{eqnarray}
\hat{S} &=& \frac{\alpha S}{4s_W^2} \ , \qquad 
\hat{T}= \alpha T \ .
\end{eqnarray}
The result of the the fit is shown in
Fig.~\ref{s06_stu_contours}. If the value of the Higgs mass increases the central value of the $S$ parameters moves to the left towards negative values. 

\begin{figure}[t]
\begin{center}
\includegraphics[width=7truecm,height=4.5truecm]{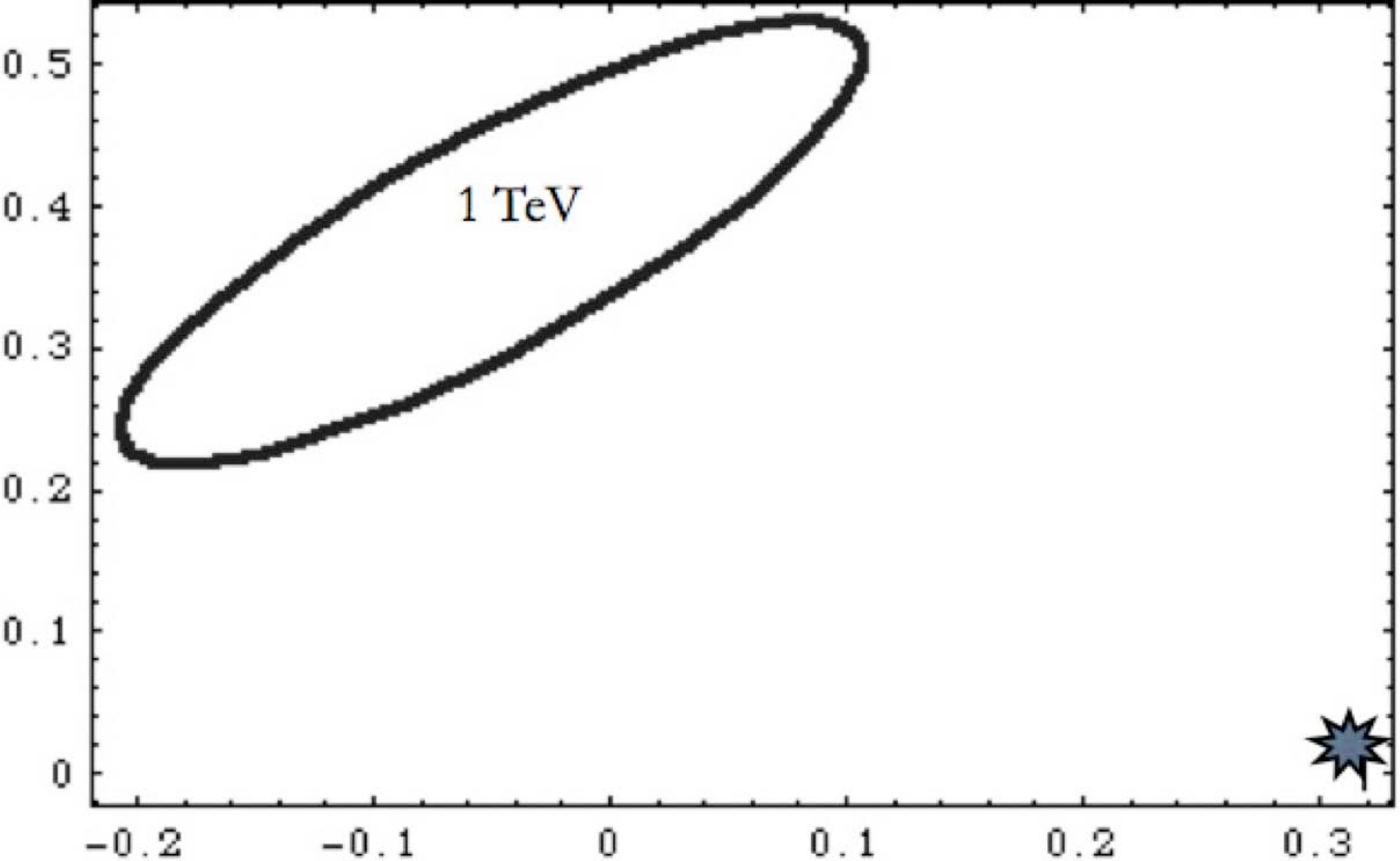}
\caption{$T$ versus $S$  for $SU(3)$ technicolor with one technifermion doublet (the full asterisk) versus precision data for a one TeV composite Higgs mass.} \label{TCSproblem}
\end{center}
\end{figure}

\noindent
In technicolor it is easy to have a vanishing $T$ parameter while typically $S$ is positive. Besides, the composite Higgs is typically heavy with respect to the Fermi scale, at least for technifermions in the fundamental representation of the gauge group and for a small number of techniflavors. The oldest technicolor models featuring QCD dynamics with three technicolors and a doublet of electroweak gauged techniflavors deviate a few sigma from the current precision tests as summarized in the figure \ref{TCSproblem}.
Clearly it is desirable to reduce the tension between the precision data and a possible dynamical mechanism underlying the electroweak symmetry breaking. It is possible to imagine different ways to achieve this goal and some of the earlier attempts have been summarized in \cite{Peskin:2001rw}. 

The computation of the $S$ parameter in technicolor theories requires the knowledge of nonperturbative dynamics rendering difficult the precise knowledge of the contribution to  $S$. {}For example, it is not clear what is the exact value of the composite Higgs mass relative to the Fermi scale and, to be on the safe side, one typically takes it to be quite large, of the order at least of the TeV. However in certain models it may be substantially lighter due to the intrinsic dynamics. We will discuss the spectrum of different strongly coupled theories in the Appendix and its relation to the electroweak parameters later in this chapter. 

 It is, however, instructive to provide a simple estimate of the contribution to $S$ which allows to guide model builders. Consider a one-loop exchange of $N_D$ doublets of techniquarks transforming according to the representation $R_{TC}$ of the underlying technicolor gauge theory and with dynamically generated mass $\Sigma_{(0)}$ assumed to be larger than the weak intermediate gauge bosons masses. Indicating with $d(R_{\rm TC})$ the dimension of the techniquark representation, and to leading order in $M_{W}/\Sigma(0)$ one finds:
 \begin{eqnarray}
S_{\rm naive} = N_D \frac{d(R_{\rm TC})}{6\pi} \ .
\end{eqnarray} 
This naive value provides, in general, only a rough estimate of the exact value of $S$. 
However, it is clear from the formula above that, the more technicolor matter is gauged under the electroweak theory the larger is the $S$ parameter and that the final $S$ parameter is expected to be positive. 

Attention must be paid to the fact that the specific model-estimate of the whole $S$ parameter, to compare with the experimental value, receives contributions also from other sectors. Such a contribution can be taken sufficiently large and negative to compensate for the positive value from the composite Higgs dynamics. To be concrete: Consider an extension of the SM in which the Higgs is composite but we also have new heavy (with a mass of the order of the electroweak) fourth family of Dirac leptons. In this case a sufficiently large splitting of the new lepton masses can strongly reduce and even offset the positive value of $S$. We will discuss this case in detail when presenting the Minimal Walking Technicolor model. The contribution of the new sector ($ S_{\rm NS}$) above, and also in  many other cases, is perturbatively under control and the total $S$ can be written as:
\begin{eqnarray}
S = S_{\rm TC} + S_{\rm NS} \ .
\end{eqnarray}
 The parameter $T$ will be, in general, modified and one has to make sure that the corrections do not spoil the agreement with this parameter.  From the discussion above it is clear that technicolor models can be constrained, via precision measurements, only model by model and the effects of possible new sectors must be properly included. We presented the constraints coming from $S$ using the underlying gauge theory information. However, in practice, these constraints apply directly to the physical spectrum. 

The classical presentation above is utterly incomplete. In fact it neglects the constraints and back-reaction on the gauge sector coming from the one giving masses to the SM fermions. To estimate these effects we have considered two simple extensions able, in an effective way, to accommodate the SM masses. In \cite{Fukano:2009zm}, to estimate these corrections,  the composite Higgs sector was coupled directly to the SM fermions \cite{Foadi:2007ue}. Here one gets relevant constraints on the $W$ parameter while the corrections do not affect the $S$-parameter. The situation changes when an entirely new sector is introduced in the flavor sector. Due to the almost inevitable interplay between the gauge and the flavor sector the back-reaction of the flavor sector is very relevant \cite{Antola:2009wq}.  Mimicking the new sector via a new (composite or not) Higgs coupling directly to the SM fermions it was observed that important corrections to the $S$ and $T$ parameters arise which can be used to compensate a possible heavy composite Higgs scenario of the technicolor sector \cite{Antola:2009wq}. To investigate these effects we adopted a straightforward and instructive model according to which we have both a composite sector and  a fundamental scalar field  (SM-like Higgs) intertwined at the electroweak scale. This idea was pioneered  in a series of papers by Simmons  \cite{Simmons:1988fu},  Dine,  Kagan and Samuel \cite{Dine:1990jd,Samuel:1990dq,Kagan:1991gh,Kagan:1992aq,Kagan:1990gi} and Carone and Georgi  \cite{Carone:1992rh,Carone:1994mx}. More recently this type of model has been investigated also in \cite{Hemmige:2001vq,Carone:2006wj,Zerwekh:2009yu}. Interesting related work can be also found in \cite{Chivukula:1990bc,Chivukula:2009ck}.

\subsection{Standard Model Fermion Masses}

Since in a purely technicolor model  the Higgs is a composite particle the Yukawa terms, when written in terms of the underlying technicolor fields, amount to four-fermion operators. The latter can be naturally interpreted as a low energy operator induced by a new strongly coupled gauge interaction emerging at energies higher than the electroweak theory. These type of theories have been termed extended technicolor interactions (ETC) \cite{Eichten:1979ah,Dimopoulos:1979es}. 

\begin{figure}[tp]
\begin{center}
\mbox{
\subfigure{\resizebox{!}{0.23\linewidth}{\includegraphics[clip=true]{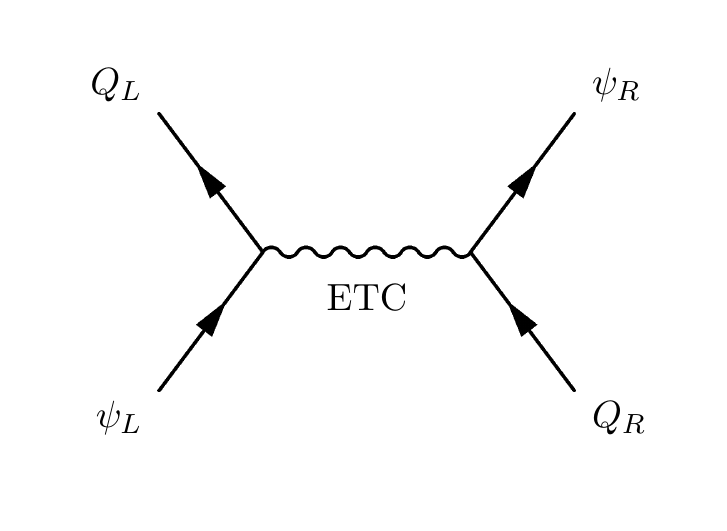}}}\qquad \qquad
\subfigure{\resizebox{!}{0.23\linewidth}{\includegraphics[clip=true]{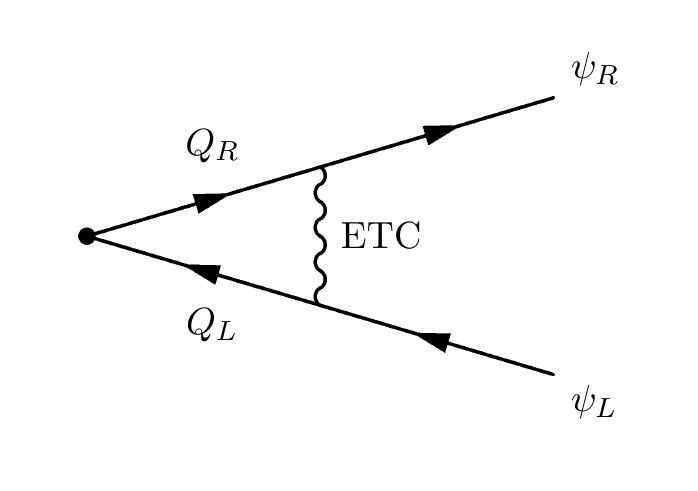}}}
}
\caption{Left Panel: ETC  gauge boson interaction involving
techniquarks and SM fermions. Right Panel: Diagram contribution to the mass to the SM fermions.}
\label{etcint}
\end{center}
\end{figure}

In the literature various extensions have been considered and we will mention them later in the text.  Here we will describe the simplest ETC model in which the ETC interactions connect the chiral symmetries of the techniquarks to those of the SM fermions (see Left panel of Fig.~\ref{etcint}).

When TC chiral symmetry breaking occurs it leads to the diagram drawn in
Fig.~\ref{etcint}b. Let's start with the case in which the ETC dynamics is represented by a $SU(N_{ETC})$ gauge group with: 
\beq N_{ETC} = N_{TC} + N_g \ , \eeq
and $N_g$ is the number of SM
generations. In order to give masses to all of the SM fermions, in this scheme, one needs a condensate for each SM fermion. This can be achieved by using as technifermion matter a complete generation of quarks and leptons (including a neutrino right) but now gauged with respect to the technicolor interactions.  

The ETC gauge group is assumed to spontaneously break $N_g$ times down to
$SU(N_{TC})$ permitting
three different mass scales, one  for each SM family. This type of technicolor with associated ETC is
termed the \emph{one family model} \cite{Farhi:1979zx}.
The heavy masses are provided by the
breaking at low energy and the light masses are provided by breaking
at higher energy scales. 
This model does not, per se, explain how the
gauge group is broken several times, neither is the breaking of weak isospin
symmetry accounted for. {}For example we cannot explain why the neutrino have masses much smaller than the associated electrons. See, however, \cite{Appelquist:2004ai} for progress on these issues. Schematically one has $SU(N_{TC} + 3)$ which breaks to  $SU(N_{TC} + 2)$ at the scale 
$\Lambda_1$ providing the first generation of fermions with a typical mass $m_1 \sim {4\pi
  (F_\pi^{TC})^3}/{\Lambda_1^2}$ at this point the gauge group breaks to $SU(N_{TC} + 1)$ with dynamical scale $\Lambda_2 $ leading to a second generation mass of the order of $m_2 \sim{4\pi
  (F_\pi^{TC})^3}/{\Lambda_2^2}$ finally the last breaking
$SU(N_{TC} )$ at scale 
$\Lambda_3$ leading to the last generation mass $m_3 \sim {4\pi
  (F_\pi^{TC})^3}/{\Lambda_3^2}$. 
 
Without specifying an ETC one can write down the most general type of four-fermion operators involving technicolor particles $Q$ and ordinary fermionic fields $\psi$.  {}Following the notation of Hill and Simmons \cite{Hill:2002ap} we write:
\beq \alpha_{ab}\frac{\bar Q\gamma_\mu T^aQ\bar\psi \gamma^\mu
  T^b\psi}{\Lambda_{ETC}^2} +
\beta_{ab}\frac{\bar Q\gamma_\mu T^aQ\bar Q\gamma^\mu
  T^bQ}{\Lambda_{ETC}^2} + 
\gamma_{ab}\frac{\bar\psi\gamma_\mu T^a\psi\bar\psi\gamma^\mu
  T^b\psi}{\Lambda_{ETC}^2} \ , \eeq
where the $T$s are unspecified ETC generators. After performing a Fierz rearrangement one has:
\beq \alpha_{ab}\frac{\bar QT^aQ\bar\psi T^b\psi}{\Lambda_{ETC}^2} +
\beta_{ab}\frac{\bar QT^aQ\bar QT^bQ}{\Lambda_{ETC}^2} +
\gamma_{ab}\frac{\bar\psi T^a\psi\bar\psi T^b\psi}{\Lambda_{ETC}^2}
+ \ldots \ , \label{etc} \eeq
The coefficients parametrize the ignorance on the specific ETC physics. To be more specific, the $\alpha$-terms, after the technicolor particles have condensed, lead to mass terms for the SM fermions
\beq m_q \approx \frac{g_{ETC}^2}{M_{ETC}^2}\langle \bar
QQ\rangle_{ETC} \ , \eeq
where $m_q$ is the mass of e.g.~a SM quark, $g_{ETC}$ is the ETC gauge 
coupling constant evaluated at the ETC scale, $M_{ETC}$ is the mass of
an ETC gauge boson and $\langle \bar QQ\rangle_{ETC}$ is the
technicolor condensate where the operator is evaluated at the ETC
scale. Note that we have not explicitly considered the different scales for the different generations of ordinary fermions but this should be taken into account for any realistic model. 

\begin{figure}[tbh]
\begin{center}
\includegraphics[width=7truecm,height=4truecm]{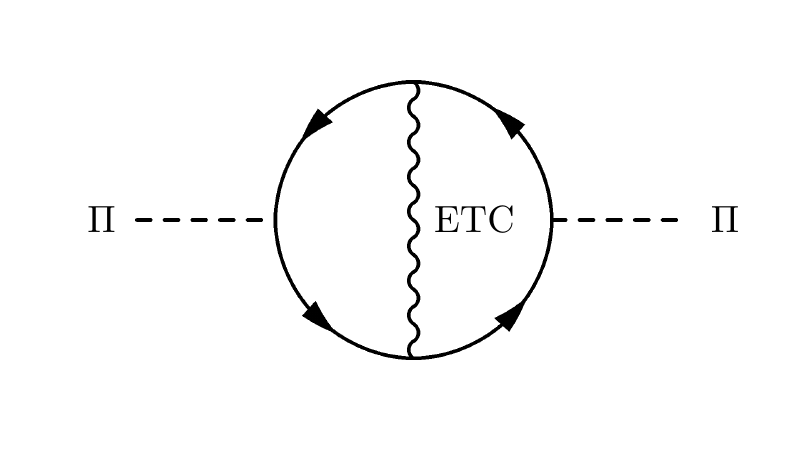}
\caption{Leading contribution to the mass of the TC pseudo
  Goldstone bosons via an exchange of an ETC gauge boson.} \label{masspgb}
\end{center}
\end{figure}

The $\beta$-terms of Eq.~(\ref{etc}) provide masses for
pseudo Goldstone bosons and also provide masses for techniaxions
\cite{Hill:2002ap}, see figure \ref{masspgb}. 
The last class of terms, namely the $\gamma$-terms of
Eq.~(\ref{etc}) induce flavor changing neutral currents. {}For example it may generate the following terms:
\beq \frac{1}{\Lambda_{ETC}^2}(\bar s\gamma^5d)(\bar s\gamma^5d) +
\frac{1}{\Lambda_{ETC}^2}(\bar \mu\gamma^5e)(\bar e\gamma^5e) + 
\ldots \ , \label{FCNC} \eeq
where $s,d,\mu,e$ denote the strange and down quark, the muon
and the electron, respectively. The first term is a $\Delta S=2$
flavor-changing neutral current interaction affecting the
$K_L-K_S$ mass difference which is measured accurately. The experimental bounds on these type of operators together with the very {\it naive} assumption that ETC will generate these operators with $\gamma$ of order one leads to a constraint on the ETC scale to be of the order of or larger than $10^3$
TeV \cite{Eichten:1979ah}. This should be the lightest ETC scale which in turn puts an upper limit on how large the ordinary fermionic masses can be. The naive estimate is that  one can account up to around 100 MeV mass for a QCD-like technicolor theory, implying that the Top quark mass value cannot be achieved.

The second term of Eq.~(\ref{FCNC}) induces flavor
changing processes in the leptonic sector such as $\mu\rightarrow e\bar ee,
e\gamma$ which are not observed.
It is clear that, both for the precision measurements  and the fermion masses, that a better theory of the flavor is needed. 

\subsection{Walking}
To better understand in which direction one should go to modify the QCD dynamics we analyze the TC condensate. 
The value of the technicolor condensate used when giving mass to the ordinary fermions should be evaluated not at the technicolor scale but at the extended technicolor one. Via the renormalization group one can relate the condensate at the two scales via:
\beq \langle\bar QQ\rangle_{ETC} =
\exp\left(\int_{\Lambda_{TC}}^{\Lambda_{ETC}}
d(\ln\mu)\gamma(\alpha(\mu))\right)\langle\bar QQ\rangle_{TC} \ ,
\label{rad-cor-tc-cond}
\eeq 
where $\gamma$ is the anomalous dimension of the techniquark mass-operator. The boundaries of the integral 
 are at the ETC scale and the TC one.
{}For TC theories with a running of the coupling constant similar to the one in QCD, i.e.
\beq \alpha(\mu) \propto \frac{1}{\ln\mu} \ , \quad {\rm for}\ \mu >
\Lambda_{TC} \ , \eeq
this implies that the anomalous dimension of the techniquark masses $\gamma \propto
\alpha(\mu)$. When computing the integral one gets
\beq \langle\bar QQ\rangle_{ETC} \sim
\ln\left(\frac{\Lambda_{ETC}}{\Lambda_{TC}}\right)^{\gamma}
\langle\bar QQ\rangle_{TC} \ , \label{QCD-like-enh} \eeq
which is a logarithmic enhancement of the operator. We can hence neglect this correction and use directly the value of the condensate at the TC scale when estimating the generated fermionic mass:
\beq m_q \approx \frac{g_{ETC}^2}{M_{ETC}^2}\Lambda_{TC}^3 \ , \qquad 
 \langle \bar QQ\rangle_{TC} \sim \Lambda_{TC}^3 \ . \eeq

The tension between having to reduce the FCNCs and at the same time provide a sufficiently large mass for the heavy fermions in the SM as well as the pseudo-Goldstones can be reduced if the dynamics of the underlying TC theory is different from the one of QCD. The computation of the TC condensate at different scales shows that  if the dynamics is such that the TC coupling does not {\it run} to the UV fixed point but rather slowly reduces to zero one achieves a net enhancement of the condensate itself with respect to the value estimated earlier.  This can be achieved if the theory has a near conformal fixed point. This kind of dynamics has been denoted of {\it walking} type. In this case 
\beq \langle\bar QQ\rangle_{ETC} \sim
\left(\frac{\Lambda_{ETC}}{\Lambda_{TC}}\right)^{\gamma(\alpha^*)}
\langle\bar QQ\rangle_{TC} \ , \label{walking-enh} \eeq
which is a much larger contribution than in QCD dynamics  \cite{Yamawaki:1985zg,Holdom:1984sk,Holdom:1981rm,Appelquist:1986an}. Here $\gamma$ is evaluated at the would be fixed point value $\alpha^*$.  Walking can help resolving the problem of FCNCs in
technicolor models since with a large enhancement of the $\langle\bar
QQ\rangle$ condensate the four-fermi operators involving SM fermions and
technifermions and the ones involving technifermions are enhanced by a factor of
$\Lambda_{ETC}/\Lambda_{TC}$ to the $\gamma$  power while the one involving only SM fermions is not enhanced.

\begin{figure}
\centering
\begin{tabular}{cc}
\resizebox{6.0cm}{!}{\includegraphics{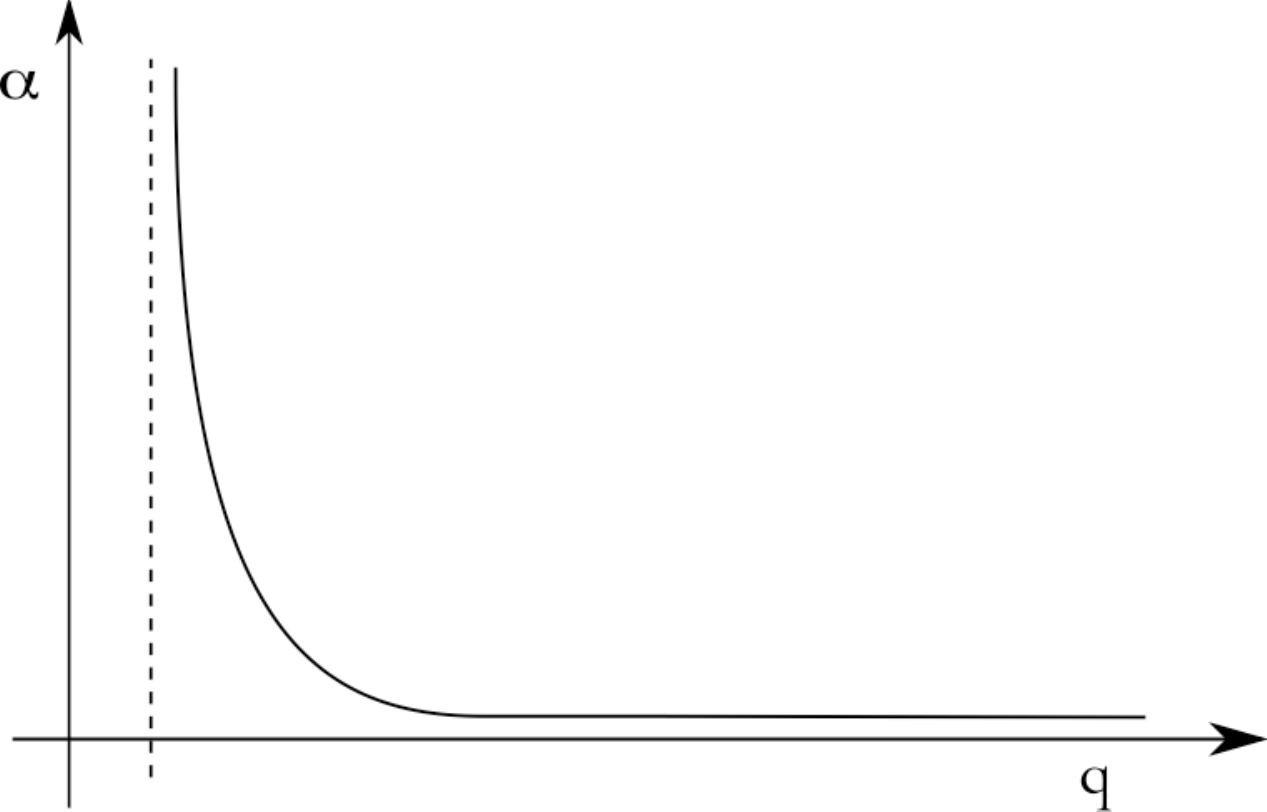}} ~~~&~~ \resizebox{6.0cm}{!}{\includegraphics{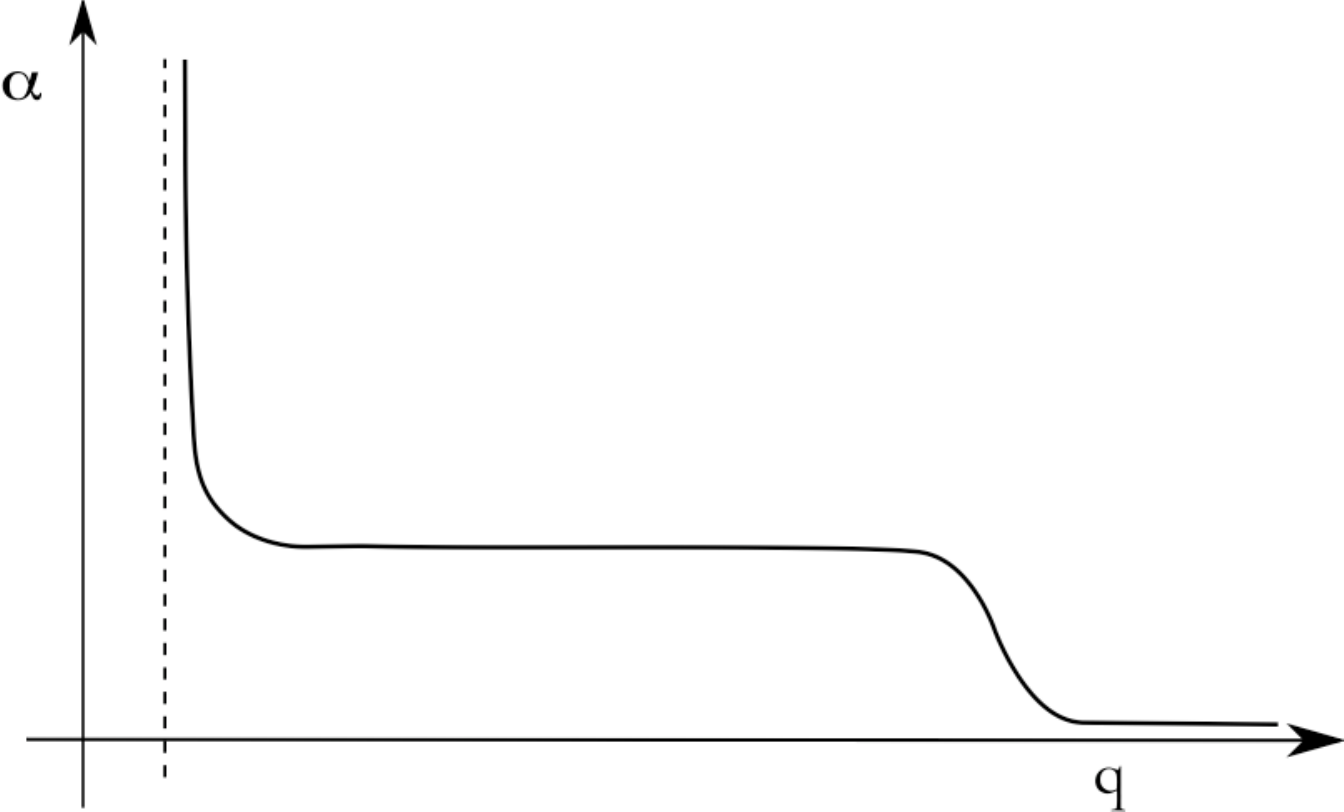}} \\&\\
&~~~~\resizebox{6.0cm}{!}{\includegraphics{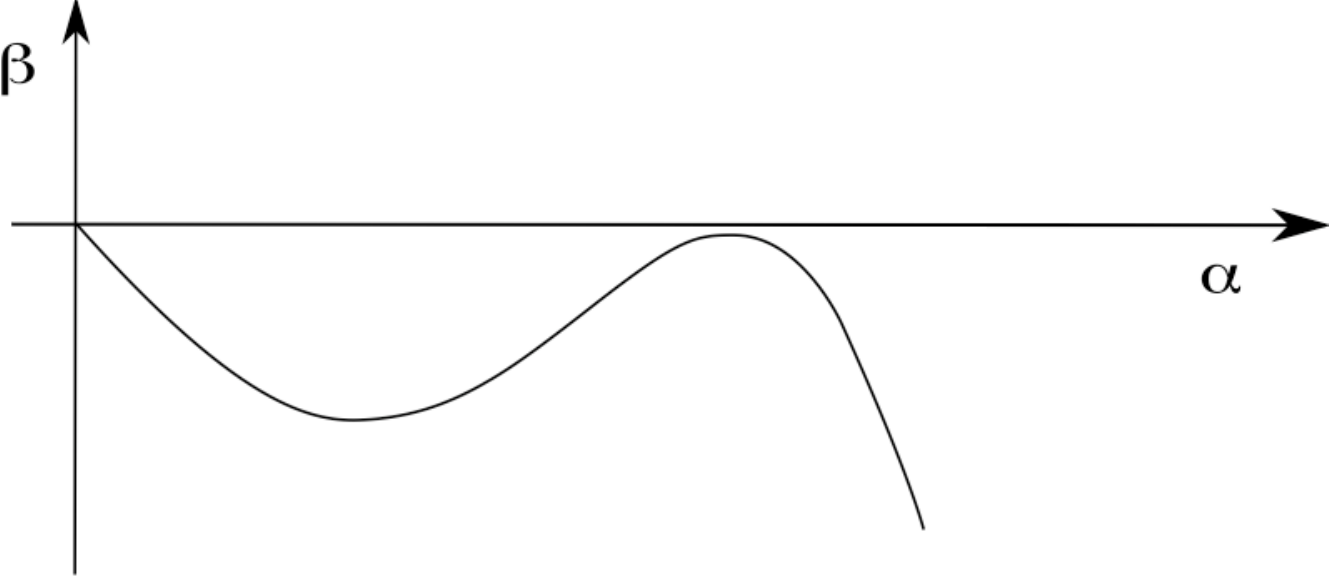}} 
\end{tabular}
\caption{Top Left Panel: QCD-like behavior of the coupling constant as function of the momentum (Running). Top Right Panel: Walking-like behavior of the coupling constant as function of the momentum (Walking). Bottom Right Panel: Cartoon of the beta function associated to a generic walking theory.}
\label{walkbeta}
\end{figure}

In the figure \ref{walkbeta} the comparison between a running and walking behavior of the coupling is qualitatively represented. 
We note tha {\it walking} is not a fundamental property for a successful model of the origin of mass of the elementary fermions featuring technicolor. In fact several alternative ideas already exist in the literature (see \cite{Antola:2009wq} and references therein).  However, a near conformal theory would still be useful to reduce the contributions to the precision data and, possibly, provide a light composite Higgs of much interest to LHC physics.

\subsection{Weinberg Sum Rules and Electroweak Parameters}\label{sec:electroweak}

Any strongly coupled dynamics, even of walking type, will generate a spectrum of resonances whose natural splitting in mass is of the order of the intrinsic scale of the theory which in this case is the Fermi scale. In order to extract predictions  for the composite vector spectrum and couplings in presence of a strongly interacting sector and an asymptotically free gauge theory, we make use of the time-honored Weinberg sum rules (WSR) \cite{Weinberg:1967kj} but we will also use the results found in \cite{Appelquist:1998xf}  allowing us to treat walking and running theories in a unified way.

\subsubsection{Weinberg sum rules}
The Weinberg sum rules (WSRs)  are linked to the  two point vector-vector minus axial-axial vacuum polarization which is known to be sensitive to chiral symmetry breaking.  We define
\begin{equation}
i\Pi_{\mu \nu}^{a,b}(q)\equiv \int\!d^4x\, e^{-i qx}
\left[<J_{\mu,V}^a(x)J_{\nu,V}^b(0)> -
 <J_{\mu,A}^a(x)J_{\nu,A}^b(0)>\right] \ ,
\label{VA}
\end{equation}
within the underlying strongly coupled gauge theory, where
\begin{equation}
\Pi_{\mu \nu}^{a,b}(q)=\left(q_{\mu}q_{\nu} - g_{\mu\nu}q^2 \right) \,
\delta^{a b} \Pi(q^2) \ .
\end{equation}
Here $a,b=1,...,N_f^2-1$, label the flavor
currents and the SU(N$_f$) generators are normalized according to
$\rm{Tr} \left[T^a T^b\right]= (1/2) \delta^{ab} $.  The
function $\Pi(q^2)$ obeys the unsubtracted dispersion relation
\begin{equation}
\frac{1}{\pi} \int_0^{\infty}\!ds\, \frac{{\rm Im}\Pi(s)}{s + Q^2}
=\Pi(Q^2) \ ,
\label{integral}
\end{equation}
where $Q^2=-q^2 >0$, and the constraint
$\displaystyle{-Q^2 \Pi(Q^2)>0}$ holds for $0 < Q^2 < \infty$~\cite{Witten:1983ut}. The discussion above is for the standard chiral symmetry breaking pattern SU(N$_f$)$\times$ SU(N$_f$)$ \rightarrow $SU(N$_f$) but it is generalizable to any breaking pattern.

Since we are taking the underlying theory to be asymptotically
free, the behavior of $\Pi(Q^2)$ at asymptotically high momenta is
the same as in ordinary QCD, i.e. it scales like $Q^{-6}$~\cite{Bernard:1975cd}. Expanding the left hand side of the dispersion relation
thus leads to the two conventional spectral function sum rules
\begin{equation}
\frac{1}{\pi} \int_0^{\infty}\!ds\,{\rm Im}\Pi(s) =0
\label{spectral1}
\quad {\rm and} \quad
\frac{1}{\pi} \int_0^{\infty}\!ds\,s \,{\rm Im}\Pi(s) =0 \ .
\end{equation}
Walking dynamics affects only the second sum rule \cite{Appelquist:1998xf} which is more sensitive to large but not asymptotically large momenta due to fact that the associated integrand contains an extra power of $s$.

We now saturate the absorptive part of the vacuum
polarization. We follow reference \cite{Appelquist:1998xf} and hence divide the
energy range of integration in three parts. The light resonance part. In this regime, the    
integral is saturated by the
Nambu-Goldstone pseudoscalar along with massive vector and
axial-vector states. If we assume, for example, that there is only a
single, zero-width vector multiplet and a single, zero-width axial
vector multiplet, then
\begin{equation}
{\rm Im}\Pi(s)=\pi F^2_V \delta \left(s -M^2_V \right) - \pi F^2_A
\delta \left(s - M^2_A \right) - \pi F^2_{\pi} \delta \left(s \right)
\ .
\label{saturation}
\end{equation}
The zero-width approximation is valid to leading order in the large
$N$ expansion for fermions in the fundamental representation of the gauge group and it is even narrower for fermions in higher dimensional representations. Since we are working near a conformal fixed point the large $N$ argument for the width is not directly applicable. We will nevertheless use this simple model for the spectrum
to  illustrate the effects of a near critical IR fixed point. 

The first WSR implies:
\begin{equation}
F^2_V - F^2_A = F^2_{\pi}\ ,
\label{1rule}
\end{equation}
where $F^2_V$ and $F^2_A$ are the vector and axial mesons decay
constants.  This sum rule holds for walking and running dynamics. {A
more general representation of the resonance spectrum would, in principle, replace
the left hand side of this relation with a sum over vector and axial
states. However the heavier resonances should not be included since in the approach of \cite{Appelquist:1998xf} the walking dynamics in the intermediate energy range is already approximated by the exchange of underlying fermions. The walking is encapsulated in the dynamical mass dependence on the momentum dictated by the gauge theory. The introduction of heavier resonances is, in practice, double counting. Note that the approach is in excellent agreement with the Weinberg approximation for QCD, since in this case, the  approximation automatically returns the known results.  }

The second sum rule receives important contributions from throughout
the near conformal region and can be
expressed in the form of:
\begin{equation}
F^2_V M^2_V - F^2_A M^2_A = a\,\frac{8\pi^2}{d(R)}\,F_{\pi}^4,
\label{2rule-2}
\end{equation}
where $a$ is expected to be positive and $O(1)$ and $d(R)$ is the dimension of the representation of the underlying fermions.  
We have generalized the result of reference \cite{Appelquist:1998xf}  to the case in which the fermions belong to a generic representation of the gauge group. In the case of running dynamics the right-hand side of the previous equation vanishes.  

{We stress that $a$ is a non-universal quantity depending on the details of the underlying gauge theory. A reasonable measure of how large $a$ can be is given by a function of the amount of walking which is the ratio of the scale above which the underlying coupling constant start running divided by the scale below which chiral symmetry breaks.} 
The fact that $a$ is positive and of order one in walking dynamics is supported, indirectly, also via the results of Kurachi and Shrock \cite{Kurachi:2006ej}. At the onset of conformal dynamics the axial and the vector will be degenerate, i.e. $M_A=M_V=M$, using the first sum rule one finds via the second sum rule $a = d({\rm R})M^2/(8\pi^2 F^2_{\pi})$ leading to a numerical value of about 4- 5 from the approximate results in \cite{Kurachi:2006ej}.  We will however use only the constraints coming from the generalized WSRs  expecting them to be less model dependent.
\subsubsection{Relating WSRs to the Effective Theory \& $S$ parameter }
The $S$ parameter is related to the absorptive part  of the
vector-vector minus axial-axial vacuum polarization as follows:
\begin{equation}
S=4\int_0^\infty \frac{ds}{s} {\rm Im}\bar{\Pi}(s)= 4\pi
\left[\frac{F^2_V}{M^2_V} - \frac{F^2_A}{M^2_A} \right] \ ,
\label{s-def}
\end{equation}
where ${\rm Im}\bar{\Pi}$ is obtained from ${\rm Im}\Pi$ by
subtracting the Goldstone boson contribution.

Other attempts to estimate the $S$ parameter for walking technicolor
theories have been made in the past \cite{Sundrum:1991rf} showing reduction of the $S$ parameter. $S$ has also been evaluated using computations inspired by the original AdS/CFT correspondence \cite{Maldacena:1997re} in \cite{Hong:2006si,Hirn:2006nt,Piai:2006vz,Agashe:2007mc,Carone:2007md,Hirayama:2007hz}. Recent attempts to use AdS/CFT inspired methods can be found in \cite{Dietrich:2009af,Dietrich:2008up,Dietrich:2008ni,Nunez:2008wi,Fabbrichesi:2008ga}.   

Kurachi, Shrock and Yamawaki \cite{Kurachi:2007at} have further confirmed the results presented in \cite{Appelquist:1998xf} with their computations tailored for describing four dimensional gauge theories near the conformal window.
The present approach \cite{Appelquist:1998xf} is more physical since it is based on the
nature of the spectrum of states associated directly to the underlying gauge theory.  

Note that we will be assuming a rather conservative approach in which the $S$ parameter, although reduced with respect to the case of a running theory, is positive and not small.  After all, other sectors of the theory such as new leptons further reduce or completely offset a positive value of $S$ due solely to the technicolor theory. 

\subsection{Naturalizing Unparticle}
It would be extremely exciting to discover new strong dynamics at the Large Collider (LHC). It is then interesting to explore the possibility to be able accommodate the unparticle scenario \cite{Georgi:2007ek} into a natural setting featuring  four dimensional strongly interacting dynamics \cite{Sannino:2008nv}.

Georgi's original idea is that at high energy there is an UV sector coupled to the SM 
through the exchange of messenger fields with a 
large mass scale $\hoch$. Below that scale 
two things happen consecutively. Firstly, the messenger sector decouples,
resulting in contact interactions between the SM and the unparticle sector. Secondly, the latter flows into a non-perturbative infrared (IR) fixed point
at a scale $\Lambda_\cU  \ll \hoch$ hence exhibiting scale invariance;
\begin{equation}
\label{eq:unscen}
{\cal L}   \sim  {\cal O}_{\rm UV} 
  {\cal O}_{\rm SM}    \to  {\cal O}_\cU {\cal O}_{\rm SM} \, .
\end{equation}
The UV unparticle operator is denoted by ${\cal O}_{\rm UV}$ and it posses integer
dimension $d_{\rm UV}$. When the IR fixed point 
is reached the operator ${\cal O}_{\rm IR} \equiv {\cal O}_\cU$ acquires a non-integer scaling dimension $\dU$ through
dimensional transmutation
\begin{equation}
\label{eq:un}
| \matel{0}{{\cal O}_\cU}{P}| \sim (\sqrt{P^2})^{\dU-1}{} \, .
\end{equation}
This defines the matrix element up to a normalization 
factor.
In the regime of exact scale invariance the spectrum of the operator ${\cal O}_\cU$ is 
continuous, does not contain isolated particle
excitations and might be regarded as one of the reasons for the name ``unparticle''. 
The unparticle propagator carries a 
CP-even phase\footnote{The resulting CP violation 
was found to be consistent with the CPT theorem 
\cite{Zwicky:2007vv}.}
\cite{Georgi:2007si,Cheung:2007zza} for space-like 
momentum. 
Effects were found to be most unconventional 
for non-integer scaling dimension $\dU$, e.g.
\cite{Georgi:2007ek,Georgi:2007si} and \cite{Cheung:2008xu}. 

The coupling of the unparticle sector to the SM \eqref{eq:unscen}  breaks the scale invariance
of the unparticle sector at a certain energy. 
Such a possibility was first investigated 
with naive dimensional analysis (NDA)
 in reference
\cite{Fox:2007sy} 
via the Higgs-unparticle coupling of the form
\begin{equation}
\label{eq:H2O}
{\cal L}^{\rm eff}  \sim {\cal O}_\cU |H|^2 \, .
\end{equation}
The dynamical interplay of the unparticle and Higgs sector in connection with the interaction 
\eqref{eq:H2O} has been studied in 
\cite{Delgado:2007dx}.
It was found, for instance, that the
Higgs VEV  induces an unparticle VEV,
which turned out to be  infrared (IR) 
divergent for their assumed range of scaling dimension and forced the authors to introduce 
various IR regulators 
\cite{Delgado:2007dx,Delgado:2008rq}.

{In work \cite{Sannino:2008nv} the unparticle scenario was elevated to a natural extension of the SM by proposing a generic framework  in which the Higgs and the unparticle sectors are both \emph{composites} of elementary
fermions.} We used four dimensional, non-supersymmetric asymptotically free gauge theories with fermionic matter. This framework allows to address, in principle, 
the dynamics beyond the use of scale invariance per se.
 
The Higgs sector is replaced by a walking  technicolor model (TC), whereas the unparticle one
corresponds to 
a gauge theory developing a nonperturbative\footnote{
We note that the Banks-Zaks \cite{Banks:1981nn} type IR points, used to illustrate  the unparticle sector in \cite{Georgi:2007ek}, are accessible in perturbation theory. This yields
anomalous dimensions of the gauge singlet  operators which are close to the pertubative ones, resulting in very small  unparticle type effects.} 
IR fixed point (conformal phase)\footnote{Strictly speaking conformal invariance is a larger symmetry than 
scale invariance but we shall use these terms interchangeably
throughout this paper. We refer the reader to reference
\cite{Polchinski:1987dy} for an investigation of the differences.} By virtue of TC there is no hierarchy problem.
We even sketched a possible unification of the two sectors,
embedding the two gauge theories in a higher 
gauge group. The model resembles the ones
of extended technicolor and leads to 
a simple explanation of the interaction between
the Higgs and the unparticle sectors. 

\subsubsection{The Higgs \& Unparticle as Composites}
\label{sec:scenario}

\begin{figure}[h]
  \begin{center}
    \includegraphics[width=0.4\textwidth]{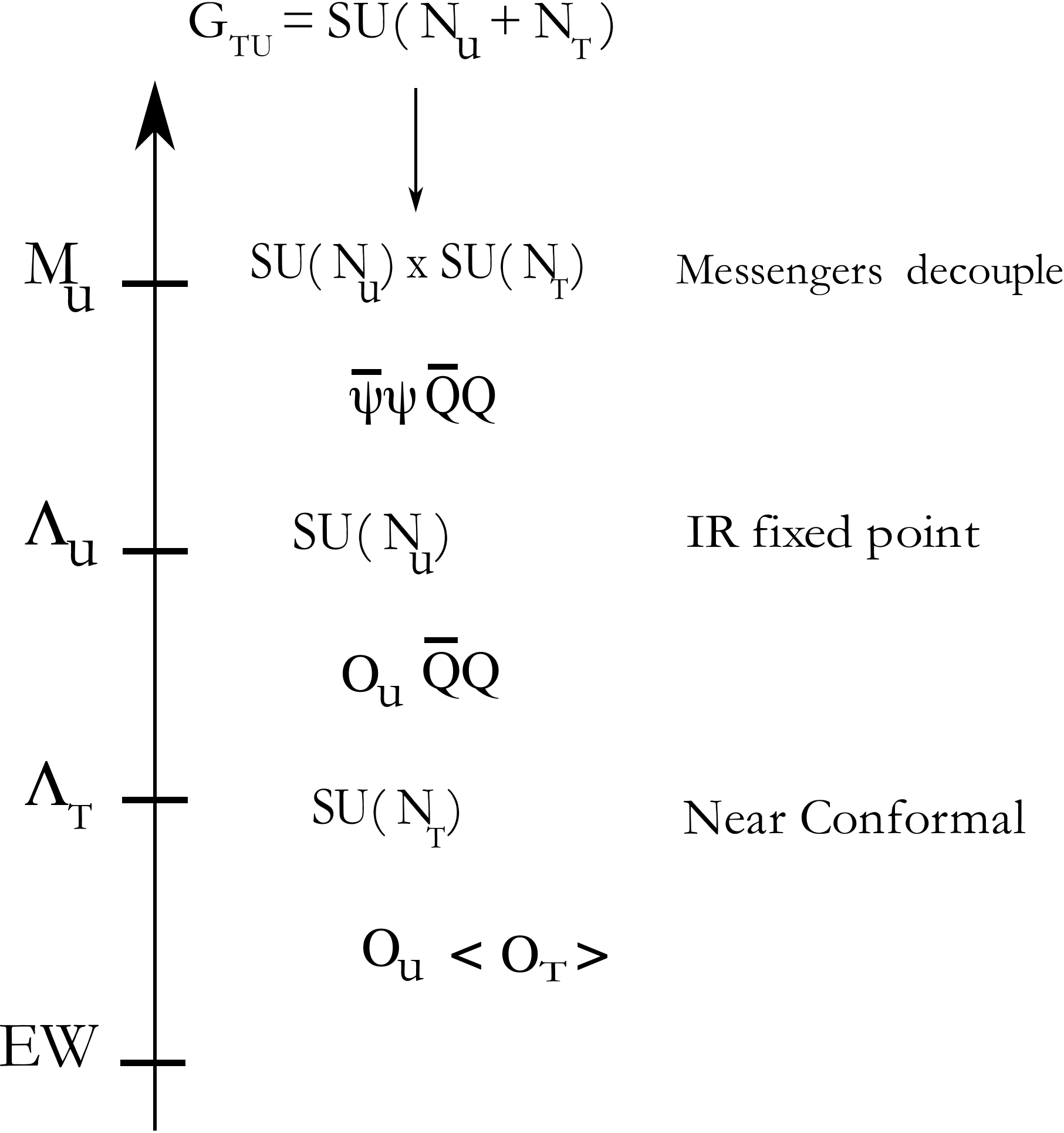}
  \end{center}
  \vspace{-14pt}
  \caption{\small Schematic scenario. The ordering of the
energy scales $\Lambda_\cU$ and $\Lambda_{\cal T}$ is not of any importance.}
\label{fig:unfig}
  \vspace{-5pt}
\end{figure}

According to \cite{Sannino:2008nv} the building block is  an extended 
$G_{T \times U} \equiv SU(N_{T})\times SU(N_{U})$ technicolor (TC)
 gauge theory.
The matter content constitutes of techniquarks  $Q^a_{f}$ charged under
the  representation $R_T$ of the TC group $SU(N_T)$ and 
Dirac techniunparticle fermions  $\Psi^A_{s}$
charged under the representation $R_U$ of the unparticle group 
$SU(N_{U})$, where  
$a/A = 1 \ldots {\rm dim}{[R_{T/U}]}$ and 
$f/s=1 \ldots F/S$ denote gauge and flavor indices
respectively. 
We will first describe the (walking) TC 
and (techni)unparticle sectors separately before
addressing  their common dynamical origin. A graphical illustration of the scenario is depicted 
in Fig.~\ref{fig:unfig} as a guidance for the reader 
throughout this section.

In the TC sector
the number of techniflavors, the matter representation and the number of colors are arranged in such a way that the dynamics is controlled by a near conformal (NC) IR fixed point. In this case the gauge coupling reaches almost a fixed point around the scale $\Lambda_{\cal T} \gg M_W$,  with $M_W$ the mass of the electroweak gauge boson.  The TC gauge coupling, at most, gently rises from this energy scale down to the electroweak one. Around the electroweak scale the TC dynamics  triggers the spontaneous breaking of the electroweak symmetry through the formation of the  technifermion condensate, which therefore has  the  quantum numbers of the SM Higgs boson. As we have explained earlier in the simplest TC models the 
technipion decay constant $F_T$ is related 
to the weak scale as $2 M_W = g  F_T$ ($g$ is the weak coupling constant) and therefore 
$F_{T} \simeq 250$~GeV. The TC scale, analogous 
to $\Lambda_{\rm QCD}$ 
for the strong force, is roughly 
$\Lambda_{TC} \sim 4 \pi F_T$.
Now we turn our attention to the unparticle sector. Here the total number of massless 
techniunparticle flavors $S$ is balanced against the
total number of colors $N_{U}$ in such a way 
that the theory, per se, is asymptotically free and 
admits a nonperturbative IR fixed point.  The energy scale around which the IR fixed point starts to set in is indicated with $\Lambda_{\cal U} \gg M_W$. 

It might be regarded as natural to assume that the unparticle and the TC sectors  have a common dynamical origin, e.g. are part of a larger gauge group at energies  above
$\Lambda_{\cal T}$  and  $\Lambda_{\cal U}$.  
Note that  the relative ordering between 
$\Lambda_{\cal T}$ and $\Lambda_{\cal U}$ 
is of no particular relevance for this scenario. The low energy relics of such a unified-type model are four-Fermi operators allowing the two 
sectors to communicate with each other at low
energy.
The unparticle sector will then be driven away from 
the fixed point due to the appearance 
of the electroweak scale in the 
TC sector.

The model
resembles models of extended technicolor (ETC)
where the techniunparticles play the role of the
SM fermions.
We refer to these type of models as 
Extended Techni-Unparticle (ETU) 
models\footnote{The work by Georgi and Kats 
\cite{Georgi:2008pq} on a two dimensional example of unparticles has triggered this work.}.  
At very high energies $E \gg M_\cU$ 
the gauge group 
$G_{T \times U}$ is thought to be 
embedded in a simple group 
$G_{\rm TU} \supset  G_{T \times U}$.
At around the scale  $\hoch$ the ETU group 
is broken to $G_{TU} \to 
G_{T \times U } $
and the heavy gauge fields receive masses 
of the order of $\hoch$ and play the role
of the messenger sector. 
Below the scale $\hoch$ the massive 
gauge fields decouple and  four-Fermi operators emerge, which corresponds to the first step
of the scenario, e.g. Eq.~\eqref{eq:unscen} 
and Fig.~\ref{fig:unfig}.
Without committing to the specific ETU dynamics
the interactions can be parametrized as: 
\begin{eqnarray}
\label{eq:ETC}
&&
{\cal L}_{< \hoch}^{\rm eff} = \alpha \frac{\bar{Q}Q \,{\bar{\Psi} \Psi}}{\Hoch{2}} + \beta \frac{{\bar{Q}Q }{\bar{Q}Q }}{\Hoch{2}} + \gamma \frac{{\bar{\Psi}\Psi }{\bar{\Psi} \Psi }}{\Hoch{2}} \,. \label{4fermi}
\end{eqnarray}
The coefficients  $\alpha$, $\beta$ and $\gamma$
(the latter should not be confused with an anomalous dimension) are  of order one, 
which can  be calculated if 
the gauge coupling $g_{\rm TU}$ is perturbative.
The Lagrangian \eqref{eq:ETC} is the relic of the
ETU(ETC) interaction and gives rise to two 
sources of dynamical chiral symmetry 
breaking in addition to the \emph{intrinsic} dynamics 
of the groups $G_{T/U}$.
These are contact interactions of the type emphasized 
in \cite{Grinstein:2008qk}.
Firstly, when one fermion pair acquires a VEV
then the $\alpha$-term turns into a tadpole and 
induces a VEV for the other fermion pair. 
{ This is what happens to the unparticle sector when the TC sector, or the SM Higgs \cite{Delgado:2007dx}, 
breaks the electroweak symmetry. 
Secondly, the $\gamma$ term 
corresponds to a Nambu$-$Jona-Lasino type 
interaction which may lead to the formation of a VEV,
for sufficiently large $\gamma$.
This mechanism 
leads to breaking of scale invariance 
even in the absence of
any other low energy scale.
Note that this mechanism
is operative in models of top condensation, 
c.f. the TC report \cite{Hill:2002ap} for an overview.
However, based on the analysis in the appendix A of \cite{Sannino:2008nv} 
we shall neglect this mechanism in the sequel of 
this paper.
We shall  refer to these two mechanism 
as $\alpha/\gamma$-induced condensates.}

At the scale $\Lambda_\cU \gg  M_W$  the unparticle gauge sector flows into an IR fixed point
and the UV operator 
${\cal O}_{\rm UV} = \bar{\Psi} \Psi$
becomes the \emph{composite unparticle}  operator $
{\cal O}_{\rm IR} \equiv {\cal O}_\cU$  with scaling 
dimension $\dU \equiv 3-\gamma_\cU$\,\footnote{The parametrization $\dU \equiv  3 -\gamma_\cU$ will be standard throughout the entire paper and in the text the scaling dimension $\dU$ and 
the anomalous dimension $\gamma_\cU$ will be used interchangeably.},
\begin{equation}
\label{eq:psis}
(\bar{\Psi} \Psi) _{UV}  \sim \Lambda_{\cal U}^{\gamma_\cU}{\cal O}_{\cU} \; ; \qquad  
M_W\ll \Lambda_{\cU} \ll  \hoch  \ .
\end{equation}
Note, the anomalous dimension $\gamma_\cU$ 
of the operator has to satisfy 
$\gamma_\cU \leq 2$ due 
to unitarity bounds of the representations of the 
conformal group \cite{Mack:1975je}. 
The Lagrangian then simply becomes
\begin{eqnarray}
\label{eq:scem}
{\cal L}^{\rm eff}_{\Lambda_\cU} = 
\alpha' \, \frac{\bar{Q}Q  \,  \Lambda_\cU^{\gamma_\cU} {\cal O}_\cU}{\Hoch{2}} + \beta' \, \frac{\bar{Q}Q \bar{Q}Q}{\Hoch{2}} + 
\gamma' \, \frac{ \Lambda_\cU^{2\gamma_\cU} \, {\cal O}_\cU {\cal O}_\cU}{\Hoch{2}} \, .
\end{eqnarray}
This realizes the second step in the scenario, 
c.f.  Fig.~\ref{fig:unfig} and
Eq.~\eqref{eq:unscen}. {The matching coefficients
$\alpha',\beta',\gamma'$ \eqref{eq:scem} 
are related to $\alpha,\beta,\gamma$ \eqref{eq:ETC} by order one coefficients.} 
The $\alpha$-term in Eq.~\eqref{eq:scem} is
similar to the unparticle-Higgs interaction in Eq.~\eqref{eq:H2O}.

The composite 
operator $\bar{Q}Q$ can be treated 
in analogy  to $\bar \Psi \Psi$ in  
\eqref{eq:psis},
\begin{equation}
\label{eq:tfcond}
(\bar{Q}Q ) _{UV}  \sim \Lambda_{\cal T}^{\gamma_{\cal T}}{\cal O}_{\cal T} \;;  \qquad  
\Lambda_{TC}\ll\Lambda_{\cal T} \ll M_{\cal U} \, ,
\end{equation}
up to  logarithmic corrections which are negligible.
Contrary to the unparticle sector the TC gauge dynamics break scale invariance through the formation of an \emph{intrinsic condensate}
\begin{equation}
\label{eq:TCcond}
\vev{ {\cal O}_{\cal T}}_{\Lambda_{\cal T}}  \simeq
 \Cw{\gamma_{\cal T}}  \Lambda_{TC}^{d_{\cal T}} 
\equiv   \Cw{\gamma_{\cal T}} \Lambda_{TC}^{3 - \gamma_{\cal T}} 
\, , \quad  \cw \equiv \Big(\frac{\Lambda_{\cal T }}{\Lambda_{ TC}} \Big) \ .
\end{equation}
The estimate of the VEV is based on
scaling from QCD and renormalization group evolution.

The relevant terms contained in the low energy effective theory around the electroweak scale 
are\footnote{
Note that in QCD-like TC models 
(the gauge coupling displays a running behavior rather than a walking one) 
one would set  $\gamma_{\cal T} \simeq 0$ in  Eqs.~\eqref{eq:TCcond} and \eqref{eq:OUint}.}:
\begin{equation}
\label{eq:OUint}
{\cal L}_{\Lambda_{\cal T}}^{\rm eff}  =  \alpha' \,  \Cw{\gamma_{\cal T}} \,
\frac{\Lambda_{TC}^{3}  \, \Lambda_\cU^{\gamma_\cU} {\cal O}_\cU}{\Hoch{2}} +  \gamma' \, 
\frac{\Lambda_\cU^{2\gamma_\cU} {\cal O}_\cU 
{\cal O}_\cU}{\Hoch{2}}  + ..   
\end{equation}
This step involves another matching procedure but
we shall not introduce further notation here
and denote the matching coefficients by simple
primes only.
As stated previously the TC condensate 
drives the TC gauge sector away from the
fixed point and the coupling increases towards the IR. The 
sector is then replaced by a low energy effective 
chiral Lagrangian featuring the relevant composite degrees of freedom \cite{Foadi:2007ue,Hill:2002ap}. The details on how the unparticle operator acquires a vev can be found in \cite{Sannino:2008nv} together with the suggestion of a UV model. Here we will simply summarize a schematic ETU model and its low energy effective description which can be useful for phenomenology.

\subsubsection{A Schematic ETU Model and its Low Energy Effective Theory}
\label{sec:ETU}

We imagine that at an energy much higher than the electroweak scale the theory is described by a 
 gauge theory
 \begin{equation}
 {\cal L}^{\rm UV} = - \frac{1}{2} {\rm Tr}\left[{\cal F}_{\mu\nu}{\cal F}^{\mu\nu} \right] +  \sum_{{\cal F} = 1}^{F} \bar \xi_{\cal F} (i \s{\partial} + g_{TU} \s{\cal A} )\xi_{\cal F} + .. \,
 \label{UV}
 \end{equation}
where ${\cal A}$ is the gauge field of the  
$SU(N_{T}  +  N_U)$ group and gauge indices
are suppressed.  
$({\xi_{\cal F}^{{\cal A}}})^T = (Q^1 ...  Q^{N_{T}}, \Psi^1 ... \Psi^{N_U} )_{\cal F}$ is the fermion field unifying the technifermion and TC matter content.  The dots in (\ref{UV}) stand for the $SU(3)\times SU(2)_L \times U(1)_Y$ gauge fields and 
their interactions to the SM fermions and technifermions. There is no elementary Higgs field in this formulation.  Unification of the TC and techniunparticle dynamics constrains the flavor symmetry of the two sectors to be identical at high energies. The matter content and the number of technifermions (TC + techniunparticles) is chosen, within the phase diagram in \cite{Sannino:2008ha}, such that the theory is asymptotically free at high energies. The non-abelian global flavor 
symmetry is  $SU_L(F)\times SU_R(F)$. 

At an intermediate scale $M_{\cal U}$, much higher than the scale where the unparticle and TC subgroup become strongly coupled, the dynamics is such that $SU(N_{T} \!+\! N_U)$ breaks to $SU(N_{T})\times SU(N_U)$. Only two flavors { (i.e. one electroweak doublet)} are gauged under the electroweak group. The global symmetry group breaks explicitly to $G_F = SU_L(2)\times SU_R(2) \times SU_L(F-2)\times SU_R(F-2)$. At this energy scale the weak interactions are, however, negligible and we can safely ignore it. 

At the scale $M_{\cal U}$ there are the $Q^i_c$ fermions  - with $i=1,\ldots, F$ and $c=1,\dots, N_T $ - as well as the $\Psi^i_u$ ones - with $i=1,\ldots, F$ and $u=1,\dots, N_U $. Assigning the indices
$i=1,2$ to the fermions gauged under the electroweak group we observe that not only the TC fermions are gauged under the electroweak but also the technunparticles. { To ensure that the 
unparticle sector is experimentally not too visible we have to 
assume a mechanism 
that provides a large mass to the charged techniunparticle fermions.} In reality this is quite a difficult task, since we do not want to break the SM weak symmetry explicitly\footnote{One could for instance 
unify the flavor symmetry of the unparticles with the technicolor gauge group into an
ETC group.
This would also produce a Lagrangian of the type \eqref{4fermi}.
The TC fermions would 
be charged under the electroweak group separately.}.
Our treatment below, however, is sufficiently general to be straightforwardly adapted to various model constructions.

As already stated in the first section, the number of
flavors and colors for the 
TC and unparticle gauge groups 
$SU(N_T)$  and $SU(N_U)$
have to be arranged such that the former is 
NC and the latter is conformal. This enforces the 
conditions:
\begin{equation}
F \leq F^{\ast}_{N_T},  \qquad  F^{\ast}_{N_U}  \leq F  -2 \ .
\label{ftc}
\end{equation}
$F^{\ast}_{N}$ denotes the critical number of flavors, for a given number of colors $N$, 
above which the theory develops an IR fixed point. Recall that two unparticle
flavors are decoupled and hence $F \to F-2$ in the
second inequality in \eqref{ftc}. The use of the phase diagram we will describe in the following chapter will be relevant to determine $F^{\ast}$.

{ Below the scale $M_{\cal U}$ all 
four-Fermi interactions have to respect
the  flavor symmetry $G_F$.} The most general four-Fermi operators  have been classified in \cite{Appelquist:1984rr} and the coefficient of the various operators depend on the specific model used to break the unified gauge theory. Upon Fierz rearrangement, the operators of greatest phenomenological relevance are, 
\begin{equation}
{\cal L}^{\rm eff} = \Big( \frac{G}{2} \bar{\Psi}_L \Sigma \Psi_R  +{\rm h.c.}\Big)+    \frac{G^{\prime}}{2M_{\cal U}^2} (\bar{\Psi}_L \Psi_R )(\bar{\Psi}_R \Psi_L) + \dots  \ ,
\end{equation}
the scalar-scalar interactions of Eq.~\eqref{4fermi}.
Here $\Sigma$ is the quark bilinear,
\begin{equation}
\Sigma_i^j  \sim ( {{Q}_L}_i {\bar{Q}_R^j})_{ UV}  \ , \qquad i=1,\ldots , F\ .
\end{equation} 
The flavor indices are contracted and the sum starts from the index value $3$;
the first two indices correspond to the $\Psi$'s charged under the electroweak force, which are decoupled at low energy. 
The fermion bilinear becomes the unparticle operator
\eqref{eq:tfcond},
\begin{equation}
\big( {\cal O_{\cal U}}\big)_i^{\;j}  =  \frac{{{\Psi}_L}_i \bar{\Psi}_R^j}{\Lambda_{\cal U}^{\gamma_{\cal U}}} \ . 
\end{equation}
The matrix $\Sigma$ at energies near the electroweak symmetry breaking scale is identified with the interpolating field for the mesonic composite operators.

{ To investigate  the coupling to the composite Higgs  we write down the low energy effective theory using linear realizations.  We parameterize 
the complex $F \times F$ matrix $\Sigma$ by
 \begin{equation}
 \Sigma = \frac{\sigma + i\,\Theta}{\sqrt{F}} + \sqrt{2}(i \Pi^a + \widetilde{\Pi}^a) {T^a} \ , 
 \end{equation}
where ($\sigma$,$\tilde \Pi$) and
($\Theta$,$\Pi$) have 
$0^{++}$ and $0^{-+}$ quantum 
numbers respectively.
The Lagrangian is given by
\begin{eqnarray}
\label{eq:sigmaeff}
{\cal L}^{\rm eff}  & = &   \frac{1}{2}{\rm Tr} \left[ (D\Sigma)^{\dagger} D\Sigma\right]  - k_1( {\rm \widehat{Tr}}\left[ \Sigma^{\dagger}\cal O_{\cal U}\right]+ {\rm h.c.}) - k_2 {\rm \widehat{Tr}}\left[ \cal O_{\cal U}  {\cal O_{\cal U}}^{\dagger} \right] \nonumber \\ &&  -  m^2_{ETC}\sum_{a=4}^{F^2-1}\frac{\Pi^a\Pi^a}{2}  - V(\Sigma, \Sigma^{\dagger}) \ ,
\end{eqnarray}
where
\begin{eqnarray}
D\Sigma = \partial \Sigma - ig W\Sigma + ig^{\prime}\Sigma BT^3_R \ ,  \quad {\rm and} \quad  W=W^a T^a_L  \ , 
\end{eqnarray}
and $ {\rm Tr}[T^a_{L/R} T^b_{L/R}] = \delta ^{ab}/2$.
The coefficients $k_1$ and $k_2$ are directly proportional to the  $\alpha^{\prime}$ and $\gamma^{\prime}$ coefficients in (\ref{eq:scem}).
The $hat$ on some of the traces indicates that the summation is only on the flavor indices from $3$ to $F$. 
Three of the Goldstone bosons play the role of the longitudinal gauge bosons and the remaining 
ones receive a mass $m^2_{ETC}$
from an ETC mechanism.
We refer the reader to reference \cite{Hill:2002ap} for discussion of different ETC models with 
mechanisms for sufficiently large mass generation.
The first term in the Lagrangian is responsible for the mass of the weak gauge bosons and 
the kinetic term for the remaining Goldstone bosons. 
The VEV's for the flavor-diagonal part of the unparticle operator, reduces to the computation performed in the previous section. 
The potential term preserves the global flavor symmetry $G_F$.
Up to dimension four, including the determinant responsible for the $\eta'$ mass in QCD, the terms  respecting the global symmetries of the TC theory are:
\begin{equation}
V(\Sigma, \Sigma^{\dagger}) = -\frac{m^2}{2} {\rm Tr} \left[ \Sigma^{\dagger}\Sigma\right] + \frac{ \lambda_1}{F} {\rm Tr} \left[ \Sigma^{\dagger} \Sigma\right]^2 + {\lambda_2}
 {\rm Tr}[( \Sigma^{\dagger}\Sigma) ^2 ] - \lambda_3( \, {\rm det}\Sigma +
  {\rm det}\Sigma^\dagger ) \ .\end{equation} 
The coefficient $m^2$ is positive to ensure chiral symmetry breaking in the TC sector. The Higgs VEV enters as follows,
\begin{equation}
\sigma = v + h \ , \qquad  {\rm with} \qquad  F_T = \sqrt{\frac{2}{F}} \, v \simeq 250 \, {\rm GeV} \ .
\end{equation}   
$F$ here is the number of flavors and $h$ the composite field with the same quantum numbers as the SM Higgs. The particles $\sigma,\Theta, \tilde \Pi$ 
all have masses of the order of $v$. The Higgs mass, the Higgs VEV and the $\Theta$ mass, for instance, are} 
\begin{eqnarray}
v^2 = \frac{m^2}{(\lambda_1 + \lambda_2 - \lambda_3)} \,,\quad 
m_h^2 =  2 m^2 \,, \qquad m_\Theta^2 = 4 v^2 \lambda_3^2
\, ,
\end{eqnarray} 
up to corrections of the order of $O(\Lambda_{TC}^2/\Hoch{2})$
due to contributions from $\alpha$-terms.

The lightest pesudoscalars of the unparticle sector
are the pseudo  Goldstone bosons
emerging from the explicit breaking of the global flavor 
symmetry in the unparticle sector. Their mass
can be read off from the linear term 
in $O_\cU$ of the effective Largrangian
\eqref{eq:sigmaeff} 
\begin{equation}
m_{ \Pi_\cU }^2 \simeq \Lambda_{TC}^2 
\left(  \frac{\Lambda_{TC}}{\Hoch{2}} \right)^2 
\left(  \frac{\Lambda_\cU}{\Lambda_{TC}} \right)^{\gamma_\cU} 
\left(  \frac{\Lambda_{\cal T}}{\Lambda_{TC}} \right)^{\gamma_{\cal T}}  \, .
\end{equation}

The unparticle propagator 
to be used for phenomenology, defined from the  
the K\"all\'en-Lehmann representation is 
\begin{equation}
\label{eq:prop2}
\Delta_\cU(q^2,\Lambda_{\rm UV}^2,\Lambda_{\rm IR}^2) = 
- \frac{B_\dU}{2 \pi} \int_{\Lambda_{\rm IR}^2}^{\Lambda_{\rm UV}^2} 
\frac{ds\, s^{\dU-2}}{s - q^2- i0}  +   s.t.    \; .
\end{equation}
The explicit form for different values of the anomalous dimension can be found in \cite{Sannino:2008nv}. We shall now turn to the question of the mixing of the 
unparticle and the Higgs \footnote{The findings in \cite{Sannino:2008nv} resemble results from extra dimensional models such as the model called HEIDI  \cite{vanderBij:2006pg}, where the continuous spectrum is mimicked by an infinite tower of narrowly spaced Kaluza-Klein modes. The difference is that this model is inherently four dimensional and that the parameters, such as the IR cut off and the strength of the unparticle-Higgs coupling, are related to each other.
This model is also different from the one in reference \cite{Delgado:2007dx} since, although both are in four dimensions, the Higgs and unparticle coupling emerges dynamically within a UV complete theory. }

The interaction term from Eq.~\eqref{eq:sigmaeff}
\begin{equation}
\label{eq:Lmix}
 {\cal L}^{\rm mix} = - g_{h {\cal O}_\cU} \, h {\cal O}_
 \cU    \,, \qquad 
g_{h {\cal O}_\cU} = \frac{k_1(F-2) }{\sqrt{F}} \,,
\end{equation}
introduces a mixing between the Higgs and the unparticle. The constant $k_1$ has mass dimension 
$\gamma_\cU$. The Higgs propagator is obtained from inverting
the combined Higgs-unparticle system
\begin{equation}
\label{eq:higgsprop}
\Delta_{hh}(q^2) = \frac{1}{q^2-m_h^2 - g_{h{\cal O}_\cU}^2 
\Delta_\cU(q^2,\Lambda_{\rm UV}^2,\Lambda_{\rm IR}^2)} \, .
\end{equation}
This, of course, results in unparticle corrections controlled by $g_{h{\cal O}_\cU}$. 
The propagator
can be rewritten in terms of a dispersion representation 
\begin{equation}
\label{eq:higgsdisperse}
\Delta_{hh}(q^2) = - \int \frac{ds\,\rho_{hh}(s)}{s - q^2- i0}  \,,
\end{equation}
where the density, 
\begin{equation}
\label{eq:norm}
\int ds \rho_{hh}(s) = 1 \, ,
\end{equation}
is automatically normalized to unity.
The non zero value of the coupling $g_{h {\cal O}_\cU}$  results solely 
in a change of basis (or poles and cuts) 
of the intermediate particles but does not change
the overall density of states. To a large extent the spectral function is
characterized by the zeros of the pole
equation and the onset of the continuum 
relative to the poles. This will depend on 
the strength of the mixing and the anomalous dimension. Somewhat exotic effects can be obtained
when the mixing term is made very large 
\cite{Delgado:2008px,vanderBij:2007um}. In the present model the mixing is determined by
$k_1$ \eqref{eq:Lmix}: 
\begin{equation}
\label{eq:k1}
k_1 \sim \alpha' \Lambda_{TC}^{\gamma_\cU}  \left(\frac{\Lambda_{TC}}{\hoch}\right)^2  \left(  \frac{\Lambda_\cU}{\Lambda_{TC}} \right)^{\gamma_\cU} \Cw{\gamma_{\cal T}} \, , \qquad 
\Cw{\phantom{x}} = \left( \frac{\Lambda_{\cal T}}{\Lambda_{TC}}   \right)
\end{equation}
which we have normalized to the TC 
scale.

The value of $k_1$ is, of course, suppressed by
the large scale $\hoch$ per se, 
but receives enhancements
from the powers of the anomalous dimensions.
For the maximal allowed anomalous dimensions 
 $\gamma_\cU \simeq \gamma_{\cal T} \simeq 2$ and a hierarchy 
of scale envisioned earlier one finds
$k_1 \Lambda_{TC}^{-\gamma_\cU}  \simeq \alpha' \cdot  O(10^{-2})$. We therefore  
expect the coupling $g_{h {\cal O}_\cU}\Lambda_{TC}^{-\gamma_\cU
} $ \eqref{eq:Lmix} to 
be considerably smaller than one.  In this case there is generally a unique solution 
to the pole equation \cite{Sannino:2008nv}. At the qualitative level 
it is an interesting question of whether the Higgs resonance is below or above the threshold \cite{vanderBij:2006pg,Delgado:2007dx}.

Here we introduced a framework  in which the Higgs and the unparticle are both composite. The underlying theories are four dimensional, asymptotically free, nonsupersymmetric gauge theories with fermionic matter.  We sketched a
possible unification of these two sectors at a scale much higher than the electroweak scale. The resulting model resembles
extended technicolor models and we termed it extended technicolor unparticle (ETU). The coupling of the unparticle sector to the SM emerges in a simple way and assumes the form of  four-Fermi interactions  below
$\hoch$. 
 
In the model  the unparticle sector
is coupled  to the composite Higgs. Another possibility
is to assume that the Higgs sector itself is unparticle-like, with a continuous mass distribution.
This UnHiggs \cite{Stancato:2008mp,Calmet:2008xe}
could find a natural setting within
walking technicolor, which is part of the present framework.
Of course it is also possible to think of an unparticle 
scenario that is not coupled to the electroweak sector,
where  scale invariance is broken at a (much) lower scale.
This could result in  interesting effects on low energy
physics as extensively studied in the literature.

With respect to this model, in the future, one can:

\begin{itemize}

\item{Study the composite Higgs production  in association with a SM gauge boson, 
both for proton-proton (LHC) and proton-antiproton 
(Tevatron) collisions via the low energy effective theory
\eqref{eq:sigmaeff}.  
In references 
\cite{Belyaev:2008yj,Zerwekh:2005wh} it has been demonstrated that such a process is enhanced with respect to the SM, due to the presence of a light composite (techni)axial resonance\footnote{
A similar analysis within an  extradimensional set up has been performed in \cite{Agashe:2008jb}.}. 
The mixing of the light composite Higgs with the unparticle sector modifies these processes in a way that can be explored at colliders.
Concretely, the transverse missing energy spectrum can be used to disentangle the unparticle sector from the TC contribution per se.}

\item{Use first principle lattice simulations 
to gain insight on the nonperturbative (near) conformal dynamics. 
It is clear from our analysis that this knowledge is crucial for describing and understanding unparticle dynamics.
As a model example in \cite{Sannino:2008nv} was considered partially gauge technicolor introduced  in \cite{Dietrich:2005jn}. These gauge theories are being studied on the lattice \cite{Appelquist:2007hu,Deuzeman:2008sc,Fodor:2008hn}.
Once the presence of a fixed point is established,
for example via lattice simulations \cite{Shamir:2008pb,Svetitsky:2008bw,DeGrand:2008dh,Fodor:2008hm}, 
 the anomalous dimension of the 
fermion mass can be determined from the
conjectured all order beta function 
\cite{Sannino:2008ha,Ryttov:2007cx}, or deduced directly via lattice simulations (see \cite{Bursa:2009we,DeGrand:2009hu,Lucini:2009an} for recent attempts) .
Moreover, on the lattice one should be able to  
directly  investigate the two-point function, i.e. the unparticle propagator.}

\item{Investigate different models at the ETU level.  For example one could adapt some models,
introduced to generate masses to the SM fermions,
in  \cite{Christensen:2006rf,Christensen:2005bt,Appelquist:2004df,Appelquist:2004ai,Appelquist:2003hn,Appelquist:2003uu,Gudnason:2006mk} to  
improve on the present ETU model.}

\item{Study possible cosmological consequences of our framework. The lightest baryon of the unparticle
gauge theory, the {\it Unbaryon}, is naturally stable (due to a  protected $U(1)$ unbaryon number) 
and therefore it is a possible dark matter candidate.
Due to the fact that we expect a closely spaced spectrum of Unbaryons and 
unparticle vector mesons, 
it shares properties in common with secluded models of dark matter \cite{Pospelov:2007mp} or previously discussed unparticle dark matter models \cite{Kikuchi:2007az}.}
\end{itemize}

Within the present framework unparticle physics emerges as a natural extension of dynamical models
of electroweak symmetry breaking.  As seen above the link opens the doors to yet unexplored collider 
phenomenology and possible new avenues 
for dark matter, such as the use of the 
{\it Unbaryon}.

\newpage
\section{Techni - Dark Cosmology: The TIMP}

The Wilkinson Microwave Anisotropy Probe (WMAP) is a NASA Explorer 
mission. By detecting the first full-sky map of the microwave sky -- and 
thus of the cosmic microwave background (CMB) -- with a resolution of under 
a degree (about the angular size of the moon), WMAP gave a wealth of 
cosmological information. 
It has produced a convincing consensus on the contents of the universe. WMAP has also determined the age, the epochs 
of the key transitions, and the geometry of the universe, while providing 
the most stringent data yet on events in the first fraction of a second 
of the universe's existence.

\begin{figure}[h] 
\center{
\includegraphics[width=10cm]{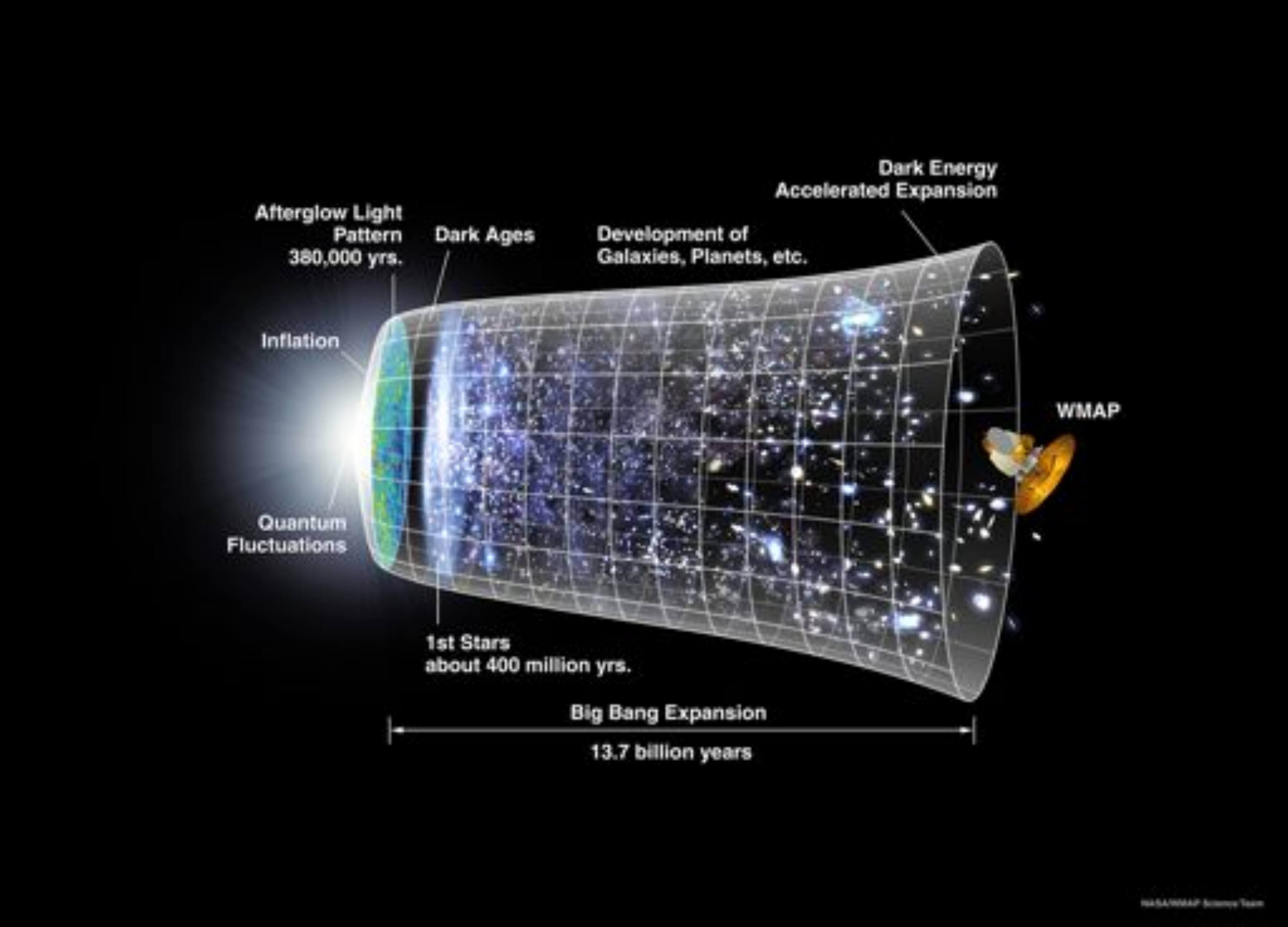}
\caption{Evolution of the universe}}
\end{figure}

More specifically, recent WMAP data \cite{Spergel:2003cb} combined with independent observations strongly indicate that  the  universe is flat and it is predominantly made of unknown forms of matter. Defining with $\Omega$ the ratio of the density to the critical density, observations indicate that the fraction of matter amounts to  $\Omega_{\rm matter} \sim 0.3$ of which the normal baryonic one is only $\Omega_{\rm baryonic} \sim 0.044$. The amount of non-baryonic matter is termed dark matter. The total $\Omega$  in the universe is dominated by dark matter and pure energy (dark energy) with the latter giving a contribution $\Omega_{\Lambda} \sim 0.7$ (see for example \cite{Frieman:2008sn,Frieman:2002wi}). Most of the  dark matter is ``cold''  (i.e. non-relativistic at freeze-out) and significant fractions of hot dark matter are hardly compatible with data. What constitutes dark matter is a question relevant for particle physics and cosmology. There is a fair chance that LHC could help to solve this puzzle by providing direct or indirect evidence for such a new type of matter.  A WIMP (Weakly Interacting Massive Particles) can be the dominant part of the non-baryonic dark matter contribution to the total $\Omega$. Axions can also be dark matter candidates but only if their mass and couplings are tuned for this purpose. If the dark matter candidate is discovered at the LHC it would be of tremendous help to cosmology since many of its properties can then be studied directly in laboratory. 

 The future Planck ESA mission will be the third generation of CMB space missions 
following the cosmic background explorer (COBE) satellite and WMAP. The primary goal of Planck is the production 
of high-sensitivity (one part per million), high-angular resolution 
(10 arcminutes) maps of the microwave sky. 
The mission goals include: 
\begin{itemize}
\item{ the precise determination of the primordial fluctuation spectrum, 
providing the information necessary for the theory of large scale 
structure formation; }
\item{ the detection of primordial gravitons/gravity waves, allowing to test 
the relationships expected from inflation; }
\item{to uncover the statistics underlying the CMB anisotropies, for instance 
by looking at non-gaussianities in the bispectrum (inflation 
generally predicts a gaussian parent distribution while topological defects 
predict deviation from gaussian distributions and rare peak fluctuations)}.
\end{itemize}

It would be theoretically very pleasing to be able to relate
the DM and the baryon energy densities in order to  explain the ratio
$\Omega_{\rm DM}/\Omega_B\sim 5$~\cite{Komatsu:2008hk}. We know that the amount of baryons in the Universe $\Omega_B\sim 0.04$ is
determined solely by the cosmic baryon asymmetry $n_B/n_\gamma \sim 6\times
10^{-10}$.  This is so since the baryon-antibaryon annihilation cross section
is so large, that virtually all antibaryons annihilate away, and only the
contribution proportional to the asymmetry remains.  
This asymmetry can be
dynamically generated after inflation.  We do not know, however, if the DM density is determined by thermal freeze-out,
by an asymmetry, or by something else.
Thermal freeze-out needs a $\sigma v \approx 3~10^{-26}\cm^3/\sec$ which is of the electroweak size,
suggesting a DM mass in the TeV range. If $\Omega_{\rm DM}$ is determined by thermal freeze-out, its proximity to $\Omega_B$
is just a fortuitous coincidence and is left unexplained.

If instead $\Omega_{\rm DM}\sim \Omega_B$ is not accidental, then the theoretical
challenge is to define a consistent scenario in which the two energy densities
are related.  Since $\Omega_B$ is a result of an asymmetry, then relating the
amount of DM to the amount of baryon matter can very well imply that
$\Omega_{\rm DM} $ is related to the same asymmetry that determines $\Omega_B$.
Such a condition is straightforwardly realized if the asymmetry for the DM
particles is fed in by the non-perturbative electroweak sphaleron transitions,
that at temperatures much larger than the temperature $T_*$ of the electroweak
phase transition (EWPT) equilibrates the baryon, lepton and DM
asymmetries.  Implementing this condition implies the following requirements: 
\begin{itemize}
\item[1.] DM must be (or must be a composite state of) a fermion, 
chiral (and thereby non-singlet) under the weak $\SU(2)_{L}$, 
and carrying an anomalous (quasi)-conserved quantum number $B'$.

\item[2.] DM (or its constituents) must have an annihilation cross section much larger
  than electroweak $\sigma_{\rm ann} \gg3~10^{-26}\cm^3/\sec$, to ensure that
  $\Omega_{\rm DM}$ is determined dominantly by the  $B'$ asymmetry.
\end{itemize}

The first condition ensures that a global quantum number corresponding to a
linear combination of $B$, $L$ and $B'$ has a weak anomaly, and thus DM
carrying $B'$ charge is produced in anomalous processes together with
left-handed quarks and leptons \cite{Kaplan:1991ah,Barr:1990ca}.  At
temperatures $T\gg T_*$ electroweak anomalous processes are in thermal
equilibrium, and equilibrate the various asymmetries $Y_{\Delta B}=c_L
Y_{\Delta L}=c_{B'}Y_{\Delta B'}\sim {\cal O}(10^{-10})$. Here the
$Y_\Delta$'s represent the difference in particle number densities $n-\bar n$
normalized to the entropy density $s$, e.g.  $Y_{\Delta B} = (n_B-\bar
n_B)/s$. These are convenient quantities since they are conserved during the
Universe thermal evolution.  

At $T\gg M_{\rm DM}$ all  particle masses can be neglected,
and $c_L$ and $c_{B'}$ are order one coefficients,  determined via chemical equilibrium conditions enforced 
by elementary reactions faster
than the Universe expansion rate~\cite{Harvey:1990qw}. These coefficients can be computed
in terms of the particle
content,  finding e.g. $c_L=-28/51$ in the SM and $c_L=-8/15$ in the MSSM.  

At $T\ll M_{\rm DM}$, the $B'$ asymmetry gets suppressed by a Boltzmann
exponential factor $e^{-M_{\rm DM}/T}$.  A key feature of sphaleron
transitions is that their rate gets suddenly suppressed at some temperature
$T_*$ slightly below the critical temperature at which $\SU(2)_L$ starts to be
spontaneously broken.  Thereby, if $M_{\rm DM}<T_*$ the $B'$ asymmetry gets
frozen at a value of ${\cal O}(Y_{\Delta B})$, while  if instead $M_{\rm DM} >
T_*$ it gets exponentially suppressed as $Y_{\Delta B'}/Y_{\Delta
  B}\sim e^{-M_{\rm DM}/T}$.

\medskip

More in detail, the sphaleron processes relate the asymmetries of the various
fermionic species with chiral electroweak interactions as follows. If
$B^{\prime}$, $B$ and $L$ are the only quantum numbers involved then the
relation is:
\begin{equation} 
\frac{Y_{\Delta B^{\prime}}}{Y_{\Delta B} } =c\cdot  {\cal S}\left(\frac{M_{\rm DM}}{T_*}\right),\qquad
c=\bar{c}_{B^{\prime}}+ \bar{c}_L \frac{Y_{\Delta L}}{Y_{\Delta B} } \ ,
\end{equation}
where the order-one $\bar{c}_{L,B^{\prime}}$ coefficients are related to the
$c_{L,B^{\prime}}$ above in a simple way.  The explicit numerical values of
these coefficients depend also on the order of the finite temperature
electroweak phase transition via the imposition or not of the weak isospin
charge neutrality. In \cite{Ryttov:2008xe,Gudnason:2006yj} the dependence on
the order of the electroweak phase transition was studied in two explicit
models, and it was found that in all cases the coefficients remain of order
one. The statistical function ${\cal S}$ is:
\begin{eqnarray}
{\cal S} (z) = \left\{ \begin{array}{rl}
\frac{6}{4\pi^2} \int_{0}^{\infty} dx\ x^2\cosh^{-2} \left( \frac{1}{2} \sqrt{x^2 + z^2 } \right) &\qquad  {\rm for~fermions} \ , \\
\frac{6}{4\pi^2} \int_{0}^{\infty} dx\ x^2\sinh^{-2} \left( \frac{1}{2} \sqrt{x^2 + z^2 } \right) &\qquad {\rm for~bosons} \ .
\end{array} \right.
\end{eqnarray}
with $S(0) = 1 (2)$ for bosons (fermions) and $S(z) \simeq 12~(z/2\pi)^{3/2} e^{-z}$ at $z\gg 1$.
We assumed the SM fields to be relativistic and checked that this is a good approximation even for the top quark \cite{Gudnason:2006yj,Ryttov:2008xe}. 
The statistic function leads to the two limiting results:
\begin{equation}
\frac{Y_{\Delta B'}}{Y_{\Delta B}} =c\times  \left\{ 
 \begin{matrix}
{\cal S}(0)\mbox{\hspace{2.3cm}}  &  \qquad  \hbox{for }M_{\rm DM}\ll  T_* \cr 
12\left({M_{\rm DM}}/{2\pi T_*}\right)^{3/2}\,e^{-M_{\rm DM}/T_*} & \qquad \hbox{for }M_{\rm DM}\gg T_* 
\end{matrix}
 \right. \quad.
\end{equation}
Under the assumption that all antiparticles carrying $B$ and $B'$ charges are
annihilated away we have $Y_{\Delta B'}/Y_{\Delta B}=n_{B'}/n_B$. The observed
DM density
\begin{equation}
\frac{\Omega_{\rm DM}}{\Omega_B}=\frac{M_{\rm DM}\,n_{B'}}{m_{p}\,n_B}\approx 5
\end{equation}
(where $m_p\approx 1\,$GeV) can be reproduced for two possible 
values of the DM mass:
\begin{itemize}
\item[i)]  $M_{\rm DM} \sim 5\,$GeV if $M_{\rm DM}\ll T_*$, times model dependent order one coefficients.
\item[ii)] $M_{\rm DM}\approx  8\, T_*\approx 2\TeV$ if $M_{\rm DM}\gg T_*$,
with a mild dependence on the model-dependent order unity coefficients.
\end{itemize}
The first solution is well known~\cite{Kaplan:1991ah} and not interesting for our purposes.
The second solution 
matches the DM mass suggested by ATIC,
in view of $T_* \sim v$~\cite{Barr:1990ca}, where $v\simeq 250\,$GeV is
the value of the electroweak breaking order parameter\footnote{More precisely,
for a Higgs  mass $m_h=120\, (300)\,$ GeV,
  Ref.~\cite{Burnier:2005hp} estimates $T_*\approx  130\, (200)\,$GeV, where the    
larger $T_*$ values  arise because of the larger values Higgs self coupling. 
For the large masses that are typical of
composite Higgs models, the self coupling is in principle 
calculable and generally large~\cite{Gudnason:2006yj}, so that 
taking $T_* \sim v$ is not unreasonable.}.

\subsection{Introducing the TIMP}

An ideal DM candidate, in agreement with PAMELA anomalies \cite{Nardi:2008ix}, and
compatible with direct DM searches,  is a TeV particle that decays 
dominantly into leptons, and that has a negligible coupling to the $Z$.

If DM is an elementary particle, this scenario needs DM to be a chiral fermion
with $\SU(2)_L$ interactions, which is very problematic.  Bounds from direct
detection are violated.  Furthermore, a Yukawa coupling $\lambda$ of DM to the
Higgs gives the desired DM mass $M_{\rm DM}\sim \lambda v\sim 2\TeV$ if
$\lambda\sim 4\pi$ is non-perturbative, hinting to a dynamically
generated mass associated to some new strongly interacting dynamics
\cite{Nussinov:1985xr,Barr:1990ca, Ryttov:2008xe,Gudnason:2006yj}.  This
assumption also solves the problem with direct detection bounds, which are
satisfied if DM is a composite $\SU(2)_L$-singlet state, made of elementary
fermions charged under $\SU(2)_L$.  

This can be realized by introducing a strongly-interacting ad-hoc `hidden'
gauge group.  A more interesting identification comes from Technicolor.  In such a
scenario, DM would be the lightest (quasi)-stable composite state carrying a
$B'$ charge of a theory of dynamical electroweak breaking featuring a spectrum
of technibaryons $(B')$ and technipions ($\Pi$). The TIMP (Technicolor Interacting Massive Particle) can have a number of phenomenologically interesting properties.  

\smallskip

\begin{itemize}
\item[i)]{
A traditional TIMP mass can be approximated by $m_{B^{\prime}} = M_{\rm DM}\approx
n_Q \Lambda_{\rm TC}$ where $n_Q$ is the number of techniquarks $Q$ bounded into
$B'$ and $\Lambda_{\rm TC}$ is the constituent mass, so that $M_{\rm DM}/m_p \approx
n_Q\Lambda_{\rm TC}/3\Lambda_{\rm QCD}$.  Denoting by $f_\pi$ ($F_\Pi$) the
(techni)pion decay constant, we have $ F_\Pi/f_\pi = \sqrt{D/3}\>
\Lambda_{\rm TC}/\Lambda_{\rm QCD}$ where $D_Q$ is the dimension of the constituent
fermions representation ($D=3$ in QCD)\footnote{The large-$N$ counting
  relevant for a generic extension of technicolor type can be found in
  Appendix F of~\cite{Sannino:2008ha} .}. Finally, the
electroweak breaking order parameter is obtained as $v^2=N_D F^2_{\Pi}$, from the sum of the
contribution of the $N_D$ electroweak techni-doublets.
Putting all together yields the estimate:
\beq
M_{\rm DM} \approx  \frac{n_Q}{\sqrt{3 D_Q N_D}} \frac{v}{f_\pi}m_p
= 2.2\TeV
\eeq
where the numerical value corresponds to the smallest number of constituents
and of techniquarks $n_Q=D_Q=2$ and $N_D=1$. }
\item[ii)]{A generic dynamical origin of the breaking
of the electroweak symmetry can lead to several natural interesting DM
candidates (see \cite{Sannino:2008ha} for a list of relevant references). 
A very interesting case is the one in which the TIMP is a pseudo-Goldstone boson \cite{Ryttov:2008xe,Gudnason:2006yj}. In this case one can observe these states also 
at colliders  experiments \cite{Foadi:2008qv}. } 

\end{itemize}

It is worth mentioning that models of
dynamical breaking of the electroweak symmetry do support the possibility of 
generating the experimentally observed baryon (and possibly also the technibaryon/DM)
asymmetry of the Universe directly at the electroweak phase
transition~\cite{Cline:2008hr,Jarvinen:2009wr,Jarvinen:2009pk}.  Electroweak baryogenesis
\cite{Shaposhnikov:1986jp} is, however, impossible in the SM
\cite{Kajantie:1995kf}.

\subsection{TIMP lifetime and decay modes}
According to~\cite{Rubakov:1984ba} the sphaleron contribution
to the  techni-baryon decay rate is negligible because exponentially
suppressed, unless the techni-baryon is heavier than several TeV.

Grand unified theories (GUTs) suggest that the baryon number $B$ is violated
by dimension-6 operators suppressed by the GUT scale $M_{\rm GUT} \sim2\cdot
10^{16}\GeV$, yielding a proton  life-time~\cite{Hisano:1992jj}
\beq
\tau(p\rightarrow\pi^0 e^+) \sim \frac{M_{\rm GUT}^4}{m_p^5}\sim
10^{41} \,{\rm sec}.
\eeq
If $B'$ is similarly violated at the same high scale $M_{\rm GUT}$,
our  TIMP would decay with life-time
\beq 
\tau \sim \frac{M_{\rm GUT}^4}{M_{\rm DM}^5}\sim 10^{26}\,{\rm sec}, 
\eeq
which falls in the ball-park required by the phenomenological analysis to explain the PAMELA anomaly \cite{Nardi:2008ix}. 
Models of unification of the SM couplings in the
presence of a dynamical electroweak symmetry breaking mechanism have been
recently explored \cite{Christensen:2005bt,Gudnason:2006mk}. Interestingly, the
scale of unification suggested by the phenomenological analysis emerges quite
naturally \cite{Gudnason:2006mk}.

\smallskip 

Low energy TIMP and nucleon (quasi)-stability imply that, in the primeval
Universe, at temperatures $T\circa{<} M_{\rm GUT}$ perturbative violation of
the $B'$ and $B$ global charges is strongly suppressed. Since this temperature
is presumably larger than the reheating temperature, it is unlikely that
$\Omega_{B}$ and $\Omega_{\rm DM}$ result directly from an asymmetry generated 
in $B'$ or  $B$. More likely, the initial seed yielding $\Omega_{\rm DM}$ and $\Omega_B$
could be an initial asymmetry in lepton number $L$ that, much along the lines 
of well studied leptogenesis scenarios~\cite{Fukugita:1986hr},  feeds the $B$ and $B'$ asymmetries 
through the sphaleron effects. 
Indeed, it has been shown that it is possible to embed seesaw-types of
scenarios in theories of dynamical symmetry breaking,
while keeping the scale of the $L$-violating Majorana masses as low as 
$\sim 10^{3}\,$TeV~\cite{Appelquist:2002me}. {In Minimal Walking Technicolor \cite{Dietrich:2005jn,Foadi:2007ue}, one additional  (techni-singlet)  $\SU(2)$-doublet  must be introduced 
to cancel the odd-number-of-doublets anomaly~\cite{Witten:1982fp}. 
An asymmetry in the $L'$ global charge associated with these new states can 
also serve as a seed for the $B$ and $B'$ asymmetries.} In \cite{Frandsen:2009fs} we have shown that it is possible to embed a low energy see-saw mechanism for the fourth family Leptons in the Minimal Walking Technicolor extension of the SM. 

\medskip

Assuming that TIMP decays is dominantly due to effective four-fermion operators,
its decay modes significantly depend on the technicolor gauge group.
In the following $L$ generically denotes any SM fermion, quark or lepton,
possibly allowed by the Lorentz and gauge symmetries of the theory.
\begin{itemize}
\item If the technicolor group is SU(3), the situation is  analogous to ordinary QCD:
the TIMP  is a fermionic $QQQ$ state, and effective $QQQL$ operators
gives $TIMP \rightarrow\Pi^- \ell^+ $ decays.  
This leads to hard leptons, but together with an excess of $\bar p$,
from the $\Pi^- \rightarrow \bar c$ decay (in view of  $\Pi^- \simeq  W^-_L$).  
\item 
If the technicolor group is SU(4) the situation is  that  the TIMP is a bosonic $QQQQ$ state, and effective $QQQQ$ operators lead to
its decay into techni-pions.
\item 
Finally, if the technicolor gauge group is SU(2) the TIMP is a bosonic $QQ$  state,  (as put forward in \cite{Ryttov:2008xe}), and effective $QQLL$ operators
lead to TIMP decays into two $L$.
Since the fundamental representation of SU(2) is pseudoreal, one actually gets 
an interesting dynamics analyzed in detail in \cite{Ryttov:2008xe}.  
Here the TIMP is a pseudo-Goldstone boson of the underlying gauge theory. 
\end{itemize}
An SU(2) technicolor model compatible with the desired features is obtained 
assuming that the left component of the Dirac field $Q$ has zero hypercharge and is a doublet under $\SU(2)_L$,
so that the TIMP is a scalar $QQ$ with no overall weak interactions, and the four-fermion operator
$(QQ)\partial_\mu (\bar L \gamma_\mu L)$ allows it to decay.
Such operator is possible for both SM leptons and quarks, so that the TIMP
branching ratios into $\ell^+ \ell^-$ and $q\bar q$ is a free parameter.

\subsection{Effective Lagrangian for the simplest TIMP and Earth Based Constraints}

{}We identify the TIMP with a complex scalar $\phi$, singlet under the SM interactions, charged under the $U(1)$ technibaryon symmetry of a generic TC theory. For example UMT includes such a state \cite{Ryttov:2008xe}.  The main differences with other models featuring an extra $U(1)$ scalar DM are: i)  The $U(1)$ is natural, i.e. it is identified with a technibaryon symmetry; ii) Compositeness requires the presence around the electroweak scale of spin one resonances leading to striking collider signatures; iii) The DM relic density is naturally linked to the baryon one via an asymmetry. 
We also introduce a (light) composite Higgs particle since it appears in several  strongly coupled theories as shown in the Appendix F of \cite{Sannino:2008ha}. 
 We consider only the TC global symmetries relevant for the electroweak sector, i.e. the  $SU(2)\times SU(2)$ spontaneously breaking to $SU(2)$. The low energy spectrum contains, besides the composite Higgs, two $SU(2)$ triplets of (axial-) vector spin one mesons.  The effective Lagrangian, before including the TIMP, has been introduced in \cite{Foadi:2007ue,Belyaev:2008yj} in the context of minimal walking theories. Here we add the TIMP  $\phi$ with the following interaction terms:
 \begin{eqnarray}
{\cal L}_{\rm DM}&=&\frac{1}{2}\ \partial_\mu\phi^\ast\partial^\mu\phi-\frac{M_\phi^2 - d_M\ v^2}{2}\ \phi^\ast\phi
+\frac{d_F}{2\Lambda^2}\ {\rm Tr}\left[F_{{\rm L}\mu\nu} F_{\rm L}^{\mu\nu}+F_{{\rm R}\mu\nu} F_{\rm R}^{\mu\nu}\right]\ \phi^\ast\phi  \nonumber \\
&+& 
d_C\ \tilde{g}^2 \ {\rm Tr}\left[C_{{\rm L}\mu}^2+C_{{\rm R}\mu}^2\right]\ \phi^\ast\phi
-\frac{d_M}{2}\ {\rm Tr}\left[M M^\dagger\right]\ \phi^\ast\phi \ .
\label{eq:Lagrangian}
\end{eqnarray}
The matrix $M$ contains the composite Higgs and the pions eaten by the electroweak bosons,
\begin{eqnarray}
M &=& \frac{1}{\sqrt{2}}\left[v+H+2\ i\ T^a\ \pi^a\right] \ ,
\end{eqnarray}
where $v$ is the vacuum expectation value, and $T^a=2\sigma^a$, $a=1,2,3$, where $\sigma^a$ are the Pauli matrices. $\Lambda $ is a scale associated to the breakdown of the low energy effective theory and it is in the TeV energy range.  $C_{{\rm L}\mu}$ and $C_{{\rm R}\mu}$ are the electroweak covariant linear combinations $\displaystyle{
C_{{\rm L}\mu} = A^a_{{\rm L}\mu}\ T^a-{g}/{\tilde{g}}\ \widetilde{W}^a\ T^a }$ and $\displaystyle{
C_{{\rm R}\mu} = A^a_{{\rm R}\mu}\ T^a-{g^\prime}/{\tilde{g}}\ \widetilde{B}\ T^3}$, 
where $\widetilde{W}^a$ and $\widetilde{B}$ are the electroweak gauge fields (before diagonalization).  $F_{\rm L}$, $F_{\rm R}$ are the fields strength tensors associated to the vector meson fields $A^a_{{\rm L}\mu}$ and $A^a_{{\rm R}\mu}$, respectively. The associated spin one mass eigenstates are indicated with $R_{1}$ and $R_2$.

In earth based experiments the TIMP interacts  with nuclei mainly via the exchange of a composite Higgs (the $d_M$ term in the above Lagrangian) and a photon. The photon interaction was considered in \cite{Bagnasco:1993st} and it is due to a nonzero electromagnetic charge radius of $\phi$. Here we stress the relevance of the composite Higgs exchange. Similar exchanges (although not from a composite theory) were studied also in \cite{McDonald:1993ex,Oikonomou:2006mh}. The Lagrangian term for the charge radius interaction is:
\beq 
\mathcal{L}_B=i e \frac{d_B}{\Lambda^2} \phi^*\overleftrightarrow{\partial_\mu} \phi \, \partial_{\nu}F^{\mu\nu} \ .
\label{eq:Bagnasco}
\eeq
 The non-relativistic cross section for scattering off a nucleon from the photon and Higgs exchanges respectively are:
\begin{equation}
\sigma_{\textrm{p}}^{\gamma} = \frac{\mu^2}{4\pi} \left[\frac{ 8\pi \, \alpha \, d_B}{\Lambda^2}\right]^2 \ , \quad
\sigma_{\textrm{nucleon}}^H = \frac{\mu^2}{4\pi}   \left[\frac{d_M  \, f m_N}{M^2_H M_{\phi}}\right]^2 
\end{equation}
with $\mu= M_{\phi} m_N /(M_{\phi} + m_N)$.
Here $m_N$ is the mass of the nucleon and the Higgs to nucleon coupling $f$ can range between $0.14$ and $0.66$ \cite{Shifman:1978zn,Andreas:2008xy}. The composite Higgs can be light in theories with higher dimensional representations \cite{Sannino:2008ha,Hong:2004td} as well as in near-conformal TC models \cite{Dietrich:2005jn}. Similarly $\phi$ can also be light~\cite{Ryttov:2008xe}. These observations justify considering both contributions for the cross section. The nucleus-dark matter cross section must be, however, still corrected by the nucleus form factor $\xi(M_{\phi})$ as described in 
\cite{Ellis:1991ef,McDonald:1993ex}. We use the same gaussian form factor as in \cite{McDonald:1993ex} which is found to be precise to within 10 $\%$ for $M_{\phi} \lesssim 100$ GeV, while for $M_{\phi} > 100$ GeV the gaussian form factor correction still overestimates the true cross-section \cite{Ellis:1991ef}. However, in the latter region CDMS and XENON currently have insufficient sensitivity even using the gaussian form factor.

\begin{figure}[htp!]
\centerline{\includegraphics[height=7cm,width=11cm]{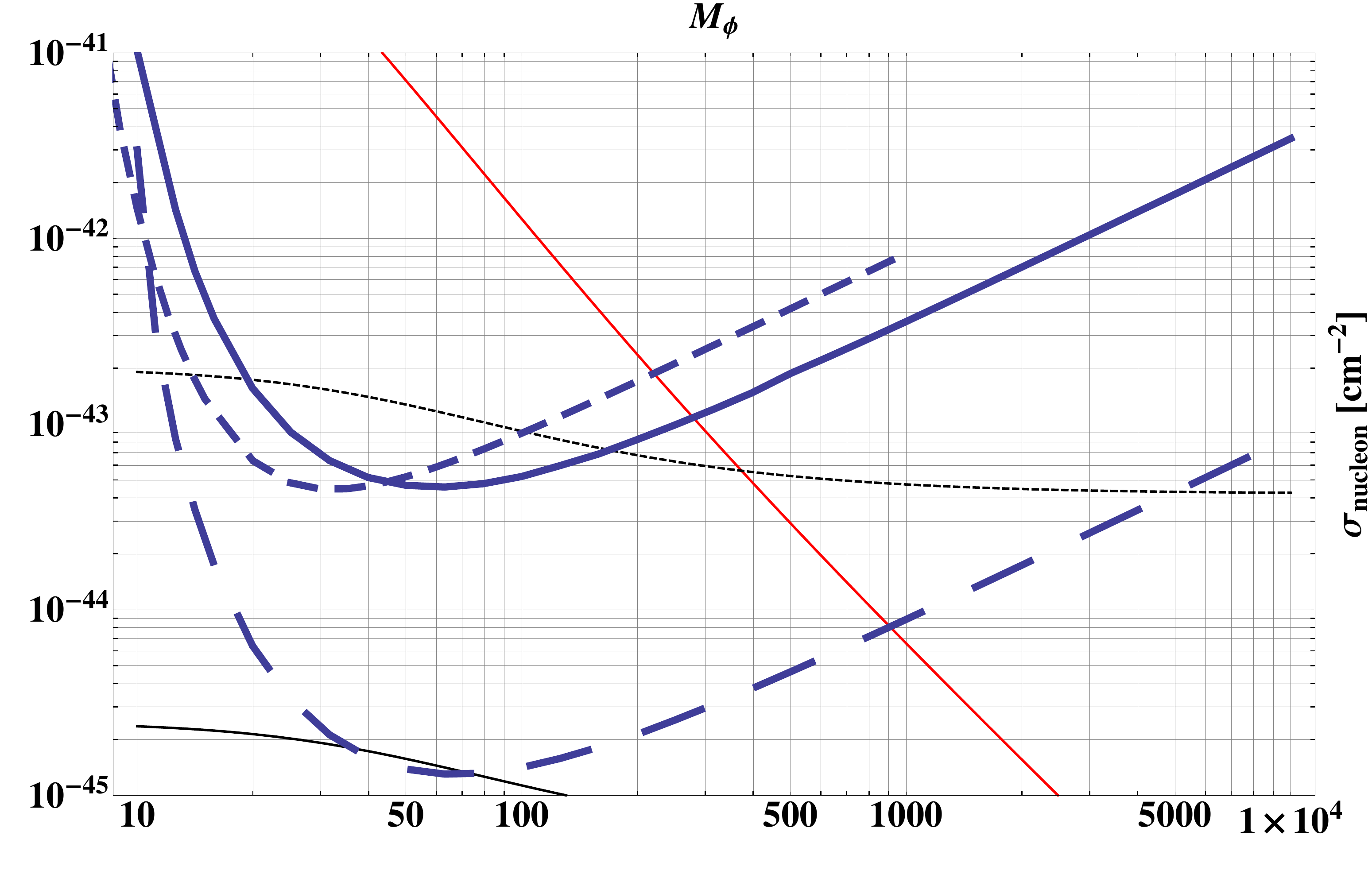}}
\caption{ TIMP -- nucleon cross section: Red solid line corresponds to the composite Higgs exchange where we used $d_M=1$ and $M_H=300$ GeV. The solid and dashed black lines correspond to the photon exchange cross section with $d_B =1$ and $\Lambda =3$ TeV (solid) and $1$ TeV (dashed). We use for reference $f=0.3$. Also plotted is the approximate exclusion limit from CDMS II Ge Combined  (solid-thick-blue) and XENON10 2007 (dashed-thick-blue).  The long-dashed-blue line is the projected superCDMS exclusion curve. The allowed region is below the CDMS and XENON curves.}\label{fig:cdms}
\end{figure}

In figure~\ref{fig:cdms} we plot the cross sections independently. The Higgs exchange is shown in solid red for a reference value of $M_H =300$ GeV and $d_M =1$.  The solid black and dashed cross sections (adjusted for $^{73}$Ge) are the photon exchange cross sections for two values of $\Lambda$, i.e.  $4\pi  F_{TC}\approx 3$ TeV  and $1$ TeV. The latter value would be a reasonable guess for low scale technicolor models \cite{Lane:1989ej}. We have also taken the fiducial value of $|d_B|=1$. In the same plot we compare our curves with the CDMS (solid-thick-blue) and XENON (dashed-thick-blue) exclusion limits. The long-dashed-blue line is the projected superCDMS exclusion curve. 

At low values of the TIMP mass the composite Higgs exchange dominates and the experimental constraints become relevant.
We also expect important interference effects which are obtained by simply combining the amplitudes of the two competing processes. The effective TIMP Lagrangian has also been used to deduce interesting collider phenomenology \cite{Foadi:2008qv}. 

There  exist several interesting models of dark matter inspired to technicolor and we refer to \cite{Kouvaris:2007iq,Kainulainen:2006wq}  for a list of relevant literature. 

 \newpage
\section{Conformal Dynamics}

We have seen that models of dynamical breaking of the electroweak symmetry are theoretically appealing and constitute one of the best motivated natural extensions of the SM.  These are also among the most challenging models to work with since they require deep knowledge of gauge dynamics in a regime where perturbation theory fails. In particular, it is of utmost importance to gain information on the nonperturbative dynamics of non-abelian four dimensional gauge theories. 
 \subsection{Phases of Gauge Theories}
Non-abelian gauge theories exist in a number of  distinct phases which can be classified according to the characteristic dependence of the potential energy on the distance between two well separated static sources.  The collection of all of these different behaviors, when represented, for example, in the  flavor-color space, constitutes the {\it phase diagram} of the given gauge theory. The phase diagram of $SU(N)$ gauge theories as functions of number of flavors, colors and matter representation has been investigated in \cite{Sannino:2004qp,Dietrich:2006cm,Ryttov:2007sr,Ryttov:2007cx,Sannino:2008ha}. 

The analytical tools we will use for such an exploration are: i) The conjectured {\it physical} all orders beta function for nonsupersymmetric gauge theories with fermionic matter in arbitrary representations of the gauge group \cite{Ryttov:2007cx}; ii) The truncated Schwinger-Dyson equation (SD) \cite{Appelquist:1988yc,Cohen:1988sq,Miransky:1996pd} (referred also as the ladder approximation in the literature);  The Appelquist-Cohen-Schmaltz (ACS) conjecture \cite{Appelquist:1999hr} which makes use of the counting of the thermal degrees of freedom at high and low temperature. 

We will show that relevant constraints can be deduced for any gauge theory and any representation via the all orders beta function and the SD methods. The ACS conjecture is not sufficiently constraining when studying theories with matter in higher dimensional representations of a generic gauge theory \cite{Sannino:2005sk, Sannino:2009aw}.

 We wish to study the phase diagram of any asymptotically free non-supersymmetric theories with fermionic matter transforming according to a generic representation of an SU($N$) gauge group as function of the number of colors and
flavors. 

We start by characterizing the possible phases via the potential $V(r)$ between
two electric test charges separated by a large distance r. The list of possible potentials is given below:
\begin{eqnarray}
 {\rm \mathbf{Coulomb:}} &~~~~~& V (r) \propto \frac{1}{r} \\
 &~~~~~& \nonumber \\
  {\rm \mathbf{Free~electric:}} &~~~~~& V (r) \propto \frac{1}{r\log(r)} \\
 &~~~~~& \nonumber \\
  {\rm \mathbf{ Free~magnetic:}} &~~~~~&  V (r) \propto \frac{\log(r)}{r}  \\
  &~~~~~& \nonumber \\
 {\rm \mathbf{ Higgs:}}&~~~~~&  V (r) \propto {\rm  constant}  \\  
 &~~~~~& \nonumber \\
   {\rm \mathbf{ Confining:}} &~~~~~&   V (r) \propto \sigma r \ .
  \end{eqnarray}
   A nice review of these phases can be found in \cite{Intriligator:1995au} which here we re-review for completeness. 
In the Coulomb phase, the electric charge $e^2(r)$  is a constant while in
the free electric phase massless electrically charged fields renormalize the charge to
zero at long distances as, i.e. $e^2(r) \sim 1/\log(r )$. QED is an abelian example of a free electric phase. The free
magnetic phase occurs when massless magnetic monopoles renormalize
the electric coupling constant at large distance with $e^2(r) \sim \log(r)$. 
  
  In the Higgs phase, the condensate of an electrically
charged field gives a mass gap to the gauge fields by the Anderson-Higgs-Kibble mechanism
and screens electric charges, leading to a potential which, up to an additive constant, has an exponential Yukawa decay to zero at long distances.
In the confining phase, there is a mass gap with electric flux confined into a thin
tube, leading to the linear potential with string tension $\sigma$.

We will be mainly interested in finding theories possessing a
non-Abelian Coulomb phase or being close in the parameter space to these theories. In this phase we have massless interacting quarks and gluons exhibiting
the Coulomb potential. This phase occurs when there is a non-trivial, infrared
fixed point of the renormalization group. These are thus non-trivial, interacting, four
dimensional conformal field theories.

To guess the behavior of the magnetic charge, at large distance separation, between two test magnetic charges  one uses the Dirac condition:
\begin{equation}
e(r) g(r) \sim 1 \ .
\end{equation}
Then it becomes clear that $g(r)$ is constant in the Coulomb phase, increases with $\log(r)$ in the free electric phase and decreases as $1/\log(r)$ in the free magnetic phase. In these three phases the potential goes like $g^2(r)/r$. A linearly rising potential in the Higgs phase for magnetic test charges corresponds to the Meissner effect in the electric charges.

Confinement does not survive the presence of massless matter in the fundamental representation, such as light quarks in QCD. This is so since it is more convenient for the underlying theory to pop from the vacuum virtual quark-antiquark pairs when pulling two electric test charges apart. The potential for the confining phase will then change and there is no distinction between Higgs and confining phase.  

Under electric-magnetic duality one exchanges electrically charged
fields with magnetic ones then the behavior in the free electric phase is
mapped in that of the free magnetic phase. The Higgs and confining phases are  also expected to be exchanged under duality. 
Confinement can then be understood as the dual Meissner effect associated with
a condensate of monopoles. 
\begin{figure}[t]
\begin{center}
\includegraphics[width=14truecm ]{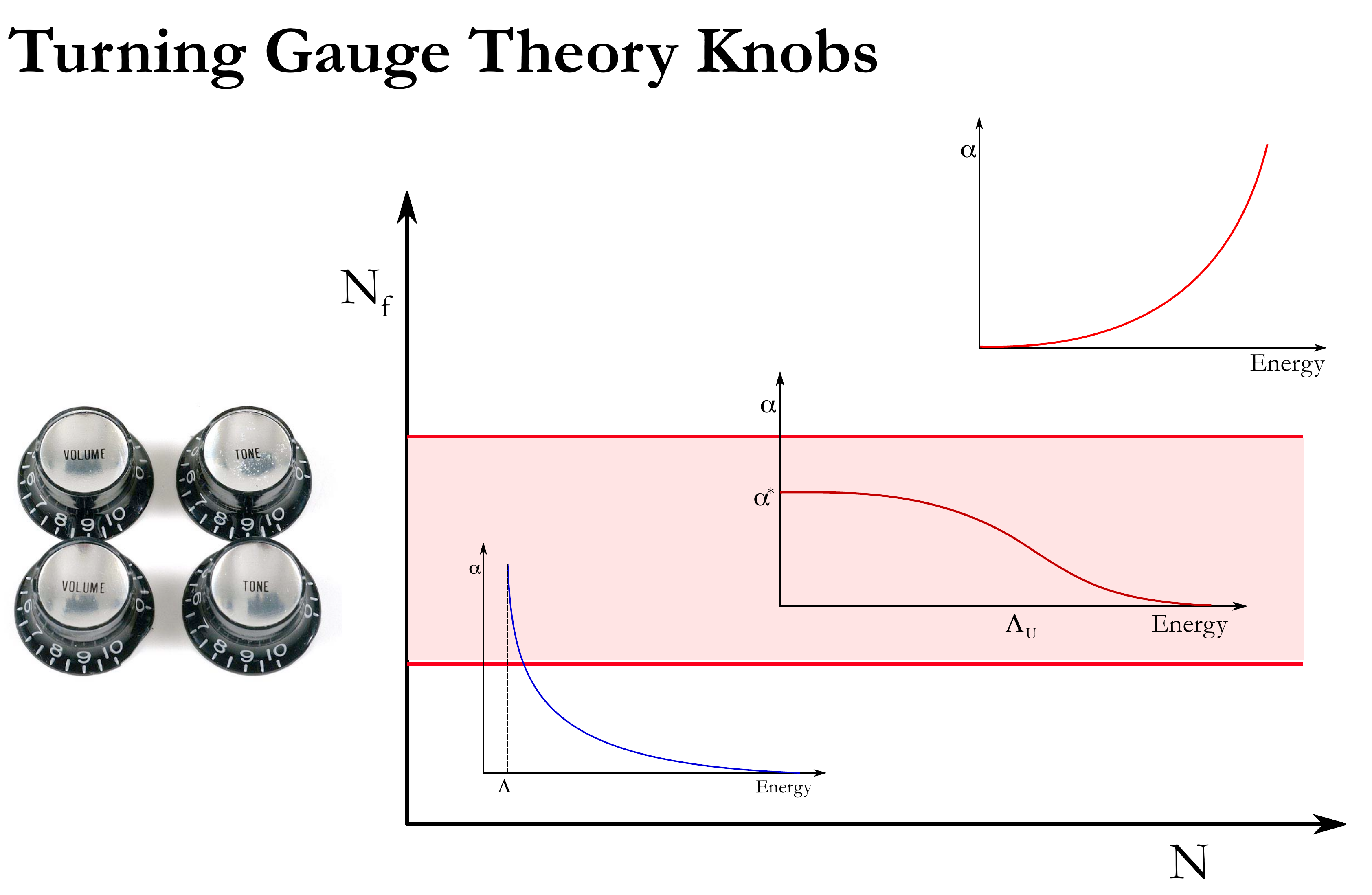}
\caption{A generic gauge theory has different Knobs one can tune. For example by changing the number of flavors one can enter in different phases. The pink region is the conformal region, i.e. the one where the coupling constant freezes at large distances (small energy). The region above the pink one corresponds to a non-abelian QED like theory and below to a QCD-like region. We have also plotted the cartoon of the running of the various coupling constants in the regions away from the boundaries of the conformal window. The diagram above is the qualitative one expected for a gauge theory with matter in the adjoint representation.} \label{knobs}
\end{center}
\end{figure}

\subsection{Analytic Methods}
\subsubsection{Physical all orders Beta Function - Conjecture}
\label{All-orders}
Recently we have conjectured an all orders beta function which allows for a bound of the conformal window \cite{Ryttov:2007cx} of $SU(N)$ gauge theories for any matter representation. 

 It is  written in a form useful for constraining the phase diagram of strongly coupled theories. It is inspired by the Novikov-Shifman-Vainshtein-Zakharov  (NSVZ) beta function for supersymmetric theories \cite{Novikov:1983uc,Shifman:1986zi} and the renormalization scheme coincides with the NSVZ one. The predictions of the conformal window coming from the above beta function are nontrivially supported by all the recent lattice results \cite{Catterall:2007yx,DelDebbio:2008wb,Catterall:2008qk,Appelquist:2007hu,
Shamir:2008pb,Deuzeman:2008sc,Lucini:2007sa}.

It reproduces the exact supersymmetric results when reducing the matter content to the one of supersymmetric gauge theories. In particular we compared our prediction for the running of the coupling constant for the pure Yang-Mills theories with the one studied via the Schroedinger functional \cite{Luscher:1992zx,Luscher:1993gh,Lucini:2007sa} and found an impressive agreement. We have also predicted that the IRFP for $SU(3)$ gauge theories could not extend below $8.25$ number of flavors. Subsequent numerical analysis  \cite{Appelquist:2007hu, Deuzeman:2008sc,Appelquist:2009ty} confirmed our prediction. 

In \cite{Sannino:2009aw} we further assumed the form of the beta function to hold for $SO(N)$ and $Sp(2N)$ gauge groups. Consider a generic gauge group with $N_f(r_i)$ Dirac flavors belonging to the representation $r_i,\ i=1,\ldots,k$ of the gauge group. The conjectured beta function reads:
\begin{eqnarray}
\beta(g) &=&- \frac{g^3}{(4\pi)^2} \frac{\beta_0 - \frac{2}{3}\, \sum_{i=1}^k T(r_i)\,N_{f}(r_i) \,\gamma_i(g^2)}{1- \frac{g^2}{8\pi^2} C_2(G)\left( 1+ \frac{2\beta_0'}{\beta_0} \right)} \ ,
\end{eqnarray}
with
\begin{eqnarray}
\beta_0 =\frac{11}{3}C_2(G)- \frac{4}{3}\sum_{i=1}^k \,T(r_i)N_f(r_i) \qquad \text{and} \qquad \beta_0' = C_2(G) - \sum_{i=1}^k T(r_i)N_f(r_i)  \ .
\end{eqnarray}
The generators $T_r^a,\, a=1\ldots N^2-1$ of the gauge group in the
representation $r$ are normalized according to
$\text{Tr}\left[T_r^aT_r^b \right] = T(r) \delta^{ab}$ while the
quadratic Casimir $C_2(r)$ is given by $T_r^aT_r^a = C_2(r)I$. The
trace normalization factor $T(r)$ and the quadratic Casimir are
connected via $C_2(r) d(r) = T(r) d(G)$ where $d(r)$ is the
dimension of the representation $r$. The adjoint
representation is denoted by $G$.

The beta function is given in terms of the anomalous dimension of the fermion mass $\gamma=-{d\ln m}/{d\ln \mu}$ where $m$ is the renormalized mass, similar to the supersymmetric case \cite{Novikov:1983uc,Shifman:1986zi,Jones:1983ip}. 
The loss of asymptotic freedom is determined by the change of sign in the first coefficient $\beta_0$ of the beta function. This occurs when
\begin{eqnarray} \label{AF}
\sum_{i=1}^{k} \frac{4}{11} T(r_i) N_f(r_i) = C_2(G) \ , \qquad \qquad \text{Loss of AF.}
\end{eqnarray}
 At the zero of the beta function we have
\begin{eqnarray}
\sum_{i=1}^{k} \frac{2}{11}T(r_i)N_f(r_i)\left( 2+ \gamma_i \right) = C_2(G) \ ,
\end{eqnarray}
Hence, specifying the value of the anomalous dimensions at the IRFP yields the last constraint needed to construct the conformal window. Having reached the zero of the beta function the theory is conformal in the infrared. For a theory to be conformal the dimension of the non-trivial spinless operators must be larger than one in order not to contain negative norm states \cite{Mack:1975je,Flato:1983te,Dobrev:1985qv}.  Since the dimension of the chiral condensate is $3-\gamma_i$ we see that $\gamma_i = 2$, for all representations $r_i$, yields the maximum possible bound
\begin{eqnarray} 
\sum_{i=1}^{k} \frac{8}{11} T(r_i)N_f(r_i) = C_2(G) \ , \qquad \gamma_i = 2 \ .
\end{eqnarray}
In the case of a single representation this constraint yields 
\begin{equation}
N_f(r)^{\rm BF} \geq \frac{11}{8} \frac{C_2(G)}{T(r)} \ , \qquad \gamma  = 2 \ .
\end{equation}
The actual size of the conformal window can be smaller than the one determined by the bound above, Eq. (\ref{AF}) and (\ref{Bound}). It may happen, in fact, that chiral symmetry breaking is triggered for a value of the anomalous dimension less than two. If this occurs the conformal window shrinks. Within the ladder approximation \cite{Appelquist:1988yc,{Cohen:1988sq}} one finds that chiral symmetry breaking occurs when the anomalous dimension is close to one. Picking $\gamma_i =1$ we find:
\begin{eqnarray}
\sum_{i=1}^{k} \frac{6}{11} T(r_i)N_f(r_i) = C_2(G) \ , \qquad  \gamma = 1 \ ,. 
\end{eqnarray}
In the case of a single representation this constraint yields 
\begin{equation}
N_f(r)^{\rm BF} \geq \frac{11}{6} \frac{C_2(G)}{T(r)} \ , \qquad \gamma =1 \ .
\end{equation}
When considering two distinct representations the conformal window becomes a three dimensional volume, i.e. the conformal {\it house} \cite{Ryttov:2007sr}.  Of course, we recover the results by Banks and Zaks \cite{Banks:1981nn} valid in the perturbative regime of the conformal window.

We note that the presence of a physical IRFP requires the vanishing of the beta function for a certain value of the coupling. The opposite however is not necessarily
true; the vanishing of the beta function is not a sufficient condition to determine if the theory has a fixed point unless the beta function is {\it physical}. By {\it physical} we mean that the beta function allows to determine simultaneously other scheme-independent quantities at the fixed point such as the anomalous dimension of the mass of the fermions. This is exactly what our beta function does.  In fact, in the case of a single representation, one finds that at the zero of the beta function one must have
\begin{eqnarray}
\gamma = \frac{11C_2(G)-4T(r)N_f}{2T(r)N_f} \ .\end{eqnarray}

 \subsubsection{Schwinger-Dyson in the Rainbow Approximation}
\label{ra}
\begin{figure}[ht]
\centerline{\includegraphics{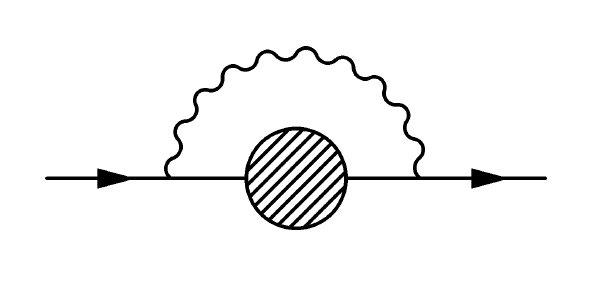}}
\caption
{Rainbow approximation for the
fermion self energy function. The boson is a gluon.} 
\label{rainbowselfenergy}
\end{figure}

{}For nonsupersymmetric theories an old way to get quantitative estimates is to use the
{\it rainbow} approximation
to the Schwinger-Dyson equation
\cite{Maskawa:1974vs,Fukuda:1976zb}, see Fig.~\ref{rainbowselfenergy}. 
\noindent
Here the full nonperturbative fermion propagator in momentum space reads
\beq iS^{-1}(p) = Z(p)\left(\slashed{p}-\Sigma(p)\right) \ , \eeq
and the Euclidianized gap equation in Landau gauge is given by
\beq \Sigma(p) =
3C_2(r)\int\frac{d^4k}{(2\pi)^4}\frac{\alpha\left((k-p)^2\right)}{(k-p)^2}\frac{\Sigma(k^2)}{Z(k^2)k^2
    + \Sigma^2(k^2)} \ , \eeq
where $Z(k^2)=1$ in the Landau gauge and we linearize
the equation by neglecting $\Sigma^2(k^2)$ in the denominator. Upon converting it into a differential equation and assuming that the coupling
$\alpha(\mu) \approx \alpha_c$ is varying slowly ($\beta(\alpha) \simeq
0$) one gets the   
approximate (WKB) 
solutions
\beq \Sigma(p) \propto p^{-\gamma(\mu)} \ , \qquad \Sigma(p) \propto
p^{\gamma(\mu) - 2} \ . \label{sol-to-gap-eq} \eeq
The critical coupling is given in terms of the quadratic Casimir
of the representation of the fermions
\beq \alpha_c \equiv \frac{\pi}{3C_2(r)} \ . \label{critical-coupling}
\eeq
The anomalous dimension of the fermion mass operator is
\beq \gamma(\mu) = 1 - \sqrt{1-\frac{\alpha(\mu)}{\alpha_c}} 
\sim \frac{3C_2(r)\alpha(\mu)}{2\pi} \ . \label{admp}
\eeq
The first solution
corresponds 
to the running of an ordinary mass term ({\it  hard} mass) of nondynamical origin and the
second solution  to a {\it soft} mass dynamically generated. In fact in the second case one observes the $1/p^2$ behavior in the limit of large momentum. 

Within this approximation spontaneous symmetry breaking occurs when $\alpha$ reaches the critical coupling $\alpha_c$ given in Eq.~(\ref{critical-coupling}). {}From Eq.~(\ref{admp}) it is clear that $ \alpha_c $ is reached when 
$\gamma$  is of order unity \cite{Yamawaki:1985zg,Cohen:1988sq,Appelquist:1988yc}. Hence the symmetry breaking occurs when the soft and the hard mass terms scale as function of the energy scale in the same way. In Ref.~\cite{Appelquist:1988yc},  it was noted that in the lowest (ladder) order, the gap equation leads to the condition $\gamma(2-\gamma)=1$ for chiral symmetry breaking to occur. To all orders in perturbation theory this condition is gauge invariant and also equivalent nonperturbatively to the condition $\gamma=1$. However, to any finite order in perturbation theory these conditions are, of course, different. Interestingly the condition $\gamma(2-\gamma)=1$ leads again to the critical coupling $\alpha_c$  when using the perturbative leading order expression for the anomalous dimension which is $\gamma=\frac{3C_2(r)}{2\pi}\alpha$ .

To summarize, the idea behind this method is simple. One simply compares the two couplings in the infrared associated to i) an infrared zero in the $\beta$ function, call it $\alpha^{\ast}$  with ii) the critical coupling, denoted with $\alpha_c$, above which a dynamical mass for the fermions generates nonperturbatively and chiral symmetry breaking occurs. If $\alpha^{\ast}$ is less than $\alpha_c$ chiral symmetry does not occur and the theory remains conformal in the infrared, viceversa if $\alpha^{\ast}$ is larger than $\alpha_c$ then the fermions acquire a dynamical mass and the theory cannot be conformal in the infrared. The condition $\alpha^{\ast} = \alpha_c$ provides the desired $N_f^{\rm SD} $ as function of $N$.  In practice to estimate $\alpha^{\ast}$ one uses the two-loop beta function while the truncated SD  equation to determine $\alpha_c$ as we have done before. This corresponds to when the anomalous dimension of the quark mass operator
becomes approximately unity.  

The two-loop fixed point value of the coupling constant
is:
\begin{eqnarray}
\frac{\alpha^*}{4\pi}=-\frac{\beta_0}{\beta_1}.
\end{eqnarray}
with the following definition of the two-loop beta function 
\begin{eqnarray}\beta (g) = -\frac{\beta_0}{(4\pi)^2} g^3 - \frac{\beta_1}{(4\pi)^4} g^5 \ ,
\label{perturbative}
\end{eqnarray}
where $g$ is the gauge coupling and the beta function coefficients are given by
\begin{eqnarray}
\beta_0 &=&\frac{11}{3}C_2(G)- \frac{4}{3}T(r)N_f \\
\beta_1 &=&\frac{34}{3} C_2^2(G)
- \frac{20}{3}C_2(G)T(r) N_f  - 4C_2(r) T(r) N_f  \ .\end{eqnarray}
To this order the two coefficients are universal,
i.e. do not depend on which renormalization group scheme one has used to determine them.
The perturbative expression for the anomalous dimension reads:
\begin{equation}
\gamma(g^2) = \frac{3}{2} C_2(r) \frac{g^2}{4\pi^2} + O(g^4) \ .
\end{equation}
With $\gamma =-{d\ln m}/{d\ln \mu}$ and $m$ the renormalized fermion mass.

For a fixed number of colors the critical number of flavors for which the
order of $\alpha^{\ast}$ and $\alpha_c$ changes is defined by imposing 
$\alpha^*{=}\alpha_c$, and it is given by
\begin{eqnarray}
{N_f^{\rm SD}} &=& \frac{17C_2(G)+66C_2(r)}{10C_2(G)+30C_2(r)}
\frac{C_2(G)}{T(r)} \ . \label{nonsusy}
\end{eqnarray}
Comparing with the previous result obtained using the all orders beta function we see that it is the coefficient of $C_2(G)/T(r)$ which is different.

\subsubsection{Thermal Counting of the Degrees of Freedom - Conjecture}
\label{ACS}
The free energy can be seen as a device to count
the relevant degrees of freedom. It can be computed, exactly, in two regimes 
of a generic asymptotically free theory: the very hot and the very cold one. 

The zero-temperature theory of interest is
characterized using the quantity $f_{IR}$, related to the free energy by
\begin{equation}  \label{eq:firdef}
f_{IR} \equiv - \lim_{T\to 0} \frac{{\cal F}(T)}{T^4}
\frac{90}{\pi^2} \ ,
\end{equation}
where $T$ is the temperature and ${\cal F}$ is the conventionally
defined free energy per unit volume. The limit is well defined if
the theory has an IRFP. 
For the special case of an
infrared-free theory 
\begin{eqnarray}
f_{IR} = \sharp~~{\rm Real~Bosons}~+ ~\frac{7}{4}~\sharp~~{\rm Weyl-Fermions} \ .
\end{eqnarray}
The corresponding expression in the large $T$ limit is
\begin{equation}  \label{eq:fuvdef}
f_{UV} \equiv - \lim_{T\to \infty} \frac{{\cal F}(T)}{T^4}
\frac{90}{\pi^2}\ .
\end{equation}
This limit is well defined if the theory has an ultraviolet
fixed point. For an asymptotically free theory $f_{UV}$ counts the
underlying ultraviolet d.o.f. in a similar way.

In terms of these quantities, the conjectured inequality \cite{Appelquist:1999hr} for any
asymptotically free theory is
\begin{equation}  \label{eq:ineq}
f_{IR} \le f_{UV}\ .
\end{equation}
This inequality has not been proven but it was
shown to be consistent with known results and then used to derive new
constraints for several strongly coupled, vector-like gauge theories.
The ACS conjecture has been used also for chiral gauge theories \cite{Appelquist:1999vs}. There it 
was also found that to make definite predictions a stronger requirement is needed \cite{Appelquist:2000qg}.

\subsection{The $SU(N)$ phase diagram}
We consider here gauge theories with fermions in any representation of the $SU(N)$ gauge group \cite{Sannino:2004qp,Dietrich:2006cm,Ryttov:2007sr,Ryttov:2007cx,Ryttov:2009yw} using the various analytic methods described above. Part of our original results have already been summarized in \cite{Sannino:2008ha}.

\begin{figure}[h]
\begin{center}\resizebox{12cm}{!}{\includegraphics{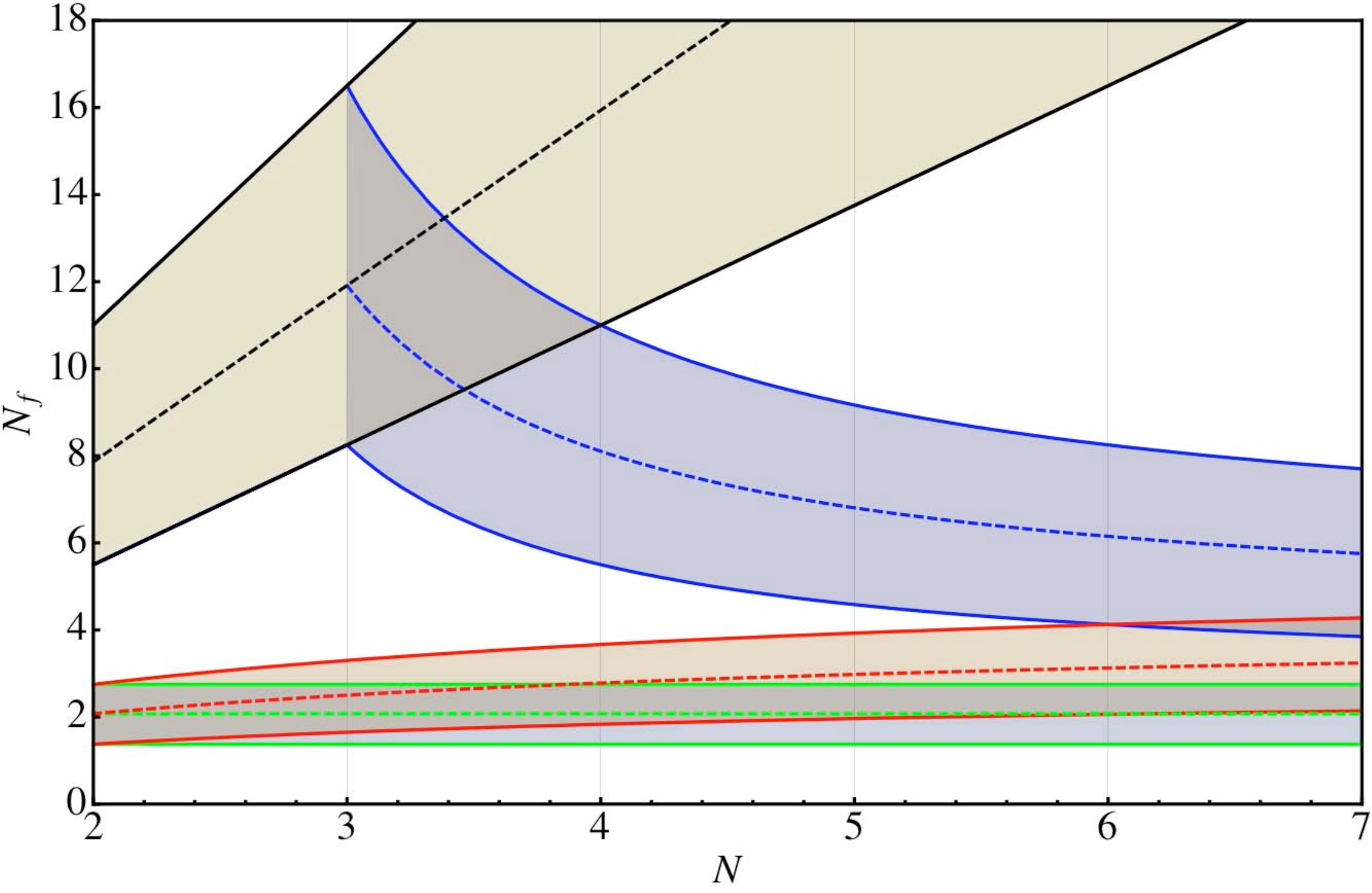}}
\caption{Phase diagram for nonsupersymmetric theories with fermions
in the: i) fundamental representation (black), ii) two-index
antisymmetric representation (blue), iii) two-index symmetric
representation (red), iv) adjoint representation (green) as a
function of the number of flavors and the number of colors. The
shaded areas depict the corresponding conformal windows. Above the
upper solid curve  the theories are no longer asymptotically free.
In between the upper and the lower solid curves the theories are
expected to develop an infrared fixed point according to the all orders
beta function. The area between the upper solid curve and
the dashed curve corresponds to the conformal window obtained in the
ladder approximation.} \label{PHComparison}\end{center}
\end{figure}

Here we simply plot in Fig.~\ref{PHComparison} the
conformal windows for various representations predicted with the physical all orders beta function and the SD approaches. 

The ladder result provides a size of the window, for every  fermion representation, smaller than the maximum bound found earlier. This is a consequence of the value of the anomalous dimension at the lower bound of the window. The unitarity constraint corresponds to $\gamma =2$ while the ladder result is closer to $\gamma \sim 1$. Indeed if we pick $\gamma =1$ our conformal window approaches the ladder result. Incidentally, a value of $\gamma$ larger than one, still allowed by unitarity, is a welcomed feature when using this window to construct walking technicolor theories. It may allow for the physical value of the mass of the top while avoiding a large violation of flavor changing neutral currents \cite{Luty:2004ye} which were investigated in  \cite{Evans:2005pu} in the case of the ladder approximation for minimal walking models.

\subsection{The $Sp(2N)$ phase diagram}
\label{sp}
$Sp(2N)$ is the subgroup of $SU(2N)$ which leaves the tensor
$J^{c_1 c_2} = ({\bf 1}_{N \times N} \otimes i \sigma_2)^{c_1 c_2}$
invariant. Irreducible tensors of $Sp(2N)$ must be traceless with respect to
$J^{c_1 c_2}$. 
Here we consider $Sp(2N)$ gauge theories with fermions transforming according to a given irreducible representation. Since $\pi^4\left[Sp(2N)\right] =Z_2$  there is a Witten topological anomaly \cite{Witten:1982fp} whenever the sum of the Dynkin indices of the various matter fields is odd. The adjoint of $Sp(2N)$ is the two-index symmetric tensor.

 \subsubsection{$Sp(2N)$ with Vector Fields}
 
Consider $2N_f$ Weyl fermions ${q_c^i}$ with ${c=1,\ldots,2N}$ and $i=1, \ldots , 2 N_f$  in the  fundamental representation of $Sp(2N)$. We have omitted the $SL(2,C)$ spinorial indices. We need an even number of flavors to avoid the Witten anomaly since the Dynkin index of the vector representation is equal to one. In the following Table we summarize the properties of the theory
\[ \begin{array}{|c|c|c|c|c|} \hline
{\rm Fields} &  \left[ Sp(2N) \right] & SU(2N_f) & T[r_i] & d[r_i] \\ \hline \hline
q &\Yfund & \Yfund & \frac{1}{2}& 2N   \\
G_{\mu}&{\rm Adj} = \Ysymm  &1&  N+1 & N(2N +1)  \\
 \hline \end{array} 
\]

\subsubsection*{\it Chiral Symmetry Breaking}

The theory is asymptotically free for $N_f \leq 11(N+1)/2$ while the relevant gauge singlet mesonic degree of freedom is:
\begin{equation}
M^{[i,j]} = \epsilon^{\alpha \beta} q_{\alpha,c_1}^{[ i} q_{\beta,c_2}^{j]} J^{c_1c_2}\ . 
\end{equation}
If the number of flavors is smaller than the critical number of flavors above which the theory develops an IRFP we expect this operator to condense and to break $SU(2N_f)$ to the maximal diagonal subgroup which is $Sp(2N_f)$ leaving behind $2N_f^2  - N_f -1$ Goldstone bosons. Also, there exist no $Sp(2N)$ stable operators constructed using the invariant tensor $\epsilon^{c_1c_2,\ldots c_{2N}}$ since they will break up into mesons $M$. This is so since  the invariant tensor $\epsilon^{c_1c_2 \ldots c_{2N}}$ breaks up into sums of products of $J^{c_1c_2}$.

\subsubsection*{\it All-orders Beta Function}

A zero in the numerator of the all orders beta function leads to the following value of the anomalous dimension of the mass operator at the IRFP: 
\begin{equation}
\gamma_{\Yfund} = \frac{11(N+1)}{N_f}  - 2 \ .
\end{equation}
Since the (mass) dimension of any scalar gauge singlet operator must be, by unitarity arguments, larger than one at the IRFP, this implies that $\gamma_{\Yfund} \leq 2$. Defining with $\gamma_{\Yfund}^{\ast}$ the maximal anomalous dimension above which the theory loses the IRFP the conformal window is:
\begin{equation}
 \frac{11}{4} (N+1) \leq  \frac{11}{2+\gamma_{\Yfund}^{\ast}} (N+1) \leq N_f \leq \frac{11}{2} (N+1) \ .
\end{equation}
{}For the first inequality we have taken the maximal value allowed for the anomalous dimension, i.e. $\gamma_{\Yfund}^{\ast} = 2$. 

\subsubsection*{\it SD}

The estimate from the truncated SD analysis yields as critical value of Weyl flavors:
\begin{equation} 
N_f^{\rm SD} = \frac{2(1+N)(67+100N)}{35+50N}  \ .
\end{equation}

\subsubsection*{\it Thermal Degrees of Freedoms}

In the UV we have $2N(2N+1)$ gauge bosons, where the extra factor of two comes from taking into account the two helicities of each massless gauge boson, and $4NN_f$ Weyl fermions. In the IR we have $2N_f^2  - N_f -1$ Goldstones and hence we have: 
\begin{equation}
f_{UV} = 2N(2N+1) + 7{NN_f} \ , \qquad f_{IR} = 2N_f^2  - N_f -1 \ .
\end{equation}
The number of flavors for which $f_{IR} = f_{UV}$ is 
\begin{equation}
N_f^{\rm Therm} = \frac{1+7N + \sqrt{3(3+10N+27N^2)}}{4} \ .
\end{equation}
No information can be obtained about the value of the anomalous dimension of the fermion bilinear at the fixed point. Assuming the conjecture to be valid the critical number of flavors  cannot exceed $N_f^{\rm Therm}$.

\begin{figure}[ht]
\centerline
{\includegraphics[height=6cm,width=18cm]{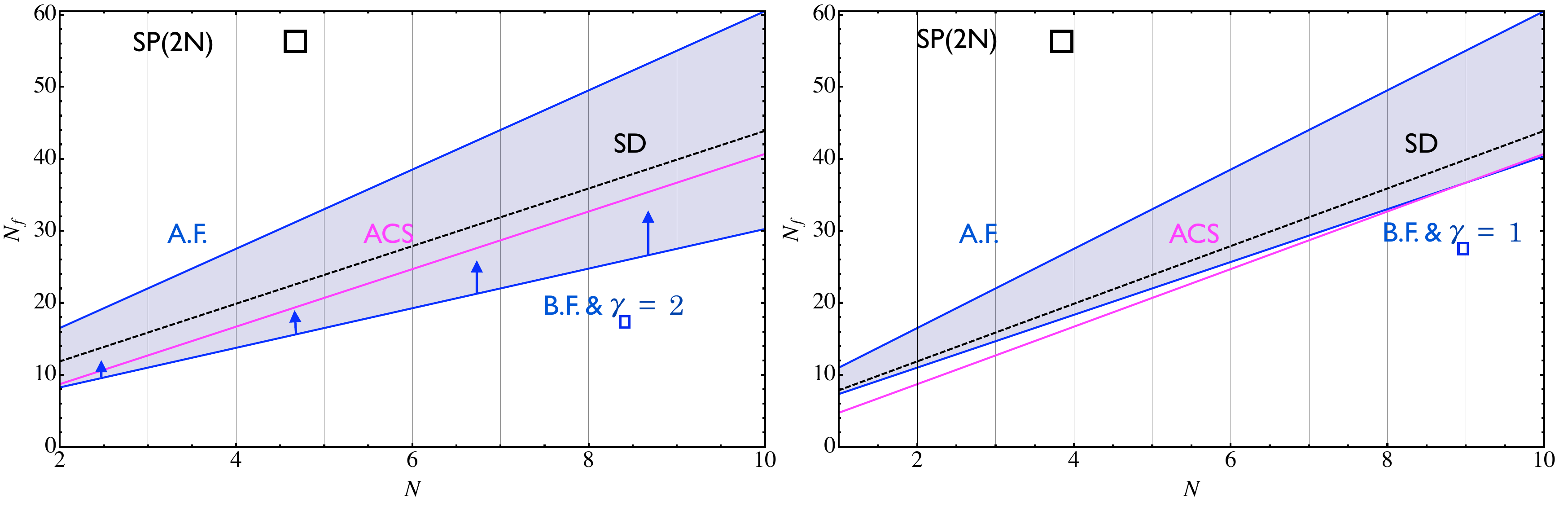}}
\caption
{Phase diagram of $Sp(2N)$ gauge theories with $2N_f$ fundamental Weyl fermions. {\it Left panel}: The upper solid (blue) line corresponds to the loss of asymptotic freedom and it is labeled by A.F.; the dashed (black) curve corresponds to the SD prediction for the breaking/restoring of chiral symmetry. The solid grey (magenta in color) line  corresponds to the ACS bound stating that the conformal region should start above this line. According to the all orders beta function (B.F.) the conformal window cannot extend below the solid (blue) line, as indicated by the arrows. This line corresponds to the anomalous dimension of the mass 
reaching the maximum value of $2$.  {\it Right panel}: The B.F. line is plotted assuming  the value of the anomalous dimension to be one.} 
\label{Sp-Fundamental}
\end{figure}

\subsubsection{On the limit  $N=1$ corresponding to $ SU(2)$}

In this case $N_f^{\rm Therm} = 2+\sqrt{\frac{15}{2}} \simeq 4.74$ and not $4\sqrt{4 - 16/81}\simeq 7.8$ as one deduces from equation (11) of \cite{Appelquist:1999hr}. The reason of the discrepancy is due to the fact that the fundamental representation of $SU(2) = Sp(2)$ is pseudoreal and hence the flavor symmetry is enhanced to $SU(2N_f)$. This enhanced symmetry is expected to break spontaneously to $Sp(2N_f)$. This yields $2N_f^2  - N_f -1$  Goldstone bosons rather than $N_f^2 - 1$ obtained assuming the global symmetry to be $SU(N_f)\times SU(N_f)\times U(1)$ spontaneously broken to $SU(N_f)\times U(1)$.  The corrected $N_f^{\rm Therm}$ value for $SU(2)$ is substantially lower than the $SD$ one which is $7.86$. The all orders beta function result is instead $5.5$ for the lowest possible value of $N_f$ below which chiral symmetry must break (corresponding to $\gamma_{\Yfund} = 2$). Imposing $\gamma_{\Yfund} =1$ (suggested by the SD approach) the all orders beta function returns $7.3$ which is closer to the $SD$ prediction. Note that there is some phenomenological interest in the $SU(2)$ gauge theory with fermionic matter in the fundamental representation. {}For example the case of $N_f=8$ has been employed in the literature as a possible template for early models of walking technicolor \cite{Appelquist:1997fp}. 
 
These results indicate that it is interesting to study the $SU(2)$ gauge theory with $N_f=5$ Dirac flavors via first principles Lattice simulation. This will allow to discriminate between the two distinct predictions, the one from the ACS and the one from the all orders beta function.

\subsubsection{$Sp(2N)$ with Adjoint Matter Fields}

Consider $N_f$ Weyl fermions ${q_{\{ c_1,c_2 \}}^i}$ with $c_1$ and $c_2$ ranging from $1$ to $2N$ and $i=1, \ldots , N_f$. This is the  adjoint representation of $Sp(2N)$ with Dynkin index $2(N+1)$. Since it is even for any $N$ there is no Witten anomaly for any $N_f$. In the following Table we summarize the properties of the theory

 \[ \begin{array}{|c|c|c|c|c|} \hline
{\rm Fields} &  \left[ Sp(2N) \right] & SU(N_f) & T[r_i] & d[r_i] \\ \hline \hline
q &\Ysymm & \Yfund &  N+1&   N(2N +1)\\
G_{\mu}&{\rm Adj} = \Ysymm  &1&  N+1 & N(2N +1) \\
 \hline \end{array} 
\]

\subsubsection*{\it Chiral Symmetry Breaking}

The theory is asymptotically free for $N_f \leq 11/2$ (recall that $N_f$ here is the number of Weyl fermions) while the relevant gauge singlet mesonic degree of freedom is:
\begin{equation}
M^{\{i,j \}} = \epsilon^{\alpha \beta} q_{\alpha, \{c_1,c_2\}}^{\{ i} q_{\beta, \{c_3,c_4 \}}^{j\}} J^{c_1c_3}J^{c_2c_4}\ . 
\end{equation}
If the number of flavors is smaller than the critical number of flavors above which the theory develops an IRFP we expect this operator to condense and to break $SU(N_f)$ to the maximal diagonal subgroup which is $SO(N_f)$ leaving behind $(N_f^2  + N_f -2)/2$ Goldstone bosons.

\subsubsection*{\it All-orders Beta Function}

Here the anomalous dimension of the mass operator at the IRFP is: 
\begin{equation}
\gamma_{\Ysymm} = \frac{11}{N_f}  - 2 \ .
\end{equation}
Since the dimension of any scalar gauge singlet operator must be larger than one at the IRFP, this implies that $\gamma_{\Ysymm} \leq 2$. Defining with $\gamma_{\Ysymm}^{\ast}$ the maximal anomalous dimension above which the theory loses the IRFP the conformal window is:
\begin{equation}
 \frac{11}{4} \leq  \frac{11}{2+\gamma_{\Ysymm}^{\ast}}  \leq N_f \leq \frac{11}{2} \ .
\end{equation}
 
\subsubsection*{\it SD}

The estimate from the truncated SD analysis yields as critical value of flavors:
\begin{equation} 
N_f^{\rm SD} =4.15  \ .
\end{equation}

\subsubsection*{\it Thermal Degrees of Freedoms}

In the ultraviolet we have $2N(2N+1)$ gauge bosons and $N(2N+1)N_f$ Weyl fermions. In the IR we have $(N_f^2  + N_f -2)/2$  Goldstone bosons. Hence: 
\begin{equation}
f_{UV} = 2N(2N+1) + \frac{7}{4}{N(2N+1)N_f} \ , \qquad f_{IR} =\frac{ N_f^2  + N_f -2}{2} \ .
\end{equation}
The number of flavors for which $f_{IR} = f_{UV}$ is 
\begin{equation}
N_f^{\rm Therm} = \frac{-2 + 7 N + 14 N^2 + \sqrt{36 + 36 N + 121 N^2 + 196 N^3 + 196 N^4}}{4} \ .
\end{equation}
This is a monotonically increasing function of $N$ which even for a value of $N$ as low as 2 yields $N_f^{\rm Therm}=35.2$ which is several times higher than the limit set by asymptotic freedom. Although this fact does not contradict the statement that the critical number of flavors is lower than $N_f^{\rm Therm}$ it shows that this conjecture does not lead to useful constraints when looking at higher dimensional representations as we observed in \cite{Sannino:2005sk} when discussing higher dimensional representations for $SU(N)$ gauge groups.

\begin{figure}[ht]
\centerline{
\includegraphics[height=6cm,width=18cm]{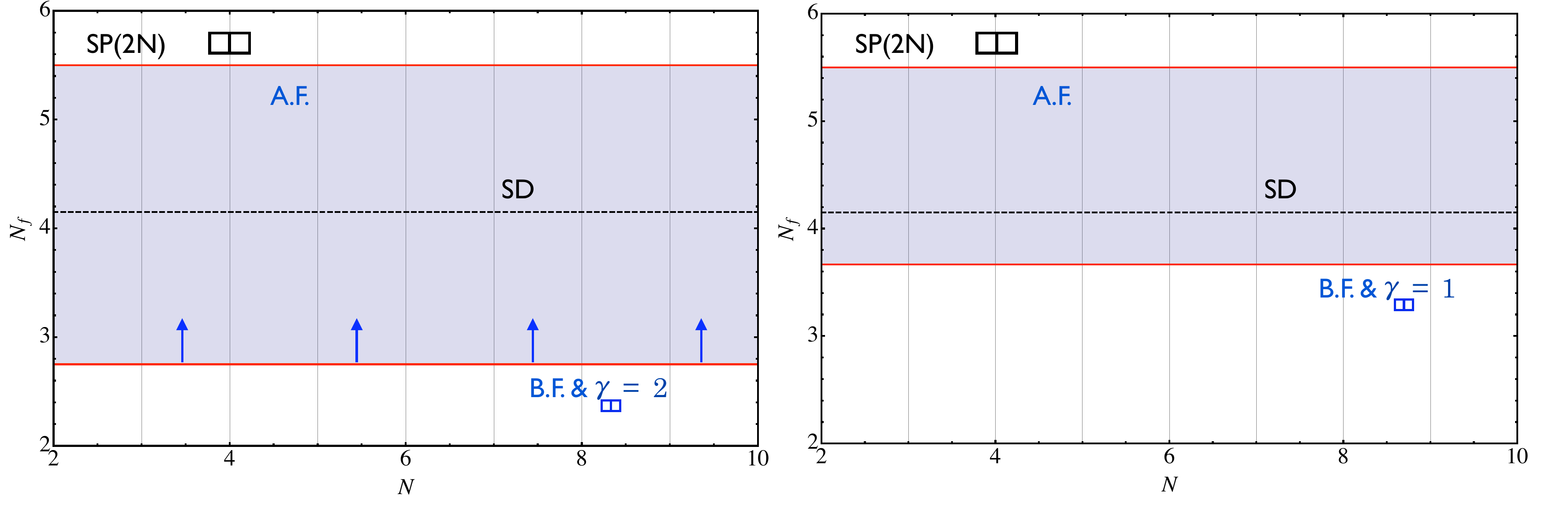}}
\caption
{Phase diagram of $Sp(2N)$ gauge theories with $N_f$ adjoint Weyl fermions. {\it Left panel}: The upper solid (red) line corresponds to the loss of asymptotic freedom and it is labeled by A.F.; the dashed (black) curve corresponds to the SD prediction for the breaking/restoring of chiral symmetry. According to the all orders beta function (B.F.) the conformal window cannot extend below the solid (red) line, as indicated by the arrows. This line corresponds to the anomalous dimension of the mass 
reaching the maximum value of $2$.  {\it Right panel}: The B.F. line is plotted assuming  the value of the anomalous dimension to be one.} 
\label{Sp-Adj}
\end{figure}

 \subsubsection{$Sp(2N)$ with  Two-Index Anti-Symmetric Representation}

Consider $N_f$ Weyl fermions ${q_{[ c_1,c_2 ]}^i}$ with $c_1$ and $c_2$ ranging from $1$ to $2N$ and $i=1, \ldots , N_f$. As for the two-index symmetric case here too the Dynkin index is even  and hence we need not to worry about the Witten anomaly. In the following Table we summarize the properties of the theory
 \[ \begin{array}{|c|c|c|c|c|} \hline
{\rm Fields} &  \left[ Sp(2N) \right] & SU(N_f) & T[r_i] & d[r_i] \\ \hline \hline
q &\Yasymm & \Yfund &  N-1&N(2N-1) -1 \\
G_{\mu}&{\rm Adj} = \Ysymm  &1&  N+1 & N(2N +1)  \\
 \hline \end{array} 
\]

\subsubsection*{\it Chiral Symmetry Breaking}

The theory is asymptotically free for $\displaystyle{N_f \leq \frac{11({N+1})}{2(N-1)}}$ with the relevant gauge singlet mesonic degree of freedom being:
\begin{equation}
M^{\{i,j \}} = \epsilon^{\alpha \beta} q_{\alpha, [c_1,c_2] }^{\{ i} q_{\beta, [c_3,c_4 ]}^{j\}} J^{c_1c_3}J^{c_2c_4}\ . 
\end{equation}
If the number of flavors is smaller than the critical number of flavors above which the theory develops an IRFP we expect this operator to condense and to break $SU(N_f)$ to the maximal diagonal subgroup which is $SO(N_f)$ leaving behind $(N_f^2  + N_f -2)/2$ Goldstone bosons.

\subsubsection*{\it All-orders Beta Function}

The anomalous dimension of the mass operator at the IRFP is: 
\begin{equation}
\gamma_{\Yasymm} = \frac{11(N+1) -2N_f(N-1)} {N_f(N-1)}  .
\end{equation}
 Defining with $\gamma_{\Yasymm}^{\ast}$ the maximal anomalous dimension above which the theory loses the IRFP the conformal window is:
\begin{equation}
 \frac{11}{4} \frac{N+1}{N-1} \leq  \frac{11}{2+\gamma_{\Yasymm}^{\ast}} \frac{N+1}{N-1} \leq N_f \leq \frac{11}{2}   \frac{N+1}{N-1} \ .
\end{equation}
The maximal value allowed for the anomalous dimension is $\gamma_{\Yasymm}^{\ast} = 2$. 

\subsubsection*{\it SD}

The SD analysis yields as critical value of flavors:
\begin{equation} 
N_f^{\rm SD} =\frac{(1+N) (83N+17)}{5(4N^2 - 3N - 1)}  \ .
\end{equation}

\subsubsection*{\it Thermal Degrees of Freedoms}

In the ultraviolet we have $2N(2N+1)$ gauge bosons and $(N(2N-1)-1)N_f$ Weyl fermions. In the IR we have $(N_f^2  + N_f -2)/2$  Goldstone bosons. Hence: 
\begin{equation}
f_{UV} = 2N(2N+1) + \frac{7}{4}{(N(2N-1)-1)N_f} \ , \qquad f_{IR} =\frac{ N_f^2  + N_f -2}{2} \ .
\end{equation}
The number of flavors for which $f_{IR} = f_{UV}$ is 
\begin{equation}
N_f^{\rm Therm} = \frac{-9 - 7 N + 14 N^2 + \sqrt{113 + 190 N - 75 N^2 - 196 N^3 + 196 N^4}}{4} \ .
\end{equation}
As explained above no useful constraint can be set with this criterion \cite{Sannino:2005sk}.

\begin{figure}[ht]
\centerline{
\includegraphics[height=6cm,width=18cm]{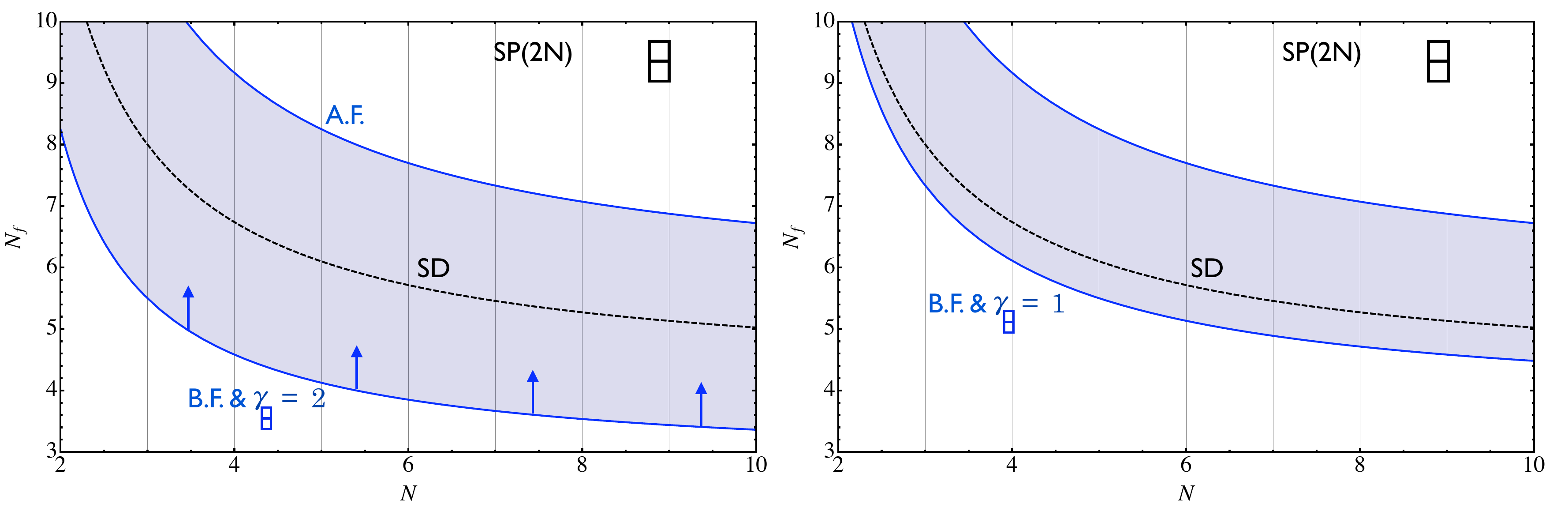}}
\caption
{Phase diagram of $Sp(2N)$ gauge theories with $N_f$ two-index antisymmetric Weyl fermions. {\it Left panel}: The upper solid (blue) curve corresponds to the loss of asymptotic freedom and it is labeled by A.F.; the dashed (black) curve corresponds to the SD prediction for the breaking/restoring of chiral symmetry. According to the all orders beta function (B.F.) the conformal window cannot extend below the solid (blue) curve, as indicated by the arrows. This curve corresponds to the anomalous dimension of the mass 
reaching the maximum value of $2$.  {\it Right panel}: The B.F. curve is plotted assuming  the value of the anomalous dimension to be one.} 
\label{Sp-Ant}
\end{figure}

\subsubsection*{\it Summary of the Results for $SP(2N)$ Gauge Theories}

\begin{figure}[ht]
\centerline{
\includegraphics[height=8cm,width=13cm]{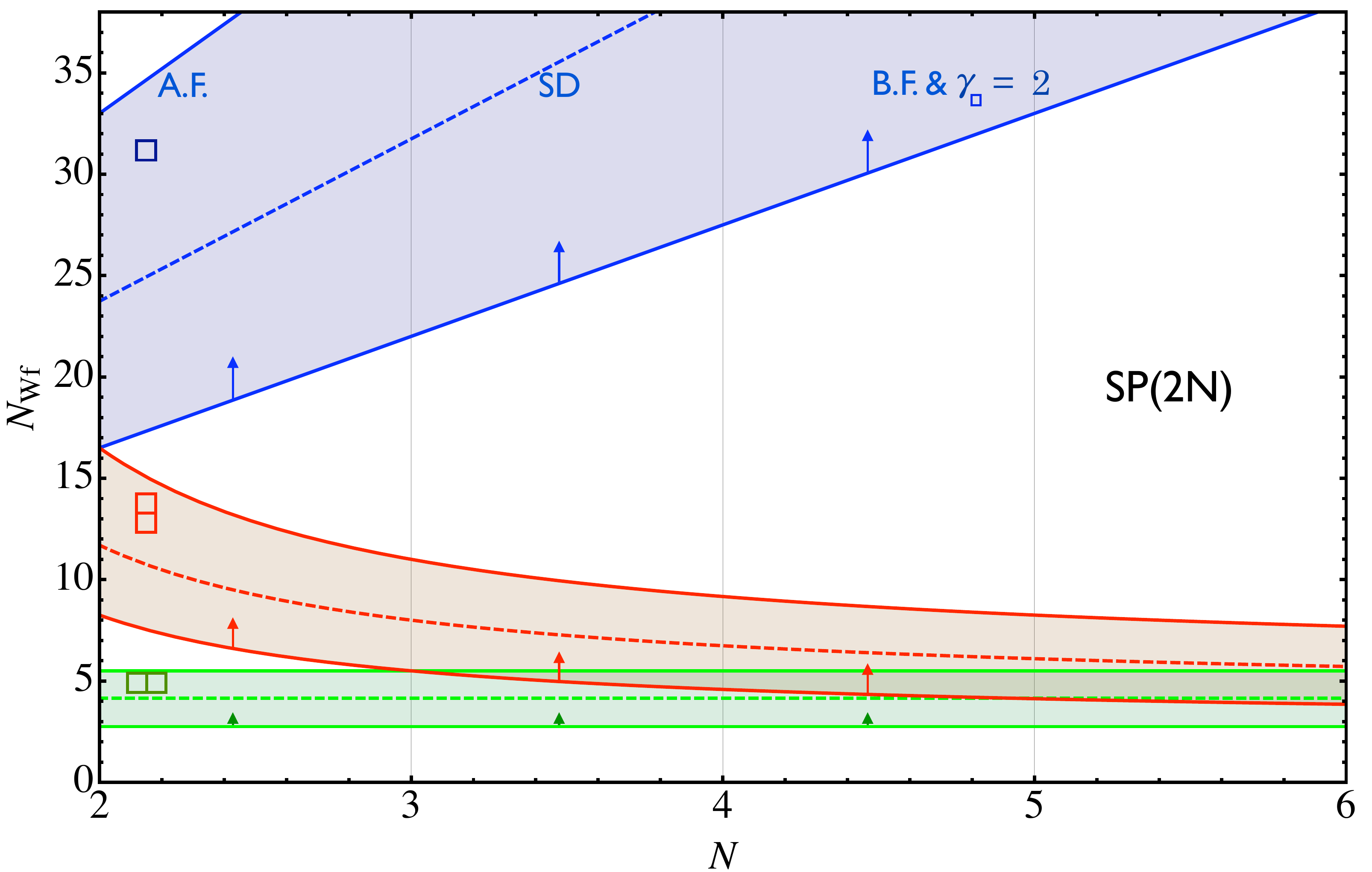}}
\caption
{Phase Diagram, from top to bottom, for $Sp(2N)$ Gauge Theories with $N_{Wf}=2N_f$  Weyl fermions in the vector representation (light blue),   $N_{Wf}=N_f$ in the  two-index antisymmetric representation (light red) and finally in the two-index symmetric (adjoint) (light green). The arrows indicate that the conformal windows can be smaller and the associated solid curves correspond to the all orders beta function prediction for the maximum extension of the conformal windows.} 
\label{Sp-PhaseDiagram}
\end{figure}

In Figure~\ref{Sp-PhaseDiagram} we summarize the relevant zero temperature and matter density phase diagram as function of the number of colors and Weyl flavors ($N_{Wf}$) for $Sp(2N)$ gauge theories. For the vector representation $N_{Wf} = 2N_f$ while for the two-index theories $N_{Wf} = N_f$. The shape of the various conformal windows are very similar to the ones for $SU(N)$ gauge theories \cite{Sannino:2004qp,Dietrich:2006cm,Ryttov:2007cx} with the difference that in this case the two-index symmetric representation is the adjoint representation and hence there is one less conformal window.

\subsection{The $SO(N)$ phase diagram}
\label{so}
We shall consider $SO(N)$ theories (for $N>5$) since they do not suffer of  a Witten
anomaly \cite{Witten:1982fp} and, besides, for $N<7$ can always be reduced to either an $SU$ or an $Sp$ theory.

 \subsubsection{$SO(N)$ with vector fields}
 
Consider $N_f$ Weyl fermions ${q_c^i}$ with ${c=1,\ldots,N}$ and $i=1, \ldots , N_f$  in the vector representation of $SO(N)$. In the following Table we summarize the properties of the theory

\[ \begin{array}{|c|c|c|c|c|} \hline
{\rm Fields} &  \left[ SO(N) \right] & SU(N_f) & T[r_i] & d[r_i] \\ \hline \hline
q &\Yfund & \Yfund & 1& N   \\
G_{\mu}&{\rm Adj} = \Yasymm  &1&  N-2 & \frac{N(N-1)}{2}  \\
 \hline \end{array} 
\]

\subsubsection*{\it Chiral Symmetry Breaking}

The theory is asymptotically free for $\displaystyle{N_f \leq \frac{11({N-2})}{2}}$.  The relevant gauge singlet mesonic degree of freedom is:
\begin{equation}
M^{\{i,j \}} = \epsilon^{\alpha \beta} q_{\alpha, c_1 }^{\{ i} q_{\beta, c_2}^{j\}} \delta^{c_1c_2}\ . 
\end{equation}
If the number of flavors is smaller than the critical number of flavors above which the theory develops an IRFP we expect this operator to condense and to break $SU(N_f)$ to the maximal diagonal subgroup which is $SO(N_f)$ leaving behind $(N_f^2  + N_f -2)/2$ Goldstone bosons.

\subsubsection*{\it All-orders Beta Function}

The anomalous dimension of the mass operator at the IRFP is: 
\begin{equation}
\gamma_{\Yfund} = \frac{11(N-2) } {N_f}  - 2 \  .
\end{equation}
Defining with $\gamma_{\Yfund}^{\ast}$ the maximal anomalous dimension above which the theory loses the IRFP the conformal window reads:
\begin{equation}
 \frac{11}{4} ({N-2}) \leq  \frac{11}{2+\gamma_{\Yfund}^{\ast}}( {N-2}) \leq N_f \leq \frac{11}{2}  ( {N-2} ) \ .
\end{equation}
The maximal value allowed for the anomalous dimension is $\gamma_{\Yfund}^{\ast} = 2$. 

\subsubsection*{\it SD}

The SD analysis yields as critical value of flavors:
\begin{equation} 
N_f^{\rm SD} =\frac{2(N-2) (50N-67)}{5(5N - 7)}  \ .
\end{equation}

\subsubsection*{\it Thermal Degrees of Freedoms}

In the ultraviolet we have $N(N- 1)$ gauge bosons and $NN_f$ Weyl fermions. In the IR we have $(N_f^2  + N_f -2)/2$  Goldstone bosons. Hence: 
\begin{equation}
f_{UV} = N(N-1) + \frac{7}{4}{NN_f} \ , \qquad f_{IR} =\frac{ N_f^2  + N_f -2}{2} \ .
\end{equation}
The number of flavors for which $f_{IR} = f_{UV}$ is 
\begin{equation}
N_f^{\rm Therm} = \frac{-2 + 7 N + \sqrt{36 - 60 N + 81 N^2}}{4} \ .
\end{equation}

This value is larger than the $SD$ result and it is larger than the asymptotic freedom constraint for $N<7$. This is not too surprising since the vector representation of $SO(N)$ for small $N$ becomes a higher representation of other groups for which we have already shown that this method is unconstraining  \cite{Sannino:2005sk}.

\begin{figure}[ht]
\centerline{
\includegraphics[height=6cm,width=18cm]{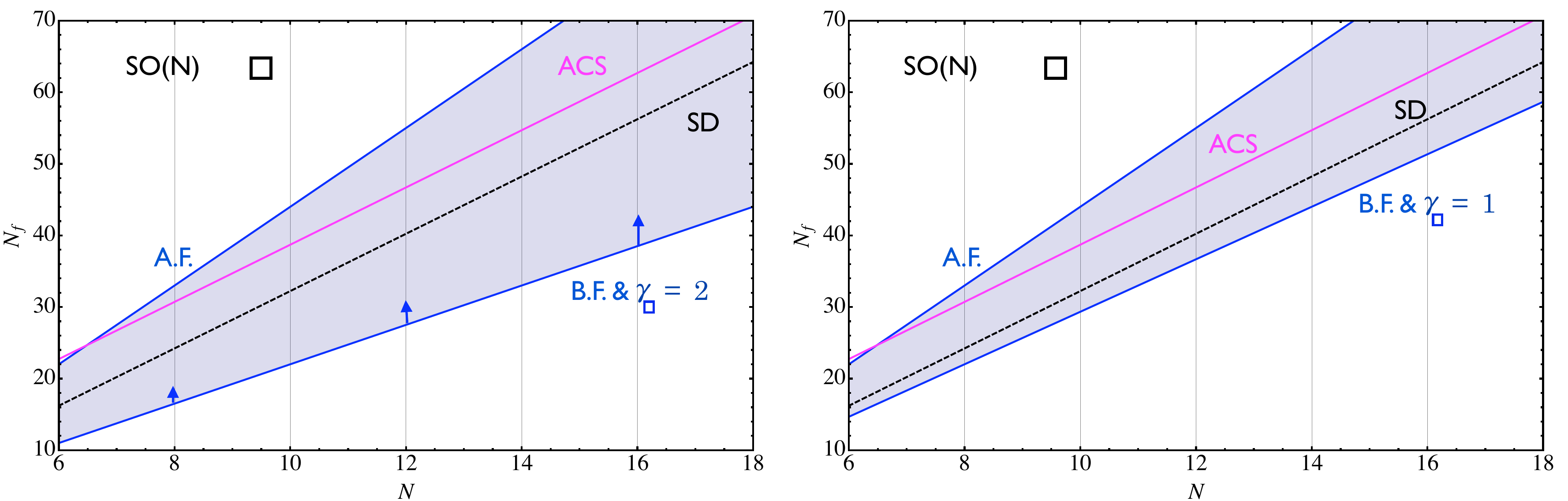}}
\caption
{Phase diagram of $SO(N)$ gauge theories with $N_f$ fundamental Weyl fermions. {\it Left panel}: The upper solid (blue) line corresponds to the loss of asymptotic freedom and it is labeled by A.F.; the dashed (black) curve corresponds to the SD prediction for the breaking/restoring of chiral symmetry. The solid grey (magenta in color) line  corresponds to the ACS bound stating that the conformal region should start above this line. According to the all orders beta function (B.F.) the conformal window cannot extend below the solid (blue) line, as indicated by the arrows. This line corresponds to the anomalous dimension of the mass 
reaching the maximum value of $2$.  {\it Right panel}: The B.F. line is plotted assuming  the value of the anomalous dimension to be one.} 
\label{SO-Fundamental}
\end{figure}

Note that the ACS line is always above the SD result. 

 \subsubsection{$SO(N)$ with Adjoint Matter Fields}
  
Consider $N_f$ Weyl fermions ${q_{[c_1,c_2]}^i}$ with $c_1$ and $c_2$ varying in the range $1,\ldots,N $ and $i=1, \ldots , N_f$. This is the  adjoint representation of $SO(N)$. In the following Table we summarize the properties of the theory

 \[ \begin{array}{|c|c|c|c|c|} \hline
{\rm Fields} &  \left[ SO(N) \right] & SU(N_f) & T[r_i] & d[r_i] \\ \hline \hline
q &\Yasymm & \Yfund &  N-2& \frac{N(N-1)}{2}  \\
G_{\mu}&{\rm Adj} = \Yasymm  &1&  N-2 & \frac{N(N-1)}{2}  \\
 \hline \end{array} 
\]

The analysis leads to a conformal window which is an identical copy of the one for the adjoint matter of the $Sp$ gauge theory which is also identical to the $SU$ case with adjoint matter. 

 \subsubsection{$SO(N)$ with  Two-Index Symmetric Representation}

Consider $N_f$ Weyl fermions ${q_{\{c_1,c_2 \}}^i}$ with $c_1$ and $c_2$ varying in the range $1,\ldots,N $ and $i=1, \ldots , N_f$, i.e. in the two-index symmetric representation of $SO(N)$.  In the following Table we summarize the properties of the theory
 
 \[ \begin{array}{|c|c|c|c|c|} \hline
{\rm Fields} &  \left[ SO(N) \right] & SU(N_f) & T[r_i] & d[r_i] \\ \hline \hline
q &\Ysymm & \Yfund &  N+2& \frac{N(N+1)}{2} -1  \\
G_{\mu}&{\rm Adj} = \Yasymm  &1&  N-2 & \frac{N(N-1)}{2}  \\
 \hline \end{array} 
\]

\subsubsection*{\it Chiral Symmetry Breaking}

The theory is asymptotically free for $\displaystyle{N_f \leq \frac{11({N-2})}{2(N+2)}}$.  The relevant gauge singlet mesonic degree of freedom is:
\begin{equation}
M^{\{i,j \}} = \epsilon^{\alpha \beta} q_{\alpha, \{c_1,c_2\} }^{\{ i} q_{\beta, \{c_3,c_4\}}^{j\}} \delta^{c_1c_3}\delta^{c_2,c_4}\ . 
\end{equation}
If the number of flavors is smaller than the critical number of flavors above which the theory develops an IRFP we expect this operator to condense and to break $SU(N_f)$ to the maximal diagonal subgroup which is $SO(N_f)$ leaving behind $(N_f^2  + N_f -2)/2$ Goldstone bosons.

\subsubsection*{\it All-orders Beta Function}

The anomalous dimension of the mass operator at the IRFP is: 
\begin{equation}
\gamma_{\Ysymm} = \frac{11(N-2) } {N_f(N+2)}  - 2 \  .
\end{equation}
Defining with $\gamma_{\Ysymm}^{\ast}$ the maximal anomalous dimension above which the theory loses the IRFP the conformal window reads:
\begin{equation}
\frac{11}{4} \frac{N-2}{N+2} \leq  \frac{11}{2+\gamma_{\Ysymm}^{\ast}} \frac{N-2}{N+2} \leq N_f \leq \frac{11}{2}  \frac{N-2}{N+2} \ .
\end{equation}
The maximal value allowed for the anomalous dimension is $\gamma_{\Ysymm}^{\ast} = 2$. 

\subsubsection*{\it SD}

The SD analysis yields as critical value of flavors:
\begin{equation} 
N_f^{\rm SD} =\frac{(N-2) (83N-34)}{10(2N^2 +3N -2)}  \ .
\end{equation}

\subsubsection*{\it Thermal Degrees of Freedoms}

In the ultraviolet we have $N(N- 1)$ gauge bosons and $(N\frac{(N+1)}{2} -1)N_f$ Weyl fermions. In the IR we have $(N_f^2  + N_f -2)/2$  Goldstone bosons. Hence: 
\begin{equation}
f_{UV} = N(N-1) + \frac{7}{4}{(N\frac{(N+1)}{2} -1)N_f} \ , \qquad f_{IR} =\frac{ N_f^2  + N_f -2}{2} \ .
\end{equation}
The number of flavors for which $f_{IR} = f_{UV}$ is 
\begin{equation}
N_f^{\rm Therm} = \frac{-18 + 7 N(1+N) + \sqrt{452 + N (-380 + N (-75 + 49 N (2 + N)))}}{8} \ .
\end{equation}

This value is several times larger than the asymptotic freedom result and hence poses no constraint  \cite{Sannino:2005sk}.

\begin{figure}[ht]
\centerline{
\includegraphics[height=6cm,width=18cm]{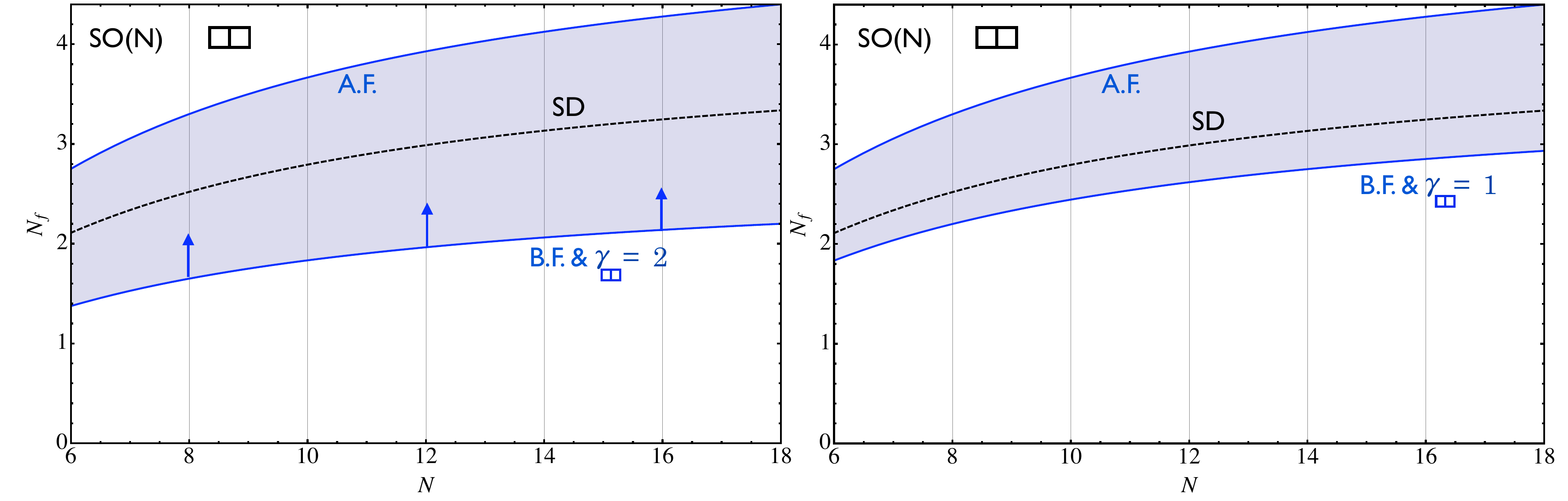}}
\caption
{Phase diagram of $SO(N)$ gauge theories with $N_f$ Weyl fermions in the two-index symmetric representation. {\it Left panel}: The upper solid (blue) curve corresponds to the loss of asymptotic freedom and it is labeled by A.F.; the dashed (black) curve corresponds to the SD prediction for the breaking/restoring of chiral symmetry. According to the all orders beta function (B.F.) the conformal window cannot extend below the solid (blue) curve, as indicated by the arrows. This curve corresponds to the anomalous dimension of the mass 
reaching the maximum value of $2$.  {\it Right panel}: The B.F. curve is plotted assuming  the value of the anomalous dimension to be one.} 
\label{SO-Sym}
\end{figure}

\subsubsection*{\it Summary for $SO(N)$ gauge theories}

\begin{figure}[t!]
\centerline{
\includegraphics[height=8cm,width=13cm]{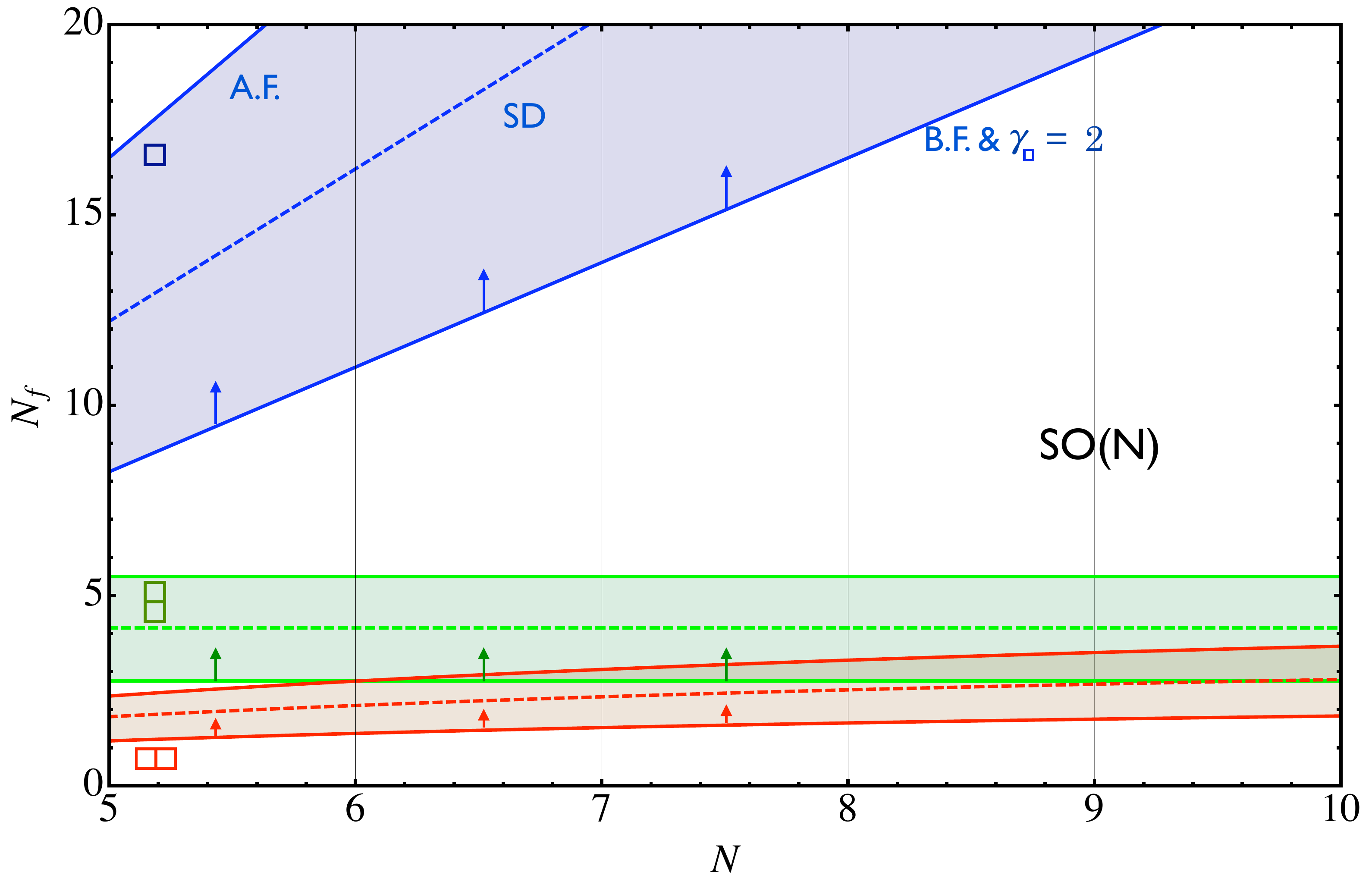}}
\caption
{Phase diagram of $SO(N)$ gauge theories with $N_f$  Weyl fermions in the vector representation, in the two-index antisymmetric (adjoint) and finally in the  two-index symmetric representation.  The arrows indicate that the conformal windows can be smaller and the associated solid curves correspond to the all orders beta function prediction for the maximum extension of the conformal windows. } 
\label{So-PhaseDiagram}
\end{figure}

In the Fig.~\ref{So-PhaseDiagram} we summarize the relevant zero temperature and matter density phase diagram as function of the number of colors and Weyl flavors ($N_{f}$) for $SO(N)$ gauge theories. The shape of the various conformal windows are very similar to the ones for $SU(N)$  and $Sp(2N)$ gauge  with the difference that in this case the two-index antisymmetric representation is the adjoint representation. We have analyzed only the theories with $N\geq 6$ since the remaining smaller $N$ theories can be deduced from $Sp$ and $SU$ using the fact that 
$SO(6)\sim SU(4)$, $SO(5)\sim Sp(4)$, 
$SO(4)\sim SU(2)\times SU(2)$, $SO(3)\sim SU(2)$, and $SO(2)\sim U(1)$.  

 At infinite $N$ it is impossible to distinguish theories with matter in the two-index symmetric representation from theories with matter in the two-index antisymmetric. This means that, in this regime, one has an obvious equivalence between theories with these two types of matter. This statement is independent on whether the gauge group is $SU$, $Sp$ or $SO(N)$. What distinguishes $SU$ from both $Sp$ and $SO$ is the fact that in these two cases one of the two two-index representations is, in fact, the adjoint representation. This simple observation automatically implies that one Weyl flavor in the two-index symmetric (antisymmetric) representation of $SO(N)$($Sp(2N)$) becomes indistinguishable from pure super Yang-Mills at large N. The original observation appeared first within the context of string theory and it is due to Sugimoto \cite{Sugimoto:1999tx} and Uranga \cite{Uranga:1999ib}. A similar comment was made in \cite{Armoni:2007jt}.

\subsection{Conformal House}

Till now the investigations dealt with fermions in a single representation of the gauge group. In fact these constitute only a small fraction of all of the possible gauge theories we can envision built out of fermions in several representations. A priori there is no reason to exclude these theories from interesting applications. In fact, we have very recently shown that one of these theories leads to a novel model of dynamical electroweak symmetry breaking possessing several interesting phenomenological features \cite{Ryttov:2008xe}.

In \cite{Ryttov:2009yw} we initiated the first systematic study of conformal gauge dynamics associated to nonsupersymmetric gauge theories featuring matter in two different representations of the undelying gauge group. The region in flavor/color space bounding the fraction of the theory developing a conformal behavior at large distances is a three-dimensional volume. Two faces of this volume correspond the the conformal areas of the gauge theory when on of the flavor numbers is set to zero. These areas are often referred as conformal {\it windows}. It is then natural to indicate the conformal volumes as the conformal {\it houses} whose {\it windows} are the two dimensional conformal areas.

\begin{figure}[h!]
\centerline{\includegraphics[height=7cm,width=10cm]{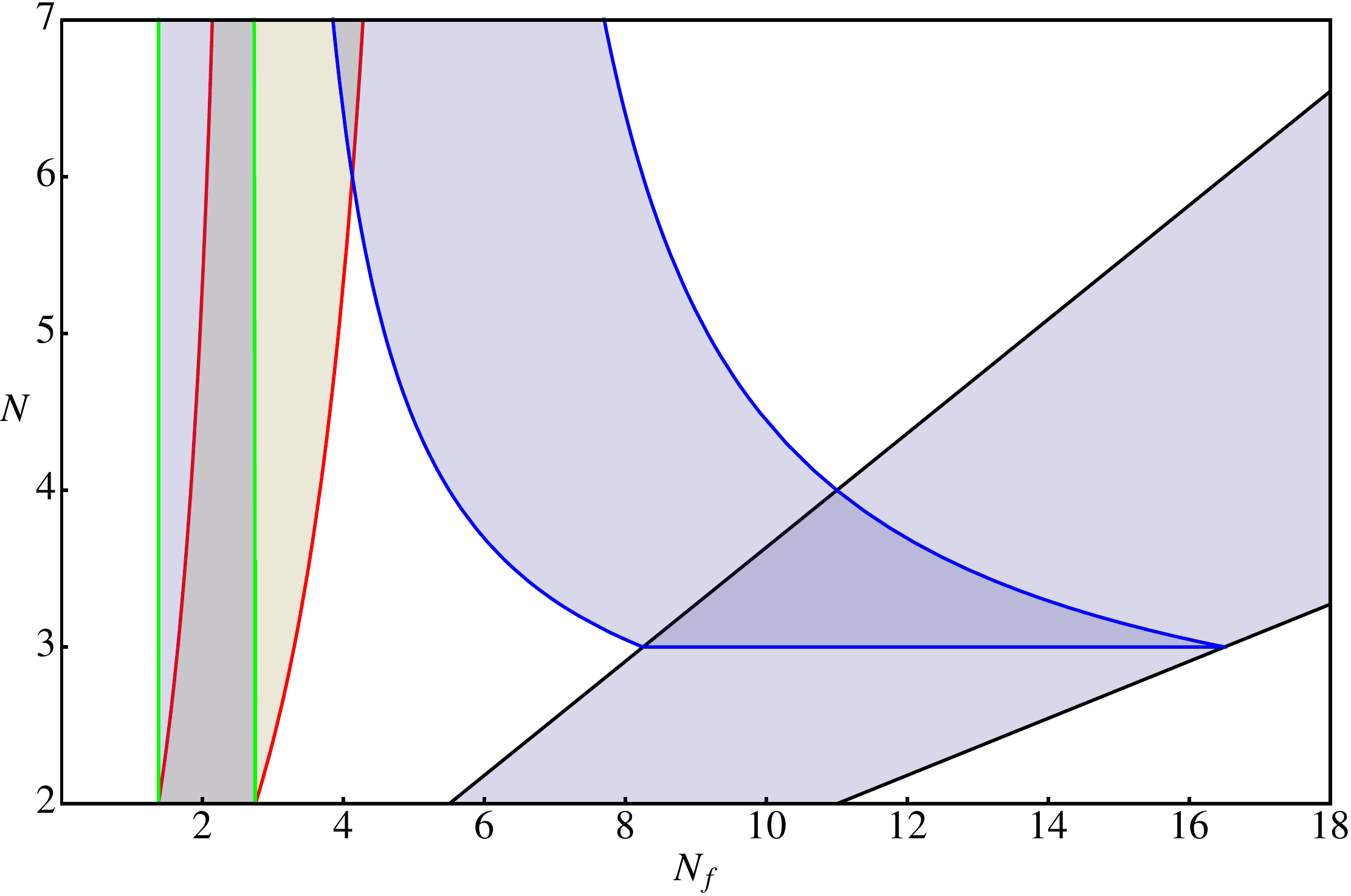}}
\caption{Phase diagram for non-supersymmetric theories with fermions in the: i) fundamental representation (black), ii) two-indexed antisymmetric representation (blue), iii) two-indexed symmetric representation (red), iv) adjoint representation (green) as a function of the number of colors and number of flavors. The shaded areas depict the corresponding conformal windows stemming from the all orders beta function. To the right of the shaded ares the theories are no longer asymptotically free while to the left of the shaded areas the theories are expected to break chiral symmetry. }\label{ReversePD}
\end{figure}

Let us review the case of a single representation but with a little twist, i.e. we draw it with the number of colors on the vertical axis. See Fig. \ref{ReversePD} where we used  the all orders beta function results. 

\begin{figure}[ht!]
\centerline{\includegraphics[height=6.5cm,width=8.82cm]{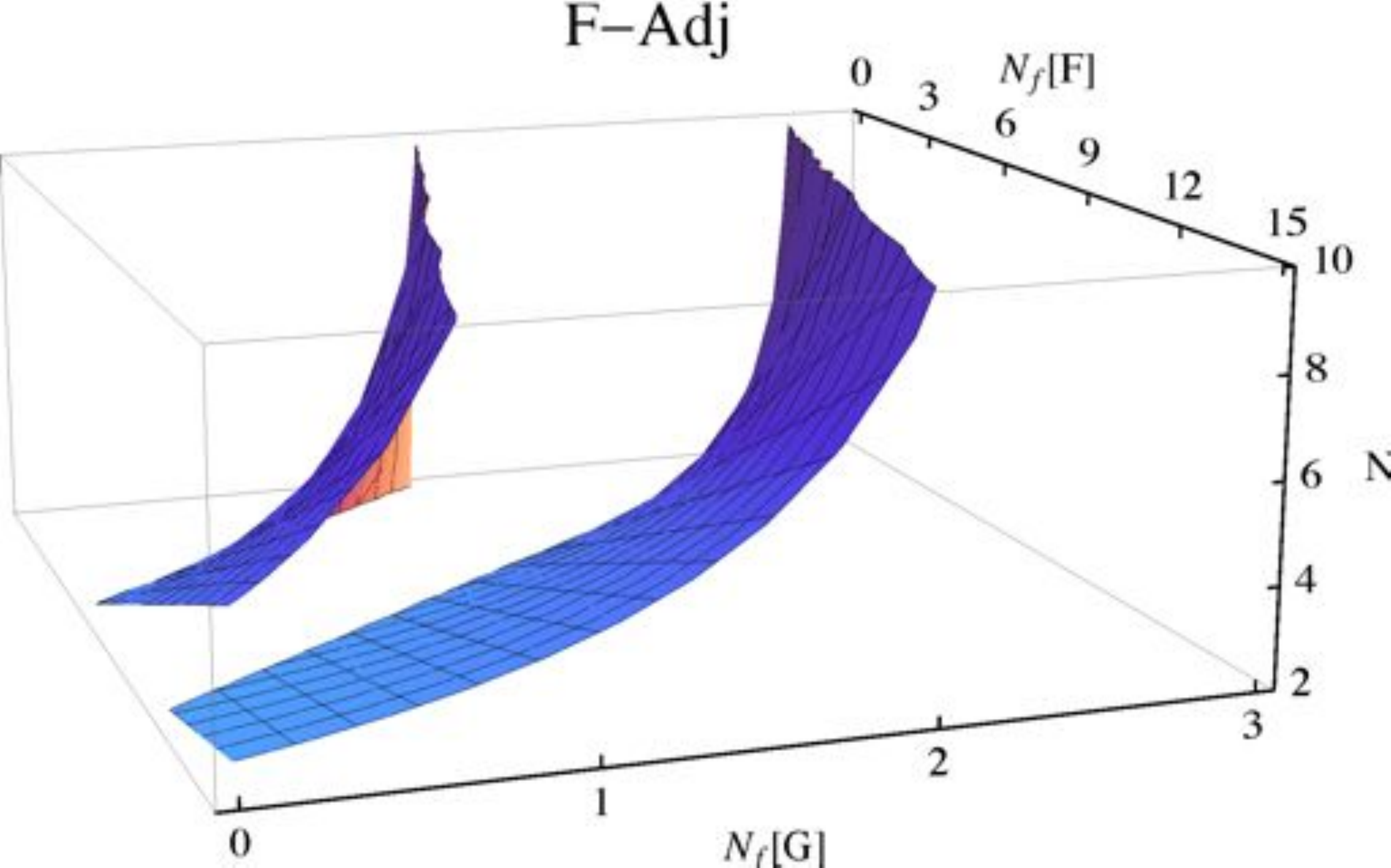}\hspace{0.8cm}}
\caption{The conformal {\it house} for a non-supersymmetric gauge theory containing fundamental and adjoint fermions. To the right of the right surface the theories are non-asymptotically free while to the left of the left surface the theories break chiral symmetry. Between the two surfaces the theories can develop an infrared fixed point.}
\label{F-Adj}
\end{figure}

Let us now generalize to multiple representations. First, the loss of asymptotic freedom is determined by the change of sign in the first coefficient of the beta function. This occurs when
\begin{eqnarray} \label{MultiAF}
\sum_{i=1}^{k} \frac{4}{11} T(r_i) N_f(r_i) = C_2(G) \ .
\end{eqnarray}
Second, we note that at the zero of the beta function we have
\begin{eqnarray}
\sum_{i=1}^{k} \frac{2}{11}T(r_i)N_f(r_i)\left( 2+ \gamma_i \right) = C_2(G) \ .
\end{eqnarray}
 Therefore specifying the value of the anomalous dimensions at the infrared fixed point yields the last constraint needed to construct the conformal region. Having reached the zero of the beta function the theory is conformal in the infrared. For a theory to be conformal the dimension of the non-trivial spinless operators must be larger than one in order to not contain negative norm states \cite{Mack:1975je,Flato:1983te,Dobrev:1985qv}.  Since the dimension of each chiral condensate is $3-\gamma_i$ we see that $\gamma_i = 2$, for all representations $r_i$, yields the maximum possible bound \footnote{Note that $\gamma \leq 2$ is an {\it exact} bound  \cite{Mack:1975je,Flato:1983te,Dobrev:1985qv}, i.e. does not depend on model computations. If it turns out that dynamically a smaller value of $\gamma$ actually delimits the conformal window this value must be less than $2$ and hence does not affect our results on the bound of the conformal windows.  }
\begin{eqnarray}\label{MultipleBound}
\sum_{i=1}^{k} \frac{8}{11} T(r_i)N_f(r_i) = C_2(G) \ .
\end{eqnarray}
For two distinct representations the conformal region is a three dimensional volume, i.e. the conformal {\it house}. The {\it windows} of the house correspond exactly to the conformal windows presented in the previous section.  In Fig. \ref{F-Adj} we plot the bound of the conformal volume in the case of fundamental and adjoint fermions.
{}For completeness we also plot below in Fig.~\ref{fig:fund} the bound on the conformal house with one species of fermions in the fundamental representation and the other in the two-index (anti)symmetric  in the (right)left panel.  We consider only two-index representations in Fig.~\ref{fig:adj}, more specifically we consider the adjoint representation together with the two-index (anti)symmetric in the \ref{fig:adj} (right) left panel. Note that to the right of the right surface the theories are non-asymptotically free while to the left of the left surface the theories break chiral symmetry. Between the two surfaces the theories can develop an infrared fixed point.
\begin{figure}[h]
\centerline{\includegraphics[height=6.5cm,width=7.70cm]{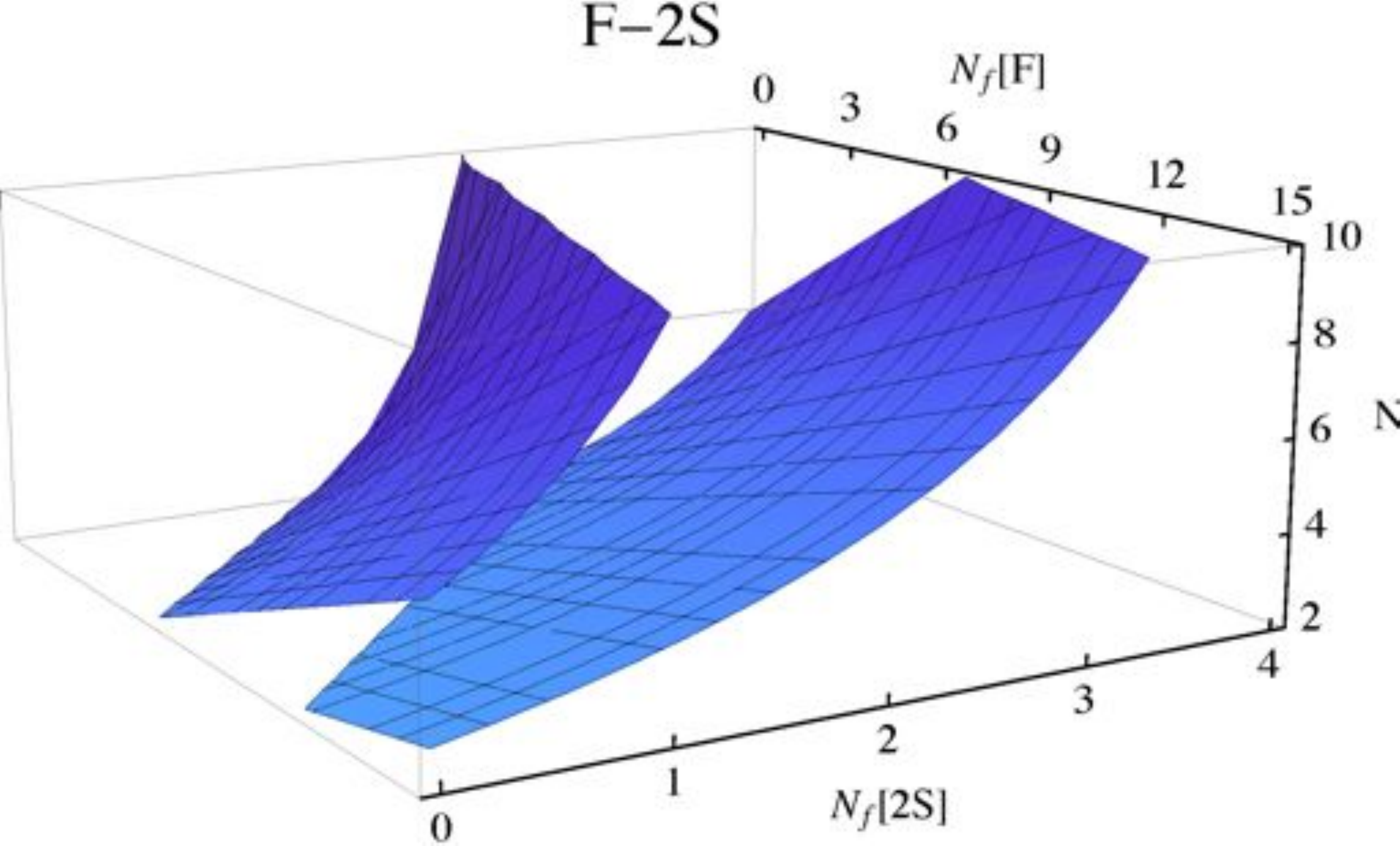}\hspace{0.8cm}\includegraphics[height=6.5cm,width=7.70cm]{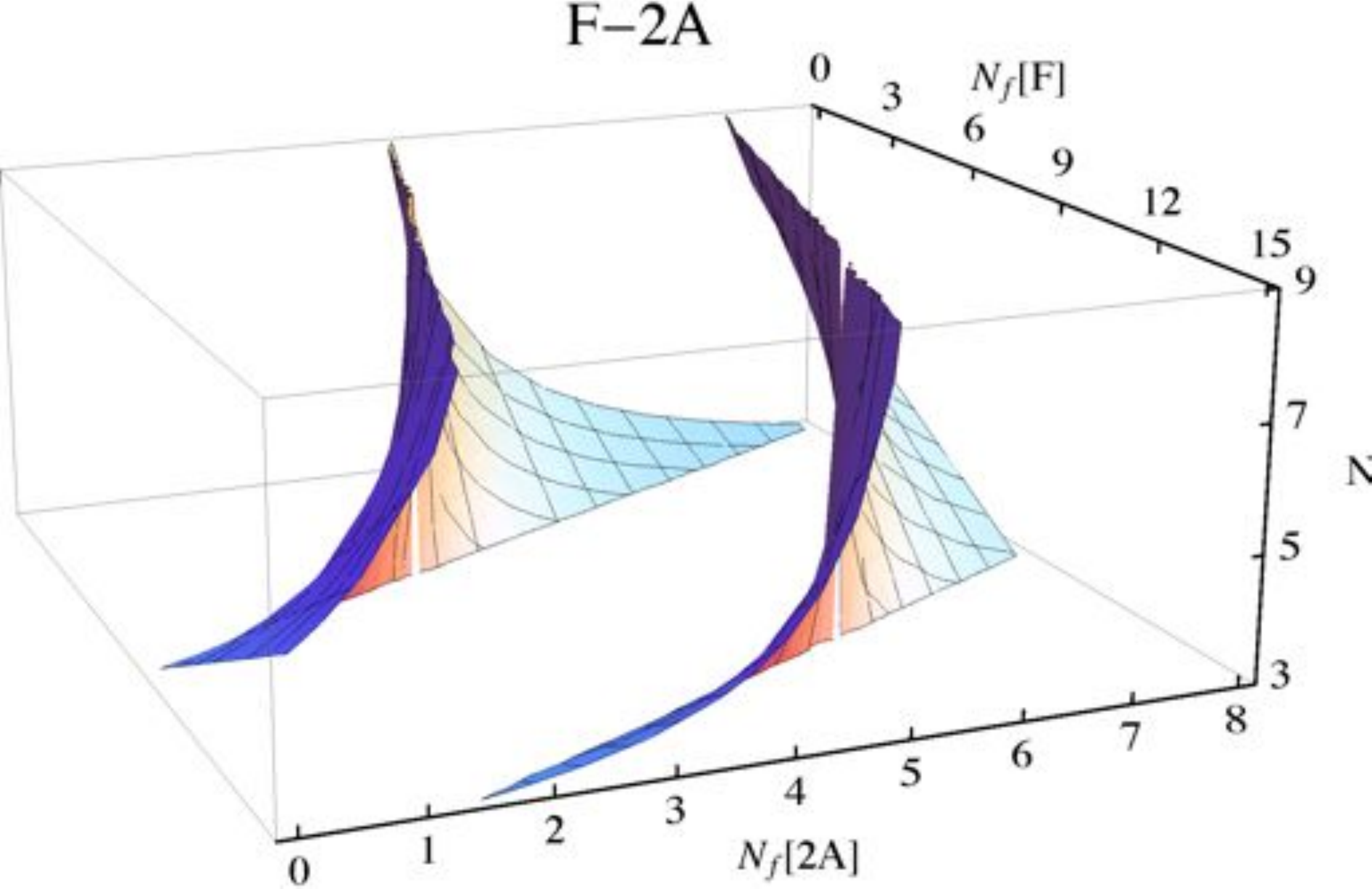}}
\caption{The conformal {\it house} for a non-supersymmetric gauge theory containing fermions in the fundamental and two-indexed symmetric representations (left) and in the fundamental and two-indexed antisymmetric representations (right).}\label{fig:fund}
\end{figure}
\begin{figure}[h]
\centerline{\includegraphics[height=6.5cm,width=7.70cm]{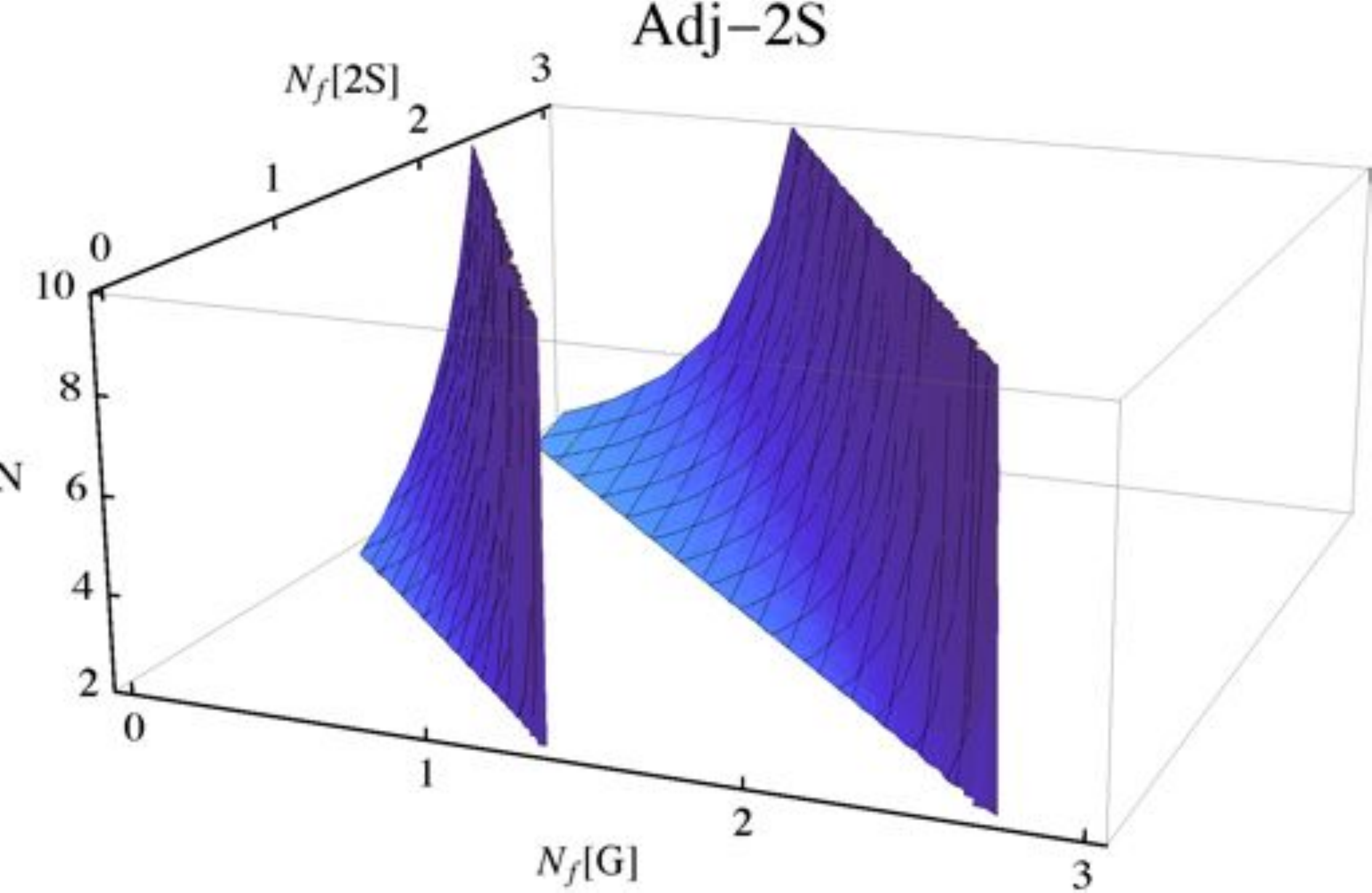}\hspace{0.8cm}\includegraphics[height=6.5cm,width=7.70cm]{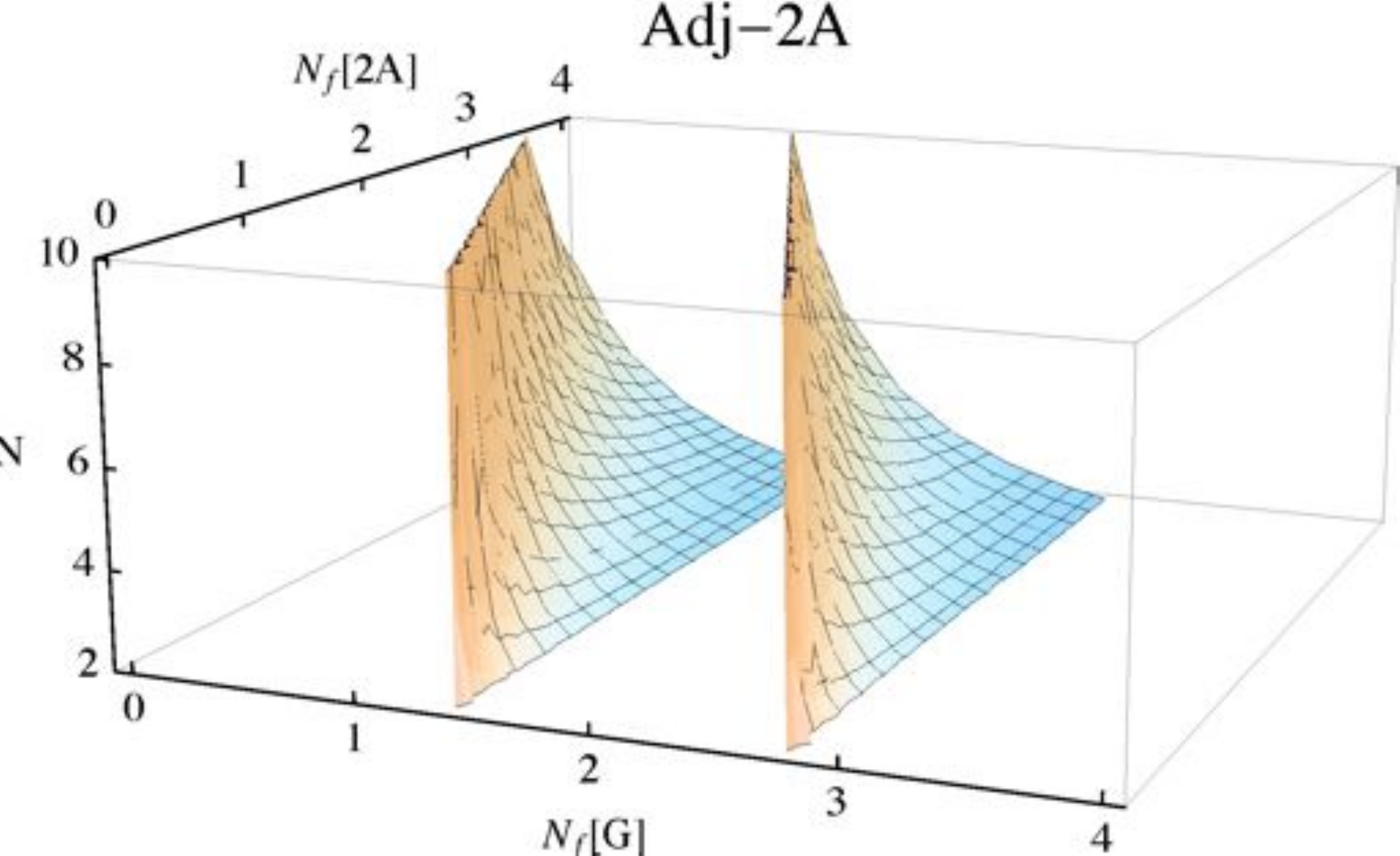}}
\caption{The conformal {\it house} for a non-supersymmetric gauge theory containing fermions in the adjoint and two-indexed symmetric representations (left) and in the adjoint and two-indexed antisymmetric representations (right).}\label{fig:adj}
\end{figure}
\begin{figure}[h]
\centerline{\includegraphics[height=6.5cm,width=7.70cm]{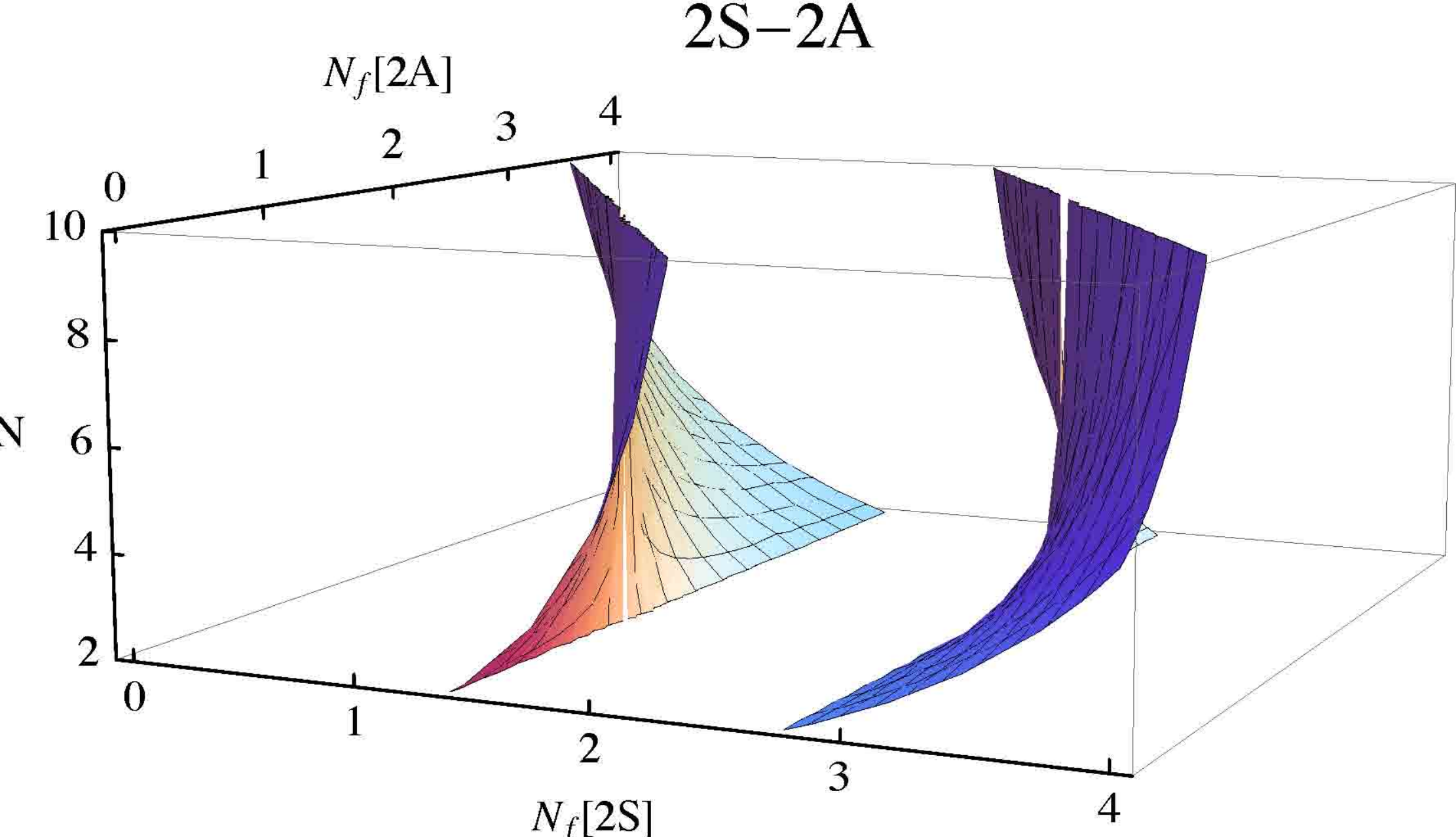}\hspace{0.8cm}}
\caption{The conformal {\it house} for a non-supersymmetric gauge theory containing fermions in the two-indexed symmetric and two-indexed antisymmetric representations.}\label{fig:two}
\end{figure}
Finally in Fig.~\ref{fig:two} we consider the last case in which one representation corresponds to the two-index symmetric and the other one is the two-index antisymmetric.

\subsection{On the behavior of physical $\beta$-functions}
Often, in the literature, one finds plotted a cartoon of the running of the coupling constant for either conformal or near conformal theories. Here we provide yet another cartoon of this running but this time using the {\it physical} form of the conjectured beta function \cite{Ryttov:2007cx} augmented with a simple ansatz for the dependence of the anomalous dimension on  the coupling constant.  The advantage is that we will be able to plot what happens when changing the number of flavors in any gauge theory. It should provide a simple framework allowing for a unified picture of the current lattice results for different representations and gauge groups. 

Let's start with the assumption that a specific theory  does posses an IRFP. In this case we have that: 
\begin{itemize}
 \item The anomalous dimension assumes a scheme independent value at the fixed point which according to the all orders beta function is:
\begin{equation}
\lim_{\mu \rightarrow 0}\gamma  = \frac{11C_2(G)  - 4 T(r) N_f}{ 2 T(r) N_f} \ .
\end{equation}  
with $r$ the fermion representation.
\item  At high energies $\gamma$ must match the perturbative expansion of the anomalous dimension. 
\end{itemize}
$\mu$ is the renormalization energy scale. It is straightforward to show that if one chooses, for the anomalous dimension, any continuous function of the coupling constant then the all orders beta function will always require, at the IRFP, that
$\gamma$ assume the value assumed above. To be able to draw the running of the coupling constant we pick as a simple approximation for $\gamma$ the second order expression for $\gamma$:
\begin{equation}
\gamma = a_0 \frac{g^2}{4\pi} +  a_1 \frac{g^4}{(4\pi)^2}  + {\cal O}(g^6)  \ ,
\label{gammapert}
\end{equation}
with 
\begin{equation}
a_0 = \frac{3}{2\pi} C_2(r) \ , \qquad   a_1= \frac{1}{16\pi^2}\left[ 3 C_2(r)^2 - \frac{10}{3} C_2(r)N_f  + \frac{97}{3} C_2(r) C_2(G)  \right] \label{as} \end{equation}
 
\begin{figure}
\begin{center}
\includegraphics[width=8cm,height=5cm]{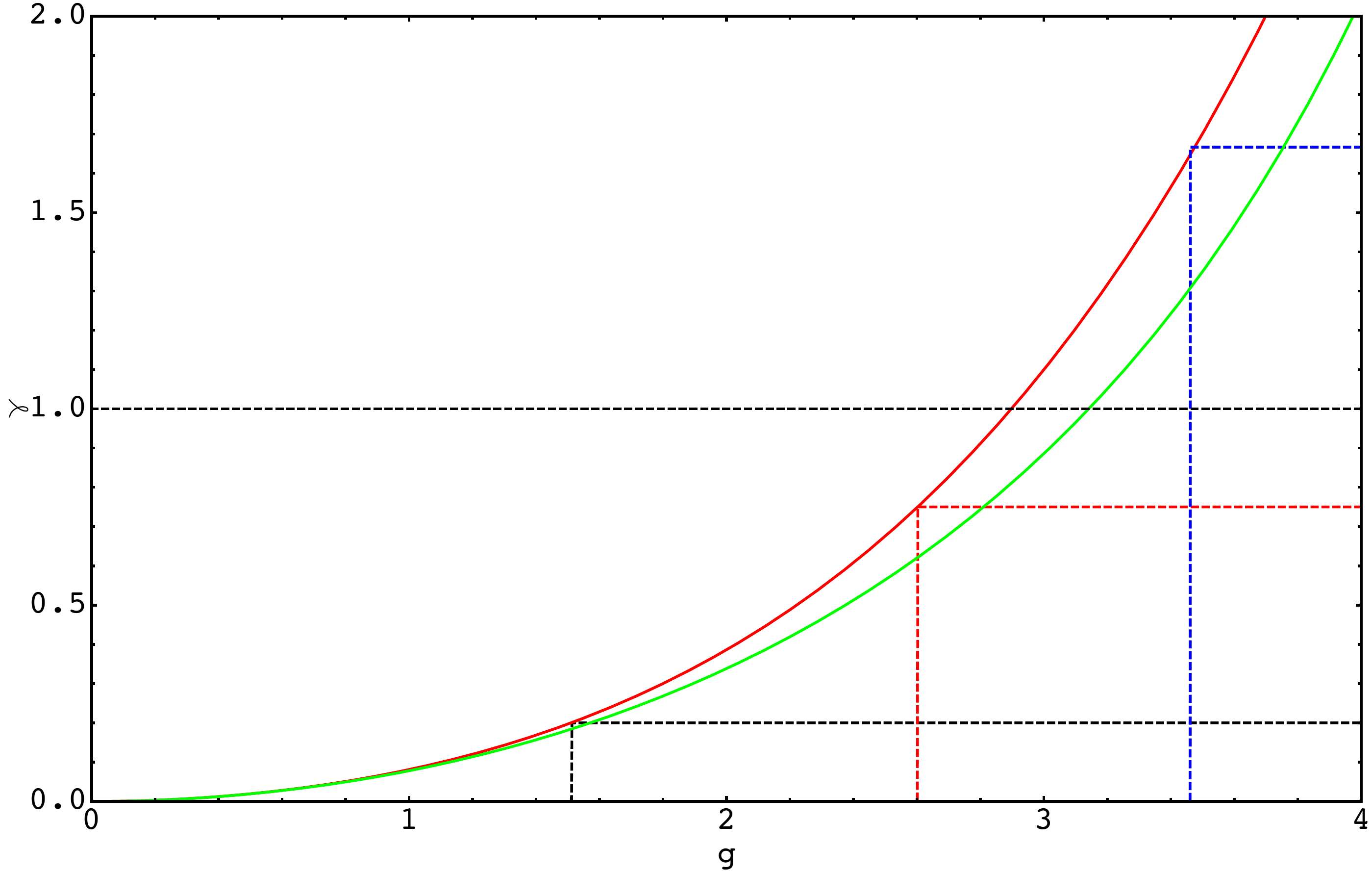}
\caption{ Anomalous dimension of the mass as function of the coupling constant  for  two Dirac flavors in the adjoint representation of $SU(2)$. To second order $\gamma$ does not differ much for different flavors in the conformal window. The horizontal blue dashed corresponds to the fixed point value for $N_f= 1.5$  Dirac flavors (i.e. 3 Weyl), the red dashed one to $N_f =2$  (four Weyl) and the black one to $N_f = 2.5$, i.e. five Weyl. The black dashed line marks the maximum value suggested by the SD equation.  The value of the coupling for which the solid and the different dashed lines meet is the fixed point value of the coupling (determined by the vertical dashed lines). The green curve is the all orders anomalous dimension for super Yang-Mills, corresponding to $N_f=0.5$, i.e. one Weyl fermion. \label{Adjoint-AD} }
\end{center}
\end{figure}

We start here by showing the case of minimal conformal theories, more specifically $SU(2)$ gauge theories with Dirac fermions in the adjoint. In Fig.~\ref{Adjoint-AD} we plot $\gamma$ up to second order, for two adjoint Dirac flavors as a red solid line. To second order $\gamma$ does not differ much for different flavors in the conformal window. The horizontal blue dashed line corresponds to the fixed point value for $N_f= 1.5$  Dirac flavors (i.e. 3 Weyl), the red dashed one to $N_f =2$  (four Weyl) and the black one to $N_f = 2.5$, i.e. five Weyl. The black dashed line marks the maximum value suggested by the SD equation.  The value of the coupling for which the solid and the different dashed lines meet is the fixed point value of the coupling (determined by the vertical dashed lines). The green curve is the all orders anomalous dimension for super Yang-Mills, corresponding to $N_f=0.5$, i.e. one Weyl fermion. 

\begin{figure}[tp]
\begin{center}
\mbox{
\subfigure{\resizebox{!}{.3\linewidth}{\includegraphics[clip=true]{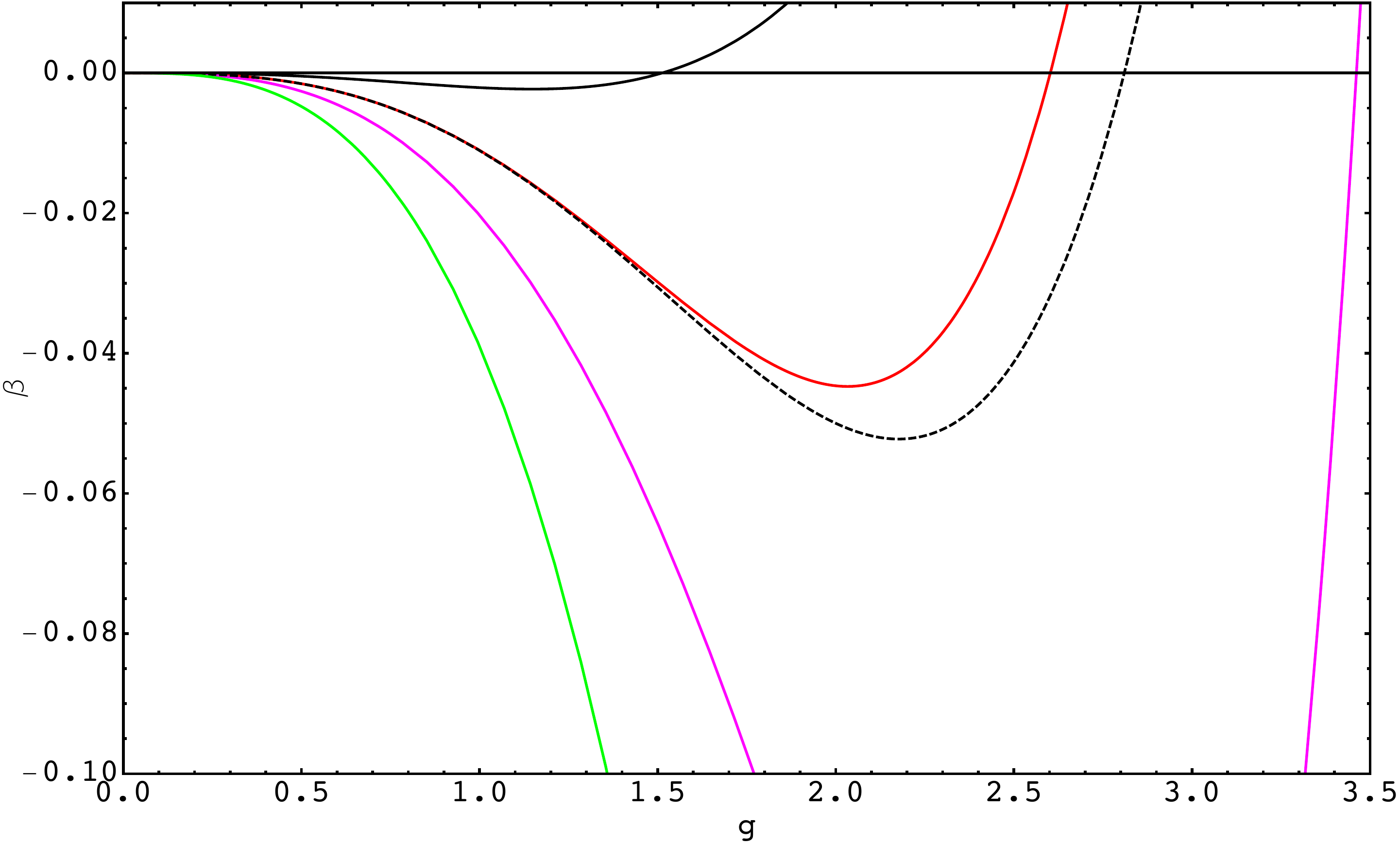}}}\qquad 
\subfigure{\resizebox{!}{0.31\linewidth}{\includegraphics[clip=true]{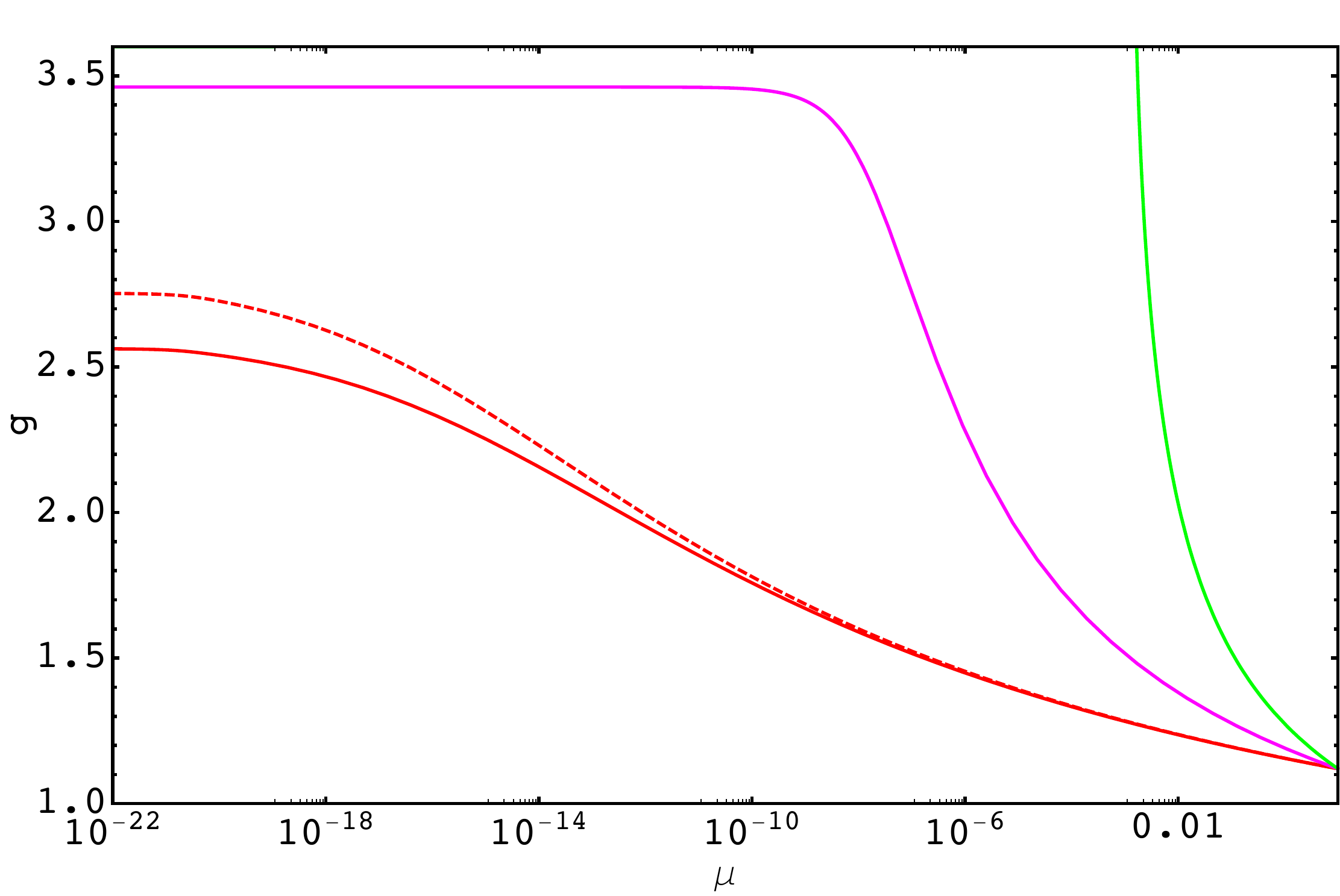}}}
}
\caption{Left Panel: Beta functions for different values of the number of Dirac flavors in the adjoint representation  of the $SU(2)$ gauge group. The black solid curve corresponds to $N_f=2.5$, the red to $N_f=2$, the dashed one is the two-loop beta function for $N_f=2$ again, while the magenta curve corresponds to $N_f=1.5$. The green curve is the beta function for super Yang-Mills.  Right Panel: Running of the coupling constants for different numbers of Dirac flavors gauged under the adjoint representation of $SU(2)$.  The red corresponds  to $N_f =2$, the magenta to $N_f = 1.5$ and the dashed red to $N_f = 1.2$ via the two loops beta function. The green line is the running of the super Yang-Mills coupling. \label{Beta-AD} }
\end{center}
\end{figure}

The analytic expression of the anomalous dimension can now be used to determine the beta functions. The different betas are shown in the left panel of Fig. \ref{Beta-AD}. The running of the coupling is, instead, plotted in the right panel of Fig.~\ref{Beta-AD}. 

 We have also plotted the two-loop beta function for $SU(2)$ with 2 Dirac flavors in the plots of the various beta functions. It is the dashed black line. Interestingly the fixed point is reached before the one obtained via the two-loop beta function. This is consistent with the recent lattice results obtained in \cite{Hietanen:2009az}. 

For completeness we also present the results for $SU(N)$ with fermions in the fundamental representation, i.e. the QCD conformal window.

\begin{figure}[h]
\begin{center}
\mbox{
\subfigure{\resizebox{!}{.3\linewidth}{\includegraphics[clip=true]{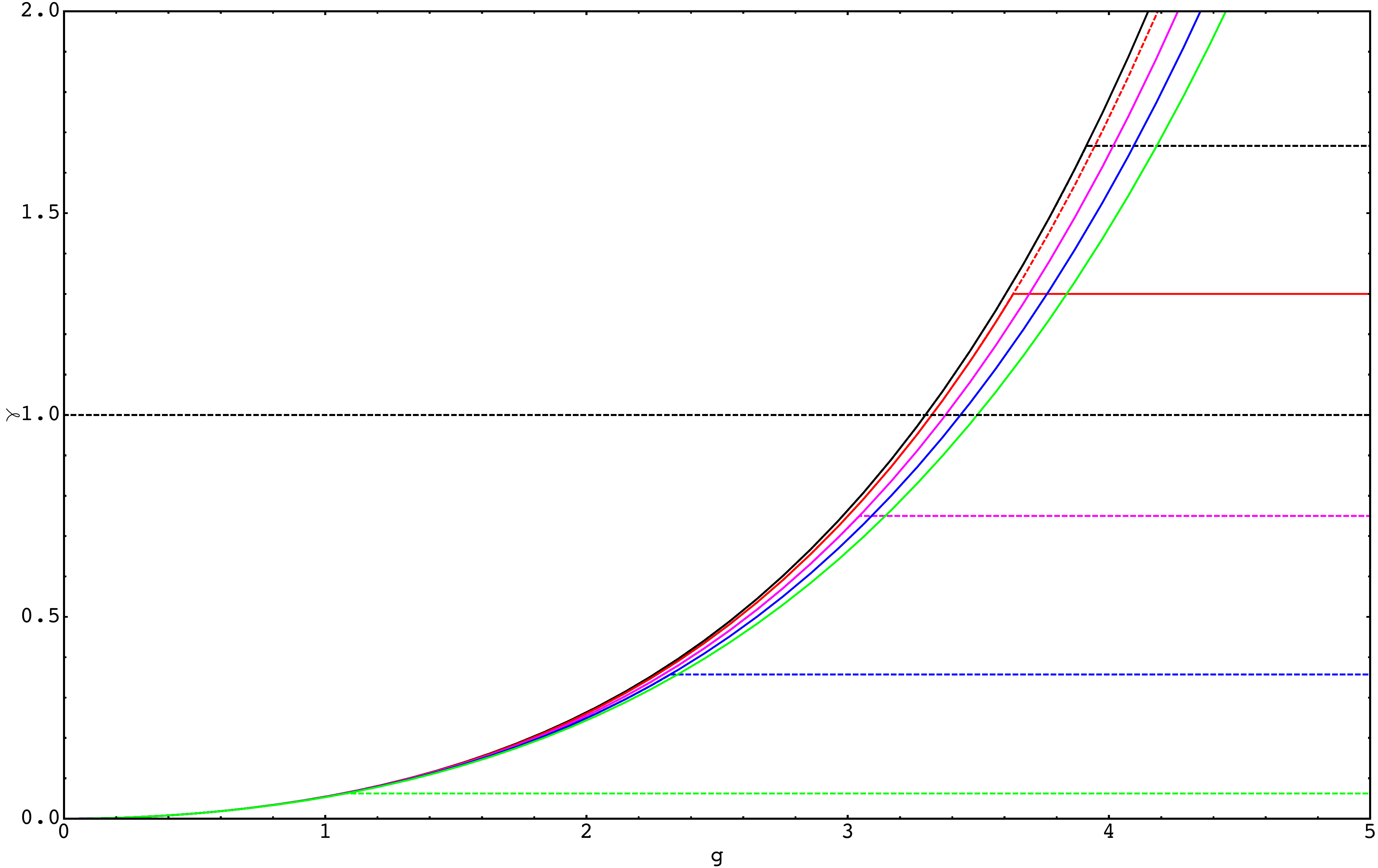}}}\qquad 
\subfigure{\resizebox{!}{0.31\linewidth}{\includegraphics[clip=true]{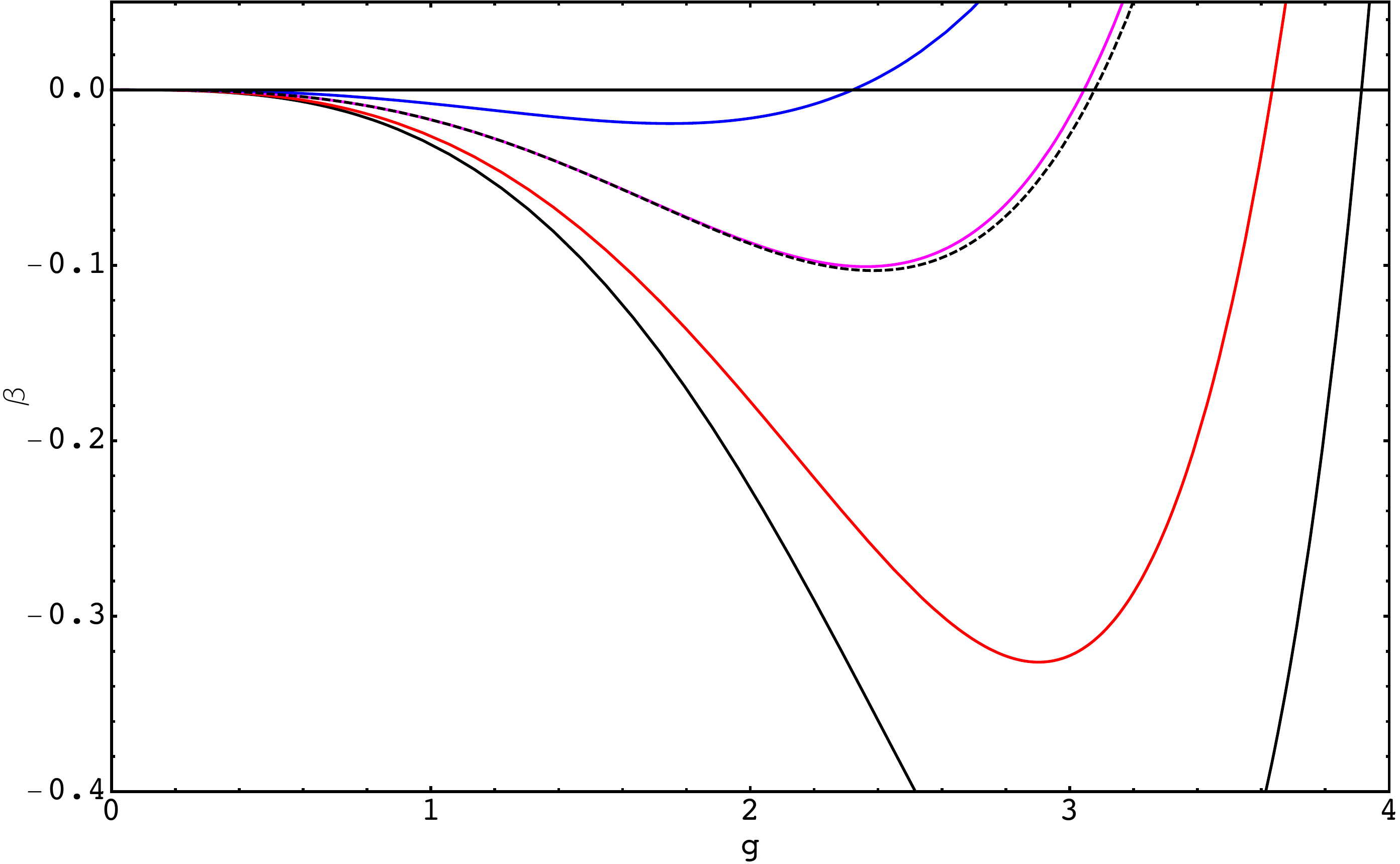}}}
}
\caption{Left Panel: Anomalous dimension of the mass as function of the coupling constant  for different number of Dirac flavors in the fundamental representation of $SU(3)$. The black solid line corresponds to $N_f=9 $,  the red to $N_f =10$ , magenta  to $N_f = 12$,  blue to $N_f=14$ and green to $N_f=16$.  The dashed black line for $\gamma=1$ marks the maximum value suggested by the SD equation.  Each dashed line parallel to the $g$ correspond to the specific value of $\gamma$ at the fixed point. The value of the coupling at the point where the dashed lines and the respective $\gamma$ meet is the fixed point one. 
Right Panel: QCD beta functions for different values of the number of Dirac flavors.   The black solid line corresponds to $N_f=9 $,  the red to $N_f =10$ , magenta  to $N_f = 12$ and blue to $N_f=14$.  The dashed line corresponds to the beta function at two loops. \label{Anomalous-F} }
\end{center}
\end{figure}
The analytic expression of the anomalous dimension can now be used to determine the beta function in between the UV fixed point and the IRFP one. The different betas are shown in right panel of Fig. \ref{Anomalous-F}.  
The general message is that one can have large values of the anomalous dimensions and yet have coupling constants at the IRFP which are small. In fact for QCD the coupling constant is always smaller than $4$ in the conformal window.  Interestingly this is true also for the case of the fermions in the adjoint representation.  Moreover for the case of $12$ flavors in QCD and two flavors in the adjoint representation we have a fixed point
coupling that is close to $3$ and a little larger than $2$ respectively.  Naively the expansion parameter in perturbation theory is $\frac{g^2}{4\pi}$. One gets for 12 flavors QCD a value close to $ 0.72$ and for MWT  $0.24$. The picture resulting from our analysis is similar to the one observed via first principle lattice simulations for QCD with 12 flavors as well as MWT and offers a possible theoretical understanding of the physics behind these lattice explorations.
 \subsection{Conformal Chiral Dynamics}

\begin{figure}[h]
\begin{center}
\includegraphics[width=6cm,]{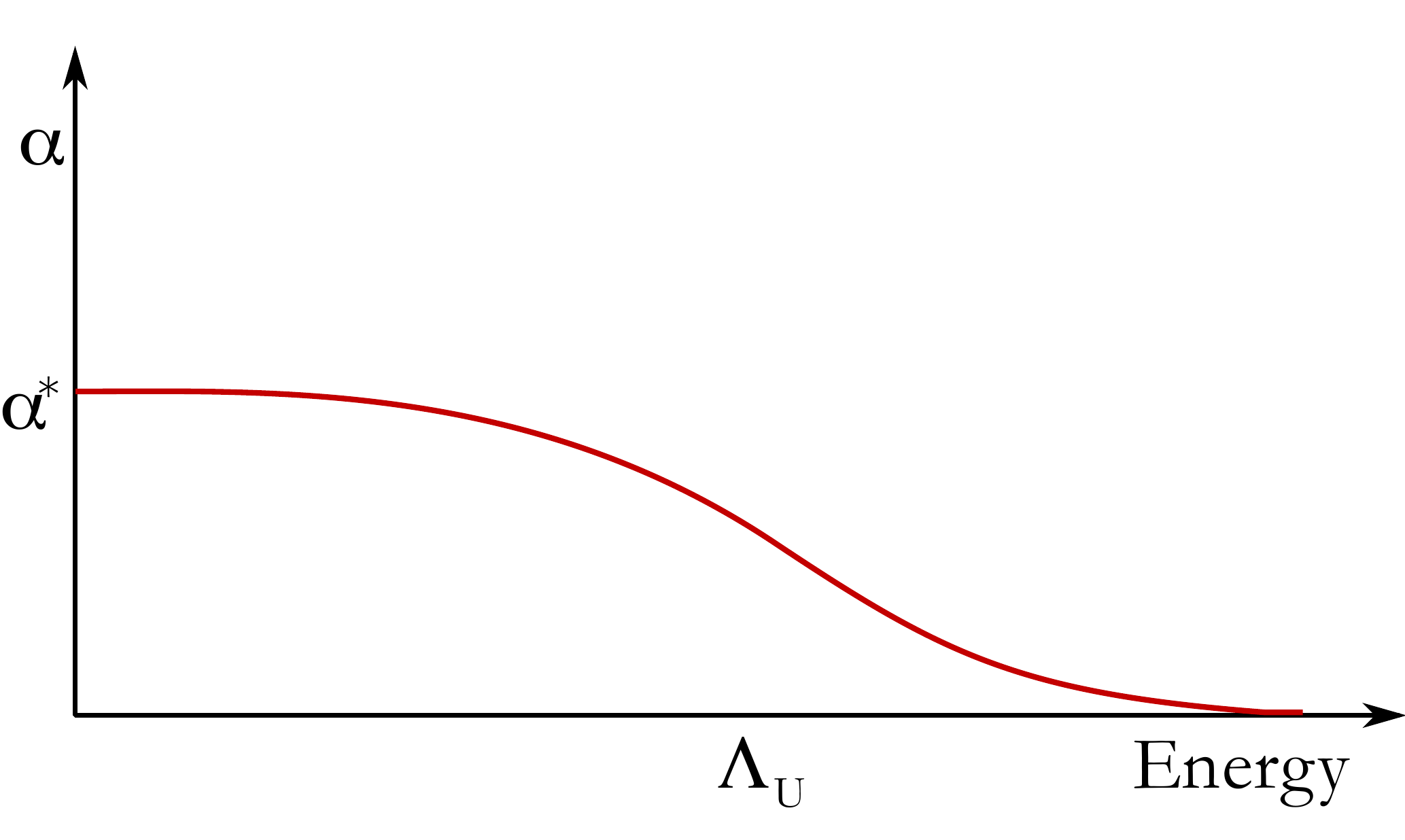}
\caption{ Running of the coupling constant in an asymptotically free gauge theory developing an infrared fixed point for a value $\alpha = \alpha^{\ast}$. }
\label{run1}
\end{center}
\end{figure}

Our starting point is a nonsupersymmetric non-abelian gauge theory with sufficient massless fermionic matter to develop a nontrivial IRFP. The cartoon of the running of the coupling constant is represented in Fig.~\ref{run1}. In the plot $\Lambda_U$ is the dynamical scale below which the IRFP is essentially reached. It can be defined as the scale for which $\alpha$ is $2/3$ of the fixed point value in a given renormalization scheme. 
If the theory possesses an IRFP the chiral condensate must vanish at large distances. Here we want to study the behavior of the condensate when a flavor singlet mass term is added to the underlying Lagrangian: 
\begin{eqnarray}
\Delta L = - m\,{\widetilde{\psi}}{\psi} + {\rm h.c.} \ ,
\end{eqnarray} 
with $m$ the fermion mass and $\psi^f_{c}$ as well as $\widetilde{\psi}_f^c$ left transforming two component spinors, $c$ and $f$ represent color and flavor indices.  The omitted color  and flavor indices, in the Lagrangian term, are contracted. 
We consider first the case of fermionic matter in the fundamental representation of the $SU(N)$ gauge group. We then generalize our results to the case of higher dimensional representations.  

The effect of such a term is to break the conformal symmetry together with some of the global symmetries of the underlying gauge theory.  The composite operator:
\begin{equation}
{{\cal O}_{\widetilde{\psi}{\psi}}}^{f^{\prime}}_f  = \widetilde{\psi}^{f^{\prime}}{\psi}_f  \ , 
\end{equation}
has mass dimension $\displaystyle{
d_{\widetilde{\psi}{\psi}} = 3 - \gamma}$ with 
$\gamma$  the anomalous dimension of the mass term. At the fixed point $\gamma$ is a positive number smaller than two \cite{Mack:1975je}. We assume $m \ll \Lambda_U$.  Dimensional analysis demands:
\begin{eqnarray}\Delta L  \rightarrow 
-m\, \Lambda_U^{\gamma} \, {\rm Tr}[\Om] + {\rm h.c.} \ .
\label{mass}
\end{eqnarray}
The mass term is a relevant perturbation around the IRFP driving the theory away from the fixed point.  It will induce a nonzero vacuum expectation value for $\Om$ itself proportional to $\delta^{f^\prime}_f$. It is convenient to define ${\rm Tr}[\Om]  = N_f  \cal O $ with $\cal O$ a flavor singlet operator.  The relevant low energy Lagrangian term is then: 
\begin{equation}
-m\, \Lambda_U^{\gamma} \, N_f \cal O + {\rm h.c.}  \ .
\end{equation}
To determine the vacuum expectation value of $\cal{O}$ we replace it, formally, with a sum over an infinite number of canonically normalized single particle states \cite{Stephanov:2007ry}:
\begin{equation}
{\cal O}(x) = \sum_{n=1}^{\infty} f_n \varphi_n (x) \ .
\end{equation}
Each state possesses a mass $M_n$ whose value is controlled by an artificial mass gap $\Delta$ and a function of $n$, call it $z(n)$, with the properties $z(n+1)>z(n)$ and $z(1)=1$. 
\begin{equation}
M^2_n = \Delta^2 z(n) \ .
\end{equation}
We also have \cite{Stephanov:2007ry}: 
\begin{equation}
f_n^2 = {\cal F}_{\dm} \frac{dz(n)}{dn}\, \Delta^2 (M^2_n)^{\dm -2} \ ,
\end{equation}
with $ {\cal F}_{\dm}$ a function depending on the scaling dimension of the operator as well as the details of the underlying dynamics \cite{Sannino:2008nv}.  
Because of the presence of the fictitious mass terms the potential reads:
\begin{eqnarray}
V &=& m\, \Lambda_U^{\gamma} \, N_f \sum_{n=1}^{\infty} f_n \varphi_n   +  \bar{m}\, \Lambda_U^{\gamma} \, N_f \sum_{n=1}^{\infty} f_n \bar{\varphi}_n  \nonumber \\
&+& \sum_{n=1}^{\infty} M^2_n \varphi_n \bar{\varphi}_n  \ .
\end{eqnarray}
The bar over the fields and the fermion mass indicates complex conjugation. The extremum condition yields:
 \begin{equation}
 \langle  \bar{\varphi}_n\rangle = - m \Lambda_U^{\gamma} N_f \frac{f_n}{M^2_n} \ ,
 \end{equation}
yielding: 
\begin{equation}
\langle {\cal O} \rangle =  - \bar{m} \Lambda_U^{\gamma} N_f \sum_{n=1}^{\infty}  \frac{f_n^2}{M^2_n}  \ .
\end{equation}
We now take the limit $\Delta^2 \rightarrow 0$ and the sum becomes an integral. {}For any specific  function $z(n)$ it is easy to show that:
\begin{equation}
\langle {\cal O} \rangle =  - \bar{m} \Lambda_U^{\gamma} N_f {\cal F}_{\dm}\Omega\left[\Lambda_{UV} ,\Lambda_{IR}\right] \ , 
\label{Ovev}
\end{equation}
 with 
 \begin{equation}
 \Omega\left[\Lambda_{UV} ,\Lambda_{IR}\right]  = \frac{1}{1-\gamma} \left[ \Lambda_{UV}^{2(1-\gamma)} - \Lambda_{IR}^{2(1-\gamma)} \right] \ .
 \end{equation}
 The  ultraviolet and infrared cutoffs are introduced to tame the integral in the respective regions. A simple physical interpretation of these cutoffs is the following. At very high energies, at scales above $\Lambda_{U}$, the underlying theory flows to the ultraviolet fixed point and we have to abandon the description in terms of the composite operator. This immediately suggests that $\Lambda_{UV}  $ is naturally identified with $\Lambda_U$. The presence of the mass term induces a mass gap, which is the quantity we are trying to determine. The induced physical mass gap is a natural infrared cutoff. We, hence, identify  $\Lambda_{IR} $ with the physical value of the condensate. We find:
 \begin{eqnarray}
 \langle \widetilde{\psi}^f_c \psi^c_f \rangle  &\propto& -m \Lambda_U^2 \ ,  \qquad ~~~~~~0 <\gamma  < 1 \ , \label{BZm} \\
 \langle \widetilde{\psi}^f_c \psi^c_f \rangle  &\propto &   -m \Lambda_U^2  \log \frac{\Lambda^2_U}{|\langle {\cal O} \rangle|}\ , ~~~   \gamma \rightarrow  1    \ , \label{SDm} \\
 \langle \widetilde{\psi}^f_c\psi^c_f \rangle   &\propto &  -m^{\frac{3-\gamma} {1+\gamma}} 
 \Lambda_U^{\frac{4\gamma} {1+\gamma}}\ , ~~~1<\gamma  \leq 2 \ .
   \label{UBm}
 \end{eqnarray}
We used  $\langle \widetilde{\psi} \psi \rangle \sim \Lambda_U^{\gamma} \langle {\cal O} \rangle $ to relate the expectation value of ${\cal O}$ to the one of the fermion condensate. Via an allowed axial rotation $m$ is now real and positive. 
It is instructive to compare these results with the ones obtained via naive dimensional analysis (NDA)  \cite{Fox:2007sy} also discussed in \cite{Luty:2008vs} and in \cite{Sannino:2008nv}. We find:
\begin{equation}
 \langle \widetilde{\psi}^f_c \psi^c_f \rangle _{\rm NDA} \propto  -m^{\frac{3-\gamma} {1+\gamma}} 
 \Lambda_U^{\frac{4\gamma} {1+\gamma}} \ . 
\end{equation}
Note that one recovers the previous scaling as function of $m$ (up to logarithmic corrections) only for $1\le \gamma \le 2$. The failure of NDA for a smaller anomalous dimension is due to the fact that  the ultraviolet physics is not captured by NDA \cite{Sannino:2008nv}. 
The effects of the Instantons on the conformal dynamics has been investigated in \cite{Sannino:2008pz}. Here it was shown that the effects of the instantons can be sizable only for a very small number of flavors given that, otherwise, the instanton induced operators are highly irrelevant.

At any nonzero value of the fermion mass the chiral and conformal symmetries are explicitly broken and single particle states emerge at low energies. A relevant set are the {\it conformal pions}, i.e. the would be Goldstones which in the limit of zero fermion mass cannot be described  via single particle states. We identify them via
\begin{eqnarray}
\langle {\Om}^{f^\prime}_{f}\rangle &=&  \langle {\cal O} \rangle \, U \quad {\rm with} \quad U=e^{i \frac{\pi}{F_{\pi}}}  \ . 
\label{U}
\end{eqnarray}
$\pi=\pi^a T^a$ and $T^a$ are the set of broken generators normalized according to ${\rm Tr} \left[T^a T^b\right] = \delta^{ab}1/2$. Substituting \eqref{U} in \eqref{mass}  and expanding up to the second order in the pion fields we have: 
\begin{eqnarray}
m^2_{\pi} F_\pi^2 =  -m\,\Lambda_U^{\gamma}\langle {\cal O}[m]\rangle \ . \label{GMOR}
\end{eqnarray}
Having determined the dependence on $m$ of $\langle O [m]\rangle$ the above generalizes the similar one in QCD \cite{Glashow:1967rx,GellMann:1968rz,Dashen:1969eg} known as the Gell-Mann Oakes Renner (GMOR)  relation. {}For example for the theories investigated above and for a very small fermion mass, $m^2_{\pi} F_\pi^2= m^2 \Lambda_U^2$. At larger masses the scaling is different  for the three cases and it can be easily deduced from our results. A similar effective Lagrangian was introduced in \cite{Sannino:2008nv}.  Assume now that the underlying gauge theory has not developed an IRFP. In this case there are only two possibilities: i) chiral symmetry breaks spontaneously yielding a condensate whose leading term in $m$ is a constant; ii) chiral symmetry is intact but a scale is still generated. Chirally paired partners emerge together with massless composite fermions appearing to saturate the 't Hooft anomaly matching conditions.  One can investigate the finite volume effects using the conformal pion lagrangian in the $\epsilon$-regime \cite{Gasser:1987ah}. However much care must be taken when constructing a low energy effective description of a near conformal gauge theory given that the standard chiral counting no needs to hold. 
Caveat lector that we assumed that a chiral Lagrangian approach is valid, in deriving Eq.~(\ref{GMOR}) in the extremely tiny m expansion over the only other scale of the theory which is $\Lambda_U$.  Another  way to think at the problem is that the mass of the fermions acts always as an {\it heavy} quark mass. Below this mass scale one has a pure confining Yang-Mills theory whose mass scale is directly controlled by the heavy quark mass. The latter observation seems to best fit current lattice results \cite{DelDebbio:2008tv,Lucini:2009an} albeit for (near) conformal theories the amount of fine-tuning needed to control the small mass approximation is harder to tame. 

\subsection {Gauge Duals and Conformal Window}

One of the most fascinating possibilities is that generic asymptotically free gauge theories have magnetic duals. In fact, in the late nineties, in a series of  ground breaking papers Seiberg  \cite{Seiberg:1994bz,Seiberg:1994pq} provided strong support for the existence of a consistent picture of such a duality within a supersymmetric framework. Supersymmetry is, however, quite special and the existence of such a duality does not automatically imply the existence of nonsupersymmetric duals. One of the most relevant  results put forward by Seiberg  has been the identification of the boundary of the conformal window for supersymmetric QCD as function of the number of flavors and colors. 
The dual theories proposed by Seiberg pass a set of mathematical consistency relations known as 't Hooft anomaly conditions (in \cite{'tHooft:1980xb}).  Another important tool has been the knowledge of the all orders supersymmetric beta function \cite{Novikov:1983uc,Shifman:1986zi,Jones:1983ip} 
 
Arguably the existence of a possible dual of a generic nonsupersymmetric asymptotically free gauge theory able to reproduce its infrared dynamics must match the 't Hooft anomaly conditions \cite{'tHooft:1980xb}.

We have exhibited several solutions of these conditions for QCD in \cite{Sannino:2009qc} and for certain gauge theories with 
higher dimensional representations in  \cite{Sannino:2009me}. An earlier exploration already appeared in the literature \cite{Terning:1997xy}. The novelty with respect to these earlier results are: i) The request that the gauge singlet operators associated to the magnetic baryons should be interpreted as bound states of ordinary baryons \cite{Sannino:2009qc}; ii) The fact that the asymptotically free condition for the dual theory matches the lower bound on the conformal window obtained using the all orders beta function  \cite{Ryttov:2007cx}. These extra constraints help restricting further the number of possible gauge duals without diminishing the exactness of the associate solutions with respect to the 't Hooft anomaly conditions.

We will briefly summarize here the novel solutions to the 't Hooft anomaly conditions for QCD and the theories with higher dimensional representations. The resulting {\it magnetic} dual allows to predict the critical number of flavors above which the asymptotically free theory, in the electric variables, enters the conformal regime as predicted using the all orders conjectured beta function \cite{Ryttov:2007cx}.

\subsubsection{QCD Duals} 
The underlying gauge group is $SU(3)$ while the
quantum flavor group is
\begin{equation}
SU_L(N_f) \times SU_R(N_f) \times U_V(1) \ ,
\end{equation}
and the classical $U_A(1)$ symmetry is destroyed at the quantum
level by the Adler-Bell-Jackiw anomaly. We indicate with
$Q_{\alpha;c}^i$ the two component left spinor where $\alpha=1,2$
is the spin index, $c=1,...,3$ is the color index while
$i=1,...,N_f$ represents the flavor. $\widetilde{Q}^{\alpha ;c}_i$
is the two component conjugated right spinor. We summarize the
transformation properties in the following table.
\begin{table}[h]
\[ \begin{array}{|c| c | c c c|} \hline
{\rm Fields} &  \left[ SU(3) \right] & SU_L(N_f) &SU_R(N_f) & U_V(1) \\ \hline \hline
Q &\Yfund &{\Yfund }&1&~~1  \\
\widetilde{Q} & \overline{\Yfund}&1 &  \overline{\Yfund}& -1   \\
G_{\mu}&{\rm Adj}   &1&1  &~~1\\
 \hline \end{array} 
\]
\caption{Field content of an SU(3) gauge theory with quantum global symmetry $SU_L(N_f)\times SU_R(N_f) \times U_V(1)$. }
\end{table}

The  global anomalies are associated to the triangle diagrams featuring at the vertices three $SU(N_f)$ generators (either all right or all left), or two 
$SU(N_f)$ generators (all right or all left) and one $U_V(1)$ charge. We indicate these anomalies for short with:
\begin{equation}
SU_{L/R}(N_f)^3 \ ,  \qquad  SU_{L/R}(N_f)^2\,\, U_V(1) \ .
\end{equation}
For a vector like theory there are no further global anomalies. The
cubic anomaly factor, for fermions in fundamental representations,
is $1$ for $Q$ and $-1$ for $\tilde{Q}$ while the quadratic anomaly
factor is $1$ for both leading to
\begin{equation}
SU_{L/R}(N_f)^3 \propto \pm 3 \ , \quad SU_{L/R}(N_f)^2 U_V(1)
\propto \pm 3 \ .
\end{equation}

 \begin{figure}[h]
\centerline{
\includegraphics[width=12cm]{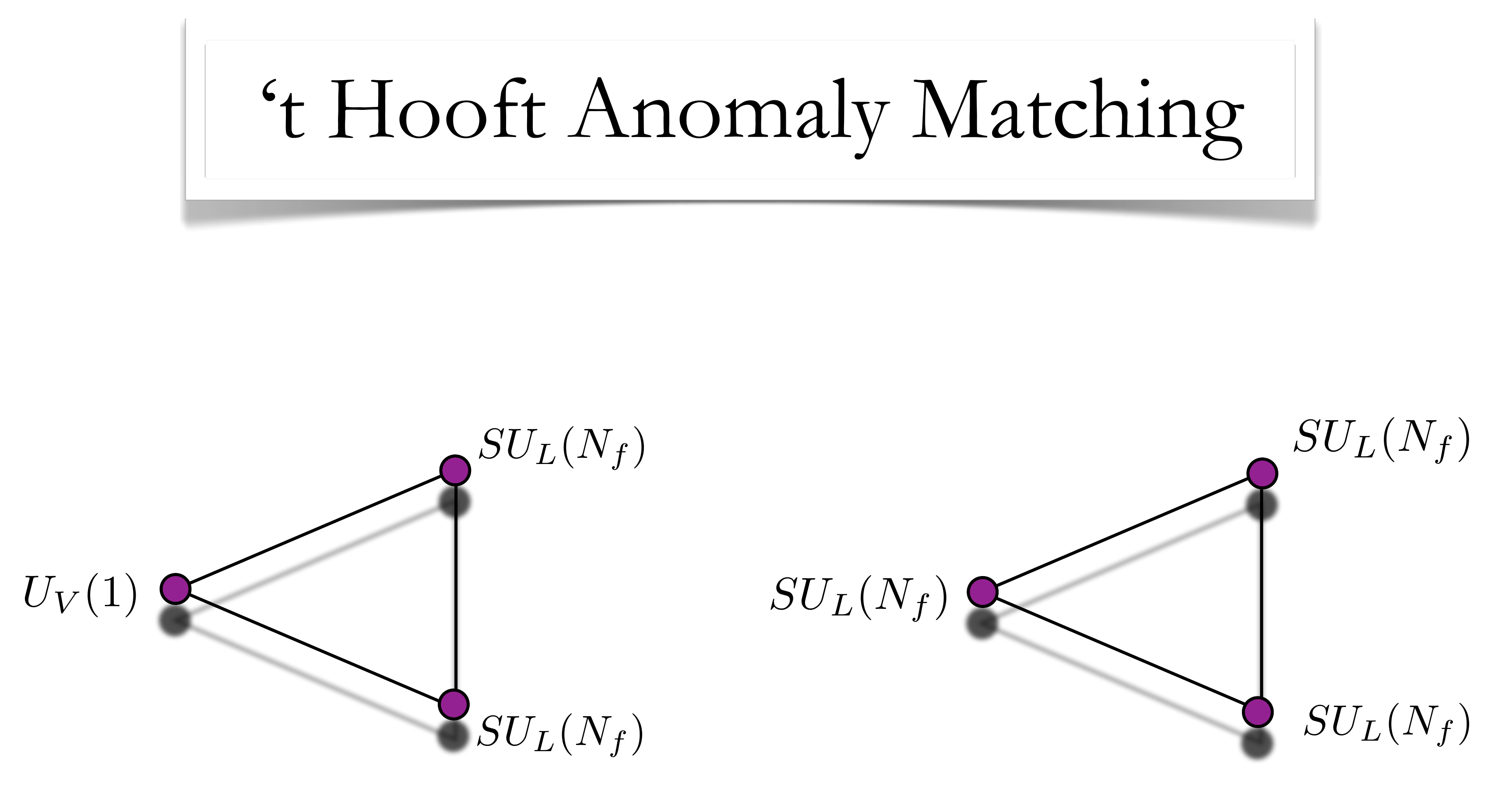}}
\caption{The 't Hooft anomaly matching conditions are related to the saturation of the global anomalies stemming out of the one-loop triangle diagrams represented, for the theory of interest, here. According to 't Hooft both theories, i.e. the electric and the magnetic ones, should yield the same global anomalies.}
\label{GA}
\end{figure}
We have mentioned already that requiring the absence of negative norm states  at the conformal point requires  $\gamma < 2$ resulting in the {\it maximum} possible extension of the conformal window bounded from below by:
\begin{equation}
N_f(r)^{\rm BF} \geq \frac{11}{8} \frac{C_2(G)}{T(r)}  \qquad { \gamma =2}\ .
\end{equation}
Specializing to three colors and fundamental representation the prediction from the all order beta function we find: 
\begin{equation}
N_f(r)^{\rm BF} \geq \frac{33}{4} = 8.25 \ , \qquad {\rm for~QCD~with}  \quad  { \gamma =2}  .
\end{equation}
The actual size of the conformal window can, however, be smaller than the one determined above without affecting the validity of the beta function. It may happen, in fact, that chiral symmetry breaking is triggered for a value of the anomalous dimension less than two. If this occurs the conformal window shrinks. As we have already mentioned the ladder approximation approach\cite{Appelquist:1988yc,{Cohen:1988sq},Appelquist:1996dq,Miransky:1996pd},  for example, predicts that chiral symmetry breaking occurs when the anomalous dimension is larger than one. Remarkably the all orders beta function encompass this possibility as well \cite{Ryttov:2007cx}. In fact, it is much more practical to quote the value predicted using the beta function by imposing $\gamma =1$:
\begin{eqnarray}
N_f(r) \geq  \frac{11}{6}  \frac{C_2(G)}{T({r})} \ ,\qquad {\gamma =1}  \ .
\end{eqnarray}
{}For QCD we have:
\begin{equation}
N_f(r)^{\rm BF} \geq   11 \ ,\qquad {\rm for~QCD~with} \quad { \gamma =1} \ .
\end{equation}
The result is very close to the one obtained using directly the ladder approximation, i.e.  $N_f \approx 4 N$, as shown in \cite{Ryttov:2007cx,Sannino:2009aw}.  

 Lattice simulations of the conformal window for various matter representations  \cite{Catterall:2007yx,Catterall:2008qk,
Shamir:2008pb,DelDebbio:2008wb,DelDebbio:2008zf, Hietanen:2008vc,Hietanen:2008mr,Appelquist:2007hu,Deuzeman:2008sc,Fodor:2008hn,DelDebbio:2008tv,DeGrand:2008kx,Appelquist:2009ty,Hietanen:2009az,Deuzeman:2009mh,DeGrand:2009et,Hasenfratz:2009ea} are in agreement with the predictions of the conformal window via the all orders beta function.

{}It would be desirable to have a novel way to determine the conformal window which makes use of exact matching conditions.  
 
\subsubsection{Dual Set Up} 
\begin{figure}[h!]
\centerline{\includegraphics[width=12cm]{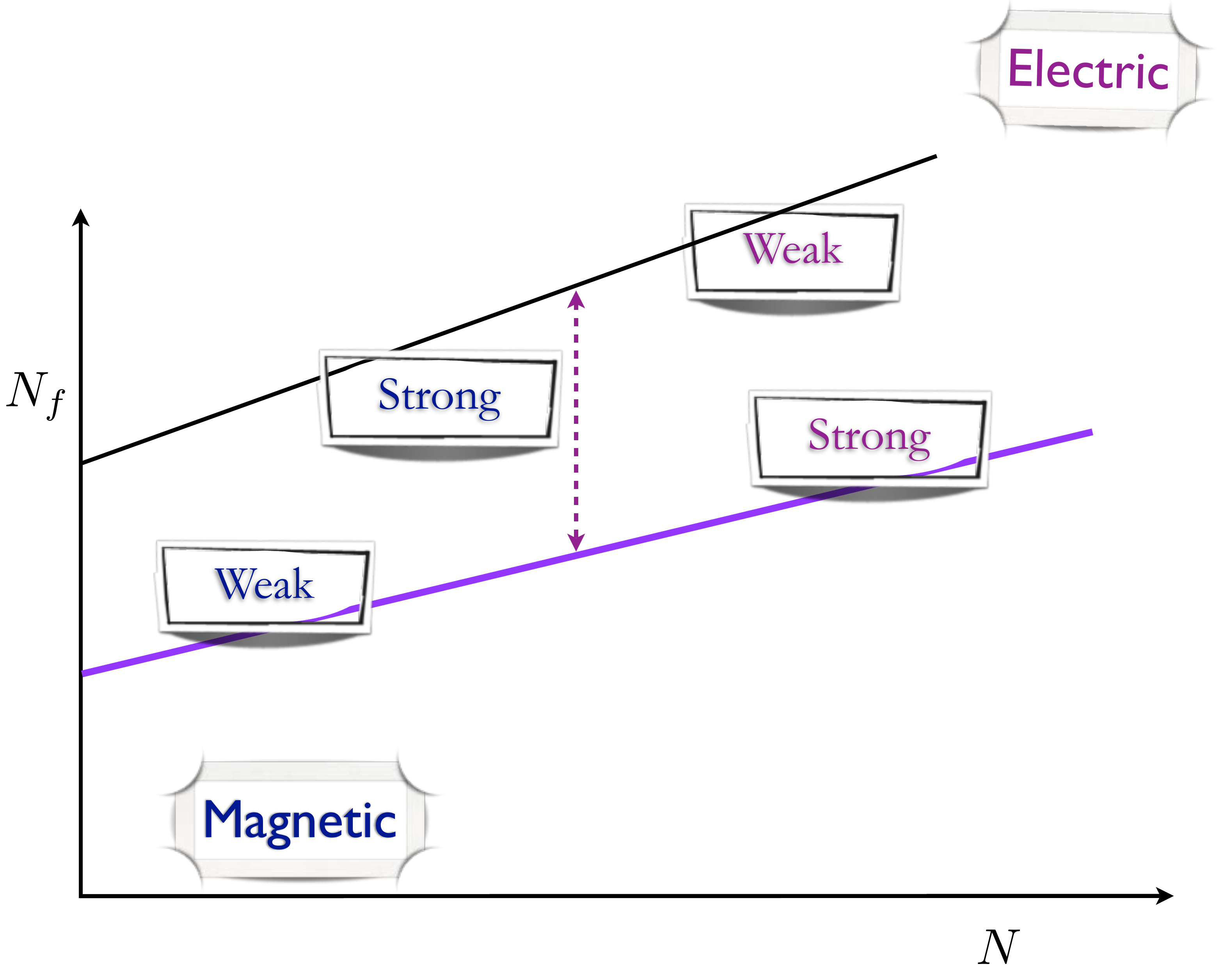}}
\caption{Schematic representation of the phase diagram as function of number of flavors and colors. For a given number of colors by increasing the number flavors within the conformal window we move from the lowest line (violet) to the upper (black) one. The upper black line corresponds to the one where one looses asymptotic freedom in the electric variables and the lower line where chiral symmetry breaks and long distance conformality is lost. In the {\it magnetic} variables the situation is reverted and the perturbative line, i.e. the one where one looses asymptotic freedom in the magnetic variables, correspond to the one where chiral symmetry breaks in the electric ones. }
\label{Duality}
\end{figure}

If a magnetic dual of QCD does exist one expects it to be weakly coupled near the critical number of flavors below which one breaks  large distance conformality in the electric variables.  This idea is depicted in Fig~\ref{Duality}. 

Determining a possible unique dual theory for QCD is, however, not simple given the few mathematical constraints at our disposal, as already observed in \cite{Terning:1997xy}. The saturation of the global anomalies is an important tool but is not able to select out a unique solution. We shall see, however, that one of the solutions, when interpreted as the QCD dual, leads to a prediction of a critical number of flavors corresponding exactly to the one obtained via the conjectured all orders beta function.

 We seek solutions of the anomaly matching conditions for a gauge theory $SU(X)$ with global symmetry group $SU_L(N_f)\times SU_R(N_f) \times U_V(1)$  featuring 
{\it magnetic} quarks ${q}$ and $\widetilde{q}$ together with $SU(X)$ gauge singlet states identifiable as baryons built out of the {\it electric} quarks $Q$. Since mesons do not affect directly global anomaly matching conditions we could add them to the spectrum of the dual theory.  We study the case in which $X$ is a linear combination of number of flavors and colors of the type $\alpha N_f + 3 \beta$ with $\alpha$ and $\beta$ integer numbers. 

We add to the {\it magnetic} quarks gauge singlet Weyl fermions which can be identified with the baryons of QCD but massless. The generic dual spectrum is summarized in table \ref{dualgeneric}.
\begin{table}[h]
\[ \begin{array}{|c| c|c c c|c|} \hline
{\rm Fields} &\left[ SU(X) \right] & SU_L(N_f) &SU_R(N_f) & U_V(1)& \# ~{\rm  of~copies} \\ \hline 
\hline 
 q &\Yfund &{\Yfund }&1&~~y &1 \\
\widetilde{q} & \overline{\Yfund}&1 &  \overline{\Yfund}& -y&1   \\
A &1&\Ythreea &1&~~~3& \ell_A \\
S &1&\Ythrees &1&~~~3& \ell_S \\
C &1&\Yadjoint &1&~~~3& \ell_C \\
B_A &1&\Yasymm &\Yfund &~~~3& \ell_{B_A} \\
B_S &1&\Ysymm &\Yfund &~~~3& \ell_{B_S} \\
{D}_A &1&{\Yfund} &{\Yasymm } &~~~3& \ell_{{D}_A} \\
{D}_S & 1&{\Yfund}  &{\Ysymm} &  ~~~3& \ell_{{D}_S} \\
\widetilde{A} &1&1&\overline{\Ythreea} &-3&\ell_{\widetilde{A}}\\
\widetilde{S} &1&1&\overline{\Ythrees} & -3& \ell_{\widetilde{S}} \\
\widetilde{C} &1&1&\overline{\Yadjoint} &-3& \ell_{\widetilde{C}} \\
 \hline \end{array} 
\]
\caption{Massless spectrum of {\it magnetic} quarks and baryons and their  transformation properties under the global symmetry group. The last column represents the multiplicity of each state and each state is a  Weyl fermion.}
\label{dualgeneric}
\end{table}
The wave functions for the gauge singlet fields $A$, $C$ and $S$ are obtained by projecting the flavor indices of the following operator
\begin{eqnarray}
\epsilon^{c_1 c_2 c_3}Q_{c_1}^{i_1} Q_{c_2}^{i_2} Q_{c_3}^{i_3}\ ,
\end{eqnarray}
over the three irreducible representations of $SU_L(N_f)$ as indicated in the table \ref{dualgeneric}. These states are all singlets under the $SU_R(N_f)$ flavor group. Similarly one can construct the only right-transforming baryons $\widetilde{A}$, $\widetilde{C}$ and $\widetilde{S}$ via $\widetilde{Q}$. The $B$ states are made by two $Q$ fields and one right field $\overline{\widetilde{Q}}$ while the $D$ fields are made by one $Q$ and two $\overline{\widetilde{Q}}$ fermions. $y$ is the, yet to be determined, baryon charge of the {\it magnetic} quarks while the baryon charge of composite states is fixed in units of the QCD quark one.The $\ell$s count the number of times the same baryonic matter representation appears as part of the spectrum of the theory. Invariance under parity and charge conjugation of the underlying theory requires $\ell_{J} = \ell_{\widetilde{J}}$~~ with $J=A,S,...,C$ and $\ell_B = - \ell_D$. 

Having defined the possible massless matter content of the gauge theory dual to QCD we compute the $SU_{L}(N_f)^3$ and $SU_{L}(N_f)^2\,\, U_V(1)$ global anomalies in terms of the new fields:   
 \begin{eqnarray}
SU_{L}(N_f)^3 &\propto &  X + \frac{(N_f-3)(N_f -6)}{2}\,\ell_A + \frac{(N_f+3)(N_f +6)}{2}\,\ell_S + (N_f^2 - 9)\,\ell_C\nonumber \\ &&   +\,(N_f-4)N_f\,\ell_{B_A} + \,(N_f+4)N_f\,\ell_{B_S}  + \frac{N_f(N_f-1)}{2}\,\ell_{D_A} \nonumber \\ 
&&+\frac{N_f(N_f+1)}{2}\,\ell_{D_S}  = 3 \ ,  \\ 
 & &\\
SU_{L}(N_f)^2\,\, U_V(1) &\propto &  y\,X +3 \frac{(N_f-3)(N_f -2)}{2}\,\ell_A + 3\frac{(N_f+3)(N_f +2)}{2}\,\ell_S + 3 (N_f^2 - 3)\,\ell_C \nonumber \\ 
 &&  +\, 3(N_f-2)N_f\,\ell_{B_A} + \,3(N_f+2)N_f\,\ell_{B_S}  + 3\frac{N_f(N_f-1)}{2}\,\ell_{D_A}\nonumber \\ &&+3\frac{N_f(N_f+1)}{2}\,\ell_{D_S} =3  \ .   \end{eqnarray}
  The right-hand side is the corresponding value of the anomaly for QCD. 
 
 \subsubsection{A Realistic QCD Dual}
We have found several solutions to the anomaly matching conditions presented above. Some were found previously in \cite{Terning:1997xy}. Here we start with a new solution in which the gauge group is $SU(2N_f - 5N)$  with the number of colors $N$  equal to $3$. It is, however, convenient to keep the dependence on $N$ explicit. 
 \begin{table}[bh]
\[ \begin{array}{|c| c|c c c|c|} \hline
{\rm Fields} &\left[ SU(2N_f - 5N) \right] & SU_L(N_f) &SU_R(N_f) & U_V(1)& \# ~{\rm  of~copies} \\ \hline 
\hline 
 q &\Yfund &{\Yfund }&1& \frac{N(2 N_f - 5)}{2 N_f - 5N} &~~~1 \\
\widetilde{q} & \overline{\Yfund}&1 &  \overline{\Yfund}& -\frac{N(2 N_f - 5)}{2 N_f - 5N}&~~~1   \\
A &1&\Ythreea &1&~~~3& ~~~2 \\
B_A &1&\Yasymm &\Yfund &~~~3& -2\\
{D}_A &1&{\Yfund} &{\Yasymm } &~~~3& ~~~2 \\
\widetilde{A} &1&1&\overline{\Ythreea} &-3&~~~2\\
 \hline \end{array} 
\]
\caption{Massless spectrum of {\it magnetic} quarks and baryons and their  transformation properties under the global symmetry group. The last column represents the multiplicity of each state and each state is a  Weyl fermion.}
\label{dual}
\end{table}
The solution above corresponds to the following value assumed by the indices and $y$ baryonic charge in table \ref{dualgeneric}.
  \begin{eqnarray}
 X=2N_f - 5N\ , \quad \ell_{A}=2\ ,  \quad \ell_{D_A} = -\ell_{B_A} =2 \ , \quad   \ell_{S}=\ell_{B_S} = \ell_{D_S} =\ell_C =0 \ , \quad y = N\,\frac{2 N_f - 5}{2 N_f - 15} \ ,\nonumber \\
 \end{eqnarray}
 with $N =3$.  $X$ must assume a value strictly larger than one otherwise it is an abelian gauge theory. This provides the first nontrivial bound on the number of flavors: 
 \begin{equation}
 N_f > \frac{5N + 1}{2}  \ , \end{equation} 
which for $N=3$ requires $N_f> 8 $. 
 \subsection*{Conformal Window from the Dual Magnetic Theory}
Asymptotic freedom of the newly found theory is dictated by the coefficient of the one-loop beta function :
 \begin{equation}
 \beta_0 = \frac{11}{3} (2N_f - 5N)  - \frac{2}{3}N_f \ . 
 \end{equation}
 To this order in perturbation theory the gauge singlet states do not affect the {magnetic} quark sector and we can hence determine  the number of flavors obtained by requiring the dual theory to be asymptotic free. i.e.: 
\begin{equation}
N_f \geq \frac{11}{4}N \qquad\qquad\qquad {\rm Dual~Asymptotic~Freedom}\ . 
\end{equation}
Quite remarkably this value {\it coincides} with the one predicted by means of the all orders conjectured beta function for the lowest bound of the conformal window, in the {\it electric} variables, when taking the anomalous dimension of the mass to be $\gamma =2 $. We recall that for any number of colors $N$ the all orders beta function requires the critical number of flavors to be larger than: 
\begin{equation}
N_f^{BF}|_{\gamma = 2} = \frac{11}{4} N \ . 
\end{equation}
{}For N=3 the two expressions yield $8.25$ \footnote{Actually given that $X$ must be at least $2$ we must have  $N_f \geq 8.5$ rather than $8.25$}. We consider this a nontrivial and  interesting result lending further support to the all orders beta function conjecture and simultaneously suggesting that this theory might, indeed, be the QCD magnetic dual. 
%
 The actual size of the conformal window matching this possible dual corresponds to setting $\gamma =2$. {}We note that although for $N_f = 9$ and $N=3$ the magnetic gauge group is $SU(3)$ the theory is not trivially QCD given that it features new massless fermions and their interactions with massless mesonic type fields.

To investigate the decoupling of each flavor at the time one needs to introduce bosonic degrees of freedom. These are not constrained by anomaly matching conditions. Interactions among the mesonic degrees of freedom and the fermions in the dual theory cannot be neglected in the regime when the dynamics is strong. The simplest mesonic operator $M_i^{j} $ transforming simultaneously according to the antifundamental representation of $SU_L(N_f)$ and the fundamental representation of  $SU_R(N_f)$  leads to the following type of interactions for the dual theory: 
\begin{eqnarray}
L_{\rm M} & = & Y_{q\widetilde{q}} \,\,q \, \, M \, \, \widetilde{q } +
Y_{A {B_A}} \,\,A\, M \overline{B}_A+ Y_{C {B_A}} \,\,C \, M \overline{B}_A +  Y_{C {B_S}} \,\,C \, M \overline{B}_S + Y_{S {B_S}} \,\,S \, M \overline{B}_S + \nonumber \\
&+&  Y_{B_A {D_A}}\,\,  B_A \, M\, \overline{D}_A +Y_{B_A {D_S}} \,\, B_A \, M\, \overline D_S + Y_{B_S {D_A}}\,\,  B_S \, M\, \overline{D}_A +Y_{B_S{D_S}} \, B_S \, M\, \overline D_S  + {\rm h.c.}  \nonumber \\
\end{eqnarray}
The coefficients of the various operators are matrices taking into account the multiplicity with which each state occurs. The number of operators drastically reduces if we consider only  the ones linear in $M$.  The dual quarks and baryons interact  via mesonic exchanges. We have considered only the meson field for the bosonic spectrum because is the one with the most obvious interpretation in terms on the electric variables.
One can also envision adding new scalars charged under the dual gauge group \cite{Terning:1997xy} and in this case one can have contact interactions between the magnetic quarks and baryons. We expect these operators to play a role near the lower bound of the conformal window of the magnetic theory where QCD is expected to become free. It is straightforward to adapt the terms above to any anomaly matching solution. 

\begin{figure}[h!]
\centerline{\includegraphics[width=12cm]{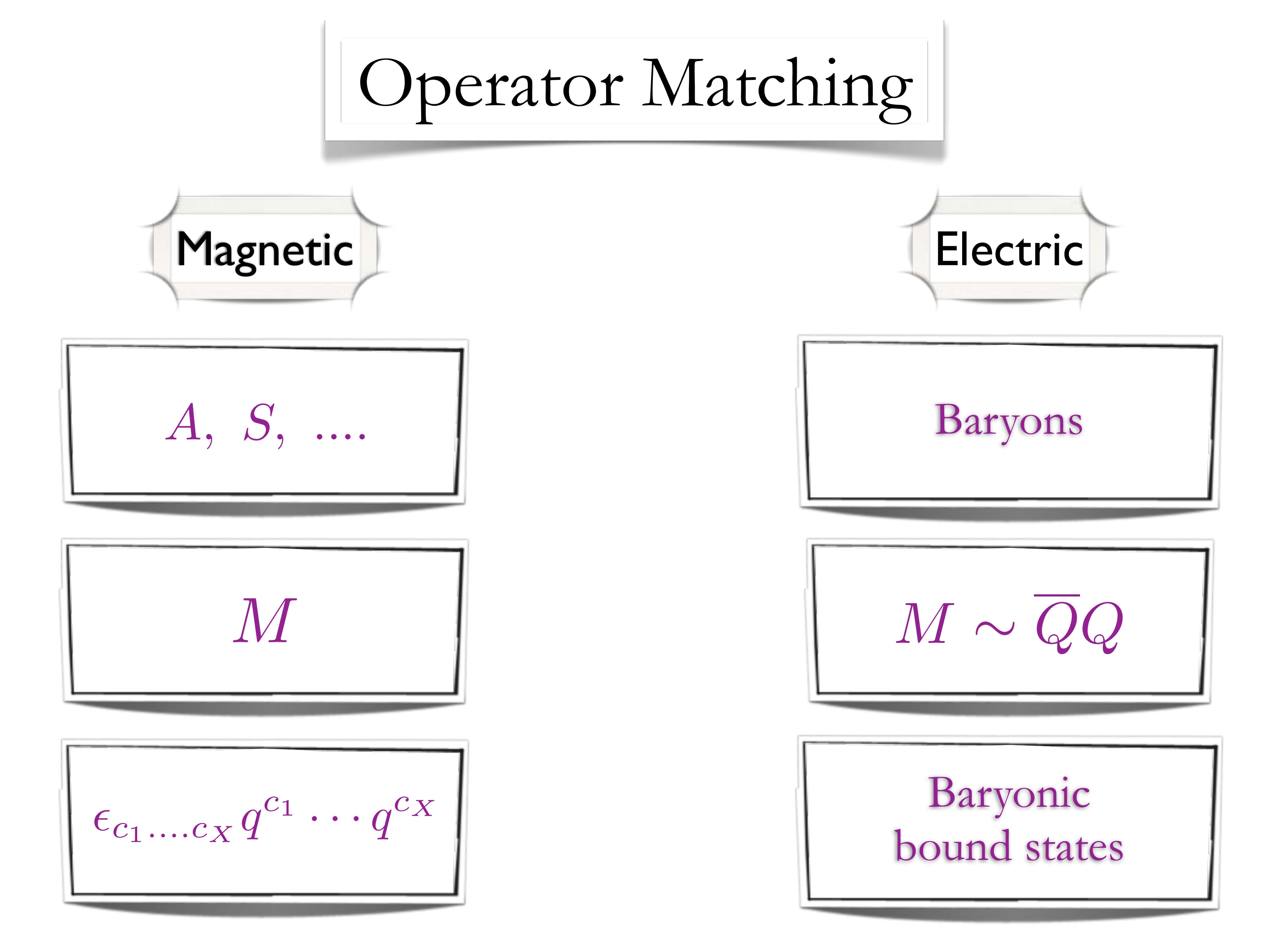}}
\caption{We propose the above correspondence between the gauge singlet operators of the magnetic theory and the electric ones. The novelty introduced in \cite{Sannino:2009qc} with respect to any of the earlier approaches is the identification of the {\it magnetic} baryons, i.e. the ones constructed via the magnetic quarks, with bound states of baryons in the electric variables.}
\label{OpM}
\end{figure}

In Seiberg's analysis it was also possible to match some of the operators of the magnetic theory with the ones of the electric theory. The situation for QCD is, in principle, more involved although it is clear that certain magnetic operators match exactly the respective ones in the electric variables. These are the meson $M$ and the massless baryons, $A$, $\widetilde{A}$, ...., $S$ shown in Table \ref{dualgeneric}.  The baryonic type operators constructed via the magnetic dual quarks have baryonic charge which is a multiple of the ordinary baryons and, hence, we propose to identify them, in the electric variables, with bound states of QCD baryons. We summarize the proposed operator matching constraints in Fig.~\ref{OpM}.

The generalization to a generic number of colors is currently under investigation. It is an interesting issue and to address it requires the knowledge of the spectrum of baryons for arbitrary number of colors.  It is reasonable to expect, however , a possible nontrivial generalization to any number of odd colors \footnote{For an even number of colors the baryons are bosons and a the analysis must modify.}. 

  \subsubsection*{Earlier solutions}
  It is worth comparing the solution above with the ones found already in the literature \cite{Terning:1997xy}.  These are: 
 \begin{eqnarray}
 X=N_f - 6\ , \quad \ell_{A}=\ell_{D_A} = -\ell_{B_A} =1 \ , \quad   \ell_{S}=\ell_{B_S} = \ell_{D_S} =\ell_C = 0 \ , \quad y = 3\,\frac{N_f - 2}{N_f - 6} \ ,\nonumber \\
 \end{eqnarray}
corresponding to $\alpha=1$ and $\beta=-2$, when taking the {\it magnetic} quark flavor symmetry assignment as in table \ref{dualgeneric}.  However, assigning the magnetic quarks $q$ to the complex representation of $SU_L(N_f)$ one has also the solution:
\begin{eqnarray}
 X=N_f  + 6\ , \quad \ell_{S}=\ell_{D_S} = -\ell_{B_S} =1 \ , \quad   \ell_{A}=\ell_{B_A} = \ell_{D_A} =\ell_C= 0 \ , \quad y = -3\,\frac{N_f + 2}{N_f +6} \ .\nonumber \\
 \end{eqnarray}
 Assuming the gauge group to be $SU(N_f\pm 6)$ the one loop coefficient of the beta function is: 
 \begin{equation}
\beta_0 = \frac{11}{3}(N_f  \mp 6 ) - \frac{2}{3}N_f \ ,
\end{equation}
where the sign corresponds to the two possibilities for $X$, i.e. $N_f \mp 6$ and we have only included the magnetic quarks \footnote{ A complex scalar gauged under the dual gauge group was also added in the dual spectrum of \cite{Terning:1997xy}. The scalar  transformed according to the fundamental representation of the left and right $SU(N_f)$ groups. The resulting $SU(N_f -6)$ gauge theory is never asymptotically free and for this reason we have not included this scalar in our discussion. }. The critical number of flavors where asymptotic freedom is lost, in the case of the $N_f-6$ gauge group, corresponds to $7.33$. On the other hand we have a stronger constraint from the fact that the gauge group must be at least $SU(2)$ and hence $N_f \geq 8$ while no useful constraint can be obtained for the $N_f + 6$ gauge group. It was argued in \cite{Terning:1997xy}
that by taking $SO$ or $Sp$ as possible gauge group rather than $SU$ one might increase the critical number of flavors to around $10$. However choosing $SO$ and $Sp$ rather than $SU$ implies that the global symmetry group is enlarged to $SU(2N_f)$ and hence it is not clear how one can still match the anomaly conditions, unless one assumes a simultaneous dynamical enhancement of the QCD global symmetries at the fixed point.

We have uncovered a novel solution of the 't Hooft anomaly matching conditions for QCD. We have shown that in the perturbative regime the new gauge theory, if interpreted as a possible QCD dual, predicts the critical number of flavors above which QCD in the nonperturbative regime, develops an infrared stable fixed point. The value is identical to the maximum bound predicted in the nonpertubative regime via the all orders conjectured beta function for nonsupersymmetric gauge theories. Recent suggestions to analyze the conformal window of nonsupersymmetric gauge theories based on different model assumptions \cite{Poppitz:2009uq} are in qualitative agreement with the precise results of the all orders beta function conjecture. It is worth noting that the combination $2N_f - 5N$ appears in the computation of the mass gap for gauge fluctuations presented in \cite{Poppitz:2009uq,Poppitz:2008hr}. It would be interesting to explore a possible link between these different approaches in the future. 

Interestingly the present solution of the anomaly matching conditions indicate a substantial larger extension of the conformal window than the one predicted using the ladder approximation \cite{Appelquist:1988yc,Cohen:1988sq,Appelquist:1996dq,Miransky:1996pd} and the thermal count of the degree of freedom \cite{Appelquist:1999hr}. 

In fact we have also find solutions for which the lower bound of the conformal window is saturated for $\gamma =1$. The predictions from the gauge duals are, however, entirely and surprisingly consistent with the maximum extension of the conformal window obtained using the all orders beta function \cite{Ryttov:2007cx}. Our main conclusion is that the  't Hooft anomaly conditions alone do not exclude the possibility that the maximum extension of the QCD conformal window is the one obtained for a large anomalous dimension of the quark mass.  

By computing the same gauge singlet correlators in QCD and its suggested dual, one can directly validate or confute this proposal via lattice simulations.

\subsubsection{Higher Representations Duals}
\subsubsection{SU(3) Gauge Theory with 2-index symmetric matter} 
The underlying gauge group is $SU(3)$ while the
quantum flavor group is
\begin{equation}
SU_L(N_f) \times SU_R(N_f) \times U_V(1) \ ,
\end{equation}
and the classical $U_A(1)$ symmetry is destroyed at the quantum
level by the Adler-Bell-Jackiw anomaly. We indicate with
$Q_{\alpha;\{c_1,c_2\}}^i$ the two component left spinor where $\alpha=1,2$
is the spin index, $c_1, c_2=1,...,3$ is the color index while
$i=1,...,N_f$ represents the flavor. $\widetilde{Q}^{\alpha ;\{c_1,c_2\}}_i$
is the two component conjugated right spinor. We summarize the
transformation properties in the following table.
\begin{table}[h]
\[ \begin{array}{|c| c | c c c|} \hline
{\rm Fields} &  \left[ SU(3) \right] & SU_L(N_f) &SU_R(N_f) & U_V(1) \\ \hline \hline
Q &\Ysymm &{\Yfund }&1&~~1  \\
\widetilde{Q} & \overline{\Ysymm}&1 &  \overline{\Yfund}& -1   \\
G_{\mu}&{\rm Adj}   &1&1  &~~1\\
 \hline \end{array} 
\]
\caption{Field content of an SU(3) gauge theory with quantum global symmetry $SU_L(N_f)\times SU_R(N_f) \times U_V(1)$. }
\end{table}

The  global anomalies are associated to the triangle diagrams featuring at the vertices three $SU(N_f)$ generators (either all right or all left), or two 
$SU(N_f)$ generators (all right or all left) and one $U_V(1)$ charge. We indicate these anomalies for short with:
\begin{equation}
SU_{L/R}(N_f)^3 \ ,  \qquad  SU_{L/R}(N_f)^2\,\, U_V(1) \ .
\end{equation}
The cubic anomaly factor, for fermions in fundamental representations,
is $1$ for $Q$ and $-1$ for $\tilde{Q}$ while the quadratic anomaly
factor is $1$ for both leading to
\begin{equation}
SU_{L/R}(N_f)^3 \propto \pm 6 \ , \quad SU_{L/R}(N_f)^2 U_V(1)
\propto \pm 6 \ .
\end{equation}

Specializing to $SU(N)$ with two-index symmetric representation the beta function we have already found that: 
\begin{equation}
N_f(r)^{\rm BF} \geq \frac{11N}{4(N+2)}  \ , \qquad SU(N)~{\rm for~Symmetric~rep.~with}  \quad  { \gamma =2}  .
\end{equation}
which for $N=3$ implies $N_f(r)^{\rm BF} \geq 1.65$.

Assuming, instead, the lower bound to occur for $\gamma =1 $ we discover that:
\begin{equation}
N_f(r)^{\rm BF} \geq \frac{11N}{3(N+2)}  \ , \qquad SU(N)~{\rm for~Symmetric~rep.~with}  \quad  { \gamma =1}  .
\end{equation}
which for $N=3$ implies: $N_f(r)^{\rm BF} \geq 2.2 $.

 We seek solutions of the anomaly matching conditions for a gauge theory $SU(X)$ with global symmetry group $SU_L(N_f)\times SU_R(N_f) \times U_V(1)$  featuring 
{\it magnetic} quarks ${q}$ and $\widetilde{q}$ together with $SU(X)$ gauge singlet states identifiable as baryons built out of the {\it electric} quarks $Q$.   We study the case in which $X$ is a linear combination of number of flavors and colors of the type $\alpha N_f + N \beta$ with $\alpha$ and $\beta$ integer numbers. In fact, in the following we will consider $N=3$. We will also require that the baryons constructed out of the {\it magnetic} quarks have integer baryonic charges with respect to the original baryon number. In this way they will be interpreted as possible bound states of the original baryons. We will see that this is an important property helping selecting possible duals.

\subsubsection{Dual Quarks in the Fundamental Representation}

In this initial investigation we search for dual electric quarks in the fundamental representation of the gauge group $X$. This choice has the virtue to keep linear in $N_f$ the asymptotic freedom condition we will investigate later. We have searched for the more complicate case of dual fermions in higher dimensional representaitons and will present this possibility in the following  section. 

We add to the {\it magnetic} quarks gauge singlet Weyl fermions which can be identified with massless baryons of the electric theory. The generic dual spectrum is summarized in table \ref{dualgeneric2indices}.
\begin{table}[h]
\[ \begin{array}{|c| c|c c c|c|} \hline
{\rm Fields} &\left[ SU(X) \right] & SU_L(N_f) &SU_R(N_f) & U_V(1)& \# ~{\rm  of~copies} \\ \hline 
\hline 
 q &\Yfund &{\Yfund }&1&~~y &1 \\
\widetilde{q} & \overline{\Yfund}&1 &  \overline{\Yfund}& -y&1   \\
A &1&\Ythreea &1&~~~3& \ell_A \\
S &1&\Ythrees &1&~~~3& \ell_S \\
C &1&\Yadjoint &1&~~~3& \ell_C \\
B_A &1&\Yasymm &\Yfund &~~~3& \ell_{B_A} \\
B_S &1&\Ysymm &\Yfund &~~~3& \ell_{B_S} \\
{D}_A &1&{\Yfund} &{\Yasymm } &~~~3& \ell_{{D}_A} \\
{D}_S & 1&{\Yfund}  &{\Ysymm} &  ~~~3& \ell_{{D}_S} \\
\widetilde{A} &1&1&\overline{\Ythreea} &-3&\ell_{\widetilde{A}}\\
\widetilde{S} &1&1&\overline{\Ythrees} & -3& \ell_{\widetilde{S}} \\
\widetilde{C} &1&1&\overline{\Yadjoint} &-3& \ell_{\widetilde{C}} \\
 \hline \end{array} 
\]
\caption{Massless spectrum of {\it magnetic} quarks and baryons and their  transformation properties under the global symmetry group. The last column represents the multiplicity of each state and each state is a  Weyl fermion.}
\label{dualgeneric2indices}
\end{table}
The wave functions for the gauge singlet fields $A$, $C$ and $S$ are obtained by projecting the flavor indices of the following operator
\begin{eqnarray}
\epsilon^{c_1c_2c_3}\epsilon^{d_1 d_2 d_3} Q_{\{c_1,d_1\}}^{i_1} Q_{\{c_2,d_2\}}^{i_2} Q_{\{c_3,d_3\}}^{i_3}\ ,
\end{eqnarray}
over the three irreducible representations of $SU_L(N_f)$ as indicated in the table \ref{dualgeneric}. These states are all singlets under the $SU_R(N_f)$ flavor group. Similarly one can construct the only right-transforming baryons $\widetilde{A}$, $\widetilde{C}$ and $\widetilde{S}$ via $\widetilde{Q}$. The $B$ states are made by two $Q$ fields and one right field $\overline{\widetilde{Q}}$ while the $D$ fields are made by one $Q$ and two $\overline{\widetilde{Q}}$ fermions. $y$ is the, yet to be determined, baryon charge of the {\it magnetic} quarks while the baryon charge of composite states is fixed in units of the electric quark one. The $\ell$s count the number of times the same baryonic matter representation appears as part of the spectrum of the theory. Invariance under parity and charge conjugation of the underlying theory requires $\ell_{J} = \ell_{\widetilde{J}}$~~ with $J=A,S,...,C$ and $\ell_B = - \ell_D$. 

Having defined the possible massless matter content of the gauge theory dual to the electric theory we compute the $SU_{L}(N_f)^3$ and $SU_{L}(N_f)^2\,\, U_V(1)$ global anomalies in terms of the new fields:   
 \begin{eqnarray}
SU_{L}(N_f)^3 &\propto &  {X } + \frac{(N_f-3)(N_f -6)}{2}\,\ell_A + \frac{(N_f+3)(N_f +6)}{2}\,\ell_S + (N_f^2 - 9)\,\ell_C\nonumber \\ &&   +\,(N_f-4)N_f\,\ell_{B_A} + \,(N_f+4)N_f\,\ell_{B_S}  + \frac{N_f(N_f-1)}{2}\,\ell_{D_A} \nonumber \\ 
&&+\frac{N_f(N_f+1)}{2}\,\ell_{D_S}  = 6 \ ,  \\ 
 & &\\
SU_{L}(N_f)^2\,\, U_V(1) &\propto &  y\, {X} +3 \frac{(N_f-3)(N_f -2)}{2}\,\ell_A + 3\frac{(N_f+3)(N_f +2)}{2}\,\ell_S + 3 (N_f^2 - 3)\,\ell_C \nonumber \\ 
 &&  +\, 3(N_f-2)N_f\,\ell_{B_A} + \,3(N_f+2)N_f\,\ell_{B_S}  + 3\frac{N_f(N_f-1)}{2}\,\ell_{D_A}\nonumber \\ &&+3\frac{N_f(N_f+1)}{2}\,\ell_{D_S} =6  \ .   \end{eqnarray}

  The left-hand expressions are identical to the ones of QCD while the right-hand side provides the corresponding value of the anomaly for the electric theory  with two-index symmetric matter.  
     
We require, as done in the QCD case  \cite{Sannino:2009qc},  that the baryonic type operators constructed via the magnetic dual quarks should have baryonic charges multiple of the ordinary baryons ones. We  identify them, in the electric variables, with bound states of ordinary baryons.   
  In the case of $N_f =2 $ the cubic anomaly vanishes identically and should not be considered. {}For three colors the electric theory looses asymptotic freedom for $3.3$ flavors and hence there is only one value of $N_f$, i.e. $N_f=3$ for which both anomalies are relevant. 
 
We have found different solutions to the anomaly matching conditions which we will present here:
\subsubsection*{First solution: $SU(2N_F -3)$ dual gauge group   }
The solutions correspond, for the case $N=3$ to the following value assumed by the indices and $y$ baryonic charge:
  \begin{eqnarray}
 X &= &2N_f- 3\ , \quad \ell_{A}= 0  \ ,  \quad \ell_{D_A} =k_1 = -\ell_{B_A}  \ , \nonumber \\
 && \nonumber \\ 
   \ell_{S}&=& -1 + 2k_1 + 5k_2 \ , \quad  \ell_{D_S} =  -2 + 4 k_1 + 9 k_2 = - \ell_{B_S} \ , \nonumber \\ 
   && \nonumber \\
  \ell_C  & = & 0 \ , \qquad   y = 3\, \frac{a + bN_f + c N_f ^2}{ 4 N_f - 6}  \ ,  \nonumber 
   \end{eqnarray}
 with $k_1$ and $k_2$ integer numbers and  \begin{equation}
 a =  10-12 k_1 - 30 k_2, \quad b= 2k_2  -  k_1  - 1  \quad   c= 3k_1 + 4 k_2 -1 \ .
 \end{equation}
We have asked that both anomaly matching conditions are satisfied for $N_f =3$  and that the solutions satisfy also the quadratic one for $N_f=2$. 
 
Of course $X$ must assume a value strictly larger than one otherwise it describes an abelian gauge theory. This provides the first nontrivial bound on the number of flavors: 
 \begin{equation}
 N_f > \frac{3+ 1}{2} = 2  \ . \end{equation} 
This value is remarkably  consistent with the maximum extension predicted using the truncated SD equation and the all orders beta function for a value of the anomalous dimension equal to one. 

Asymptotic freedom of the newly found theory is dictated by the coefficient of the one-loop beta function :
 \begin{equation}
 \beta_0 = \frac{11}{3} (2N_f- 3)  - \frac{2}{3}N_f\ . 
 \end{equation}
 To this order in perturbation theory the gauge singlet states do not affect the {magnetic} quark sector and we can hence determine  the number of flavors obtained by requiring the dual theory to be asymptotic free. i.e.: 
\begin{equation}
N_f \geq \frac{33}{20} = 1.65 \ , \qquad\qquad\qquad {\rm Dual~Asymptotic~Freedom}  \qquad {\rm I}\ . 
\end{equation}
This value {\it coincides} with the one predicted by means of the all orders conjectured beta function for the lowest bound of the conformal window, in the {\it electric} variables, when taking the anomalous dimension of the mass to be $\gamma =2 $. We recall that for any number of colors $N$ the all orders beta function requires the critical number of flavors to be larger than: 
\begin{equation}
N_f^{BF}|_{\gamma = 2} = \frac{11N}{4(N+2)}  \ . 
\end{equation}
{}For N=3 the two expressions yield $1.65$. Actually given that $X$ must be larger than one this solution requires $N_f > 2$ rather than $1.65$. This last feature was also observed for the QCD dual case. We simply consider this as a signal that we cannot arrive at the maximum value of $\gamma$, nevertheless we can still arrive at a value for the anomalous dimension larger than one according to this solution. If one requires an even more stringent constraint $X \geq 2$ we then find $N_f>2.5$ which is very close to the result obtained setting $\gamma=1$ in the all orders beta function.

The baryon charge of the magnetic baryons is: 
\begin{equation}
B[q^X] = X \times y = \frac{3}{2} (a+bN_f + c N_f^2)   = {\rm Operator~matching} = 3\, n  \ ,
\end{equation}
with $n$ an integer requiring  $a+bN_f + c N_f^2$ to be an even number. This extra constraints is easily satisfied by choosing, for example, $k_1 =1$ and $k_2 = 0$ yielding $B[q^X] = 3\, (N_f^2 - N_f  -1)$. {}Intriguingly for $N_f=2$ one recovers the standard baryonic charge.

\subsubsection*{Second solution: SU(7Nf - 15)}

The solutions correspond to the following value assumed by the indices and $y$ baryonic charge: 
  \begin{eqnarray}
 X &= &7N_f- 15\ , \quad \ell_{A}=0 \ ,  \quad \ell_{D_A} = k_1  = -\ell_{B_A}  \ , \nonumber \\
 && \nonumber \\ 
   \ell_{S}&=& 2k_1 + 5k_2 \ , \quad  \ell_{D_S} =    4 k_1 + 9 k_2  = - \ell_{B_S} \ , \nonumber \\ 
   && \nonumber \\
   \ell_C  & = & 0 \ , \qquad 
  y = 3 \frac{a+ bN_f +c N_f^2}{14 N_f  - 30}  \ .\nonumber \\
   && \ \ 
 \end{eqnarray}
 with $k_1$ and $k_2$ integer numbers and 
 \begin{equation}
a = 4- 12 k_1 - 30 k_2 \ , \quad b= 2 k_2  - k_1 \ , \quad c= 3 k_1 +4 k_2  \ .
\end{equation}
The baryon charge of the magnetic baryons is: 
\begin{equation}
B[q^X] = X \times y = \frac{3}{2} (a+bN_f + c N_f^2)   = {\rm Operator~matching} = 3\, n  \ ,
\end{equation}
with $n$ an integer requiring  $a+bN_f + c N_f^2$ to be an even number. This extra constraints is also easily satisfied by choosing, for example, $k_1 =k_2 =0$  yielding $B[q^X] = 6$ for any $N_f$ corresponding to a di-baryon charge. One can also consider the case $k_2 = 1$ and $k_1=0$. 

The condition $X>1$ yields: 
 \begin{equation}
 N_f > \frac{16}{7} \simeq 2.29  \ . \end{equation} 
This value is also remarkably  consistent with the maximum extension predicted using the truncated SD equation and the all orders beta function for a value of the anomalous dimension equal to one. 

Asymptotic freedom of the newly found theory is dictated by the coefficient of the one-loop beta function :
 \begin{equation}
 \beta_0 = \frac{11}{3} (7N_f- 15)  - \frac{2}{3}N_f\ , 
 \end{equation}
 yielding
 \begin{equation}
N_f \geq \frac{11}{5} = 2.2\ , \qquad\qquad\qquad {\rm Dual~Asymptotic~Freedom} \qquad {\rm II}\ . 
\end{equation}
This value {\it coincides} with the one predicted by means of the all orders conjectured beta function for the lowest bound of the conformal window, in the {\it electric} variables, when taking the anomalous dimension of the mass to be $\gamma =1 $. We recall that for any number of colors $N$ the all orders beta function requires the critical number of flavors to be larger than: 
\begin{equation}
N_f^{BF}|_{\gamma = 1} = \frac{11N}{3(N+2)}  \ . 
\end{equation}
{}For N=3 the two expressions yield $2.2$.  This value is even closer to the one obtained imposing the condition $X>1$ which is  circa $2.29$. 

Interestingly the two class of solutions suggest that the electric theory is not conformal but walking. We observe that the predictions from the dual theory does {\it not} depend on the all orders beta function. However it is remarkable that the predictions are very close to the ones predicted using the beta function ansatz.  

An interesting property of this solution is that one can saturate the anomaly matching conditions directly via the presumed magnetic quarks. It is, in fact, sufficient to set $k_1 = k_2 =0$ to see this. If we apply the all order beta function we can investigate when chiral symmetry is restored. Setting $\gamma=1$ for the dual theory one finds that $N_f$ should be less than or equal to about $2.29$ which is lower than the value for which the electric theory looses asymptotic freedom. This seems to indicate  that more matter is needed and the solution $k_1=k_2=0$ is not an exact dual according to the all orders beta function. However one can investigate the nonperturbative dynamics of this theory via first principle lattice simulations and test the duality independently.   

\subsubsection*{Third solution: $SU(\alpha N_f (N+2)  - \beta N + \delta)$ and two-index symmetric magnetic quarks. }

We have investigated also the case in which the magnetic quarks are in the same two-index representation of the gauge group $X = \alpha N_f (N+2)  - \beta N + \delta$. In this case the first coefficient of both anomalies must be modified according to $X \rightarrow X(X+1)/2$ to take into account of the change of the representation of the dual quarks.  We have found several solutions for different integer values of the coefficients $\alpha$, $\beta$ and $\delta$. We present two examples here for $N=3$: 

\vskip .2cm
{$\alpha =2$, $\beta=5$ and $\delta=0$}

 \begin{eqnarray}
 X &= &10 N_f- 15 \ , \quad \ell_{A}=0 \ ,  \quad \ell_{D_A} = k_1  = -\ell_{B_A}  \ , \nonumber \\
 && \nonumber \\ 
   \ell_{S}&=&  -2 + 2k_1 + 5k_2 \ , \quad  \ell_{D_S} =  4 +  4 k_1 + 9 k_2  = - \ell_{B_S} \ , \nonumber \\ 
   && \nonumber \\
   \ell_C  & = & 0 \ , \qquad 
  y = 3 \frac{a+ bN_f +c N_f^2}{10(2 N_f  - 3)(5N_f - 7)}  \ .\nonumber \\
   && \ \ 
 \end{eqnarray}
 with $k_1$ and $k_2$ integer numbers and  
 \begin{equation}
a = 16- 12 k_1 - 30 k_2 \ , \quad b= 22 + 2 k_2  - k_1 \ , \quad c= 6 + 3 k_1 +4 k_2  \ .
\end{equation}

The one-loop coefficient of the beta function is: 
 \begin{eqnarray}
  \beta_0 = \frac{11}{3}X   - \frac{2}{3}N_f  (X +2) \ . 
 \end{eqnarray}
 Asymptotic freedom requires the previous coefficient to be positive which means: 
 \begin{eqnarray}
 1.58 \leq  N_f   \leq 5.22, \quad  {\rm Dual~asymptotic~freedom~condition}\ .
 \end{eqnarray}
 The lower bound is now close to the value of the critical number of flavors corresponding to the maximum extension ($\gamma =2$) value where the all orders beta function requires the electric theory to start developing an IRFP. This time the condition $X>1$ yields a weaker constraint, i.e. $N_f > 1.5$, with respect to the asymptotic freedom constraint on the lowest value for $N_f$. The trend is different with respect to the case in which we considered magnetic quarks transforming according to the fundamental representation of the $SU(X)$ gauge group. 

\vskip .2cm
{$\alpha =4 $, $\beta = 11$ and $\delta =2$}

In this case the solution is:
 \begin{eqnarray}
 X &= & 20N_f- 31 \ , \quad \ell_{A}=0 \ ,  \quad \ell_{D_A} = k_1  = -\ell_{B_A}  \ , \nonumber \\
 && \nonumber \\ 
   \ell_{S}&=&  -2 + 2k_1 + 5k_2 \ , \quad  \ell_{D_S} =  25 +  4 k_1 + 9 k_2  = - \ell_{B_S} \ , \nonumber \\ 
   && \nonumber \\
   \ell_C  & = & 0 \ , \qquad 
  y = 3 \frac{a+ bN_f +c N_f^2}{10(20N_f  - 31)(2N_f - 3)}  \ .\nonumber \\
   && \ \ 
 \end{eqnarray}
 with $k_1$ and $k_2$ integer numbers and 
 \begin{equation}
a = 16- 12 k_1 - 30 k_2 \ , \quad b= 85 -  k_1 + 2k_2 \ , \quad c= 27 + 3 k_1 +4 k_2  \ .
\end{equation}
 Asymptotic freedom requires the previous coefficient to be positive which means: 
 \begin{eqnarray}
 1.59 \leq  N_f   \leq 5.36, \quad  {\rm Dual~asymptotic~freedom~condition}\ .
 \end{eqnarray}
 The lower bound is close again to the value of the critical number of flavors corresponding to the maximum extension ($\gamma =2$) value where the all orders beta function requires the electric theory to start developing an IRFP. The condition $X>1$ yields a  constraint, i.e. $N_f > 1.6$ consistent with the lowest value of the asymptotic freedom window. Note that we have arranged $X$ in such a way that for $X\geq 2$ we recover identically the all order beta function bound for $\gamma=2$.  
 
We were able to find a solution for different values of $\alpha$, in particular for $\alpha =3$ the condition $X\geq 2$  is consistent with the bound of the all orders beta function but for $\gamma =1$ and asymptotic freedom requires $2.15 < N_f < 5.28 $.

{} For duals  with {\it magnetic} quarks in the two index symmetric representation we find more difficult to have a reasonable interpretation of the magnetic baryons, i.e. possessing $B[q^X] = 3n$.  Our findings suggest that duals with fermions in the fundamental representation are, actually, privileged. 

We have found solutions matching the predictions coming from the conjectured all orders beta function also in the case of theories with fermions in the two-index symmetric representation of the $SU(3)$ gauge group. Moreover if one uses dual quarks in the fundamental representation the typical size of the allowed conformal window is consistent with the $\gamma =1$ condition. On the other hand, when using dual quarks in the two-index symmetric representation the size of the conformal window compatible with 't Hooft anomaly matching can extend to match the one obtained using $\gamma =2 $ in the all orders beta function. However the latter case is disfavored by the operator matching conditions given that the $U_V(1)$ charge of magnetic baryons is typically not an integer number of ordinary baryons.

\subsubsection{Minimal Conformal Theories: SU(N) with Adjoint Weyl Matter} 
We considered till now only a fixed number of colors since the spectrum of possible composite fermions increases when increasing the number of colors. We turn our attention now to another class of two-index theories for which the dependence on the number of colors, spectrum-wise, is trivial. These are theories with a generic number of Weyl fermions transforming according to the adjoint representation of the underlying $SU(N)$ gauge group.  The associated quantum flavor group is simply $ SU(N_f)$.
We indicate with
$\lambda_{\alpha;a}^i$ the two component left spinor where $\alpha=1,2$
is the spin index, $a=1,...,N^2 -1 $ is the color index while
$i=1,...,N_f$ represents the flavor. We summarize the
transformation properties in the following table:
\begin{table}[h]
\[ \begin{array}{|c| c | c |} \hline
{\rm Fields} &  \left[ SU(N) \right] & SU(N_f)  \\ \hline \hline
\lambda &{\rm Adj} &{\Yfund }  \\
G_{\mu}&{\rm Adj}   &1\\
 \hline \end{array} 
\]
\caption{Field content of an SU(N) gauge theory with quantum global symmetry $SU(N_f)$. }
\end{table}

The  global anomalies are associated to the triangle diagrams featuring at the vertices three $SU(N_f)$ generators. We indicate these anomalies for short with:
\begin{equation}
SU(N_f)^3   \ .
\end{equation}
For a vector like theory there are no further global anomalies. The
cubic anomaly factor, for fermions in the fundamental representation,
is one leading to 
\begin{equation}
SU(N_f)^3 \propto N^2 - 1 \ .
\end{equation}
   We seek solutions of the anomaly matching conditions for a possible dual gauge theory $SO(X)$  featuring 
{\it magnetic} Weyl quarks ${q}$ transforming according to the vector representation of the gauge group. The global symmetry group is then $SU(N_f)$. 
We also add gauge singlet fields built out of the {\it electric} quarks $\lambda$.  The dual spectrum is summarized in table \ref{dualgeneric4}.
\begin{table}[h]
\[ \begin{array}{|c| c|c |c|} \hline
{\rm Fields} &\left[ SO(X) \right] & SU(N_f) & \# ~{\rm  of~copies} \\ \hline 
\hline 
 q &\Yfund &{\Yfund } &1 \\
\Lambda &1&{\Yfund}& \ell_\Lambda \\
 \hline \end{array} 
\]
\caption{Massless spectrum of {\it magnetic} quarks and baryons and their  transformation properties under the global symmetry group. The last column represents the multiplicity of each state and each state is a  Weyl fermion.}
\label{dualgeneric4}
\end{table}
The gauge singlet state $\Lambda$ is nothing but the gauge singlet built out of the gauge field strength and ${\lambda}$. We can have several copies of $\Lambda$. 

Having defined the possible massless matter content of the gauge theory dual we compute the relevant anomaly:   
 \begin{eqnarray}
SU(N_f)^3 &\propto &  X +  \ell_{\Lambda}= N^2 - 1  \ .  \end{eqnarray}
  The right-hand side is the corresponding value of the anomaly for the electric theory. {}For any $X$ we have a solution which is: 
  \begin{equation} 
  \ell_{\Lambda} =  N^2-1 -X \ . 
  \end{equation}
  The one-loop coefficient of the beta function is: 
 \begin{eqnarray}
  \beta_0 = \frac{11}{3}(X-2)   - \frac{2}{3}N_f \ . 
 \end{eqnarray}
 We find that for $X= N_f -1$ asymptotical freedom is lost for: 
 \begin{eqnarray}
  N_{f} \geq \frac{11}{3}  \ , \quad  {\rm Dual~asymptotic~freedom~ and~}N_f~{\rm Weyl~fermions} \ ,
 \end{eqnarray}
 in total agreement with the lower bound of the conformal window obtained by imposing $\gamma=1$ in the all orders beta function. 
 In fact $N_f$ must be larger or equal than four for the dual $SO(N_f-1)$ theory to be a non-abelian gauge theory. Since $N_f$ counts the number of Weyl fermions we have found that the number of Dirac flavors above which we expect any $SU(N)$ gauge theory to develop an infrared fixed point must be equal or larger than two. This is an extremely interesting result since it agrees with earlier analytical expectations obtained using several different analytic methods as well as recent first principle lattice results \cite{Catterall:2007yx,Catterall:2008qk,Hietanen:2008mr,DelDebbio:2008tv,Hietanen:2009az,DelDebbio:2009fd}.

 We have also explored the possibility to introduce dual fermions in the adjoint representation of a SU dual gauge group. Although solutions to the anomaly conditions are straightforward we find that the solution above is the one which better fits the numerical and analytical results. 
 
\subsubsection*{Higher Representations Duals summary}
We provided the first investigation of the conformal window of nonsuperymmetric gauge theories with sole fermionic matter transforming according to higher dimensional representation of the underlying gauge group. We argued that, if the duals exist, they are gauge theories with fermions transforming according to the defining representation of the dual gauge group. The resulting conformal windows match the one stemming from the all orders beta function results when taking the anomalous dimension of the fermion mass to be unity. 
 In particular our results for the adjoint representation indicate that for two Dirac flavors any $SU(N)$ gauge theory should enter the conformal window. These results are in excellent agreement with numerical and previous analytical results \cite{Sannino:2004qp,Dietrich:2006cm,Ryttov:2007cx,Sannino:2009aw,Poppitz:2009uq}.  The mapping of higher dimensional representations into duals with fermions in the fundamental representation can be the source of the observed universality of the {\it size} of the various phase diagrams for different representations noted in \cite{Ryttov:2007sr}. 

\subsection{Phases of Chiral Gauge Theories}

Chiral gauge theories, in which at least part of the matter field
content is in complex representations of the gauge group, play an
important role in efforts to extend the SM. These
include grand unified theories, dynamical breaking of symmetries,
and theories of quark and lepton substructure. An important
distinction from vector-like theories such as QCD is that since at
least some of the chiral symmetries are gauged, mass terms that
would explicitly break these chiral symmetries are forbidden in the
Lagrangian. Another key feature is that the fermion content is
subject to a constraint not present in vectorial gauge theories,
the cancellation of gauge and gravitational anomalies.

Chiral theories received much attention in the 1980's~\cite{Ball:1988xg},
focusing on their strong coupling behavior in the infrared. One
possibility is confinement with the gauge symmetry as well as
global symmetries unbroken, realized by the formation of gauge
singlet, massless composite fermions. Another is confinement with
intact gauge symmetry but with some of the global symmetries broken
spontaneously, leading to the formation of gauge-singlet Goldstone
bosons. It is also possible for these theories to exist in the
Higgs phase, dynamically breaking their own gauge
symmetries~\cite{Raby:1979my}. Depending on particle content, they might
even remain weakly coupled. This will happen if the theory has an
interacting but weak infrared fixed point. The symmetries will then
remain unbroken, and the infrared and underlying degrees of freedom
will be the same.

Supersymmetric (SUSY) chiral theories have also received
considerable attention over the years, since most of the known
examples of dynamical supersymmetry breaking involve these kinds of
theories \cite{Poppitz:1998vd}.

 Studies of chiral gauge theories have typically made use of
the 't Hooft global anomaly matching conditions~\cite{'tHooft:1980xb} along
with $1/N$ expansion, and not-so-reliable most attractive channel
(MAC)  analysis and instanton computations. Direct
approaches using strong coupling lattice methods are
still difficult.

Here we confront the  results obtained in Ref.~\cite{Appelquist:1999vs,Appelquist:2000qg} using the thermal degree of count freedom with the generalization of the all orders beta function useful to constrain these extremely interesting phase diagrams. The two important class of theories we are going to investigate are the Bars-Yankielowicz (BY) \cite{Bars:1981se} model involving fermions in the two-index symmetric tensor
representation, and the other is a generalized Georgi-Glashow (GGG)
model involving fermions in the two-index antisymmetric tensor
representation. In each case, in addition to fermions in complex
representations, a set of $p$ anti fundamental-fundamental pairs
are included and the allowed phases are considered as a function of
$p$. Several possible phases emerge, consistent with global anomaly
matching and the thermal inequality. It was noted in \cite{Appelquist:2000qg} that in the real world case of two-flavor
QCD (a vector-like theory with {\it all} fermions in a real
representation) nature prefers to minimize $f_{IR}$. Neglecting the
small bare quark masses, global anomaly matching admits two
possible low energy phases, broken chiral symmetry through the
formation of the bilinear $<{\bar\psi}\psi
>$ condensate, or unbroken chiral symmetry through the formation of confined
massless baryons. Both effective low energy theories are infrared
free. The three Goldstone bosons of the former (chosen by nature)
lead to $ f_{IR}= 3$, and the two massless composite Dirac fermions
of the latter lead to $f_{IR}= 7$.

We focus almost completely on symmetry breaking patterns
corresponding to the formation of bilinear condensates. We suggested in \cite{Appelquist:2000qg}
that in general, the phase corresponding to confinement with all
symmetries unbroken, where all the global anomalies are matched by
massless composite fermions, is not preferred. Instead, the global
symmetries associated with fermions in real representations break
spontaneously via bilinear condensate formation as in QCD. With
respect to the fermions in complex representations, however, the
formation of bilinear condensates is suggested to be disfavored
relative to confinement and the preservation of the global
symmetries via massless composite fermion formation.

Bilinear condensate formation is of course not the only possibility
in a strongly coupled gauge field theory. We extended
our discussion to include general condensate formation for one
simple example, the $SU(5)$ Georgi-Glashow model, which has
fermions in only complex representations and has only a $U(1)$
global symmetry. This symmetry can be broken via only higher
dimensional condensates. Interestingly, this breaking pattern, with
confinement and unbroken gauge symmetry, leads to the minimum value
of $f_{IR}$. This highlights the important general question of the
pattern of symmetry breaking in chiral theories when arbitrary
condensate formation is considered. Higher dimensional condensates
might play an important role, for example, in the dynamical
breaking of symmetries in extensions of the SM
\cite{Eichten:1985fs}.

\subsubsection{All-orders beta function for Chiral Gauge Theories}
A generic chiral gauge theory has always a set of matter fields for which one cannot provide a mass term, but it can also contain vector-like matter. We hence suggest the following minimal modification of the all orders beta function \cite{Ryttov:2007cx} for any nonsupersymmetric chiral gauge theory: 
\begin{equation}
\beta_{\chi} (g)= -\frac{g^3}{(4\pi)^2} \frac{\beta_0 - \frac{2}{3}\sum_{i=1}^{k}T(r_i)p(r_i) \gamma_i (g^2)} 
{1 - \frac{g^2}{8\pi^2}C_2(G)\left(1 + \frac{2\beta^{\prime}_{ \chi}}{\beta_0} \right)} \ ,
\end{equation}
where $p_i$ is the number of vector like pairs of fermions in the representation $r_i$ for which an anomalous dimension of the mass $\gamma_i$ can be defined.  $\beta_0$ is the standard one loop coefficient of the beta function while $\beta^{\prime}_{\chi}$ expression is readily obtained by imposing that when expanding  $\beta_{\chi}$ one recovers the two-loop coefficient correctly and its explicit expression is not relevant here. Note that the previous expression readily reproduces our original beta function when setting to zero the number of true chiral fermions. At this point one can investigate the conformal window for chiral gauge theories in a manner analogous to the case of purely vector matter. We will use this method to determine the conformal window of the most relevant chiral theories, already used already in particle physics, and described in much detail below. According to the new beta function gauge theories without vector-like matter but featuring several copies of purely chiral matter will be conformal when the number of copies is such that the first coefficient of the beta function vanishes identically.  Using topological excitations an analysis of this case was performed in \cite{Poppitz:2009uq}. 
 
\subsection{ The Bars Yankielowicz (BY) Model}
\label{due}

This model is based on the single gauge group $SU(N\geq 3) $ and
includes fermions transforming as a symmetric tensor
representation, $S=\psi
_{L}^{\{ab\}}$, $a,b=1,\cdots ,N$; $\ N+4+p$ conjugate fundamental
representations: $\bar{F}_{a,i}=\psi _{a,iL}^{c}$, where $i=1,\cdots ,N+4+p$%
; and $p$ fundamental representations, $F^{a,i}=\psi _{L}^{a,i},\ i=1,\cdots
,p$. The $p=0$ theory is the basic chiral theory, free of gauge
anomalies by virtue of cancellation between the antisymmetric
tensor and the $N+4$ conjugate fundamentals. The additional $p$
pairs of fundamentals and conjugate fundamentals, in a real
representation of the gauge group, lead to no gauge anomalies.

The global symmetry group is
\begin{equation}
G_{f}=SU(N+4+p)\times SU(p)\times U_{1}(1)\times U_{2}(1)\ .
\label{gglobal3}
\end{equation}
Two $U(1)$'s are the linear combination of the original $U(1)$'s generated
by $S\rightarrow e^{i\theta _{S}}S$ , $\bar{F}\rightarrow e^{i\theta _{\bar{F%
}}}\bar{F}$ and $F\rightarrow e^{i\theta _{F}}F$ that are left invariant by
instantons, namely that for which $\sum_{j}N_{R_{j}}T(R_{j})Q_{R_{j}}=0$,
where $Q_{R_{j}}$ is the $U(1)$ charge of $R_{j}$ and $N_{R_{j}}$ denotes
the number of copies of $R_{j}$.

Thus the fermionic content of the theory is
\begin{table}[h]
\[ \begin{array}{c |  cc  c c c } \hline
{\rm Fields} &\left[ SU(N) \right] & SU(N+4+p) & SU(p) & U_1(1) &U_2(1) \\ \hline 
\hline 
 S &\Ysymm &1 &1 & N+4 &2p \\
  \bar{F} &\bar{\Yfund}  &\bar{\Yfund} & 1  & -(N+2) & - p \\
   {F} &\Yfund  &1 & \Yfund  & N+2 & - (N - p) \\
 \hline \end{array} 
\]
\caption{The Bars Yankielowicz (BY) Model}
\end{table}
where the first $SU(N)$ is the gauge group, indicated by the square brackets.

The perturbative beta function (trivially related to the one defined via the derivative of  $g$) is
\begin{equation}
\beta =\mu \frac{d\alpha }{d\mu }=-\beta _{1}{\LARGE (}\frac{\alpha ^{2}}{
2\pi }{\Large )}-\beta _{2}{\large (}\frac{\alpha ^{3}}{4\pi
^{2}}{\large )} +O(\alpha ^{4}) \ , \label{betafun}
\end{equation}
where the terms of order $\alpha ^{4}$ and higher are scheme-dependent. For
the present model, we have $\beta _{1}=3N-2-(2/3)p$ and $\beta
_{2}=(1/4)\{13N^{2}-30N+1+12/N-2p((13/3)N-1/N)\}$. Thus the theory is
asymptotically free for
\begin{equation}
p<(9/2)N-3 \ . \label{afby}
\end{equation}
We shall restrict $p$ so that this condition is satisfied.

Because of asymptotic freedom, the thermodynamic free-energy may be computed
in the $T\rightarrow \infty $ limit. An enumeration of the degrees of
freedom leads to
\begin{equation}
f_{UV}=2(N^{2}-1)+\frac{7}{4}{\large
[}\frac{N(N+1)}{2}+(N+4)N+2pN{\Large ]} \ .
\label{finfby}
\end{equation}

The infrared realization of this theory will vary depending on the number $p$
of conjugate fundamental-fundamental pairs. We begin by discussing the $p=0$
theory and then map out the phase structure as function of $p$.

\subsubsection*{The $p=0$ Case}

For $p=0$, the fermions are in complex representations of the
$SU(N)$ gauge group and the global symmetry group is
$G_{f}=SU(N+4)\times U_{1}(1)$. The theory is strongly coupled at
low energies, so it is expected either to confine or to break some
of the symmetries, consistent with global anomaly matching.

All the global anomalies of the
underlying theory may be matched at low energies providing that the
massless spectrum is composed of gauge singlet composite fermions
transforming according to the antisymmetric second-rank tensor
representation of $SU(N+4)$. They are described by the composite
operators $\bar{F}_{[i}S\bar{F}_{j]}$ and have charge $-N$ under
the $U_{1}(1)$ global symmetry.

With only these massless composites in the low energy spectrum, there are no
dimension-four interactions, so the composites are noninteracting in the
infrared. Therefore the thermodynamic free energy may be computed in the
limit $T\rightarrow 0$. Enumerating the degrees of freedom gives
\begin{equation}
f_{IR}^{sym}(p=0) = \frac{7}{4}\frac{(N+4)(N+3)}{2}\ ,
\label{fzerosym}
\end{equation}
where the superscript indicates that the full global symmetry is
intact. Clearly $f_{IR}^{sym}(p=0)< f_{UV}(p=0)$, satisfying the
inequality of Eq.~(\ref{eq:ineq})~\cite{Appelquist:1999vs}.

While the formation of confined massless composite fermions and the
preservation of $G_{f}$ is consistent with anomaly matching and the
thermal inequality, the same can be seen to be true of broken
symmetry channels. We consider first the Higgs phase corresponding
to the maximally attractive channel.
\begin{equation}
\Ysymm \times \bar{\Yfund} \rightarrow \Yfund \ ,
\end{equation}
leading to the formation of the $S\bar{F}$ condensate
\begin{equation}
\varepsilon ^{\gamma \delta }S_{\gamma }^{ai}\bar{F}_{\{a,i\},\delta }\ ,
\label{gauge breaking}
\end{equation}
where $\gamma ,\delta =1,2$ are spin indices and $a,i=1,\cdots ,N$
are gauge and flavor indices. This condensate breaks $U_{1}(1)$ and
all the gauge symmetries, and it breaks $SU(N+4)$ to $SU(4)$. But
the $SU(N)$ subgroup of $SU(N+4)$ combines with the gauge group,
leading to a new global symmetry $SU^{\prime }(N)$. For this group,
$\bar{F}_{a,i\leq N}$ is reducible, to the symmetric
$\bar{F}^{S}=\bar{F}_{\{a,i\}}$ and the anti-symmetric
$\bar{F}^{A}=\bar{F}_{[a,i]}$ representations.

The broken $SU(N+4)$ generator
\[
Q_{(N+4)}=\left(
\begin{tabular}{ccc|ccc}
$4$ &  &  &  &  &  \\ & $\ddots $ &  &  &  &  \\ &  & $4$ &  &  &
\\ \hline &  &  & $-N$ &  &  \\ &  &  &  & $\ddots $ &  \\ &  &  &
&  & $-N$
\end{tabular}
\right) \ ,
\]
combines with $Q_{1}$ giving a residual global symmetry $U_{1}^{\prime }(1)=$
$\frac{1}{N+4}(2Q_{1}-Q_{(N+4)})$. The breakdown pattern thus is
\begin{equation}
\left[ SU(N)\right] \times SU(N+4)\times U_{1}(1)\rightarrow SU^{\prime
}(N)\times SU(4)\times U_{1}^{\prime }(1) \ .
\label{rim}
\end{equation}

The gauge bosons have become massive as have some fermions. The fermionic
spectrum, with respect to the residual global symmetry is
\begin{equation}
\begin{tabular}{c|cccc}
&  & $SU^{\prime }(N)$ & $SU(4)$ & $U_{1}^{\prime }(1)$ \\ \hline \\
& $S$ & $ \Ysymm $& $1$ & $2$ \\
massive &  &  &  &  \\
& $\bar{F}^{S}$ &$ \overline{\Ysymm}$ & $1$ & $-2$ \\
&&&&\\
 \cline{1-3}\cline{2-5}
 &&&&\\
& $\bar{F}^{A}$ & $\overline{\Yasymm
}$ & $1$ & $-2$ \\
massless &  &  &  &  \\
& $\bar{F}_{i>N}$ & $\overline{\Yfund}$ & $\overline{
\Yfund}$ & $-1$
\end{tabular}
\end{equation}
This breaking pattern gives $N^{2}+8N$ Goldstone bosons, $N^{2}-1$ of which
are eaten by the gauge bosons. So only $8N+1$ remain as part of the massless
spectrum along with the massless fermions. The global anomalies are again
matched by this spectrum. Those associated with the unbroken group $%
SU^{\prime }(N)\times SU(4)\times U_{1}^{\prime }(1)$ are matched by the
massless fermions, while those associated with the broken global generators
are matched by the Goldstone bosons. Since the Goldstone bosons do not
couple singly to the massless fermions (no dimension-four operators), the
effective zero-mass theory is free at low energies.

 It follows that the thermodynamic free energy may be computed at
$T
\rightarrow 0$ by counting the degrees of freedom. The result is
\begin{equation}
f_{IR}^{Higgs}(p=0) = (8N+1)+\frac{7}{4}[\frac{1}{2}N(N-1)+4N] \ ,
\end{equation}
where the superscript indicates that the gauge symmetry is
(partially) broken. Just as in the case of the symmetric phase, the
inequality Eq.~(\ref{eq:ineq}) is satisfied: $f_{IR}^{Higgs}(p=0)<
f_{UV}(p=0)$.

As an aside, we note that according to the idea of complementarity
this low energy phase may be thought of as having
arisen from confining gauge forces rather than the Higgs
mechanism. Confinement then would partially break the
global symmetry to the above group forming the necessary Goldstone
bosons. It would also produce gauge singlet massless composite
fermions to replace precisely the massless elementary fermions in
the above table.

We have identified  \cite{Appelquist:2000qg} two possible phases of this theory consistent
with global anomaly matching and the inequality
Eq.~(\ref{eq:ineq}). One confines and breaks no symmetries. The
other breaks the chiral symmetry according to Eq.~(\ref{rim}). For
any finite value of $N$, $f_{IR}^{sym}(p=0)< f_{IR}^{Higgs}(p=0)$.
The symmetric phase is thus favored if the number of degrees of
freedom, or the entropy of the system near freeze-out, is
minimized. In the limit $N\rightarrow
\infty$, the Goldstone bosons do not contribute to leading order,
and $f_{IR}^{sym}(p=0)
\rightarrow f_{IR}^{Higgs}(p=0)$. We return to a discussion
of the infinite $N$ limit after describing the general ($p > 0$)
model.

What about other symmetry breaking phases of the $p=0$ theory
corresponding to bilinear condensate formation? In addition to
$S\bar{F}$ condensates, there are also $SS$ and $\bar{F}\bar{F}$
possibilities. Several of these correspond to attractive channels,
although not maximally attractive, due to gluon exchange. We have
considered all of them for the case $N=3$, and have shown that the
effective low energy theory is infrared free and that the number of
low energy degrees of freedom is larger than the symmetric phase.

\subsubsection*{The General Case}

We next consider the full range of $p$ allowed by asymptotic
freedom: $0 < p< (9/2)N-3$. For $p$ near $(9/2)N - 3$, an infrared
stable fixed point exists, determined by the first two terms in the
$\beta$ function. This can be arranged by taking both $N$ and $p$
to infinity with the difference $(9/2)N - p$ fixed, or at finite
$N$ by continuing to nonintegral $p$. The infrared coupling is then
weak and the theory neither confines nor breaks symmetries. The
fixed point leads to an approximate, long-range conformal symmetry.
As $p$ is reduced, the screening of the long range force decreases,
the coupling increases, and confinement and/or symmetry breaking
set in. We consider three strong-coupling possibilities, each
consistent with global anomaly matching.

\subsubsection*{Confinement with no symmetry Breaking}

It was observed by Bars and Yankielowicz~\cite{Bars:1981se} that
confinement without chiral symmetry breaking is consistent with
global anomaly matching provided that the spectrum of the theory
consists of massless composite fermions transforming under the
global symmetry group as follows:

\begin{table}[h]
\[ \begin{array}{c |  cc  c c c } \hline
{\rm Fields} &\left[ SU(N) \right] & SU(N+4+p) & SU(p) & U_1(1) &U_2(1) \\ \hline 
\bar{F}^{+}S^{+} \bar{F} &1 &\overline{\Yasymm}&1 & - N & 0 \\
  F^{+}S^{+}F &1  &{\Yfund} & \Yfund  & N & - N \\
 F^{+}SF^{+} &1 &1 &\overline{\Ysymm}  & - N & 2 N  \\
 \hline \end{array} 
\]
\end{table}
%

The effective low-energy theory is free. In Ref.~\cite{Appelquist:1999vs}, the
thermodynamic free energy for this phase was computed, giving
\begin{equation}
f_{IR}^{sym}=\frac{7}{4}[\frac{1}{2}(N+4+p)(N+3+p)+p(N+4+p)+\frac{1}{2}
p(p+1)] \ .
\end{equation}
The inequality $f_{IR}^{sym} < f_{UV}$ was then invoked to argue
that this phase is possible only if $p$ is less than a certain
value (less than the asymptotic freedom bound). For large $N$, the
condition is $p<(15/14)^{1/2}N$.

\subsubsection*{Chiral symmetry breaking}

Since this theory is vector-like with respect to the $p$
$F$-$\bar{F}$ pairs, it may be anticipated that these pairs
condense according to
\begin{equation}
\overline{\Yfund }\times
\Yfund
\rightarrow 1\ ,
\end{equation}
leading to a partial breaking of the chiral symmetries. The gauge-singlet
bilinear condensate (fermion mass) is of the form
\begin{equation}
\varepsilon ^{\gamma \delta }F_{\gamma }^{a,i}\bar{F}_{a,N+4+i,\delta}\ ,
\label{chiral breaking}
\end{equation}
where $i=1,...,p$.

This leads to the symmetry breaking pattern $SU(N+4+p)\times SU(p)\times
U_{1}(1)\times U_{2}(1)\rightarrow SU(N+4)\times SU_{V}(p)\times
U_{1}^{\prime }(1)\times U_{2}^{\prime }(1)$, producing $2pN+p^{2}+8p$ gauge
singlet Goldstone bosons. The $U^{\prime }(1)^{\prime}s$ are combinations of
the $U(1)^{\prime}s$ and the broken generator of $SU(N+4+p)$
\begin{equation}
Q_{(N+4+p)}=\left(
\begin{tabular}{ccc|ccc}
$-p$ &  &  &  &  &  \\
& $\ddots $ &  &  &  &  \\
&  & $-p$ &  &  &  \\ \hline
&  &  & $N+4$ &  &  \\
&  &  &  & $\ddots $ &  \\
&  &  &  &  & $N+4$%
\end{tabular}
\right).
\end{equation}

At this stage, the remaining massless theory is the $p=0$ theory described
above, together with the $2pN+p^{2}+8p$ gauge-singlet Goldstone bosons.
Since the Goldstone bosons are associated with the broken symmetry, there
will be no dimension-four interactions between them and the $p=0$ theory.
This theory may therefore be analyzed at low energies by itself, leading to
the possible phases described above. Two possible phases of the $p=0$ theory
were discussed in detail. One corresponds to confinement and massless
composite fermion formation with no chiral symmetry breaking. For the
general theory, this corresponds to

\begin{itemize}
\item  Partial chiral symmetry breaking but no gauge symmetry breaking. The
vector-like $p$ pairs of $F$ and $\bar{F}$ condense, and others
form composite fermions.
\end{itemize}

The massless spectrum consists of the $2pN+p^{2}+8p$ Goldstone
bosons together with the $(N+4)(N+3)/2$ composite fermions of the
$p=0$ sector. All are confined. The final global symmetry is
$SU(N+4)\times SU_{V}(p)\times U_{1}^{\prime }(1)\times
U_{2}^{\prime }(1)$. Global anomalies are matched partially by the
massless composites and partially by the Goldstone bosons. Since
both theories are infrared free, the free energy may be computed in
the $T \rightarrow 0$ limit to give

\begin{equation}
f_{IR}^{brk+sym}=(2pN+p^{2}+8p)+\frac{7}{4}[\frac{1}{2}(N+4)(N+3)]
\ .
\end{equation}
The inequality Eq.~(\ref{eq:ineq}) thus allows this phase for $p$
less than a certain value below the asymptotic freedom bound but
above the value at which the symmetric phase becomes possible. For
large N, the limit is $p/N$ less than $\simeq 2.83$.

\bigskip

The other phase of the $p=0$ theory considered above, corresponds
to the MAC for symmetry breaking and the Higgsing of the gauge
group with a further breaking of the chiral symmetry. For the
general theory ($p > 0$), it leads to

\begin{itemize}
\item Further chiral symmetry breaking and gauge symmetry breaking.
\end{itemize}

The final global symmetry is $SU^{\prime}(N) \times SU(4) \times
SU_{V}(p)\times U_{1}^{\prime }(1)\times U_{2}^{\prime }(1)$. The
massless spectrum consists of the $2pN+p^{2}+8p$ Goldstone bosons
associated with the $p$ $F$-$\bar{F}$ pairs, together with the
$8N+1$ Goldstone bosons and $N(N+1)/2 + 4N$ massless elementary
fermions of the $p=0$ sector. Global anomalies are matched
partially by Goldstone bosons and partially by the remaining
massless fermions. The effective low energy theories are infrared
free, and we have

\begin{eqnarray}
f_{IR}^{brk+Higgs} &=&(2pN+p^{2}+8p)+(8N+1)  \nonumber \\
&&+\frac{7}{4}[\frac{1}{2}N(N-1)+4N] \ .
\end{eqnarray}

\begin{figure}[htb]
\center{
\includegraphics{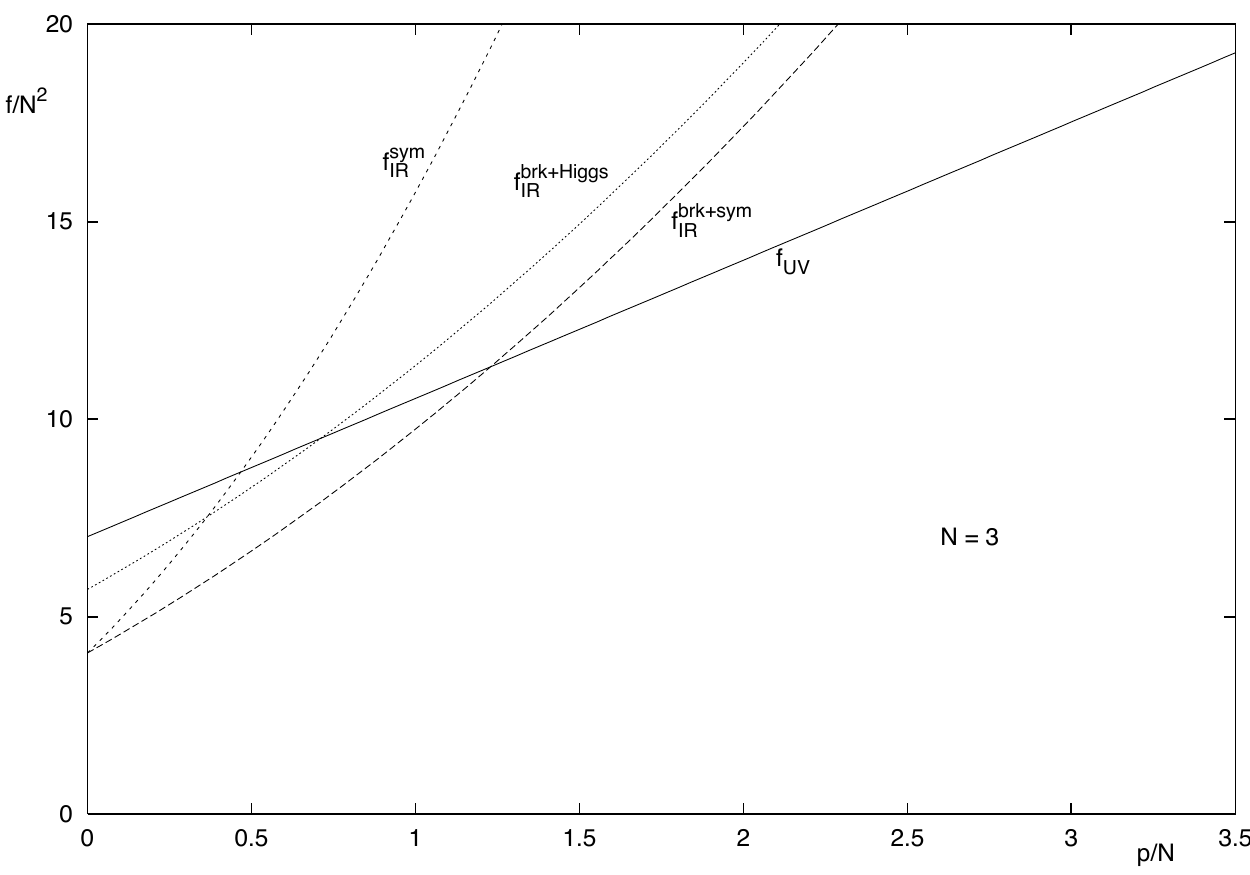}
\caption{BY
Model: Degree of freedom count $f$ (normalized to $N^2$) for
different phases as function of the number $p$ of $F-\bar{F}$ pairs
for the choice $N=3$. $f_{IR}^{sym}$ indicates confinement with
intact chiral global symmetry while $f_{IR}^{brk+sym}$ indicates
confinement with partial chiral symmetry breaking.
$f_{IR}^{brk+Higgs}$ indicates partial chiral symmetry breaking
with gauge symmetry breaking. $f_{UV}$ counts the underlying
degrees of freedom. As $N$ increases, the $f_{IR}^{brk+sym}$ and
$f_{IR}^{brk+Higgs}$ curves approach each other.}
\label{BYfp}}
\end{figure}

\bigskip

Three possible phases of the general Bars-Yankielowicz model have
now been identified. In Fig.~\ref{BYfp}, we summarize the
computation of $f_{IR}/N^{2}$ for each phase and compare with
$f_{UV}$ for the choice $N=3$. Other choices are qualitatively the
same. Each phase satisfies the inequality Eq.~(\ref{eq:ineq}) for
$p/N$ small enough. As $p$ is reduced, the first phase allowed by
the inequality corresponds to confinement with condensation of the
$p$ fermions in the real representation of the gauge group and the
breaking of the associated chiral symmetry, along with unbroken
chiral symmetry and massless composite fermion formation in the
$p=0$ sector. The degree of freedom count is denoted by
$f^{brk+sym}_{IR}$.

 The two other phases are also allowed by the inequality as $p$ is
reduced further. But for any finite value of $N$ and for any value
of $p>0$, the curve for $f_{IR}^{brk+sym}$ is the lowest of the
$f_{IR}$ curves. Thus the lowest infrared degree-of-freedom count
corresponds to a complete breaking of the chiral symmetry
associated with the $p$ $F$-$\bar{F}$ pairs (the vector- like part
of the theory with the fermions in a real representation of the
gauge group), and no breaking of the chiral symmetry associated
with the $p=0$ sector (the part of the theory with the fermions in
complex representations).

It is instructive to examine this model in the infinite $N$ limit.
If the limit is taken with $p/N$ fixed, the curves for $
f_{IR}^{brk+sym}/N^{2}$ and $f_{IR}^{brk+Higgs}/N^{2}$ become
degenerate for all values of $p/N$, and are below the curve for
$f_{IR}^{sym}$. If the limit $N \rightarrow \infty$ is taken with
$p$ fixed, all the curves become degenerate, and the phases are not
distinguished by the number of degrees of freedom. The authors of
Ref.~\cite{Eichten:1985fs} analyzed the model in the $N \rightarrow
\infty$ limit with confinement assumed and
 noted that the $U_{1}(1)$ symmetry cannot break
because no appropriate order parameter can form in this limit. This
is consistent with the above discussion since each of the phases
preserves the $U_{1}(1)$ for any $N$.

As with the $p=0$ theory, there are other possible symmetry
breaking phases corresponding to bilinear condensate formation.
Some of these are attractive channels, although not maximally
attractive, due to gluon exchange. We have considered several
possibilities. Each leads to an effective low energy theory that is
infrared free, and each gives a larger value of $f_{IR}$ than the
phase corresponding to the lowest curve in Fig.~\ref{BYfp}:
complete breaking of the chiral symmetry associated with $p$
additional $F$-$\bar{F}$ pairs and no breaking of the chiral
symmetry associated with the sector of the theory with the fermions
in complex representations.


\subsubsection*{Generalized BY - Conformal Window from the chiral beta function}
\begin{figure}[ht]
\centerline
{\includegraphics[height=6cm,width=18cm]{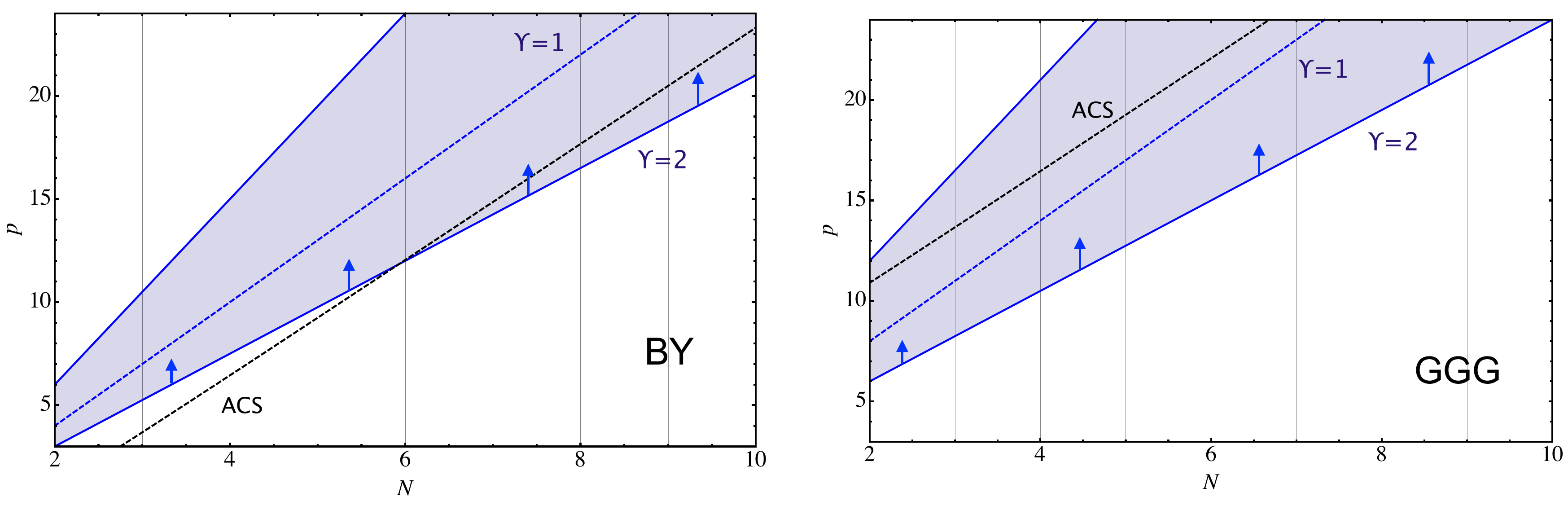}}
\caption
{ {\it Left panel}: Phase diagram of  the BY generalized model. The upper solid (blue) line corresponds to the loss of asymptotic freedom; the dashed (blue) curve corresponds to the chiral beta function prediction for the breaking/restoring of chiral symmetry. The dashed black line  corresponds to the ACS bound stating that the conformal region should start above this line. We have augmented the ACS method with the Appelquist-Duan-Sannino \cite{Appelquist:2000qg} extra requirement that the phase with the lowest number of massless degrees of freedom wins among all the possible phases in the infrared a chiral gauge theory can have. We hence used $f^{\rm brk + sym}_{IR}$ and $f_{UV}$ to determine this curve.  According to the all orders beta function (B.F.) the conformal window cannot extend below the solid (blue) line, as indicated by the arrows. This line corresponds to the anomalous dimension of the mass 
reaching the maximum value of $2$.  {\it Right panel}: The same plot for the GGG model.} 
\label{Chiral}
\end{figure}

From the numerator of the chiral beta function and the knowledge of the one-loop coefficient of the BY perturbative beta function the predicted conformal window is: 
\begin{equation}
 3\frac{(3N-2)}{2+\gamma^{\ast}}\leq p \leq \frac{3}{2}(3N-2) \ ,  
\end{equation}
with $\gamma^{\ast}$ the largest possible value of the anomalous dimension of the mass. The maximum value of the number of $p$ flavors is obtained by setting $\gamma^{\ast} = 2$: 
\begin{equation}
\frac{3}{4} (3N-2) \leq p \leq \frac{3}{2}(3N-2)  \ , \qquad \gamma^{\ast} = 2 \ ,
\end{equation}
while for $\gamma^{\ast} =1$ one gets: 
\begin{equation}
(3N-2) \leq p \leq \frac{3}{2} (3N-2)  \ , \qquad \gamma^{\ast} = 1 \ .
\end{equation}
The chiral beta function predictions for the conformal window are compared with the thermal degree of freedom investigation provided above as shown in the left panel of Fig.~\ref{Chiral}.
In order to derive a prediction from the ACS method we augmented it with the Appelquist-Duan-Sannino \cite{Appelquist:2000qg} extra requirement that the phase with the lowest number of massless degrees of freedom wins among all the possible phases in the infrared a chiral gauge theory can have. We hence used the condition $f^{\rm brk + sym}_{IR}=f_{UV}$ to determine this curve.
The thermal critical number is: 
\begin{eqnarray}
p^{\rm Therm} = \frac{1}{4}\left[-16 + 3N + \sqrt{208  - 196N + 69 N^2} \right] \ .  
\end{eqnarray}

\subsection{ The Generalized Georgi-Glashow (GGG) Model}

This model is similar to the BY model just considered. It is an
$SU(N\geq 5)$ gauge theory, but with fermions in the
anti-symmetric, rather than symmetric, tensor representation. The
complete fermion content is $A=\psi _{L}^{[ab]},$ $a,b=1,\cdots
,N$; an additional $N-4+p$ fermions in the conjugate fundamental
representations: $\bar{F}_{a,i}=\psi
_{a,iL}^{c},$ $i=1,\cdots ,N-4+p$; and $p$ fermions in the fundamental
representations, $F^{a,i}=\psi _{L}^{a,i}$, $i=1,\cdots ,p$.

The global symmetry is
\begin{equation}
G_{f}=SU(N-4+p)\times SU(p)\times U_{1}(1)\times U_{2}(1) \ .
\end{equation}
where the two $U(1)$'s are anomaly free. With respect to this symmetry, the
fermion content is

\begin{table}[h]
\[ \begin{array}{c |  cc  c c c } \hline
{\rm Fields} &\left[ SU(N) \right] & SU(N -4+p) & SU(p) & U_1(1) &U_2(1) \\ \hline 
\hline &&&&&\\
 A &\Yasymm &1 &1 & N- 4 &2p \\
  \bar{F} &\bar{\Yfund}  &\bar{\Yfund} & 1  & -(N - 2) & - p \\
   {F} &\Yfund  &1 & \Yfund  & N - 2 & - (N - p) \\
 \hline \end{array} 
\]
\caption{The Generalized Georgi-Glashow (GGG) Model}
\end{table}

%


\bigskip

For the two-loop $\beta $-function, we have $\beta _{1}=3N+2-(2/3)p$ and $\beta
_{2}=(1/4)\{13N^{2}+30N+1+12/N-2p((13/3)N-1/N)\}$. Thus the theory is
asymptotically free if
\begin{equation}
p<(9/2)N+3 \ .
\end{equation}
We restrict $p$ so that this condition is satisfied. Because of
asymptotic freedom, the thermodynamic free-energy may be computed
in the $T\rightarrow \infty $ limit. We have
\begin{equation}
f_{UV}=2(N^{2}-1)+\frac{7}{4}{\large
[}\frac{N(N-1)}{2}+(N-4)N+2pN{\Large ]}\ .
\end{equation}
As with the BY model, we first discuss the $p=0$ theory and then
consider the general case.

\subsubsection*{The $p=0$ Case}

The global symmetry group is $G_{f}=SU(N-4)\times U_{1}(1)$. The
theory is strongly coupled at low energies, so it is expected
either to confine or to break some of the symmetries, consistent
with global anomaly matching~\cite{'tHooft:1980xb}.

In the case of complete confinement and unbroken symmetry, to
satisfy global anomaly matching the massless spectrum consists of
gauge singlet composite fermions $\bar{F}_{\{i}A\bar{F}_{j\}}$
transforming according to the symmetric second-rank tensor
representation of $SU(N-4)$ with charge $-N$ under the $ U_{1}(1)$
global symmetry~\cite{Bars:1981se}. The composites are noninteracting in the
infrared. Therefore the thermodynamic free energy may be computed
in the limit $T\rightarrow 0$. Enumerating the degrees of freedom
gives
\begin{equation}
f_{IR}^{sym}(p=0)=\frac{7}{4}\frac{(N-4)(N-3)}{2} \ .
\end{equation}
Clearly $f_{IR}^{sym}(p=0)<f_{UV}(p=0)$, satisfying the inequality
Eq.~(\ref{eq:ineq}).

We next consider symmetry breaking due to bilinear condensate
formation by first examining the maximally attractive channel:
\begin{equation}
\Yasymm
\times \overline{
\Yfund
}\rightarrow
\Yfund
\ ,
\end{equation}
leading to the formation of the $A\bar{F}$ condensate
\begin{equation}
\varepsilon ^{\gamma \delta }A_{\gamma }^{ai}
\bar{F}_{a,i, \delta } \ ,
\end{equation}
where $\gamma ,\delta =1,2$ are spin indices, $a=1,\cdots ,N$, is a
gauge index and $i=1,\cdots ,N-4$ is a flavor index. This
condensate breaks the $ U_{1}(1)$ symmetry and breaks the gauge
symmetry $SU(N)$ to $SU(4)$. The broken gauge subgroup $SU(N-4)$
combines with the flavor group, leading to a new global symmetry
$SU^{\prime }(N-4)$, while the broken gauge $SU(N)$ generator
\[
Q_{(N)}=\left(
\begin{tabular}{ccc|ccc}
$4$ &  &  &  &  &  \\
& $\ddots $ &  &  &  &  \\
&  & $4$ &  &  &  \\ \hline
&  &  & $4-N$ &  &  \\
&  &  &  & $\ddots $ &  \\
&  &  &  &  & $4-N$%
\end{tabular}
\right) \ ,
\]
combines with $U_{1}(1)$ to form a residual global symmetry
$U^{\prime }(1)$. The remaining symmetry is thus $[SU(4)]\times
SU^{\prime }(N-4)\times U_{1}^{\prime }(1)$. \ All Goldstone bosons
are eaten by gauge bosons.

We have
\begin{equation}
\bar{F}_{a,i}=\left(
\begin{tabular}{c}
$\bar{F}_{j,i}\rightarrow \bar{F}_{[j,i]}+\bar{F}_{\{j,i\}}$ \\
\hline $\bar{F}_{c,i}$
\end{tabular}
\right)
\end{equation}
and
\begin{equation}
A^{ab}=\left(
\begin{tabular}{c|c}
$A^{ij}$ & $A^{ic}$ \\ \hline & $A^{cd}$
\end{tabular}
\right) ,
\end{equation}
where $a,b=1,\cdots ,N$, \ $i,j=1,\cdots ,N-4$, and $c,d=N-3,\cdots
,N$. The $A\bar{F}$ condensate pairs $\bar{F}_{[j,i]}$ with
$A^{ij}$ and $\bar{F}
_{c,i}$ with $A^{ic}$. This leaves only $A^{cd}$,  which is neutral under $
U^{\prime }(1)$, as the fermion content of the $SU(4)$ gauge
theory.

This $SU(4)$ theory is also strongly coupled in the infrared and we
expect it to confine. The most attractive channel for condensate
formation, for example, is
\begin{equation}
\Yasymm
\times
\Yasymm
\rightarrow
1 \ ,
\end{equation}
leading to the bilinear condensate
\begin{equation}
\varepsilon ^{\gamma \delta }A_{\gamma }^{ab}A_{\delta }^{cd}\varepsilon
_{1\cdots (N-4)abcd} \ ,  \label{AA}
\end{equation}
a singlet under the gauge group. Thus, in the infrared, the only
massless fermions are the $\bar{F}_{\{j,i\}}'s$ in the symmetric
two-index tensor representation of $SU^{\prime }(N-4)$.
Interestingly, the massless fermion content and the low energy
global symmetry are precisely the same for the symmetric and Higgs
phases. Therefore,
\begin{equation}
f_{IR}^{higgs}(p=0)= f_{IR}^{sym}(p=0)
= \frac{7}{4}[\frac{1}{2}(N-4)(N-3)]\ .
\end{equation}
The fermions are composite in the first case and elementary in the
second. This is another example of the complementarity
idea.  While the two phases are not distinguished by the
low energy considerations used here, they {\it are} different
phases. However, other ideas involving energies on the order of the
confinement and/or breaking scales will have to be employed to
distinguish them.

A general study of the phases of chiral gauge theories should
include higher dimensional as well as bilinear condensate
formation. We have done this for one case, the $p=0$ $SU(N=5)$
model, which possesses only a $U(1)$ global symmetry. Among the
various phases that may be considered is one that confines but
breaks the global $U(1)$. This corresponds to the formation of
gauge invariant higher dimensional condensates, for example of the
type $\left(\bar{F}A\bar{F}\right)^2$. There is no bilinear
condensate for this breaking pattern. Global anomaly matching is
satisfied by the appearance of a single massless Goldstone boson
and no other massless degrees of freedom. This phase clearly
minimizes the degree of freedom count (the entropy near
freeze-out), among the phases described by infrared free effective
theories. The unbroken phase, by contrast, must include a massless
composite fermion for anomaly matching, and therefore gives a
larger $f_{IR}$. This suggests that higher dimensional condensate
formation may indeed be preferred in this model. It will be
interesting to study this possibility in more detail and to see
whether higher dimensional condensate formation plays an important
role in the larger class of chiral theories considered here and in
other theories.

\subsubsection*{The General Case}

The full range of $p$ allowed by asymptotic freedom may be
considered just as it was for the BY model. For $p$ near
$(9/2)N+3$, an infrared stable fixed point exists, determined by
the first two terms in the $\beta $ function. The infrared coupling
is then weak and the theory neither confines nor breaks symmetries.
As $p$ decreases, the coupling strengthens, and confinement and/or
symmetry breaking set in. We consider two possibilities consistent
with global anomaly matching.

\subsubsection*{Confinement with no symmetry breaking}

It is known~\cite{Bars:1981se} that confinement without chiral symmetry
breaking is consistent with global anomaly matching provided that
the spectrum of the theory consists of gauge singlet massless
composite fermions transforming under the global symmetry group as
follows:

\begin{table}[h]
\[ \begin{array}{c |  cc  c c c } \hline
{\rm Fields} &\left[ SU(N) \right] & SU(N-4+p) & SU(p) & U_1(1) &U_2(1) \\ \hline 
\bar{F}^{+}A \bar{F} &1 &\overline{\Ysymm}&1 & - N & 0 \\
  F^{+}A^{+}F &1  &{\Yfund} & \Yfund  & N & - N \\
 F^{+}AF^{+} &1 &1 &\overline{\Yasymm}  & - N & 2 N  \\
 \hline \end{array} 
\]
\end{table}

%

The effective low energy is free. Thus the thermodynamic free
energy may be computed in the limit $T \rightarrow 0$ to give
\begin{equation}
f_{IR}^{sym}=\frac{7}{4}[\frac{1}{2}(N-4+p)(N-3+p)+p(N-4+p)+\frac{1}{2}
p(p-1)] \ .
\end{equation}
The inequality Eq.~(\ref{eq:ineq}) allows this phase when $p/N$ is
less than $\simeq 2.83$, for large $N$.

\subsubsection*{Chiral symmetry breaking}

As in the BY model it may be expected that the fermions in a real
representation of the gauge group (the $p$ $F$-$\bar{F}$ pairs)
will condense in the pattern
\begin{equation}
\overline{
\Yfund
}\times
\Yfund
\rightarrow 1.
\end{equation}
The gauge-singlet bilinear condensate (fermion mass) is of the form
\begin{equation}
\varepsilon ^{\gamma \delta }F_{\gamma }^{a,i}\bar{F}_{a,N-4+i,\delta} \ ,
\end{equation}
where $i=1,...,p$, leading to the symmetry breaking pattern
\begin{equation}
\begin{tabular}{c}
$SU(N-4+p)\times SU(p)\times U_{1}(1)\times U_{2}(1)$ \\
$\rightarrow SU(N-4)\times SU_{V}(p)\times U_{1}^{\prime }(1)\times
U_{2}^{\prime }(1),$
\end{tabular}
\end{equation}
and producing $2pN+p^{2}-8p$ gauge singlet Goldstone bosons.

The $U^{\prime }(1)^{\prime }s$ are combinations of the
$U(1)^{\prime }s$ and the broken generator of $SU(N-4+p)$
\begin{equation}
Q_{(N-4+p)}=\left(
\begin{tabular}{ccc|ccc}
$-p$ &  &  &  &  &  \\
& $\ddots $ &  &  &  &  \\
&  & $-p$ &  &  &  \\ \hline
&  &  & $N-4$ &  &  \\
&  &  &  & $\ddots $ &  \\
&  &  &  &  & $N-4$%
\end{tabular}
\right)\ .
\end{equation}

The remaining massless theory is the $p=0$ theory described above,
together with the $2pN+p^{2}-8p$ gauge-singlet Goldstone bosons.
Since the Goldstone bosons are associated with the broken symmetry,
there will be no dimension-four (Yukawa) interactions between them
and the $p=0$ fields. The $p=0$ theory may therefore be analyzed by
itself, leading to the possible phases described above. Two phases
were considered, one symmetric and the other broken by the
maximally attractive bilinear condensate, and they were seen to
lead to identical low energy theories.

Thus, in either case, the degree-of-freedom count for the general
theory, corresponding to the breaking of the chiral symmetry
associated with the $p$ $F$-$\bar{F}$ pairs, gives
\begin{equation}
f_{IR}^{brk}=(2pN+p^{2}-8p)+\frac{7}{4}[\frac{1}{2}(N-4)(N-3)] \ .
\end{equation}

\begin{figure}[htb]
\center{
 \includegraphics{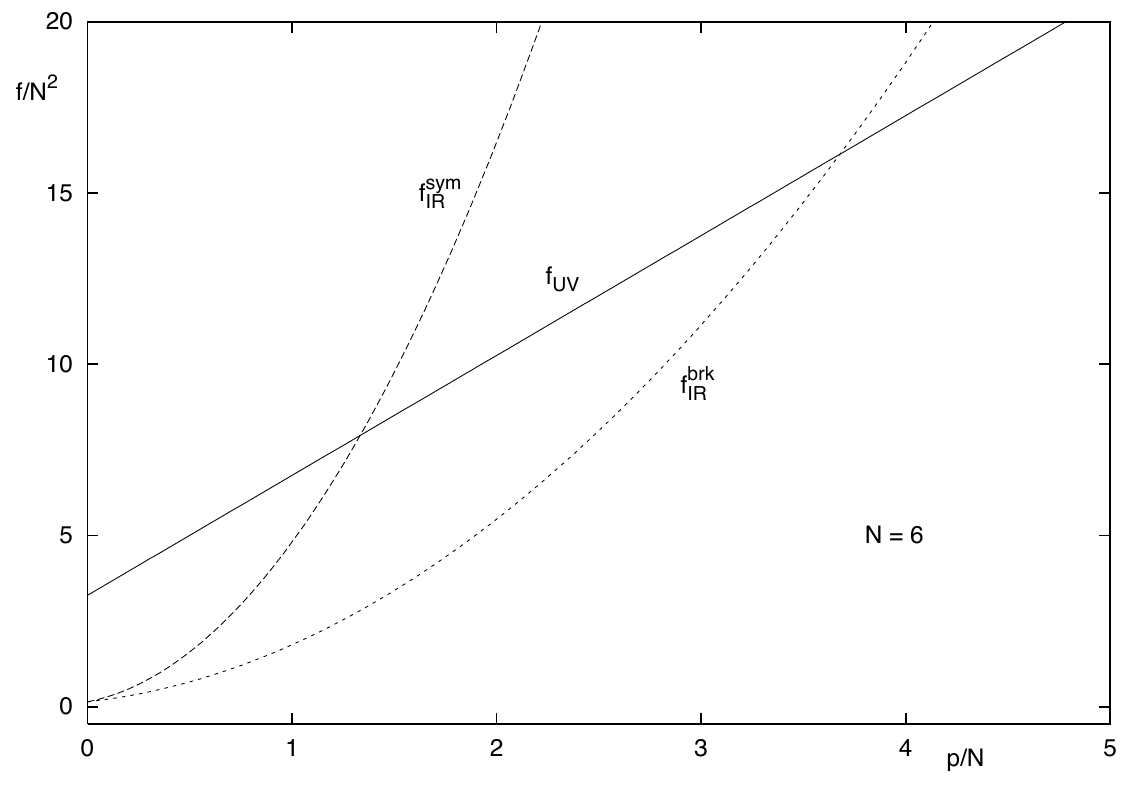}
\caption{ GGG Model:
Degree of freedom count $f$ (normalized to $N^2$) for different
phases as function of the number $p$ of $F-\bar{F}$ pairs for the
choice $N=6$. Other choices are qualitatively the same.
$f_{IR}^{sym}$ indicates confinement with intact chiral global
group while $f_{IR}^{brk}$ indicates either confinement or gauge
symmetry breaking, with partial chiral symmetry breaking. $f_{UV}$
counts the underlying degrees of freedom.}
\label{GGGfp}}
\end{figure}

\bigskip

  To summarize, two possible phases of the general GGG model have been
considered. In Fig.~\ref{GGGfp}, we plot the two computations of
$f_{IR}/N^{2}$ for the choice $N=6$ and compare them with $f_{UV}$.
Each phase satisfies the inequality Eq.~(\ref{eq:ineq}) for $p$
below some value. As $p$ is reduced, the first phase allowed by the
inequality corresponds to partial chiral symmetry breaking. For any
$p$, $f_{IR}^{brk}$ is the lower of the $f_{IR}$ curves. Thus the
lower infrared degree-of-freedom count corresponds to a complete
breaking of the chiral symmetry associated with $p$ additional
$F$-$\bar{F}$ pairs (the vector like part of the theory) and no
breaking of the chiral symmetry associated with the fermions in a
complex representation of the gauge group. Whether the latter
behavior is due to confinement or the Higgsing of the gauge group
has not been determined. These conclusions remain valid in the
infinite $N$ limit with $p/N$ fixed. If the limit $N
\rightarrow \infty$, is taken with $p$ fixed, the two curves become degenerate.

As we have already noted, for the $p=0$ $SU(5)$ theory
there is a still lower degree of freedom count  when higher
dimensional condensates are considered. This will therefore also be
true of the general-$p$ case for $SU(5)$. This even lower count
corresponds to complete confinement along with breaking of the
chiral symmetry associated with the $p$ $F$-$\bar{F}$ pairs {\it
and} breaking of the remaining global $U(1)$ symmetry. It will be interesting
to see whether a preference for this phase can be confirmed by
a dynamical study of this model and whether
similar higher dimensional condensate formation is favored in a
more general class of models.

\subsubsection*{GGG- Conformal Window from the chiral beta function}
From the numerator of the chiral beta function and the knowledge of the one-loop coefficient of the GGG perturbative beta function the predicted window is: 
\begin{equation}
 3\frac{(3N+2)}{2+\gamma^{\ast}}\leq p \leq \frac{3}{2}(3N+2) \ .  
\end{equation}
It maximum lower bound is obtained setting $\gamma^{\ast} = 2$ and one obtains: 
\begin{equation}
\frac{3}{4} (3N+ 2) \leq p \leq \frac{3}{2}(3N + 2)  \ , \qquad \gamma^{\ast} = 2 \ ,
\end{equation}
while for $\gamma^{\ast} =1$ one gets: 
\begin{equation}
(3N + 2) \leq p \leq \frac{3}{2} (3N+ 2)  \ , \qquad \gamma^{\ast} = 1 \ .
\end{equation}

The chiral beta function predictions for the conformal window are compared with the thermal degree of freedom investigation provided above and the result is shown in the right panel of Fig.~\ref{Chiral}.

As for the BY case, in order to derive a prediction from the ACS method we augmented it with the Appelquist-Duan-Sannino \cite{Appelquist:2000qg} extra requirement that the phase with the lowest number of massless degrees of freedom wins among all the possible phases in the infrared a chiral gauge theory can have. We hence used the condition $f^{\rm brk}_{IR}=f_{UV}$ to determine this curve. We expect that chiral symmetry should break below this black-dashed line according to this method.
The thermal critical number is: 
\begin{eqnarray}
p^{\rm Therm} = \frac{1}{4}\left[16 + 3N + \sqrt{56 + 68N + 69 N^2} \right] \ .  
\end{eqnarray}
 Interestingly in this case we have not yet reached $\gamma=1$ while the opposite is true for the generalized BY model. At large N we expect the two models to lead to identical results. 

\subsection{Two Chiral SUSY Models}

 Although this review is devoted principally to non-SUSY chiral models,
 we briefly describe two chiral SUSY models \cite{Appelquist:2000qg}: the supersymmetric
generalization of the one generation $SU(5)$ Georgi-Glashow model
\cite{Affleck:1983vc} and the related ($3-2$) model (see \cite{Poppitz:1998vd} for a
review of this model and relevant references).

The $SU(5)$ model contains a single antisymmetric tensor chiral
superfield $A$ and an antifundamental chiral superfield $\bar{F}$.
The vector superfield $W_{\alpha}$ includes the standard vector
boson and the associated gluino in the adjoint representation of
$SU(5)$. The global symmetry is the anomaly-free $U_R(1)\times
U_A(1)$, and the charge assignments are:

\begin{equation}
\begin{tabular}{cccccc}
& $[SU(5)]$ & $U_{R}(1)$ & $U_{A}(1)$ \\ & & & & &
\\ $A$ &
\Yasymm & $-1$ & $-1$ \\ &  &  &  &  &  \\ $\bar{F}$ & $\overline{%
\Yfund
}$ &
 $+ 9$ & $+3$ \\ &  &  &  &  &  \\ $W_{\alpha}$ &
$Adj$ & $-1$ & $ 0$%
\end{tabular}
\end{equation}

\bigskip
A special feature of this model is that the classical vacuum is
unique. The absence of flat directions is due to the fact that
there exists no holomorphic gauge invariant polynomial constructed
out of the supersymmetric fields. This feature guarantees that when
comparing phases through their degree-of-freedom count, we know
that we are considering a single underlying theory. By contrast, in
SUSY gauge theories with flat directions, non-zero condensates
associated with the breaking of global symmetries correspond to
different points in moduli space and therefore to different
theories.

This model was studied long ago \cite{Affleck:1983vc} and various possible phases
were seen to be consistent with global anomaly matching. One
preserves supersymmetry along with the global symmetries. This
requires composite massless fermions to saturate the global
anomalies. It was shown that there are several, rather complicated,
solutions, with at least five Weyl fermions (which for
supersymmetry to hold must be cast in five chiral superfields). The
charge assignments for one of them is \cite{Affleck:1983vc}:
\begin{equation}
(-5,-26),~(5,20),~(5,24),~(0,-1),~(0,9) \ ,
\end{equation}
where the first entry is the $U_A(1)$ charge and the second is the
$U_R(1)$ charge of each chiral superfield.

Other possibilities are that SUSY breaks with the global symmetries
unbroken or that one or both of the global symmetries together with
supersymmetry break spontaneously. It is expected \cite{Affleck:1983vc} that
in a supersymmetric theory without classical flat directions, the
spontaneous breaking of global symmetries also signals spontaneous
supersymmetry breaking. In these cases, the only massless fields
will be the Goldstone boson(s) associated with the broken global
symmetries and/or some massless fermions transforming under the
unbroken chiral symmetries, together with the Goldstone Weyl
fermion associated with the spontaneous supersymmetry breaking.

In Reference~\cite{Affleck:1983vc} it was suggested on esthetic grounds that
the supersymmetric solution seems less plausible. Additional
arguments that supersymmetry is broken are based on investigating
correlators in an instanton background \cite{Meurice:1984ai}. However a firm
solution to this question is not yet available.

Since all the above phases are non interacting in the infrared we
may reliably compute $f_{IR}$ and note that the phase that
minimizes the degree-of-freedom count is the one that breaks
supersymmetry and both of the global symmetries. This phase
consists of two $U(1)$ Goldstone bosons and a single Weyl Goldstino
associated with the breaking of SUSY. Thus $f_{IR} = 15/4$. SUSY
preserving phases and those that leave one or both of the $U(1)'s$
unbroken lead to more degrees of freedom.  It will be interesting to
see whether further dynamical studies confirm that the
maximally broken phase is indeed preferred

This phase is similar to the minimal-$f_{IR}$ phase in the
nonsupersymmetric $SU(5)$ model in that both correspond to
higher dimensional condensate formation. In the SUSY case, one can
construct two independent order parameters. The one for $U_R(1)$ is
the gluino condensate (scalar component of the chiral superfield
$W^{\alpha} W_{\alpha}$) while the one for $U_A(1)$ can be taken to
be the scalar component of the chiral superfield
$\bar{F}_a\bar{F}_b A^{ac} (W^{\alpha} W_{\alpha})^b_c$.

Finally we comment on a well known and related chiral model for
dynamical supersymmetry breaking: the ($3-2$) model. Unlike the
models considered so far, this model involves multiple couplings,
i.e. two gauge couplings and a Yukawa one. Without the Yukawa
interaction the theory posses a run-away vacuum. The model has an
$SU(3)\times SU(2)$ gauge symmetry and a $U_{Y}(1)\times U_{R}(1)$
anomaly free global symmetry. As above, the low energy phase that
minimizes the number of degrees of freedom is the one that breaks
supersymmetry along with both of the global symmetries. The
massless spectrum is the same as in the parent chiral $SU(5)$ case.
In the ($3-2$) model, however, the low energy spectrum has been
computed \cite{Bagger:1994hh} in a self-consistent weak-Yukawa coupling
approximation, where it was noted that the $U_{R}(1)$ breaks along
with supersymmetry, leaving intact the $U_{Y}(1)$. The spectrum
consists of two massless fermions (a Goldstino and the fermion
associated with the unbroken $U_{Y}(1)$) and the $U_{R}(1)$
Goldstone boson. If this is indeed the ground state, then the
number of infrared degrees of freedom is not minimized in this weak
coupling case.

\subsubsection*{GGG \& BY in brief}
 We have considered the low energy structure of two chiral gauge
 theories, the Bars-Yankielowicz (BY) model and the generalized
 Georgi-Glashow (GGG) model. Each contains a core of fermions
 in complex representation of the gauge group, along with a set
 of $p$ additional fundamental-anti-fundamental pairs. In each case,
 for $p$ near but not above the value for which asymptotic freedom is lost,
the model will have a weak infrared fixed point and exist in the
non-abelian Coulomb phase.

  As $p$ drops, the infrared coupling strengthens and one or more phase
transitions to strongly coupled phases are expected. Several
possible phases have been identified that are consistent with
global anomaly matching, and that satisfy the inequality
Eq.~(\ref{eq:ineq}) for low enough $p$. One is confinement with the
gauge symmetry and additional global symmetries unbroken. Another
is confinement with the global symmetry broken to that of the $p=0$
theory. Still another is a Higgs phase, with both gauge and chiral
symmetries broken. Both symmetry breaking phases correspond to
bilinear condensate formation. The infrared degree of freedom count
$f_{IR}$ for each of these phases is shown in Figs.~\ref{BYfp} and \ref{GGGfp},
along with the corresponding ultraviolet count $f_{UV}$.

In \cite{Appelquist:2000qg} we suggested that at each value of $p$, these theories will
choose the phase that minimizes the degree of
freedom count as defined by $f_{IR}$, or equivalently
the phase that minimizes the entropy near freeze-out ($S(T)\approx
(2\pi^{2}/45) T^{3} f_{IR}$). As may be seen from
 Figs.~\ref{BYfp} and \ref{GGGfp}, this idea leads to the following picture. As $p$
drops below some critical value, the $p$
fundamental-anti-fundamental pairs condense at some scale
$\Lambda$, breaking the full global symmetry to the symmetry of the
$p=0$ theory and producing the associated Goldstone bosons. For the
remaining theory with fermions in only complex representations, the
phase with the global symmetry unbroken and the global anomalies
matched by massless fermions is preferred to phases with further
global symmetry breaking via bilinear condensates. We have not yet
shown that this is true relative to all bilinear condensate
formation. Also, this does not exclude the possibility that some
strongly coupled infrared phase (such as a strong non abelian
coulomb phase) leads to the smallest value for $f_{IR}$ and is
still consistent with global anomaly matching.

We extended our discussion to include general condensate formation
for one simple example, the $SU(5)$ Georgi-Glashow model with
fermions in only complex representations and a single $U(1)$ global
symmetry. This symmetry can be broken via only a higher dimensional
condensate. For this model, interestingly, we noted that the
breaking of the $U(1)$ with confinement and unbroken gauge symmetry
leads to the minimum value of $f_{IR}$ among phases that are
infrared free. This highlights the important question of the
pattern of symmetry breaking in general chiral theories (or any
theories for that matter) when arbitrary condensate formation is
considered. Higher dimensional condensates could play an important
role in the dynamical  breaking of symmetries in extensions of the
SM \cite{Roux:1999dq}. The enumeration of degrees of freedom in
the effective infrared theory is a potentially useful guide to
discriminate among the possibilities.

Finally, we commented on two supersymmetric chiral models: the
supersymmetric version of the $SU(5)$ Georgi-Glashow model and the
closely related ($3-2$) model. Both have a $U_{R}(1) \times
U_{Y}(1)$ global symmetry. In each case, the phase that minimizes
the number of massless degrees of freedom corresponds to the
breaking of SUSY and both of its global symmetries. In the case of
the ($3-2$) model, however, an analysis in the case of a weak
Yukawa coupling (see \cite{Poppitz:1998vd} for a discussion and relevant
references) leads to the conclusion that the $U_{Y}(1)$ is not
broken. If this truly represents the ground state in the case of
weak coupling, then the degree of freedom count is not the minimum
among possible phases that respect global anomaly matching.

For the nonsupersymmetric chiral gauge theories
discussed here, we identified a variety of possible
zero-temperature phases and conjectured \cite{Appelquist:2000qg} that the theories will
choose from among them the one that minimizes the infrared degree
of freedom count. 

However the degree of freedom count is not sufficient to provide a complete insight on the conformal window of chiral gauge theories. We have, hence, introduced a novel all orders chiral beta function which naturally extends the one for vector like theories we introduced earlier. 

We note that the determination of the conformal window of the generalized BY and GGG models should be taken into account when constructing ETC models featuring chiral gauge theories given that the these theories will not be able to break any symmetry if the number of vector-like fermions is within the window determined above.

\subsection{Comparison Chart}

We investigated the conformal windows for $SU$, $SO$ and $Sp$ nonsupersymmetric gauge theories with fermions in any representation of the underlying gauge group using four independent analytic methods. One observes a universal value, i.e. independent of the representation, of the ratio of the area of the maximum extension of the conformal window, predicted using the all orders beta function, to the asymptotically free one, as defined in \cite{Ryttov:2007sr}. It is easy to check from the results presented that this ratio is not only independent on the representation but also on the particular gauge group chosen. 

The four  methods we used to unveil the conformal windows are the all orders beta function (BF), the SD truncated equation, the thermal degrees of freedom method and least but not the last Gauge - Duality. Have vastly different starting points and there was no, a priori, reason to agree with each other.

In the Table below we compare directly the various analytical methods. 
\begin{table}[ht]
\caption{Direct comparison among the various analytic methods}
\centering
\begin{tabular}
{|c|c|c|c|c|c|c|c|} \hline
{Method} &  ~~~ Fund.  ~~~& ~Higher~&~Multiple~&~~Susy~~& ~~~$\gamma$~~~&AMC&$\chi$  \\ \hline \hline
BF & +  & +  &+ & + & + &+&+ \\
SD &  + & +  & - & - & -&-&-\\
ACS & + & - & - & + &  -&-&+ \\
 \hline 
 \end{tabular} 
 \label{comparison}
\end{table}
 The three plus signs in the second column indicate that the three analytic methods do constrain the conformal window of $SU$, $Sp$ and $SO$ gauge theories with fermions in the fundamental representation. Only BF and SD provide useful constraints in the case of the higher dimensional representations as summarized in the third column. {}When multiple representations participate in the gauge dynamics the BF constraints can be used directly \cite{Ryttov:2007cx,Ryttov:2008xe} to determine the extension of the conformal (hyper)volumes while extra dynamical information and approximations are required in the $SD$ approach. Since gauge theories with fermions in several representations of the underlying gauge group must contain higher dimensional representations the ACS is expected to be less efficient in this case \footnote{We do not consider super QCD a theory with higher dimensional representations.}. These results are summarized in the fourth column. The all orders beta function reproduces the supersymmetric exact results when going over the super Yang-Mills case, the ACS conjecture was  proved successful when tested against the supersymmetric conformal window results \cite{Appelquist:1999hr}. However the SD approximation does not reproduce any supersymmetric result \cite{Appelquist:1997gq}.  The results are summarized in the fifth column. Finally, it is of  theoretical and phenomenological interest -- for example to construct sensible UV completions of models of dynamical electroweak symmetry breaking and unparticles --  to compute the anomalous dimension of the mass of the fermions at the (near) conformal fixed point.  Only the all orders beta function provides a simple closed form expression as it is summarized in the sixth column. 

We have also suggested that it is interesting to study the $SU(2)$ gauge theory with $N_f=5$ Dirac flavors via first principles Lattice simulations since it will discriminate between the two distinct predictions, the one from the ACS conjecture and the one from the all orders beta function.

We have presented a comprehensive analysis of the phase diagram of non-supersymmetric vector-like and strongly coupled SU($N$) gauge theories 
with matter in various representations of the gauge group. 

Other approaches, such as the instanton-liquid model \cite{Schafer:1995pz} or the one developed in \cite{Grunberg:2000ap,Gardi:1998ch,Grunberg:1996hu} have also been used to investigate the QCD chiral phase transition as function of the number of flavors. According to the instanton-liquid model one expects for QCD the transition to occur for a very small number of flavors. This result is at odds with the bound for the conformal window found with our novel beta function \cite{Ryttov:2007cx} as well lattice data \cite{Appelquist:2007hu}. 

A more recent approach makes use of certain topological excitations \cite{Poppitz:2009uq,Poppitz:2009tw}, whose link to the conformal window is yet to be proved, leads to predictions very close to the ones of the all orders beta function. However within this approach there is no prediction for the anomalous dimension at the fixed point. 

The all orders beta function method represents, however,  a much more direct way to estimate the conformal window and to capture some of its salient properties. Besides, the form of the beta function is highly consistent with the {\it exact} results stemming from the solutions of the 't Hooft Anomaly conditions leading to interesting possible gauge duals. In the table above the 't Hooft Anomaly matching conditions have been abbreviated via (AMC). We have seen that is even possible to predict the conformal window of phenomenologically relevant chiral gauge theories. In the table we have used the symbol $\chi$ to refer to these kind of theories. 

Several questions remain open such as what happens on the right hand side of the infrared fixed point as we increase further the coupling. Does a generic strongly coupled theory develop a new UV fixed point as we increase the coupling beyond the first IR value \cite{Kaplan:2009kr}? If this were the case our beta function would still be a valid description of the running of the coupling of the constant in the region between the trivial UV fixed point and the neighborhood of the first IR fixed point. One might also consider extending our beta function to take into account of this possibility as done in \cite{Antipin:2009wr}. It is also possible that no non-trivial UV fixed point forms at higher values of the coupling constant for any value of the number of flavors within the conformal window. Gauge-duals seem to be in agreement with the simplest form of the beta function. The extension of the all orders beta function to take into account fermion masses has appeared in \cite{Dietrich:2009ns}.

We briefly summarize the current status of the lattice results. The $SU(2)$ gauge theory with four adjoint Weyl fermions, known as Minimal Walking Technicolor, seems to be (near) conformal \cite{Catterall:2007yx,Catterall:2008qk,DelDebbio:2008zf,Hietanen:2008vc,Hietanen:2009az,Pica:2009hc,Catterall:2009sb} as predicted using the all orders beta function. There are also interesting early results for the anomalous dimension of the mass of these theories \cite{Pica:2009hc,Bursa:2009we,Lucini:2009an}. However the uncertainties on this quantity are still quite large with a quoted result of $ 0.05 \leq \gamma \leq 0.56$. It is worth mentioning that the predicted value of $\gamma$ from the all orders beta function is $\gamma = 3/4=0.75$ was already known to be less then one. This of course does not diminish the value of Minimal Walking Theories nor renders them less relevant for particle physics phenomenology. In fact the relevant point is that near conformal behavior reduces the corrections to the precision data and it was already pointed out in \cite{Evans:2005pu} that another mechanism for the generation of the fermion masses is needed along the lines of the one presented in \cite{Antola:2009wq}.   

Next to Minimal Walking Technicolor, i.e. $SU(3)$ gauge theory with 2 Dirac flavors in the two-index symmetric representation has also been investigated in \cite{Shamir:2008pb,DeGrand:2008kx,Fodor:2009ar} and there are very preliminary indications that this theory is (near) conformal. Searches for the conformal
window in  $SU(3)$ gauge theories with fundamental representation
quarks have also received recent attention 
\cite{Appelquist:2009ty,Appelquist:2009ka,Fodor:2009wk,Fodor:2008hn,
Deuzeman:2009mh}.

Our analysis  substantially increases the number of asymptotically free gauge theories which can be used to construct SM extensions making use of (near) conformal dynamics. Current Lattice simulations can test our predictions and lend further support or even disprove  the emergence of a universal picture possibly relating the phase diagrams of gauge theories of fundamental interactions.


\newpage
\section{Minimal Conformal Models}

The simplest technicolor model has $N_{T f}$ Dirac fermions in
the fundamental representation of $SU(N)$. These models, when
extended to accommodate the fermion masses through the extended technicolor interactions,
suffer from large flavor changing neutral currents. This problem is
alleviated if the number of flavors is sufficiently large
such that the theory is almost conformal. This is estimated to happen
for $N_{T f} \sim 4 N$ \cite{Yamawaki:1985zg} as also summarized in the section dedicated to the Phase Diagram of strongly interacting theories. This, in turn, implies a large
contribution to the oblique parameter $S$ (within naive estimates) \cite{Hong:2004td}. Although near the conformal window \cite{
Appelquist:1998xf,Sundrum:1991rf} the $S$ parameter is reduced due to
non-perturbative corrections, it is still too large if the
model has a large particle content. In addition, such models may
have a large number of pseudo Nambu-Goldstone bosons. By
choosing a higher dimensional technicolor representation for the new technifermions one
can overcome these problems \cite{Sannino:2004qp,Hong:2004td}. 

To have a very low $S$ parameter one would ideally have a technicolor theory which with only one doublet breaks dynamically the electroweak theory but at the same time being walking to reduce the $S$ parameter. The walking nature then also enhances the scale responsible for the fermion mass generation. 
 
According to the phase diagram exhibited earlier the promising candidate theories with the properties required are either theories with fermions in the adjoint representation or two index symmetric one. In Table \ref{symmetric} we present the
generic S-type theory.
\begin{table}[h]
\begin{center}
\begin{tabular}{c||ccccc }
 & $SU(N)$ & $SU_L(N_{T f})$& $SU_R(N_{T f})$&$U_V(1)$ & $U_A(1)$  \\
  \hline \hline \\
${Q_L}$& $\symm$ & $\fund$ & $1$ & $1$ & $1$  \\
 &&&\\
 $\bar{Q}_R$ & $\bsymm $ & $1$& $\bfund$& $-1$ & $1$ \\
 &&&\\
$G_{\mu}$ & {\rm Adj} & $0$&$0$ &$0$ & $0$    \\
\end{tabular}
\end{center}
\caption{Schematic representation of a generic nonsupersymmetric vector like $SU(N)$ gauge theory with matter content 
in the two-index representation.  Here $Q_{L(R)}$ are Weyl fermions.}
\label{symmetric}
\end{table}

The relevant feature, found first in \cite{Sannino:2004qp} using the ladder approximation, is that the
S-type theories can be near conformal already at $N_{T f}=2$ when $N=2$ or $3$. This
should be contrasted with theories in which the fermions are in the fundamental
representation for which the minimum number of flavors required to
reach the conformal window is eight for $N=2$. This last statement is supported by the all order beta function results \cite{Ryttov:2007cx} as well as lattice simulations \cite{Catterall:2007yx,Shamir:2008pb,Appelquist:2007hu}. The critical value of flavors increases with the number of colors
for the gauge theory with S-type matter: the limiting value is
$4.15$ at large $N$. 

The situation is different for the theory with A-type matter. Here the critical number of flavors increases when decreasing the
number of colors. The maximum value of about $N_{Tf}=12$  is obtained - in the ladder approximation - for
$N=3$, i.e. standard QCD.  In reference \cite{Hong:2004td} it has been argued that 
the nearly conformal A-type theories have, already 
at the perturbative level, a very large $S$ parameter with respect to the experimental data. These theories can be re-considered if one gauges under the electroweak symmetry only a part of the flavor symmetries as we shall see in the section dedicated to {\it partially gauged} technicolor.

\subsection{Minimal Walking Technicolor (MWT)}

The dynamical sector we consider, which underlies the Higgs mechanism, is an SU(2) technicolor gauge theory with two adjoint
technifermions \cite{Sannino:2004qp}. The theory is asymptotically free if the number of flavors $N_f$ is less than $2.75$ according to the ladder approximation. Lattice results support the conformal or near conformal behavior of this theory. In any event the symmetries and properties of this model make it ideal for a comprehensive study for LHC physics. The all order beta function prediction is that this gauge theory is, in fact, conformal. In this case we can couple another non-conformal sector to this gauge theory and push it away from the fixed point. 

The two adjoint fermions are conveniently written as \beq Q_L^a=\left(\begin{array}{c} U^{a} \\D^{a} \end{array}\right)_L , \qquad U_R^a \
, \quad D_R^a \ ,  \qquad a=1,2,3 \ ,\eeq with $a$ being the adjoint color index of SU(2). The left handed fields are arranged in three
doublets of the SU(2)$_L$ weak interactions in the standard fashion. The condensate is $\langle \bar{U}U + \bar{D}D \rangle$ which
correctly breaks the electroweak symmetry as already argued for ordinary QCD in (\ref{qcd-condensate}).

The model as described so far suffers from the Witten topological anomaly \cite{Witten:1982fp}. However, this can easily be solved by
adding a new weakly charged fermionic doublet which is a technicolor singlet \cite{Dietrich:2005jn}. Schematically: 
\beq L_L =
\left(
\begin{array}{c} N \\ E \end{array} \right)_L , \qquad N_R \ ,~E_R \
. \eeq In general, the gauge anomalies cancel using the following
generic hypercharge assignment
\begin{align}
Y(Q_L)=&\frac{y}{2} \ ,&\qquad Y(U_R,D_R)&=\left(\frac{y+1}{2},\frac{y-1}{2}\right) \ , \label{assign1} \\
Y(L_L)=& -3\frac{y}{2} \ ,&\qquad
Y(N_R,E_R)&=\left(\frac{-3y+1}{2},\frac{-3y-1}{2}\right) \ \label{assign2} ,
\end{align}
where the parameter $y$ can take any real value \cite{Dietrich:2005jn}. In our notation
the electric charge is $Q=T_3 + Y$, where $T_3$ is the weak
isospin generator. One recovers the SM hypercharge
assignment for $y=1/3$.
\begin{figure}
\centerline{\includegraphics[width=8cm]{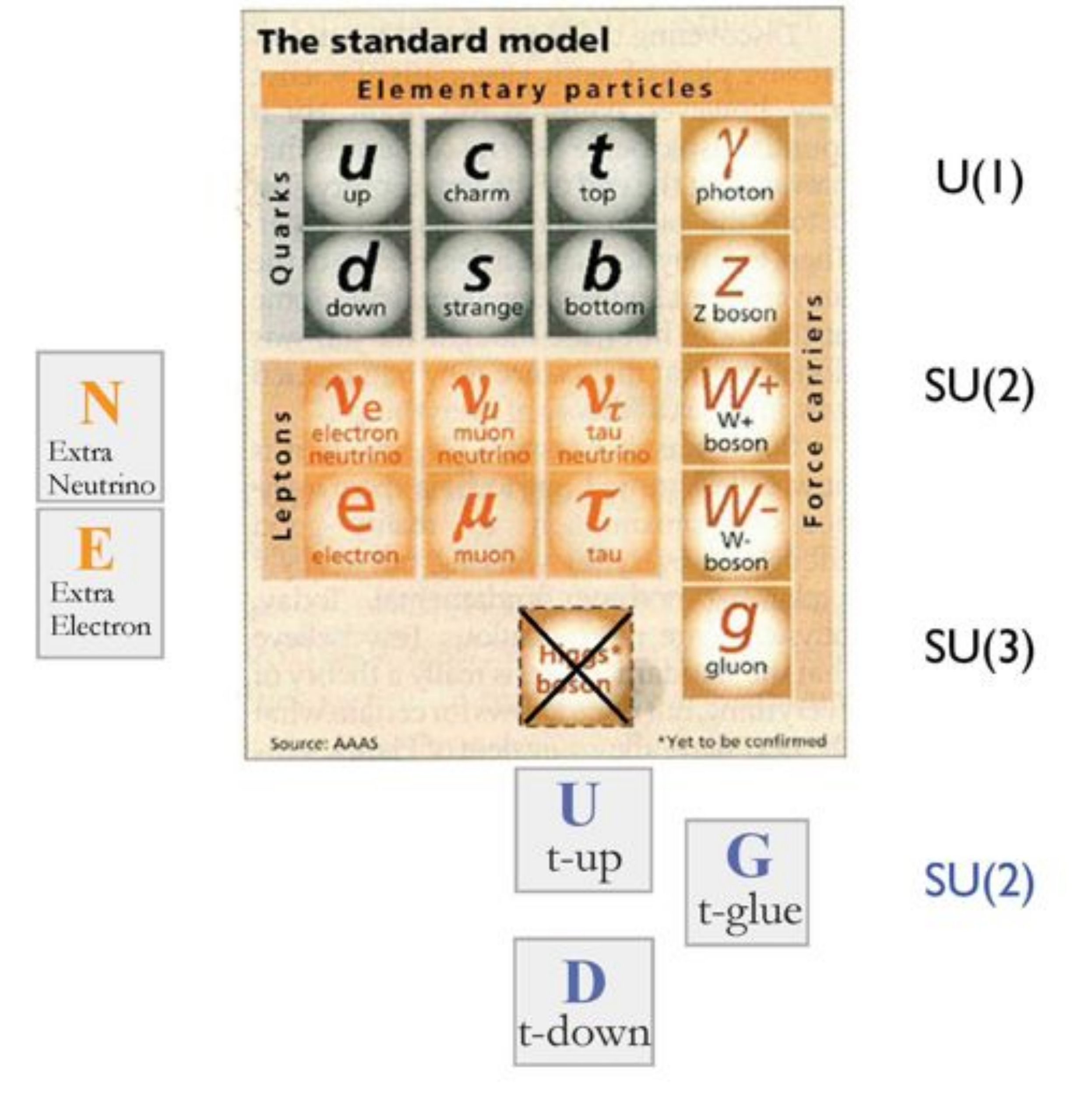}}
\label{MWTposter}
\caption{Cartoon of the Minimal Walking Technicolor Model extension of the SM.}
\end{figure}
To discuss the symmetry properties of the
theory it is
convenient to use the Weyl basis for the fermions and arrange them in the following vector transforming according to the
fundamental representation of SU(4)
\beq Q= \begin{pmatrix}
U_L \\
D_L \\
-i\sigma^2 U_R^* \\
-i\sigma^2 D_R^*
\end{pmatrix},
\label{SU(4)multiplet} \eeq where $U_L$ and $D_L$ are the left
handed techniup and technidown, respectively and $U_R$ and $D_R$ are
the corresponding right handed particles. Assuming the standard
breaking to the maximal diagonal subgroup, the SU(4) symmetry
spontaneously breaks to $SO(4)$. Such a breaking is driven by the
following condensate \beq \langle Q_i^\alpha Q_j^\beta
\epsilon_{\alpha \beta} E^{ij} \rangle =-2\langle \overline{U}_R U_L
+ \overline{D}_R D_L\rangle \ , \label{conde}
 \eeq
where the indices $i,j=1,\ldots,4$ denote the components
of the tetraplet of $Q$, and the Greek indices indicate the ordinary
spin. The matrix $E$ is a $4\times 4$ matrix defined in terms
of the 2-dimensional unit matrix as
 \beq E=\left(
\begin{array}{cc}
0 & \mathbbm{1} \\
\mathbbm{1} & 0
\end{array}
\right) \ . \eeq

Here 
$\epsilon_{\alpha \beta}=-i\sigma_{\alpha\beta}^2$ and $\langle
 U_L^{\alpha} {{U_R}^{\ast}}^{\beta} \epsilon_{\alpha\beta} \rangle=
 -\langle  \overline{U}_R U_L
 \rangle$. A similar expression holds for the $D$ techniquark.
The above condensate is invariant under an $SO(4)$ symmetry. This leaves us with nine broken  generators with associated Goldstone bosons.

Replacing the Higgs sector of the SM with the MWT the Lagrangian now
reads:
\begin{eqnarray}
\mathcal{L}_H &\rightarrow &  -\frac{1}{4}{\cal F}_{\mu\nu}^a {\cal F}^{a\mu\nu} + i\bar{Q}_L
\gamma^{\mu}D_{\mu}Q_L + i\bar{U}_R \gamma^{\mu}D_{\mu}U_R +
i\bar{D}_R \gamma^{\mu}D_{\mu}D_R \nonumber \\
&& +i \bar{L}_L \gamma^{\mu} D_{\mu} {L}_L +
i\bar{N}_R \gamma^{\mu}D_{\mu}N_R + i\bar{E}_R
\gamma^{\mu}D_{\mu}E_R
\end{eqnarray}
with the technicolor field strength ${\cal F}_{\mu\nu}^a =
\partial_{\mu}{\cal A}_{\nu}^a - \partial_{\nu}{\cal A}_{\mu}^a + g_{TC} \epsilon^{abc} {\cal A}_{\mu}^b
{\cal A}_{\nu}^c,\ a,b,c=1,\ldots,3$.
For the left handed techniquarks the covariant derivative is:

\begin{eqnarray}
D_{\mu} Q^a_L &=& \left(\delta^{ac}\partial_{\mu} + g_{TC}{\cal
A}_{\mu}^b \epsilon^{abc} - i\frac{g}{2} \vec{W}_{\mu}\cdot
\vec{\tau}\delta^{ac} -i g'\frac{y}{2} B_{\mu} \delta^{ac}\right)
Q_L^c \ .
\end{eqnarray}
${\cal A}_{\mu}$ are the techni gauge bosons, $W_{\mu}$ are the
gauge bosons associated to SU(2)$_L$ and $B_{\mu}$ is the gauge
boson associated to the hypercharge. $\tau^a$ are the Pauli matrices
and $\epsilon^{abc}$ is the fully antisymmetric symbol. In the case
of right handed techniquarks the third term containing the weak
interactions disappears and the hypercharge $y/2$ has to be replaced
according to whether it is an up or down techniquark. For the
left-handed leptons the second term containing the technicolor
interactions disappears and $y/2$ changes to $-3y/2$. Only the last
term is present for the right handed leptons with an appropriate
hypercharge assignment.

\subsubsection{Low Energy Theory for MWT}

We construct the effective theory for MWT including composite scalars and vector bosons, their self interactions, and their interactions with the electroweak gauge fields and the SM fermions

\subsubsection{Scalar Sector} \label{sec:scalar}
The relevant effective theory for the Higgs sector at the electroweak scale consists, in our model, of a composite Higgs and its pseudoscalar partner, as well as nine pseudoscalar Goldstone bosons and their scalar partners. These
can be assembled in the matrix
\begin{eqnarray}
M = \left[\frac{\sigma+i{\Theta}}{2} + \sqrt{2}(i\Pi^a+\widetilde{\Pi}^a)\,X^a\right]E \ ,
\label{M}
\end{eqnarray}
which transforms under the full SU(4) group according to
\begin{eqnarray}
M\rightarrow uMu^T \ , \qquad {\rm with} \qquad u\in {\rm SU(4)} \ .
\end{eqnarray}
The $X^a$'s, $a=1,\ldots,9$ are the generators of the SU(4) group which do not leave  the vacuum expectation value (VEV) of $M$ invariant
\begin{eqnarray}
\langle M \rangle = \frac{v}{2}E
 \ .
\end{eqnarray}
Note that the notation used is such that $\sigma$ is a \emph{scalar}
while the $\Pi^a$'s are \emph{pseudoscalars}. It is convenient to
separate the fifteen generators of SU(4) into the six that leave the
vacuum invariant, $S^a$, and the remaining nine that do not, $X^a$.
Then the $S^a$ generators of the SO(4) subgroup satisfy the relation
\begin{eqnarray}
S^a\,E + E\,{S^a}^{T} = 0 \ ,\qquad {\rm with}\qquad  a=1,\ldots  ,  6 \ ,
\end{eqnarray}
so that $uEu^T=E$, for $u\in$ SO(4). The explicit realization of the generators is shown in appendix \ref{appgen}.
With the tilde fields included, the matrix $M$ is invariant in form under U(4)$\equiv$SU(4)$\times$U(1)$_{\rm
A}$, rather than just SU(4). However the U(1)$_{\rm A}$ axial symmetry is anomalous, and is therefore broken at the quantum level.

The connection between the composite scalars and the underlying techniquarks can be derived from the transformation properties under SU(4), by observing that the elements of the matrix $M$ transform like techniquark bilinears:
\begin{eqnarray}
M_{ij} \sim Q_i^\alpha Q_j^\beta \varepsilon_{\alpha\beta} \quad\quad\quad {\rm with}\ i,j=1\dots 4.
\label{M-composite}
\end{eqnarray}
Using this expression, and the basis matrices given in appendix \ref{appgen}, the scalar fields can be related to the wavefunctions of the techniquark bound states. This gives the following charge eigenstates:
\begin{eqnarray}
\begin{array}{rclcrcl}
v+H & \equiv & \sigma \sim  \overline{U}U+\overline{D}D  &,~~~~ &
\Theta  &\sim& i \left(\overline{U} \gamma^5 U+\overline{D} \gamma^5 D\right) \ ,  \\
A^0 & \equiv & \widetilde{\Pi}^3  \sim  \overline{U}U-\overline{D}D &,~~~~ &
\Pi^0 & \equiv & \Pi^3 \sim i \left(\overline{U} \gamma^5 U-\overline{D} \gamma^5 D\right) \ , \\
A^+ & \equiv & {\displaystyle \frac{\widetilde{\Pi}^1 - i \widetilde{\Pi}^2}{\sqrt{2}}} \sim \overline{D}U &,~~~~&
\Pi^+ & \equiv & {\displaystyle \frac{\Pi^1 - i \Pi^2}{\sqrt{2}}} \sim i \overline{D} \gamma^5 U \ , \\
A^- & \equiv & {\displaystyle \frac{\widetilde{\Pi}^1 + i \widetilde{\Pi}^2}{\sqrt{2}}} \sim \overline{U}D &,~~~~&
\Pi^- & \equiv & {\displaystyle \frac{\Pi^1 + i \Pi^2}{\sqrt{2}}} \sim i \overline{U} \gamma^5 D \ ,
\end{array}
\label{TM-eigenstates}
\end{eqnarray}
for the technimesons, and
\begin{eqnarray}
\begin{array}{rcl}
\Pi_{UU} & \equiv & {\displaystyle \frac{\Pi^4 + i \Pi^5 + \Pi^6 + i \Pi^7}{2}} \sim U^T C U \ , \\
\Pi_{DD} & \equiv & {\displaystyle \frac{\Pi^4 + i \Pi^5 - \Pi^6 - i \Pi^7}{2}} \sim D^T C D \ , \\
\Pi_{UD} & \equiv & {\displaystyle \frac{\Pi^8 + i \Pi^9}{\sqrt{2}}} \sim U^T C D \ , \\
\widetilde{\Pi}_{UU} & \equiv &
{\displaystyle \frac{\widetilde{\Pi}^4 + i \widetilde{\Pi}^5 + \widetilde{\Pi}^6 + i \widetilde{\Pi}^7}{2}} \sim i U^T C \gamma^5 U \ , \\
\widetilde{\Pi}_{DD} & \equiv &
{\displaystyle \frac{\widetilde{\Pi}^4 + i \widetilde{\Pi}^5 - \widetilde{\Pi}^6 - i \widetilde{\Pi}^7}{2}} \sim i D^T C \gamma^5 D  \ , \\
\widetilde{\Pi}_{UD} & \equiv & {\displaystyle \frac{\widetilde{\Pi}^8 + i \widetilde{\Pi}^9}{\sqrt{2}}} \sim i U^T C \gamma^5 D \ ,
\end{array}
\label{TB-eigenstates}
\end{eqnarray}
for the technibaryons, where $U\equiv (U_L,U_R)^T$ and $D\equiv (D_L,D_R)^T$ are Dirac technifermions, and $C$ is the charge conjugation matrix, needed to form Lorentz-invariant objects. To these technibaryon charge eigenstates we must add the corresponding charge conjugate states ({\em e.g.} $\Pi_{UU}\rightarrow \Pi_{\overline{U}\overline{U}}$).

It is instructive to split the scalar matrix into four two by two blocks as follows:
\begin{equation}  
M=\begin{pmatrix} {\cal X}  & {\cal O}  \\ {\cal O}^T & {\cal Z} \end{pmatrix} \ ,
\end{equation}
with ${\cal X}$ and ${\cal Z}$ two complex symmetric matrices accounting for six independent degrees of freedom each and ${\cal O}$ is a generic complex two by two matrix featuring eight real bosonic fields. ${\cal O}$ accounts for the SM like Higgs doublet and a second copy as well as for the three Goldstones which upon electroweak gauging will become the longitudinal components of the intermediate massive vector bosons. 

The electroweak subgroup can be embedded in SU(4), as explained in detail in \cite{Appelquist:1999dq}. Here SO(4) is the subgroup to which SU(4) is maximally broken. The generators $S^a$, with $a=1,2,3$, form an SU(2) subgroup of SU(4), which we denote by SU(2)$_{\rm V}$, while $S^4$ forms a U(1)$_{\rm V}$ subgroup. The $S^a$ generators, with $a=1,..,4$, together with the $X^a$ generators, with $a=1,2,3$, generate an SU(2)$_{\rm L}\times$SU(2)$_{\rm R}\times$U(1)$_{\rm V}$ algebra. This is easily seen by changing genarator basis from $(S^a,X^a)$ to $(L^a,R^a)$, where
\begin{eqnarray}
L^a \equiv \frac{S^a + X^a}{\sqrt{2}} = \begin{pmatrix}\frac{\tau^a}{2}\ \ \  & 0 \\ 0 & 0\end{pmatrix} \ , \ \
{-R^a}^T \equiv \frac{S^a-X^a}{\sqrt{2}}  = \begin{pmatrix}0 & 0 \\ 0 & -\frac{{\tau^a}^T}{2}\end{pmatrix} \ ,
\end{eqnarray}
with $a=1,2,3$. The electroweak gauge group is then obtained by gauging ${\rm SU(2)}_{\rm L}$, and the ${\rm U(1)}_Y$ subgroup of ${\rm SU(2)}_{\rm R}\times {\rm U(1)}_{\rm V}$, where
\begin{eqnarray}
Y =  -{R^3}^T + \sqrt{2}\ Y_{\rm V}\ S^4 \ ,
\end{eqnarray}
and $Y_{\rm V}$ is the U(1)$_{\rm V}$ charge. For example, from Eq.~(\ref{assign1}) and Eq.~(\ref{assign2}) we see that $Y_{\rm V}=y$ for the techniquarks, and $Y_{\rm V}=-3y$ for the new leptons. As SU(4) spontaneously breaks to SO(4), ${\rm SU(2)}_{\rm L}\times {\rm SU(2)}_{\rm R}$ breaks to ${\rm SU(2)}_{\rm V}$. As a consequence, the electroweak symmetry breaks to ${\rm U(1)}_Q$, where
\begin{eqnarray}
Q = \sqrt{2}\ S^3 + \sqrt{2}\ Y_{\rm V} \ S^4 \ .
\end{eqnarray}
Moreover the ${\rm SU(2)}_{\rm V}$ group, being entirely contained in the unbroken SO(4), acts as a custodial isospin, which insures that the $\rho$ parameter is equal to one at tree-level.

The electroweak covariant derivative for the $M$ matrix is
\begin{eqnarray}
D_{\mu}M =\partial_{\mu}M - i\,g \left[G_{\mu}(y)M + MG_{\mu}^T(y)\right]  \
, \label{covariantderivative}
\end{eqnarray}
where
\begin{eqnarray}
g\ G_{\mu}(Y_{\rm V}) & = & g\ W^a_\mu \ L^a + g^{\prime}\ B_\mu \ Y  \nonumber \\
& = & g\ W^a_\mu \ L^a + g^{\prime}\ B_\mu \left(-{R^3}^T+\sqrt{2}\ Y_{\rm V}\ S^4\right) \ .
\label{gaugefields}
\end{eqnarray}
Notice that in the last equation $G_\mu(Y_{\rm V})$ is written for a general U(1)$_{\rm V}$ charge $Y_{\rm V}$, while in Eq.~(\ref{covariantderivative}) we have to take the U(1)$_{\rm V}$ charge of the techniquarks, $Y_{\rm V}=y$, since these are the constituents of the matrix $M$, as explicitly shown in Eq.~(\ref{M-composite}).

Three of the nine Goldstone bosons associated with the broken generators become the longitudinal degrees of freedom of
the massive weak gauge bosons, while the extra six Goldstone bosons will acquire a mass due to extended technicolor interactions (ETC) as well as the
electroweak interactions per se. Using a bottom up approach we will not commit to a specific ETC theory but limit ourself to introduce the minimal low energy operators  needed to construct a phenomenologically viable theory. The new Higgs Lagrangian is
\begin{eqnarray}
{\cal L}_{\rm Higgs} &=& \frac{1}{2}{\rm Tr}\left[D_{\mu}M D^{\mu}M^{\dagger}\right] - {\cal V}(M) + {\cal L}_{\rm ETC} \ ,
\label{Letc}
\end{eqnarray}
where the potential reads
\begin{eqnarray}
{\cal V}(M) & = & - \frac{m^2}{2}{\rm Tr}[MM^{\dagger}] +\frac{\lambda}{4} {\rm Tr}\left[MM^{\dagger} \right]^2 
+ \lambda^\prime {\rm Tr}\left[M M^{\dagger} M M^{\dagger}\right] \nonumber \\
& - & 2\lambda^{\prime\prime} \left[{\rm Det}(M) + {\rm Det}(M^\dagger)\right] \ ,
\label{Vdef}
\end{eqnarray}
and ${\cal L}_{\rm ETC}$ contains all terms which are generated by the ETC interactions, and not by the chiral symmetry breaking sector. Notice that the determinant terms (which are renormalizable) explicitly break the U(1)$_{\rm A}$ symmetry, and give mass to $\Theta$, which would otherwise be a massless Goldstone boson. While the potential has a (spontaneously broken) SU(4) global symmetry, the largest global symmetry of the kinetic term is SU(2)$_{\rm L}\times$U(1)$_{\rm R}\times$U(1)$_{\rm V}$ (where U(1)$_{\rm R}$ is the $\tau^3$ part of SU(2)$_{\rm R}$), and becomes SU(4) in the $g,g^\prime\rightarrow 0$ limit. Under electroweak gauge transformations, $M$ transforms like
\begin{eqnarray}
M(x) \rightarrow u(x;y) \ M(x) \ u^T(x;y) \ ,
\label{transf-M}
\end{eqnarray}
where
\begin{eqnarray}
u(x;Y_{\rm V}) = \exp{\left[i\alpha^a(x)L^a+i\beta(x)\left(-{R^3}^T+\sqrt{2}\ Y_{\rm V}\ S^4\right)\right]} \ ,
\label{u}
\end{eqnarray}
and $Y_{\rm V}=y$.
We explicitly break the SU(4) symmetry in order to provide mass to the Goldstone bosons which are not eaten by the weak gauge bosons. We, however, preserve the full 
SU(2)$_{\rm L}\times$SU(2)$_{\rm R}\times$U(1)$_{\rm V}$ subgroup of SU(4), since breaking SU(2)$_{\rm R}\times$U(1)$_{\rm V}$ to U(1)$_Y$ would result in a potentially dangerous violation of the custodial isospin symmetry. Assuming parity invariance we write: 
\begin{eqnarray}
{\cal L}_{\rm ETC} = \frac{m_{\rm ETC}^2}{4}\ {\rm Tr}\left[M B M^\dagger B + M M^\dagger \right] + \cdots \ , \label{VETCdef}
\end{eqnarray}
where the ellipses represent possible higher dimensional operators, and $B\equiv 2\sqrt{2}S^4$ commutes with the SU(2)$_{\rm L}\times$SU(2)$_{\rm R}\times$U(1)$_{\rm V}$ generators.

The potential ${\cal V}(M)$ is SU(4) invariant. It produces a VEV
which parameterizes the techniquark condensate, and spontaneously
breaks SU(4) to SO(4). In terms of the model parameters the VEV is
\begin{eqnarray}
v^2=\langle \sigma \rangle^2 = \frac{m^2}{\lambda + \lambda^\prime - \lambda^{\prime\prime} } \ ,
\label{VEV}
\end{eqnarray}
while the Higgs mass is
\begin{eqnarray}
M_H^2 = 2\ m^2 \ .
\end{eqnarray}
The linear combination $\lambda + \lambda^{\prime} -
\lambda^{\prime\prime}$ corresponds to the Higgs self coupling in
the SM. The three pseudoscalar mesons $\Pi^\pm$, $\Pi^0$ correspond
to the three massless Goldstone bosons which are absorbed by the
longitudinal degrees of freedom of the $W^\pm$ and $Z$ boson. The
remaining six uneaten Goldstone bosons are technibaryons, and all
acquire tree-level degenerate mass through, not yet specified, ETC interactions:
\begin{eqnarray}
M_{\Pi_{UU}}^2 = M_{\Pi_{UD}}^2 = M_{\Pi_{DD}}^2 = m_{\rm ETC}^2  \ .
\end{eqnarray}
The remaining scalar and pseudoscalar masses are
\begin{eqnarray}
M_{\Theta}^2 & = & 4 v^2 \lambda^{\prime\prime} \nonumber \\
M_{A^\pm}^2 = M_{A^0}^2 & = & 2 v^2 \left(\lambda^{\prime}+\lambda^{\prime\prime}\right)
\end{eqnarray}
for the technimesons, and
\begin{eqnarray}
M_{\widetilde{\Pi}_{UU}}^2 = M_{\widetilde{\Pi}_{UD}}^2 = M_{\widetilde{\Pi}_{DD}}^2 =
m_{\rm ETC}^2 + 2 v^2 \left(\lambda^{\prime} + \lambda^{\prime\prime }\right) \ ,
\end{eqnarray}
for the technibaryons. 
To gain insight on some of the mass relations one can use \cite{Hong:2004td}.

\subsubsection{Vector Bosons}
The composite vector bosons of a theory with a global SU(4) symmetry are conveniently described by the four-dimensional traceless Hermitian matrix
\begin{eqnarray}
A^\mu = A^{a\mu} \ T^a \ ,
\end{eqnarray}
where $T^a$ are the SU(4) generators: $T^a=S^a$, for $a=1, \dots ,6$, and $T^{a+6}=X^a$, for $a=1, \dots ,9$. Under an arbitrary SU(4) transformation, $A^\mu$ transforms like
\begin{equation}
A^\mu \ \rightarrow \ u\ A^\mu \ u^\dagger \ ,\ \ \ {\rm where} \ u\in {\rm SU(4)} \ .
\label{vector-transform}
\end{equation}
Eq.~(\ref{vector-transform}), together with the tracelessness of the matrix $A_\mu$, gives the connection with the techniquark bilinears:
\begin{equation}
A^{\mu,j}_{i}  \sim \ Q^{\alpha}_i  \sigma^{\mu}_{\alpha \dot{\beta}}  \bar{Q}^{\dot{\beta},j}
- \frac{1}{4} \delta_{i}^j Q^{\alpha}_k  \sigma^{\mu}_{\alpha \dot{\beta}} \bar{Q}^{\dot{\beta},k} \ .
\end{equation}
Then we find the following relations between the charge eigenstates and the wavefunctions of the composite objects:
\begin{eqnarray}
\begin{array}{rclcrcl}
v^{0\mu} & \equiv & A^{3\mu} \sim \bar{U} \gamma^\mu U - \bar{D} \gamma^\mu D & , & 
a^{0\mu} & \equiv & A^{9\mu} \sim \bar{U} \gamma^\mu \gamma^5 U - \bar{D} \gamma^\mu \gamma^5 D \\
v^{+\mu} & \equiv & {\displaystyle \frac{A^{1\mu}-i A^{2\mu}}{\sqrt{2}}} \sim \bar{D} \gamma^\mu U & , &
a^{+\mu} & \equiv & {\displaystyle \frac{A^{7\mu}-i A^{8\mu}}{\sqrt{2}}} \sim  \bar{D} \gamma^\mu  \gamma^5 U \\
v^{-\mu} & \equiv & {\displaystyle \frac{A^{1\mu}+i A^{2\mu}}{\sqrt{2}}} \sim  \bar{U} \gamma^\mu D & , &  
a^{-\mu} & \equiv & {\displaystyle \frac{A^{7\mu}+i A^{8\mu}}{\sqrt{2}}} \sim  \bar{U} \gamma^\mu  \gamma^5 D \\
v^{4\mu} & \equiv & A^{4\mu} \sim \bar{U} \gamma^\mu U + \bar{D} \gamma^\mu D  & , & & &
\end{array}
\label{TMV-eigenstates}
\end{eqnarray}
for the vector mesons, and
\begin{eqnarray}
\begin{array}{rcl}
x_{UU}^\mu & \equiv & {\displaystyle \frac{A^{10\mu}+i A^{11\mu}+A^{12\mu}+ i A^{13\mu}}{2}} \sim   U^T C \gamma^\mu \gamma^5 U \ , \\
x_{DD}^\mu & \equiv & {\displaystyle \frac{A^{10\mu}+i A^{11\mu}-A^{12\mu}- i A^{13\mu}}{2}} \sim   D^T C \gamma^\mu \gamma^5 D \ , \\
x_{UD}^\mu & \equiv & {\displaystyle \frac{A^{14\mu}+i A^{15\mu}}{\sqrt{2}}} \sim  D^T C \gamma^\mu \gamma^5 U \ , \\
s_{UD}^\mu & \equiv & {\displaystyle \frac{A^{6\mu}-i A^{5\mu}}{\sqrt{2}}} \sim U^T C \gamma^\mu  D \ ,
\end{array}
\label{TBV-eigenstates}
\end{eqnarray}
for the vector baryons.

There are different approaches on how to introduce vector mesons at the effective Lagrangian level. At the tree level they are all equivalent. The main differences emerge when exploring quantum corrections.

In the appendix we will show how to introduce the vector mesons in a way that renders the following Lagrangian amenable to loop computations.  
Based on these premises, the minimal kinetic Lagrangian is:
\begin{eqnarray}
{\cal L}_{\rm kinetic} = -\frac{1}{2}{\rm Tr}\Big[\widetilde{W}_{\mu\nu}\widetilde{W}^{\mu\nu}\Big] - \frac{1}{4}B_{\mu\nu}B^{\mu\nu}
-\frac{1}{2}{\rm Tr}\Big[F_{\mu\nu}F^{\mu\nu}\Big] + m_A^2 \ {\rm Tr}\Big[C_\mu C^\mu\Big] \ ,
\label{massterm}
\end{eqnarray}
where $\widetilde{W}_{\mu\nu}$ and $B_{\mu\nu}$ are the ordinary field strength tensors for the electroweak gauge fields. Strictly speaking the terms above are not only kinetic ones since the Lagrangian contains a mass term as well as self interactions. The tilde on $W^a$ indicates  that the associated states are not yet the SM weak triplets: in fact these states mix with the composite vectors to form mass eigenstates corresponding to the ordinary $W$ and $Z$ bosons. $F_{\mu\nu}$ is the field strength tensor for the new SU(4) vector bosons,
\begin{eqnarray}
F_{\mu\nu} & = & \partial_\mu A_\nu - \partial_\nu A_\mu - i\tilde{g}\left[A_\mu,A_\nu\right]\ ,
\label{strength}
\end{eqnarray}
and the vector field $C_\mu$ is defined by
\begin{eqnarray}
C_\mu \ \equiv \ A_\mu \ - \ \frac{g}{\tilde{g}}\ G_\mu (y) \ .
\end{eqnarray}
As shown in the appendix this is the appropriate linear combination to take which transforms homogeneously under the electroweak symmetries:
\begin{eqnarray}
C_\mu(x) \ \rightarrow \ u(x;y)\ C_\mu(x) \ u(x;y)^\dagger \ ,
\label{transf-C}
\end{eqnarray}
where $u(x;Y_{\rm V})$ is given by Eq.~(\ref{u}). (Once again, the specific assignment $Y_{\rm V}=y$, due to the fact that the composite vectors are built out of techniquark bilinears.) The mass term in Eq.~(\ref{massterm}) is gauge invariant (see the appendix), and gives a degenerate mass to all composite vector bosons, while leaving the actual gauge bosons massless. (The latter acquire mass as usual from the covariant derivative term of the scalar matrix $M$, after spontaneous symmetry breaking.)

The $C_\mu$ fields couple with $M$ via gauge invariant operators. Up
to dimension four operators the Lagrangian is (see the appendix for a more general treatment):
\begin{eqnarray}
{\cal L}_{\rm M-C} & = & \tilde{g}^2\ r_1 \ {\rm Tr}\left[C_\mu C^\mu M M^\dagger\right]
+ \tilde{g}^2\ r_2 \ {\rm Tr}\left[C_\mu M {C^\mu}^T M^\dagger \right] \nonumber \\
& + & i \ \tilde{g}\ r_3 \ {\rm Tr}\left[C_\mu \left(M (D^\mu M)^\dagger - (D^\mu M) M^\dagger \right) \right]
+ \tilde{g}^2\ s \ {\rm Tr}\left[C_\mu C^\mu \right] {\rm Tr}\left[M M^\dagger \right] \ . \nonumber \\
\end{eqnarray}
The dimensionless parameters $r_1$, $r_2$, $r_3$, $s$ parameterize
the strength of the interactions between the composite scalars and
vectors in units of $\tilde{g}$, and are therefore naturally
expected to be of order one. However, notice that for
$r_1=r_2=r_3=0$ the overall Lagrangian possesses two independent
SU(2)$_{\rm L}\times$U(1)$_{\rm R}\times$U(1)$_{\rm V}$ global
symmetries. One for the terms involving $M$ and one for the terms
involving $C_\mu$~\footnote{The gauge fields explicitly break the
original SU(4) global symmetry to SU(2)$_{\rm L}\times$U(1)$_{\rm
R}\times$ U(1)$_{\rm V}$, where U(1)$_{\rm R}$ is the $T^3$ part of
SU(2)$_{\rm R}$, in the SU(2)$_{\rm L}\times$SU(2)$_{\rm
R}\times$U(1)$_{\rm V}$ subgroup of SU(4).}. The Higgs potential
only breaks the symmetry associated with $M$, while leaving the
symmetry in the vector sector unbroken. This {\em enhanced symmetry}
guarantees that all $r$-terms are still zero after loop corrections.
Moreover if one chooses $r_1$, $r_2$, $r_3$ to be small the near enhanced symmetry will protect these values against large corrections \cite{Casalbuoni:1995qt,Appelquist:1999dq}.

We can also construct dimension four operators including only
$C_{\mu}$ fields.  

\subsubsection{Fermions and Yukawa Interactions}
The fermionic content of the effective theory consists of the SM quarks and leptons, the new lepton doublet $L=(N,E)$ introduced to cure the Witten anomaly, and a composite techniquark-technigluon doublet. 

We now consider the limit according to which the SU(4) symmetry is, at first, extended to ordinary quarks and leptons. Of course, we will need to break this symmetry to accommodate the SM phenomenology. We start by arranging the SU(2) doublets in SU(4) multiplets as we did for the techniquarks in Eq.~(\ref{SU(4)multiplet}). We therefore introduce the four component vectors $q^i$ and $l^i$,
\begin{eqnarray} 
q^i= \begin{pmatrix}
u^i_L \\
d^i_L \\
-i\sigma^2 {u^i_R}^* \\
-i\sigma^2 {d^i_R}^*
\end{pmatrix}\ , \quad
l^i= \begin{pmatrix}
\nu^i_L \\
e^i_L \\
-i\sigma^2 {\nu^i_R}^* \\
-i\sigma^2 {e^i_R}^*
\end{pmatrix}\ ,
\end{eqnarray}
where $i$ is the generation index. Note that such an extended SU(4) symmetry automatically predicts the presence of a right handed neutrino for each generation. In addition to the SM fields there is an SU(4) multiplet for the new leptons,
\begin{eqnarray}
L = \begin{pmatrix}
N_L \\
E_L \\
-i\sigma^2 {N_R}^* \\
-i\sigma^2 {E_R}^*
\end{pmatrix}\ ,
\end{eqnarray}
and a multiplet for the techniquark-technigluon bound state,
\begin{eqnarray} 
\widetilde{Q}= \begin{pmatrix}
\widetilde{U}_L \\
\widetilde{D}_L \\
-i\sigma^2 {\widetilde{U}_R}^* \\
-i\sigma^2 {\widetilde{D}_R}^*
\end{pmatrix}\ .
\end{eqnarray}
With this arrangement, the electroweak covariant derivative for the fermion fields can be written
\begin{eqnarray}
D_\mu \ = \  \partial_\mu \  - \  i \ g \ G_\mu (Y_{\rm V})  \ , 
\end{eqnarray}
where $Y_{\rm V}=1/3$ for the quarks, $Y_{\rm V}=-1$ for the leptons, $Y_{\rm V}=-3y$ for the new lepton doublet, and $Y_{\rm V}=y$ for the techniquark-technigluon bound state. One can check that these charge assignments give the correct electroweak quantum numbers for the SM fermions. In addition to the covariant derivative terms, we should add a term coupling $\widetilde{Q}$ to the vector field $C_\mu$, which transforms globally under electroweak gauge transformations. Such a term naturally couples the composite fermions to the composite vector bosons which otherwise would only feel the week interactions. Based on this, we write the following gauge part of the fermion Lagrangian:
\begin{eqnarray}
{\cal L}_{\rm fermion} & = & i\ \overline{q}^i_{\dot{\alpha}}  \overline{\sigma}^{\mu,\dot{\alpha} \beta} D_\mu  q^i_\beta 
+ i\ \overline{l}^i_{\dot{\alpha}}  \overline{\sigma}^{\mu,\dot{\alpha} \beta} D_\mu  l^i_\beta 
+ i\ \overline{L}_{\dot{\alpha}}  \overline{\sigma}^{\mu,\dot{\alpha} \beta} D_\mu  L_\beta 
+ i\ \overline{\widetilde{Q}}_{\dot{\alpha}}  \overline{\sigma}^{\mu,\dot{\alpha} \beta} D_\mu  \widetilde{Q}_\beta  \nonumber \\
& + & x\ \overline{\widetilde{Q}}_{\dot{\alpha}}  \overline{\sigma}^{\mu,\dot{\alpha} \beta} C_\mu  \widetilde{Q}_\beta
\label{fermion-kinetic}
\end{eqnarray}
The terms coupling the SM fermions or the new leptons to $C_\mu$ are in general not allowed. In fact under electroweak gauge transformations any four-component fermion doublet $\psi$ transforms like
\begin{eqnarray}
\psi \rightarrow  u(x;Y_{\rm V}) \ \psi \ ,
\label{transf-psi}
\end{eqnarray}
and from Eq.~(\ref{transf-C}) we see that a term like $\psi^\alpha  \sigma^{\mu}_{\alpha \dot{\beta}} C_\mu  \overline{\psi}^{\dot{\beta}}$ is only invariant if $Y_{\rm V}=y$. Then we can distinguish two cases. First, we can have
$y\neq 1/3$ and $y\neq -1$, in which case $\psi^\alpha  \sigma^{\mu}_{\alpha \dot{\beta}} C_\mu  \overline{\psi}^{\dot{\beta}}$ is only invariant for $\psi=\widetilde{Q}$. Interaction terms of the SM fermions with components of $C_\mu$ are still possible, but these would break the SU(4) chiral simmetry even in the limit in which the electroweak gauge interactions are switched off. Second, we can have $y=1/3$ or $y=-1$. Then $\psi^\alpha  \sigma^{\mu}_{\alpha \dot{\beta}} C_\mu  \overline{\psi}^{\dot{\beta}}$ is not only invariant for $\psi=\widetilde{Q}$, but also for either $\psi=q^i$ or $\psi=l^i$, respectively. In the last two cases, however, the corresponding interactions are highly suppressed, since these give rise to anomalous couplings of the light fermions with the SM gauge bosons, which are tightly constrained by experiments.

We now turn to the issue of providing masses to ordinary fermions.  In the first chapter the simplest ETC model has been briefly reviewed. Many extensions of technicolor have been suggested in the literature to address this problem. Some of the extensions use another strongly coupled gauge dynamics,  others introduce fundamental scalars. Many variants of the schemes presented above exist and a review of the major models is the one by Hill and Simmons \cite{Hill:2002ap}. At the moment there is not yet a consensus on which is the correct ETC. To keep the number of fields minimal we make the most economical ansatz, i.e. we parameterize our ignorance about a complete ETC theory by simply coupling the fermions to our low energy effective Higgs. This simple construction minimizes the flavor changing neutral currents problem. It is worth mentioning that it is possible to engineer a schematic ETC model proposed first by Randall in \cite{Randall:1992vt} and adapted for the MWT in \cite{Evans:2005pu} for which the effective theory presented in the main text can be considered a minimal description. Another non minimal way to give masses to the ordinary fermions is to (re)introduce a new Higgs doublet as already done many times in the literature \cite{Simmons:1988fu,Dine:1990jd,Kagan:1990az,Kagan:1991gh,Carone:1992rh,Carone:1993xc,Gudnason:2006mk}. 

Depending on the value of $y$ for the techniquarks, we can write different Yukawa interactions which couple the SM fermions to the matrix $M$. Let $\psi$ denote either $q^i$ or $l^i$. If $\psi$ and the techniquark multiplets $Q^a$ have the same U(1)$_{\rm V}$ charge, then the Yukawa term
\begin{eqnarray}
- \psi^T M^* \psi + {\rm h.c.} \ ,
\label{yukawa1}
\end{eqnarray}
is gauge invariant, as one can check explicitly from Eq.~(\ref{transf-M}) and Eq.~(\ref{transf-psi}). Otherwise, if $\psi$ and $Q^a$ have different U(1)$_{\rm V}$ charges, we can only write a gauge invariant Lagrangian with the off-diagonal terms of $M$, which contain the Higgs and the Goldstone bosons:
\begin{eqnarray}
- \psi^T M_{\rm off}^*\ \psi + {\rm h.c.} \ .
\label{yukawa2}
\end{eqnarray}
In fact $M_{\rm off}$ has no U(1)$_{\rm V}$ charge, since
\begin{eqnarray}
S^4 M_{\rm off} + M_{\rm off} {S^4}^T = 0 \ ,
\end{eqnarray}
The last equation implies that the U(1)$_{\rm V}$ charges of $\psi^T$ and $\psi$ cancel in Eq.~(\ref{yukawa2}). The latter is actually the only viable Yukawa Lagrangian for the new leptons, since the corresponding U(1)$_{\rm V}$ charge is $Y_{\rm V}=-3y \neq y$, and for the ordinary quarks, since Eq.~(\ref{yukawa1}) contains $qq$ terms which are not color singlets. 

We notice however that neither Eq.~(\ref{yukawa1}) nor Eq.~(\ref{yukawa2}) are phenomenologically viable yet, since they leave the SU(2)$_{\rm R}$ subgroup of SU(4) unbroken, and the corresponding Yukawa interactions do not distinguish between the up-type and the down-type fermions. In order to prevent this feature, and recover agreement with the experimental input, we break the SU(2)$_{\rm R}$ symmetry to U(1)$_{\rm R}$ by using the projection operators $P_U$ and $P_D$, where
\begin{eqnarray}
P_U = \begin{pmatrix} 1 & 0 \\ 0 & \frac{1+\tau^3}{2} \end{pmatrix} \ , \quad
P_D = \begin{pmatrix} 1 & 0 \\ 0 & \frac{1-\tau^3}{2} \end{pmatrix} \ .
\end{eqnarray}
Then, for example, Eq.~(\ref{yukawa1}) should be replaced by
\begin{eqnarray}
- \psi^T \left(P_U M^* P_U\right) \psi - \psi^T \left(P_D M^* P_D\right) \psi + {\rm h.c.} \ .
\label{yukawa3}
\end{eqnarray}

 {}For illustration we distinguish two different cases for our analysis, $y\neq -1$ and $y= -1$, and write the corresponding Yukawa interactions:
 \newline
(i) $y\neq -1$. In this case we can only form gauge invariant terms with the SM fermions by using the off-diagonal $M$ matrix. Allowing for both $N-E$ and $\widetilde{U}-\widetilde{D}$ mass splitting, we write
\begin{eqnarray}
{\cal L}_{\rm Yukawa} &=& -\ y_u^{ij}\ q^{i T} \left(P_U M_{\rm off}^* P_U\right) q^j
- y_d^{ij}\ q^{i T} \left(P_D M_{\rm off}^* P_D\right) q^j \nonumber \\
&& -\ y_\nu^{ij}\ l^{i T} \left(P_U M_{\rm off}^* P_U\right) l^j
- y_e^{ij}\ l^{i T} \left(P_D M_{\rm off}^* P_D\right) l^j \nonumber \\
&& -\ y_N\ L^T \left(P_U M_{\rm off}^* P_U\right) L
-\ y_E\ L^T \left(P_D M_{\rm off}^* P_D\right) L \nonumber \\
&& -\ \ y_{\widetilde{U}} \widetilde{Q}^T \left(P_U M^* P_U\right) \widetilde{Q} 
- \ y_{\widetilde{D}} \widetilde{Q}^T \left(P_D M^* P_D\right) \widetilde{Q} \ + {\rm h.c.} \ ,
\label{yukawa-1}
\end{eqnarray}
where $y_u^{ij}$, $y_d^{ij}$, $y_\nu^{ij}$, $y_e^{ij}$ are arbitrary complex matrices, and $y_N$, $y_E$, $y_{\widetilde{U}}$, $y_{\widetilde{D}}$ are complex numbers.
\newline
Note that the underlying strong dynamics already provides a dynamically generated mass term for $\widetilde{Q}$ of the type: 
\begin{equation}
{k}\,  \widetilde{Q}^T  M^* \widetilde{Q} + {\rm h.c.} \ ,
\end{equation}
with ${k}$ a dimensionless coefficient of order one and entirely fixed within the underlying theory.  The splitting between the up and down type techniquarks is due to physics beyond the technicolor interactions \footnote{Small splittings with respect to the electroweak scale will be induced by the SM corrections per se.}. Hence the Yukawa interactions for $\widetilde{Q}$ must be interpreted as already containing the dynamical generated mass term.  

(ii) $y= -1$. In this case we can form gauge invariant terms with the SM leptons and the full $M$ matrix:
\begin{eqnarray}
{\cal L}_{\rm Yukawa} &=& -\ y_u^{ij}\ q^{i T} \left(P_U M_{\rm off}^* P_U\right) q^j
- y_d^{ij}\ q^{i T} \left(P_D M_{\rm off}^* P_D\right) q^j \nonumber \\
&& -\ y_\nu^{ij}\ l^{i T} \left(P_U M^* P_U\right) l^j
- y_e^{ij}\ l^{i T} \left(P_D M^* P_D\right) l^j \nonumber \\
&& -\ y_N\ L^T \left(P_U M_{\rm off}^* P_U\right) L
-\ y_E\ L^T \left(P_D M_{\rm off}^* P_D\right) L \nonumber \\
&& -\ \ y_{\widetilde{U}} \widetilde{Q}^T \left(P_U M^* P_U\right) \widetilde{Q} 
- \ y_{\widetilde{D}} \widetilde{Q}^T \left(P_D M^* P_D\right) \widetilde{Q} \ + {\rm h.c.} \ .
\label{yukawa-2}
\end{eqnarray}
Here we are assuming Dirac masses for the neutrinos, but we can easily add also Majorana mass terms. At this point one can exploit the symmetries of the kinetic terms to induce a GIM mechanism, which works out exactly like in the SM. Therefore, in both Eq.~(\ref{yukawa-1}) and Eq.~(\ref{yukawa-2}) we can assume $y_u^{ij}$, $y_d^{ij}$, $y_\nu^{ij}$, $y_e^{ij}$ to be diagonal matrices, and replace the $d^i_L$ and $\nu^i_L$ fields, in the kinetic terms, with $V_q^{ij} d^j_L$ and $V_l^{ij} \nu^j_L$, respectively, where $V_q$ and $V_l$ are the mixing matrices.  

When $y=-1$ $\widetilde{Q}$ has the same quantum numbers of the ordinary leptons, except for the technibaryon number. If the technibaryon number is violated they can mix with the ordinary leptons behaving effectively as a  fourth generation leptons (see Eq.~(\ref{yukawa-2})). However this will reintroduce, in general, anomalous couplings with intermediate gauge bosons for the ordinary fermions and hence we assume the mixing to be small.

\subsection{Constraining the MWT effective Lagrangian via WSRs  \& $S$ parameter }

In our effective theory the $S$ parameter is directly proportional to the parameter $r_3$ via:
\begin{eqnarray}
S= \frac{8\pi}{\tilde{g}^2}\, \chi \,(2- \chi)   \ , \quad {\rm with} \quad \chi = \frac{v^2\tilde{g}^2}{2M^2_A} \, r_3 \ ,
\label{S}
\end{eqnarray}
where we have expanded in $g/\tilde{g}$ and kept only the leading order. The full expression can be found in the appendix D of \cite{Foadi:2007ue} and it is also reported in the appendix here.
We can now use the  sum rules to relate $r_3$ to other parameters in the theory for the running and the walking case.  Within the effective theory we deduce:
\begin{eqnarray}
F^2_V = \left(1 - \, \chi \frac{r_2}{r_3}\right)\, \frac{2M^2_A}{\tilde{g}^2} = \frac{2M^2_V}{\tilde{g}^2}\ , \quad F^2_A = 2\frac{M^2_A}{\tilde{g}^2}(1 - \chi)^2 \ , \quad F^2_{\pi} = v^2 ( 1 -\chi \, r_3 ) \ .\end{eqnarray}
Hence the first WSR reads:
\begin{eqnarray}
1+r_2 - 2 r_3 = 0 \ ,
\end{eqnarray}
while the second:
\begin{eqnarray}
 (r_2 - r_3) (v^2\tilde{g}^2 (r_2 + r_3) - 4 M^2_A) = a \frac{16\pi^2}{d(R)}  v^2\left( 1 - \chi \, r_3\right)^2 \ .
\end{eqnarray}

To gain analytical insight we consider the limit in which $\tilde{g} $ is small while $g/\tilde{g}$ is still much smaller than one. To leading order in $\tilde{g}$ the second sum rule simplifies to:
\begin{eqnarray}
r_3 - r_2  = a \frac{4\pi^2}{d(R)}\frac{v^2}{M^2_A} \ ,
\end{eqnarray}
Together with the first sum rule we find:
\begin{eqnarray}
r_2 = 1 - 2 t \ , \qquad r_3 = 1 - t \ ,
\end{eqnarray}
with
\begin{eqnarray}
t= a \frac{4\pi^2}{d(R)}\frac{v^2}{M^2_A} \ .
\end{eqnarray}
The approximate $S$ parameter reads.
\begin{eqnarray}
S= 8\pi\, \frac{v^2}{M^2_A} (1-t)   \ . \end{eqnarray}
A positive $a$  renders $S$ smaller than expected in a running theory for a given value of the axial mass. In the next subsection we will make a similar analysis without taking the limit of small $\tilde{g}$.

 \subsubsection{Axial-Vector Spectrum via WSRs}

It is is interesting to determine the relative vector to axial
spectrum as function of one of the two masses, say the axial one,
for a fixed value of the $S$ parameter meant to be associated to a given underlying gauge theory. 

\begin{figure}[!t]
\centering
\resizebox{10cm}{!}{\includegraphics{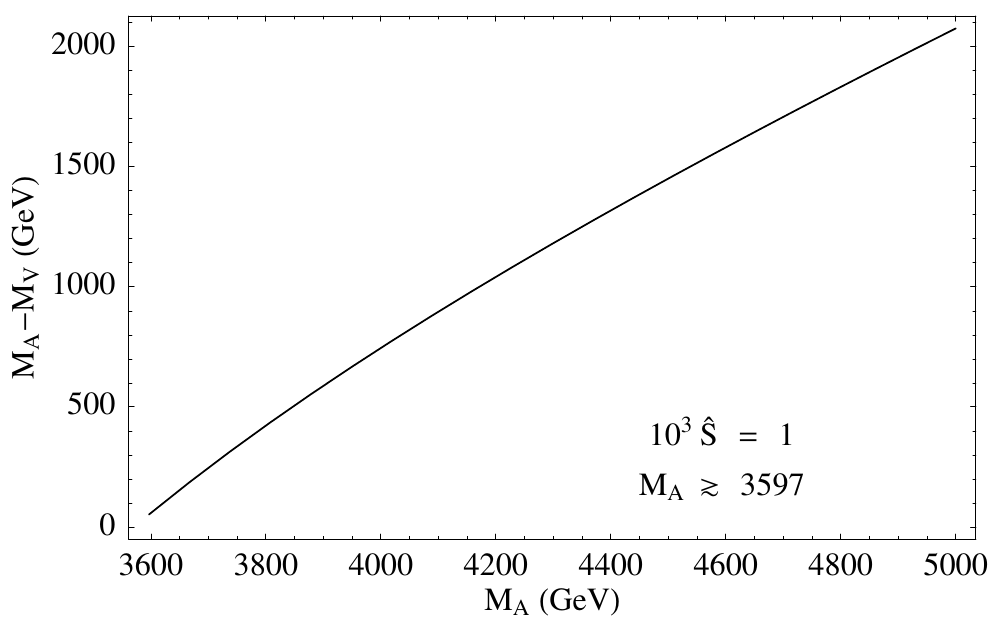}}
\caption{In the picture above we  have set $10^3 \hat{S}=1$, corresponding to $S\simeq 0.11$.  In the appendix we have provided the relation between $\hat{S}$ and the traditional S. Here we have imposed the first and the second WSR for $a=0$. This corresponds to an underlying gauge theory with a standard running behavior of the coupling constant. }
\label{runningwsr}
\end{figure}

For a running type dynamics (i.e. $a=0$) the two WSRs force the vector mesons to be quite heavy (above 3 TeV) in order to have a relatively low $S$ parameter ($S\simeq$ 0.1). This can be seen directly from Eq.~(\ref{S}) in the running regime, where $r_2=r_3=1$. This leads to
\begin{eqnarray}
M_A^2 \ \gtrsim \ \frac{8\pi v^2}{S} \ ,
\label{lowMA}
\end{eqnarray}
which corresponds to $M_A\gtrsim$ 3.6 TeV, for $S\simeq$ 0.11. Perhaps a more physical way to express this is to say that it is hard to have an intrinsically small $S$ parameter for running type theories. By small we mean smaller than the scaled up technicolor version of QCD with two techniflavors, in which $S\simeq$ 0.3. In Figure \ref{runningwsr} we plot the difference between the axial and vector mass as function of the axial mass, for $S\simeq$0.11. Since Eq.~(\ref{lowMA}) provides a lower bound for $M_A$, this plot shows that in the running regime the axial mass is always heavier than the vector mass. In fact the $M_A^2-M_V^2$ difference is proportional to $r_2$, with a positive proportionality factor (see the appendix), and $r_2=1$ in the running regime.

\begin{figure}[!t]
\centering
\begin{tabular}{cc}
\hskip -1.2cm\resizebox{8.5cm}{!}{\includegraphics{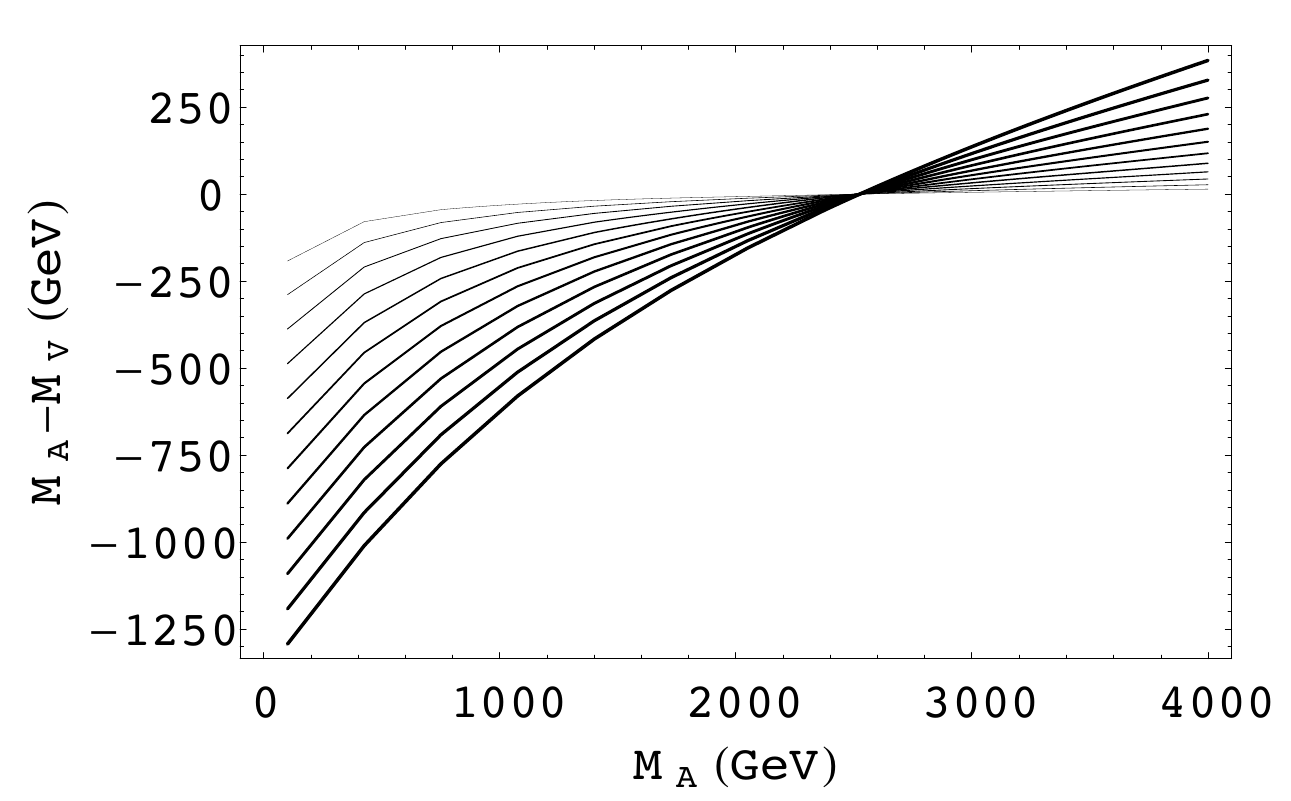}} & \resizebox{8.5cm}{!}{\includegraphics{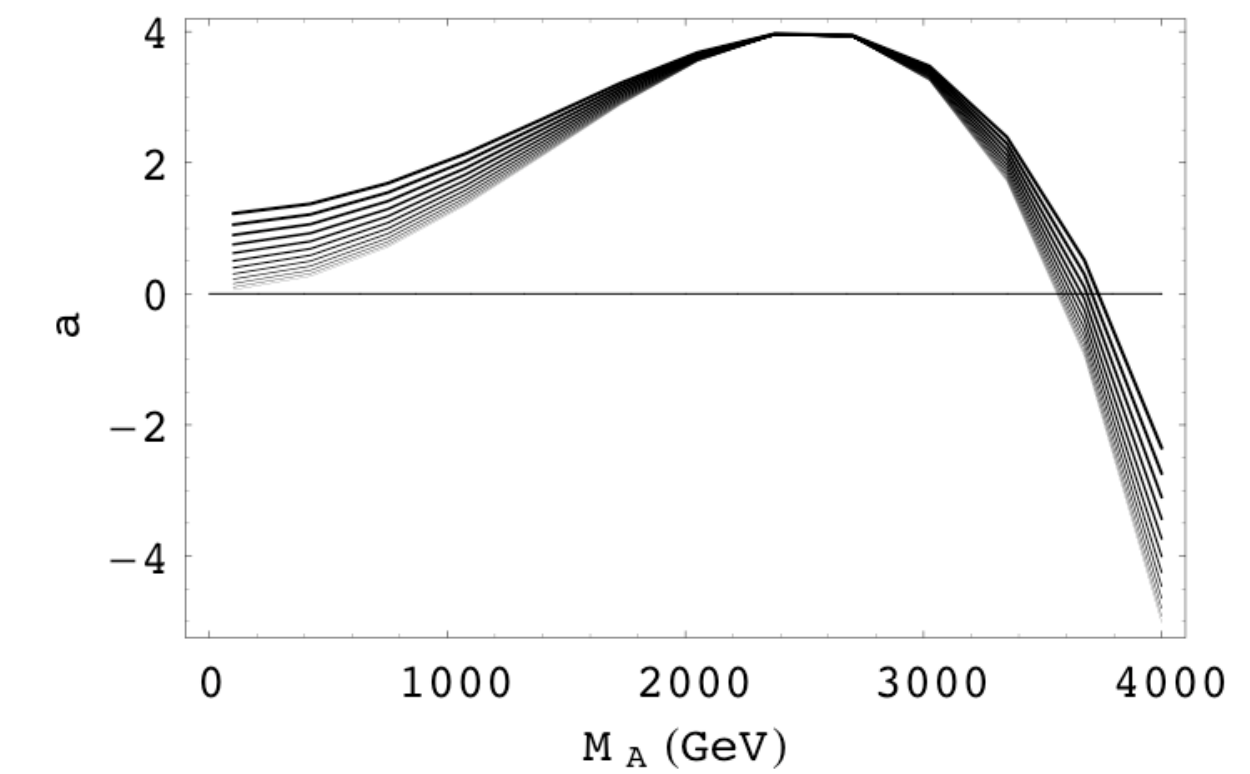}}
\end{tabular}
\caption{In the two pictures above we have set $10^3 \hat{S}=1$, corresponding to $S\simeq 0.11$, and the different curves are obtained by varying $\tilde{g}$ from one (the thinnest curve)  to eight  (the thickest curve). We have imposed the first WSR. {\it Left Panel}: We plot the allowed values of $M_A-M_V$ as function of $M_A$. {\it Right Panel}: We compute the value that $a$ should assume as function of $M_A$ in order for the second WSR to be satisfied in the walking regime. Note that $a$ is expected to be positive or zero. }
\label{walkwsr}
\end{figure}

When considering the second WSR modified by the walking dynamics, we observe that it is possible to have quite light spin one  vector mesons compatible with a small $S$ parameter. We numerically solve the first and second WSR in presence of the contribution due to walking in the second sum rule. The results are summarized in Figure \ref{walkwsr}. As for the running case we set again $S\simeq 0.11$. This value is close to the estimate in the underlying MWT \footnote{For the MWT we separate the contribution due to the new leptonic sector (which will be dealt with later in the main text) and the one due to the underlying strongly coupled gauge theory which is expected to be well represented by the perturbative contribution and is of the order of $1/2\pi$. When comparing with the $S$ parameter from the vector meson sector of the effective theory we should subtract from the underlying $S$ the one due to the new fermionic composite states $\tilde{U}$ and $\tilde{D}$. This contribution is very small since it is $1/6\pi$ in the limit when these states are degenerate and heavier than the $Z$ gauge boson.}. The different curves are obtained by varying $\tilde{g}$ from one (the thinnest curve)  to eight (the thickest curve). We plot the allowed values of $M_A-M_V$ as function of $M_A$ in the left panel, having imposed only the first sum rule. In the right panel we compute the corresponding value that $a$ should assume as function of $M_A$ in order for the second WSR to be satisfied in the walking regime as well.
Our computation shows that it is possible to have walking theories with light vector mesons and a small $S$ parameter. Such a scenario, however, needs a positive value of $a$, together with an exotic vector spectrum according to which the axial is lighter than its associated vector meson\footnote{The sum rules do not contain sufficient information to predict the relative spectral ordering of the vector and axial vector mesons. We have hence studied both orderings.}. However for spin one fields heavier than roughly 2.5 TeV and with still a positive $a$ one has an axial meson heavier than the vector one.  A degenerate spectrum allows for a small $S$ but with  relatively large values of $a$ and spin one masses around 2.5 TeV. We observe that $a$ becomes zero (and eventually negative) when the vector spectrum becomes sufficiently heavy. In other words we recover the running behavior for large masses of spin-one fields. Although in the plot we show negative values of $a$ one should stop the analysis after having reached a zero value of $a$. In fact, for masses heavier than roughly 3.5 TeV the second WSR for the running behavior, i.e. $a=0$, is enforced.  

Our results are general and help elucidating how different underlying dynamics will manifest itself at LHC. Any four dimensional strongly interacting theory replacing the Higgs mechanism, with two Dirac techniflavors gauged under the electroweak theory, is expected to have a spectrum of the low lying vector resonances like the one presented above. We note that the resonance inversion does not imply a negative or zero S parameter. 

\subsubsection{Reducing $S$ via New Leptons}
\begin{figure}[!t]
\centering
\begin{tabular}{cc}
\resizebox{6.5cm}{!}{\includegraphics{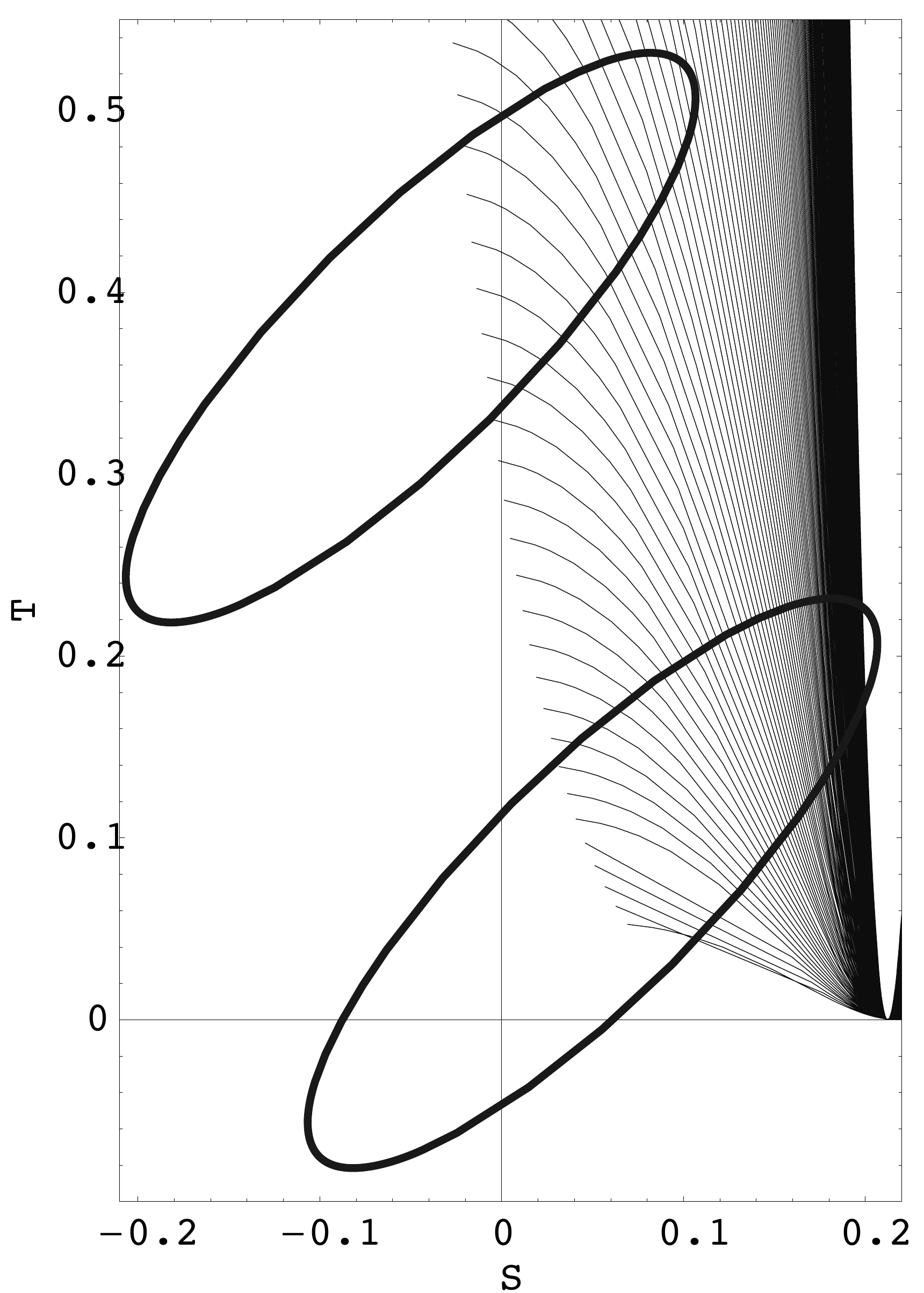}}
~~&~~~~
\resizebox{6.39cm}{!}{\includegraphics{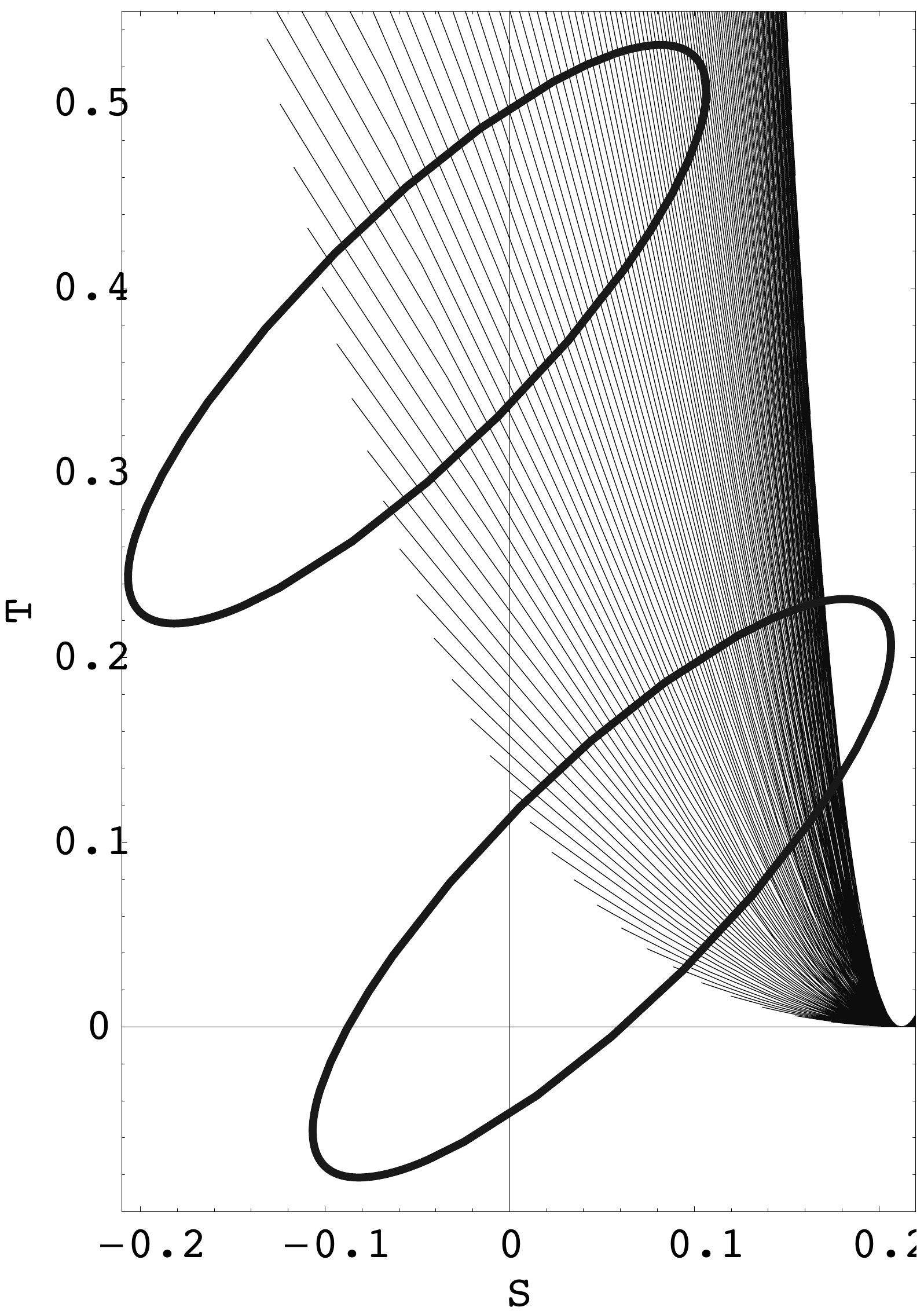}}
\end{tabular}
\caption{The ellipses represent the 68\% confidence region for the $S$ and $T$ parameters. The upper ellipse is for  a reference Higgs mass of the order of a TeV, the lower curve is for a light Higgs with mass around 114 GeV. The contribution from the MWT theory per se and of the leptons as function of the new lepton masses is expressed by the dark grey region. The left panel has been obtained using a SM type hypercharge assignment while the right hand graph is for $y=1$. }
\label{ST}
\end{figure}

We have studied the effects of the lepton family on the
electroweak parameters in~\cite{Dietrich:2005jn}, we summarize here the main results
in Figure \ref{ST}.  The ellipses represent the one standard deviation
confidence region for the $S$ and $T$ parameters. The upper ellipse
is for  a reference Higgs mass of the order of 1 TeV while the lower
curve is for a light Higgs with mass around 114 GeV. The
contribution from the MWT theory per se and of the leptons as
function of the new lepton masses is expressed by the dark grey
region. The left panel has been obtained using a SM type hypercharge assignment while the right hand graph is for $y=1$. In both pictures the regions of overlap between the theory and the precision contours are achieved when the upper component of the weak isospin doublet is lighter than the lower component. The opposite case leads to a total $S$ which is larger than the one predicted within the new strongly coupled dynamics per se.  This is due to the sign of the hypercharge for the new leptons. The mass range used in the plots, in the case of the SM hypercharge assignment is $100-1000$~GeV for the new electron and $50-800$~GeV for the new Dirac neutrino, while it is $100-800$ and $100-1000$~GeV respectively for the $y=1$ case. The plots have been obtained assuming a Dirac mass for the new neutral lepton (in the case of a SM hypercharge assignment).  {The range of the masses for which the theory is in the ellipses, for a reference Higgs mass of a TeV, is $100-400$~GeV for the new neutrino and about twice the mass of the neutrino for the new electron.}

The analysis for the Majorana mass case has been performed in \cite{Kainulainen:2006wq} where one can again show that it is possible to be within the 68\% contours. 

The contour plots we have drawn take into account the values of the top
mass which has dropped dramatically since our first comparison of our model 
theory in \cite{Dietrich:2005jn} to the experimental data. 

We have provided a comprehensive extension of the SM which embodies (minimal) walking technicolor theories and their interplay with the SM particles. Our extension of the SM features all of the relevant low energy effective degrees of freedom linked to our underlying minimal walking theory. These include scalars, pseudoscalars as well as spin one fields. The bulk of the Lagrangian has been spelled out in \cite{Foadi:2007ue}. The link with underlying strongly coupled gauge theories has been achieved via the time-honored Weinberg sum rules. The modification of the latter according to walking has been taken into account. We have also analyzed the case in which the underlying theory behaves like QCD rather than being near an infrared fixed point. This has allowed us to gain insight on the spectrum of the spin one fields which is an issue of phenomenological interest. In the appendix we have: i) provided the explicit construction of all of the SU(4) generators, ii) shown how to construct the effective Lagrangian in a way which is amenable to quantum corrections,  iii) shown the explicit form of the mass matrices for all of the particles, iv) provided a summary of all of the relevant electroweak parameters and their explicit dependence on the coefficients of our effective theory. The vacuum alignment problem associated to the weak corrections has been investigated in \cite{Dietrich:2009ix} and the results support the breaking pattern envisioned here.

We have introduced the model in a format which is useful for collider phenomenology, as well as for computing corrections due to (walking) technicolor for different observable of the SM, even in the flavor sector. 

\subsubsection{ Beyond $S$ and $T$: New Constraints for Walking Technicolor}
In \cite{Foadi:2007se} we investigated the effects of the electroweak precision measurements beyond the time honored $S$ and $T$ ones. 
Once the hypercharge of the underlying technifermions is fixed all of the derived precision parameters defined in \cite{Barbieri:2004qk} are function solely of the gauge couplings, masses of the gauge bosons and the first excited spin-one states and one more parameter $\chi$:
\begin{eqnarray}
\hat{S} &=& \frac{(2 - \chi )\chi g^2}{2\tilde{g}^2}
      \ , \\ 
W &=& \frac{g^2}{2\tilde{g}^2}\frac{M_W^2 }{M_A^2M_V^2}{(M_A^2+(\chi-1)^2M_V^2)}
\ ,  \\
Y &=&  \frac{g'^2}{2\tilde{g}^2}\frac{M_W^2}{M_A^2M_V^2} {((1+4y^2)M_A^2+(\chi -1)^2M_V^2)}
 \ , \\
X &=& \frac{g\,g'}{2\tilde{g}^2} \, \frac{M_W^2}{M^2_A M^2_V}{ (M_A^2 - (\chi -1)^2M_V^2 )} \ .
\end{eqnarray}
$\hat{T}=\hat{U}=V=0$.  $g$ and $g^{\prime}$ are the weak and hypercharge couplings, $M_W$ the gauge boson mass, $y$ the coefficient parameterizing different hypercharge choices of the underlying technifermions \cite{Foadi:2007ue}, $\tilde{g}$ the technistrong vector mesons coupling to the Goldstones in the technicolor limit, i.e. $a=0$. It was realized in \cite{Appelquist:1999dq, Duan:2000dy} and further explored in \cite{Foadi:2007ue} that for walking theories, i.e. $a\neq 0$, the WSRs allow for a new parameter $\chi$ which in the technicolor limit reduces to $\chi_0=\tilde{g}^2 v^2/2M^2_A$ 
with $F^2_{\pi}=v^2(1-\chi^2/\chi_0)$ the electroweak vacuum expectation value and $M_{A(V)}$ the mass of the axial(vector) lightest spin-one field. To make direct contact with the WSRs and for the reader's convenience we recall the relations:\begin{eqnarray}
F^2_V =  \frac{2M^2_V}{\tilde{g}^2}\ , \quad F^2_A = 2\frac{M^2_A}{\tilde{g}^2}(1 - \chi)^2 \ .\end{eqnarray}
We have kept the leading order in the electroweak couplings over the technistrong coupling $\tilde{g}$ in the expressions above while we used the full expressions \cite{Foadi:2007ue} in making the plots.

How do we study the constraints? From the expressions above we have four independent parameters, $\tilde{g}$, $\chi$, $M_V$ and $M_A$ at the effective Lagrangian level. Imposing the first WSR and assuming a fixed value of $\hat{S}$ leaves two independent parameters which we choose to be $\tilde{g}$ and $M_A$. From the modified second WSR we read off the value of $a/d(R)$.

In \cite{Foadi:2007se} we first constrained the spectrum and couplings of theories of  WT with a positive value of the $\hat{S}$ parameter compatible with the associated precision measurements at the one sigma level. More specifically we will take $\hat{S}\simeq 0.0004$ which is the highest possible value compatible with precision data for a very heavy Higgs \cite{Barbieri:2004qk}. Of course the possible presence of another sector can allow for a larger intrinsic $\hat{S}$. We are interested in the constraints coming from W and Y after having fixed $\hat{S}$.  The analysis can easily be extended to take into account sectors not included in the new strongly coupled dynamics.
\begin{figure*}[htb] 
\begin{tabular}{ccc}
{\includegraphics[height=9cm,width=5cm]{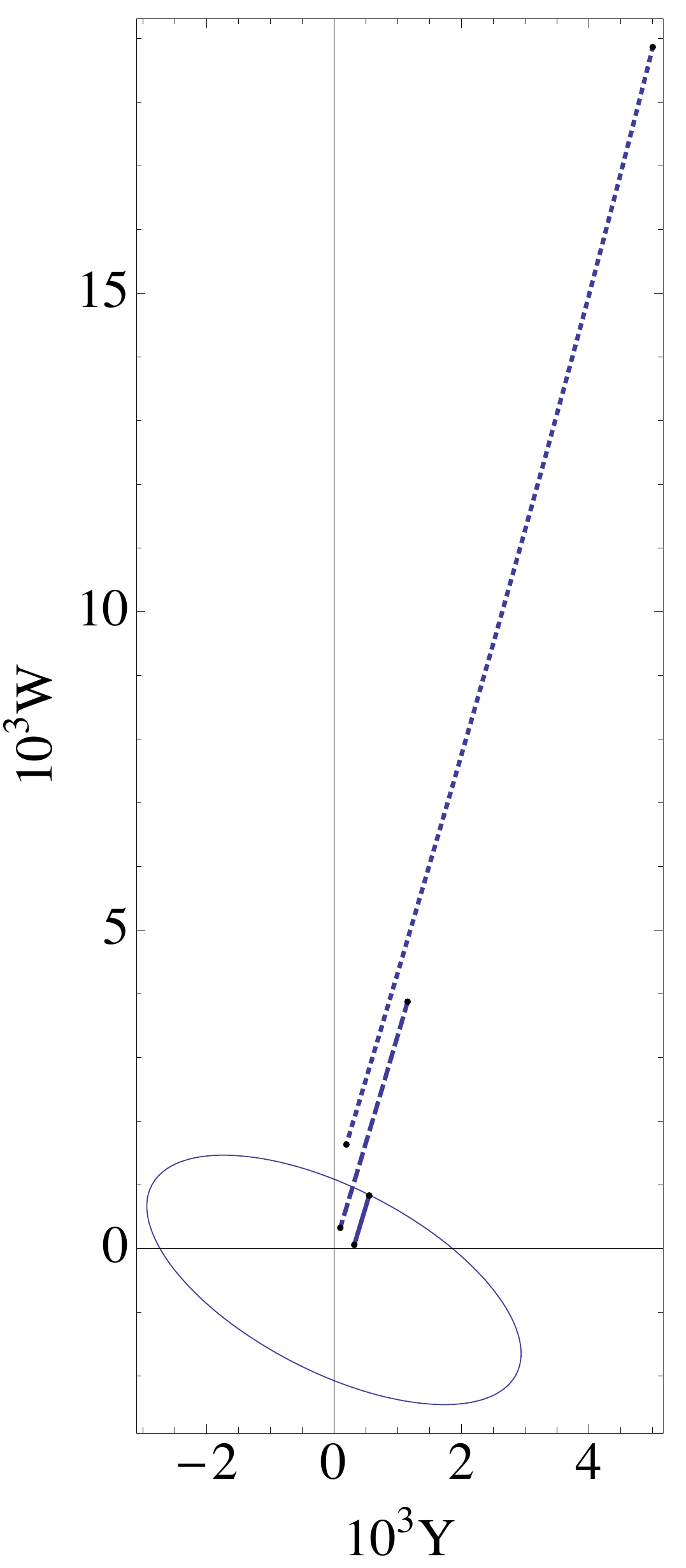}} ~~~&~~~ 
{\includegraphics[height=9cm,width=5cm]{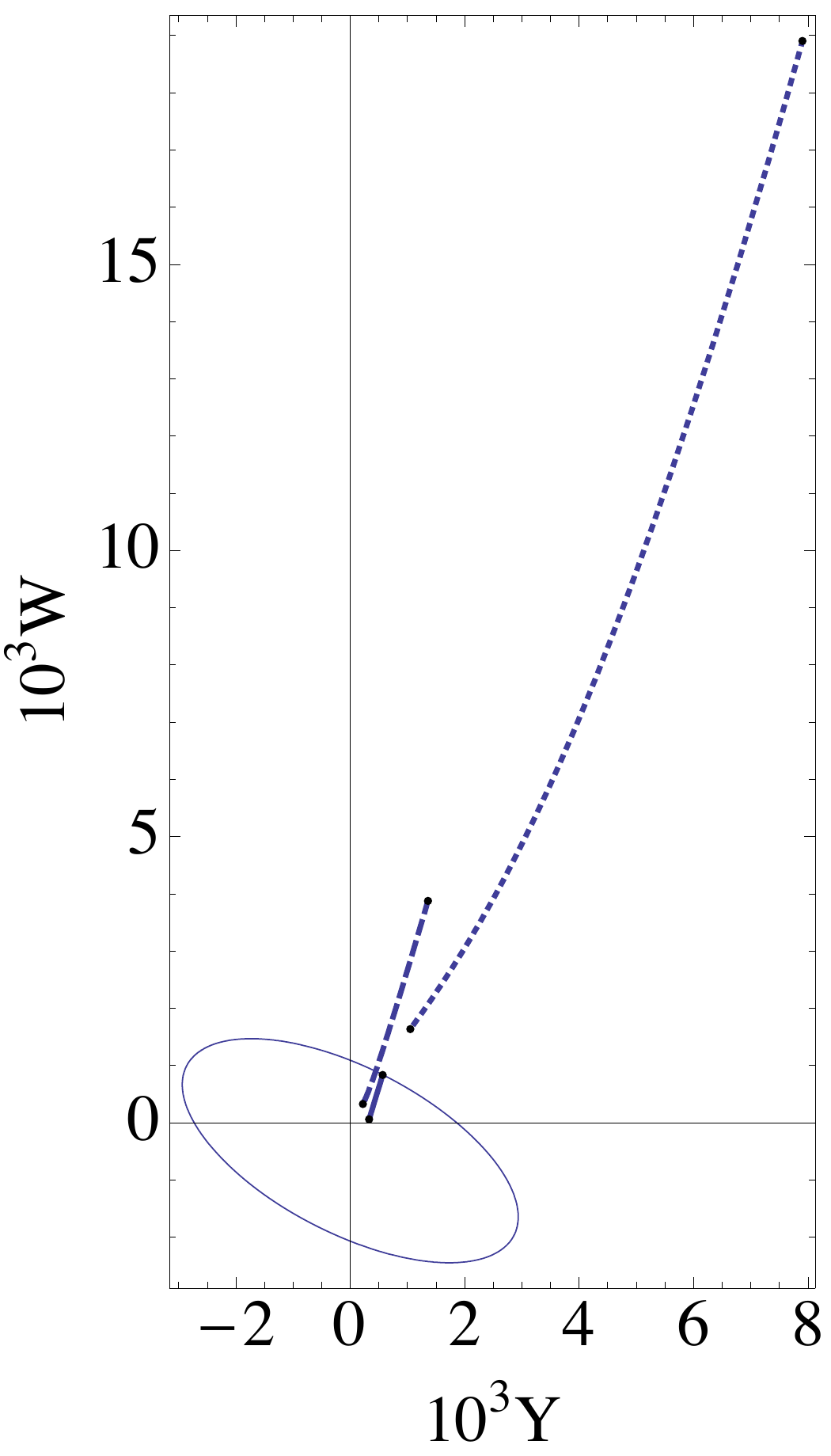}}~~~&~~~{\includegraphics[height=9cm,width=5cm]{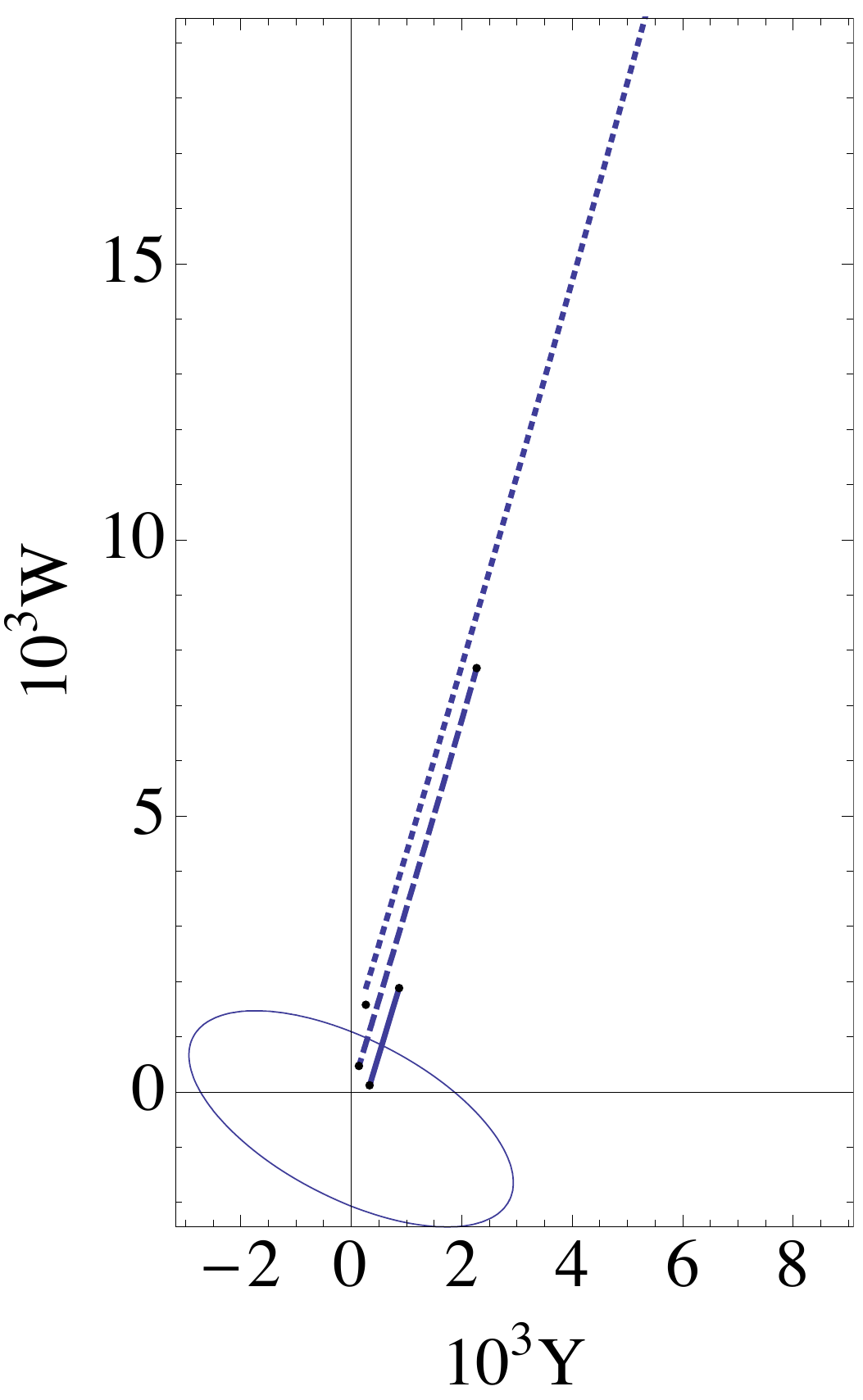}}
\\ 
~~~{WT with $y=0$}&~~~{WT with $y=1$ }&~~~CT with $y=0$
\end{tabular}
\caption{\label{fig:label1}  The ellipses in the WY plane corresponds to the 95\% confidence level obtained scaling the standard error ellipse axis by a 2.447 factor. The three segments, meant to be all on the top of each other,  in each plot correspond to different values of $\tilde{g}$. The  solid line corresponds to $\tilde{g}=8$, the dashed line to $\tilde{g}=4$ and the dotted one to $\tilde{g}=2$. The lines are drawn as function of $M_A$ with the point closest to the origin obtained for $M_A = 600$ GeV while the further away corresponds to $M_A=150$ GeV. We assumed $\hat{S} = 0.0004$ for WT while $\hat{S}$ is $0$ in CT by construction.  
}
\end{figure*}
A light spin-one spectrum can be achieved only if the axial is much lighter than the associated vector meson. The second is that WT models, even with small $\hat{S}$, are sensitive to the  W-Y constraints as can be seen from the plots in Fig.~\ref{fig:label1}. Since X is a higher derivative of $\hat{S}$ it is not constraining. We find that WT dynamics with a small $\tilde{g}$ coupling and a light axial vector boson is not preferred by electroweak data. Only for values of $\tilde{g}$ larger than or about $8$ the axial vector meson can be light, i.e. of the order of $200$ GeV. However WT dynamics with a small intrinsic S parameter does not allow the spin-one vector partner to be degenerate with the light axial but  predicts it to be much heavier Fig.~\ref{fig:label2}. If the spin-one masses are very heavy then the spectrum has a standard ordering pattern, i.e. the vector meson lighter than the axial meson. We also show in Fig.~\ref{fig:label2} the associated value of $a$. We were the first to make the prediction of a very light axial vector mesons in \cite{Foadi:2007ue} on the base of the modified WSRs, even lighter than the associated vector mesons.  Eichten and Lane put forward a similar suggestion in \cite{Eichten:2007sx}.  We find that a WT dynamics alone compatible with precision electroweak data can accommodate a light spin-one axial resonance only if the associated vector partner is much heavier and in the regime of a strong $\tilde{g}$ coupling.   $a$.  We find tension with the data at a level superior to the 95\% confidence level for: a) WT models featuring $M_A \simeq M_V$ spectrum with a common and very light mass; b) WT models with an axial vector meson lighter than $300$ GeV and $\tilde{g}$ smaller than $4$, an axial vector meson with a mass lighter than or around $600$ GeV and $\tilde{g}$ smaller than $2$.  

\subsubsection
{Introducing and constraining Custodial Technicolor}
We now constrain also models proposed in \cite{Appelquist:1999dq, Duan:2000dy} which, at the effective Lagrangian level, possess an explicit {\it custodial} symmetry for the S parameter. We will refer to this class of models as custodial technicolor (CT) \cite{Foadi:2007se} . The new custodial symmetry is  present in the BESS 
models \cite{Casalbuoni:1988xm,Casalbuoni:1995yb,Casalbuoni:1995qt} which will therefore be constrained as well. In this case we expect our constraints to be similar to the ones also discussed in \cite{Casalbuoni:2007dk}.

Custodial technicolor corresponds to the case for which $M_A=M_V=M$ and $\chi=0$. The effective Lagrangian acquires a new symmetry, relating a vector and an axial field, which can be interpreted as a custodial symmetry for the S parameter \cite{Appelquist:1999dq, Duan:2000dy}.  The only non-zero parameters are now:
\begin{eqnarray}
W &=& \frac{g^2}{\tilde{g}^2}\frac{M_W^2 }{M^2}
\ ,  \\
Y &=&  \frac{g'^2}{2\tilde{g}^2}\frac{M_W^2}{M^2} {(2+4y^2)} \ .
\end{eqnarray}
A CT model cannot be achieved in walking dynamics and must be interpreted as a new framework. In other words CT does not respect the WSRs and hence it can only be considered as a phenomenological type model in search of a fundamental strongly coupled theory. To make our point clearer note that a degenerate spectrum of light spin-one resonances (i.e. $M<4\pi F_{\pi}$) leads to  a very large $\hat{S}=g^2 F^2_{\pi}/4M^2$. We needed only the first sum rule together with the statement of degeneracy of the spectrum to derive this $\hat{S}$ parameter. This statement is universal and it is true for WT and ordinary technicolor. 
The Eichten and Lane \cite{Eichten:2007sx} scenario of almost degenerate and very light spin-one states can only be achieved within a near CT models. A very light vector meson with a small number of techniflavors fully gauged under the electroweak can  be accommodated in CT. This scenario was considered in \cite{ Zerwekh:2005wh,Zerwekh:2007pw} and our constraints apply here.

We find that in CT it is possible to have a very light and degenerate spin-one spectrum  if $\tilde{g}$ is sufficiently large, of the order say of 8 or larger as in the WT case.

We constrained the electroweak parameters intrinsic to WT or CT, however, in general other sectors may contribute to the electroweak observables, an explicit example is the new heavy lepton family introduced above \cite{Dietrich:2005jn}. 

To summarize we have suggested in \cite{Foadi:2007se} a way to constrain WT theories with any given S parameter. We have further constrained relevant models featuring a custodial symmetry protecting the S parameter.  When increasing the value of the S parameter while reducing the amount of walking we recover the technicolor constraints \cite{Peskin:1990zt}. We found bounds on the lightest spectrum of WT and CT theories with an intrinsically small S parameter. Our results are applicable to {\it any} dynamical model of electroweak symmetry breaking featuring near conformal dynamics \'a la walking technicolor.

\begin{figure*}[htb] 
\begin{center}
\begin{tabular}{cc}
{\includegraphics[height=5cm,width=7cm]{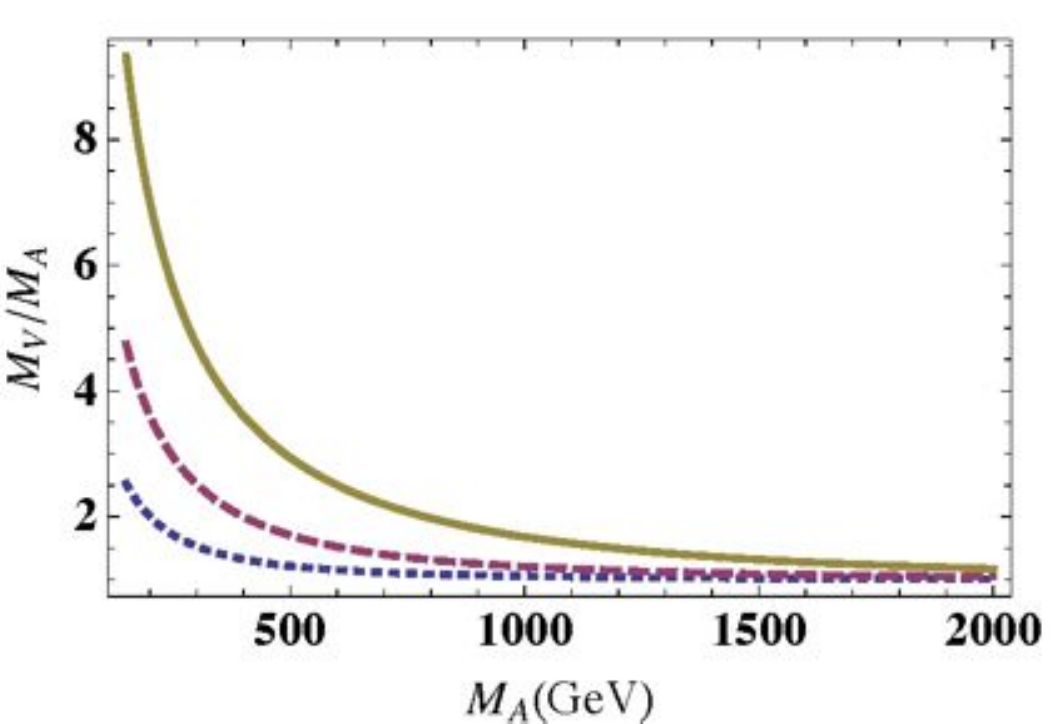}}~~~~&~~~~ 
{\includegraphics[height=4.8cm,width=7.2cm]{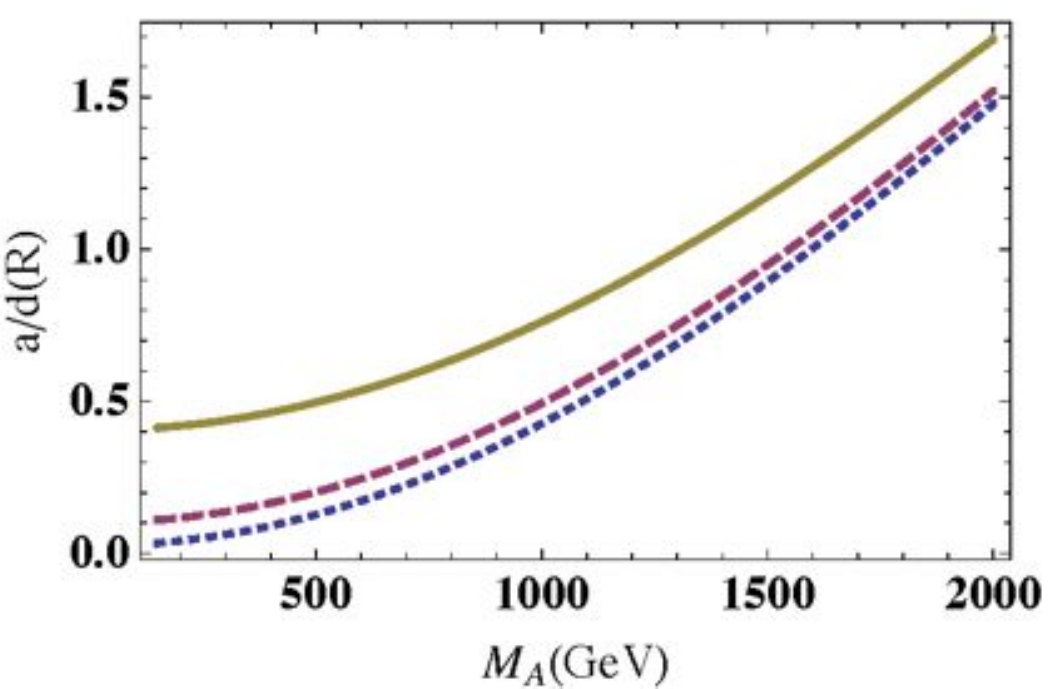}}
\end{tabular}
\end{center}
\caption{\label{fig:label2}  
In the left panel we plot the ratio of the vector over axial mass as function of the axial mass for a WT theory with an intrinsic small S parameter. The vector and axial spectrum is close only when their masses are of the order of the TeV scale and around 2 TeV and onwards the vector is lighter than the axial. The right panel shows the value $a/d(R)$ as function of the axial mass. In both plots the solid, dashed and dotted lines corresponds respectively to  $\tilde{g}=8,4,2$.
}
\end{figure*}%

\subsubsection{An ETC example for MWT and The Top Mass}

It is instructive to present a simple model \cite{Evans:2005pu} which shows how one can embed MWT in an extended technicolor model
capable of generating the top quark mass.  

When the techni-quarks are not in the fundamental representation
of the technicolor group it can be hard to feed
down the electroweak symmetry breaking condensate to generate the
SM fermion masses
\cite{Lane:1989ej,Christensen:2005cb}.  Here, following \cite{Evans:2005pu}, we
wish to highlight that the minimal model can be recast as an SO(3)
theory with fundamental representation techni-quarks. The model
can therefore rather easily be enlarged to an extended technicolor
theory \cite{Lane:1989ej} in the spirit of many examples in the
literature. We will concentrate on the top quark sector - the ETC
gauge bosons in this sector violate weak isospin and one must be
careful to compute their contribution to the T
parameter \cite{Chivukula:1995dc}.

We start by recognizing that adjoint multiplets of SU(2) can be written as
fundamental representations of SO(3). This trick will now allow us
to enact a standard ETC pattern from the literature - it is
particularly interesting that for this model of higher dimensional
representation techniquarks there is a simple ETC model. We will
follow the path proposed in \cite{Randall:1992vt} where we gauge
the full flavor symmetry of the fermions. 

If we were simply interested in the fourth family then the
enlarged ETC symmetry is a Pati-Salam type
unification. We stack the doublets

\beq \left[\left( \begin{array}{c} U^a \\ D^a \end{array}
\right)_L \ , \quad \left( \begin{array}{c} N \\ E \end{array}
\right)_L \right],\hspace{1.5cm} \left[U_R^a\ ,~N_R \right],
\hspace{1.5cm} \left[D_R^a \ ,~E_R \right]\eeq into 4 dimensional
multiplets of SU(4). One then invokes some symmetry breaking
mechanism at an ETC scale (we will not speculate on the mechanism
here though see figure 1) 

\beq SU(4)_{ETC} \rightarrow SO(3)_{TC} \times U(1)_Y \eeq

The technicolor dynamics then proceeds to generate a techniquark
condensate $\langle \bar{U} U \rangle = \langle \bar{D} D \rangle
\neq 0$. The massive gauge bosons associated with the broken ETC
generators can then feed the symmetry breaking condensate down to
generate fourth family lepton masses

\beq m_N = m_E \simeq {\langle \bar{U}U \rangle \over
\Lambda_{ETC}^2} \eeq 

One could now naturally proceed to include the third (second,
first) family by raising the ETC symmetry group to SU(8) (SU(12),
SU(16)) and a series of appropriate symmetry breakings. This would
generate masses for all the SM fermions but no isospin
breaking mass contributions within fermion doublets. The simplest
route to generate such splitting is to make the ETC group chiral
so that different ETC couplings determine the isospin +1/2 and
-1/2 masses. Let us only enforce such a pattern for the top quark
and fourth family here since the higher ETC scales are far beyond
experimental probing.

We can, for example, have the SU(7) multiplets
\beq \left[\left( \begin{array}{c} U^a \\ D^a \end{array}
\right)_L\ , \quad \left( \begin{array}{c} N \\ E \end{array}
\right)_L\ , \quad \left( \begin{array}{c} t^c \\ b^c \end{array}
\right)_L \right], \hspace{1.5cm} \left[U_R^a\ ,~ N_R\ ,~t^c_R
\right] \eeq here $a$ will become the technicolor index and $c$
the QCD index. We also have a right handed SU(4) ETC group that
only acts on \beq \left[D_R^a\ , \quad E_R \right]\eeq

The right handed bottom quark is left out of the ETC dynamics and
only has proto-QCD SU(3) dynamics. The bottom quark will thus be
left massless. The symmetry breaking scheme at, for example, a
single ETC scale would then be

\beq SU(7) \times SU(4) \times SU(3) \rightarrow SO(3)_{TC} \times SU(3)_{QCD}  \eeq

The top quark now also acquires a mass from the broken gauge
generators naively equal to the fourth family lepton multiplet.
Walking dynamics has many features though that one would expect to
overcome the traditional small size of the top mass in ETC models.
Firstly the enhancement of the techniquark self energy at high
momentum enhances the ETC generated masses by a factor potentially
as large as  $\Lambda_{ETC}/ \Sigma(0)$. In
\cite{Appelquist:2003hn} it is argued that this effect alone may
be sufficient to push the ETC scale to 4 TeV and still maintain
the physical top mass.

The technicolor coupling is near conformal and strong
so the ETC dynamics will itself be quite strong at its breaking
scale which will tend to enhance light fermion masses
\cite{Evans:1994fb}. In this ETC model the top quark will also feel
the effects of the extra massive octet of axial gluon-like gauge
fields that may induce a degree of top condensation a l\`{a} top
color models \cite{Miransky:1989nu}. We conclude that a 4-8 TeV ETC scale for generating the top mass is possible. In this model the fourth family lepton would then have a mass of the same order and well in excess of the current search limit $M_Z/2$.

\subsubsection{The Next to Minimal Walking Technicolor Theory (NMWT)}
\label{4}

The theory with three technicolors contains an even number of electroweak doublets, and hence
it is not subject to a Witten anomaly.  
The doublet of technifermions, is then represented again as:
\bear
Q_L^{\{C_1,C_2 \}}
=
\left(\begin{array}{l}U^{\{C_1,C_2 \}}\\ D^{\{C_1,C_2 \}}\end{array}\right)_L \ ,
\qquad 
Q_R^{\{C_1,C_2\}}&=&\left(U_R^{\{C_1,C_2\}},~ D_R^{\{C_1,C_2\}}\right) \ .
\nn
\eear
Here $C_i=1,2,3$ is the technicolor index and $Q_{L(R)}$ is a doublet (singlet) with respect 
to the weak interactions. 
Since the two-index symmetric representation of $SU(3)$ is complex the flavor symmetry is $SU(2)_L\times SU(2)_R\times U(1)$. 
Only three Goldstones emerge and
are absorbed in the longitudinal components of the weak vector bosons.

Gauge anomalies are absent with the choice $Y=0$ for the hypercharge of the left-handed technifermions:
\bear
Q_L^{(Q)}
=
\left(\begin{array}{l} U^{(+1/2)}\\ D^{(-1/2)}\end{array}\right)_L \ .
\eear
Consistency requires for the right-handed technifermions (isospin singlets):
\bear
Q_R^{(Q)}&=&\left(U_R^{(+1/2)},~D_R^{-1/2}\right)
\nn
Y&=&~~+1/2,~-1/2 \ .
\eear 
All of these states will be bound into hadrons. There is no need for an associated fourth family of leptons, and hence it is not expected to be observed in the experiments.

Here the low-lying technibaryons are fermions constructed with three techniquarks in the following way:
\begin{eqnarray}
B_{f_1,f_2,f_3;\alpha} = Q^{\{C_1,C_2 \}}_{L;\alpha,
f_1}Q^{\{ C_3,C_4 \}}_{L;\beta,f_2} Q^{\{ C_5,C_6 \}}_{L;\gamma,f_3}\epsilon^{\beta \gamma}
\epsilon_{C_1 C_3 C_5}\epsilon_{C_2 C_4 C_6} \ .
\end{eqnarray}
where $f_i=1,2$ corresponds to $U$ and $D$ flavors, and we are not specifying the flavor symmetrization which in any 
event will have to be such that the full technibaryon wave function is fully antisymmetrized in technicolor, flavor and spin. 
$\alpha$, $\beta$, and $\gamma$ assume the values of one or 
two and represent the ordinary spin. Similarly we can construct different technibaryons using only right fields or a mixture of left and right.

\subsection{Beyond Minimal Walking Technicolor (BMWT)}

When going beyond MWT one finds new and interesting theories able to break the electroweak symmetry while featuring a walking dynamics and yet not at odds with precision measurements, at least when comparing with the naive $S$ parameter.  A compendium of these theories can be found in \cite{Dietrich:2006cm}. Here we will review only the principla type of models one can construct.


\subsubsection{Partially gauged technicolor\label{pgt}}

A small modification of the traditional technicolor approach, which neither
involves additional particle species nor more complicated gauge groups, 
allows constructing several other viable candidates. It consists in letting 
only one doublet of techniquarks transform non-trivially under the electroweak
symmetries with the rest being electroweak singlets, as first suggested in 
\cite{Dietrich:2005jn} and later also used in \cite{Christensen:2005cb}.
Still, all techniquarks transform under the technicolor gauge group. Thereby only one techniquark doublet contributes directly\footnote{Via Technicolor interactions all of the matter content of the theory will affect physical observables associated to the sector coupled to the electroweak symmetry.} to the oblique 
parameter which is thus kept to a minimum for theories which need more
than one family of techniquarks to be quasi-conformal. It is the condensation 
of that first electroweakly charged family that breaks the electroweak 
symmetry. The techniquarks which are uncharged under the electroweak gauge group are
natural building blocks for components of dark matter.

Among the partially gauged cases one of the interesting candidates \cite{Dietrich:2006cm} is the theory with eight techniflavors
in the two-index antisymmetric representation of SU(4). The techniquarks of 
one of the four families carry electroweak charges while the others are 
electroweak singlets. Gauge anomalies are avoided if the two electrically 
charged techniquarks possess half-integer charges.
The technihadron spectrum contains technibaryons made of only two
techniquarks because in the two-index
antisymmetric representation of SU(4) a singlet can already be formed in
that case.
Otherwise technisinglets can also be formed from four techniquarks. All 
technihadrons formed from techniquarks without electrical charges
can contribute to dark matter.
Due to the special charge assignment of the electrically charged particles
(opposite half-integer charges) certain combinations of those can also be 
contained in electrically uncharged technibaryons. For instance we can 
construct the following completely neutral technibaryon:
\bear
\epsilon_{t_1t_2t_3t_4}Q_L^{{t_1t_2},f}Q_L^{{t_3t_4},f^{\prime}}
{\epsilon_{ff^{\prime}}} \ ,
\eear
where $\epsilon_{ff^{\prime}}$ saturates the SU(2)$_L$ indices of the two
gauged techniquarks and the first antisymmetric tensor $\epsilon$ is summed 
over the technicolor indices. We have suppressed the spin indices. This particle is an interesting candidate for 
dark matter and it is hardly detectable in any earth based experiment
\cite{Gudnason:2006yj}.

Since the two index antisymmetric representation of $SU(4)$ is real the model's flavor symmetry is enhanced
to SU(2$N_f$=16)\footnote{It is slightly explicitly broken by the
electroweak interactions.  Here, additionally, there is a difference, on
the electroweak level, between gauged and ungauged techniquark families.}, 
which, when it breaks to SO(16), induces 135 Goldstone
bosons\footnote{Obviously the Goldstone bosons must receive a sufficiently large
mass. This is usually achieved in extended technicolor. Still, they could be
copious at LHC.}. 

It is worth recalling that the centre group symmetry left invariant
by the fermionic matter is a Z$_2$
symmetry. Hence there is a well defined order parameter for confinement
\cite{Sannino:2005sk} which can play a
role in the early Universe.

\subsubsection{Split technicolor}

We summarize here also another
possibility \cite{Dietrich:2005jn} according to which we keep the technifermions gauged under the electroweak theory in the fundamental 
representation of the $SU(N)$ technicolor group while still reducing the number of techniflavors needed to be
near the conformal window. Like for the partially gauged case described above 
this can be achieved by adding matter uncharged under the weak interactions. 
The difference to section \ref{pgt} is that this part of matter transforms 
under a different representation of the technicolor gauge group than the 
part coupled directly to the electroweak sector. {}For example, for definiteness let's choose it to be a 
massless Weyl fermion in the adjoint representation of the technicolor gauge 
group. The resulting theory has the same matter content as $N_f$-flavor super QCD but without the scalars;
hence the name "split technicolor." We expect the critical number of flavors above which one enters the
conformal window $N_f^\mathrm{II}$ to lie within the range
\bear
\frac{3}{2}<\frac{N^\mathrm{II}_f}{N}<\frac{11}{2} \ .
\eear
The lower bound is the exact supersymmetric value for a non-perturbative
conformal fixed point \cite{Intriligator:1995au}, while the upper bound is
the one expected in the theory without a technigluino. The matter content of
"split technicolor" lies between that of super QCD and QCD-like theories with 
matter in the fundamental representation.

For two colors the number of (techni)flavors needed to be near the
conformal window in the split case is at least three, while for three
colors more than five flavors are required. These values are still larger
than the ones for theories with fermions in the two-index
symmetric representation. It is useful to remind the reader that in
supersymmetric theories the critical number of flavors needed to enter the
conformal window does not coincide with the critical number of flavors
required to restore chiral symmetry. The scalars in supersymmetric theories
play an important role from this point of view. We note that a split
technicolor-like theory has been used in \cite{Hsu:2004mf}, to
investigate the strong CP problem.

Split technicolor shares some features with theories of split
supersymmetry advocated and studied in
\cite{ArkaniHamed:2004fb,Giudice:2004tc} as possible extensions of the
SM. Clearly, we have introduced split technicolor---differently
from split supersymmetry---to address the hierarchy problem. This is why we
do not expect new scalars to appear at energy scales higher than the one of
the electroweak theory.

\subsection{Ultra Minimal Walking Technicolor}

Here we provide an explicit example \cite{Ryttov:2008xe} of (near) conformal (NC) technicolor with two types of technifermions, i.e. transforming according to two different representations of the underlying technicolor gauge group \cite{Dietrich:2006cm,Lane:1989ej}. The model possesses a number of interesting properties to recommend it over the earlier models of dynamical electroweak symmetry breaking:
\begin{itemize}
\item
 Features the lowest possible value of the naive $S$ parameter \cite{Peskin:1990zt,{Peskin:1991sw}} while possessing a dynamics which is NC.

 \item Contains, overall, the lowest possible number of fermions.

\item Yields natural DM candidates.
\end{itemize}
Due to the above properties we term this model {\it Ultra Minimal near conformal Technicolor} (UMT). It is constituted by an $SU(2)$ technicolor gauge group with two Dirac flavors in the fundamental representation also carrying electroweak charges, as well as, two additional Weyl fermions in the adjoint representation but singlets under the SM gauge groups.

By arranging the additional fermions in higher dimensional representations, it is possible to construct models which have a particle content smaller than the one of partially gauged technicolor theories. In fact instead of considering additional fundamental flavors we shall consider adjoint flavors. Note that for two colors there exists only one distinct two-indexed representation.

How many adjoint fermions are needed to build the above NC model? Information on the conformal window for gauge theories containing fermions transforming according to distinct representations is vital. We used the all orders beta-function \cite{Ryttov:2007cx} extended to the multiple representation case presented in the {\it Conformal House} section and found that for a two color theory with two fundamental flavors the critical number of adjoint Weyl fermions above which one looses asymptotic freedom is $4.50$. Second, we noted that at the zero of the beta function we have
\begin{eqnarray}
\sum_{i=1}^{k} \frac{2}{11}T(r_i)N_f(r_i)\left( 2+ \gamma_i \right) = C_2(G) \ .
\end{eqnarray}
 Therefore specifying the value of the anomalous dimensions at the infrared fixed point yields the last constraint needed to construct the conformal window. Having reached the zero of the beta function the theory is conformal in the infrared. For a theory to be conformal the dimension of the non-trivial spinless operators must be larger than one in order to not contain negative norm states \cite{Mack:1975je,Flato:1983te,Dobrev:1985qv}.  Since the dimension of the chiral condensate is $3-\gamma_i$ we see that $\gamma_i = 2$, for all representations $r_i$, yields the maximum possible bound
\begin{eqnarray}\label{Bound}
\sum_{i=1}^{k} \frac{8}{11} T(r_i)N_f(r_i) = C_2(G) \ .
\end{eqnarray}
This implies, for example, that for a two technicolor theory with two fundamental Dirac flavors the critical number of adjoint Weyl fermions needed to reach the bound above on the conformal window  is $1.75$. Naively then a two technicolor theory with two fundamental Dirac flavors and one adjoint Weyl fermion would be a good candidate for a NC technicolor theory. However this is hardly the case since the theory equals two flavor supersymmetric QCD but without the scalars. For two colors and two flavors supersymmetric QCD is known to exhibit confinement with chiral symmetry breaking \cite{Intriligator:1995au}. Since the critical number of flavors above which one enters the conformal window is three it will most likely not exhibit NC dynamics. Throwing away the scalars only drives it further away from the NC scenario. The actual size of the conformal window can be smaller than the one determined by the bound above. It may happen, in fact, that chiral symmetry breaking is triggered for a value of the anomalous dimension less than two. If this occurs the conformal window shrinks. Within the ladder approximation \cite{Appelquist:1988yc,{Cohen:1988sq}} one finds that chiral symmetry breaking occurs when the anomalous dimension is close to one. Picking $\gamma_i =1$ we find:
\begin{eqnarray}
\sum_{i=1}^{k} \frac{6}{11} T(r_i)N_f(r_i) = C_2(G) \ .
\end{eqnarray}
In this case when considering a two color theory with two fundamental Dirac flavors the critical number of adjoint Weyl flavors is $2.67$. Hence, our candidate for a NC theory with a minimal $S$ parameter has two colors, two fundamental Dirac flavors charged under the electroweak symmetries and two adjoint Weyl fermions. This is the Ultra Minimal NC Technicolor model (UMT).

If it turns out that the anomalous dimension above which chiral symmetry breaking occurs is larger than one we can still use the model just introduced. We will simply break its conformal dynamics by adding masses (anyway needed for phenomenological reasons) for the adjoint fermions.

The fermions transforming according to the fundamental representation are arranged into electroweak doublets in the standard way and may be written as:

\begin{eqnarray}
T_L = \left( \begin{array}{c} U \\ D  \end{array}  \right)_L \ , \qquad U_R \ , \ \  D_R
\end{eqnarray}

The additional adjoint Weyl fermions needed to render the theory quasi conformal are denoted as $\lambda^f$ with $f=1,2$. They are not charged under the electroweak symmetries. Also we have suppressed technicolor indices. The theory is anomaly free using the following hypercharge assignment
\begin{eqnarray}
Y(T_L) = 0 \ , \qquad Y(U_R) = \frac{1}{2} \ , \qquad Y(D_R) = -\frac{1}{2} \ , \qquad Y(\lambda^f) = 0 \ ,
\end{eqnarray}
Our notation is such that the electric charge is $Q = T_3 + Y$. Replacing the Higgs sector of the SM with the above technicolor theory the Lagrangian reads:
\begin{eqnarray}
\mathcal{L}_H \rightarrow  -\frac{1}{4}{F}_{\mu\nu}^a {F}^{a\mu\nu} + i\overline{T}_L
\gamma^{\mu}D_{\mu}T_L + i\overline{U}_R \gamma^{\mu}D_{\mu}U_R +
i\overline{D}_R \gamma^{\mu}D_{\mu}D_R + i \overline{\lambda} \overline{\sigma}^{\mu} D_{\mu} \lambda \ ,
\end{eqnarray}
with the technicolor field strength ${F}_{\mu\nu}^a =
\partial_{\mu}{A}_{\nu}^a - \partial_{\nu}{A}_{\mu}^a + g_{TC} \epsilon^{abc} {A}_{\mu}^b
{A}_{\nu}^c,\ a,b,c=1,\ldots,3$. The covariant derivatives for the various fermions are
\begin{eqnarray}
D_{\mu} T_L &=& \left( \partial_{\mu} -i g_{TC} {A}_{\mu}^a \frac{\tau^a}{2} -i g W_{\mu}^a \frac{L^a}{2} \right) T_L \ , \\
D_{\mu}U_R &=& \left( \partial_{\mu} - i g_{TC} A_{\mu}^{a} \frac{\tau^a}{2} - i \frac{g'}{2} B_{\mu} \right)U_R \ , \\
 D_{\mu}D_R &=& \left( \partial_{\mu} - i g_{TC} A_{\mu}^{a} \frac{\tau^a}{2} + i \frac{g'}{2} B_{\mu} \right)D_R \\
 D_{\mu}\lambda^{a,f} &=& \left( \delta^{ac} \partial_{\mu} + g_{TC} A_{\mu}^b \epsilon^{abc} \right) \lambda^{c,f} \ ,
\end{eqnarray}

Here $g_{TC}$ is the technicolor gauge coupling, $g$ is the electroweak gauge coupling and $g'$ is the hypercharge gauge coupling. Also $W_{\mu}^a$ are the electroweak gauge bosons while $B_{\mu}$ is the gauge boson associated to the hypercharge. Both $\tau^a$ and $L^a$ are Pauli matrices and they are the generators of the technicolor and weak gauge groups respectively.

The global symmetries of the theory are most appropriately handled by first arranging the fundamental fermions into a quadruplet of $SU(4)$

\begin{eqnarray}
Q &=& \left( \begin{array}{c}
U_L \\
D_L \\
-i\sigma^2U_R^* \\
-i\sigma^2D_R^*
\end{array} \right) \ .
\end{eqnarray}

\begin{figure}
\centerline{\includegraphics[width=8cm]{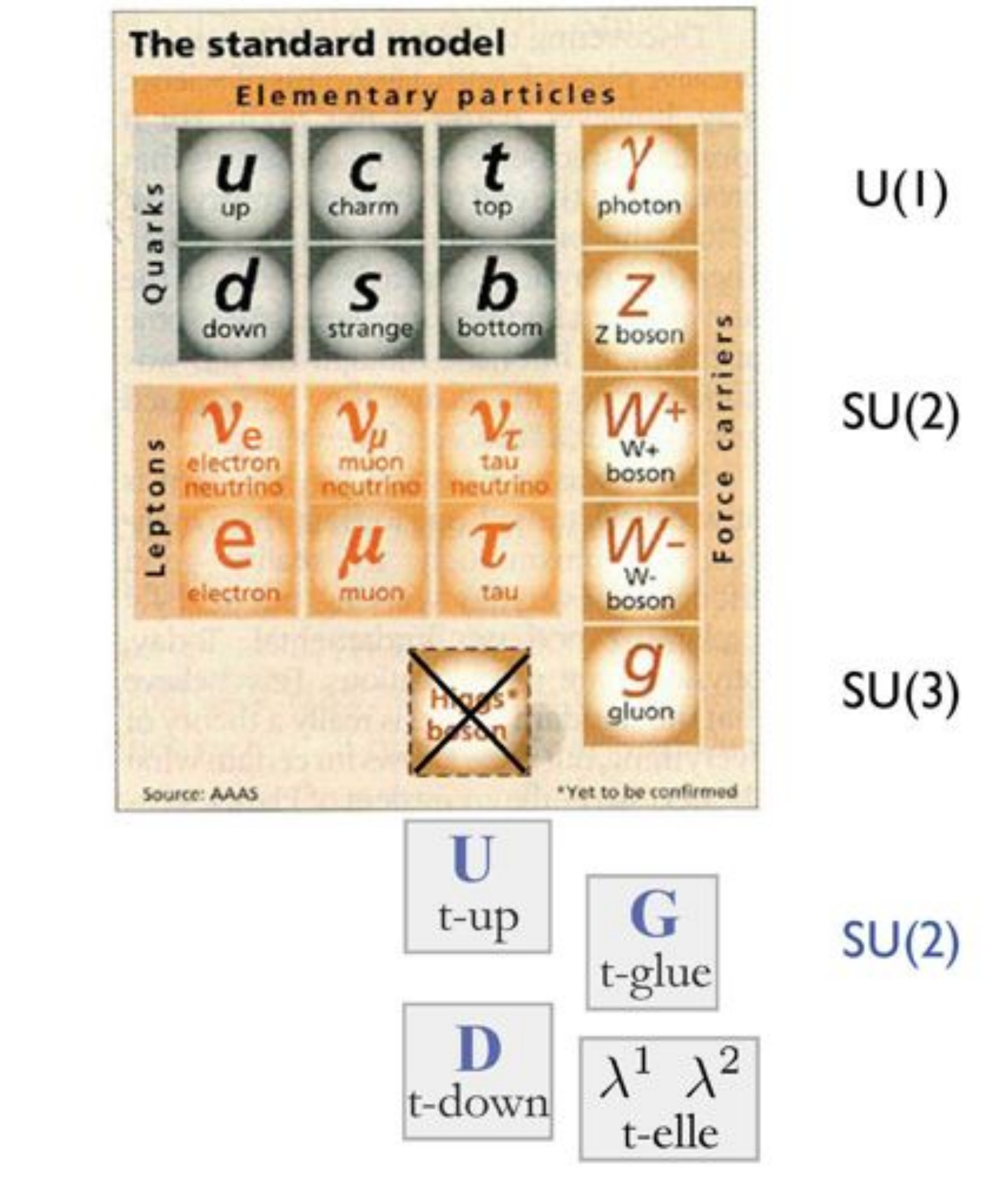}}
\label{UMTposter}
\caption{Cartoon of the Ultra Minimal Walking Technicolor extension of the SM.}
\end{figure}

Since the fermions belong to pseudo-real and real representations of the gauge group the global symmetry of the theory is enhanced and can be summarized as
\begin{eqnarray}
\begin{array}{c||ccc}
 &\qquad SU(4) &\qquad SU(2) &\qquad U(1) \\ \hline\hline
Q &\qquad \fund &\qquad 1 &\qquad -1  \\
\lambda &\qquad 1 &\qquad \fund &\qquad \frac{1}{2}
\end{array}
\end{eqnarray}
The abelian symmetry is anomaly free. Following Ref. \cite{Raby:1979my} the characteristic chiral symmetry breaking scale of the adjoint fermions is larger than that of the fundamental ones since the dimension of the adjoint representation is larger than the dimension of the fundamental representation. We expect, however, the two scales to be very close to each other since the number of fundamental flavors is rather low. In the two-scale technicolor models \cite{Lane:1989ej} the dynamical assumption is instead, that the different scales of the condensates are very much apart from each other.

The global symmetry group $G=SU(4)\times  SU(2) \times U(1)$ breaks to $H= Sp(4) \times SO(2)\times Z_2$. The stability group $H$ is dictated by the (pseudo)reality of the fermion representations and the breaking is triggered by the formation of the following two condensates
\begin{eqnarray}\label{condensate1}
\langle Q_F^{\alpha, c} Q_{F'}^{\beta, c'} \epsilon_{\alpha\beta} \epsilon_{cc'} E_4^{FF'} \rangle &=& -2 \langle \overline{U}_R U_L + \overline{D}_R D_L \rangle \\
\langle \lambda_{f}^{\alpha, k}\lambda_{f'}^{\beta, k'} \epsilon_{\alpha\beta} \delta_{kk'} E_2^{ff'} \rangle &=& -2 \langle \lambda^1 \lambda^2 \rangle \label{condensate2}
\end{eqnarray}
where
\begin{eqnarray}
E_4 = \left( \begin{array}{cc}
0_{2\times2} & \mathbf{1}_{2\times2} \\
-\mathbf{1}_{2\times2} & 0_{2\times2}
\end{array} \right) \ , \qquad E_2 = \left( \begin{array}{cc}
0 & 1 \\
1 & 0
\end{array} \right)
\end{eqnarray}
The flavor indices are denoted with $F,F'=1,\ldots,4$ and $f,f'=1,2$, the spinor indices as $\alpha,\beta=1,2$ and the color indices as $c,c'=1,2$ and $k,k'=1,\ldots,3$. Also the notation is such that $U_L^{\alpha} U_R^{*\beta} \epsilon_{\alpha\beta} = -\overline{U}_RU_L$ and $\lambda^{1,\alpha} \lambda^{2, \beta} \epsilon_{\alpha\beta} = \lambda^1 \lambda^2$. Under the $U(1)$ symmetry $Q$ and $\lambda$ transform as
\begin{equation}
Q \rightarrow e^{-i\alpha}Q \ , \qquad {\rm and } \qquad \lambda \rightarrow e^{-i\frac{\alpha}{2}} \lambda \ ,
\end{equation}
and the two condensates are simultaneously invariant if
\begin{equation}
\alpha = 2 k \pi \ , \qquad {\rm with}~ k ~{\rm an~integer}\ .
\end{equation}
Only the $\lambda$ fields will transform nontrivially under the remaining $Z_2$, i.e.  $\lambda \rightarrow - \lambda$.

The relevant degrees of freedom are efficiently collected in two distinct matrices, $M_4$ and $M_2$, which transform as $M_4 \rightarrow g_4M_4g_4^T$ and $M_2 \rightarrow g_2M_2g_2^T$ with $g_4 \in SU(4)$ and $g_2 \in SU(2)$. Both $M_4$ and $M_2$ consist of a composite iso-scalar and its pseudoscalar partner together with the Goldstone bosons and their scalar partners:
\begin{eqnarray}
M_4 &=& \left[ \frac{\sigma_4 + i \Theta_4}{2} + \sqrt{2}\left( i \Pi_4^i+ \tilde{\Pi}_4^i \right) X_4^i \right] E_4 \ , \qquad i=1,\ldots,5 \ , \\
M_2 &=& \left[ \frac{\sigma_2 + i \Theta_2}{\sqrt{2}} + \sqrt{2} \left( i \Pi_2^i+ \tilde{\Pi}_2^i \right) X_2^i \right] E_2 \ , \qquad i=1,2 \ .
\end{eqnarray}

The notation is such that $X_{4}$ and $X_{2}$ are the broken generators of $SU(4)$ and $SU(2)$ respectively. An explicit realization can be found in the Appendix of \cite{Ryttov:2008xe}. Also $\sigma_{4}$ and $\Theta_{4}$ are the composite Higgs and its pseudoscalar partner while $\Pi_{4}^i$ and $\tilde{\Pi}_{4}^i$ are the Goldstone bosons and their associated scalar partners. {}For SU(2) one simply substitutes the index $4$ with the index $2$.   With the above normalization of the $M$ matrices the kinetic term of each component field is canonically normalized. Under an infinitesimal global symmetry transformation we have:
\begin{eqnarray}
\delta M = i \alpha^a \left( T^a M + M T^{aT} \right) \ .\end{eqnarray}
Here $T$ is the full set of generators of the unbroken group (either SU(4) or SU(2)). With the $\Theta$ and $\tilde{\Pi}^i$ states included the matrices are actually form invariant under $U(4)$ and $U(2)$ with the abelian parts being broken by anomalies. We construct our Lagrangian by considering only the terms preserving the anomaly free U(1) symmetry. As we will see this implies that $\Theta_4$ and $\Theta_2$ are not mass eigensates. In the diagonal basis we will find one massless and one massive state. The massless state corresponds to the $U(1)$ Goldstone boson.

The relation between the composite scalars and the underlying degrees of freedom can be found by first noting that $M_4$ and $M_2$ transform as:
\begin{eqnarray}
M_4^{FF'} \sim Q^FQ^{F'} \ , \qquad M_2^{ff'} \sim \lambda^f \lambda^{f'}
\end{eqnarray}
where both color and spin indices have been contracted. It then follows that the composite states transform as:

\begin{equation}
\begin{array}{rclcrcl}\label{mesons}
\nu_4 +H_4  & \equiv & \sigma_4 \sim \overline{U}U + \overline{D}D & ,~~~~ & \Theta_4 & \sim & i \left( \overline{U}\gamma^5 U + \overline{D} \gamma^5 D \right) \ , \\
\Pi^0 & \equiv & \Pi^3 \sim i \left( \overline{U} \gamma^5 U - \overline{D} \gamma^5 D \right) & ,~~~~ & \tilde{\Pi}^0 & \equiv & \tilde{\Pi}^3  \sim \overline{U}U - \overline{D}D \ , \\
\Pi^+ & \equiv & {\displaystyle \frac{\Pi^1 - i \Pi^2}{\sqrt{2}} \sim i \overline{D} \gamma^5 U} & ,~~~~ & \tilde{\Pi}^+ & \equiv & \frac{\tilde{\Pi}^1 - i \tilde{\Pi}^2}{\sqrt{2}} \sim \overline{D} U\ , \\
\Pi^- & \equiv & \frac{\Pi^1 + i \Pi^2}{\sqrt{2}} \sim i\overline{U} \gamma^5 D & ,~~~~ & \tilde{\Pi}^- & \equiv & \frac{\tilde{\Pi}^1 + i \tilde{\Pi}^2}{\sqrt{2}} \sim \overline{U}D \ , \\
\Pi_{UD} & \equiv & \frac{\Pi^4 + i \Pi^5}{\sqrt{2}} \sim U^T C D & ,~~~~ & \tilde{\Pi}_{UD} & \equiv & \frac{\tilde{\Pi}^4 + i \tilde{\Pi}^5}{\sqrt{2}} \sim iU^T C \gamma^5 D \ , \\
\Pi_{\overline{U}\overline{D}} & \equiv & \frac{\Pi^4 - i \Pi^5}{\sqrt{2}} \sim \overline{U} C \overline{D}^T & ,~~~~ & \tilde{\Pi}_{\overline{U}\overline{D}} & \equiv & \frac{\tilde{\Pi}^4 - i \tilde{\Pi}^5}{\sqrt{2}} \sim i \overline{U} C \gamma^5  \overline{D}^T \ , \\
\end{array}
\end{equation}
and
\begin{equation}
\begin{array}{rclcrcl}
\nu_2 +H_2  & \equiv & \sigma_2 \sim \overline{\lambda}_D \lambda_D & ,~~~~ & \Theta_2 & \sim & i\overline{\lambda}_D \gamma^5 \lambda_D \ , \\
\Pi_{\lambda \lambda} & \equiv & \frac{\Pi^6-i\Pi^7}{\sqrt{2}} \sim \lambda_D^T C \lambda_D & ,~~~~ & \tilde{\Pi}_{\lambda \lambda} & \equiv & \frac{\tilde{\Pi}^6-i\tilde{\Pi}^7}{\sqrt{2}}  \sim i \lambda_D^T C \gamma_5 \lambda_D\ , \\
\Pi_{\overline{\lambda}\overline{\lambda}} & \equiv & \frac{\Pi^6+i\Pi^7}{\sqrt{2}} \sim \overline{\lambda}_D C \overline{\lambda}_D^T & ,~~~~ & \tilde{\Pi}_{\overline{\lambda}\overline{\lambda}} & \equiv & \frac{\tilde{\Pi}^6+i\tilde{\Pi}^7}{\sqrt{2}}  \sim i \overline{\lambda}_D C \gamma_5 \overline{\lambda}_D^T \ , \\
\end{array}
\end{equation}

Here $U=(U_L,U_R)^T$, $D=(D_L,D_R)^T$ and $\lambda_D=(\lambda^1, - i\sigma^2 {\lambda^2}^{\ast})^T$. Another set of states are the composite fermions
\begin{equation}
\Lambda^f = \lambda^{a,f}  \sigma^{\mu\nu} F_{\mu\nu}^a  \ , \qquad \qquad {f=1,2} \ , \qquad {a=1,2,3} \ .
\end{equation}
where $F_{\mu\nu}^a$ is the technicolor field strength.

To describe the interaction with the weak gauge bosons we embed the electroweak gauge group in $SU(4)$ as done in \cite{Appelquist:1999dq}.  First we note that the following generators
\begin{eqnarray}
L^a = \frac{S^a_4+X^a_4}{\sqrt{2}} = \left( \begin{array}{cc}
\frac{\tau^a}{2} &  \\
 & 0
\end{array} \right) \ , \qquad
R^a = \frac{X^{aT}_4-S^{aT}_4}{\sqrt{2}} = \left( \begin{array}{cc}
0 & \\
 & \frac{\tau^a}{2}
\end{array} \right)
\end{eqnarray}
with $a=1,2,3$ span an $SU(2)_L\times SU(2)_R$ subalgebra. By gauging $SU(2)_L$ and the third generator of $SU(2)_R$ we obtain the electroweak gauge group where the hypercharge is $Y=-R^3$. Then as $SU(4)$ breaks to $Sp(4)$ the electroweak gauge group breaks to the electromagnetic one with the electric charge given by $Q=\sqrt{2}S^3$.

Due to the choice of the electroweak embedding the weak interactions explicitly reduce the $SU(4)$ symmetry to $SU(2)_L \times U(1)_Y \times U(1)_{TB}$ which is further broken to $U(1)_{\rm em} \times U(1)_{TB}$ via the technicolor interactions. $U(1)_{TB}$ is the technibaryon number and its generator corresponds to the $S_4^4$ diagonal generator. The remaining $SU(2)\times U(1)$ spontaneously break, only via the (techni)fermion condensates, to $SO(2)\times Z_2$. We prefer to indicate $SO(2)$ with $U(1)_{T\lambda}$. We summarize some of the relevant low-energy technihadronic states according to the final unbroken symmetries in Table \ref{symmetries}.
\begin{table}
\begin{eqnarray}
\begin{array}{c||ccccc}
 &\qquad SU(2)_L &\qquad U(1)_{\rm em} &\qquad U(1)_{ TB} &\qquad U(1)_{T\lambda}& \qquad  Z_2 \\ \hline\hline
H_4,\ \Theta_4 & \qquad  1 &\qquad 0  &\qquad 0 & \qquad 0 & \qquad 0 \\
\stackrel{\rightarrow}{\Pi},\  \stackrel{\rightarrow}{\tilde{\Pi}} &\qquad 3 &\qquad +1,0,-1 &\qquad 0 & \qquad 0 & \qquad 0  \\
\Pi_{UD},\  \tilde{\Pi}_{UD}&\qquad  1 &\qquad 0  &\qquad \frac{1}{\sqrt{2}} & \qquad 0 & \qquad 0 \\
\Pi_{\overline{U}\overline{D}} ,\  \tilde{\Pi}_{\overline{U}\overline{D}} & \qquad  1 &\qquad 0  &\qquad -\frac{1}{\sqrt{2}} & \qquad 0 & \qquad 0 \\
H_2,\ \Theta_2 & \qquad  1 &\qquad 0  &\qquad 0 & \qquad 0 & \qquad 0 \\
\Pi_{\lambda\lambda},\ \tilde{\Pi}_{\lambda\lambda} & \qquad  1 &\qquad 0  &\qquad 0 & \qquad 1 & \qquad 0 \\
\Pi_{\overline{\lambda}\overline{\lambda}},\ \tilde{\Pi}_{\overline{\lambda}\overline{\lambda}} & \qquad  1 &\qquad 0  &\qquad 0 & \qquad -1 & \qquad 0 \\
\Lambda_D&\qquad  1 &\qquad 0  &\qquad 0 & \qquad \frac{1}{2} & \qquad -1
\end{array} \nonumber
\end{eqnarray}
\caption{Summary table of the relevant low-energy technihadronic states for UMT. We display their $SU(2)_L$ weak interaction charges together with their electromagnetic ones. We also show the remaining global symmetries.}
\label{symmetries}
\end{table}
We have arranged the composite fermions into a Dirac fermion
\begin{equation}
\Lambda_D=
\left(
\begin{array}{c}
  \Lambda^1     \\
  -i\sigma^2 {\Lambda^2}^{\ast}
\end{array}
\right) \ .
\end{equation}

Except for the triplet of Goldstone bosons charged under the electroweak symmetry the rest of the states are electroweak neutral.  In the unitary gauge the $\vec{\Pi}$ states become the longitudinal components of the massive electroweak gauge bosons. $\Pi_{UD}$ ($\tilde{\Pi}_{UD}$) is a pseudoscalar(scalar) diquark charged under the technibaryon number $U(1)_{TB}$ while $\Pi_{\lambda\lambda}$ (${\tilde\Pi}_{\lambda\lambda}$) is charged under the $U(1)_{T\lambda}$. $\Lambda_D$ is the composite fermionic state charged under
both $U(1)_{T\lambda}$ and $Z_2$.

The technibaryon number $U(1)_{TB}$  is anomalous due to the presence of the weak interactions:
\begin{equation}
\partial_{\mu} J^{\mu}_{TB} = \frac{1}{2\sqrt{2}} \frac{g^2}{32 \pi^2} \epsilon_{\mu\nu\rho\sigma} W^{\mu \nu} {W}^{\rho\sigma}  \ , \quad {\rm and} \quad
J^{\mu}_{TB} =  \frac{1}{2\sqrt{2}}\left( \bar{U} \gamma^{\mu}U +
 \bar{D} \gamma^{\mu}D \right) \ .
\end{equation}

With the above discussion of the electroweak embedding the covariant derivative for $M_4$ is:
\begin{eqnarray}
D_{\mu} M_4 &=& \partial_{\mu} M_4 - i \left[ G_{\mu} M_4 + M_4 G_{\mu}^T  \right] \ , \qquad G_{\mu} =
\left( \begin{array}{cc}
g W_{\mu}^a \frac{\tau^a}{2} & 0 \\
0 & -g' B_{\mu} \frac{\tau^3}{2}
\end{array} \right)  \ .
\end{eqnarray}

We are now in a position to write down the effective Lagrangian. It contains the kinetic terms and a potential term:
\begin{eqnarray}
\mathcal{L} &=& \frac{1}{2} \text{Tr} \left[ D_{\mu} M_4 D^{\mu} M_4^{\dagger} \right] + \frac{1}{2} \text{Tr} \left[ \partial_{\mu} M_2 \partial^{\mu} M_2^{\dagger} \right] - \mathcal{V} \left( M_4, M_2 \right)
\end{eqnarray}
where the potential is:
\begin{eqnarray}
\mathcal{V} \left( M_4, M_2 \right) &=& -\frac{m_4^2}{2} \text{Tr}\left[ M_4M_4^{\dagger} \right] + \frac{\lambda_4}{4}\text{Tr}\left[ M_4M_4^{\dagger} \right]^2 + \lambda_4' \text{Tr} \left[ M_4M_4^{\dagger}M_4M_4^{\dagger} \right] \nonumber \\
&&
-\frac{m_2^2}{2} \text{Tr}\left[ M_2M_2^{\dagger} \right] + \frac{\lambda_2}{4}\text{Tr}\left[ M_2M_2^{\dagger} \right]^2 + \lambda_2' \text{Tr} \left[ M_2M_2^{\dagger}M_2M_2^{\dagger} \right] \\
&&
+ \frac{\delta}{2}\text{Tr}\left[ M_4M_4^{\dagger} \right] \text{Tr}\left[ M_2M_2^{\dagger} \right] +4\delta' \left[ \left( \det M_2 \right)^2 \text{Pf}\ M_4 + \text{h.c.} \right] \ . \nonumber
\end{eqnarray}

Once $M_4$ develops a vacuum expectation value the electroweak symmetry breaks and three of the eight Goldstone bosons - $\Pi^0,\ \Pi^+$ and $\Pi^-$ - will be eaten by the massive gauge bosons. One can find the actual minimization of the potential and mass spectrum in \cite{Ryttov:2008xe}.

It is also useful to show hot to construct the non-linear effective theory of the associated Goldstone bosons. We shall consider the elements of the global symmetry $G$ as $6\times 6$ matrices.
The generators of $SU(4)$ sit in the upper left corner while the generators of $SU(2)$ sit in the lower right corner. The generator of $U(1)$ is diagonal. We divide the nineteen generators of $G$ into the eleven that leave the vacuum invariant $S$ and the eight that do not $X$. An explicit realization of $S$ and $X$ can be found in the Appendix of Ref. \cite{Ryttov:2008xe}.

An element of the coset space $G/H$ is parameterized by
\begin{eqnarray}
\mathcal{V}(\xi) = \exp \left( i \xi^iX^i \right) E \ ,
\end{eqnarray}
where
\begin{eqnarray}
E= \left( \begin{array}{cc}
E_4 & \\
  & E_2
\end{array}\right) \ , \qquad  \xi^i X^i = \sum_{i=1}^{5} \frac{\Pi^iX^i}{F_{\pi}} + \sum_{i=6}^{7} \frac{\Pi^iX^i}{\tilde{F}_{\pi}} + \frac{\Pi^8 X^8}{\hat{F}_{\pi}} \ .
\end{eqnarray}
The Goldstone bosons are denoted as $\Pi^i,\ i=1,\ldots,8$ and $F_{\pi}, \tilde{F}_{\pi}$ and $\hat{F}_{\pi}$ are the related Goldstone boson decay constants. Since the entire global symmetry $G$ is expected to break approximately at the same scale we also expect the three decay constants to have close values. The element $\mathcal{V}$ of the coset space transforms non-linearly
\begin{eqnarray} \label{trans}
\mathcal{V}(\xi) \rightarrow g \mathcal{V}(\xi) h^{\dagger}(\xi, g)
\end{eqnarray}
where $g$ is an element of $G$ and $h$ is an element of $H$. To describe the Goldstone bosons interaction with the weak gauge bosons we embed the electroweak gauge group in $SU(4)$ as done above and also in \cite{Appelquist:1999dq}.
With the embedding of the electroweak gauge group in hand it is appropriate to introduce the hermitian, algebra valued, Maurer-Cartan one-form
\begin{eqnarray}
\omega_{\mu} = i \mathcal{V}^{\dagger} D_{\mu} \mathcal{V}
\end{eqnarray}
where the electroweak covariant derivative is
\begin{eqnarray}
D_{\mu} \mathcal{V} = \partial_{\mu} \mathcal{V} -iG_{\mu} \mathcal{V} \ , \qquad G_{\mu} = \left( \begin{array}{ccc}
gW_{\mu}^{a} \frac{\tau^a}{2} & & \\
 & -g'B_{\mu}\frac{\tau^3}{2}  &  \\
 & & 0
\end{array} \right) \ .
\end{eqnarray}
From the above transformation properties of $\mathcal{V}$ it is clear that $\omega_{\mu}$ transforms as
\begin{eqnarray}
\omega_{\mu} \rightarrow h(\xi, g) \omega_{\mu} h^{\dagger}(\xi, g) + h(\xi, g) \partial_{\mu} h^{\dagger}(\xi, g) \ .
\end{eqnarray}
With $\omega_{\mu}$ taking values in the algebra of $G$ we can decompose it into a part $\omega_{\mu}^{\parallel}$ parallel to $H$ and a part $\omega_{\mu}^{\perp}$ orthogonal to $H$
\begin{eqnarray}
\omega_{\mu}^{\parallel} = 2S^a \text{Tr}\left[ S^a \omega_{\mu} \right] \ , \qquad \omega_{\mu}^{\perp} = 2X^i \text{Tr} \left[ X^i \omega_{\mu} \right] \ .
\end{eqnarray}
It is clear that $\omega_{\mu}^{\parallel}$ ($\omega_{\mu}^{\perp}$) is an element of the algebra of $H$ ($G/H$) since it is a linear combination of $S^a$ ($X^i$). They have the following transformation properties
\begin{eqnarray}
\omega_{\mu}^{\parallel} \rightarrow h(\xi, g) \omega_{\mu}^{\parallel} h^{\dagger}(\xi, g) + h(\xi, g) \partial_{\mu} h^{\dagger}(\xi, g) \ , \qquad \omega_{\mu}^{\perp} \rightarrow h(\xi, g) \omega_{\mu}^{\perp} h^{\dagger}(\xi, g)
\end{eqnarray}

We are now in a position to construct the non-linear Lagrangian. We shall only consider terms containing at most two derivatives. By noting that the generator $X^8$ corresponding to the broken $U(1)$ is not traceless we can also write a double-trace term besides the standard one-trace term:
\begin{eqnarray}
\mathcal{L} = \text{Tr} \left[a \omega_{\mu}^{\perp} \omega^{\mu\perp}  \right] + b \text{Tr} \left[ \omega_{\mu}^{\perp} \right] \text{Tr} \left[ \omega^{\mu \perp} \right] \ ,
\end{eqnarray}
The coefficients $a=\text{diag}\left(F_{\pi}^2, F_{\pi}^2, F_{\pi}^2, F_{\pi}^2, \tilde{F}_{\pi}^2, \tilde{F}_{\pi}^2 \right)$ and $b=\frac{\hat{F}_{\pi}^2}{2} - \frac{4F_{\pi}^2}{9} - \frac{\tilde{F}_{\pi}^2}{18}$ are chosen such that the kinetic term is canonically normalized:
\begin{eqnarray}
\mathcal{L} = \frac{1}{2} \sum_{i=1}^8 \partial_{\mu} \Pi^i \partial^{\mu} \Pi^i + \ldots \ .
\end{eqnarray}
We conclude this section by connecting the linear and non-linear theories
\begin{eqnarray}
F_{\pi}^2 = \frac{v_4^2}{2} \ , \qquad \tilde{F}_{\pi}^{2} = v_2^2 \ , \qquad \hat{F}_{\pi}^2 = \frac{1}{9} \left( 4v_4^2 + v_2^2 \right) \ .
\end{eqnarray}
For this model much is still left to be done. {}For example one should include the massive vector states and provide a detailed computation of the higher-loop corrections on the effective Lagrangians introduced above similar to what has been done for the case of a single representation \cite{Bijnens:2009qm}. The interest here resides in the fact that the interplay between the two sectors will inevitably lead to interesting effects never explored before.

The UMT provided a {\it novel} type of TIMP, i.e. a  di-techniquark,
with the following unique features:
\begin{itemize}
\item It is a quasi-Goldstone of the underlying gauge
theory receiving a mass term only from interactions not present in the
technicolor theory per se.
\item The lightest technibaryon is a {\it singlet}
with respect to weak interactions.
\item Its relic density can be related to
the SM lepton number over the baryon number if the asymmetry is produced above the
eletroweak phase transition.
\end{itemize}

In appendix  of \cite{Ryttov:2008xe} we provided a much detailed model
computation of the ratio $TB/B$ making use of
the chemical equilibrium conditions and the sphaleron processes active
around the electroweak phase transition.

In the approximation where also the top quark is considered massless around
the electroweak phase transition (we have also checked that the effects of the top mass do not change our results) the $TB/B$ is independent of
the order of the electroweak phase transition and reads
\begin{eqnarray}
- \frac{\sqrt{2}\cdot TB}{B} &=& \frac{\sigma}{2} \left( 3 + \xi \right) \ ,
\end{eqnarray}
where  $\sigma\equiv \sigma_U = \sigma_D$ is the statistical function for the techniquarks.  The $U$ and $D$ constituent-type masses are assumed to be dynamically generated and equal.  $\xi = L/B$ is the SM lepton over the baryon number.

If DM is identified with the lightest technibaryon in our model the
ratio of the dark to baryon mass of the universe is
\begin{equation}
\frac{\Omega_{TB}}{\Omega_B} = \frac{m_{TB}}{m_p}  \frac{\widetilde{TB}}{B} \ ,
\end{equation}
with $m_{TB}$ the technibaryon mass and $\widetilde{TB}=- {\sqrt{2}
TB}$ the technibaryon number normalized in such a way that it is minus one for the lightest state.

 The bulk of the mass of the lightest technibaryon is not due to the technicolor interactions as it was in the original proposal  \cite{Nussinov:1985xr,Barr:1990ca}. This is similar to the
case studied in \cite{Gudnason:2006yj}. The interactions providing mass to the techibaryon are the SM interactions per se and ETC. The main effect of these interactions
will be in the strength and the order of the electroweak phase transition as shown in
\cite{Cline:2008hr}.

What it was found is that, differently from the case in which the technibaryon acquires mass only due to technicolor interactions, one achieves the desired phenomenological ratio of DM to baryon matter with a light technibaryon mass with respect to the weak interaction scale. In fact the mass can be even lower than $100$~{\rm GeV}. This DM candidate can be produced at the Large Hadron Collider experiment. A general investigation of the TIMP detection properties and LHC phenomenology can be found in \cite{Foadi:2008qv}.  

This TIMP is a template for a more general class of models according to which the lightest one is neutral under the SM interactions. Models belonging to this class are, for example, partially gauged technicolor.

\subsection*{A comment}
We have shown above several of the still many available models for dynamical electroweak symmetry breaking. If any of the theories above are actually conformal rather than near conformal then one can couple it to yet another sector which is not conformal and this interaction will lift the conformality in a manner similar to the one we outlined in the {\it Naturalized Unparticle} section. As for the anomalous dimension of the mass of some of these theories, if it turns out not to be very large, we will then require another mechanism to give mass to the SM fermions. Some of these mechanisms exists and have been already hinted to in the $S$-parameter section. 

 The models provide concrete examples of models of dynamical electroweak symmetry breaking with small corrections to the precision observables and having a chance to be observed at the LHC. It would certainly be interesting, in the near future, to extend the number of possible models and investigate their experimental signatures \cite{Belyaev:2008yj,Zerwekh:2005wh,Frandsen:2009fs,Antipin:2009ks,Antipin:2009ch}.

 \newpage
\section{Electroweak Phase Transition for Technicolor}

The experimentally observed baryon asymmetry of the universe may be
generated at the electroweak phase transition (EWPT)
\cite{Shaposhnikov:1986jp,Shaposhnikov:1987tw,Shaposhnikov:1987pf,
Farrar:1993sp,Farrar:1993hn,Gavela:1993ts,Gavela:1994ds}. {}For the
mechanism to be applicable it requires the presence of new physics 
beyond the SM
\cite{Nelson:1991ab,Joyce:1994bi,Joyce:1994fu,Joyce:1994zn,
Joyce:1994zt,Cline:1995dg}. {An essential condition for electroweak
baryogenesis is that the  baryon-violating interactions induced by
electroweak sphalerons are sufficiently slow immediately after the
phase transition to avoid the destruction of the baryons that have
just been created.  This is achieved when the thermal average of the
Higgs field evaluated on the ground state,  in the broken phase of
the electroweak symmetry, is large enough compared to the critical
temperature at the time of the transition (see for example
ref.~\cite{Cline:2006ts} and references therein),
\beq
        \phi_c/ T_c   > 1.
\label{cond} 
\eeq 
In the SM, the bound (\ref{cond}) was believed to be satisfied
only for very light Higgs
bosons \cite{Carrington:1991hz,Arnold:1992fb,Arnold:1992rz,
Anderson:1991zb,Dine:1992wr}.  However, this was before the mass
of the top quark was known.  With $m_t=175$ GeV, nonperturbative studies
of the phase transition \cite{Kajantie:1995kf,Kajantie:1996mn,Rummukainen:1998as} show that
the bound (\ref{cond}) cannot be satisfied for {\it any} value of the
Higgs mass (see also \cite{Gynther:2005av,Gynther:2005dj}).
In addition to the
difficulties with producing a large enough initial baryon asymmetry,
the impossibility of satisfying the sphaleron constraint (\ref{cond})
in the SM provides an incentive for seeing whether the situation
improves in various extensions of the SM \cite{CQW,DR,CM,LR}. 

In \cite{Cline:2008hr,Jarvinen:2009wr,Jarvinen:2009pk}  we  explored  the electroweak phase transition in  models
in which the electroweak symmetry is broken dynamically
\cite{Weinberg:1979bn,Susskind:1978ms}. 

As mentioned earlier, on the astrophysical side, technicolor models are capable of
providing interesting dark matter candidates, since the  new strong 
interactions confine techniquarks in technimeson and technibaryon
bound states.  Technibaryons are therefore natural dark matter candidates 
\cite{Nussinov:1985xr,Barr:1990ca,Gudnason:2006yj}. In fact it is
possible to {naturally} understand the observed ratio of the dark to
luminous matter mass fraction of the universe if the technibaryon
possesses an asymmetry 
\cite{Nussinov:1985xr,Barr:1990ca,Gudnason:2006yj}. If the latter is
due to a net $B-L$ generated at some high energy scale, then this
would be subsequently distributed among {\em all} electroweak
doublets by fermion-number violating processes in the SM at
temperatures above the electroweak scale
\cite{Shaposhnikov:1991cu,Kuzmin:1991ft,Shaposhnikov:1991wi}, thus
naturally generating a technibaryon asymmetry as well.  

The order of the electroweak
phase transition (EWPT) depends on the underlying type of strong dynamics
and  plays an important role for baryogenesis
\cite{Cline:2002aa,Cline:2006ts}. The technicolor chiral phase
transition at finite temperature is mapped onto the electroweak one.
Attention must be paid to the way in which the electroweak symmetry
is embedded into the global symmetries of the underlying technicolor
theory.  An interesting preliminary analysis dedicated to earlier
models of technicolor has been performed in \cite{Kikukawa:2007zk}. 

Here we wish to briefly review recent investigations on the EWPT in a class of realistic
and viable technicolor models. An explicit phenomenological
realization of walking models consistent with the electroweak
precision data  is the MWT model or alike\cite{Foadi:2007ue}. 

 The effective theory
contains composite scalars and spin-one vectors. Compatibility
between the electroweak precision constraints and tree-level
unitarity of $WW$-scattering was demonstrated  in
\cite{Foadi:2008ci}.

 The study of
longitudinal WW scattering unitarity versus precision measurements
within the effective Lagrangian approach demonstrated that it is
possible to pass the precision tests while simultaneously delaying
the onset of unitarity \cite{Foadi:2008ci,Foadi:2008xj}.


The tree-level effective potential is obtained by evaluating the
potential in (\ref{Vdef}) and (\ref{VETCdef}) in the background where
the (composite) Higgs fields assumes the vacuum expectation value $\sigma$, 
{\it i.e.,}
$M = \sigma E/2$. It has the SM form
\begin{eqnarray} \label{Vtree}
 V^{(0)} = \frac{1}{4}\left(\lambda+\lambda'-\lambda''\right) \left(\sigma^2-v^2\right)^2 =\frac{M_H^2}{8v^2} \left(\sigma^2-v^2\right)^2 \ .
\end{eqnarray}
The effective potential at one loop can be naturally divided into
zero- and nonzero-temperature contributions.

We begin by constructing the one-loop effective potential at zero
temperature.  We fix the counterterms so as to  preserve the
tree-level definitions of the VEV and the Higgs mass, {\it i.e.,}
$M^2_H = 2\bar{\lambda} v^2$ with $\bar{\lambda}=\lambda +
\lambda^{\prime} -\lambda^{\prime \prime}$. The one loop contribution
to the potential then reads:
\begin{eqnarray} \label{VT0}
V^{(1)}_{T=0} = \frac{1}{64\pi^2} \sum_{i} n_i\, f_{i}(M_i(\sigma)) +
V_{\rm GB} \ ,
\end{eqnarray}
where the index $i$ runs over all of the mass eigenstates, 
except for the Goldstone bosons (GB), and $n_i$ is the multiplicity 
factor for a given scalar particle while for Dirac fermions is $-4$ 
times the multiplicity factor of the specific fermion.  
The function $f_i$ is:
\begin{equation} \label{fdef}
f_i = M^4_i(\sigma) \left[\log\frac{M^2_{i}(\sigma)}{M^2_i(v)}  - \frac{3}{2}\right] + 2M^2_i(\sigma) \, M^2_i(v)  \ ,
\end{equation}
where $M^2_i(\sigma)$ is the background dependent mass term of the
$i$-th particle. This prescription would lead to  infrared
divergences in the 't Hooft-Landau gauge for $V_{\rm GB}$, the GB
contribution, when evaluated at the tree-level VEV, due to the
vanishing of the GB masses. Different ways of dealing with this
problem have been discussed in the literature. One possibility is to
regularize the infrared divergence by replacing $M^2_i(v)$ with some
characteristic mass scale.   However with this prescription the
tree-level VEV and Higgs mass get shifted by the presence of the
one-loop correction.  A simpler approach is to neglect the GB
contribution, since in practice it never has a strong effect on the
phase transition.  We tried both methods and found that they give
essentially indistinguishable results.

To explicitly evaluate the potential above it is useful to split the
scalar matrix into four $2\times 2$ blocks as follows:
\begin{equation}  
M=\begin{pmatrix} {\cal X}  & {\cal O}  \\ {\cal O}^T & {\cal Z} \end{pmatrix} \ ,
\end{equation}
with ${\cal X}$ and ${\cal Z}$ two complex symmetric matrices
accounting for six independent degrees of freedom each and ${\cal O}$
a generic complex $2\times 2$ matrix representing eight real bosonic
fields. ${\cal O}$ accounts for the SM-like Higgs doublet, a second
doublet, and the three GB's absorbed by the
longitudinal  gauge bosons. We find $n_{\cal X} = n_{\cal Z} = 6$
while the two weak doublets split into two SU(2)$_V$ isoscalars, {\it i.e.,}
the Higgs ($n_H=1$) and ${\Theta}$ ($n_{\Theta}=1$) with different
masses and two independent triplets, {\it i.e.,} $n_{GB}=3$ and $n_{A}=3$.

For the contribution of the gauge bosons we have $n_W=6$ and $n_Z=3$.
In the fermionic sector we will consider only the heaviest
particles, {\it i.e.,} the top for which $n_T = -12$ and the two new
leptons $n_N = n_E = -4$.  

The one-loop, ring-improved, finite-temperature effective potential 
can be divided into fermionic, scalar and vector contributions,
\begin{eqnarray}
V_T^{(1)} = {V_T^{(1)}}_{\rm f}+ {V_T^{(1)}}_{\rm b}+ {V_T^{(1)}}_{\rm gauge}\ .
\end{eqnarray}
The fermionic contribution at high temperature reads:
\begin{eqnarray} \label{VTf}
{V_T^{(1)}}_{\rm f} = 2\frac{T^2}{24} \sum_{f} n_{f} M_{f}^2(\sigma) +\frac{1}{16 \pi^2}\sum_fn_{f} M_{f}^4(\sigma)\left[\log\frac{M_{f}^2(\sigma)}{T^2}-c_f\right]
\end{eqnarray}
where $c_f \simeq 2.63505$, $n_{\rm Top}=3$, $n_{N}=n_{E}=1$, and we 
have neglected ${\cal O}\left(1/T^2\right)$ terms.
The field-dependent masses are
\begin{equation}
M_{\rm Top} (\sigma) = m_{\rm Top}\frac{\sigma}{v} \ , \quad  M_{\rm N}(\sigma) = m_{ N}\frac{\sigma}{v} \ , \quad M_{\rm E} = m_{E}\frac{\sigma}{v} \ ,
\end{equation}
with $m_{\rm Top}$, $m_{N}$ and $m_{E}$ the physical masses. 
Notice that the logarithmic term in (\ref{VTf}) combines with a 
similar term in the zero-temperature potential (\ref{VT0}) so that their sum is analytic in the masses $M_{f}^2(\sigma)$.
 
For the scalar part of the thermal potential one must resum the
contribution of the ring diagrams. Following Arnold and Espinosa
\cite{Arnold:1992rz} we write
\begin{eqnarray} \label{VTb}
{V_T^{(1)}}_{\rm b} =\frac{T^2}{24} \sum_{b} n_{b} M_{b}^2(\sigma)  - \frac{T}{12\pi} \sum_b\,n_b \,M_{b}^3(\sigma,T) \nonumber\\ 
-\frac{1}{64 \pi^2}\sum_bn_{b} M_{b}^4(\sigma)\left[\log\frac{M_{b}^2(\sigma)}{T^2}-c_b\right]
\ ,
\end{eqnarray}
where $c_b\simeq 5.40762 $ and $M_b(\sigma,T)$ the thermal mass
which follows from the 
tree-level plus one-loop thermal contribution to the potential. {}For the gauge bosons, 
\begin{eqnarray} \label{VTgb}
{V_T^{(1)}}_{\rm gauge} =\frac{T^2}{24} \sum_{gb} 3 M_{gb}^2(\sigma)  - \frac{T}{12\pi} \sum_{gb}\left[ 2 M_{T,gb}^3(\sigma)+ M_{L,gb}^3(\sigma,T)\right] \nonumber \\
-\frac{1}{64 \pi^2}\sum_{gb}n_{gb} M_{gb}^4(\sigma)\left[\log\frac{M_{gb}^2(\sigma)}{T^2}-c_b\right] \ .
\end{eqnarray}
Here $M_{T,gb}$ ($M_{L,gb}$ ) is the transverse (longitudinal) mass
of a given gauge boson and we have $M_{T,gb}(\sigma)=
M_{L,gb}(\sigma,T=0) = M_{gb}(\sigma)$. Only the longitudinal gauge
bosons acquire a thermal mass squared at the leading order, 
$O(g^2T^2)$. The transverse bosons acquire instead a 
magnetic mass squared  of order $g^4T^2$ which we have neglected. 

The explicit form of the transverse and longitudinal gauge boson 
mass matrix is given in the Appendix of \cite{Cline:2008hr}.

We used the one-loop high temperature approximation together with the
summation of the ring-diagrams  to evaluate the effective potential
in our numerical calculations. The full expression of the finite
temperature potential is given as a sum of the tree level potential
(\ref{Vtree}), the zero-temperature one-loop contribution
(\ref{VT0}), and the one-loop thermal corrections at high
temperature, (\ref{VTf}), (\ref{VTb}), and (\ref{VTgb}). We assumed
that the phase transition takes place when the two minima are
degenerate. This then defines the critical value of the thermal
average of the composite Higgs field $\phi_c$, in the broken phase,
at the critical temperature $T_c$. Above the critical temperature the
ground state is the one at the origin of the Higgs field. {}For
convenience we subtracted from the potential a temperature-dependent
constant which is defined in such a way that $V(\sigma,T)=0$ for
$\sigma=0$.

The relevant input parameters are the zero-temperature masses of the
Higgs ($M_H$) and its pseudoscalar partner $\Theta$ ($M_\Theta$). The
phase transition also depends on the masses of the scalar partners of
the Goldstone bosons $A^{0,\pm}$ ($M_A$), on the mass scale of the
scalar baryons $m_{\rm ETC}$, and on the masses of the heavy
fermions. For simplicity, we chose the masses of the new fermions to
be equal, \begin{eqnarray} M_{E}^2 = M_{N}^2 \equiv M_{\rm f}^2 \ .
\end{eqnarray} This choice does not seem to have a strong effect on
the phase transition; for example we checked that using instead $M_E
\simeq 2 M_N$, very similar results were obtained. We have neglected
the heavy composite vectors of MWT since they are expected to
decouple at the scale of the EWPT.   At this scale, the couplings to
the SM gauge bosons are simply $g$, $g'$. We set the parameter $y$ to
$y=1/3$ so that the MWT hypercharge assignment equals the SM one.
Notice that $y$ appears only in the longitudinal Debye mass of the
$Z$ boson. Since the effective potential terms are proportional to
$M_i^{2}(\sigma)$ or $M_i^{4}(\sigma)$, the contributions of the
fermions and the composite scalars typically dominate over that of
the relatively light $Z$ boson, whence the dependence of the phase
transition on $y$ is negligible.

The details of the investigation and results can be found in \cite{Cline:2008hr} here we briefly summarize the general results. We find that in the parameter region where a strong first order transition is
observed the composite Higgs and its pseudoscalar partner $\Theta$
are light enough to be produced at the LHC. Moreover one expects, for
this range of parameters of the effective Lagrangian, sizable
deviations from the SM predictions at the LHC \cite{Belyaev:2008yj}.  We emphasize that the spectrum is completely
fixed by the underlying gauge theory and that first principle lattice
simulations can test our results.

\subsection{Electroweak Deconfinig and Chiral Phase Transitions}
\label{sect4.d}
As suggested in \cite{Sannino:2008ha}, an intriguing possibility can
 emerge in that one can have {\em two independent} phase transitions
 at nonzero temperature in technicolor theories, whenever the theory
 possesses a nontrivial center symmetry.  The two phase transitions
 are the chiral one, directly related to the electroweak phase
 transition, and a confining one at lower temperatures.  During the
 history of the universe one predicts a phase transition around the
 electroweak scale and another one at lower temperatures with a jump
 in the entropy proportional to the number of degrees of freedom
 liberated (or gapped) when increasing (decreasing) the temperature
 (see \cite{Mocsy:2003qw} for a simple explanation of this phenomenon
 and a list of relevant references). This may have very interesting
 cosmological consequences. Here we have concentrated on the
 chiral one alone. The interplay with the confining one, expected to
 occur at lower temperatures, can be studied by coupling the
 effective Lagrangian presented here to the Polyakov-loop effective
 degree of freedom as done in \cite{Mocsy:2003qw}.  

\subsection{Extra Electroweak Chiral Phase Transitions}

In \cite{Jarvinen:2009wr} were presented generic models of DEWB possessing a surprisingly rich finite temperature phase diagram structure. The basic ingredients are: i) At least two different composite Higgs sectors; ii) One charged under the EW symmetry; iii) An underlying strong dynamics mixing the two sectors. An explicit realization  is UMT \cite{Ryttov:2008xe}. 

Consider an asymptotically free gauge theory having sufficient matter to posses, at least, two independent non-abelian global symmetries spontaneously breaking, in the infrared, to two subgroups. One of the initial symmetries (or both) must contain the EW one in order to drive EW symmetry breaking. The Goldstones which are not eaten by the longitudinal components of the weak gauge bosons receive masses from other, unspecified, sectors.  The analysis done in \cite{Jarvinen:2009wr} is sufficiently general that one needs not to specify such sectors.

Let us denote with $\rm I$ and ${\rm I\,I}$ the two non-abelian global symmetries. They are broken at low temperatures and restored at very high temperatures. The restoration of each symmetry will typically happen at two different critical temperatures. We indicate with $\langle H_{\rm I}\rangle$ and  $\langle H_{\rm I\,I}\rangle$ the thermal average of the two condensates. The zero temperature physical masses $M_{{\rm I}}$ and $M_{{\rm I\,I}}$ of the two composite Higgses together with $\beta$ (measuring the mixing between the two), as well as the collection of all the other couplings mixing the two sectors constitute the parameters allowing us to make a qualitative picture of the complex phase structure. 

In figure ~\ref{one} we present three possible versions of the two-dimensional phase diagram as function of the temperature as well as one of the zero-temperature masses of one of the Higgses (holding fixed the other). The three plots are meant for three different strengths $\beta$ of the mixing while keeping the other relevant parameters fixed.  Four distinct regions are classified via the broken versus unbroken number of global symmetries. To simplify the discussion we are taking $\beta$ to be the parameter controlling the mixing between the two sectors. In fact, one should use the entire ensemble of parameters whose associated operators mix the different sectors. 

Let us describe the situation before embedding the EW symmetry within any of the two non-abelian global symmetries. We envision the following possibilities: i) The two sectors do not talk to each other ($\beta=0$). In this case the two PTs happen at different temperatures and do not interfere (left panel). ii) The two sectors do feel each other when $\beta\neq 0$. Possible phase diagrams are depicted in the central and right panel of Fig.~\ref{one}. 
\begin{figure}[htp!]
\centerline{\includegraphics[height=4.5cm,width=4.5cm]{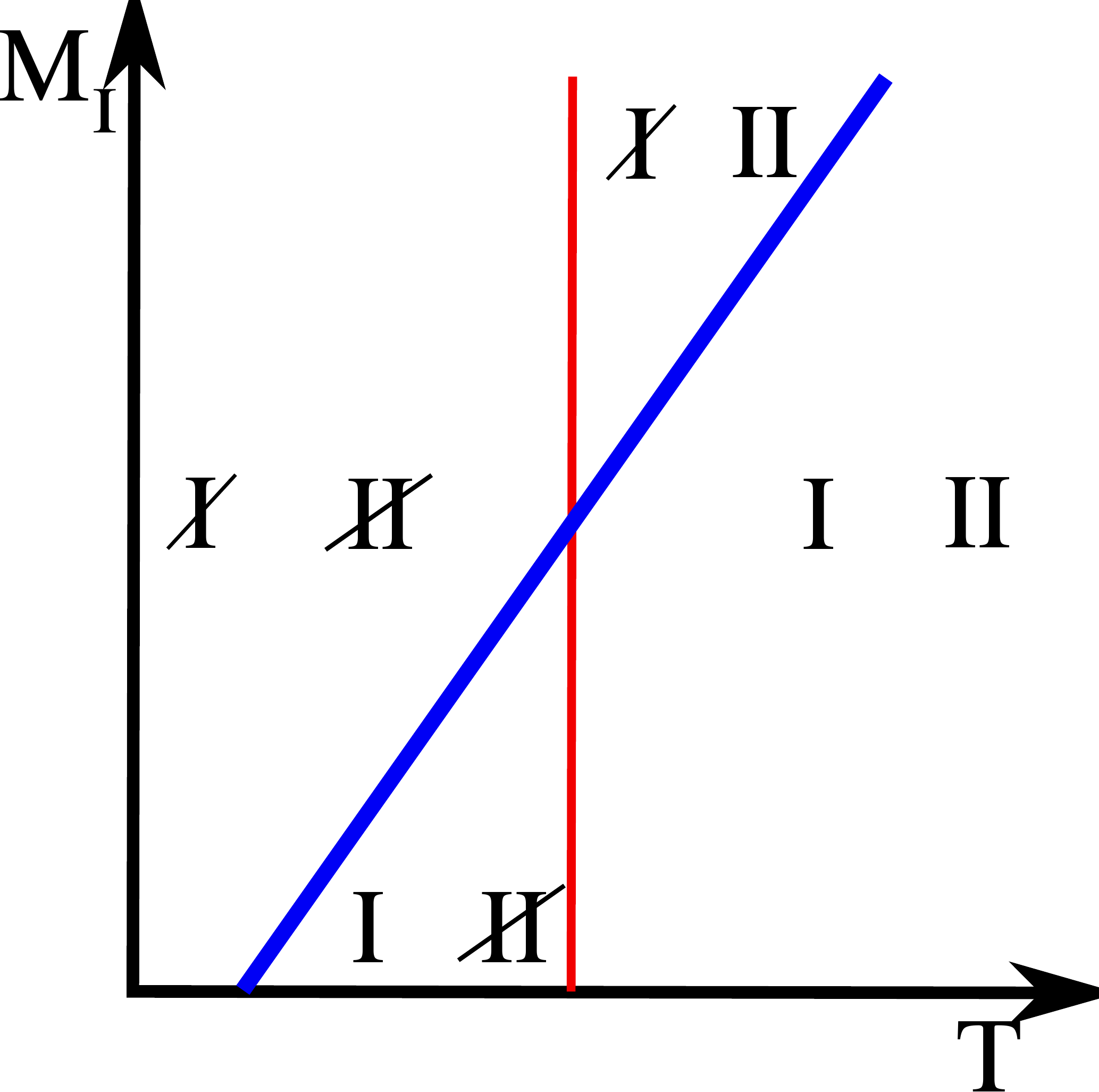} \hskip 1.3cm\includegraphics[height=4.5cm,width=4.5cm]{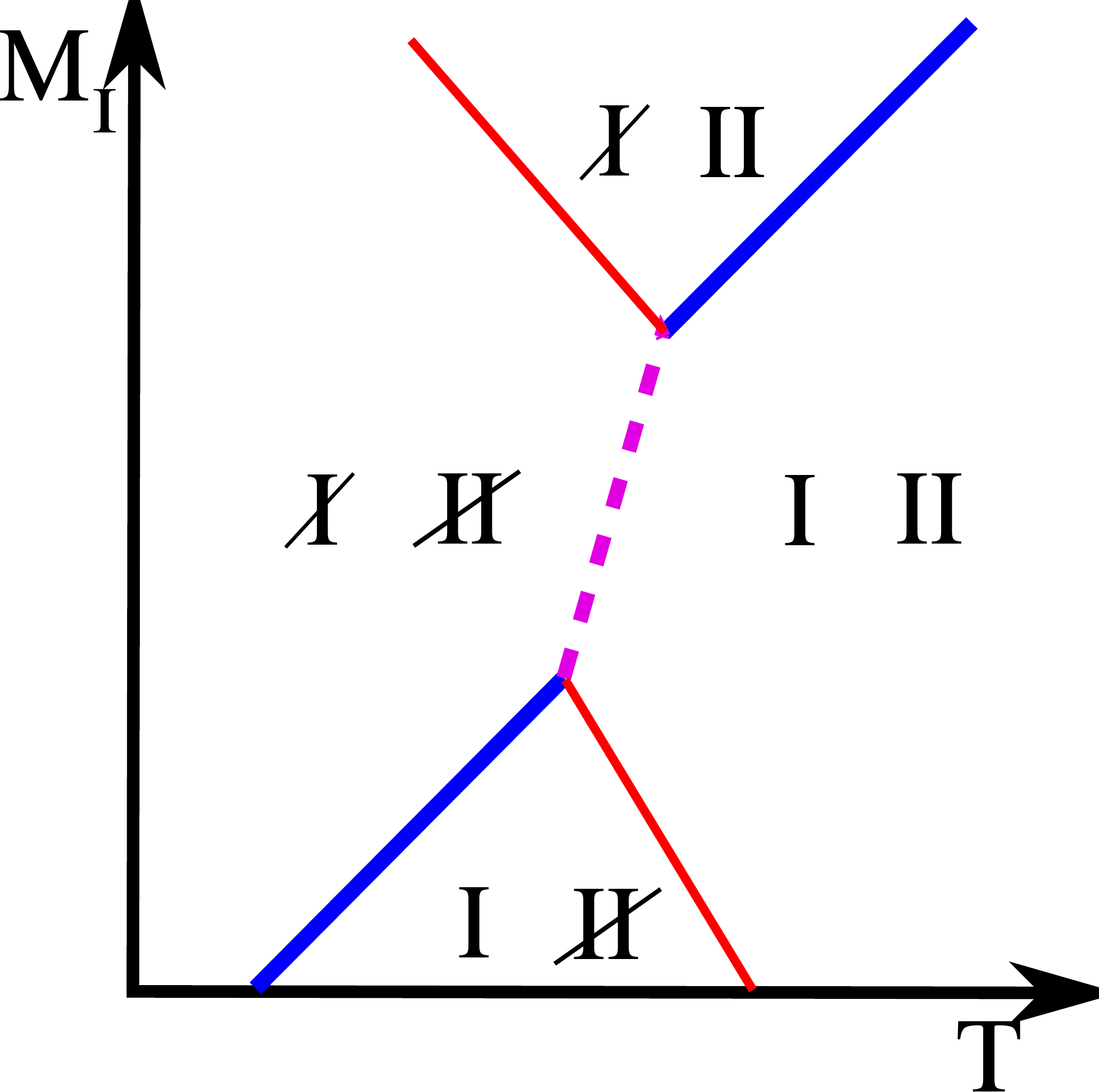}\hskip 1.3cm\includegraphics[height=4.5cm,width=4.5cm]{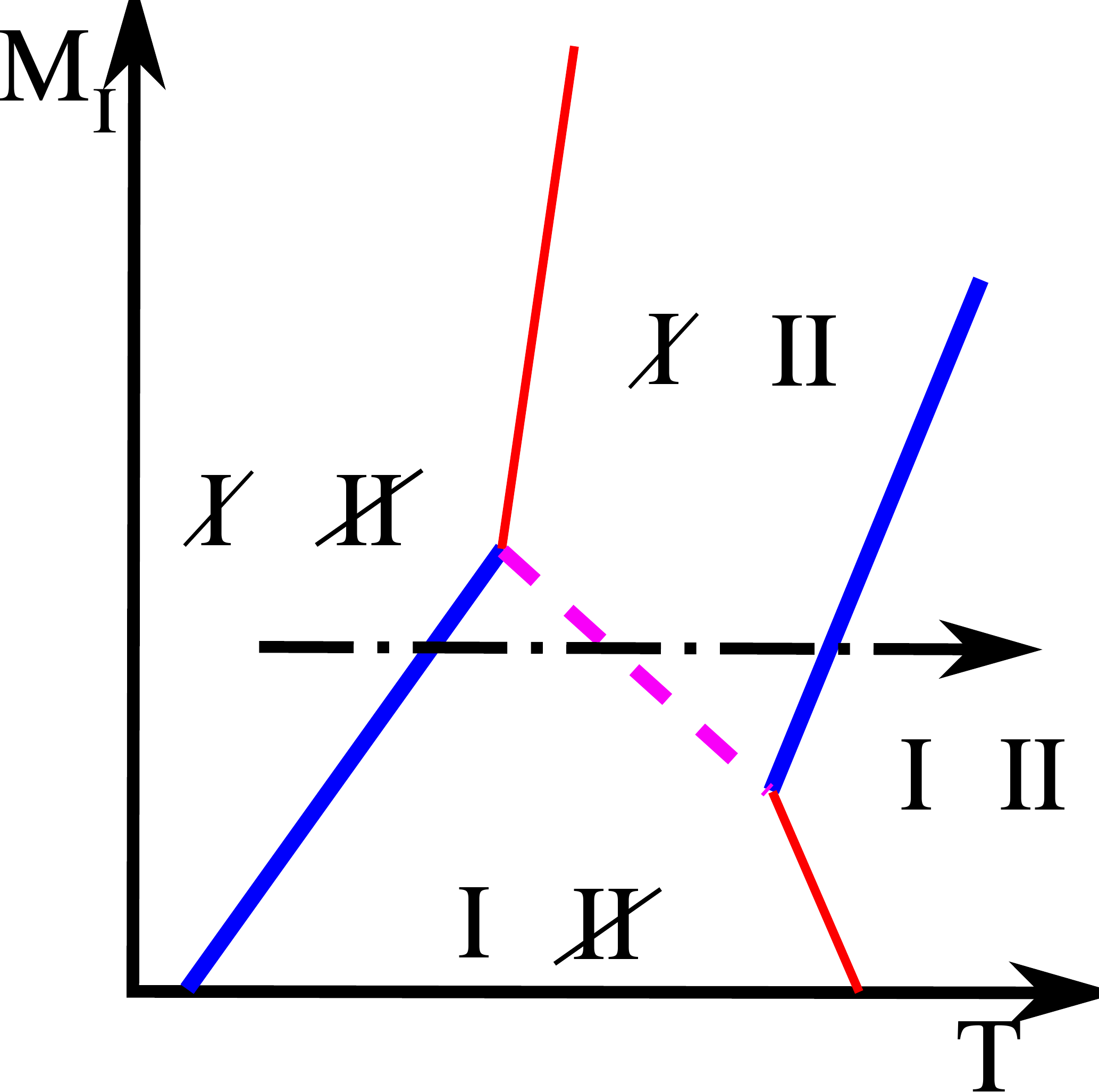}}
\caption{Possible Phase Diagrams: {\bf Left Panel:} The two transitions do not feel each other ($\beta=0$). {\bf Central and Right Panels:} The two transitions do interfere with each other  ($\beta \neq 0$).  }\label{one}
\end{figure}
In a generic strongly coupled theory the two global symmetries are bound to talk to each other and hence the second possibility is the one expected. A new line can develop (the dashed one depicted in the central and right panel) entirely due to the interactions between the two sectors. This line allows several new possible PTs. {}For example, according to the central phase diagram the transition between two broken to two unbroken phases can occur at the same critical temperature along the dashed line in the $M_{\rm I}-T$ plane. What strikes us as a very intriguing possibility is the pattern of PTs one can encounter following the right panel phase diagram. Along the horizontal dashed arrow line we have three subsequent PTs constituted by the first condensate being melted twice and re-generated once while the second one melts only once. We could also plot a diagram similar to the one in the right panel but with the first vertex lower than the second one (with respect to $M_{\rm I}$).  In fact, more sophisticated PTs can occur.

Let us turn on the EW fields by gauging the relevant symmetries within, for definitiveness, the first sector. There is no elementary SM Higgs but we require the new strong dynamics to drive EW symmetry breaking.  The units of the dynamically generated scale of the new strong dynamics are now fixed by the mass of the weak gauge bosons. Would the {\it extra} transition associated to the right-panel diagram survive? What are the main effects of the SM on the phase diagram?  It would be very interesting if a complex PT structure appears in technicolor-like extensions of the SM when the universe reaches temperatures near the EW scale. Similar possibilities have been investigated earlier in the case of the two Higgs doublet model \cite{Land:1992sm}. The EW fields will impinge on the PTs and the relevant degrees of freedom are the weak gauge bosons and the SM fermions. The gauge fields couple via covariant derivatives while the fermions communicate by means of effective Yukawa-type interactions  as proposed in Minimal Walking Technicolor (MWT) \cite{Sannino:2004qp,Foadi:2007ue}. In fact, the top quark has the major impact on the phase diagram due to its very large Yukawa coupling.

Interestingly these generic feature are seen to appear when  tested again specific TC models of the type envisioned above \cite{Jarvinen:2009pk}. 

Future analysis include the possible experimental observation of the new type of electroweak phase transitions proposed above via, for example, gravitational waves experiments as well as first principle lattice simulations. 

\section{Conclusion}

We  introduced  different topics related to the dynamical breaking of the electroweak symmetry. We have also reported on the status of the phase diagram for  generic nonsupersymmetric gauge theories with fermionic matter obtained via analytic methods. We have also provided a novel analysis of chiral gauge theories relevant for particle physics phenomenology making use of a newly introduced chiral beta function.  As a relevant example for breaking the electroweak symmetry dynamically we  introduced  different types of minimal conformal models of dynamical electroweak symmetry breaking. We have also summarized a simple model of unparticle physics. We discussed possible astroparticle physical applications and discussed the electroweak phase transition in technicolor theories. A number of appendices have been added to support the reader with group theoretical results, notation and some needed technical information.

 \subsection*{Acknowledgments}
 I thank the organizers of the 49$^{th}$ Cracow School of Theoretical Physics, the Yukawa Institute for Theoretical Physics in Kyoto and the XXX Elementary Particles and Fields Meeting in Brasil for providing a very nice scientific environment. I am deeply indebted to M. Antola, S. Catterall, L. Del Debbio,  S. Di Chiara, D. D. Dietrich, R. Foadi, J.~Giedt, M.~Heikinheimo, M. T. Frandsen, C. Kouvaris, M. J\"{a}rvinen, I. Masina, T.~A.  Ryttov, J. Schechter, K. Tuominen and R. Zwicky, for pleasant fruitful collaborations on the various topics presented in this review, comments and/or careful reading of the manuscript. 

\newpage

\appendix 
\section{Exploring Unification within Technicolor}
 
 Although it is not a fundamental prerequisite of any specific extension of the SM, it is a fact that unification of the SM couplings is an attractive feature. It is, hence, instructive to investigate what happens to the SM couplings when the Higgs sector is replaced with a new strongly coupled theory a la' technicolor \cite{Gudnason:2006mk}.  

 We start by investigating the  one-loop evolution of the SM couplings
 once the SM Higgs is replaced by the MWT model.   
The evolution of the coupling constant $\alpha_n$, at the
one-loop level, of a gauge theory
is controlled by
\begin{equation}\label{running}
{\alpha_{n}^{-1}(\mu) = \alpha_{n}^{-1}(M_Z) - \frac{b_n}{2\pi}\ln
\left(\frac{\mu}{M_Z}\right) \ ,}
\end{equation}
where $n$ refers to the gauge group being $SU(n),$ for $n\geq 2$ or $U(1),$
 for $n=1\ $.

The first coefficient of the beta function $b_n$ is
\begin{eqnarray}
b_n =  \frac{2}{3}T(r) N_{wf} + \frac{1}{3} T(r')N_{cb} -
\frac{11}{3}C_2(G) \ ,
\end{eqnarray}
where $T(r)$ is the Casimir of the
representation $r$ to which the
fermions belong, $T(r')$ is the
Casimir of the representation $r'$
to which the bosons belong. $N_{wf}$ and $N_{cb}$ are respectively the
number of Weyl fermions and the number of complex scalar bosons. $C_2(G)$ is the
quadratic Casimir of the adjoint representation of the gauge group.

The SM gauge group is $SU(3)\times SU(2)\times U(1)$. We have three
associated coupling constants which one can imagine to unify at some
very high energy scale $M_{GUT}$. This means that the three
couplings are all equal at the scale $M_{GUT}$, i.e.
$\alpha_3(M_{GUT})= \alpha_2(M_{GUT})=\alpha_1(M_{GUT})$ with
$\alpha_1= \alpha/(c^2\cos^2 \theta_w)$ and $\alpha_2 = \alpha/
\sin^2\theta_w$, where $c$ is a normalization constant to be
determined shortly.

Assuming one-loop unification using
Eq.~(\ref{running}) for $n=1,2,3,$ one
finds the following relation
\begin{eqnarray} \label{unification}
\frac{b_3-b_2}{b_2-b_1} & = & \frac{\alpha_3^{-1} -\alpha^{-1}\sin^2
\theta_w}{(1+c^2)\alpha^{-1} \label{1}
\sin^2\theta_w-c^2\alpha^{-1}} \ .
\end{eqnarray}
In the above expressions the
Weinberg angle
$\theta_w$, the electromagnetic coupling constant $\alpha$ and the
strong coupling constant $\alpha_3$ are all evaluated at the
$Z$ mass. For a
given particle content we shall denote the LHS of
Eq.~(\ref{unification}) by $B_{\rm theory}$ and the RHS by
$B_{\rm
  exp}$. Whether $B_{\rm theory}$ and $B_{\rm exp}$ agree is a simple way to
check if
the coupling constants unify. We shall use the experimental
values $\sin^2 \theta_w (M_Z) = 0.23150\pm 0.00016$,
$\alpha^{-1}(M_Z) = 128.936 \pm 0.0049$, $\alpha_3(M_Z) =
0.119\pm 0.003$ and $M_Z = 91.1876(21)$ GeV \cite{Yao:2006px}.
The unification scale is given by the expression
\begin{eqnarray}
{M_{GUT} = M_Z
\exp
\left[{{2\pi}\frac{\alpha_2^{-1}(M_Z)-\alpha_1^{-1}(M_Z)}{b_2-b_1}}\right]
\ . }
\end{eqnarray}

While the normalizations of the
coupling constants of the two
non-Abelian gauge groups are fixed by
the appropriately normalized
generators of the gauge groups, the
normalization of the Abelian
coupling constant is a priori arbitrary. The normalization of the
Abelian coupling constant can be
fixed by a rescaling of the
hypercharge $Y \rightarrow cY$ along with
$g\to g/c\ $. The normalization
constant $c$ is
chosen by imposing that all three coupling constants have a common
normalization
\begin{eqnarray}
\text{Tr}\,(c^2Y^2) = \text{Tr}\,(T_3^2) \ ,
\end{eqnarray}
where $T_3$ is the generator of the weak
  isospin and
the trace is over all the relevant fermionic particles on which the
generators act. It is sufficient to fix it for a given fermion
  generation (in a complete multiplet of the unification group).

The previous normalization is consistent with
an $SU(5)$-type normalization for the generators of $U(1)$
of hypercharge,
$SU(2)_L$ and $SU(3)_c\ $.

As well explained in the paper by Li
and Wu \cite{Li:2003zh}:
{\it  At one-loop a contribution to $b_3 - b_2$ or $b_2 - b_1$ emerges
  only from particles not forming complete representations
  \footnote{Such as the five and the ten dimensional representation of
    the unifying gauge group $SU(5)$.} of the unified gauge
  group}. {}For example the gluons, the weak gauge bosons and the
Higgs particle of the SM do not form complete
representations of $SU(5)$ but ordinary quarks and leptons do. Here we
mean that these particles form complete representations of $SU(5)$, all
the way from the unification scale
    down to the electroweak scale.
The particles not forming complete representations will presumably
join at the unification scale with new particles and together then
form complete representations of the unified gauge group. Note, that
although there is no contribution to the unification point of the
particles forming complete
representations, the running of each
coupling constant is affected by all of the particles present at low
energy.

\subsection{Dis-unification in SM}
As a warm up, we consider the SM with
$N_g$ generations. In
this case we find $c=\sqrt{3/5}$, which is the same value one finds
when the hypercharge is upgraded to one of the generators of $SU(5)$,
and therefore the beta function
coefficients are

\begin{eqnarray}
b_3 & = & \frac{4}{3}N_g -11 \ ,\\
b_2 & = & {\frac{4}{3} N_g - \frac{22}{3} +
\underbrace{\frac{1}{6}}_{\rm Higgs} \ , } \\
b_1 & = & {\frac{3}{5}\left( \frac{20}{9}N_g +\frac{1}{6} \right) =
\frac{4}{3}N_g +\underbrace{\frac{1}{10}}_{\rm Higgs} \ .}
\end{eqnarray}

Here $N_g$ is the number of
generations. It is clear that the
SM does not unify since $B_{\rm theory} \sim 0.53$ while
$B_{\rm exp} \sim 0.72\ $.

Note that the spectrum relevant for computing
$B_{\rm theory}$  is constituted by the gauge bosons and the standard
model
Higgs. The contribution of quarks and leptons drops out in agreement
with the fact that they form complete representations of the unifying
gauge group which, given the present normalization for $c$, is at
least $SU(5)$. Hence the predicted value of $B_{\rm theory}$ is
independent of the number of
generations. However the overall running
for the three couplings is dependent on the number of
  generations and in
Fig.~\ref{SM} we show the behavior of the three couplings with $N_g
= 3$.

\begin{figure}[htbp]
\begin{center}
\includegraphics[width=0.6\linewidth]{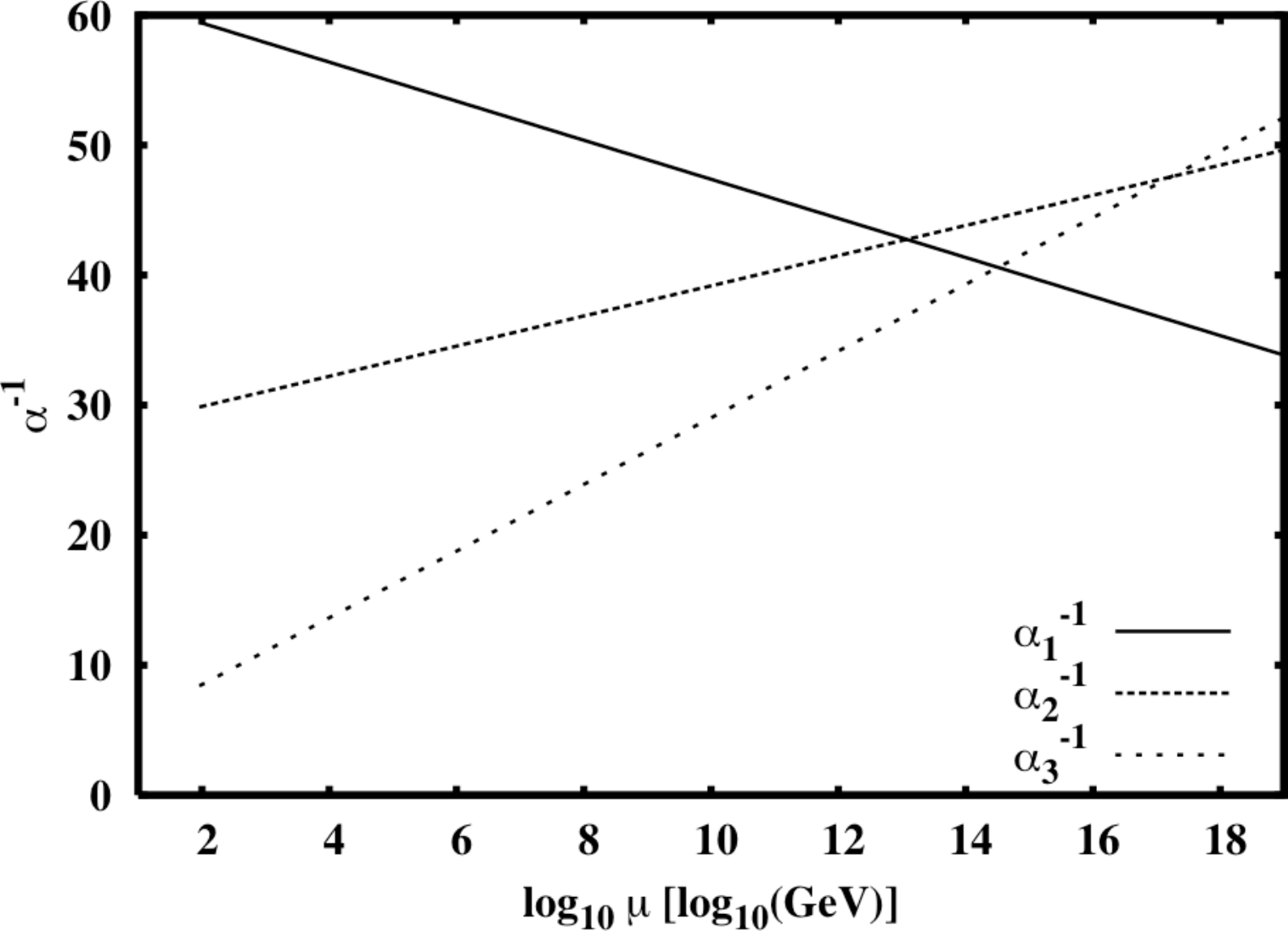}
\end{center}
\caption{The running of the three SM gauge couplings.}
\label{SM}
\end{figure}

\subsection{Studying SU(3)$\times$SU(2)$\times$U(1) Unification in  MWT  }
Here we compare a few examples in which the SM Higgs is
replaced by a technicolor-like theory. A similar analysis was performed in \cite{Christensen:2005bt}. In this section we press on phenomenological successful technicolor models with technimatter in
higher dimensional representations and demonstrate that the simplest
model helps unifying the SM couplings while other more
traditional approaches are less successful. We also show that by a
small modification of the technicolor
dynamics, all of the four
couplings can unify \footnote{Since the technicolor dynamics is
  strongly coupled at the electroweak scale the last point on the
  unification of all of the couplings is meant to be only
  illustrative.}.

We examine what happens to the running of the SM couplings when the Higgs sector is replaced by the
MWT theory introduced earlier. This model has
technicolor group $SU(2)$ with two techniflavors in the two-index
symmetric representation of the technicolor group. As already
mentioned to avoid Witten's $SU(2)$
anomaly, the minimal solution is to
add a
new lepton family. We still assume an $SU(5)$-type
unification leading to $c^2=3/5$.
The beta function
coefficients will be those of
  the SM minus the Higgs plus the extra
contributions from the techniparticles, ergo
\begin{eqnarray}
b_3 & = & \frac{4}{3}N_g -11 \ ,\\
b_2 & = & \frac{4}{3} N_g - \frac{22}{3} + \frac{2}{3}
\frac{1}{2}\left( \frac{2(2+1)}{2} + 1 \right) = \frac{4}{3} \left(
N_g + 1 \right) - \frac{22}{3} \ ,\\
b_1 & = & \frac{3}{5} \left( \frac{20}{9} N_g + \frac{20}{9} \right)
= \frac{4}{3}\left( N_g+1 \right) \ ,
\end{eqnarray}
where $N_g$ is the number of ordinary SM generations. From this we
see that $B_{\rm theory}=0.68$ and $B_{\rm exp} = 0.72\ $ and hence argue that we have a better
unification than in the SM
with an elementary Higgs. The running of the SM couplings is shown in
Fig.~\ref{TC} for three ordinary SM generations.
\begin{figure}[!tbp]
\begin{center}
\includegraphics[scale=0.6]{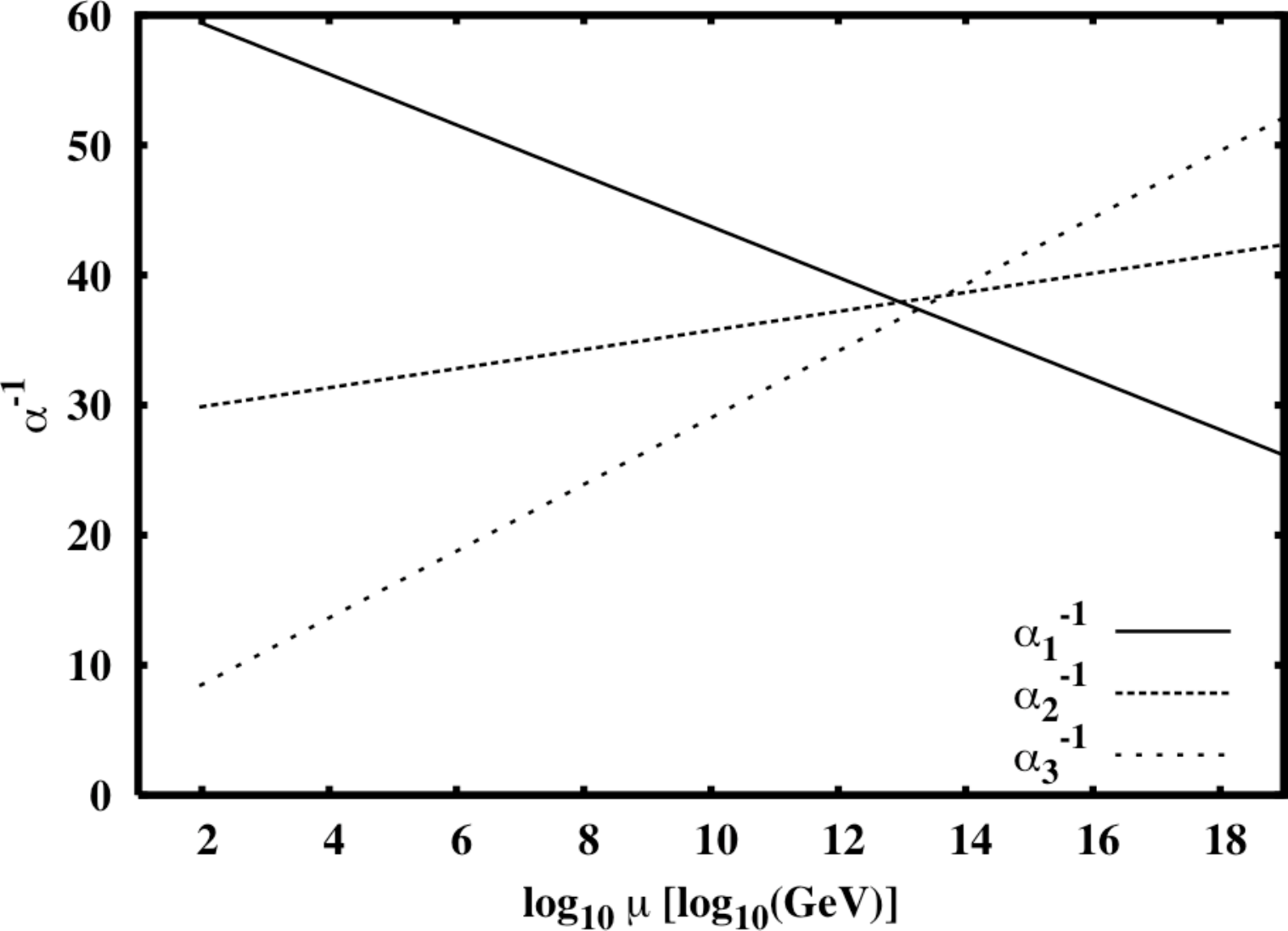}
\end{center}
\caption{The running of the SM gauge couplings in the
    presence of adjoint technifermions (the technicolor
coupling is not included here).}\label{TC}
\end{figure}
Note that the increase in $B_{\rm theory}$ with respect to the
SM is due to the fact that,
typically, bosonic contributions
are numerically suppressed with respect to fermionic
ones and
that,
while $b_1-b_2 = 22/3$ receives only a contribution from the gauge
sector, $b_2-b_3 = 11/3 + 4/3$ has
two contributions, a gauge one and a
fermionic one. These results are a
direct consequence of the fact that we have no
ordinary quarks related to the new leptonic family.

\subsubsection{The Technicolor Coupling Constant}

Until now we have not discussed the technicolor coupling
constant $\alpha_{TC}$.
It is possible that the technicolor
interaction does not unify with the other three forces or unifies
later. A single step unification is though esthetically more appealing to
us.  Remembering that the Casimir of the
two-index symmetric representation of $SU(N_{TC})$ is
$(N_{TC}+2)/2$ the first coefficient
of the beta function
$b_{TC}$ is easily found to be
\begin{eqnarray}
b_{TC} = \frac{2}{3} (N_{TC}+2)  N_{f} - \frac{11}{3}N_{TC} \ ,
\end{eqnarray}
where $N_{TC}$ is the number of
technicolors and $N_f$ is the
number of techniflavors. For two
colors and two flavors we find
$b_{TC}=-2$. Observing that, somewhat accidentally, also $b_2=-2$ for
three ordinary SM
generations, we
conclude that the technicolor coupling constant cannot unify with the
other three couplings at the same point. We are assuming, quite
naturally, that the low energy starting points of $\alpha_2$ and
$\alpha_{TC}$ are different.

Insisting that the technicolor coupling constant must
unify with the other coupling constants at
$M_{GUT}$, we need to modify
at a given scale
$ {X}<M_{GUT}$ either the
overall running of the SM couplings or the one of technicolor. To make less
steep the running of the SM couplings one could add new
generations. To avoid the loss of asymptotic freedom for the week
coupling we find that at most only one entire new SM like
generation can be added at an intermediate
scale.
If we, however, choose not to modify the running of the SM coupling constants, the
running of the technicolor coupling constant must at some point $X < M_{GUT}$ become
steeper. This can be achieved by enhancing the number of technigluons
and lowering the contribution due to the
techniquarks at the
scale $X$. An elegant way to implement this idea is to imagine
that the techniquarks - belonging to
the three dimensional
two-index symmetric representation of $SU(2)$ - are
embedded in the
fundamental representation of $SU(3)$ at the scale $X$. At energies
below $X$ we have $b^{<
  X}_{TC}=-2$ and for energies larger than
$X$ we have $b_{TC}^{> X} = -29/3$. If we take the technicolor
coupling to start running at the electroweak scale $M_{EW}\sim 246 \
\text{GeV}$ and unifying with the three SM couplings at the
unification scale we find an expression for the intermediate scale
$X$
\begin{equation}
\ln X = \frac{1}{b_{TC}^{< X} - b_{TC}^{>
 X}}\bigg\{2\pi \bigg(\alpha_{TC}^{-1}(M_{EW}) - \alpha_{TC}^{-1}(M_{GUT})\bigg) +
 b_{TC}^{< X} \ln M_{EW} - b_{TC}^{> X}\ln M_{GUT} \bigg\} \ .
 \label{XX}
\end{equation}
If we take the starting point of the running of the technicolor
coupling to be the critical coupling close to the conformal
window
we have ${\alpha_{TC}(M_{EW}) =
\pi/(3C_2(\square\hspace{-1pt}\square)) = \pi/6}\ $. Also
using the numbers
$\alpha_{TC}(M_{GUT}) = {\alpha_i(M_{GUT}) \sim 0.026\ ,
  i=1,2,3\ }$, $M_{GUT} \sim 9.45 \times 10^{12}
\,\text{GeV}$ we find the intermediate scale
to be $X \sim 830$ GeV.

\subsubsection{Proton Decay}
Grand Unified Theories lead, generally, to proton decay. Gauge bosons
of mass ${M_{V} < M_{GUT}}$ are
responsible for the decay of the proton into $\pi^0$ and $e^+$. The
lifetime of the proton is estimated to be \cite{Giudice:2004tc}
\begin{eqnarray}
\tau &=& {\frac{4f^2_{\pi}M^4_{V}}{\pi m_p \alpha_{GUT}^2 \left(1+D+F\right)^2
\alpha_N^2\left[A_R^2+\left(1+|V_{ud}|^2 \right)^2A_L^2\right]}} \\
& = &{\left( \frac{M_{GUT}}{10^{16}\ \text{GeV}} \right)^4 \left(
\frac{\alpha_{GUT}^{-1}}{35} \right)^2 \left( \frac{0.015\
\text{GeV}^3}{\alpha_N} \right)^2 \left(\frac{2}{\mathcal{A}} \right)^2 2.7
\times 10^{35}\ \text{yr} \ , }
\end{eqnarray}
where we have used $f_{\pi} = 0.131\ \text{GeV}$, the chiral
Lagrangian factor $1+D+F = 2.25$, the operator renormalization
factors $\mathcal{A} \equiv A_L=A_R$ and the hadronic matrix element
is taken from lattice results \cite{Aoki:1999tw} to be $\alpha_N
=-0.015\ \text{GeV}^3$. Following Ross \cite{Ross:1985ai}, we have
estimated $\mathcal{A} \sim 2$ but a larger value $\sim 5$ is quoted in \cite{Giudice:2004tc} .
The lower bound on the unification scale comes from
the Super-Kamiokande limit $\tau > 5.3 \times 10^{33}$
{yr}
\cite{Suzuki:2001rb}
\begin{eqnarray}
M_{GUT} > M_{V} & > & {\left( \frac{35}{\alpha_{GUT}^{-1}} \right)^{1/2}
\left( \frac{\alpha_N}{-0.015\ \text{GeV}^3}\right)^{1/2} \left(
\frac{\mathcal{A}}{2} \right)^{1/2}\ 3.7\times 10^{15}\ \text{GeV} \ .}
\end{eqnarray}

In the MWT model extension of the SM we find ${\alpha^{-1}_{GUT} \sim
  37.5}$ and $M_{GUT} \sim  10^{13} \,\text{GeV}$ yielding too fast
proton decay.

\subsection{Constructing an Unifying Group}
We provide a simple embedding of our matter content into a unifying gauge group. To construct this group we first summarize the charge assignments in table \ref{single}.
\begin{table}[h]
\caption{Quantum Numbers of the MWT + One SM Family}
\begin{center}
\begin{tabular}{c|c|c|c|c}
&$SO_{TC}(3)$&$SU_c(3)$&$SU_L(2)$&$U_Y(1)$ \\
\hline \hline
$q_L$ &1&3&2&1/6 \\
$u_R$ & 1 &3 &1 & 2/3 \\
$d_R$ & 1&3 &1&-1/3\\
$L $& 1&1&2&-1/2 \\
$e_R$ & 1 &1&1&-1 \\
\hline
$Q_L$ &3&1&2&1/6 \\
$U_R$ & 3 &1 &1 & 2/3 \\
$D_R$ & 3&1 &1&-1/3\\
${\cal L}_L $& 1&1&2&-1/2 \\
$\zeta_R$ & 1 &1&1&-1 \\
\end{tabular}
\end{center}
\label{single}
\end{table}
For simplicity we have considered right transforming leptons only
for the charged ones. Also, the techniquarks are classified as being
fundamentals of $SO(3)$ rather than adjoint of $SU(2)$. Except for
topological differences, linked to the center group of the two
groups, there is no other difference. This choice allows us to show
the resemblance of the technicolor fermions with ordinary quarks. We
can now immediately arrange each SM family within an ordinary
$SU(5)$ gauge theory. The relevant question is how to incorporate
the technicolor sector (here we mean also the new Lepton family). An
easy way out is to double the weak and hypercharge gauge groups as
described in table \ref{tabledouble}.
\begin{table}[h]
\caption{MWT + One SM Family enlarged gauge group}
\begin{center}
\begin{tabular}{c|c|c|c|c|c|c}
&$SO_{TC}(3)$&$SU_1(2)$&$U_1(1)$&$SU_c(3)$&$SU_2(2)$&$U_{2}(1)$ \\
\hline \hline
$q_L$ &1&1&0&3&2&1/6 \\
$u_R$& 1&1&0 &3 &1 & 2/3 \\
$d_R$ &1&1& 0&3 &1&-1/3\\
$L $&1&1 &0&1&2&-1/2 \\
$e_R$ &1&1& 0 &1&1&-1 \\
\hline
$Q_L$ &3&2&1/6&1&1&0 \\
$U_R$ & 3 &1 & 2/3&1&1&0 \\
$D_R$ & 3&1 &-1/3&1&1&0\\
${\cal L}_L $& 1&2&-1/2&1&1&0 \\
$\zeta_R$ & 1 &1&-1&1&1&0 \\
\end{tabular}
\end{center}
\label{tabledouble}
\end{table}
This assignment allows us to arrange the low energy matter  fields
into complete representations of $SU(5)\times SU(5)$. To recover the
low energy assignment one invokes a spontaneous breaking of the
group down to $SO(3)_{TC}\times SU_c(3)\times SU_L(2) \times U_Y(1)$
\footnote{To achieve such as a spontaneous breaking of the gauge
group one needs new matter fields around or slightly above the grand
unified scale transforming with respect to both the gauge groups. }.
We summarize in table \ref{GUT} the technicolor and SM fermions
transformation properties with respect to the grand unified group.
\begin{table}[h]
\caption{GUT}
\begin{center}
\begin{tabular}{c|c|c}
&$SU(5)$&$SU(5)$ \\
\hline \hline
$\bar{A}_{SM}$ &1&$\overline{10}$\\
$F_{SM}$ & 1&5 \\
\hline
$\bar{A}_{MWT}$ &$\overline{10}$&1\\
$F_{MWT}$& 5&1 \\
\end{tabular}
\end{center}
\label{GUT}
\end{table}
Here the fields $A$ and $F$ are standard Weyl fermions and the gauge
couplings of the two $SU(5)$ groups need to be the same. We have
shown here that it is easy to accommodate all of the matter fields
in a single semi-simple gauge group. This is a minimal embedding and
others can be envisioned. New fields must be present at the grand
unified scale (and hence will not affect the running at low energy)
guaranteeing the desired symmetry breaking pattern.

 We have not yet considered the problem of how the ordinary fermions
acquire mass.   
We parametrize our ETC, or better our ignorance about a complete ETC theory, with the (re)introduction of a single Higgs type doublet on the top of the minimal walking theory whose main purpose is to give mass to the ordinary fermions. This simple construction leads to no flavor changing neutral currents and does not upset the agreement with the precision tests which our MWT theory already passes brilliantly. We are able to give mass to all of the fermions and the contribution to the beta functions reads:
\begin{eqnarray}
b_3 & = & \frac{4}{3}N_g -11 \ ,\\
b_2 & = & \frac{4}{3} \left(
N_g + 1 \right) - \frac{22}{3}  + \frac{1}{6}\ ,\\
b_1 & = &  \frac{4}{3}\left( N_g+1 \right) +\frac{1}{10} \ ,
\end{eqnarray}
leading to
\begin{equation}
B_{theory}=0.71 \ ,
\end{equation}
a value which, at the one loop level, is even closer to the
experimental value of $0.72$ than the original MWT theory alone. The
unification scale is also slightly higher than in MWT alone and it
is of the order of $1.2 \times 10^{13}$~ GeV.  The ETC construction
presented above has already been used many times in the literature
\cite{Simmons:1988fu,Dine:1990jd,Kagan:1990az,Kagan:1991gh,Carone:1992rh,Carone:1993xc}.
We find the results very encouraging.  We wish to add that the need
for walking dynamics in the gauge sector is important since it helps
reducing the value of the S-parameter which is typically large even
before taking into account the problems due to the introduction of
an ETC sector.

 We wish to improve on the unification point (before taking into account of possible ETC type corrections) and delay it, energy-wise, to
avoid the experimental bounds on the proton decay.

We hence need a minimal modification of our extension of the SM with
the following properties: i) it is natural, i.e. it does not
reintroduce the hierarchy problem, ii) it does not affect the working
technicolor sector, iii) it allows for a straightforward unification
with a resulting theory which is asymptotically free, iv) it yields a
phenomenologically viable proton decay rate and possibly leads also
to dark matter candidates.

Point i) forces us to add new fermionic-type matter while ii) can be
satisfied by modifying the matter content of the SM per se. A simple
thing to do is to explore the case in which we consider adjoint
fermionic matter for the strong and weak interactions. We will show
that this is sufficient to greatly improve the proton decay problem
while also improving unification with respect to the MWT theory. To be
more specific, we add one colored
Weyl fermion transforming solely
according to the adjoint representation of $SU(3)$ and a Weyl fermion
transforming according to the adjoint
representation of $SU_L(2)$. These
fermions can
be identified with the gluino and wino in supersymmetric extensions of
the SM.  The big
hierarchy is still under control in the present model.

Since our theory is not supersymmetric the introduced fermions need
not be degenerate with the associated gauge bosons. Their masses can
be of the order of, or larger than, the electroweak scale. Finally,
naturality does not forbid the presence of a fermion associated to
the hypercharge gauge boson and hence this degree of freedom may
occur in the theory. Imagining a unification of the value of the
masses at the unification scale also requires the presence of such a
${U(1)}$ bino-type fermion.

In this case
the one-loop beta function coefficients are
\begin{eqnarray}
b_3 &=& \frac{4}{3}N_g -11 + 2 \ , \\
b_2 &=& \frac{4}{3} \big( N_g +1 \big) - \frac{22}{3} + \frac{4}{3} \ ,
\\
b_1 &=& \frac{4}{3} \big( N_g +1 \big) \ .
\end{eqnarray}
This gives $B_{\rm theory} = 13/18 \sim
0.72{(2)}$ which is in
excellent agreement with the experimental value. Note also that the
unification scale is ${M_{GUT} \sim 2.65 \times 10^{15} \text{GeV}}$
which brings the proton decay within the correct order of magnitude set by
experiments.
\begin{figure}[ht]
\begin{center}
\includegraphics[width=0.6\linewidth]{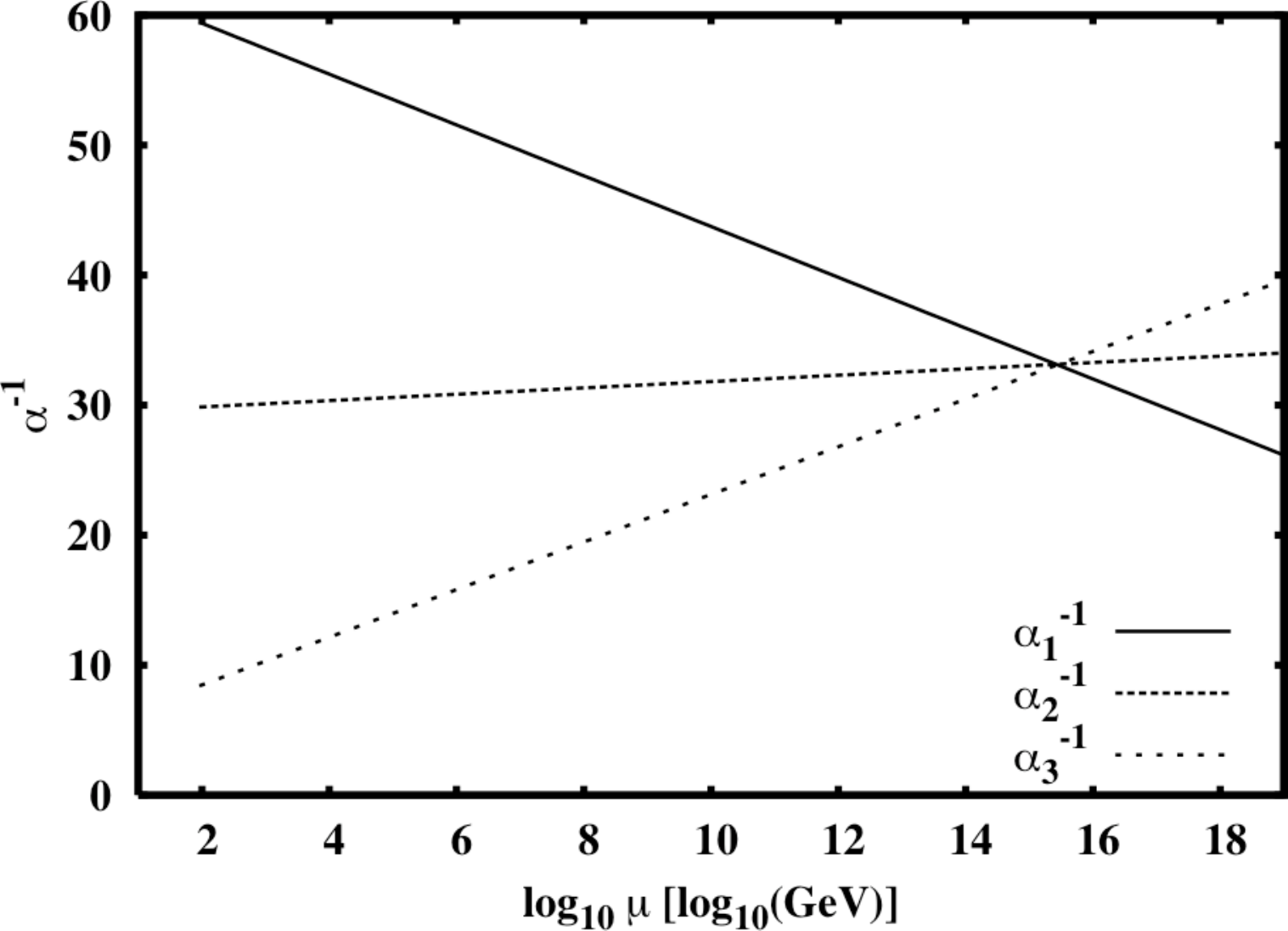}
\end{center}
\caption{The running of the three SM gauge couplings in the new model
  with also adjoint fermionic matter for the SM gauge groups.}
\label{SM-Improved}
\end{figure}

\begin{figure}[ht]
\begin{center}
\mbox{\subfigure{\resizebox{!}{5.3cm}{\includegraphics{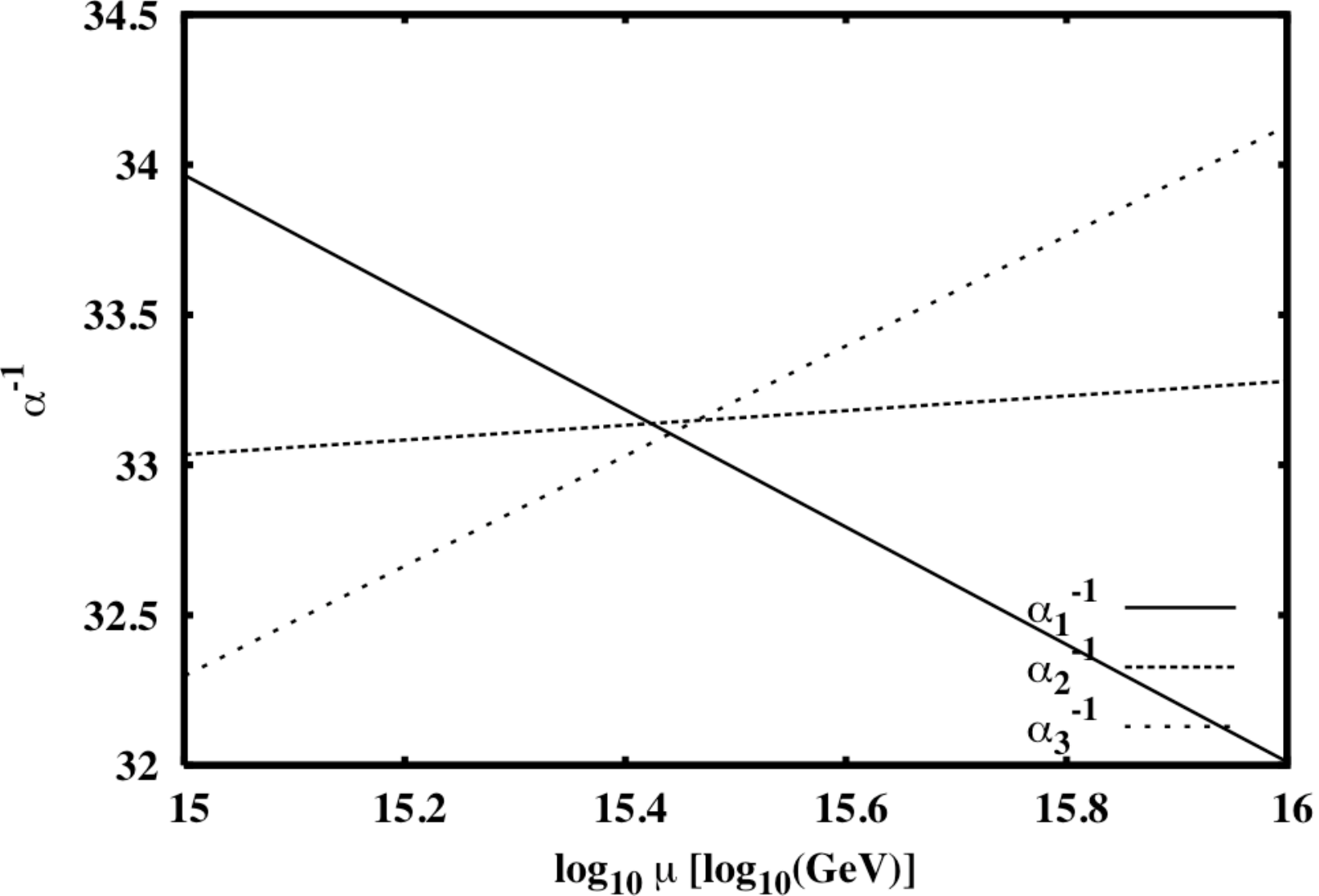}}}
\quad
\subfigure{\resizebox{!}{5.3cm}{\includegraphics{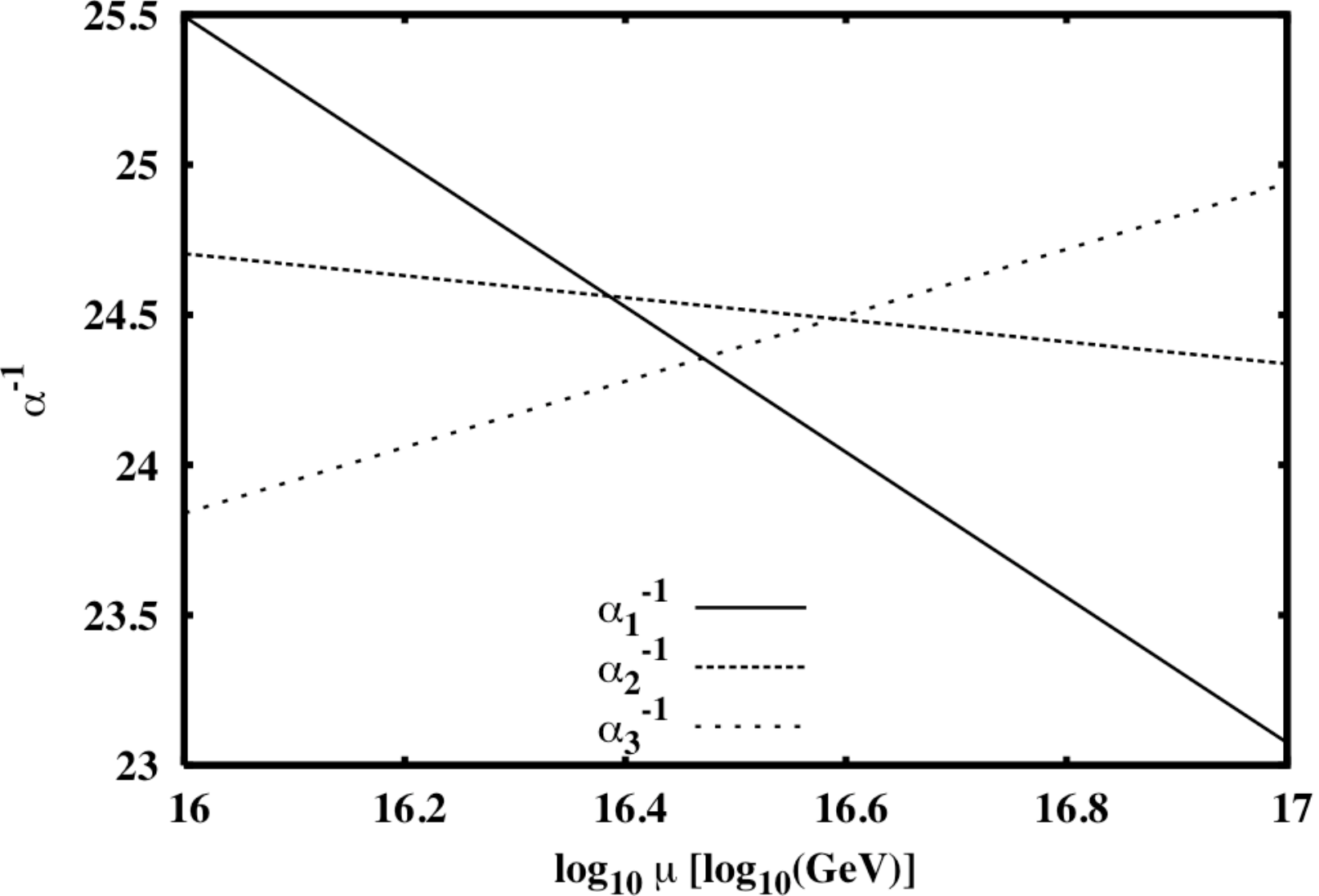}}}}
\end{center}
\caption{\textit{Left Panel}: A zoom around the unification point of the running of the
  three SM gauge couplings in the new model with extra fermionic adjoint matter
  for the SM gauge groups. \textit{Right Panel}: A zoom around the
  unification point for the couplings in the MSSM. }
\label{ZoomCompare}
\end{figure}

We can make the technicolor coupling unify with the SM couplings, as
done in the MWT section.  Using
Eq.~(\ref{XX}), we find now $X\sim
10^8$ GeV. We recall here that $X$ is the scale above which our
technicolor theory becomes an ${SU(3)}$ gauge theory with the fermions
transforming according to the fundamental representation.

It is phenomenologically appealing that the scale $X$ is much higher than the electroweak scale. This allows our technicolor coupling to walk for a sufficiently large range of energy to allow for the introduction of extended technicolor interactions needed to give masses to the SM particles.
\begin{figure}[htbp]
\begin{center}
\includegraphics[width=0.6\linewidth]{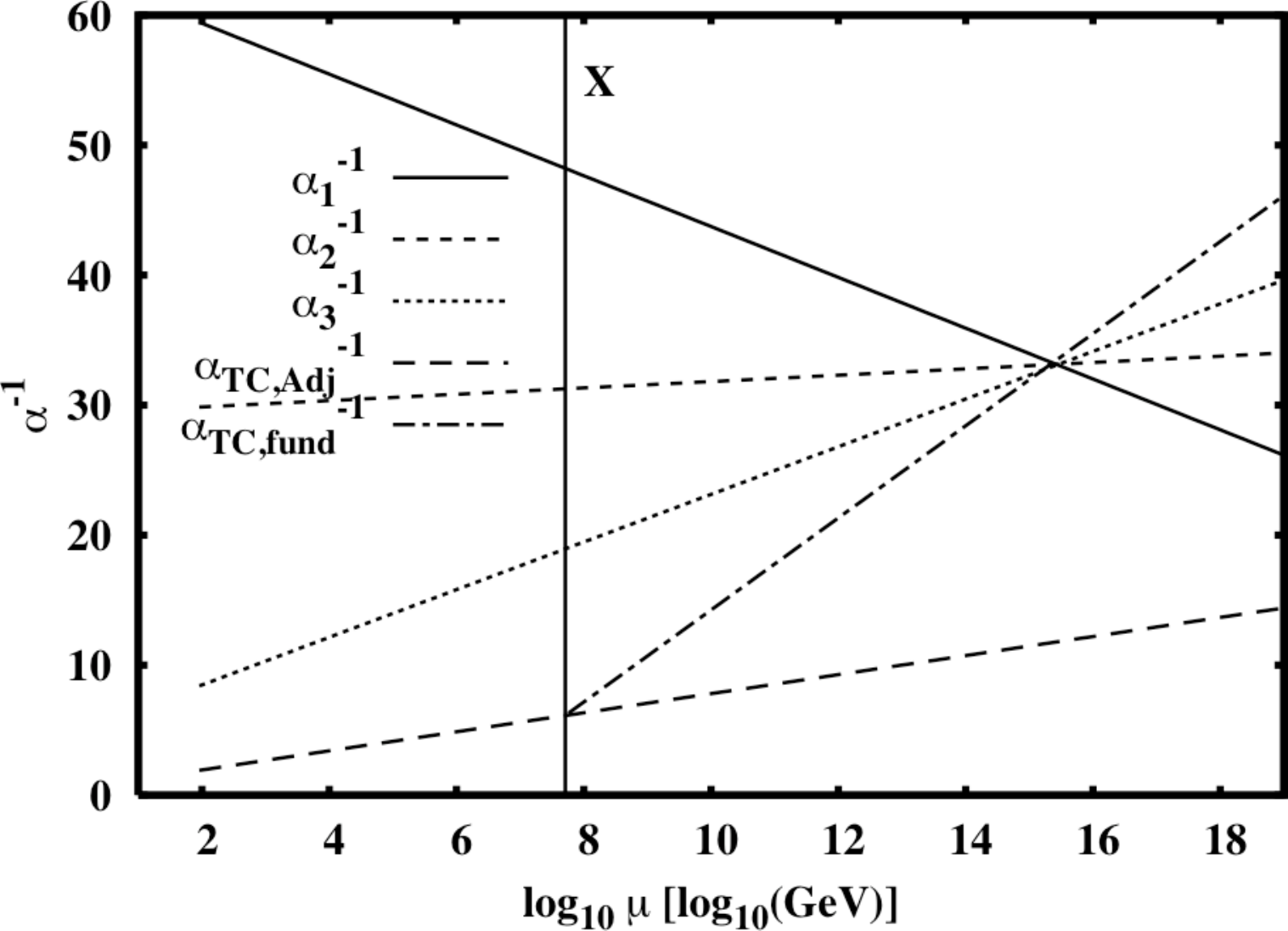}
\end{center}
\caption{The running of the three SM gauge couplings as well as the
  technicolor one. MWT is made to unify with the other three couplings
  by enhancing the gauge group from ${SU(2)}$ of technicolor
  to ${SU(3)}$ while
  keeping the same fermionic matter content. We see that the scale
  where this enhancement of the gauge group should dynamically occur
  to obtain complete unification is around $10^8$ GeV. }
\label{SM-ImprovedX}
\end{figure}
 
Unification of the SM gauge couplings is considered one of the
strongest points in favor of a supersymmetric extension of the SM
and hence it is reasonable to compare our results with the {SUSY}
ones. In {SUSY}, one finds $B_{\rm theory}=0.714$ which is
remarkably close to the experimental value $B_{\rm exp}\sim 0.72$
but it is not better than the value predicted in the present model
which is $0.72(2)$. Obviously this comparison must be taken with a
grain of salt since we still need to provide mass to the SM fermions
and take care of the threshold corrections. {}For example according
to the model introduced in section III, subsection G, to give masses
to all of the fermions yields a theoretical value for the
unification which is around $0.76$.

There are three possible candidates for dark matter here, depending on
which one is the lightest one and on the extended technicolor
interactions which we have not yet specified: The chargeless fermion in the adjoint
representation of $SU(2)_L$, i.e. the wino-like object as well as the
bino-type one.  The third possibility is the heavy neutrino-like fermion whose dark matter potential features are being currently investigated \cite{Kainulainen:2006wq}.

We have introduced a technicolor model which leads to the unification of the SM gauge couplings. At the one-loop level the model provides a higher degree of unification when compared to other technicolor models and to the minimal supersymmetric extension of the SM.

The phenomenology, both for collider experiments and cosmology, of the present extension of the SM is very
rich and needs to be explored in much detail.

The model has many
features in common with split and non-split supersymmetry \cite{ArkaniHamed:2004fb,Giudice:2004tc} and also with models proposed in \cite{Bajc:2006ia,Dorsner:2006fx} while others in common with
technicolor.

\newpage
\section{Basic group theory relations }
\label{eleven}
The Dynkin indices label the highest weight
of an irreducible representation and uniquely characterise the
representations. The Dynkin indices for some of the most common
representations are given in Table \ref{dynkin}. For details on the concept
of Dynkin indices see, for example \cite{Dynkin:1957um,Slansky:1981yr}.

\begin{table}
\begin{center}
\begin{tabular}{r|c}
representation&Dynkin indices\\
\hline
singlet&(000\dots 00)\\
fundamental (F)&(100\dots 00)\\
antifundamental ($\bar{\mathrm{F}}$)&(000\dots 01)\\
adjoint (G)&(100\dots 01)\\
$n$-index symmetric (S$_n$)&(n00\dots 00)\\
2-index antisymmetric (A$_2$)&(010\dots 00)\\
\end{tabular}
\caption{Examples for Dynkin indices for some common representations.}
\label{dynkin}
\end{center}
\end{table}

For a representation, R, with the Dynkin indices
$(a_1,a_2,\dots,a_{N-2},a_{N-1})$ the
quadratic Casimir operator reads \cite{White:1992aa}
\bear 2N \,C_2(\mathrm{R}) &=& \sum_{m=1}^{N-1}[ N(N-m)ma_m + m(N-m){a_m}^2 +\nonumber \\&&
\sum_{n=0}^{m-1}2n(N-m)a_na_m ] \label{C} \eear
and the dimension of R is given by
\bear
d(\mathrm{R})
=
\prod_{p=1}^{N-1}
\left\{
\frac{1}{p!}
\prod_{q=p}^{N-1}
\left[
\sum_{r=q-p+1}^p(1+a_r)
\right]
\right\},
\label{d}
\eear
which gives rise to the following structure
\bear d(\mathrm{R}) &=& (1+a_1)(1+a_2)\dots(1+a_{N-1}) \times
\nn && \times (1+\sfr{a_1+a_2}{2}) \dots
(1+\sfr{a_{N-2}+a_{N-1}}{2}) \times
\nn && \times (1+\sfr{a_1+a_2+a_3}{3}) \dots
(1+\sfr{a_{N-3}+a_{N-2}+a_{N-1}}{3}) \times
\nn && \times \dots \times
\nn && \times
(1+\sfr{a_1+\dots+a_{N-1}}{N-1}). \eear
The Young tableau associated to a given Dynkin index
$(a_1,a_2,\dots,a_{N-2},a_{N-1})$ is easily constructed. The length of
row $i$ (that is the number of boxes per row) is given in terms of the Dynkin
indices by the expression $r_i = \sum_{i}^{N-1} a_{i}$. The length of
each column is indicated by $c_k$; $k$ can assume any positive integer value.
Indicating the total number of boxes associated to a given Young tableau with 
$b$ one has another compact expression for $C_2(\mathrm{R})$,
\bear
2N \, C_2(\mathrm{R})
= 
N\left[ bN +\sum_i r^2_i  -\sum_ic_i^2 - \frac{b^2}{N} \right]\ ,
\eear
and the sums run over each column and row.

\subsection{Realization of the generators for MWT \label{appgen}}

It is convenient to use the following representation of SU(4)
\beq S^a = \begin{pmatrix} \bf A & \bf B \\ {\bf B}^\dag & -{\bf A}^T
\end{pmatrix} \ , \qquad X^i = \begin{pmatrix} \bf C & \bf D \\ {\bf
    D}^\dag & {\bf C}^T \end{pmatrix} \ , \eeq
where $A$ is hermitian, $C$ is hermitian and traceless, $B = -B^T$ and
$D = D^T$. The ${S}$ are also a representation of the $SO(4)$
generators, and thus leave the vacuum invariant $S^aE + E {S^a}^T = 0\ $.
Explicitly, the generators read
\beq S^a = \frac{1}{2\sqrt{2}}\begin{pmatrix} \tau^a & \bf 0 \\ \bf 0 &
  -\tau^{aT} \end{pmatrix} \ , \quad a = 1,\ldots,4 \ , \eeq
where $a = 1,2,3$ are the Pauli matrices and $\tau^4 =
\mathbbm{1}$. These are the generators of SU$_V$(2)$\times$ U$_V$(1).
\beq S^a = \frac{1}{2\sqrt{2}}\begin{pmatrix} \bf 0 & {\bf B}^a \\
{\bf B}^{a\dag} & \bf 0 \end{pmatrix} \ , \quad a = 5,6 \ , \eeq
with
\beq B^5 = \tau^2 \ , \quad B^6 = i\tau^2 \ . \eeq
The rest of the generators which do not leave the vacuum invariant are
\beq X^i = \frac{1}{2\sqrt{2}}\begin{pmatrix} \tau^i & \bf 0 \\
\bf 0 & \tau^{iT} \end{pmatrix} \ , \quad i = 1,2,3 \ , \eeq
and
\beq X^i = \frac{1}{2\sqrt{2}}\begin{pmatrix} \bf 0 & {\bf D}^i \\
{\bf D}^{i\dag} & \bf 0 \end{pmatrix} \ , \quad i = 4,\ldots,9 \ ,
\eeq
with
\beq\begin{array}{r@{\;}c@{\;}lr@{\;}c@{\;}lr@{\;}c@{\;}l}
D^4 &=& \mathbbm{1} \ , & \quad D^6 &=& \tau^3 \ , & \quad D^8 &=& \tau^1 \ , \\
D^5 &=& i\mathbbm{1} \ , & \quad D^7 &=& i\tau^3 \ , & \quad D^9 &=& i\tau^1
\ .
\end{array}\eeq

The generators are normalized as follows
\beq {\rm Tr}\left[S^aS^b\right] =\frac{1}{2}\delta^{ab}\ , \qquad \ , {\rm Tr}\left[X^iX^j\right] =
\frac{1}{2}\delta^{ij} \ , \qquad {\rm Tr}\left[X^iS^a\right] = 0 \ . \eeq

\newpage
\section{Vector Mesons as Gauge Fields} \label{sec:hidden}
We show how to rewrite the vector meson Lagrangian in a gauge invariant way.  We assume the scalar sector to transform according to a given but otherwise arbitrary representation of the flavor symmetry group $G$. This is a straightforward generalization of the Hidden Local Gauge symmetry idea \cite{Bando:1984ej,Bando:1987br}, used in a similar context for the BESS models \cite{Casalbuoni:1995qt}. At the tree approximation this approach is identical to the one introduced first in \cite{Kaymakcalan:1984bz,Kaymakcalan:1983qq}.

\subsection{Introducing Vector Mesons}
Let us start with a generic flavor symmetry group $G$ under which a scalar field $M$ transforms globally in a given, but generic, irreducible representation $R$.  We also introduce an algebra valued  one-form $A=A^{\mu}dx_{\mu}$ taking values in a copy of the algebra of the group $G$, call it $G^{\prime}$, i.e.
\begin{eqnarray}
 A_{\mu}=A^{a}_{\mu} T^a \ , \qquad {\rm with } \qquad T^a \in {\cal{A}}(G^{\prime}) \ .
 \end{eqnarray}
At this point the full group structure is the semisimple group $G\times G^{\prime}$. $M$ does not transform under $G^{\prime}$. Given that $M$ and $A$ belong to two different groups we need another field to connect the two. We henceforth introduce a new scalar field $N$ transforming according to the fundamental of $G$ and to the antifundamental of $G^{\prime}$. We then upgrade $A$ to a gauge field over $G^{\prime}$.
\begin{table}[t]
\caption{Field content}
\begin{center}
\begin{tabular}{c||cc}
&\,\,\,$G$\,\,\,&\,\,\,$G^{\prime}$\,\,\, \\
\hline
&&\\
$M$\,\,\,&\,\,\,$R$\,\,\,&\,\,\,${\mathbf 1}$\,\,\, \\
&&\\
$N$\,\,\,&\,\,\,$\fund$\,\,\,&\,\,\,$\overline{\fund}$\,\,\, \\
&&\\
$A_{\mu}$\,\,\,&\,\,\,${\mathbf 1}$\,\,\,&\,\,\,${\rm Adj}$\,\,\, \\
\end{tabular}
\end{center}
\label{default}
\end{table}%
The covariant derivative for $N$ is:
\begin{eqnarray}
D_{\mu}N=\partial_{\mu} N + i\,\tilde{g} \, N\,A_{\mu} \ .
\end{eqnarray}
We now force $N$ to acquire the following vacuum expectation value
\begin{eqnarray}
\langle N^i_j \rangle = \delta^i_j \, v^{\prime} \ ,
\end{eqnarray}
which leaves the diagonal subgroup - denoted with $G_{V}$ - of $G \times G^{\prime}$ invariant. Clearly $G_V$ is a copy of $G$. Note that it is always possible to arrange a suitable potential term for $N$ leading to the previous pattern of symmetry breaking. $v/v^{\prime}$ is expected to be much less than one and the {\it unphysical} massive degrees of freedom associated to the fluctuations of $N$ will have to be integrated out. The would-be Goldstone bosons associated to $N$ will become the longitudinal components of the massive vector mesons.

To connect $A$ to $M$ we define the one-form transforming only under $G$ via $N$ which - in the deeply spontaneously broken phase of $N$ - reads:
\begin{eqnarray}
\frac{ {\rm Tr} [N N^{\dagger}]}{{\rm dim}(F)} \, P_{\mu} =\frac{D_{\mu}N N^{\dagger} - N D_{\mu}N^{\dagger}} {2\, i \tilde{g}} \ , \qquad P_{\mu} \rightarrow u P_{\mu} u^{\dagger} \ ,
\end{eqnarray}
with $u$ being an element of $G$ and dim($F$) the dimension of the fundamental representation of $G$. When evaluating $P_{\mu}$ on the vacuum expectation value for $N$ we recover $A_{\mu}$:
\begin{eqnarray}
\langle P_{\mu } \rangle =  A_{\mu} \ .
\end{eqnarray}
  At this point it is straightforward to write the Lagrangian containing $N$, $M$ and $A$ and their self-interactions.
 Being in the deeply broken phase of $G\times G^{\prime}$ down to $G_{V}$ we count $N$ as a dimension zero field. This is consistent with the normalization for $P_{\mu}$.

The simplest\footnote{Another nonminimal term is ${\rm Tr} \left[NFN^{\dagger}M (NFN^{\dagger})^{T} M^{\dagger}\right]$.} kinetic term of the Lagrangian is:
\begin{eqnarray}
L_{kinetic} = -\frac{1}{2}{\rm Tr} \left[F_{\mu\nu}F^{\mu\nu}\right]  + \frac{1}{2}{\rm Tr} \left[DN DN^{\dagger}\right] +\frac{1}{2}{\rm Tr} \left[ \partial M \partial M^{\dagger} \right] \ .
\end{eqnarray}
The second kinetic term will provide a mass to the vector mesons.
Besides the potential terms for $M$ and $N$ there is another part of the Lagrangian which is of interest to us. This is the one mixing $P$ and $M$.  Up to dimension four and containing at most two powers of $P$ and $M$ this is:
\begin{eqnarray}
L_{P-M} & = & \tilde{g}^2\ r_1 \ {\rm Tr}\left[P_\mu P^\mu M M^\dagger\right]
+ \tilde{g}^2\ r_2 \ {\rm Tr}\left[P_\mu M {P^\mu}^T M^\dagger \right] \nonumber \\
& + & i \ \tilde{g}\ r_3 \ {\rm Tr}\left[P_\mu \left(M (D^\mu M)^\dagger - (D^\mu M) M^\dagger \right) \right]
+ \tilde{g}^2\ s \ {\rm Tr}\left[P_\mu P^\mu \right] {\rm Tr}\left[M M^\dagger \right] \ . \nonumber \\
\end{eqnarray}
The dimensionless parameters $r_1$, $r_2$, $r_3$, $s$ parameterize the strength of the interactions between the composite scalars and vectors in units of $\tilde{g}$, and are therefore expected to be of order one. We have assumed $M$ to belong to the two index symmetric representation of a generic G= SU(N). It is straightforward to generalize the previous terms to the case of an arbitrary representation $R$ with respect to any group G.
Further higher derivative interactions including $N$ can be included systematically.

\subsection{Further Gauging of G}
In this case we add another gauge field $G_{\mu}$ taking values in the algebra of $G$. We then define the correct covariant derivatives for $M$ and $N$. {}For $N$, for example, we have:
\begin{eqnarray}
D_{\mu}N=\partial_{\mu} N -i\,g\,G_{\mu}\,N+ i\,\tilde{g} \, N\,A_{\mu} \ .
\label{covariant-N}
\end{eqnarray}
Evaluating the previous expression on the vacuum expectation value of $N$ we recover the field $C_{\mu}$ introduced in the text. To be more precise we need to use $P_{\mu}$ again but with the covariant derivative for $N$ replaced by the one in the equation above.
\begin{table}[hb]
\caption{Field content}
\begin{center}
\begin{tabular}{c||cc}
&\,\,\,$G$\,\,\,&\,\,\,$G^{\prime}$\,\,\, \\
\hline
&&\\
$M$\,\,\,&\,\,\,$R$\,\,\,&\,\,\,${\mathbf 1}$\,\,\, \\
&&\\
$N$\,\,\,&\,\,\,$\fund$\,\,\,&\,\,\,$\overline{\fund}$\,\,\, \\
&&\\
$A_{\mu}$\,\,\,&\,\,\,${\mathbf 1}$\,\,\,&\,\,\,${\rm Adj}$\,\,\, \\
&&\\
$G_{\mu}$\,\,\,&\,\,\,${\rm Adj}$\,\,\,&\,\,\,${\mathbf 1}$\,\,\, \\
\end{tabular}
\end{center}
\label{default2}
\end{table}%

\newpage
\section{The Topological Terms and Massive Spin One States}

In the previous section we introduced the vector mesons as gauge bosons of a {\it fake} new gauge symmetry and provided a mass term resorting to an Higgsing procedure. In fact this symmetry does not exist  and there is {\it no} notion of a {\it minimal way} to break it. If all the terms are included correctly one recovers, {\it de facto}, a non-renormalizable Lagrangian for vector mesons preserving only the correct global flavor symmetries of the problem. This, of course, is true also for the terms involving vectors, pions and the space-time $\epsilon_{\mu\nu\rho\sigma}$ structure.  This correct way to proceed was already suggested some time ago in \cite{Duan:2000dy}. We will review here the salient points on the analysis done in \cite{Duan:2000dy}.

\subsection{The ${\protect\epsilon}$ terms for   $SU(N_f)\times SU(N_f)$  }

We construct an effective Lagrangian which manifestly possesses the
global symmetry $SU_L(N_f)\times SU_R(N_f)$ of the underlying
theory. We assume that chiral symmetry is broken according to the
standard pattern $SU_L(N_f)\times SU_R(N_f)\rightarrow SU_V(N_f)$.
The $N_f^2 - 1$ Goldstone bosons are encoded in the $N_f\times N_f$
matrix $U$ transforming linearly under a chiral rotation \begin{equation}
U\rightarrow u_L U u_R^{\dagger} \ ,
\end{equation}
with $u_{L/R} \in SU_{L/R}(N_f)$. $U$ satisfies the non linear realization
constraint $U U^{\dagger} =1 $. We also require ${\rm det} U =1$. In this
way we avoid discussing the axial $U_A(1)$ anomaly at the effective
Lagrangian level (see Ref.~{\cite{Kaymakcalan:1984bz,Kaymakcalan:1983qq,Jain:1987sz}} for a general discussion of
anomalies). We have
\begin{equation}
{U=e^{i \frac{\Phi}{v}}} \ ,
\end{equation}
with $\Phi=\sqrt{2}\Phi^a T^a$ representing the $N^2_f - 1$
Goldstone bosons. $T^{a}$ are the generators of $SU(N_{f})$, with
$a=1,...,N_{f}^{2}-1$ and $\displaystyle{{\rm Tr}\left[
T^{a}T^{b}\right] = \frac{1}{2}\delta ^{ab}}$. $v$ is the vacuum
expectation value.

As done above we enlarge the spectrum of massive particles including vector and
axial-vector fields $A_{L/R}^{\mu }=A_{L/R}^{\mu ,a}T^{a}$\footnote{
We rescale $A$ by the coupling constant $\tilde{g}$.}. 

The Wess-Zumino \cite{Wess:1971yu} action is the first example of $\epsilon$ term. It can be
compactly written using the language of differential forms. It is useful to
introduce the Maurer-Cartan one forms:
\begin{equation}
\alpha=\left(\partial_{\mu}U\right)U^{-1}\, dx^{\mu}\equiv
\left(dU\right)U^{-1}\ , \quad \beta=U^{-1}dU=U^{-1}\alpha U \ .  \label{MC}
\end{equation}
$\alpha$ and $\beta$ are algebra valued one forms and transform,
respectively, under the left and right $SU(N_f)$ flavor group. The
Wess-Zumino effective action is
\begin{equation}
\Gamma_{WZ}\left[U\right]=C\, \int_{M^5} {\rm Tr} \left[\alpha^5\right] \ .
\label{WZ}
\end{equation}
The price to pay in order to make the action local is to augment by one the
space dimensions. Hence the integral must be performed over a
five-dimensional manifold whose boundary ($M^4$) is the ordinary Minkowski
space. The constant $C$ is fixed to be
\begin{equation}
C=-i\frac{N}{240\pi^2} \ ,
\end{equation}
by comparing the current algebra prediction for the time honored
process $\pi^{0}\rightarrow 2\gamma$ with the amplitude predicted
using Eq.~(\ref{WZ}) once we gauge the electromagnetic sector of
the Wess-Zumino term, and $N$ is the number of colors.

We now consider $\epsilon$ type terms involving the vector and
axial vector particles. As for the non $\epsilon$ part of the
Lagrangian we first gauge the WZ
term under the $SU_L(N_f)\times SU_R(N_f)$ chiral symmetry group.
This procedure automatically induces new $\epsilon$ terms
\cite{Witten:1983tw,Witten:1983tx,Kaymakcalan:1984bz,Kaymakcalan:1983qq,Jain:1987sz}, leading to the following
Lagrangian,
\begin{eqnarray}
\Gamma _{WZ}\left[ U,\;A_{L},\;A_{R}\right] &=&\Gamma _{WZ}\left[ U\right]
\,+\,5Ci\,\int_{M^{4}}{\rm Tr}\left[ A_{L}\alpha ^{3}+A_{R} \beta ^{3}\right]
\nonumber \\
&&-5C\,\int_{M^{4}}{\rm Tr}\left[ (dA_{L}A_{L}+A_{L}dA_{L})\alpha
+(dA_{R}A_{R}+A_{R}dA_{R})\beta \right]  \nonumber \\
&&+5C\,\int_{M^{4}}{\rm Tr}\left[
dA_{L}dUA_{R}U^{-1}-dA_{R}dU^{-1}A_{L}U \right]  \nonumber \\
&&+5C\,\int_{M^{4}}{\rm Tr}\left[ A_{R}U^{-1}A_{L}U\beta
^{2}-A_{L}UA_{R}U^{-1}\alpha ^{2}\right]  \nonumber \\
&&+\frac{5C}{2}\,\int_{M^{4}}{\rm Tr}\left[ (A_{L}\alpha
)^{2}-(A_{R}\beta )^{2}\right] +5Ci\,\int_{M^{4}}{\rm Tr}\left[
A_{L}^{3}\alpha +A_{R}^{3}\beta \right]  \nonumber \\
&&+5Ci\,\int_{M^{4}}{\rm Tr}\left[
(dA_{R}A_{R}+A_{R}dA_{R})U^{-1}A_{L}U-(dA_{L}A_{L}
+A_{L}dA_{L})UA_{R}U^{-1}
\right]  \nonumber \\
&&+5Ci\,\int_{M^{4}}{\rm Tr}\left[ A_{L}UA_{R}U^{-1}A_{L}\alpha
+A_{R}U^{-1}A_{L}UA_{R}\beta \right]  \nonumber \\
&&+5C\,\int_{M^{4}}{\rm Tr}\left[
A_{R}^{3}U^{-1}A_{L}U-A_{L}^{3}UA_{R}U^{-1}+\frac{1}{2}
(UA_{R}U^{-1}A_{L})^{2}\right]  \nonumber \\
&&-5Cr\,\int_{M^{4}}{\rm Tr}\left[ F_{L}UF_{R}U^{-1}\right] \ .  \label{GWZ1}
\end{eqnarray}
Here the two-forms $F_{L}$ and $F_{R}$ are defined as
$F_{L}=dA_{L}-iA_{L}^{2}$ and $F_{R}=dA_{R}-iA_{R}^{2}$ with the
one form $A_{L/R}=A^{\mu}_{L/R}dx_{\mu}$. The previous Lagrangian,
when identifying the vector fields with true gauge vectors,
correctly saturates the underlying global anomalies.

The last term in Eq.~(\ref{GWZ1}) is a gauge covariant term which can always
be added if parity is not imposed. The last term in Eq.(\ref{GWZ1}) is not invariant under parity, so the
parameter $r$ must vanish. All the other terms are related by gauge
invariance.

Imposing just global chiral invariance, together with $P$ and $C$, the
previous Lagrangian has ten unrelated terms \cite{Duan:2000dy}:
\begin{eqnarray}
\Gamma _{WZ}\left[ U,\;A_{L},\;A_{R}\right] &=&\Gamma _{WZ}\left[ U\right]
\,+\,5c_{1}\,i\int_{M^{4}}{\rm Tr}\left[ A_{L}\alpha ^{3}+A_{R}\beta ^{3}
\right]  \nonumber \\
&&+5c_{2}\,\int_{M^{4}}{\rm Tr}\left[
(dA_{L}A_{L}+A_{L}dA_{L})\alpha +(dA_{R}A_{R}+A_{R}dA_{R})\beta
\right]  \nonumber \\ &&-5c_{3}\int_{M^{4}}{\rm Tr}\left[
dA_{L}dUA_{R}U^{-1}-dA_{R}dU^{-1}A_{L}U
\right]  \nonumber \\
&&-5c_{4}\,\int_{M^{4}}{\rm Tr}\left[ A_{R}U^{-1}A_{L}U\beta
^{2}-A_{L}UA_{R}U^{-1}\alpha ^{2}\right]  \nonumber \\
&&-\frac{5c_{5}}{2}\int_{M^{4}}{\rm Tr}\left[ (A_{L}\alpha
)^{2}-(A_{R}\beta )^{2}\right] +5c_{6}\,i\int_{M^{4}}{\rm Tr}\left[
A_{L}^{3}\alpha +A_{R}^{3}\beta \right]  \nonumber \\
&&+5c_{7}\,i\int_{M^{4}}{\rm Tr}\left[
(dA_{R}A_{R}+A_{R}dA_{R})U^{-1}A_{L}U-(dA_{L}A_{L}+A_{L}dA_{L})UA_{R}U^{-1}
\right]  \nonumber \\
&&+5c_{8}\,i\int_{M^{4}}{\rm Tr}\left[ A_{L}UA_{R}U^{-1}A_{L}\alpha
+A_{R}U^{-1}A_{L}UA_{R}\beta \right]  \nonumber \\
&&-5c_{9}\,\int_{M^{4}}{\rm Tr}\left[
A_{R}^{3}U^{-1}A_{L}U-A_{L}^{3}UA_{R}U^{-1}\right]  \nonumber \\
&&- \frac{5c_{10}}{2} \,\int_{M^{4}}{\rm Tr}\left[
(UA_{R}U^{-1}A_{L})^{2}
\right] \ ,  \label{GWZ2}
\end{eqnarray}
\noindent where the $c$-coefficients are imaginary. We see that while the gauging procedure of the Wess Zumino term automatically generates
a large number of $\epsilon $ terms, it does not guarantee that we have
uncovered all terms consistent with chiral, $P$ and $C$ invariance. Indeed
there is still one new single trace term \cite{Duan:2000dy} to add
to the action:
\begin{equation}
c_{11}i\int_{M^{4}}{\rm Tr}\left[ A_{L}^{2}\left( UA_{R}U^{-1}\alpha -\alpha
UA_{R}U^{-1}\right) +A_{R}^{2}\left( U^{-1}A_{L}U\beta -\beta
U^{-1}A_{L}U\right) \right] \ ,  \label{undici}
\end{equation}
and $c_{11}$ is an imaginary coefficient. Imposing invariance under $CP$ has
been very useful to reduce the number of possible $\epsilon $ terms. For
example it is easy to verify that a term of the type ${\rm Tr}\left[
dA_{L}\left( UA_{R}U^{-1}\right) ^{2}\right] $ is $CP$ odd.

In Appendix A of \cite{Duan:2000dy} we provided a general proof that all the dimension four (i.e.
4-derivative) terms involving the Lorentz tensor $\epsilon_{\mu \nu \rho
\sigma}$, which are consistent with global chiral symmetries as well as $C$
and $P$ invariance, are the ones presented in Eq.~(\ref{GWZ2}) and
Eq.~(\ref {undici}).

\subsection{The ${\protect\epsilon}$ terms for $SU(2N_{f})$}

We consider now fermions in a pseudoreal representation, for example $SU(2)$ technicolor with $N_f$ fermions in the fundamental representation. The global symmetry group is $SU(2N_f)$ and if chiral symmetry breaking occurs we expect it to break to $Sp(2N_f)$. 
We divide the generators ${T}$ of $SU(2N_f)$, normalized according
to $\displaystyle{{\rm
Tr}\left[T^aT^b\right]}=\frac{1}{2}\delta^{ab}$, into two classes.
We call the generators of $Sp(2N_f)$ $\{S^{a}\}$ with $a=1,\ldots
,2N_f^2+N_f$, and the remaining $SU(2N_{f})$ generators
(parameterizing the quotient space $SU(2N_{f})/Sp(2N_{f})$)
$\{X^{i}\}$ with $i=1,\ldots ,2N_f^2-N_f-1$.

This breaking pattern gives $2N^2_f - N_f -1$ Goldstone bosons, encoded in
the antisymmetric matrix $U^{ij}$ and $i,j = 1,\ldots,2N_f$ as follows:
\begin{equation}
U=e^{i\frac{\Pi ^{i}X^{i}}{v}}\,E\ ,
\end{equation}
where the $N_f\times N_f$ matrix $E$ is
\begin{equation}
E=\left(
\begin{array}{cc}
{\bf 0} & {\bf 1} \\
-{\bf 1} & {\bf 0}
\end{array}
\right) \ .
\end{equation}
$U$ transforms linearly under a chiral rotation
\begin{equation}
U\rightarrow u\, U \, u^{T} \ ,
\end{equation}
with $u\in SU(2N_f)$. The non linear realization constraint,
$\displaystyle {UU^{\dagger }=1}$, is automatically satisfied.

The generators of the $Sp(2N_f)$ satisfy the following relation,
\begin{equation}
S^{T}\,E+E\,S=0\ ,
\end{equation}
while the $X^i$ generators obey,
\begin{equation}
X^{T}=E\,X\,E^{T}\ ,
\end{equation}
Using this last relation we can easily demonstrate that $U^{T}=-U$. We also
require
\begin{equation}
{\rm Pf}\,U=1\ ,
\end{equation}
avoiding in this way to consider the explicit realization of the underlying
axial anomaly at the effective Lagrangian level.

We define the following vector field
\begin{equation}
A_{\mu }=A_{\mu }^{a}T^{a}\ ,
\end{equation}
which formally transforms under a $SU(2N_{f})$ rotation as
\begin{equation}
A_{\mu }\rightarrow uA_{\mu }u^{\dagger }-i\partial _{\mu }uu^{\dagger }\ .
\end{equation}

We generate the $\epsilon $ terms following the same
procedure used for the $SU_{L}(N_{f})\times SU_{R}(N_{f})$ global symmetry
case. First we introduce the one form
\begin{equation}
\alpha =\left( dU\right) U^{-1}\ .
\end{equation}
It is sufficient to define only $\alpha $ since the analog of $\beta
=U^{-1}dU=\alpha ^{T}$ is now not an independent form. The Wess-Zumino
action term is:
\begin{equation}
\tilde{\Gamma}_{WZ}\left[ U\right] =C\,\int_{M^{5}}{\rm Tr}\left[ \alpha
^{5} \right] \ ,  \label{WZSp}
\end{equation}
where again we are integrating on a five dimensional manifold and $C=-i\frac{
2}{240\pi ^{2}}$ for $N=2$. We are considering here an $SU(2)$ underlying gauge theory with fermions in the fundamental representation. 

We now gauge the Wess-Zumino action under the $SU(2N_f)$ chiral symmetry
group. This procedure provides single trace $\epsilon$-terms involving
vector, axial and Goldstones with an universal coupling $C$. The gauged $CP$
invariant Wess-Zumino term is
\begin{eqnarray}
\tilde{\Gamma}_{WZ}\left[ U,\;A\right] &=&\tilde{\Gamma}_{WZ}\left[ U\right]
\,+\,10Ci\int_{M^{4}}{\rm Tr}\left[ A\alpha ^{3}\right]  \nonumber \\
&&-10C\,\int_{M^{4}}{\rm Tr}\left[ (dAA+AdA)\alpha \right]  \nonumber \\
&&-5C\,\int_{M^{4}}{\rm Tr}\left[ dAdUA^{T}U^{-1}-dA^{T}dU^{-1}AU \right]
\nonumber \\
&&-5C\,\int_{M^{4}}{\rm Tr}[UA^{T}U^{-1}(A\alpha ^{2}+\alpha ^{2}A)]
\nonumber \\
&&+\,5C\int_{M^{4}}{\rm Tr}\left[ (A\alpha )^{2}\right] +10C\,i\int_{M^{4}}
{\rm Tr}\left[ A^{3}\alpha \right]  \nonumber \\
&&+10C\,i\int_{M^{4}}{\rm Tr}\left[ (dAA+AdA)UA^{T}U^{-1}\right]  \nonumber
\\
&&-10C\,i\int_{M^{4}}{\rm Tr}\left[ A\alpha AUA^{T}U^{-1}\right]  \nonumber
\\
&&+10C\,\int_{M^{4}}{\rm Tr}\left[ A^{3}UA^{T}U^{-1}+\frac{1}{4}
(AUA^{T}U^{-1})^{2}\right] \ ,
\end{eqnarray}
where $A=A^{\mu}dx_{\mu}$. 
The previous Lagrangian must be generalized to be only globally invariant under a chiral rotation and invariant
under $CP$ and one obtains \cite{Duan:2000dy}.
\begin{eqnarray}
\tilde{\Gamma}_{WZ}\left[ U,\;A\right] &=&\tilde{\Gamma}_{WZ}\left[ U\right]
\,+\,C_{1}i\int_{M^{4}}{\rm Tr}\left[ A\alpha ^{3}\right]  \nonumber \\
&&-C_{2}\,\int_{M^{4}}{\rm Tr}\left[ (dAA+AdA)\alpha \right]  \nonumber \\
&&-C_{3}\int_{M^{4}}{\rm Tr}\left[ dAdUA^{T}U^{-1}-dA^{T}dU^{-1}AU \right]
\nonumber \\
&&-C_{4}\,\int_{M^{4}}{\rm Tr}[UA^{T}U^{-1}(A\alpha ^{2}+\alpha ^{2}A)]
\nonumber \\
&&+\,C_{5}\int_{M^{4}}{\rm Tr}\left[ (A\alpha )^{2}\right]
+C_{6}\,i\int_{M^{4}}{\rm Tr}\left[ A^{3}\alpha \right]  \nonumber \\
&&+C_{7}\,i\int_{M^{4}}{\rm Tr}\left[ (dAA+AdA)UA^{T}U^{-1}\right]  \nonumber
\\
&&-C_{8}i\int_{M^{4}}{\rm Tr}\left[ A\alpha AUA^{T}U^{-1}\right]
+C_{9}\,\int_{M^{4}}{\rm Tr}[A^{3}UA^{T}U^{-1}]  \nonumber \\
&&+C_{10}\int_{M^{4}}{\rm Tr}[(AUA^{T}U^{-1})^{2}]  \nonumber \\
&&+C_{11}i\int_{M^{4}}{\rm Tr}[A^{2}(\alpha UA^{T}U^{-1}-UA^{T}U^{-1}\alpha
)] \ ,
\end{eqnarray}
where $C_{i}$ are imaginary. The last term is a new term not
generated by gauging the Wess-Zumino effective action. 

At this point the application to extensions of the SM featuring chiral dynamics is straightforward. Summarizing, the SM gauge bosons, being true gauge fields, must be introduced via the correct gauging of the Wess-Zumino term. Any other spin one field which is not a gauge degree of freedom must be introduced in the manner presented above, i.e. allowing for a very general form of the interactions with the Goldstone bosons featuring an $\epsilon$ tensor. Often, in literature, spin-one non-gauge degrees of freedom are introduced again as gauge degrees of freedom (see for example \cite{Lane:2009ct}). This latter procedure can be considered as a simple phenomenological approach.  

\newpage
\section{Spectrum of Strongly Coupled Theories: Higgsless versus Higgsful theories}
\label{7}
Often, in the literature, a number of incorrect statements are made when discussing the spectrum of technicolor theories. Here we  will try to clarify first the situation in QCD and then show how to use new analytic means to gain control over the spectrum of strongly coupled theories with fermions in higher dimensional representations. 

 One approach 
 is based on studying the theory in the large number
of colors (N) limit \cite{'tHooft:1973jz,Witten:1979kh}. At the same time one may
obtain more information by requiring the theory to model the 
(almost) spontaneous breakdown of chiral
 symmetry \cite{Nambu:1961tp,GellMann:1960np}. A standard test case, for ordinary QCD, is pion pion
scattering in the energy range up to about 1 GeV. Some time ago,
an attempt was made \cite{Sannino:1995ik,Harada:1995dc} to implement this
combined scenario.  We used pion pion scattering to provide some insight on the low lying hadronic spectrum of QCD. 
 
 Before turning to the spectrum of the lightest composite states in QCD we offer  a simple definition of Higgsless theory: {\it If the composite state with the same quantum numbers of the Higgs is not the lightest  particle in the spectrum after the Goldstones then the theory is Higgsless. } 
 In practice we  will use the massive spin one states to compare the mass of the composite Higgs with. 
 
 \subsection{The lightest composite scalars in QCD}
The scalar sector of QCD and any technicolor theory constitutes a complicated sector. {}For QCD, in \cite{Harada:2003em},  using the 't Hooft large N limit, chiral dynamics and unitarity constraints
the $f_0(600)$ resonance mass was found to be around 550 MeV. Other authors \cite{Pelaez:2003dy,Oller:1997ti,Uehara:2003ax} have found similar results.
 Such a low
value would make it different from a p-wave quark-antiquark state,
which is expected to be in the 1000-1400 MeV range.
We assume then that 
it is a four quark state (glueball states
 are expected to be in the 1.5
GeV range from lattice investigations).
 Four quark
states of diquark-quark type \cite{Jaffe:1976ig,Jaffe:1976ih} and meson-meson type
\cite{Weinstein:1982gc} have been
 discussed in the literature for many years.
Accepting this picture, however,  
poses a problem for the accuracy of the large N
inspired description of the scattering since
four quark states are  
predicted not to exist in the large N limit of QCD.
We shall take the point of view that a four quark 
type state is present since it allows a natural fit to
the low energy data.
 In practice, since the 
parameters
of the pion contact and rho exchange contributions are fixed, the 
sigma is the most important one for fitting and fits may 
even be achieved \cite{Harada:1996wr} if the vector meson
 piece is neglected. However
the well established, presumably four
 quark type, $f_0$(980) resonance
 must be included to achieve a fit in the region just
 around 1 GeV.

 There is by now a fairly large literature
 on the effect of light ``exotic" scalars in low energy meson meson
 scattering. There seems to be a consenesus, arrived at using rather
different approaches (keeping however, unitarity), that the sigma exists.

Here we  use two large N limits of QCD as well as our information on the low lying spectrum of QCD to extract information on the spectrum of the lightest states for strongly coupled theories with fermions in various representations of the underlying strongly coupled gauge group. Lifting the strongly coupled scale to the electroweak one for theories with underlying fermions in two index representations we  will show that the light scalar with the same quantum numbers of the Higgs is lighter than the lightest techni-vector meson.

\subsection{Scalars in the 't Hooft Large N:  Higgsless theories}

We concentrate on the lightest scalar $f_0(600)$ and on the vector meson $\rho(770)$. The $q\bar{q}$ nature of the vector meson is clear. This means that its mass does not scale with the number of colors while its width decrease as $1/N$. We argued above that $f_0(600)$ is a multiquark state. In this case its mass scales with a positive power of N and its width remains constant or grows with N. In formulae: 
\begin{eqnarray}
m_{\rho} ^2&\sim &~~~~\Lambda_{QCD}^2 \ , \qquad \Gamma_{\rho} \sim \frac{1}{N} \\
m_{f_0}^2 &\sim &N^p \Lambda_{QCD}^2 \ , \qquad \Gamma_{f_0} \sim {N}^q \ ,
\end{eqnarray}
with $p>0$ and $q>-1$. 

Scaling up these results to the electroweak theory is straightforward. We first generalize the number of technidoublets gauged under the electroweak theory as well the number of technicolors $N_{TC}$, holding fixed the weak scale we have:
\begin{eqnarray}
M_{T\rho} &=& \frac{\sqrt{2}v_{weak}}{F_{\pi}}
\frac{\sqrt{3}}
{\sqrt{N_D\, N_{TC}}} m_{\rho}
 \\
M_{T f_{0}} &=& \frac{\sqrt{2}v_{weak}}{F_{\pi}\sqrt{N_D}} 
\left(\frac{N_{TC}}{\sqrt{3}}\right)^{\frac{p-1}{2}}m_{f_0}  \ ,
\end{eqnarray} 
where $N_D$ is the number of doublets, $v_{weak}$ is the electroweak scale and the extra $\sqrt{2}$ is due to our normalization of the pion decay constant. Note that for $p=0$ and $q=-1$  the $f_0(600)$ would scale like the $\rho$ and would then be regarded as a quark-antiquark meson at large $N$. However, as we mentioned, there are, by now, strong indications that this state  is not of $q\bar{q}$ nature and hence $p>0$ and $q>-1$. 

Let us choose for definitiveness $p=1$. Already for $N_{TC}\sim 6$, for any $N_D$ the scalar is heavier than the vector meson. Hence for fermions in the fundamental representation of the technicolor theory we expect {\it no scalars} lighter than the respective vector mesons for any $N_{TC}$ larger than or about $6$ technicolors. It is hence fair to call these theories Higgsless. Note that the previous statements may be altered if the theory features walking dynamics.

\subsection{Alternative Large N limits}
The previous results are in agreement with the common lore about the light spectrum of QCD-like theories. Interestingly even if for N=3 one has a scalar state lighter than the lightest vector meson it becomes heavier already for N$>$6.  Clearly the reason behind this is that, due to its multiquark nature, the lightest state possesses different scaling properties than the vector meson.  
The situation changes when we consider alternative extensions of QCD using higher dimensional representations.  At large $N$ different extensions capture different dynamical properties of QCD.  

\subsubsection{The Two Index Antisymmetric Fermions - Link to QCD}    
   Consider redefining the $N=3$ quark field with
 color index A (and flavor
index not written) as
\begin{equation}
        q_A= \frac{1}{2} \epsilon_{ABC}q^{\left[B,C\right]}\ ,\qquad  q^{\left[B,C\right]}=-q^{\left[C,B\right]},
\label{redefine}
\end{equation}
so that, for example, $q_1=q^{23}$ and similarly for the adjoint field,
${\bar q}^1={\bar q}_{23}$ etc. This is just a
 trivial change of variables. However for $N>3$ the resulting theory will be
 different since the two index
 antisymmetric quark representation has $N(N-1)/2$ rather
 than $N$ color components.
As was pointed out by Corrigan and Ramond
 \cite{Corrigan:1979xf}, who were
mainly interested in the problem of the
 baryons at large N, this shows
that the extrapolation of QCD to higher $N$ is not unique.
 Further investigation
of the properties of the alternative extrapolation model
introduced in \cite{Corrigan:1979xf} was 
carried out 
 by Kiritsis and Papavassiliou \cite{Kiritsis:1989ge}.
 
     It may be worthwhile to remark that gauge theories with
two index quarks have gotten a great deal of attention.
 Armoni, Shifman and Veneziano \cite{Armoni:2003gp} have
 proposed an interesting
 relation between certain sectors of the two index antisymmetric
 (and symmetric) theories at
 large number of colors and
sectors of super Yang-Mills (SYM).
Using a supersymmetric inspired
effective Lagrangian approach $1/N$ corrections were
investigated in \cite{Sannino:2003xe}. 

 Besides these two limits a third one for
 massless one-flavor QCD, which
 is in between the 't Hooft and Corrigan Ramond
ones, has been been
 proposed in \cite{Ryttov:2005na}.
 Here one first splits the
 QCD Dirac fermion into the two elementary Weyl fermions
 and afterwards assigns one of them
 to transform according to a rank-two antisymmetric
 tensor while the other
 remains in the fundamental representation of the
 gauge group. For three
 colors one reproduces
one-flavor QCD and for a generic number of colors the
theory is chiral. {}The generic $N$ is
 a particular case of the 
generalized
Georgi-Glashow (gGG) model \cite{Georgi:1985hf}.
The finite temperature phase transition and its relation with chiral symmetry has been investigated in \cite{Sannino:2005sk} while the effects of a nonzero
 baryon chemical potential were pioneered in \cite{Frandsen:2005mb}. More recent work in this direction has appeared in the literature \cite{Cherman:2009fh,Buchoff:2009za}.  In particular in \cite{Buchoff:2009za} the authors have shown that one of the high density QCD phases investigated  in \cite{Frandsen:2005mb}, i.e. the color superconductive one, seem to be favored at large N. This is a very interesting result which modifies and improves on the results in  \cite{Frandsen:2005mb}. 
On the validity of the large N equivalence between different theories we refer the reader to \cite{Unsal:2006pj,Kovtun:2005kh}.

   To illustrate the large N counting when quarks are designated
to transform according to the two index antisymmetric representation
of color SU(3) one may employ \cite{'tHooft:1973jz} the mnemonic where
each tensor index of this group is represented by a directed line.
Then the quark-quark gluon interaction is pictured as in Fig. \ref{FigA}. 
\begin{figure}[htbp]
\centering 
{
\includegraphics[width=
7.5cm,clip=true]{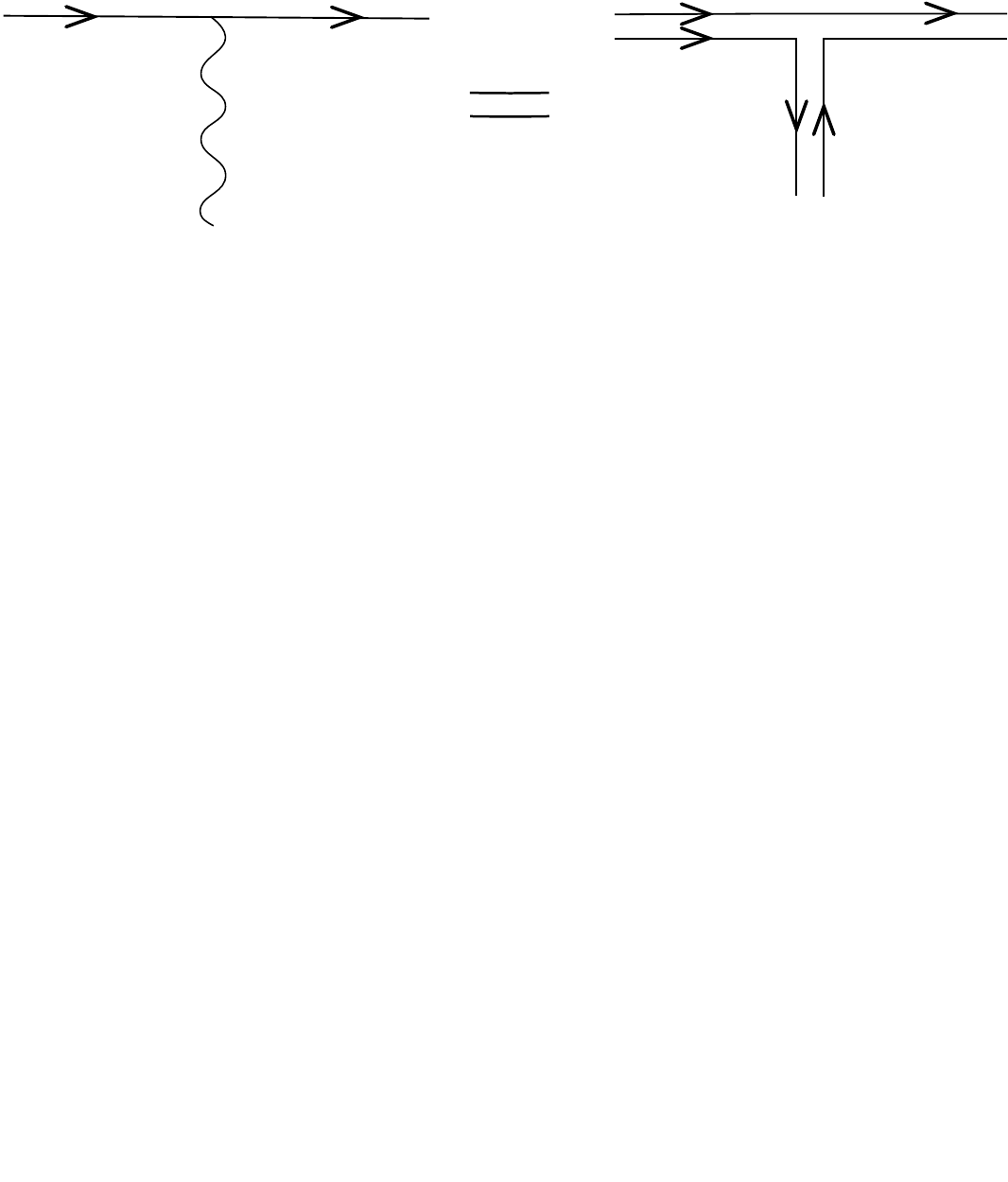}
}
\vskip -7cm
\caption[]
{Two index fermion - gluon vertex.} \label{FigA}
\end{figure}
The two index quark is pictured as two lines with arrows pointing in the 
same direction, as opposed to the gluon which has two lines with arrows 
pointing in opposite directions. The coupling constant representing this 
vertex is taken to be $g_t/\sqrt{N}$, where $g_t$ does not depend on $N$ and is kept fixed.

    A ``one point function", like the pion decay constant, $F_\pi$
has as it's simplest diagram, Fig.~\ref{FigB}
          
\begin{figure}[htbp]
\centering 
{
\includegraphics[width=
3.0cm,clip=true]{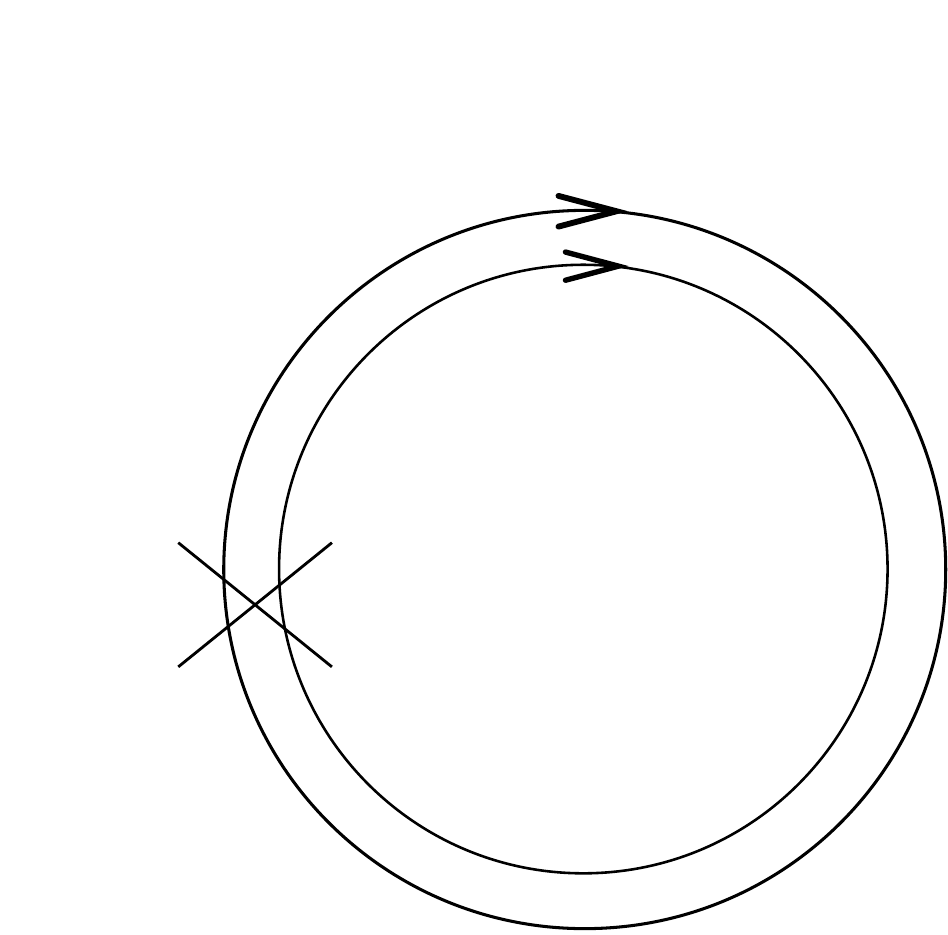}
}
\caption[]
{Diagram for $F_\pi$ for the two index quark.} \label{FigB}
\end{figure}

The X represents a pion insertion and is associated with a
normalization factor for the color part of the pion's wavefunction,
\begin{eqnarray}
            \frac{\sqrt{2}}{\sqrt{N(N-1)}},
\label{wf}
\end{eqnarray}        
which scales for large N as $1/N$. The two circles each
carry a quark index so their factor scales as $N^2$ for
 large N; more
precisely, taking the antisymmetry into account, the factor is
\begin{equation}          
 \frac{N(N-1)}{2}.
\label{loop}
\end{equation}
The product of Eqs. (\ref{wf}) and (\ref{loop})
 yields the $N$ scaling for
$F_{\pi}$:
\begin{equation}
F_{\pi}^2(N)= \frac{N(N-1)}{6}F_{\pi}^2(3).
\label{fpiscaling}
\end{equation}
For large N, $F_\pi$ scales proportionately to $N$ rather
than to $\sqrt{N}$ as in the case of the 't Hooft extrapolation.

    Using this scaling the $\pi \pi$ scatttering amplitude, $A$ scales as,
\begin{equation}
A(N)=\frac{6}{N(N-1)}A(3),
\label{Ascaling}
\end{equation}
which, for large N scales as $1/N^2$ rather than
 as $1/N$  in
the 't Hooft extrapolation. This scaling law for
 large N may be verified
from the mnemonic in Fig.~\ref{FigC}, where there is
 an $N^2$ factor from the two loops
multiplied by four factors of $1/N$ from the X's.

\begin{figure}[htbp]
\centering 
{
\includegraphics[width=
4.0cm,clip=true]{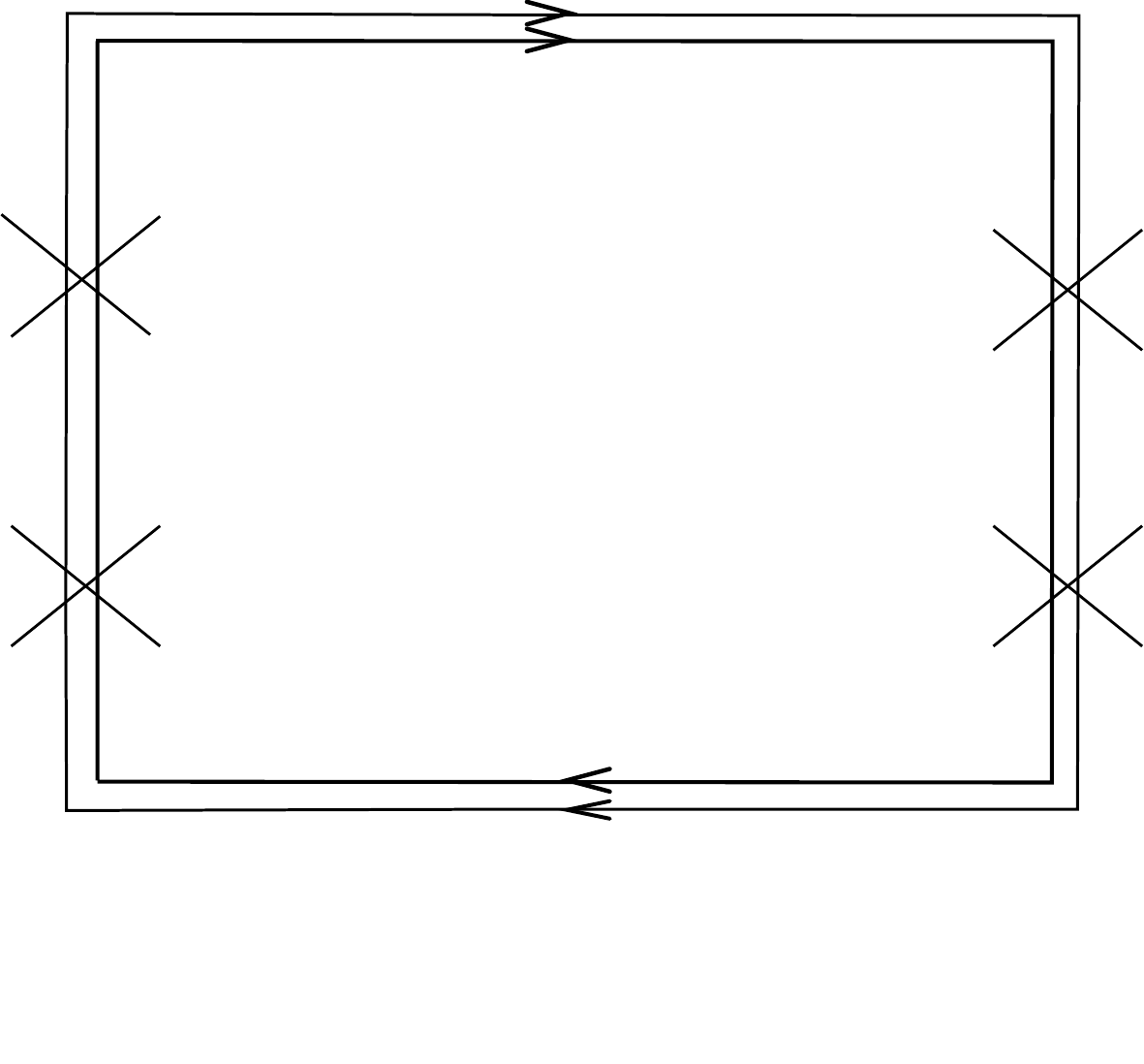}
}
\vskip -1cm
\caption[]
{Diagram for the scattering amplitude, A with the 2 index quark. } \label{FigC}
\end{figure}

    There is still another different feature with respect to the 't Hooft expansion; consider the typical
$\pi \pi$ scattering diagram with an extra internal (two index)
quark loop, as shown in Fig.~\ref{FigD}.
\begin{figure}[htbp]
\centering 
{
\includegraphics[width=
8.5cm,clip=true]{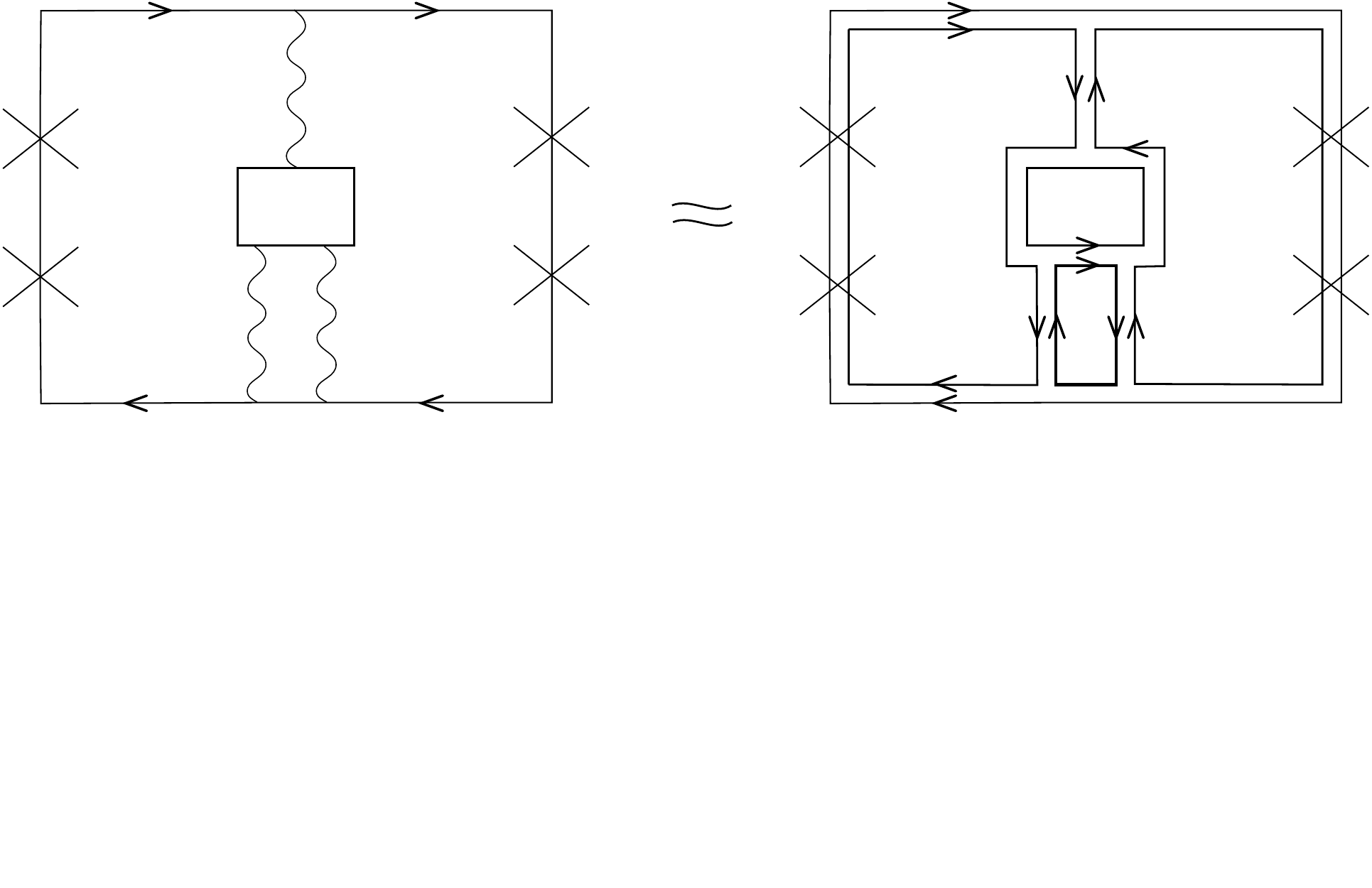}
}
\vskip -3cm
\caption[]
{Diagram for the scattering amplitude, A including an internal 2 index quark loop. } \label{FigD}
\end{figure}

 In this diagram there are 
four X's (factor from Eq.(\ref{wf})), five index loops (factor 
from Eq.(\ref{loop})) and six gauge coupling constants. These
combine to give a large N scaling behavior proportional
to $1/N^2$ for the $\pi \pi$ scattering amplitude. We see that
diagrams with an extra internal 2 index quark loop are not suppressed
compared to the leading diagrams. This is analogous,
as pointed out in \cite{Kiritsis:1989ge}, to the 
behavior of diagrams with an extra gluon loop in the 't Hooft
extrapolation scheme. Now, Fig.~\ref{FigD} is a diagram which can describe
a sigma particle exchange. Thus in the 2 index quark scheme,
``exotic" four quark resonances can appear at the leading order
in addition to the usual two quark resonances. The possibility of a sigma-type state appearing at leading order
means that one can construct a unitary $\pi \pi$ amplitude  
already at $N$ = 3 in the 2 antisymmetric index scheme. From the
 point of view of low energy $\pi \pi$ scattering, it seems to be unavoidable to say
that the 2 index scheme is  more realistic than the 't Hooft
scheme given the existence of a four quark type sigma.

  Of course, the usual 't Hooft extrapolation has a number of
 other things to recommend it. These include the fact that nearly all
meson resonances seem to be of the quark- antiquark type, the OZI rule
predicted holds to a good approximation and baryons emerge
 in an elegant way as solitons in the model. 

 A fair statement is that each extrapolation emphasizes
different aspects of $N$ = 3 QCD.
 In particular, the usual scheme
is not really a replacement for the true theory. That appears to be the
 meaning of the fact that the continuation to $N>3$ is not unique.  

\subsubsection{Quarks in two index symmetric color representation}

    Clearly the assignment of femions to the two index
 symmetric representation of
color SU(3) is very similar to the previous case.
 We denote the fields
as,
\begin{equation}
q_{\{AB \}}=q_{\{BA \}} \ .
\label{symquarks}
\end{equation}
There will be $N(N+1)/2$ different color states for the two index
 symmetric quarks. This means that
there is no value of $N$ for which
 the symmetric theory can be made to
 correspond to true QCD. On the other hand, for large N we can make the approximation 
\begin{equation}
A^{sym}(N) \approx A^{asym}(N),
\label{symasym}
\end{equation}
for the $\pi \pi$ scattering amplitude.

    As far as the large N counting goes,
 the mnemonics in Figs. \ref{FigA}-\ref{FigD}
are still applicable to the case of quarks
 in the two index symmetric color
representation. For not so large N, the scaling
 factor for the pion insertion is
\begin{eqnarray}
            \frac{\sqrt{2}}{\sqrt{N(N+1)}},
\label{symwf}
\end{eqnarray}        
and the pion decay constant scales as
\begin{equation}
F_{\pi}^{sym}(N)\propto \sqrt{\frac{N(N+1)}{2}}.
\label{symfpiscaling}
\end{equation}.

     With the identification $A^{QCD}=A^{asym}(3)$, the use of
Eq.(\ref{symasym})
enables us to estimate the large N
scattering amplitude as,
\begin{equation}
A^{sym}(N) \approx \frac{6}{N^2}A^{QCD}.
\label{bigN}
\end{equation}

In 
applications to minimal walking technicolor theories
this formula 
is useful for making estimates involving weak gauge bosons via
the Goldstone boson equivalence theorem \cite{Lee:1977eg}.

     Finally we remark on the large N scaling rules for meson 
and glueball masses and decays in either the two index
antisymmetric or two index symmetric schemes. Both meson and
 glueball masses scale
as $(N)^0$. Furthermore, all six reactions of the type
\begin{equation}
a\rightarrow b + c,
 \label{abc}
\end{equation}
where a,b and c can stand for either a meson or a glueball,
scale as $1/N$. This is illustrated in Fig. \ref{FigE} for the case
of a meson decaying into two glueballs; note that the
glueball insertion scales as $1/N$ and that two interaction
 vertices are involved.

\begin{figure}[htbp]
\centering 
{
\includegraphics[width=
4.5cm,clip=true]{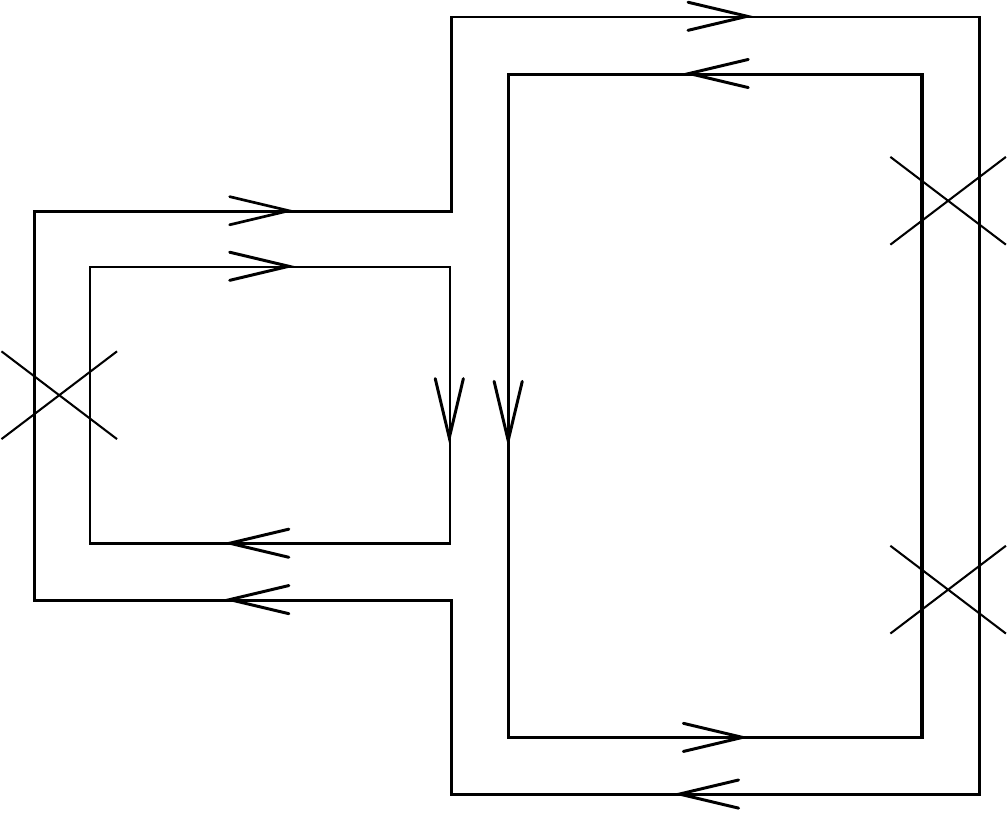}
}
\caption[]
{Diagram for meson decay into two glueballs.} \label{FigE}
\end{figure}
\vskip -.5 cm
 
 \subsubsection{Spectrum for Higher Dimensional Representations: Higgsful theories}
Combining our knowledge of the QCD spectrum together with the rules above for the two index antisymmetric representation we deduce the following large N scaling:
\begin{eqnarray}
m_{\rho} ^2&\sim &\Lambda_{QCD}^2 \ , \qquad \Gamma_{\rho} \sim \frac{2}{N(N - 1)} \\
m_{f_0}^2 &\sim & \Lambda_{QCD}^2 \ , \qquad \Gamma_{f_0} \sim\frac{2}{N(N - 1)}  \ .
\end{eqnarray}
 The fact that in QCD  the state $f_0(600)$ is not narrow indicates that the unknown coefficient in the expression for the width, expected to be order one, is large. However, as we increase the number of colors we expect this state to become quickly narrow.  
Scaling up these results for a technicolor theory with $N_{TC}$  colors and fermions in the two index antisymmetric representation we have:
\begin{eqnarray}
M_{T\rho} &=& \frac{\sqrt{2}v_{weak}}{F_{\pi}}
\frac{\sqrt{3}\sqrt{2}}
{\sqrt{N_D\, N_{TC}(N_{TC}-1) }} m_{\rho}
 \\
M_{T f_{0}} &=&\frac{\sqrt{2}v_{weak}}{F_{\pi}}
\frac{\sqrt{3}\sqrt{2}}
{\sqrt{N_D\, N_{TC}(N_{TC}-1) }} m_{f_0}  \ .
\end{eqnarray} 
The input values here are the QCD masses for $f_0(600)$ and $\rho(770)$. Differently from the 't Hooft case the scalar will remain lighter than the associate technivector meson for any number of technicolors. Finally, increasing the number of technicolors and techniflavors we can achieve a very light scalar, lighter then its own technivector. Since in these theories one cannot differentiate a fermion-antifermion state from a multi fermion states we map the lightest scalar into the composite Higgs. 

So, even without invoking walking dynamics, higher dimensional representations provide a composite Higgs lighter than the technivector meson. These theories are Higgsful for any number of colors. 

One can pass from the two index antisymmetric to the two index symmetric by replacing $N_{TC}-1$ with $N_{TC}+1$ in the expressions above and matching the result at infinite number of colors. In Fig.~\ref{levelsplittingthooft} the physical spectrum of spin one vector bosons and the lightest scalar is reported in TeV units in the case of two doublets ($N_D=2$ ) of technifermions for different number of colors. At $N=3$ we match the spectrum to QCD  for the two index antisymmetric representation. On the left panel we draw the spectrum for the two index antisymmetric extension of QCD while on the right we consider the two index symmetric representation normalized at large N with the two index antisymmetric one. {}For any $N_D$ and $N_{TC}$ the scalar is always lighter than the associated vector meson. In the case of the two index symmetric on approaches light masses a little faster when increasing the number of colors. 
\begin{figure}[htbp]
\centering \leavevmode
{
\includegraphics[width=
7.5cm,clip=true]{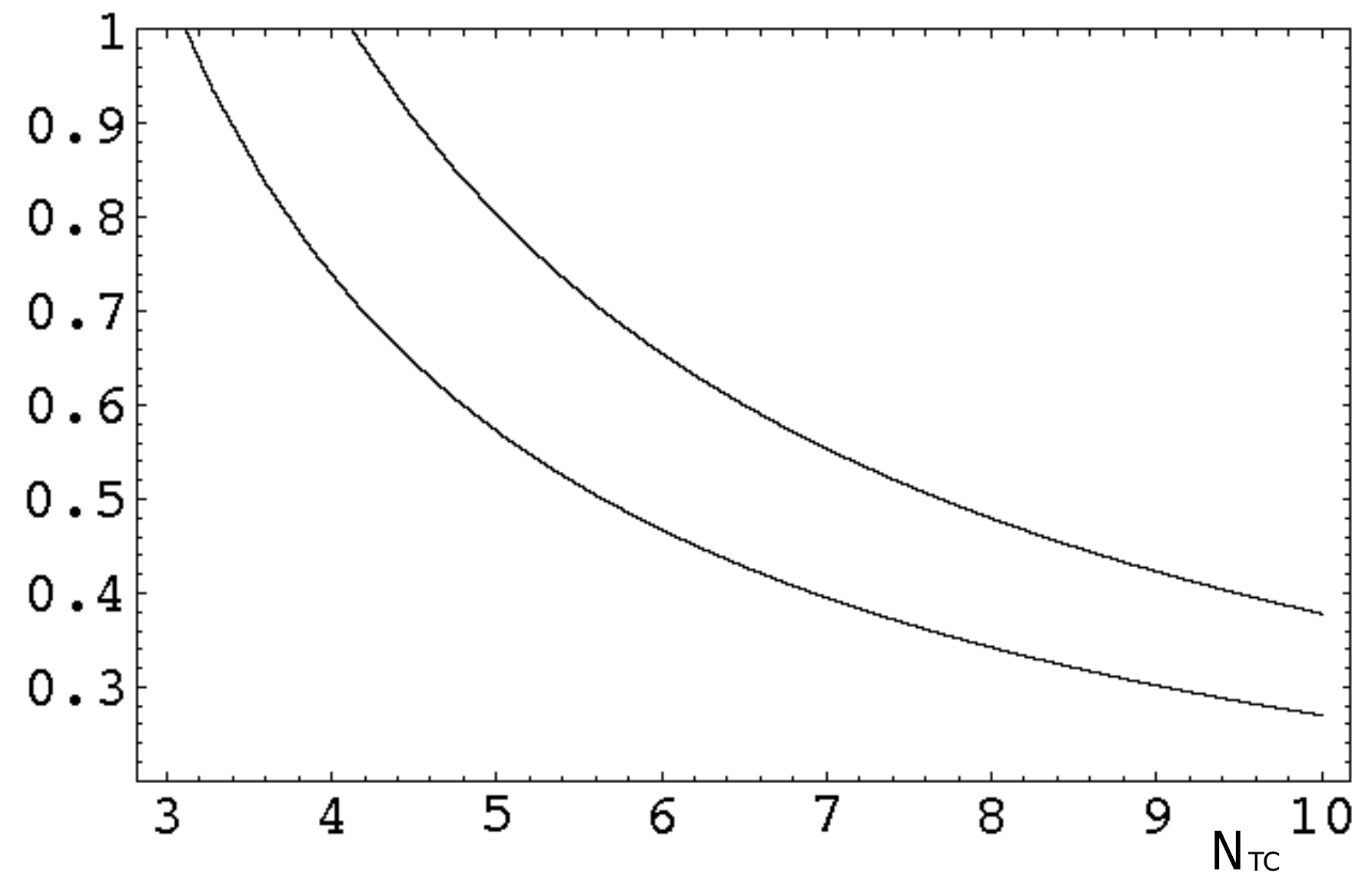}}\hskip .5cm 
\leavevmode{
\includegraphics[width=7.5cm,clip=true]{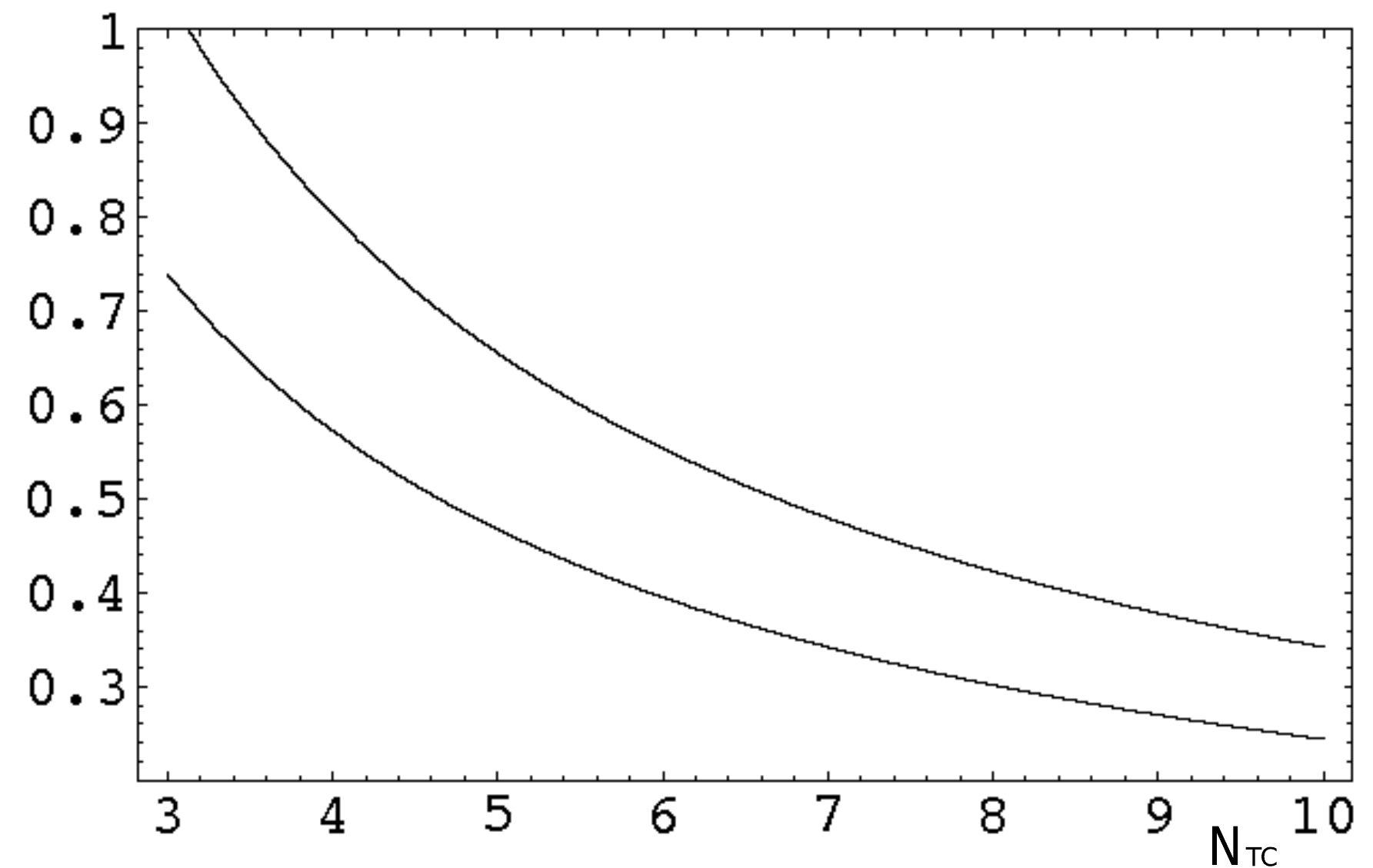}}
\caption[Corrigan-Ramond Scaling]
{Mass of the lightest vector meson (higher curve) and scalar meson (lower curve) as function of the number of colors in TeV units. At $N=3$ we match the spectrum to QCD for the two index antisymmetric representation. Here we use $N_D=2$. On the left panel we draw the spectrum for the two index antisymmetric extension of QCD while on the right we consider the two index symmetric representation. Note that now for any $N_D$ and $N_{TC}$ the scalar is always lighter than the associated vector meson. } \label{levelsplittingthooft}
\end{figure}

Above we  demonstrated that i) It is possible to have composite theories which are Higgsful ii) the resulting composite Higgs is light with respect to the TeV scale. The comparison with precision data must then be revised for these theories since the associated $S$ parameter constraint changes. Note that in the proof we  used only a straightforward geometrical scaling.

What happens to the mass of the composite Higgs in the case of walking?  By increasing the number of flavors all of the composite states from the chiral-symmetric broken side become massless when reaching the fixed point since the only invariant scale of the theory vanishes there \cite{Chivukula:1996kg}. This is supported by lattice simulations \cite{Catterall:2007yx}. We are, however, interested in the ratio between the masses of the various states to the pion decaying constant which is fixed to be the electroweak scale. Simple arguments suggest that if the transition is second order then there will be a light composite Higgs or else its mass to decay constant ratio will not vanish near the conformal point. In any event one can write a low energy effective action for the composite scalar with the quantum numbers of the Higgs -- treating it as a dilaton -- using trace and axial anomaly as well as chiral symmetry as done in \cite{Sannino:1999qe}. A similar analysis using trace anomaly has been also discussed in \cite{Goldberger:2007zk}. The resulting action contains, by construction, non-analitc powers of the composite Higgs field \cite{Sannino:1999qe} and must be treated as generating functional for the anomalous transformations of the underlying dynamics. 

The possibility of a light composite Higgs  in (walking) technicolor was first advocated in \cite{Hong:2004td,Dietrich:2005jn,Dietrich:2005wk,Dietrich:2006cm} and also proposed in \cite{Goldberger:2007zk} and \cite{Doff:2008xx}.  Since, as shown above using standard scaling arguments, it is possible to construct technicolor theories with a light composite Higgs it is relevant to study its phenomenological signatures \cite{Zerwekh:2005wh,Fan:2008jk}.

\newpage



\begin{thebibliography}{300}

\bibitem{Gambino:2003xc}
 P.~Gambino,
 Int.\ J.\ Mod.\ Phys.\  A {\bf 19}, 808 (2004)
 [arXiv:hep-ph/0311257].



\bibitem{Ellis:2009pz}
  J.~Ellis,
  Nucl.\ Phys.\  A {\bf 827}, 187C (2009)
  [arXiv:0902.0357 [hep-ph]].

\bibitem{Altarelli:2009bz}
  G.~Altarelli,
  arXiv:0902.2797 [hep-ph].


\bibitem{Giudice:2007qj}
 G.~F.~Giudice,
 J.\ Phys.\ Conf.\ Ser.\  {\bf 110}, 012014 (2008)
 [arXiv:0710.3294 [hep-ph]].

\bibitem{Mangano:2008ag}
  M.~L.~Mangano,
  Int.\ J.\ Mod.\ Phys.\  A {\bf 23}, 3833 (2008)
  [arXiv:0802.0026 [hep-ph]].

  
\bibitem{Altarelli:2008yi}
  G.~Altarelli,
  arXiv:0805.1992 [hep-ph].


\bibitem{Barbieri:2008zz}
  R.~Barbieri,
  Nuovo Cim.\  {\bf 123B}, 485 (2008).



\bibitem{Grojean:2009fd}
  C.~Grojean,
  arXiv:0910.4976 [Unknown].

\bibitem{DeRoeck:2009id}
  A.~De Roeck {\it et al.},
  arXiv:0909.3240 [Unknown].



\bibitem{Accomando:2006ga}
  E.~Accomando {\it et al.},
  arXiv:hep-ph/0608079.


\bibitem{Weinberg:1979bn}
 S.~Weinberg,
 Phys.\ Rev.\  D {\bf 19}, 1277 (1979).

\bibitem{Susskind:1978ms}
 L.~Susskind,
 Phys.\ Rev.\  D {\bf 20}, 2619 (1979).

\bibitem{Hill:2002ap}
 C.~T.~Hill and E.~H.~Simmons,
 Phys.\ Rept.\  {\bf 381}, 235 (2003)
 [Erratum-ibid.\  {\bf 390}, 553 (2004)]
 [arXiv:hep-ph/0203079].

\bibitem{Sannino:2008ha}
 F.~Sannino,
 arXiv:0804.0182 [hep-ph].

\bibitem{Lane:2002wv}
 K.~Lane,
 arXiv:hep-ph/0202255.

\bibitem{Sannino:2004qp}
 F.~Sannino and K.~Tuominen,
 Phys.\ Rev.\  D {\bf 71}, 051901 (2005)
 [arXiv:hep-ph/0405209].

\bibitem{Dietrich:2005jn}
 D.~D.~Dietrich, F.~Sannino and K.~Tuominen,
 Phys.\ Rev.\  D {\bf 72}, 055001 (2005)
 [arXiv:hep-ph/0505059].

\bibitem{Dietrich:2005wk}
 D.~D.~Dietrich, F.~Sannino and K.~Tuominen,
 Phys.\ Rev.\  D {\bf 73}, 037701 (2006)
 [arXiv:hep-ph/0510217].

\bibitem{Foadi:2007ue}
 R.~Foadi, M.~T.~Frandsen, T.~A.~Ryttov and F.~Sannino,
 Phys.\ Rev.\  D {\bf 76}, 055005 (2007)
 [arXiv:0706.1696 [hep-ph]].

\bibitem{Dietrich:2006cm}
 D.~D.~Dietrich and F.~Sannino,
 Phys.\ Rev.\  D {\bf 75}, 085018 (2007)
 [arXiv:hep-ph/0611341].

\bibitem{Ryttov:2007cx}
 T.~A.~Ryttov and F.~Sannino,
 Phys.\ Rev.\  D {\bf 78}, 065001 (2008)
 [arXiv:0711.3745 [hep-th]].

\bibitem{Shrock:2007km}
 R.~Shrock,
 arXiv:hep-ph/0703050.

\bibitem{Sarkar:1995dd}
 S.~Sarkar,
 Rept.\ Prog.\ Phys.\  {\bf 59}, 1493 (1996)
 [arXiv:hep-ph/9602260].

\bibitem{Chanowitz:1988ae}
 M.~S.~Chanowitz,
 Ann.\ Rev.\ Nucl.\ Part.\ Sci.\  {\bf 38}, 323 (1988).

\bibitem{Farhi:1980xs}
 E.~Farhi and L.~Susskind,
 Phys.\ Rept.\  {\bf 74}, 277 (1981).

\bibitem{Kaul:1981uk}
 R.~K.~Kaul,
 Rev.\ Mod.\ Phys.\  {\bf 55}, 449 (1983).

\bibitem{Chivukula:2000mb}
 R.~S.~Chivukula,
 arXiv:hep-ph/0011264.


\bibitem{'tHooft:1980xb}
  G.~'t Hooft, C.~Itzykson, A.~Jaffe, H.~Lehmann, P.~K.~Mitter, I.~M.~Singer and R.~Stora,
{\it  New York, Usa: Plenum ( 1980) 438 P. ( Nato Advanced Study Institutes Series: Series B, Physics, 59)}


\bibitem{'tHooft:1973jz}
  G.~'t Hooft,
  Nucl.\ Phys.\  B {\bf 72}, 461 (1974).


\bibitem{Witten:1979kh}
 E.~Witten,
 Nucl.\ Phys.\  B {\bf 160}, 57 (1979).

\bibitem{Peskin:1990zt}
 M.~E.~Peskin and T.~Takeuchi,
 Phys.\ Rev.\ Lett.\  {\bf 65}, 964 (1990).

\bibitem{Peskin:1991sw}
 M.~E.~Peskin and T.~Takeuchi,
 Phys.\ Rev.\  D {\bf 46}, 381 (1992).

\bibitem{Kennedy:1990ib}
 D.~C.~Kennedy and P.~Langacker,
 Phys.\ Rev.\ Lett.\  {\bf 65}, 2967 (1990)
 [Erratum-ibid.\  {\bf 66}, 395 (1991)].

\bibitem{Altarelli:1990zd}
 G.~Altarelli and R.~Barbieri,
 Phys.\ Lett.\  B {\bf 253}, 161 (1991).

\bibitem{hep-ph/9306267}
 I.~Maksymyk, C.~P.~Burgess and D.~London,
 Phys.\ Rev.\  D {\bf 50}, 529 (1994)
 [arXiv:hep-ph/9306267].

\bibitem{Barbieri:2004qk}
 R.~Barbieri, A.~Pomarol, R.~Rattazzi and A.~Strumia,
 Nucl.\ Phys.\  B {\bf 703}, 127 (2004)
 [arXiv:hep-ph/0405040].

\bibitem{Chivukula:2004af}
 R.~S.~Chivukula, E.~H.~Simmons, H.~J.~He, M.~Kurachi and M.~Tanabashi,
 Phys.\ Lett.\  B {\bf 603}, 210 (2004)
 [arXiv:hep-ph/0408262].

\bibitem{Peskin:2001rw}
 M.~E.~Peskin and J.~D.~Wells,
 Phys.\ Rev.\  D {\bf 64}, 093003 (2001)
 [arXiv:hep-ph/0101342].

\bibitem{Fukano:2009zm}
 H.~S.~Fukano and F.~Sannino,
 arXiv:0908.2424 [hep-ph].

\bibitem{Antola:2009wq}
 M.~Antola, M.~Heikinheimo, F.~Sannino and K.~Tuominen,
 arXiv:0910.3681 [hep-ph].

\bibitem{Simmons:1988fu}
 E.~H.~Simmons,
 Nucl.\ Phys.\  B {\bf 312}, 253 (1989).

\bibitem{Dine:1990jd}
 M.~Dine, A.~Kagan and S.~Samuel,
 Phys.\ Lett.\  B {\bf 243}, 250 (1990).

\bibitem{Samuel:1990dq}
 S.~Samuel,
 Nucl.\ Phys.\  B {\bf 347}, 625 (1990).

\bibitem{Kagan:1991gh}
 A.~Kagan and S.~Samuel,
 Phys.\ Lett.\  B {\bf 270}, 37 (1991).

\bibitem{Kagan:1992aq}
 A.~Kagan and S.~Samuel,
 Phys.\ Lett.\  B {\bf 284}, 289 (1992).

\bibitem{Kagan:1990gi}
 A.~Kagan and S.~Samuel,
 Int.\ J.\ Mod.\ Phys.\  A {\bf 7}, 1123 (1992).

\bibitem{Carone:1992rh}
 C.~D.~Carone and E.~H.~Simmons,
 Nucl.\ Phys.\  B {\bf 397}, 591 (1993)
 [arXiv:hep-ph/9207273].

\bibitem{Carone:1994mx}
 C.~D.~Carone, E.~H.~Simmons and Y.~Su,
 Phys.\ Lett.\  B {\bf 344}, 287 (1995)
 [arXiv:hep-ph/9410242].

\bibitem{Hemmige:2001vq}
 V.~Hemmige and E.~H.~Simmons,
 Phys.\ Lett.\  B {\bf 518}, 72 (2001)
 [arXiv:hep-ph/0107117].

\bibitem{Carone:2006wj}
 C.~D.~Carone, J.~Erlich and J.~A.~Tan,
 Phys.\ Rev.\  D {\bf 75}, 075005 (2007)
 [arXiv:hep-ph/0612242].

\bibitem{Zerwekh:2009yu}
 A.~R.~Zerwekh,
 arXiv:0907.4690 [hep-ph].

\bibitem{Chivukula:1990bc}
 R.~S.~Chivukula, A.~G.~Cohen and K.~D.~Lane,
 Nucl.\ Phys.\  B {\bf 343}, 554 (1990).

\bibitem{Chivukula:2009ck}
 R.~Sekhar Chivukula, N.~D.~Christensen, B.~Coleppa and E.~H.~Simmons,
 Phys.\ Rev.\  D {\bf 80}, 035011 (2009)
 [arXiv:0906.5567 [hep-ph]].

\bibitem{Eichten:1979ah}
 E.~Eichten and K.~D.~Lane,
 Phys.\ Lett.\  B {\bf 90}, 125 (1980).

\bibitem{Dimopoulos:1979es}
 S.~Dimopoulos and L.~Susskind,
 Nucl.\ Phys.\  B {\bf 155}, 237 (1979).

\bibitem{Farhi:1979zx}
 E.~Farhi and L.~Susskind,
 Phys.\ Rev.\  D {\bf 20}, 3404 (1979).

\bibitem{Appelquist:2004ai}
 T.~Appelquist, N.~D.~Christensen, M.~Piai and R.~Shrock,
 Phys.\ Rev.\  D {\bf 70}, 093010 (2004)
 [arXiv:hep-ph/0409035].

\bibitem{Yamawaki:1985zg}
 K.~Yamawaki, M.~Bando and K.~i.~Matumoto,
 Phys.\ Rev.\ Lett.\  {\bf 56}, 1335 (1986).

\bibitem{Holdom:1984sk}
 B.~Holdom,
 Phys.\ Lett.\  B {\bf 150}, 301 (1985).

\bibitem{Holdom:1981rm}
 B.~Holdom,
 Phys.\ Rev.\  D {\bf 24}, 1441 (1981).

\bibitem{Appelquist:1986an}
 T.~W.~Appelquist, D.~Karabali and L.~C.~R.~Wijewardhana,
 Phys.\ Rev.\ Lett.\  {\bf 57}, 957 (1986).

\bibitem{Weinberg:1967kj}
 S.~Weinberg,
 Phys.\ Rev.\ Lett.\  {\bf 18}, 507 (1967).

\bibitem{Appelquist:1998xf}
 T.~Appelquist and F.~Sannino,
 Phys.\ Rev.\  D {\bf 59}, 067702 (1999)
 [arXiv:hep-ph/9806409].

\bibitem{Witten:1983ut}
 E.~Witten,
 Phys.\ Rev.\ Lett.\  {\bf 51}, 2351 (1983).

\bibitem{Bernard:1975cd}
 C.~W.~Bernard, A.~Duncan, J.~LoSecco and S.~Weinberg,
 Phys.\ Rev.\  D {\bf 12}, 792 (1975).

\bibitem{Kurachi:2006ej}
 M.~Kurachi and R.~Shrock,
 JHEP {\bf 0612}, 034 (2006)
 [arXiv:hep-ph/0605290].

\bibitem{Sundrum:1991rf}
 R.~Sundrum and S.~D.~H.~Hsu,
 Nucl.\ Phys.\  B {\bf 391}, 127 (1993)
 [arXiv:hep-ph/9206225].

\bibitem{Maldacena:1997re}
 J.~M.~Maldacena,
 Adv.\ Theor.\ Math.\ Phys.\  {\bf 2}, 231 (1998)
 [Int.\ J.\ Theor.\ Phys.\  {\bf 38}, 1113 (1999)]
 [arXiv:hep-th/9711200].

\bibitem{Hong:2006si}
 D.~K.~Hong and H.~U.~Yee,
 Phys.\ Rev.\  D {\bf 74}, 015011 (2006)
 [arXiv:hep-ph/0602177].

\bibitem{Hirn:2006nt}
 J.~Hirn and V.~Sanz,
 Phys.\ Rev.\ Lett.\  {\bf 97}, 121803 (2006)
 [arXiv:hep-ph/0606086].

\bibitem{Piai:2006vz}
 M.~Piai,
 arXiv:hep-ph/0609104.

\bibitem{Agashe:2007mc}
 K.~Agashe, C.~Csaki, C.~Grojean and M.~Reece,
 JHEP {\bf 0712}, 003 (2007)
 [arXiv:0704.1821 [hep-ph]].



\bibitem{Carone:2007md}
 C.~D.~Carone, J.~Erlich and M.~Sher,
 Phys.\ Rev.\  D {\bf 76}, 015015 (2007)
 [arXiv:0704.3084 [hep-th]].
  C.~D.~Carone, J.~Erlich and J.~A.~Tan,
  Phys.\ Rev.\  D {\bf 75}, 075005 (2007)
  [arXiv:hep-ph/0612242].

\bibitem{Hirayama:2007hz}
  T.~Hirayama and K.~Yoshioka,
  JHEP {\bf 0710}, 002 (2007)
  [arXiv:0705.3533 [hep-ph]].


\bibitem{Dietrich:2009af}
  D.~D.~Dietrich, M.~Jarvinen and C.~Kouvaris,
  arXiv:0908.4357 [hep-ph].
\bibitem{Dietrich:2008ni}
  D.~D.~Dietrich and C.~Kouvaris,
   ``Constraining vectors and axial-vectors in walking technicolour by a
  Phys.\ Rev.\  D {\bf 78}, 055005 (2008)
  [arXiv:0805.1503 [hep-ph]].


\cite{Dietrich:2008up}
\bibitem{Dietrich:2008up}
  D.~D.~Dietrich and C.~Kouvaris,
  Phys.\ Rev.\  D {\bf 79}, 075004 (2009)
  [arXiv:0809.1324 [hep-ph]].
\bibitem{Hirn:2008tc}
  J.~Hirn, A.~Martin and V.~Sanz,
  Phys.\ Rev.\  D {\bf 78}, 075026 (2008)
  [arXiv:0807.2465 [hep-ph]].
\bibitem{Fabbrichesi:2008ga}
  M.~Fabbrichesi, M.~Piai and L.~Vecchi,
   ``Dynamical electro-weak symmetry breaking from deformed AdS: vector mesons
  Phys.\ Rev.\  D {\bf 78}, 045009 (2008)
  [arXiv:0804.0124 [hep-ph]].
\bibitem{Nunez:2008wi}
  C.~Nunez, I.~Papadimitriou and M.~Piai,
  arXiv:0812.3655 [hep-th].



\bibitem{Kurachi:2007at}
 M.~Kurachi, R.~Shrock and K.~Yamawaki,
 Phys.\ Rev.\  D {\bf 76}, 035003 (2007)
 [arXiv:0704.3481 [hep-ph]].

\bibitem{Georgi:2007ek}
 H.~Georgi,
 Phys.\ Rev.\ Lett.\  {\bf 98}, 221601 (2007)
 [arXiv:hep-ph/0703260].

\bibitem{Sannino:2008nv}
 F.~Sannino and R.~Zwicky,
 Phys.\ Rev.\  D {\bf 79}, 015016 (2009)
 [arXiv:0810.2686 [hep-ph]].

\bibitem{Zwicky:2007vv}
 R.~Zwicky,
 Phys.\ Rev.\  D {\bf 77}, 036004 (2008)
 [arXiv:0707.0677 [hep-ph]].

\bibitem{Georgi:2007si}
 H.~Georgi,
 Phys.\ Lett.\  B {\bf 650}, 275 (2007)
 [arXiv:0704.2457 [hep-ph]].

\bibitem{Cheung:2007zza}
 K.~Cheung, W.~Y.~Keung and T.~C.~Yuan,
 Phys.\ Rev.\ Lett.\  {\bf 99}, 051803 (2007)
 [arXiv:0704.2588 [hep-ph]].

\bibitem{Cheung:2008xu}
 K.~Cheung, W.~Y.~Keung and T.~C.~Yuan,
 AIP Conf.\ Proc.\  {\bf 1078}, 156 (2009)
 [arXiv:0809.0995 [hep-ph]].

\bibitem{Fox:2007sy}
 P.~J.~Fox, A.~Rajaraman and Y.~Shirman,
 Phys.\ Rev.\  D {\bf 76}, 075004 (2007)
 [arXiv:0705.3092 [hep-ph]].

\bibitem{Delgado:2007dx}
 A.~Delgado, J.~R.~Espinosa and M.~Quiros,
 JHEP {\bf 0710}, 094 (2007)
 [arXiv:0707.4309 [hep-ph]].

\bibitem{Delgado:2008rq}
 A.~Delgado, J.~R.~Espinosa, J.~M.~No and M.~Quiros,
 JHEP {\bf 0804}, 028 (2008)
 [arXiv:0802.2680 [hep-ph]].

\bibitem{Banks:1981nn}
 T.~Banks and A.~Zaks,
 Nucl.\ Phys.\  B {\bf 196}, 189 (1982).

\bibitem{Polchinski:1987dy}
 J.~Polchinski,
 Nucl.\ Phys.\  B {\bf 303}, 226 (1988).

\bibitem{Georgi:2008pq}
 H.~Georgi and Y.~Kats,
 Phys.\ Rev.\ Lett.\  {\bf 101}, 131603 (2008)
 [arXiv:0805.3953 [hep-ph]].

\bibitem{Grinstein:2008qk}
 B.~Grinstein, K.~A.~Intriligator and I.~Z.~Rothstein,
 Phys.\ Lett.\  B {\bf 662}, 367 (2008)
 [arXiv:0801.1140 [hep-ph]].

\bibitem{Mack:1975je}
 G.~Mack,
 Commun.\ Math.\ Phys.\  {\bf 55}, 1 (1977).

\bibitem{Appelquist:1984rr}
 T.~Appelquist, M.~J.~Bowick, E.~Cohler and A.~I.~Hauser,
 Phys.\ Rev.\  D {\bf 31}, 1676 (1985).

\bibitem{vanderBij:2006pg}
 J.~J.~van der Bij and S.~Dilcher,
 Phys.\ Lett.\  B {\bf 638}, 234 (2006)
 [arXiv:hep-ph/0605008].

\bibitem{Delgado:2008px}
 A.~Delgado, J.~R.~Espinosa, J.~M.~No and M.~Quiros,
 JHEP {\bf 0811}, 071 (2008)
 [arXiv:0804.4574 [hep-ph]].

\bibitem{vanderBij:2007um}
 J.~J.~van der Bij and S.~Dilcher,
 Phys.\ Lett.\  B {\bf 655}, 183 (2007)
 [arXiv:0707.1817 [hep-ph]].

\bibitem{Stancato:2008mp}
 D.~Stancato and J.~Terning,
 arXiv:0807.3961 [hep-ph].

\bibitem{Calmet:2008xe}
 X.~Calmet, N.~G.~Deshpande, X.~G.~He and S.~D.~H.~Hsu,
 Phys.\ Rev.\  D {\bf 79}, 055021 (2009)
 [arXiv:0810.2155 [hep-ph]].

\bibitem{Belyaev:2008yj}
 A.~Belyaev, R.~Foadi, M.~T.~Frandsen, M.~Jarvinen, F.~Sannino and A.~Pukhov,
 Phys.\ Rev.\  D {\bf 79}, 035006 (2009)
 [arXiv:0809.0793 [hep-ph]].

\bibitem{Zerwekh:2005wh}
 A.~R.~Zerwekh,
 Eur.\ Phys.\ J.\  C {\bf 46}, 791 (2006)
 [arXiv:hep-ph/0512261].

\bibitem{Agashe:2008jb}
 K.~Agashe, S.~Gopalakrishna, T.~Han, G.~Y.~Huang and A.~Soni,
 Phys.\ Rev.\  D {\bf 80}, 075007 (2009)
 [arXiv:0810.1497 [hep-ph]].

\bibitem{Appelquist:2007hu}
 T.~Appelquist, G.~T.~Fleming and E.~T.~Neil,
 Phys.\ Rev.\ Lett.\  {\bf 100}, 171607 (2008)
 [Erratum-ibid.\  {\bf 102}, 149902 (2009)]
 [arXiv:0712.0609 [hep-ph]].

\bibitem{Deuzeman:2008sc}
 A.~Deuzeman, M.~P.~Lombardo and E.~Pallante,
 Phys.\ Lett.\  B {\bf 670}, 41 (2008)
 [arXiv:0804.2905 [hep-lat]].

\bibitem{Fodor:2008hn}
 Z.~Fodor, K.~Holland, J.~Kuti, D.~Nogradi and C.~Schroeder,
 arXiv:0809.4890 [hep-lat].

\bibitem{Shamir:2008pb}
 Y.~Shamir, B.~Svetitsky and T.~DeGrand,
 Phys.\ Rev.\  D {\bf 78}, 031502 (2008)
 [arXiv:0803.1707 [hep-lat]].

\bibitem{Svetitsky:2008bw}
 B.~Svetitsky, Y.~Shamir and T.~DeGrand,
 arXiv:0809.2885 [hep-lat].

\bibitem{DeGrand:2008dh}
 T.~DeGrand, Y.~Shamir and B.~Svetitsky,
 arXiv:0809.2953 [hep-lat].

\bibitem{Fodor:2008hm}
 Z.~Fodor, K.~Holland, J.~Kuti, D.~Nogradi and C.~Schroeder,
 arXiv:0809.4888 [hep-lat].

\bibitem{Bursa:2009we}
 F.~Bursa, L.~Del Debbio, L.~Keegan, C.~Pica and T.~Pickup,
 arXiv:0910.4535 [hep-ph].

\bibitem{Lucini:2009an}
  B.~Lucini,
  arXiv:0911.0020 [hep-ph].


\bibitem{DeGrand:2009hu}
 T.~DeGrand,
 arXiv:0910.3072 [hep-lat].

\bibitem{Christensen:2006rf}
 N.~D.~Christensen and R.~Shrock,
 Phys.\ Rev.\  D {\bf 74}, 015004 (2006)
 [arXiv:hep-ph/0603149].

\bibitem{Christensen:2005bt}
 N.~D.~Christensen and R.~Shrock,
 Phys.\ Rev.\  D {\bf 72}, 035013 (2005)
 [arXiv:hep-ph/0506155].

\bibitem{Appelquist:2004df}
 T.~Appelquist and R.~Shrock,
 AIP Conf.\ Proc.\  {\bf 721}, 261 (2004).

\bibitem{Appelquist:2003hn}
 T.~Appelquist, M.~Piai and R.~Shrock,
 Phys.\ Rev.\  D {\bf 69}, 015002 (2004)
 [arXiv:hep-ph/0308061].

\bibitem{Appelquist:2003uu}
 T.~Appelquist and R.~Shrock,
 Phys.\ Rev.\ Lett.\  {\bf 90}, 201801 (2003)
 [arXiv:hep-ph/0301108].

\bibitem{Gudnason:2006mk}
 S.~B.~Gudnason, T.~A.~Ryttov and F.~Sannino,
 Phys.\ Rev.\  D {\bf 76}, 015005 (2007)
 [arXiv:hep-ph/0612230].

\bibitem{Pospelov:2007mp}
 M.~Pospelov, A.~Ritz and M.~B.~Voloshin,
 Phys.\ Lett.\  B {\bf 662}, 53 (2008)
 [arXiv:0711.4866 [hep-ph]].

\bibitem{Kikuchi:2007az}
 T.~Kikuchi and N.~Okada,
 Phys.\ Lett.\  B {\bf 665}, 186 (2008)
 [arXiv:0711.1506 [hep-ph]].

\bibitem{Spergel:2003cb}
 D.~N.~Spergel {\it et al.}  [WMAP Collaboration],
 Astrophys.\ J.\ Suppl.\  {\bf 148}, 175 (2003)
 [arXiv:astro-ph/0302209].

\bibitem{Frieman:2008sn}
 J.~Frieman, M.~Turner and D.~Huterer,
 Ann.\ Rev.\ Astron.\ Astrophys.\  {\bf 46}, 385 (2008)
 [arXiv:0803.0982 [astro-ph]].

\bibitem{Frieman:2002wi}
 J.~A.~Frieman, D.~Huterer, E.~V.~Linder and M.~S.~Turner,
 Phys.\ Rev.\  D {\bf 67}, 083505 (2003)
 [arXiv:astro-ph/0208100].

\bibitem{Komatsu:2008hk}
 E.~Komatsu {\it et al.}  [WMAP Collaboration],
 Astrophys.\ J.\ Suppl.\  {\bf 180}, 330 (2009)
 [arXiv:0803.0547 [astro-ph]].

\bibitem{Kaplan:1991ah}
 D.~B.~Kaplan,
 Phys.\ Rev.\ Lett.\  {\bf 68}, 741 (1992).

\bibitem{Barr:1990ca}
 S.~M.~Barr, R.~S.~Chivukula and E.~Farhi,
 Phys.\ Lett.\  B {\bf 241}, 387 (1990).

\bibitem{Harvey:1990qw}
 J.~A.~Harvey and M.~S.~Turner,
 Phys.\ Rev.\  D {\bf 42}, 3344 (1990).

\bibitem{Ryttov:2008xe}
 T.~A.~Ryttov and F.~Sannino,
 Phys.\ Rev.\  D {\bf 78}, 115010 (2008)
 [arXiv:0809.0713 [hep-ph]].

\bibitem{Gudnason:2006yj}
 S.~B.~Gudnason, C.~Kouvaris and F.~Sannino,
 Phys.\ Rev.\  D {\bf 74}, 095008 (2006)
 [arXiv:hep-ph/0608055].


\bibitem{Burnier:2005hp}
 Y.~Burnier, M.~Laine and M.~Shaposhnikov,
 JCAP {\bf 0602}, 007 (2006)
 [arXiv:hep-ph/0511246].

\bibitem{Nardi:2008ix}
 E.~Nardi, F.~Sannino and A.~Strumia,
 JCAP {\bf 0901}, 043 (2009)
 [arXiv:0811.4153 [hep-ph]].

\bibitem{Nussinov:1985xr}
 S.~Nussinov,
 Phys.\ Lett.\  B {\bf 165}, 55 (1985).

\bibitem{Foadi:2008qv}
 R.~Foadi, M.~T.~Frandsen and F.~Sannino,
 Phys.\ Rev.\  D {\bf 80}, 037702 (2009)
 [arXiv:0812.3406 [hep-ph]].


\bibitem{Kouvaris:2007iq}
  C.~Kouvaris,
  Phys.\ Rev.\  D {\bf 76}, 015011 (2007)
  [arXiv:hep-ph/0703266].
  C.~Kouvaris,
  Phys.\ Rev.\  D {\bf 77}, 023006 (2008)
  [arXiv:0708.2362 [astro-ph]].
  C.~Kouvaris,
  Phys.\ Rev.\  D {\bf 78}, 075024 (2008)
  [arXiv:0807.3124 [hep-ph]].
  M.~Y.~Khlopov and C.~Kouvaris,
  Phys.\ Rev.\  D {\bf 77}, 065002 (2008)
  [arXiv:0710.2189 [astro-ph]].
\bibitem{Khlopov:2008ty}
  M.~Y.~Khlopov and C.~Kouvaris,
  Phys.\ Rev.\  D {\bf 78}, 065040 (2008)
  [arXiv:0806.1191 [astro-ph]].
  K.~Belotsky, M.~Khlopov and C.~Kouvaris,
  Phys.\ Rev.\  D {\bf 79}, 083520 (2009)
  [arXiv:0810.2022 [astro-ph]].

\bibitem{Kainulainen:2006wq}
 K.~Kainulainen, K.~Tuominen and J.~Virkajarvi,
 Phys.\ Rev.\  D {\bf 75}, 085003 (2007)
 [arXiv:hep-ph/0612247].


\bibitem{Cline:2008hr}
 J.~M.~Cline, M.~Jarvinen and F.~Sannino,
 Phys.\ Rev.\  D {\bf 78}, 075027 (2008)
 [arXiv:0808.1512 [hep-ph]].

\bibitem{Jarvinen:2009wr}
 M.~Jarvinen, T.~A.~Ryttov and F.~Sannino,
 Phys.\ Lett.\  B {\bf 680}, 251 (2009)
 [arXiv:0901.0496 [hep-ph]].

\bibitem{Jarvinen:2009pk}
 M.~Jarvinen, T.~A.~Ryttov and F.~Sannino,
 Phys.\ Rev.\  D {\bf 79}, 095008 (2009)
 [arXiv:0903.3115 [hep-ph]].

\bibitem{Shaposhnikov:1986jp}
 M.~E.~Shaposhnikov,
 JETP Lett.\  {\bf 44}, 465 (1986)
 [Pisma Zh.\ Eksp.\ Teor.\ Fiz.\  {\bf 44}, 364 (1986)].



\bibitem{Kajantie:1995kf}
  K.~Kajantie, M.~Laine, K.~Rummukainen and M.~E.~Shaposhnikov,
  Nucl.\ Phys.\  B {\bf 466}, 189 (1996)
  [arXiv:hep-lat/9510020].


\bibitem{Rubakov:1984ba}
 V.~A.~Rubakov,
 Nucl.\ Phys.\  B {\bf 256}, 509 (1985).

\bibitem{Hisano:1992jj}
 J.~Hisano, H.~Murayama and T.~Yanagida,
 Nucl.\ Phys.\  B {\bf 402}, 46 (1993)
 [arXiv:hep-ph/9207279].

\bibitem{Fukugita:1986hr}
 M.~Fukugita and T.~Yanagida,
 Phys.\ Lett.\  B {\bf 174}, 45 (1986).

\bibitem{Appelquist:2002me}
 T.~Appelquist and R.~Shrock,
 Phys.\ Lett.\  B {\bf 548}, 204 (2002)
 [arXiv:hep-ph/0204141].

\bibitem{Witten:1982fp}
 E.~Witten,
 Phys.\ Lett.\  B {\bf 117}, 324 (1982).

\bibitem{Frandsen:2009fs}
 M.~T.~Frandsen, I.~Masina and F.~Sannino,
 arXiv:0905.1331 [hep-ph].

\bibitem{Bagnasco:1993st}
 J.~Bagnasco, M.~Dine and S.~D.~Thomas,
 Phys.\ Lett.\  B {\bf 320}, 99 (1994)
 [arXiv:hep-ph/9310290].

\bibitem{McDonald:1993ex}
 J.~McDonald,
 Phys.\ Rev.\  D {\bf 50}, 3637 (1994)
 [arXiv:hep-ph/0702143].

\bibitem{Oikonomou:2006mh}
 V.~K.~Oikonomou, J.~D.~Vergados and C.~C.~Moustakidis,
 Nucl.\ Phys.\  B {\bf 773}, 19 (2007)
 [arXiv:hep-ph/0612293].

\bibitem{Shifman:1978zn}
 M.~A.~Shifman, A.~I.~Vainshtein and V.~I.~Zakharov,
 Phys.\ Lett.\  B {\bf 78}, 443 (1978).

\bibitem{Andreas:2008xy}
 S.~Andreas, T.~Hambye and M.~H.~G.~Tytgat,
 JCAP {\bf 0810}, 034 (2008)
 [arXiv:0808.0255 [hep-ph]].

\bibitem{Hong:2004td}
 D.~K.~Hong, S.~D.~H.~Hsu and F.~Sannino,
 Phys.\ Lett.\  B {\bf 597}, 89 (2004)
 [arXiv:hep-ph/0406200].

\bibitem{Ellis:1991ef}
 J.~R.~Ellis and R.~A.~Flores,
 Phys.\ Lett.\  B {\bf 263}, 259 (1991).

\bibitem{Lane:1989ej}
 K.~D.~Lane and E.~Eichten,
 Phys.\ Lett.\  B {\bf 222}, 274 (1989).

\bibitem{Ryttov:2007sr}
 T.~A.~Ryttov and F.~Sannino,
 Phys.\ Rev.\  D {\bf 76}, 105004 (2007)
 [arXiv:0707.3166 [hep-th]].

\bibitem{Appelquist:1988yc}
 T.~Appelquist, K.~D.~Lane and U.~Mahanta,
 Phys.\ Rev.\ Lett.\  {\bf 61}, 1553 (1988).

\bibitem{Cohen:1988sq}
 A.~G.~Cohen and H.~Georgi,
 Nucl.\ Phys.\  B {\bf 314}, 7 (1989).

\bibitem{Miransky:1996pd}
 V.~A.~Miransky and K.~Yamawaki,
 Phys.\ Rev.\  D {\bf 55}, 5051 (1997)
 [Erratum-ibid.\  D {\bf 56}, 3768 (1997)]
 [arXiv:hep-th/9611142].

\bibitem{Appelquist:1999hr}
 T.~Appelquist, A.~G.~Cohen and M.~Schmaltz,
 Phys.\ Rev.\  D {\bf 60}, 045003 (1999)
 [arXiv:hep-th/9901109].

\bibitem{Sannino:2005sk}
 F.~Sannino,
 Phys.\ Rev.\  D {\bf 72}, 125006 (2005)
 [arXiv:hep-th/0507251].

\bibitem{Sannino:2009aw}
 F.~Sannino,
 Phys.\ Rev.\  D {\bf 79}, 096007 (2009)
 [arXiv:0902.3494 [hep-ph]].

\bibitem{Intriligator:1995au}
 K.~A.~Intriligator and N.~Seiberg,
 Nucl.\ Phys.\ Proc.\ Suppl.\  {\bf 45BC}, 1 (1996)
 [arXiv:hep-th/9509066].

\bibitem{Novikov:1983uc}
 V.~A.~Novikov, M.~A.~Shifman, A.~I.~Vainshtein and V.~I.~Zakharov,
 Nucl.\ Phys.\  B {\bf 229}, 381 (1983).

\bibitem{Shifman:1986zi}
 M.~A.~Shifman and A.~I.~Vainshtein,
 Nucl.\ Phys.\  B {\bf 277}, 456 (1986)
 [Sov.\ Phys.\ JETP {\bf 64}, 428 (1986\ ZETFA,91,723-744.1986)].


\bibitem{Catterall:2007yx}
 S.~Catterall and F.~Sannino,
 Phys.\ Rev.\  D {\bf 76}, 034504 (2007)
 [arXiv:0705.1664 [hep-lat]].

\bibitem{DelDebbio:2008wb}
 L.~Del Debbio, M.~T.~Frandsen, H.~Panagopoulos and F.~Sannino,
 JHEP {\bf 0806}, 007 (2008)
 [arXiv:0802.0891 [hep-lat]].

\bibitem{Catterall:2008qk}
 S.~Catterall, J.~Giedt, F.~Sannino and J.~Schneible,
 JHEP {\bf 0811}, 009 (2008)
 [arXiv:0807.0792 [hep-lat]].

\bibitem{Lucini:2007sa}
 B.~Lucini and G.~Moraitis,
 PoS {\bf LAT2007}, 058 (2007)
 [arXiv:0710.1533 [hep-lat]].

\bibitem{Luscher:1992zx}
 M.~Luscher, R.~Sommer, U.~Wolff and P.~Weisz,
 Nucl.\ Phys.\  B {\bf 389}, 247 (1993)
 [arXiv:hep-lat/9207010].

\bibitem{Luscher:1993gh}
 M.~Luscher, R.~Sommer, P.~Weisz and U.~Wolff,
 Nucl.\ Phys.\  B {\bf 413}, 481 (1994)
 [arXiv:hep-lat/9309005].

\bibitem{Appelquist:2009ty}
 T.~Appelquist, G.~T.~Fleming and E.~T.~Neil,
 Phys.\ Rev.\  D {\bf 79}, 076010 (2009)
 [arXiv:0901.3766 [hep-ph]].

\bibitem{Jones:1983ip}
 D.~R.~T.~Jones,
 Phys.\ Lett.\  B {\bf 123}, 45 (1983).

\bibitem{Flato:1983te}
 M.~Flato and C.~Fronsdal,
 Lett.\ Math.\ Phys.\  {\bf 8}, 159 (1984).

\bibitem{Dobrev:1985qv}
 V.~K.~Dobrev and V.~B.~Petkova,
 Phys.\ Lett.\  B {\bf 162}, 127 (1985).

\bibitem{Maskawa:1974vs}
 T.~Maskawa and H.~Nakajima,
 Prog.\ Theor.\ Phys.\  {\bf 52}, 1326 (1974).

\bibitem{Fukuda:1976zb}
 R.~Fukuda and T.~Kugo,
 Nucl.\ Phys.\  B {\bf 117}, 250 (1976).

\bibitem{Appelquist:1999vs}
 T.~Appelquist, A.~G.~Cohen, M.~Schmaltz and R.~Shrock,
 Phys.\ Lett.\  B {\bf 459}, 235 (1999)
 [arXiv:hep-th/9904172].

\bibitem{Appelquist:2000qg}
 T.~Appelquist, Z.~y.~Duan and F.~Sannino,
 Phys.\ Rev.\  D {\bf 61}, 125009 (2000)
 [arXiv:hep-ph/0001043].

\bibitem{Ryttov:2009yw}
 T.~A.~Ryttov and F.~Sannino,
 arXiv:0906.0307 [hep-ph].

\bibitem{Luty:2004ye}
 M.~A.~Luty and T.~Okui,
 JHEP {\bf 0609}, 070 (2006)
 [arXiv:hep-ph/0409274].

\bibitem{Evans:2005pu}
 N.~Evans and F.~Sannino,
 arXiv:hep-ph/0512080.

\bibitem{Appelquist:1997fp}
 T.~Appelquist, J.~Terning and L.~C.~R.~Wijewardhana,
 Phys.\ Rev.\ Lett.\  {\bf 79}, 2767 (1997)
 [arXiv:hep-ph/9706238].

\bibitem{Sugimoto:1999tx}
 S.~Sugimoto,
 Prog.\ Theor.\ Phys.\  {\bf 102}, 685 (1999)
 [arXiv:hep-th/9905159].

\bibitem{Uranga:1999ib}
 A.~M.~Uranga,
 JHEP {\bf 0002}, 041 (2000)
 [arXiv:hep-th/9912145].

\bibitem{Armoni:2007jt}
 A.~Armoni,
 JHEP {\bf 0704}, 046 (2007)
 [arXiv:hep-th/0703229].

\bibitem{Hietanen:2009az}
 A.~J.~Hietanen, K.~Rummukainen and K.~Tuominen,
 arXiv:0904.0864 [hep-lat].

\bibitem{Stephanov:2007ry}
 M.~A.~Stephanov,
 Phys.\ Rev.\  D {\bf 76}, 035008 (2007)
 [arXiv:0705.3049 [hep-ph]].

\bibitem{Luty:2008vs}
 M.~A.~Luty,
 JHEP {\bf 0904}, 050 (2009)
 [arXiv:0806.1235 [hep-ph]].

\bibitem{Sannino:2008pz}
 F.~Sannino,
 Phys.\ Rev.\  D {\bf 80}, 017901 (2009)
 [arXiv:0811.0616 [hep-ph]].

\bibitem{Glashow:1967rx}
 S.~L.~Glashow and S.~Weinberg,
 Phys.\ Rev.\ Lett.\  {\bf 20}, 224 (1968).

\bibitem{GellMann:1968rz}
 M.~Gell-Mann, R.~J.~Oakes and B.~Renner,
 Phys.\ Rev.\  {\bf 175}, 2195 (1968).

\bibitem{Dashen:1969eg}
 R.~F.~Dashen,
 Phys.\ Rev.\  {\bf 183}, 1245 (1969).

\bibitem{Gasser:1987ah}
 J.~Gasser and H.~Leutwyler,
 Phys.\ Lett.\  B {\bf 188}, 477 (1987).

\bibitem{Seiberg:1994bz}
 N.~Seiberg,
 Phys.\ Rev.\  D {\bf 49}, 6857 (1994)
 [arXiv:hep-th/9402044].

\bibitem{Seiberg:1994pq}
 N.~Seiberg,
 Nucl.\ Phys.\  B {\bf 435}, 129 (1995)
 [arXiv:hep-th/9411149].

\bibitem{Sannino:2009qc}
 F.~Sannino,
 Phys.\ Rev.\  D {\bf 80}, 065011 (2009)
 [arXiv:0907.1364 [hep-th]].

\bibitem{Sannino:2009me}
 F.~Sannino,
 arXiv:0909.4584 [hep-th].

\bibitem{Terning:1997xy}
 J.~Terning,
 Phys.\ Rev.\ Lett.\  {\bf 80}, 2517 (1998)
 [arXiv:hep-th/9706074].

\bibitem{Appelquist:1996dq}
 T.~Appelquist, J.~Terning and L.~C.~R.~Wijewardhana,
 Phys.\ Rev.\ Lett.\  {\bf 77}, 1214 (1996)
 [arXiv:hep-ph/9602385].

\bibitem{DelDebbio:2008zf}
 L.~Del Debbio, A.~Patella and C.~Pica,
 arXiv:0805.2058 [hep-lat].

\bibitem{Hietanen:2008vc}
 A.~Hietanen, J.~Rantaharju, K.~Rummukainen and K.~Tuominen,
 PoS {\bf LATTICE2008}, 065 (2008)
 [arXiv:0810.3722 [hep-lat]].

\bibitem{Hietanen:2008mr}
 A.~J.~Hietanen, J.~Rantaharju, K.~Rummukainen and K.~Tuominen,
 JHEP {\bf 0905}, 025 (2009)
 [arXiv:0812.1467 [hep-lat]].

\bibitem{DelDebbio:2008tv}
 L.~Del Debbio, A.~Patella and C.~Pica,
 arXiv:0812.0570 [hep-lat].

\bibitem{DeGrand:2008kx}
 T.~DeGrand, Y.~Shamir and B.~Svetitsky,
 Phys.\ Rev.\  D {\bf 79}, 034501 (2009)
 [arXiv:0812.1427 [hep-lat]].

\bibitem{Deuzeman:2009mh}
 A.~Deuzeman, M.~P.~Lombardo and E.~Pallante,
 arXiv:0904.4662 [hep-ph].

\bibitem{DeGrand:2009et}
 T.~DeGrand,
 arXiv:0906.4543 [hep-lat].

\bibitem{Hasenfratz:2009ea}
 A.~Hasenfratz,
 Phys.\ Rev.\  D {\bf 80}, 034505 (2009)
 [arXiv:0907.0919 [hep-lat]].

\bibitem{Poppitz:2009uq}
 E.~Poppitz and M.~Unsal,
 JHEP {\bf 0909}, 050 (2009)
 [arXiv:0906.5156 [hep-th]].

\bibitem{Poppitz:2008hr}
 E.~Poppitz and M.~Unsal,
 JHEP {\bf 0903}, 027 (2009)
 [arXiv:0812.2085 [hep-th]].

\bibitem{DelDebbio:2009fd}
 L.~Del Debbio, B.~Lucini, A.~Patella, C.~Pica and A.~Rago,
 Phys.\ Rev.\  D {\bf 80}, 074507 (2009)
 [arXiv:0907.3896 [hep-lat]].

\bibitem{Ball:1988xg}
 R.~D.~Ball,
 Phys.\ Rept.\  {\bf 182}, 1 (1989).

\bibitem{Raby:1979my}
 S.~Raby, S.~Dimopoulos and L.~Susskind,
 Nucl.\ Phys.\  B {\bf 169}, 373 (1980).

\bibitem{Poppitz:1998vd}
 E.~Poppitz and S.~P.~Trivedi,
 Ann.\ Rev.\ Nucl.\ Part.\ Sci.\  {\bf 48}, 307 (1998)
 [arXiv:hep-th/9803107].

\bibitem{Bars:1981se}
 I.~Bars and S.~Yankielowicz,
 Phys.\ Lett.\  B {\bf 101}, 159 (1981).

\bibitem{Eichten:1985fs}
 E.~Eichten, R.~D.~Peccei, J.~Preskill and D.~Zeppenfeld,
 Nucl.\ Phys.\  B {\bf 268}, 161 (1986).

\bibitem{Affleck:1983vc}
 I.~Affleck, M.~Dine and N.~Seiberg,
 Phys.\ Lett.\  B {\bf 137}, 187 (1984).

\bibitem{Meurice:1984ai}
 Y.~Meurice and G.~Veneziano,
 Phys.\ Lett.\  B {\bf 141}, 69 (1984).

\bibitem{Bagger:1994hh}
 J.~Bagger, E.~Poppitz and L.~Randall,
 Nucl.\ Phys.\  B {\bf 426}, 3 (1994)
 [arXiv:hep-ph/9405345].

\bibitem{Roux:1999dq}
 F.~S.~Roux, T.~Torma and B.~Holdom,
 Phys.\ Rev.\  D {\bf 61}, 056009 (2000)
 [arXiv:hep-ph/9907540].

\bibitem{Appelquist:1997gq}
 T.~Appelquist, A.~Nyffeler and S.~B.~Selipsky,
 Phys.\ Lett.\  B {\bf 425}, 300 (1998)
 [arXiv:hep-th/9709177].

\bibitem{Schafer:1995pz}
 T.~Schafer and E.~V.~Shuryak,
 Phys.\ Rev.\  D {\bf 53}, 6522 (1996)
 [arXiv:hep-ph/9509337].

\bibitem{Grunberg:2000ap}
 G.~Grunberg,
 Phys.\ Rev.\  D {\bf 65}, 021701 (2002)
 [arXiv:hep-ph/0009272].

\bibitem{Gardi:1998ch}
 E.~Gardi and G.~Grunberg,
 JHEP {\bf 9903}, 024 (1999)
 [arXiv:hep-th/9810192].

\bibitem{Grunberg:1996hu}
 G.~Grunberg,
 arXiv:hep-ph/9608375.

\bibitem{Poppitz:2009tw}
 E.~Poppitz and M.~Unsal,
 arXiv:0910.1245 [hep-th].

\bibitem{Kaplan:2009kr}
 D.~B.~Kaplan, J.~W.~Lee, D.~T.~Son and M.~A.~Stephanov,
 arXiv:0905.4752 [hep-th].

\bibitem{Antipin:2009wr}
 O.~Antipin and K.~Tuominen,
 arXiv:0909.4879 [hep-ph].

\bibitem{Dietrich:2009ns}
 D.~D.~Dietrich,
 Phys.\ Rev.\  D {\bf 80}, 065032 (2009)
 [arXiv:0908.1364 [hep-th]].

\bibitem{Pica:2009hc}
 C.~Pica, L.~Del Debbio, B.~Lucini, A.~Patella and A.~Rago,
 arXiv:0909.3178 [hep-lat].

\bibitem{Catterall:2009sb}
 S.~Catterall, J.~Giedt, F.~Sannino and J.~Schneible,
 arXiv:0910.4387 [hep-lat].

\bibitem{Fodor:2009ar}
 Z.~Fodor, K.~Holland, J.~Kuti, D.~Nogradi and C.~Schroeder,
 arXiv:0908.2466 [hep-lat].

\bibitem{Appelquist:2009ka}
 T.~Appelquist {\it et al.},
 arXiv:0910.2224 [hep-ph].

\bibitem{Fodor:2009wk}
 Z.~Fodor, K.~Holland, J.~Kuti, D.~Nogradi and C.~Schroeder,
 arXiv:0907.4562 [hep-lat].

\bibitem{Appelquist:1999dq}
 T.~Appelquist, P.~S.~Rodrigues da Silva and F.~Sannino,
 Phys.\ Rev.\  D {\bf 60}, 116007 (1999)
 [arXiv:hep-ph/9906555].

\bibitem{Casalbuoni:1995qt}
 R.~Casalbuoni, A.~Deandrea, S.~De Curtis, D.~Dominici, R.~Gatto and M.~Grazzini,
 Phys.\ Rev.\  D {\bf 53}, 5201 (1996)
 [arXiv:hep-ph/9510431].

\bibitem{Randall:1992vt}
 L.~Randall,
 Nucl.\ Phys.\  B {\bf 403}, 122 (1993)
 [arXiv:hep-ph/9210231].

\bibitem{Kagan:1990az}
 A.~Kagan and S.~Samuel,
 Phys.\ Lett.\  B {\bf 252}, 605 (1990).

\bibitem{Carone:1993xc}
 C.~D.~Carone and H.~Georgi,
 Phys.\ Rev.\  D {\bf 49}, 1427 (1994)
 [arXiv:hep-ph/9308205].



\bibitem{Dietrich:2009ix}
  D.~D.~Dietrich and M.~Jarvinen,
  Phys.\ Rev.\  D {\bf 79}, 057903 (2009)
  [arXiv:0901.3528 [hep-ph]].

\bibitem{Foadi:2007se}
 R.~Foadi, M.~T.~Frandsen and F.~Sannino,
 Phys.\ Rev.\  D {\bf 77}, 097702 (2008)
 [arXiv:0712.1948 [hep-ph]].

\bibitem{Duan:2000dy}
 Z.~y.~Duan, P.~S.~Rodrigues da Silva and F.~Sannino,
 Nucl.\ Phys.\  B {\bf 592}, 371 (2001)
 [arXiv:hep-ph/0001303].

\bibitem{Eichten:2007sx}
 E.~Eichten and K.~Lane,
 Phys.\ Lett.\  B {\bf 669}, 235 (2008)
 [arXiv:0706.2339 [hep-ph]].

\bibitem{Casalbuoni:1988xm}
 R.~Casalbuoni, S.~De Curtis, D.~Dominici, F.~Feruglio and R.~Gatto,
 Int.\ J.\ Mod.\ Phys.\  A {\bf 4}, 1065 (1989).

\bibitem{Casalbuoni:1995yb}
 R.~Casalbuoni, A.~Deandrea, S.~De Curtis, D.~Dominici, F.~Feruglio, R.~Gatto and M.~Grazzini,
 Phys.\ Lett.\  B {\bf 349}, 533 (1995)
 [arXiv:hep-ph/9502247].

\bibitem{Casalbuoni:2007dk}
 R.~Casalbuoni, F.~Coradeschi, S.~De Curtis and D.~Dominici,
 Phys.\ Rev.\  D {\bf 77}, 095005 (2008)
 [arXiv:0710.3057 [hep-ph]].

\bibitem{Zerwekh:2007pw}
 A.~R.~Zerwekh, C.~O.~Dib and R.~Rosenfeld,
 Phys.\ Rev.\  D {\bf 75}, 097702 (2007)
 [arXiv:hep-ph/0702167].

\bibitem{Christensen:2005cb}
 N.~D.~Christensen and R.~Shrock,
 Phys.\ Lett.\  B {\bf 632}, 92 (2006)
 [arXiv:hep-ph/0509109].

\bibitem{Chivukula:1995dc}
 R.~S.~Chivukula, B.~A.~Dobrescu and J.~Terning,
 Phys.\ Lett.\  B {\bf 353}, 289 (1995)
 [arXiv:hep-ph/9503203].

\bibitem{Evans:1994fb}
 N.~J.~Evans,
 Phys.\ Lett.\  B {\bf 331}, 378 (1994)
 [arXiv:hep-ph/9403318].

\bibitem{Miransky:1989nu}
 V.~A.~Miransky, T.~Nonoyama and K.~Yamawaki,
 Mod.\ Phys.\ Lett.\  A {\bf 4}, 1409 (1989).

\bibitem{Hsu:2004mf}
 S.~D.~H.~Hsu and F.~Sannino,
 Phys.\ Lett.\  B {\bf 605}, 369 (2005)
 [arXiv:hep-ph/0408319].

\bibitem{ArkaniHamed:2004fb}
 N.~Arkani-Hamed and S.~Dimopoulos,
 JHEP {\bf 0506}, 073 (2005)
 [arXiv:hep-th/0405159].

\bibitem{Giudice:2004tc}
 G.~F.~Giudice and A.~Romanino,
 Nucl.\ Phys.\  B {\bf 699}, 65 (2004)
 [Erratum-ibid.\  B {\bf 706}, 65 (2005)]
 [arXiv:hep-ph/0406088].

\bibitem{Bijnens:2009qm}
 J.~Bijnens and J.~Lu,
 arXiv:0910.5424 [hep-ph].

\bibitem{Antipin:2009ks}
 O.~Antipin, M.~Heikinheimo and K.~Tuominen,
 JHEP {\bf 0910}, 018 (2009)
 [arXiv:0905.0622 [hep-ph]].

\bibitem{Antipin:2009ch}
 O.~Antipin and K.~Tuominen,
 Phys.\ Rev.\  D {\bf 79}, 075011 (2009)
 [arXiv:0901.4243 [hep-ph]].

\bibitem{Shaposhnikov:1987tw}
 M.~E.~Shaposhnikov,
 Nucl.\ Phys.\  B {\bf 287}, 757 (1987).

\bibitem{Shaposhnikov:1987pf}
 M.~E.~Shaposhnikov,
 Nucl.\ Phys.\  B {\bf 299}, 797 (1988).

\bibitem{Farrar:1993sp}
 G.~R.~Farrar and M.~E.~Shaposhnikov,
 Phys.\ Rev.\ Lett.\  {\bf 70}, 2833 (1993)
 [Erratum-ibid.\  {\bf 71}, 210 (1993)]
 [arXiv:hep-ph/9305274].

\bibitem{Farrar:1993hn}
 G.~R.~Farrar and M.~E.~Shaposhnikov,
 Phys.\ Rev.\  D {\bf 50}, 774 (1994)
 [arXiv:hep-ph/9305275].

\bibitem{Gavela:1993ts}
 M.~B.~Gavela, P.~Hernandez, J.~Orloff and O.~Pene,
 Mod.\ Phys.\ Lett.\  A {\bf 9}, 795 (1994)
 [arXiv:hep-ph/9312215].

\bibitem{Gavela:1994ds}
 M.~B.~Gavela, M.~Lozano, J.~Orloff and O.~Pene,
 Nucl.\ Phys.\  B {\bf 430}, 345 (1994)
 [arXiv:hep-ph/9406288].

\bibitem{Nelson:1991ab}
 A.~E.~Nelson, D.~B.~Kaplan and A.~G.~Cohen,
 Nucl.\ Phys.\  B {\bf 373}, 453 (1992).

\bibitem{Joyce:1994bi}
 M.~Joyce, T.~Prokopec and N.~Turok,
 Phys.\ Lett.\  B {\bf 338}, 269 (1994)
 [arXiv:hep-ph/9401352].

\bibitem{Joyce:1994fu}
 M.~Joyce, T.~Prokopec and N.~Turok,
 Phys.\ Rev.\ Lett.\  {\bf 75}, 1695 (1995)
 [Erratum-ibid.\  {\bf 75}, 3375 (1995)]
 [arXiv:hep-ph/9408339].

\bibitem{Joyce:1994zn}
 M.~Joyce, T.~Prokopec and N.~Turok,
 Phys.\ Rev.\  D {\bf 53}, 2930 (1996)
 [arXiv:hep-ph/9410281].

\bibitem{Joyce:1994zt}
 M.~Joyce, T.~Prokopec and N.~Turok,
 Phys.\ Rev.\  D {\bf 53}, 2958 (1996)
 [arXiv:hep-ph/9410282].

\bibitem{Cline:1995dg}
 J.~M.~Cline, K.~Kainulainen and A.~P.~Vischer,
 Phys.\ Rev.\  D {\bf 54}, 2451 (1996)
 [arXiv:hep-ph/9506284].

\bibitem{Cline:2006ts}
 J.~M.~Cline,
 arXiv:hep-ph/0609145.

\bibitem{Carrington:1991hz}
 M.~E.~Carrington,
 Phys.\ Rev.\  D {\bf 45}, 2933 (1992).

\bibitem{Arnold:1992fb}
 P.~Arnold,
 Phys.\ Rev.\  D {\bf 46}, 2628 (1992)
 [arXiv:hep-ph/9204228].

\bibitem{Arnold:1992rz}
 P.~Arnold and O.~Espinosa,
 Phys.\ Rev.\  D {\bf 47}, 3546 (1993)
 [Erratum-ibid.\  D {\bf 50}, 6662 (1994)]
 [arXiv:hep-ph/9212235].

\bibitem{Anderson:1991zb}
 G.~W.~Anderson and L.~J.~Hall,
 Phys.\ Rev.\  D {\bf 45}, 2685 (1992).

\bibitem{Dine:1992wr}
 M.~Dine, R.~G.~Leigh, P.~Y.~Huet, A.~D.~Linde and D.~A.~Linde,
 Phys.\ Rev.\  D {\bf 46}, 550 (1992)
 [arXiv:hep-ph/9203203].



\bibitem{Kajantie:1996mn}
  K.~Kajantie, M.~Laine, K.~Rummukainen and M.~E.~Shaposhnikov,
  Phys.\ Rev.\ Lett.\  {\bf 77}, 2887 (1996)
  [arXiv:hep-ph/9605288].


\bibitem{Rummukainen:1998as}
  K.~Rummukainen, M.~Tsypin, K.~Kajantie, M.~Laine and M.~E.~Shaposhnikov,
  Nucl.\ Phys.\  B {\bf 532}, 283 (1998)
  [arXiv:hep-lat/9805013].


\bibitem{Gynther:2005av}
  A.~Gynther and M.~Veps\"al\"ainen,
  JHEP {\bf 0603}, 011 (2006)
  [arXiv:hep-ph/0512177].

\bibitem{Gynther:2005dj}
  A.~Gynther and M.~Veps\"al\"ainen,
  JHEP {\bf 0601}, 060 (2006)
  [arXiv:hep-ph/0510375].

\bibitem{CQW}
  M.~Carena, M.~Quiros and C.~E.~M.~Wagner,
  Phys.\ Lett.\ B {\bf 380}, 81 (1996)
  [arXiv:hep-ph/9603420].

\bibitem{DR}
J.~M.~Cline and K.~Kainulainen,
  Nucl.\ Phys.\ B {\bf 482}, 73 (1996)
  [arXiv:hep-ph/9605235].

  M.~Laine,
  Nucl.\ Phys.\ B {\bf 481}, 43 (1996)
  [Erratum-ibid.\ B {\bf 548}, 637 (1999)]
  [arXiv:hep-ph/9605283].

M.~Losada,
  Phys.\ Rev.\ D {\bf 56}, 2893 (1997)
  [arXiv:hep-ph/9605266].

\bibitem{CM}
 J.~M.~Cline and G.~D.~Moore,
  Phys.\ Rev.\ Lett.\  {\bf 81}, 3315 (1998)
  [arXiv:hep-ph/9806354].

\bibitem{LR}
  M.~Laine and K.~Rummukainen,
  Phys.\ Rev.\ Lett.\  {\bf 80}, 5259 (1998)
  [arXiv:hep-ph/9804255];
  Nucl.\ Phys.\ B {\bf 535}, 423 (1998)
  [arXiv:hep-lat/9804019].




\bibitem{Shaposhnikov:1991cu}
 M.~E.~Shaposhnikov,
 Phys.\ Lett.\  B {\bf 277}, 324 (1992)
 [Erratum-ibid.\  B {\bf 282}, 483 (1992)].

\bibitem{Kuzmin:1991ft}
 V.~A.~Kuzmin, V.~A.~Rubakov and M.~E.~Shaposhnikov,
{\it  In *Moscow 1991, Proceedings, Sakharov memorial lectures in physics, vol. 2* 779-789}

\bibitem{Shaposhnikov:1991wi}
 M.~E.~Shaposhnikov,
 Nucl.\ Phys.\ Proc.\ Suppl.\  {\bf 26}, 78 (1992).

\bibitem{Cline:2002aa}
 J.~M.~Cline,
 arXiv:hep-ph/0201286.

\bibitem{Kikukawa:2007zk}
 Y.~Kikukawa, M.~Kohda and J.~Yasuda,
 Phys.\ Rev.\  D {\bf 77}, 015014 (2008)
 [arXiv:0709.2221 [hep-ph]].

\bibitem{Foadi:2008ci}
 R.~Foadi and F.~Sannino,
 Phys.\ Rev.\  D {\bf 78}, 037701 (2008)
 [arXiv:0801.0663 [hep-ph]].

\bibitem{Foadi:2008xj}
 R.~Foadi, M.~Jarvinen and F.~Sannino,
 Phys.\ Rev.\  D {\bf 79}, 035010 (2009)
 [arXiv:0811.3719 [hep-ph]].

\bibitem{Mocsy:2003qw}
 A.~Mocsy, F.~Sannino and K.~Tuominen,
 Phys.\ Rev.\ Lett.\  {\bf 92}, 182302 (2004)
 [arXiv:hep-ph/0308135].

\bibitem{Land:1992sm}
 D.~Land and E.~D.~Carlson,
 Phys.\ Lett.\  B {\bf 292}, 107 (1992)
 [arXiv:hep-ph/9208227].

\bibitem{Yao:2006px}
 W.~M.~Yao {\it et al.}  [Particle Data Group],
 J.\ Phys.\ G {\bf 33}, 1 (2006).

\bibitem{Li:2003zh}
 L.~F.~Li and F.~Wu,
 Int.\ J.\ Mod.\ Phys.\  A {\bf 19}, 3217 (2004)
 [arXiv:hep-ph/0304238].

\bibitem{Aoki:1999tw}
 S.~Aoki {\it et al.}  [JLQCD Collaboration],
 Phys.\ Rev.\  D {\bf 62}, 014506 (2000)
 [arXiv:hep-lat/9911026].

\bibitem{Ross:1985ai}
 G.~G.~Ross,
{\it  Reading, Usa: Benjamin/cummings ( 1984) 497 P. ( Frontiers In Physics, 60)}

\bibitem{Suzuki:2001rb}
 Y.~Suzuki {\it et al.}  [TITAND Working Group],
 arXiv:hep-ex/0110005.

\bibitem{Bajc:2006ia}
 B.~Bajc and G.~Senjanovic,
 JHEP {\bf 0708}, 014 (2007)
 [arXiv:hep-ph/0612029].

\bibitem{Dorsner:2006fx}
 I.~Dorsner and P.~Fileviez Perez,
 JHEP {\bf 0706}, 029 (2007)
 [arXiv:hep-ph/0612216].

\bibitem{Dynkin:1957um}
 E.~B.~Dynkin,
 Trans.\ Am.\ Math.\ Soc.\  {\bf 6}, 111 (1957).

\bibitem{Slansky:1981yr}
 R.~Slansky,
 Phys.\ Rept.\  {\bf 79}, 1 (1981).

\bibitem{White:1992aa}
 P.~L.~White,
 Nucl.\ Phys.\  B {\bf 403}, 141 (1993)
 [arXiv:hep-ph/9207231].

\bibitem{Bando:1984ej}
 M.~Bando, T.~Kugo, S.~Uehara, K.~Yamawaki and T.~Yanagida,
 Phys.\ Rev.\ Lett.\  {\bf 54}, 1215 (1985).

\bibitem{Bando:1987br}
 M.~Bando, T.~Kugo and K.~Yamawaki,
 Phys.\ Rept.\  {\bf 164}, 217 (1988).

\bibitem{Kaymakcalan:1984bz}
 O.~Kaymakcalan and J.~Schechter,
 Phys.\ Rev.\  D {\bf 31}, 1109 (1985).

\bibitem{Kaymakcalan:1983qq}
 O.~Kaymakcalan, S.~Rajeev and J.~Schechter,
 Phys.\ Rev.\  D {\bf 30}, 594 (1984).

\bibitem{Jain:1987sz}
 P.~Jain, R.~Johnson, U.~G.~Meissner, N.~W.~Park and J.~Schechter,
 Phys.\ Rev.\  D {\bf 37}, 3252 (1988).

\bibitem{Wess:1971yu}
 J.~Wess and B.~Zumino,
 Phys.\ Lett.\  B {\bf 37}, 95 (1971).

\bibitem{Witten:1983tw}
 E.~Witten,
 Nucl.\ Phys.\  B {\bf 223}, 422 (1983).

\bibitem{Witten:1983tx}
 E.~Witten,
 Nucl.\ Phys.\  B {\bf 223}, 433 (1983).

\bibitem{Lane:2009ct}
  K.~Lane and A.~Martin,
  arXiv:0907.3737 [hep-ph].

\bibitem{Nambu:1961tp}
 Y.~Nambu and G.~Jona-Lasinio,
 Phys.\ Rev.\  {\bf 122}, 345 (1961).

\bibitem{GellMann:1960np}
 M.~Gell-Mann and M.~Levy,
 Nuovo Cim.\  {\bf 16}, 705 (1960).

\bibitem{Sannino:1995ik}
 F.~Sannino and J.~Schechter,
 Phys.\ Rev.\  D {\bf 52}, 96 (1995)
 [arXiv:hep-ph/9501417].

\bibitem{Harada:1995dc}
 M.~Harada, F.~Sannino and J.~Schechter,
 Phys.\ Rev.\  D {\bf 54}, 1991 (1996)
 [arXiv:hep-ph/9511335].

\bibitem{Harada:2003em}
 M.~Harada, F.~Sannino and J.~Schechter,
 Phys.\ Rev.\  D {\bf 69}, 034005 (2004)
 [arXiv:hep-ph/0309206].

\bibitem{Pelaez:2003dy}
 J.~R.~Pelaez,
 Phys.\ Rev.\ Lett.\  {\bf 92}, 102001 (2004)
 [arXiv:hep-ph/0309292].

\bibitem{Oller:1997ti}
 J.~A.~Oller and E.~Oset,
 Nucl.\ Phys.\  A {\bf 620}, 438 (1997)
 [Erratum-ibid.\  A {\bf 652}, 407 (1999)]
 [arXiv:hep-ph/9702314].

\bibitem{Uehara:2003ax}
 M.~Uehara,
 arXiv:hep-ph/0308241.

\bibitem{Jaffe:1976ig}
 R.~L.~Jaffe,
 Phys.\ Rev.\  D {\bf 15}, 267 (1977).

\bibitem{Jaffe:1976ih}
 R.~L.~Jaffe,
 Phys.\ Rev.\  D {\bf 15}, 281 (1977).

\bibitem{Weinstein:1982gc}
 J.~D.~Weinstein and N.~Isgur,
 Phys.\ Rev.\ Lett.\  {\bf 48}, 659 (1982).

\bibitem{Harada:1996wr}
 M.~Harada, F.~Sannino and J.~Schechter,
 Phys.\ Rev.\ Lett.\  {\bf 78}, 1603 (1997)
 [arXiv:hep-ph/9609428].

\bibitem{Corrigan:1979xf}
 E.~Corrigan and P.~Ramond,
 Phys.\ Lett.\  B {\bf 87}, 73 (1979).

\bibitem{Kiritsis:1989ge}
 E.~B.~Kiritsis and J.~Papavassiliou,
 Phys.\ Rev.\  D {\bf 42}, 4238 (1990).

\bibitem{Armoni:2003gp}
 A.~Armoni, M.~Shifman and G.~Veneziano,
 Nucl.\ Phys.\  B {\bf 667}, 170 (2003)
 [arXiv:hep-th/0302163].

\bibitem{Sannino:2003xe}
 F.~Sannino and M.~Shifman,
 Phys.\ Rev.\  D {\bf 69}, 125004 (2004)
 [arXiv:hep-th/0309252].

\bibitem{Ryttov:2005na}
 T.~A.~Ryttov and F.~Sannino,
 Phys.\ Rev.\  D {\bf 73}, 016002 (2006)
 [arXiv:hep-th/0509130].

\bibitem{Georgi:1985hf}
 H.~Georgi,
 Nucl.\ Phys.\  B {\bf 266}, 274 (1986).

\bibitem{Frandsen:2005mb}
 M.~T.~Frandsen, C.~Kouvaris and F.~Sannino,
 Phys.\ Rev.\  D {\bf 74}, 117503 (2006)
 [arXiv:hep-ph/0512153].

\bibitem{Cherman:2009fh}
  A.~Cherman, T.~D.~Cohen and R.~F.~Lebed,
  Phys.\ Rev.\  D {\bf 80}, 036002 (2009)
  [arXiv:0906.2400 [hep-ph]].

\bibitem{Buchoff:2009za}
  M.~I.~Buchoff, A.~Cherman and T.~D.~Cohen,
  arXiv:0910.0470 [hep-ph].

\bibitem{Unsal:2006pj}
 M.~Unsal and L.~G.~Yaffe,
 Phys.\ Rev.\  D {\bf 74}, 105019 (2006)
 [arXiv:hep-th/0608180].

\bibitem{Kovtun:2005kh}
 P.~Kovtun, M.~Unsal and L.~G.~Yaffe,
 Phys.\ Rev.\  D {\bf 72}, 105006 (2005)
 [arXiv:hep-th/0505075].

\bibitem{Lee:1977eg}
 B.~W.~Lee, C.~Quigg and H.~B.~Thacker,
 Phys.\ Rev.\  D {\bf 16}, 1519 (1977).

\bibitem{Chivukula:1996kg}
 R.~S.~Chivukula,
 Phys.\ Rev.\  D {\bf 55}, 5238 (1997)
 [arXiv:hep-ph/9612267].

\bibitem{Sannino:1999qe}
 F.~Sannino and J.~Schechter,
 Phys.\ Rev.\  D {\bf 60}, 056004 (1999)
 [arXiv:hep-ph/9903359].

\bibitem{Goldberger:2007zk}
 W.~D.~Goldberger, B.~Grinstein and W.~Skiba,
 Phys.\ Rev.\ Lett.\  {\bf 100}, 111802 (2008)
 [arXiv:0708.1463 [hep-ph]].

\bibitem{Doff:2008xx}
 A.~Doff, A.~A.~Natale and P.~S.~Rodrigues da Silva,
 Phys.\ Rev.\  D {\bf 77}, 075012 (2008)
 [arXiv:0802.1898 [hep-ph]].

\bibitem{Fan:2008jk}
 J.~Fan, W.~D.~Goldberger, A.~Ross and W.~Skiba,
 Phys.\ Rev.\  D {\bf 79}, 035017 (2009)
 [arXiv:0803.2040 [hep-ph]].

\end{thebibliography}
\end{document}